\documentclass[prl,english,reprint,nofootinbib,aps,superscriptaddress,showpacs,showkeys]{revtex4-1}

\usepackage{babel,calc,amsmath,amsthm,amssymb,graphicx,subfigure,xcolor,mathtools,
      longtable,braket,bm,extarrows}

\usepackage[T1]{fontenc}
\usepackage[unicode=true]{hyperref}
\hypersetup{
     colorlinks=true,       		
     linkcolor=blue,          	
     citecolor=red,            
     urlcolor=magenta,           	
}
\newtheorem*{theorem*}{Theorem}
\newtheorem{theorem}{Theorem}

\newtheorem*{corollary*}{Corollary}

\newtheorem*{lemma*}{Lemma}

\newtheorem*{proposition*}{Proposition}

\newtheorem*{conjecture*}{Conjecture}
\theoremstyle{definition}

\newtheorem*{definition*}{Definition}
\theoremstyle{remark}
\newtheorem{remark}{Remark}
\newtheorem*{remark*}{Remark}

\newcommand{\Eq}[1]{Eq.~\eqref{#1}}

\newcommand{\p}[2]{\dfrac{\partial\,#1}{\partial #2}}

\begin{document}

\title{A Possible Mechanism to Alter Gyromagnetic Factor}


\author{Jing-Ling Chen}
\email{chenjl@nankai.edu.cn}
\affiliation{Theoretical Physics Division, Chern Institute of Mathematics, Nankai
      University, Tianjin 300071, People's Republic of China}

\author{Xing-Yan Fan}
\affiliation{Theoretical Physics Division, Chern Institute of Mathematics, Nankai
      University, Tianjin 300071, People's Republic of China}

\author{Xiang-Ru Xie}
\affiliation{School of Physics, Nankai University, Tianjin 300071, People's Republic
      of China}

\date{\today}

\begin{abstract}
 Dirac has predicted that the $g$ factor of an electron is strictly equal to 2 in the framework of relativistic quantum mechanics. However, later physicists have found that this factor can be slightly deviated from 2 (i.e., the problem of anomalous magnetic moments of leptons) when they consider quantum filed theory. This fact thus renders the $g$ factors of free leptons serving as precision tests for quantum electrodynamics, the standard model and beyond. In this work, we re-examine the problem of $g$ factor within the framework of relativistic quantum mechanics. We propose a possible mechanism called the ``electron-braidon mixing'', such that the $g$ factor of an electron can be visibly altered. Our results are hopeful to be verified in experiments and also shed new light to the problem of the anomalous magnetic moments of leptons.
\end{abstract}


\maketitle
\emph{Introduction}.---The gyromagnetic factor (i.e., the $g$ factor) is a meaningful measurable quantity involving angular momentum. It serves as not only a vital parameter characterising different types of particles, but also a promising tool probing ``new physics''. For example, Debierre \emph{et al}. utilized the $g$ factor of few-electron ions and their isotope shifts to investigate the ``fifth fundamental force'' in nature \cite{2020PLBDebierre}.

      In quantum mechanics, the $g$ factors appear in Dirac's equation of a lepton moving in a weak and homogeneous magnetic field $\vec{B}$ under the nonrelativistic limiting~\cite{GreinRQM,GreineQMIntroM}. Precisely, the term $\bigl[e/(2\,m\,c)\bigr]\big(\vec{\ell}+2\,\vec{S}\big)\cdot\vec{B}$ in Pauli's Hamiltonian implies the interactive energy between angular momenta and magnetic field $\vec{B}$, where $e$ represents the unit electric charge, $m$ is the mass of lepton, $c$ expresses the speed of light in vacuum, $\vec{\ell}=\vec{r}\times \vec{p}$ is the orbit angular momentum, $\vec{S}=(\hbar/2)\vec{\sigma}$ is the spin-$1/2$ angular momentum, $\hbar=h/(2\pi)$, $h$ is Planck's constant, and $\vec{\sigma}=(\sigma_x, \sigma_y, \sigma_z)$ is the vector of Pauli matrices. Define the $g$ factors of lepton with respect to the orbital angular momentum as $g_{\rm\ell}$ and the spin as $g_{\rm s}$, then one obtains $g_{\rm\ell}=1$ and $g_{\rm s}=2\,g_{\rm\ell}$, which is just Dirac's prediction in 1928~\cite{1928Dirac}. Actually one may rewrite the interactive energy stemming from spin above in terms of the spin magnetic moment $\vec{\mu}_{\rm m} =-g_{\rm s} \mu_0 \vec{\sigma}/2$, with $\mu_0 \equiv e\,\hbar/(2\,m\,c)$ marking the Bohr magneton. Thus experimentally, one can determine $g_{\rm s}$ via measuring the spin magnetic moment $\vec{\mu}_{\rm m}$. Following this idea, Kinsler and Houston performed the first test of $g_{\rm s} \approx 2$, while their data were comparatively coarse in those days \cite{1934PPREAMM}. However afterwards, more and more experiments (including the muon $g-2$ experiments) confirmed physicists that $g_{\rm s} >2$ rigorously~\cite{1956SciKusch,1966PTKusch}.

      The first theoretical breakthrough of predicting the anomalous $g$ factor of lepton came from quantum electrodynamics (QED). In 1948, Schwinger considered the radiative correction to calculate the ``anomalous'' $a_e :=(g_{\rm s} -2)/2 =\alpha/(2\,\pi)\approx 0.001162$, which means that $g_{\rm s} \approx 2.002324$ \cite{1948PRSchwinger}; here $\alpha\approx 1/137$ indicates the fine structure constant. Noting the best experiment at that time showed $g_{\rm s} \approx 2.00238(10)$ \cite{1948PRKusch}, namely the relative error was
      $28.4669$ ppm exclusively.

      With the development of quantum field theory (QFT), especially the Standard Model (SM) \cite{1961NPGlashow,1967PRLWeinberg,Salam}, physicists are able to compute the anomalous $g$ factor to a new level for one thing, through $g_{\rm s} =2+a_l =2+O(\alpha)$ ($l=e,\mu,\tau$ presenting one of the three charged leptons); for another, pay more attention to the anomalous $g$ factor of the heavier charged lepton, such as muon \cite{2020PRAoyama,2022NRPLehner}. In the light of SM, the sensitivity of $a_l$, described via $\delta\,a_l /a_l$, is proportional to $m_l /M$, where $m_l$ refers to the mass of the lepton $l$; while $M\gg m_l$ can be chosen, e.g., the mass of another heavier particle falling into the category of SM, or the mass of a postulated heavy state beyond SM \cite{2017AMMu}. Hence the heavier a lepton is, the more sensitive it is for exploring ``new physics'' transcending SM. In 1960, Garwin \emph{et al}. realized the first measurement for the anomalous $g$ factor of muon \cite{1960PRGarwin}, with the accuracy $0.007\%=70$ ppm. Up to now, the newest experimental error fell to $0.2$ ppm \cite{2023PRLG2}, which paves an enticing way for physicists to find ``new physics''.

      Nevertheless, SM cannot explain the anomalous $g$ factors of leptons perfectly, since the genesis and direct outcome of quantum fluctuation remain outstanding, which impels physicists to seek other rationalization about anomalous $g$ factor, e.g., (i). uncontrollable external fields. Steinmetz \emph{et al}. studied the Dirac and Klein-Gordon equations with an additional Pauli term regarding electromagnetic dipole moment, which led to an alternation of $g$ factor \cite{2019EPJASteinmetz}. (ii). Finite radius treatment for lepton, rather than ``point-like'' particle \cite{1993HIDehmelt,2018Maruani}. Whilst the two ways mentioned above set a lepton as a single particle in a typical state.

      Since all charged leptons share the same traits except for mass merely, we take the electron as a typical example thereafter, and our computation is also applicable to the other two charged leptons. It is well-known that Dirac predicted $g_{\rm s}=2$ in the framework of relativistic quantum mechanics (RQM), while later physicists found that it was slightly deviated from 2 when one considers QFT. This gives rise to a scientific question: Can one alter the $g$ factor just in the framework of RQM? The purpose of this work is to propose a possible mechanism to alter the $g$ factor in the framework of RQM. For convenience, we would like to call the novel mechanism as the ``electron-braidon mixing'', which is inspired by the successes of quark mixing \cite{QuarkMix} and neutrino mixing \cite{NeutriMix} in particle physics. By means of this extraordinary mechanism, we may visibly alter the $g$ factor of an electron, and thus shedding a new light to the problem of the anomalous magnetic moments of leptons.

      The work is organized as follows. First, we briefly review Dirac's prediction of $g_{\rm s}=2$. Second, we introduce the new concept of ``braidon''. Third, we propose a possible mechanism of ``electron-braidon mixing'' to alter the $g$ factor of an electron.
      Conclusion and discussion are made in the end.

      \emph{Dirac's Prediction}.---In modern physics, special relativity and quantum mechanics are two cornerstones. As for special relativity, a fundamental relation is undoubtedly the relativistic energy-momentum (REM) relation, which is given by
      $E^{2}=p^{2}c^2+m^{2}c^4$,
and $E$ is the energy, $p=|\vec{p}|$ is the magnitude of the linear momentum $\vec{p}=(p_x, p_y, p_z)$,  and $m$ is the rest mass of the particle.
When one turns to quantum mechanics, the REM relation is recast to an operator-form as~\cite{2021ModernQM}
\begin{equation}\label{eq:EP2}
H^2=\vec{p}^{\,2} c^2+m^2c^4,
\end{equation}
where the energy $E$ is regarded as the eigenvalue of the Hamiltonian operator $H$.

However, Eq. (\ref{eq:EP2}) is not applicable directly to Schr{\" o}dinger's equation.
The reason is that Schr{\" o}dinger's equation requires a linear operator $H$, instead of a square form $H^2$ as shown in Eq. (\ref{eq:EP2}).
In 1928, to reconcile the REM relation with Schr{\" o}dinger's equation, Dirac successfully formulated a relativistic Hamiltonian operator for a free electron~\cite{1928Dirac}; namely by linearizing the RME relation, Dirac creatively presented that
 \begin{eqnarray}\label{eq:DiracH}
         H_{\rm e}&=& c\vec{\alpha}\cdot\vec{p}+\beta\,m c^2,
 \end{eqnarray}
where $\vec{\alpha}=\sigma_x\otimes \vec{\sigma}$, $\beta=\sigma_z\otimes\openone $ are Dirac's matrices, with $\openone$ being the $2\times 2$ identity operator.
For convenience, we call $ H_{\rm e}$ as the Dirac Hamiltonian operator of a \emph{Dirac's electron}, and $m\equiv m_{\rm e}$ represents the rest mass of electron. After squaring the Hamiltonian operator $H_{\rm e}$ in Eq. (\ref{eq:DiracH}), one easily recovers the REM relation (\ref{eq:EP2}).

The relativistic Hamiltonian of a Dirac's electron in an electromagnetic field is given by
      \begin{equation}\label{eq:Hev-1}
            \mathcal{H}_{\rm e} =c\,\vec{\alpha}\cdot\vec{\Pi}
                  +\beta\,m_{\rm e}\,c^2 +q\,\phi,
      \end{equation}
where the momentum operator $\vec{p}$ has been replaced by the canonical momentum operator
\begin{equation}\label{eq:Hev-1}
\vec{\Pi}= \vec{p}-\dfrac{q}{c} \vec{A},
      \end{equation}
with $\vec{A}$ and $\phi$ being the vector and scalar potentials respectively, and $q=-e$ represents the charge of electron.

Dirac's equation reads
      \begin{equation}\label{eq:DiracEqE}
            {\rm i}\hbar \frac{\partial \,\Psi(t)}{\partial t} =H \Psi(t),
      \end{equation}
where $H=\mathcal{H}_{\rm e}$ is the Hamiltonian of Dirac's electron in presence of an electromagnetic field.  By setting
      \begin{eqnarray}
            \Psi (t)=\begin{bmatrix}
                  \tilde{\eta} \\ \tilde{\chi}
            \end{bmatrix}=\begin{bmatrix}
                  \eta \\ \chi
            \end{bmatrix}{\rm e}^{-{\rm i}\frac{m_{\rm e}\,c^2}{\hbar} t},
      \end{eqnarray}
      and considering the non-relativistic limit, one arrives at
      the Pauli equation in quantum mechanics as~\cite{GreinRQM,GreineQMIntroM}
      \begin{equation}\label{eq:pauli-1}
            {\rm i}\hbar\p{\eta}{t} = \bigg[
                        \dfrac{1}{2\,m_{\rm e}} \Bigl(\vec{p}-\dfrac{q}{c} \vec{A}\Bigr)^2
                        -\dfrac{q\,\hbar}{2\,m_{\rm e}\,c} \vec{\sigma}\cdot\vec{B}+q\,\phi\bigg]\eta,
      \end{equation}
      or
      %
%
%
      \begin{equation}\label{eq:pauli-3}
            {\rm i}\hbar\p{\eta}{t} = \bigg[
                        \dfrac{\vec{p}^{\, 2}}{2\,m_{\rm e}}
                        -\dfrac{q}{2\,m_{\rm e}\,c} \left(\vec{\ell}+2 \vec{S}\right)\cdot\vec{B}+q\,\phi\bigg]\eta.
      \end{equation}
     This form explicitly shows Dirac's prediction of $g_{s}=2$.

\emph{Dirac's Braidon}.---Braiding is one of the most common phenomena in daily life. For instance, the bamboo baskets are woven with the bamboo strips and the sweaters are knitted with the wool. Mathematically, the basic braiding relation is given by \cite{1999braid}
      \begin{equation}\label{eq:bra-1b}
            \mathbb{A}\mathbb{B}\mathbb{A}=\mathbb{B}\mathbb{A}\mathbb{B},
      \end{equation}
      where $\mathbb{A}$ and $\mathbb{B}$ are braiding operations. The  braiding relation has been applied expansively in physics. For example, it has been generalized to the Yang-Baxter equation~\cite{1989Jimbo}, serving as an indispensable key to solve the quantum integrable models~\cite{1992Mattis}, such as the one dimensional many-body problem with repulsive delta-function interaction~\cite{1967Yang}, the Heisenberg spin chains~\cite{1999ChenGeXue}, and so on~\cite{1994YangGe,2008PRABHandDH,2019ToPhy}. Furthermore, due to braiding operations, the non-abelian anyon with fault-tolerant features has become a competitive candidate for building topological quantum computer~\cite{2008Nayak}.

      The braid relation (\ref{eq:bra-1b}) was presented by Artin in 1925, even before the establishment of quantum mechanics. It gives rise to an interesting question: Does the fundamental braiding relation \eqref{eq:bra-1b} have a connection with the well-known angular momentum operator in quantum mechanics? The answer is positive, which is related to the following theorem:
      \begin{theorem}\label{thm:Braid}
            Let $\Gamma_x$, $\Gamma_y$ be two generalized Pauli matrices, i.e., $\Gamma_x^2=\Gamma_y^2=\mathbb{I}$, and $\left\{\Gamma_x, \Gamma_y\right\}=\Gamma_x \Gamma_y+\Gamma_y \Gamma_x=0$, with $\mathbb{I}$ being the identity operator, and then the two operators
            \begin{eqnarray}\label{eq:bra-1d}
                 \mathbb{A}={\rm e}^{\frac{{\rm i}\pi}{4}\Gamma_x},\qquad
                  \mathbb{B}= {\rm e}^{\frac{{\rm i}\pi}{4}\Gamma_y}
            \end{eqnarray}
            fulfil the braid relation (\ref{eq:bra-1b}).
      \end{theorem}
      The proof of Theorem \ref{thm:Braid} can be found in the Supplementary Information (SI) \cite{SI}. Based on the first two generalized Pauli matrices $\Gamma_x$ and $\Gamma_y$, one can generate the third generalized Pauli matrix as $\Gamma_z=-{\rm i}\,\Gamma_x \Gamma_y$. Accordingly, we can define a generalized spin-$1/2$ angular momentum operator as
      \begin{eqnarray}
            \vec{\tau} =\frac{\hbar}{2} \,\vec{\Gamma}=\frac{\hbar}{2} \left(
                  \Gamma_x, \Gamma_y, \Gamma_z\right),
      \end{eqnarray}
      which satisfies $\vec{\tau}\times\vec{\tau}={\rm i}\hbar \vec{\tau}$. Furthermore, when one sets $\vec{\Gamma}=\vec{\sigma}$, then $\vec{\tau}$ becomes the usual spin-$1/2$ angular momentum operator $\vec{S}=(\hbar/2)\vec{\sigma}$.

      Based on Dirac's matrices $\vec{\alpha}$ and $\beta$, one may realize the three components of $\vec{\Gamma}$ as
      \begin{subequations}
       \begin{eqnarray}
            && {\Gamma}_x =\dfrac{\kappa_1 (\vec\alpha\cdot\hat{n})
                  +\kappa_2\,\beta}{\sqrt{\kappa_1^2 +\kappa_2^2}},\label{eq:s-3e}\\
           && {\Gamma}_y =\dfrac{-\kappa_2 (\vec\alpha\cdot\hat{n})+\kappa_1\beta}{
                  \sqrt{\kappa_1^2 +\kappa_2^2}},  \label{eq:s-3e1}\\
            && {\Gamma}_z ={\rm i}\,\beta(\vec\alpha\cdot\hat{n}).\label{eq:s-3e2}
      \end{eqnarray}
      \end{subequations}
      Here $\hat{n}=(n_x, n_y, n_z)$ is a unit vector,
      $\kappa_1$ and $\kappa_2$ are two parameters.
      Following Theorem \ref{thm:Braid}, if one defines $\mathbb{C}={\rm e}^{\frac{{\rm i}\pi}{4}\Gamma_z}$, then he has
      \begin{eqnarray}\label{eq:bra-2a}
            &&\mathbb{A}\mathbb{B}\mathbb{A}=\mathbb{B}\mathbb{A}\mathbb{B},\;\;\;
            \mathbb{B}\mathbb{C}\mathbb{B}=\mathbb{C}\mathbb{B}\mathbb{C},\;\;\;
            \mathbb{C}\mathbb{A}\mathbb{C}=\mathbb{A}\mathbb{C}\mathbb{A}.
      \end{eqnarray}
      In other words, any two elements in the set $\{\mathbb{A}, \mathbb{B}, \mathbb{C}\}$ satisfy the basic braiding relation.

      Now, an interesting thing comes.
      Let us compare the Dirac Hamiltonian operator (\ref{eq:DiracH}) with ${\Gamma}_x$ in \Eq{eq:s-3e}. Via direct observation, if let
      $\hat{n}=\hat{p}=\vec{p}/|\vec{p}|$, $\kappa_1 =p\,c$, and $\kappa_2 =m\,c^2$, then we arrive at
      \begin{eqnarray}\label{eq:DiracH-3}
            {\Gamma}_x =\dfrac{H_{\rm e}}{\sqrt{\vec{p}^{\,2} c^2+m^2c^4}},
      \end{eqnarray}
      which is nothing but \emph{the normalized Dirac Hamiltonian operator} of Dirac's electron.
      However, this immediately leads to a significant question: If ${\Gamma}_x $ corresponds to the Dirac Hamiltonian of a Dirac's electron, then what will ${\Gamma}_y$ and ${\Gamma}_z$ correspond to?

\begin{figure}[b]
            \centering
            \includegraphics[width=80mm]{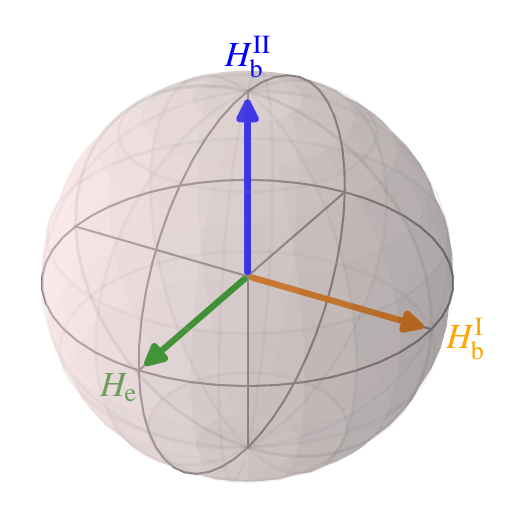}
            \caption{Illustration of Dirac's electron and Dirac's braidons in the space of angular momentum. From the viewpoint of the generalized spin-$1/2$ angular momentum operator $\vec{\tau}=(\tau_x, \tau_y, \tau_z)$, the Hamiltonian $H_{\rm e}$ of a Dirac's electron corresponds to the $x$-component of spin operator $\vec{\tau}$, while the Hamiltonians $H_{\rm b}^{\rm I}$ and $H_{\rm b}^{\rm II}$ of Dirac's braidons correspond to the other two components of $\vec{\tau}$.}\label{fig:spin}
      \end{figure}

      Similarly, after substituting $\hat{n}$, $\kappa_1$, and $\kappa_2$ into Eq. (\ref{eq:s-3e1}) and Eq. (\ref{eq:s-3e2}), we attain
      \begin{subequations}
            \begin{eqnarray}
                  && {\Gamma}_y =\dfrac{H_{\rm b}^{\rm I}}{
                        \sqrt{\vec{p}^{\,2} c^2+m^2c^4}}, \\
                  && {\Gamma}_z =\dfrac{H_{\rm b}^{\rm II}}{
                        \sqrt{\vec{p}^{\,2} c^2+m^2c^4}},
            \end{eqnarray}
      \end{subequations}
      with
      \begin{subequations}
            \begin{eqnarray}
                  && H_{\rm b}^{\rm I} =-m\,c^2 (\vec\alpha\cdot\hat{p})+\beta\,p\,c,
                        \\
                  && H_{\rm b}^{\rm II} =\sqrt{\vec{p}^2 c^2+m^2c^4}\,{\rm i}\,\beta(
                        \vec\alpha\cdot\hat{p}).
            \end{eqnarray}
      \end{subequations}      
      Perceive that the exponential maps of ${\Gamma}_x$, ${\Gamma}_y$ and ${\Gamma}_z$ obey the braiding rules in (\ref{eq:bra-2a}), and hence we would like to call $H_{\rm b}^{\rm I}$ (and $H_{\rm b}^{\rm II}$) as the Dirac Hamiltonian operators of \emph{type-I (and type-II) Dirac's braidons}.
      In Fig. \ref{fig:spin}, we have illustrated Dirac's electron and Dirac's braidons in the space of spin-$1/2$ angular momentum.
      One finds that the Hamiltonian operator $H_{\rm e}$ of Dirac's electron plays a role as the $x$-component of the generalized spin-$1/2$ angular momentum operator $\vec{\tau}$.

      Certainly, based on the theory of angular momentum, one can transform $\tau_j$ to $\tau_k$ by a rotation around the $l$-axis with an angle $\pi/2$, where distinct $j,k,l\in\{x,y,z\}$. Mathematically, it is expressed by $\tau_k ={\rm e}^{
            -\frac{{\rm i}}{\hbar} \frac{\pi}{2} \tau_l} \tau_j {\rm e}^{
                  \frac{{\rm i}}{\hbar} \frac{\pi}{2} \tau_l}$,
      or $\Gamma_k ={\rm e}^{-{\rm i} \frac{\pi}{4} \Gamma_l} \Gamma_j {\rm e}^{
            {\rm i} \frac{\pi}{4} \Gamma_l}$,
      where two of its components are equivalent to
      \begin{subequations}
            \begin{eqnarray}
                  && H_{\rm b}^{\rm I} =\mathcal{C} H_{\rm e} \mathcal{C}^\dagger, \;\;\;\;\;\; \mathcal{C}={\rm e}^{-{\rm i} \frac{\pi}{4} \Gamma_z}, \label{eq:hh} \\
                  && H_{\rm b}^{\rm II} =\mathcal{D} H_{\rm e} \mathcal{D}^\dagger, \;\;\; \mathcal{D}={\rm e}^{{\rm i} \frac{\pi}{4} \Gamma_y}.
                        \label{eq:hhh}
            \end{eqnarray}
      \end{subequations}
      The physical meaning of Eq. (\ref{eq:hh}) is that the Hamiltonian operator $H_{\rm b}^{\rm I}$ can be obtained from the Hamiltonian operator $H_{\rm e}$ via the unitary transformation $\mathcal{C}$, similar for Eq. (\ref{eq:hhh}). Thus, in our work, Dirac's braidon is not a new particle. Namely, Dirac's braidon is still an electron, but a unitary-transformed ``electron''.
      
      \vspace{5mm}
      
      \onecolumngrid
      \begin{widetext}
      \begin{table}[h]
	\centering
\caption{Dirac' electron versus Dirac's braidons. Here $|\Psi_j \rangle$, $|\Psi'_j \rangle$ and $|\Psi''_j \rangle$ $(j=1, 2, 3, 4)$ are eigenstates of Dirac's electron and Dirac's braidons, respectively. $\hat{\mathcal{Z}}^r$ and $\hat{\mathcal{Z}}^s$ are position Zitterbewegung operator and spin Zitterbewegung operator, respectively. $E_{\rm p}=\sqrt{p^{2}c^2+m^{2}c^4}$, $\Delta_1=\frac{-{\rm i}\hbar c}{2 E_{\rm p}} \left({\rm e}^{{\rm i}\,2\,E_{\rm p} t/\hbar} {-1} \right)$, $\Delta_2=\frac{-mc}{p}\Delta_1$, $\Delta_3=\left(1+{\rm i}\,\frac{m\,c}{p}\right)\Delta_1$,
      $\vec{F}_1 =\frac{1}{p(p+p_z)} \big[(p^2 -p_x^2)+p\,p_z\big]\hat{e}_x
            -\frac{p_x p_y}{p(p+p_z)} \hat{e}_y -\frac{p_x}{p} \hat{e}_z$, and
      $\vec{F}_2 =\vec{F}_1 \times\hat{p}$.}
\begin{tabular}{lccc}
\hline\hline
 & Dirac's Electron  \;\;\;\;\;\;\;& Type-I Dirac's Braidon  \;\;\;\;\;\;\;& Type-II Dirac's Braidon  \\
  \hline
Hamiltonian & $H_{\rm e}=c\vec{\alpha}\cdot\vec{p}+\beta\,m c^2$ & $H_{\rm b}^{\rm I} =-m\,c^2 (\vec\alpha\cdot\hat{p})+\beta\,p\,c$  & $H_{\rm b}^{\rm II} =\sqrt{\vec{p}^2 c^2+m^2c^4}\,{\rm i}\,\beta(
                        \vec\alpha\cdot\hat{p})$\\
 \hline
 Mass & $m$ & $m$ & $m$\\
 \hline
Magnitude of $\vec{p}$ &	$p$ & $p$ & $p$\\
  \hline
REM relation & \;\;$H_{\rm e}^2=\vec{p}^{\,2}c^2+m^2c^4$\;\; &  $(H_{\rm b}^{\rm I})^2=\vec{p}^{\,2}c^2+m^2c^4$ & $(H_{\rm b}^{\rm II})^2=\vec{p}^{\,2}c^2+m^2c^4$\\
  \hline
 $\vec{J}$ \; & $[H_{\rm e}, \vec{J}]=0$ & $[H_{\rm b}^{\rm I}, \vec{J}]=0$ & $[H_{\rm b}^{\rm II}, \vec{J}]=0$\\
  \hline
Wavefunction & $|\Psi\rangle_{\rm e}\equiv |\Psi\rangle$ &  $|\Psi\rangle_{\rm b}^{\rm I}
      =\mathcal{C}|\Psi\rangle_{\rm e}\equiv |\Psi'\rangle$ & $|\Psi\rangle_{\rm b}^{\rm II}=\mathcal{D}|\Psi\rangle_{\rm e}\equiv |\Psi''\rangle$\\
  \hline
PZ & $\langle\Psi_j|\hat{\mathcal{Z}}_{\rm e}^r|\Psi_k \rangle
            =0$, $(j=k)$ & $\langle\Psi'_j|\hat{\mathcal{Z}}_{\rm b}^{{\rm I},r}|\Psi'_k \rangle
            =0$, $(j=k)$ & $\langle\Psi''_j|\hat{\mathcal{Z}}_{\rm b}^{{\rm II},r}|\Psi''_k \rangle
            =0$, $(j=k)$ \\
  \hline
& $\langle\Psi_1|\hat{\mathcal{Z}}_{\rm e}^r|\Psi_3 \rangle
            =\Delta_1\,\dfrac{-m\,c^2}{E_{\rm p}}\, \hat{p}$  & $\langle\Psi'_1|\hat{\mathcal{Z}}_{\rm b}^{{\rm I},r}|\Psi'_3 \rangle
            =\Delta_1\,\dfrac{-m\,c^2}{E_{\rm p}}\, \hat{p}$ & $\langle\Psi''_1|\hat{\mathcal{Z}}_{\rm b}^{{\rm II},r}|\Psi''_3 \rangle
            =0$ \\
  \hline
& $\langle\Psi_1|\hat{\mathcal{Z}}_{\rm e}^r|\Psi_4 \rangle
            =\Delta_1\,(\vec{F}_1 +{\rm i}\,\vec{F}_2)$  & $\langle\Psi'_1|\hat{\mathcal{Z}}_{\rm b}^{{\rm I},r}|\Psi'_4 \rangle
            =\Delta_2\,(\vec{F}_1 +{\rm i}\,\vec{F}_2)$ & $\langle\Psi''_1|\hat{\mathcal{Z}}_{\rm b}^{{\rm II},r}|\Psi''_4 \rangle
            =\Delta_3\,(\vec{F}_1 +{\rm i}\,\vec{F}_2)$  \\
  \hline
SZ & $\langle\Psi_j|\hat{\mathcal{Z}}_{\rm e}^s|\Psi_k \rangle
            =0$, $(j=k)$ & $\langle\Psi'_j|\hat{\mathcal{Z}}_{\rm b}^{{\rm I},s}|\Psi'_k \rangle
            =0$, $(j=k)$ & $\langle\Psi''_j|\hat{\mathcal{Z}}_{\rm b}^{{\rm II},s}|\Psi''_k \rangle
            =0$, $(j=k)$ \\
  \hline
& $\langle\Psi_1|\hat{\mathcal{Z}}_{\rm e}^s|\Psi_3 \rangle
            =0$  & $\langle\Psi'_1|\hat{\mathcal{Z}}_{\rm b}^{{\rm I},s}|\Psi'_3 \rangle
            =0$ & $\langle\Psi''_1|\hat{\mathcal{Z}}_{\rm b}^{{\rm II},s}|\Psi''_3 \rangle
            =0$ \\
  \hline
& $\langle\Psi_1|\hat{\mathcal{Z}}_{\rm e}^s|\Psi_4 \rangle
            =-\Delta_1(\vec{F}_1 +{\rm i}\,\vec{F}_2)\times \vec{p}$  \, & $\langle\Psi'_1|\hat{\mathcal{Z}}_{\rm b}^{{\rm I},s}|\Psi'_4 \rangle
            =-\Delta_2(\vec{F}_1 +{\rm i}\,\vec{F}_2)\times \vec{p}$ \,& $\langle\Psi''_1|\hat{\mathcal{Z}}_{\rm b}^{{\rm II},s}|\Psi''_4 \rangle
            =-\Delta_3(\vec{F}_1 +{\rm i}\,\vec{F}_2)\times \vec{p}$  \\
 \hline\hline
\end{tabular}\label{tab:compare}
\end{table}
\end{widetext}
\twocolumngrid

      In Table \ref{tab:compare}, we have compared some physical properties of Dirac's electron and Dirac's braidons.
      They have the same mass, the same magnitude of linear momentum, and the same REM relation. If one considers the total angular momentum $\vec{J}=\vec{\ell}+\vec{S}$, where $\vec{S}=(\hbar/2)\vec{\Sigma}=(\hbar/2)(\openone\otimes\vec{\sigma})$ is the spin-$1/2$ angular momentum operator in the $4\times 4$ matrix-representation, then $[H_{\rm e},\ \vec{J}]=[H_{\rm b}^{\rm I},\ \vec{J}]=[H_{\rm b}^{\rm II},\ \vec{J}]=0$, which means that $\vec{J}$ is a conservation quantity for both Dirac's electron and Dirac's braidons. The above analysis shows that one cannot distinguish Dirac's electron from Dirac's braidons simply based on the mass, the linear momentum, the REM relation, and the conservation of total angular momentum. Suppose $|\Psi\rangle_{\rm e}$ is the wavefunction of a free Dirac's electron, and then based on Eq. (\ref{eq:hh}) and Eq. (\ref{eq:hhh}), one knows that $|\Psi\rangle_{\rm b}^{\rm I}
      =\mathcal{C}|\Psi\rangle_{\rm e}$ $\big($and $|\Psi\rangle_{\rm b}^{\rm II}=\mathcal{D}|\Psi\rangle_{\rm e}\big)$ are the wavefunctions of the type-I (and type-II) Dirac braidons, respectively. However, the behaviors of Dirac's electron and Dirac's braidons are not always identical. For example, they behave differently in the relativistic effects of position Zitterbewegung (PZ) \cite{1931Zitter} and spin Zitterbewegung (SZ), which we have explicitly shown in SI~\cite{SI}.


\emph{A Possible Mechanism to Alter the $g$ Factor}.---For a spin operator $\vec{S}=(S_x, S_y, S_z)$, its three components play the similar roles. Thus in the textbooks of quantum mechanics, it is easy to encounter the following spin operator $\vec{S}\cdot\hat{n}=S_x n_x+ S_y n_y+ S_z n_z$, i.e., $\vec{S}\cdot\hat{n}$ represents a mixture of $S_x$, $S_y$, and $S_z$. Since $\vec{\tau}=(\tau_x, \tau_y, \tau_z)$ is also a kind of spin operator, thus one naturally has the following mixture of ${H}_{\rm e}$, ${H}_{\rm b}^{\rm I}$ and ${H}_{\rm b}^{\rm II}$: $ H_{\rm mix}=\cos\vartheta\,{H}_{\rm e} +\sin\vartheta\,(
                  \cos\varphi\,{H}_{\rm b}^{\rm I}
                  +\sin\varphi\,{H}_{\rm b}^{\rm II})$,
which denotes the mixed Hamiltonian. We put the mixed physical system into the electromagnetic filed with potential $(\vec{A}, \phi)$, and then $H_{\rm mix}$ turns to
   \begin{eqnarray}\label{eq:mix-2}
            \mathcal{H}_{\rm mix}&=&\cos\vartheta\,\mathcal{H}_{\rm e} +\sin\vartheta(
                  \cos\varphi\,\mathcal{H}_{\rm b}^{\rm I}
                  +\sin\varphi\,\mathcal{H}_{\rm b}^{\rm II})\nonumber\\
                  &&+q\phi,
      \end{eqnarray}
where
      \begin{subequations}
            \begin{eqnarray}
            && \mathcal{H}_{\rm e} = c(\vec{\alpha}\cdot\vec{\Pi})
                        +\beta\,m_{\rm e}\,c^2, \\
            && \mathcal{H}_{\rm b}^{\rm I} =-\dfrac{m_{\rm e}\,c^2}{2} \Big\{
                        (\vec{\alpha}\cdot\vec{\Pi}),\ \dfrac{1}{\sqrt{\vec{\Pi}^2}}
                        \Big\}+c\,\beta\sqrt{\vec{\Pi}^2}, \\
            && \mathcal{H}_{\rm b}^{\rm II} = {\rm i}\dfrac{c}{2} \Bigg\{\beta(
                              \vec{\alpha}\cdot\vec{\Pi}),\
                        \sqrt{1+\dfrac{m_{\rm e}^2 c^2}{\vec{\Pi}^2}}\Bigg\},
            \end{eqnarray}
      \end{subequations}
      and $\{X, Y\}=XY+YX$ denotes the anti-commutator. One may verify that $\mathcal{H}_{\rm mix}$ is hermitian, and it reduces to ${H}_{\rm mix}$ if the electromagnetic filed vanishes.

      Based on the mechanism of \emph{electron-braidon mixing} as shown in Eq. (\ref{eq:mix-2}), we reexamine Dirac's equation (\ref{eq:DiracEqE}), where $H$ takes $\mathcal{H}_{\rm mix}$.
      Following a similar computational procedure~\cite{SI}, we finally obtain the $g$ factor as
      \begin{eqnarray}\label{eq:g-1a}
                 g_{\rm s}=2\left(1- \dfrac{\cos\vartheta \sin\vartheta\,\cos\varphi}{\cos^2\vartheta+{\sin^2}\vartheta\,
                              {\sin^2}\varphi }\dfrac{m_{\rm e}c}{p}\right),
            \end{eqnarray}
      with $ \vartheta \in [-\pi/4, \pi/4]$. This means that the mixture system of $H_{\rm e}$-${H}_{\rm b}^{\rm I}$-${H}_{\rm b}^{\rm II}$ can indeed alter the $g$ factor by adjusting the mixing angles of $\vartheta$ and $\varphi$.

     \begin{remark}One may perform an experiment to observe the changeable $g$ factor of an electron.
     For simplicity, let us consider the the mixture of $H_{\rm e}$-${H}_{\rm b}^{\rm I}$. In this case $\varphi=0$, and Eq. (\ref{eq:g-1a}) becomes
     \begin{eqnarray}\label{eq:g-1b}
                 g_{\rm s}=2\left(1- \tan\vartheta\,\dfrac{m_{\rm e}c}{p}\right).
            \end{eqnarray}
    There is a clear physical picture of obtaining $\mathcal{H}_{\rm mix}$ from $H_{\rm e}$. Let us start from the Hamiltonian $H_{\rm e}$. Quantum mechanically we perform a unitary transformation $ \mathcal{C}(\vartheta) = {\rm e}^{-{\rm i} \frac{\vartheta}{2} \Gamma_z}$
     on Dirac's electron, then it becomes a mixture of $H_{\rm e}$ and ${H}_{\rm b}^I$, i.e., $ H_{\rm e} \mapsto   H_{\rm mix} = \mathcal{C}(\vartheta) H_{\rm e} \mathcal{C}(\vartheta)^\dagger= \cos\vartheta H_{\rm e} +\sin\vartheta H_{\rm b}^I$.
       After putting the mixed physical system into an external electromagnetic field, we have the mixed Hamiltonian as
       $\mathcal{H}_{\rm mix}= \cos\vartheta\,\mathcal{H}_{\rm e} +\sin\vartheta\,\mathcal{H}_{\rm b}^{\rm I}+q\, \phi$.
      Note that $\mathcal{C}(\vartheta)$ does not depend on mass $m_{\rm e}$, and therefore it is feasible to be realized in experiment.
      \end{remark}

\emph{Conclusion and Discussion}.---In this work, we have proposed a possible mechanism of ``electron-braidon mixing'' to alter the $g$ factor of an electron in the framework of RQM. The approach is also applicable to other charged spin-1/2 particles, like muon, tauon, proton and so on. Alternatively, one may interpret the anomalous magnetic moments of electron through the \emph{electron-braidon mixing}. For instance, for the electron $g_{\rm e}\approx 2.0023$, which corresponds to mixing angle $\vartheta_{\rm e}\approx -\arctan[(g_{\rm e}-2)p/(2m_{\rm e}c)]$; for the muon $g_{\mu}\approx 2.0023$, which corresponds to mixing angle $\vartheta_{\mu}\approx -\arctan[(g_{\mu}-2)p/(2m_{\mu}c)]$; and for the proton $g_{\rm p}\approx 5.5856$ \cite{2003PLBGp,2000RMPCDat}, which corresponds to mixing angle $\vartheta_{\rm p}\approx -\arctan[(g_{\rm p}-2)p/(2 m_{\rm p}c)]$.

Based on Eq. (\ref{eq:g-1b}), one can evidently alter the $g$ factor in a quantum-mechanical experiment. In an extreme situation, one may set $g_{\rm s}=0$ in Eq. (\ref{eq:g-1b}),
and then obtains a magic mixing angle as
\begin{eqnarray}
                        && \vartheta_{\rm magic}=\arctan\left(\dfrac{p}{m_{\rm e}\,c}\right).
\end{eqnarray}
This is a theoretical prediction of our work, and the zero $g$ factor can be possibly observed in the lab.
Coincidentally, in the mixed system of $H_{\rm e}$-${H}_{\rm b}^{\rm I}$, the relativistic effect of ``spin Zitterbewegung'' vanishes when one chooses the magic mixing angle. Eventually, although the possible mechanism is proposed in the framework of RQM, it is also applicable to QFT and SM. We anticipate ``new physics'' will emerge when one applies the extraordinary idea of ``electron-braidon mixing'' to QFT and SM.

\begin{acknowledgments}
   J.L.C. is supported by the National Natural Science Foundation of China (Grants No. 12275136 and 12075001). X.Y.F. is supported by the Nankai Zhide Foundation.
\end{acknowledgments}

\vspace{3mm}
\noindent\textbf{Author contributions}\\
J.L.C supervised the project. All authors read the paper and discussed the results.
\vspace{3mm}

\noindent\textbf{Data Availability}\\
All data that support the findings of this study are included within the article (and any supplementary files).
\vspace{3mm}

\noindent\textbf{Competing interests}\\
The authors declare no competing interests.

\end{document}



\title{Supplemental Material for\\
``A Possible Mechanism to Alter Gyromagnetic Factor''}

\author{Jing-Ling Chen}
\email{chenjl@nankai.edu.cn}
\affiliation{Theoretical Physics Division, Chern Institute of Mathematics, Nankai
      University, Tianjin 300071, People's Republic of China}

\author{Xing-Yan Fan}
\affiliation{Theoretical Physics Division, Chern Institute of Mathematics, Nankai
      University, Tianjin 300071, People's Republic of China}

\author{Xiang-Ru Xie}
\affiliation{School of Physics, Nankai University, Tianjin 300071, People's Republic
      of China}

\date{\today}

%

   \maketitle

\tableofcontents

\newpage

\part{The Hamiltonion Operators of Dirac's Electron and Dirac's Braidons}

\section{Motivation}

In quantum mechanics, physical quantities are represented by operators. Given two operators $A$ and $B$, if they have nothing to do with each other, then they satisfy the following simplest commutation relation
\begin{equation}\label{eq:bra-1a}
[A, B]=AB-BA=0,
\end{equation}
i.e., the operators $A$ and $B$ are mutually commutative. However, in the more general case, operators are not commutative, thus motivating people to investigate the commutation relations among them.

As it is well-known, the angular momentum operator is very significant in quantum theory. Usually the angular momentum operator is denoted by a three-dimensional vector
\begin{equation}
\vec{J}=(J_x, J_y, J_z),
\end{equation}
whose three components satisfy the following elegant commutation relations
\begin{eqnarray}\label{eq:ang-1a}
&&[J_x, J_y]={\rm i}\hbar\, J_z, \nonumber\\
&&[J_y, J_z]={\rm i}\hbar\, J_x, \nonumber\\
&&[J_z, J_x]={\rm i}\hbar\, J_y,
\end{eqnarray}
or in a vector form as
\begin{eqnarray}
            \vec{J} \times \vec{J} = {\rm i} \hbar \vec{J},
      \end{eqnarray}
with $\hbar=h/2\pi$, and $h$ is Planck's constant.
There are two typical examples for the angular momentum operator. The first one is the orbital angular momentum operator
\begin{equation}
\vec{\ell}=\vec{r}\times \vec{p}=(\ell_x, \ell_y, \ell_z),
\end{equation}
which is defined in the external space (i.e., the space of position operator $\vec{r}$), and
\begin{eqnarray}
&&\vec{p}=(p_x, p_y, p_z)=-{\rm i}\hbar\,\left(\frac{\partial}{\partial x}, \frac{\partial}{\partial y}, \frac{\partial}{\partial z}\right)
\end{eqnarray}
is the linear momentum operator. The second one is the spin-$1/2$ angular momentum operator
\begin{eqnarray}\label{eq:s-1}
            \vec{s} =\frac{\hbar}{2} \vec{\sigma},
      \end{eqnarray}
which is defined in the intrinsic space of spin. Here
\begin{eqnarray}
\vec{\sigma}=(\sigma_x, \sigma_y, \sigma_z)
\end{eqnarray}
is the vector of Pauli's matrices, which are represented by the following $2\times 2$ matrices
\begin{eqnarray}\label{eq:Pauli-a}
 {\sigma}_x=\left(
                  \begin{array}{cc}
                    0 & 1 \\
                    1 & 0 \\
                  \end{array}
                \right), \;
                {\sigma}_y=\left(
                  \begin{array}{cc}
                    0 & -{\rm i} \\
                    {\rm i} & 0 \\
                  \end{array}
                \right), \;
                {\sigma}_z=\left(
                  \begin{array}{cc}
                    1 & 0 \\
                    0 & -1 \\
                  \end{array}
                \right).
\end{eqnarray}

Obviously, Eq. (\ref{eq:bra-1a}) is a simple relation between two operators $A$ and $B$. Nevertheless, a more nontrivial relation between them can be given as follows
\begin{equation}\label{eq:bra-1b}
ABA=BAB,
\end{equation}
which is nothing but the \emph{fundamental braid relation} presented by Emil Artin in 1925, even before the establishment of quantum mechanics. If the relation (\ref{eq:bra-1a}) describes a phenomenon that the operation $B$ does not disturb the other operation $A$, then the relation (\ref{eq:bra-1b}) describes a phenomenon called \emph{braiding}~\cite{1999braid}.

Braiding is one of the most common phenomena in human's daily life. For instance, the bamboo baskets are woven with the bamboo strips and the sweaters are knitted with the wool. The braiding operations can be described by mathematics via Eq. (\ref{eq:bra-1b}).
Actually, such a braid relation can have significant applications in mathematical physics, e.g., it has been generalized to be the famous Yang-Baxter equation~\cite{1989Jimbo}, which serves as an indispensable key to solve the quantum integrable models~\cite{1992Mattis}, such as the many-body problem in one dimension with repulsive delta-function interaction~\cite{1967Yang}, the Heisenberg spin chains~\cite{1999ChenGeXue}, and so on~\cite{1994YangGe,2019ToPhy}. Moreover, due to the braiding operations, the non-Abelian anyon with fault-tolerant feature has become a competitive candidate for building the topological quantum computer~\cite{2008Nayak}.

Now, it gives rise to an interesting question: Does the braid relation have any connection with the angular momentum operator? The answer is positive. Based on the Pauli matrices, after defining the following two operators
\begin{eqnarray}\label{eq:bra-1c}
  && A= {\rm e}^{\frac{{\rm i}\pi}{4}\sigma_x}={\rm e}^{\frac{{\rm i}\pi}{2\hbar}s_x}, \nonumber\\
  && B= {\rm e}^{\frac{{\rm i}\pi}{4}\sigma_y}={\rm e}^{\frac{{\rm i}\pi}{2\hbar}s_y},
\end{eqnarray}
 one may easily verify that the braid relation (\ref{eq:bra-1b}) holds. Such a fact implies that the fundamental braid relation can be realized by the spin-$1/2$ angular momentum operators.

In modern physics, special relativity and quantum mechanics are two cornerstones. As for special relativity, a fundamental relation is undoubtedly the relativistic energy-momentum (REM) relation of a particle, which is given by
 \begin{equation}\label{eq:EP1}
E^{2}=p^{2}c^2+m^{2}c^4,
 \end{equation}
where $E$ is the energy, $c$ is the speed of light in vacuum, $p=|\vec{p}|$ is the magnitude of the linear momentum $\vec{p}$,  and $m$ is the rest mass of the particle.
When one turns to quantum mechanics, the REM relation is recast to an operator-form as~\cite{2021ModernQM}
\begin{equation}\label{eq:EP2}
H^2=\vec{p}^{\,2} c^2+m^2c^4,
\end{equation}
where the energy $E$ is regarded as the eigenvalue of the Hamiltonian operator $H$.

However, Eq. (\ref{eq:EP2}) is not applicable directly to Schr{\" o}dinger's equation
\begin{eqnarray}
\mathrm{i} \hbar \dfrac{\partial |\Psi(t)\rangle}{\partial t} =H |\Psi(t)\rangle,
\end{eqnarray}
where $|\Psi(t)\rangle$ is the wave-function. The reason is that Schr{\" o}dinger's equation requires a linear operator $H$, instead of a square operator $H^2$ as shown in Eq. (\ref{eq:EP2}).
In 1928, to reconcile the REM relation and Schr{\" o}dinger's equation, Dirac successfully formulated a relativistic Hamiltonian operator for a free electron~\cite{1928Dirac}. Namely, by linearizing the RME relation (\ref{eq:EP2}) Dirac creatively presented that
 \begin{eqnarray}\label{eq:DiracH}
         H_{\rm e}&=& c\vec{\alpha}\cdot\vec{p}+\beta\,m c^2,
 \end{eqnarray}
where
      \begin{equation}\label{eq:EP3}
         \vec{\alpha}=\left(\begin{array}{cc}
                    0 & \vec{\sigma} \\
                  \vec{\sigma} & 0 \\
                  \end{array}
                \right),
         \quad \beta=\left(\begin{array}{cc}
                  \openone & 0 \\
                  0 & -\openone \\
                  \end{array}
                \right),
      \end{equation}
are Dirac's matrices, with $\openone$ being the $2\times 2$ identity matrix.
In this work, for convenience, we shall call $ H_{\rm e}$ as the Dirac Hamiltonian operator of a \emph{Dirac's electron}. After squaring the Hamiltonian operator $H_{\rm e}$ in Eq. (\ref{eq:DiracH}), one easily recovers the REM relation (\ref{eq:EP2}).

Then, it gives rise to another interesting question: Does the fundamental braid relation have any connection with the fundamental Hamiltonian of a Dirac's electron? Our motivation is to investigate the deep connection between the fundamental braid relation and the relativistic quantum mechanics. We find that the spin-$1/2$ angular momentum operator serves as a natural bridge to connect them. We shall expound this issue in next sections.

\section{Pauli Matrices and the Braid Relation}

Pauli's matrices as shown in Eq. (\ref{eq:Pauli-a}) satisfy the following two basic relations
\begin{subequations}
\begin{eqnarray}
  && {\sigma}_x^2={\sigma}_y^2={\sigma}_z^2=\openone, \label{eq:Pauli-b-1}\\
       && {\sigma}_x {\sigma}_y={\rm i} {\sigma}_z, \;\;\; {\sigma}_y {\sigma}_z={\rm i} {\sigma}_x, \;\;\;
        {\sigma}_z {\sigma}_x={\rm i} {\sigma}_y. \label{eq:Pauli-b-2}
\end{eqnarray}
\end{subequations}
Eq. (\ref{eq:Pauli-b-1}) indicates that the square of any Pauli's matrix gives the identity matrix, while Eq. (\ref{eq:Pauli-b-2}) implies that
each Pauli's matrix can be generated by the other two Pauli's matrices, e.g., $\sigma_z=-{\rm i} \sigma_x\sigma_y$.

Based on Eq. (\ref{eq:Pauli-b-2}), one can have the following anti-commutation relation
\begin{eqnarray}\label{eq:Pauli-b-3}
   && \{{\sigma}_x, {\sigma}_y\}=0, \;\;\; \{{\sigma}_y, {\sigma}_z\}=0, \;\;\; \{{\sigma}_z, {\sigma}_x\}=0,
\end{eqnarray}
with $\{X, Y\}=XY+YX$ being the anti-commutator. Eq. (\ref{eq:Pauli-b-3}) means that two different Pauli's matrices are anti-commutative. Similarly, due to Eq. (\ref{eq:Pauli-b-2}), one can have the following commutation relation
\begin{eqnarray}
       && [{\sigma}_x, {\sigma}_y]= {\rm i} 2\, {\sigma}_z, \;  [{\sigma}_y, {\sigma}_z]= {\rm i}2\, {\sigma}_x, \;
       [{\sigma}_z, {\sigma}_x]= {\rm i}2\, {\sigma}_y,
\end{eqnarray}
which leads to
\begin{eqnarray}\label{eq:Pauli-b-4}
       [s_x, s_y]= {\rm i} \hbar s_z, \;\;\;  [s_y, s_z]= {\rm i} \hbar s_x, \;\;\;   [s_z, s_x]= {\rm i} \hbar s_y.
\end{eqnarray}
Based on Eq. (\ref{eq:Pauli-b-1}), it is easy to prove that
\begin{eqnarray}\label{eq:Pauli-b-5}
       \vec{s}^{\, 2}=s(s+1)\hbar^2 \openone=\frac{1}{2}\left(\frac{1}{2}+1\right)\hbar^2 \openone.
\end{eqnarray}
Eqs. (\ref{eq:Pauli-b-4}) and (\ref{eq:Pauli-b-5}) indicate that $\vec{s}$ is indeed an angular momentum operator with the spin value $s=1/2$, hence Eq. (\ref{eq:s-1}) is the simplest realization of the spin-$1/2$ angular momentum operator based on Pauli's matrices.

Now, we show that the fundamental braid relation can be established based on the $2\times 2$ Pauli matrices. Explicitly, we need to prove that two operators $A$ and $B$ in Eq. (\ref{eq:bra-1c}) indeed satisfy the braid relation (\ref{eq:bra-1b}). The following is the proof.

\begin{proof}
From Eq. (\ref{eq:bra-1c}) one has
\begin{eqnarray}
  && A= {\rm e}^{\frac{{\rm i}\pi}{4}\sigma_x}=\frac{1}{\sqrt{2}}\left(\openone+{\rm i} \sigma_x\right), \nonumber\\
  && B= {\rm e}^{\frac{{\rm i}\pi}{4}\sigma_y}=\frac{1}{\sqrt{2}}\left(\openone+{\rm i} \sigma_y\right).
\end{eqnarray}
Because
\begin{eqnarray}\label{eq:bra-1d}
 ABA & = & \frac{1}{\sqrt{2}}\left(\openone+{\rm i} \sigma_x\right)\frac{1}{\sqrt{2}}\left(\openone+{\rm i} \sigma_y\right)\frac{1}{\sqrt{2}}\left(\openone+{\rm i} \sigma_x\right), \nonumber\\
  &&=\frac{1}{2\sqrt{2}} \left[\left(\openone-\sigma_x^2-\{\sigma_x, \sigma_y\}\right)+{\rm i}\left(2\sigma_x+\sigma_y-\sigma_x\sigma_y\sigma_x\right)\right]\nonumber\\
  &&=\frac{{\rm i}}{\sqrt{2}} \left(\sigma_x+\sigma_y\right),
\end{eqnarray}
and similarly
\begin{eqnarray}
  && BAB= \frac{{\rm i}}{\sqrt{2}} \left(\sigma_x+\sigma_y\right).
\end{eqnarray}
Thus the braid relation (\ref{eq:bra-1b}) holds.
\end{proof}

\begin{remark}
The derivation of the above proof is very simple, however, we are not able to find the original reference stating this point in the literature. $\blacksquare$
\end{remark}

Furthermore, one may notice that, if two basic relations shown in (\ref{eq:Pauli-b-1}) and (\ref{eq:Pauli-b-2}) are satisfied, then the braid relation (\ref{eq:bra-1b}) is automatically proved. Consequently, we have the following more general theorem, whose proof is similar.

\begin{theorem}
            Let $\Gamma_x$, $\Gamma_y$ be two generalized Pauli's matrices, i.e., they satisfy $\Gamma_x^2=\Gamma_y^2=\mathbb{I}$, and
            $\left\{\Gamma_x, \Gamma_y\right\}=0$, with $\mathbb{I}$ being the identity operator, then the following two operators
\begin{eqnarray}\label{eq:bra-1dd}
  && A= {\rm e}^{\frac{{\rm i}\pi}{4}\Gamma_x}, \nonumber\\
  && B= {\rm e}^{\frac{{\rm i}\pi}{4}\Gamma_y},
\end{eqnarray}
satisfy the braid relation (\ref{eq:bra-1b}).
\end{theorem}

\begin{remark}
Based on the first two generalized Pauli matrices $\Gamma_x$ and $\Gamma_y$, one can generate the third generalized Pauli matrix as
\begin{eqnarray}
  && \Gamma_z=-{\rm i}\, \Gamma_x\Gamma_y.
\end{eqnarray}
Accordingly one can define a generalized spin-$1/2$ angular momentum operator as
\begin{eqnarray}
            \vec{J} =\frac{\hbar}{2} \,\vec{\Gamma}=\frac{\hbar}{2} \left(\Gamma_x, \Gamma_y, \Gamma_z\right).
\end{eqnarray}
Furthermore, it is easy to see that for any two operators in the set
$\{A= {\rm e}^{\frac{{\rm i}\pi}{4}\Gamma_x}, B= {\rm e}^{\frac{{\rm i}\pi}{4}\Gamma_y}, C= {\rm e}^{\frac{{\rm i}\pi}{4}\Gamma_z}\}$,
they form a braid relation, i.e.,
\begin{eqnarray}
&& ABA=BAB,\nonumber\\
&& BCB=CBC, \nonumber\\
&& CAC=ACA.
\end{eqnarray}
$\blacksquare$
\end{remark}

\section{Parameter-Dependent Spin-$1/2$ Angular Momentum Operators}

If one focuses on Eq. (\ref{eq:s-1}), he may notice that the spin-$1/2$ operator $\vec{S}$ does not depend on any parameter.  In other words, the realization of the spin-$1/2$ operator given in Eq. (\ref{eq:s-1}) is parameter-independent. It immediately raises an interesting question: Can one have the generalized spin-$1/2$ operator that depend on some specific parameters? The answer is yes. In this section, we come to construct such a generalized spin-$1/2$ angular momentum operator.

Let us look at the $4 \times 4$ Dirac matrices as shown in Eq. (\ref{eq:EP3}),
\begin{equation}
            \vec{\alpha}=
            \left(\begin{array}{cc}
                    0 & \vec{\sigma} \\
                  \vec{\sigma} & 0 \\
                  \end{array}
                \right)=\sigma_x\otimes\vec{\sigma},\;\;\;\;\;
             \beta=
             \left(\begin{array}{cc}
                  \openone & 0 \\
                  0 & -\openone \\
                  \end{array}
                \right)
           =\sigma_z\otimes\openone,
      \end{equation}
which satisfy the following relations
\begin{subequations}
\begin{eqnarray}
  && {\alpha}_x^2={\alpha}_y^2={\alpha}_z^2=\beta^2 =\mathbb{I}, \\
       && \{{\alpha}_x, \beta\}=\{{\alpha}_y, \beta\}=\{{\alpha}_z, \beta\}=0,
\end{eqnarray}
\end{subequations}
with $\mathbb{I}=\openone\otimes\openone$ being the $4\times 4$ identity matrix.
Now let us introduce a unit vector
\begin{eqnarray}
  && \hat{n}=(n_x, n_y, n_z),
\end{eqnarray}
with $|\hat{n}|= \sqrt{n_x^2+n_y^2+n_z^2}=1$, and then we define a combined operator
\begin{eqnarray}
  && \vec{\alpha}\cdot \hat{n}= \alpha_x n_x+ \alpha_y n_y+\alpha_z n_z,
\end{eqnarray}
which satisfies
\begin{subequations}
\begin{eqnarray}
  && (\vec{\alpha}\cdot \hat{n})^2=\beta^2 =\mathbb{I}, \\
       && \{\vec{\alpha}\cdot \hat{n}, \beta\}=0.
\end{eqnarray}
\end{subequations}

We now introduce the generalized  spin-$1/2$ angular momentum operator as
\begin{eqnarray}\label{eq:s-2}
            \vec{\tau} =\frac{\hbar}{2}\, \vec{\Gamma}= \frac{\hbar}{2}\, ({\Gamma}_x, {\Gamma}_y, {\Gamma}_z),
      \end{eqnarray}
where the first two components of $\vec{\Gamma}$ (i.e., ${\Gamma}_x$ and ${\Gamma}_y$) are generated from $\vec{\alpha}\cdot \hat{n}$ and $\beta$ through a rotation as
\begin{equation}
            \left(
              \begin{array}{c}
                {\Gamma}_x \\
                {\Gamma}_y \\
              \end{array}
            \right)
            =
            \left(\begin{array}{cc}
                    \cos\varphi & \sin\varphi \\
                  -\sin\varphi & \cos\varphi \\
                  \end{array}
                \right) \left(
                          \begin{array}{c}
                            \vec{\alpha}\cdot \hat{n} \\
                            \beta \\
                          \end{array}
                        \right),
      \end{equation}
i.e.,
\begin{eqnarray}\label{eq:s-2a}
  &&  {\Gamma}_x= \cos\varphi\, (\vec{\alpha}\cdot \hat{n}) + \sin\varphi\, \beta, \nonumber\\
  && {\Gamma}_y= -\sin\varphi\, (\vec{\alpha}\cdot \hat{n}) + \cos\varphi\, \beta.
\end{eqnarray}
It is easy to prove that
\begin{subequations}
\begin{eqnarray}
  &&  {\Gamma}_x^2={\Gamma}_y^2=\mathbb{I}, \\
       && \{{\Gamma}_x, {\Gamma}_y\}={\Gamma}_x{\Gamma}_y+{\Gamma}_y{\Gamma}_x=0,
\end{eqnarray}
\end{subequations}
thus ${\Gamma}_x$ and ${\Gamma}_y$ can be viewed as two generalized Pauli's matrices. After performing the direct calculation, the third generalized Pauli matrix can be determined as
\begin{eqnarray}\label{eq:s-2b}
   {\Gamma}_z&=& -{\rm i}\, {\Gamma}_x {\Gamma}_y = {\rm i}\, \beta(\vec{\alpha}\cdot \hat{n}),
\end{eqnarray}
which does not depend on the parameter $\varphi$. \Eq{eq:s-2}, Eq. (\ref{eq:s-2a}), and Eq. (\ref{eq:s-2b}) lead to
\begin{eqnarray}
            \vec{\tau} \times \vec{\tau} = {\rm i} \hbar \, \vec{\tau},
      \end{eqnarray}
thus we have successfully constructed the generalized spin-$1/2$ angular momentum operator $\vec{\tau}$, which are $(\varphi, \hat{n})$-dependent.

\begin{remark}
Alternatively, one may write
\begin{eqnarray}\label{eq:s-3a}
   && \cos\varphi=\dfrac{\kappa_1}{\sqrt{\kappa_1^2+\kappa_2^2}}, \;\;\;  \sin\varphi=\dfrac{\kappa_2}{\sqrt{\kappa_1^2+\kappa_2^2}},
      \end{eqnarray}
where $\kappa_1$ and $\kappa_2$ are two parameters. One then has three generalized Pauli's matrices as
\begin{eqnarray}\label{eq:s-3e}
          && {\Gamma}_x =  \dfrac{\kappa_1\vec{\alpha}\cdot\hat{n}+\kappa_2\,\beta}{
                  \sqrt{\kappa_1^2+\kappa_2^2}},\nonumber\\
          && {\Gamma}_y = \dfrac{-\kappa_2\,\vec{\alpha}\cdot\hat{n}+\kappa_1\beta}{
                  \sqrt{\kappa_1^2+\kappa_2^2}},\nonumber\\
          && {\Gamma}_z =  {\rm i}\, \beta(\vec{\alpha}\cdot \hat{n}),
      \end{eqnarray}
and three components of the generalized spin-$1/2$ angular momentum operator as
      \begin{eqnarray}\label{eq:s-3f}
          &&  {\tau}_x = \frac{\hbar}{2}\, \dfrac{(\kappa_1\vec{\alpha}\cdot\hat{n}+\kappa_2\,\beta)}{
                  \sqrt{\kappa_1^2+\kappa_2^2}},\nonumber\\
          && {\tau}_y =\frac{\hbar}{2}\, \dfrac{(-\kappa_2\,\vec{\alpha}\cdot\hat{n}+\kappa_1\beta)}{
                  \sqrt{\kappa_1^2+\kappa_2^2}},\nonumber\\
          && {\tau}_z = \frac{\hbar}{2}\, {\rm i}\, \beta(\vec{\alpha}\cdot \hat{n}).
      \end{eqnarray}
$\blacksquare$
\end{remark}

\section{Dirac's Electron and Dirac's Braidons}

In this section, we come to link the Dirac Hamiltonian operator with the generalized Pauli matrices. To do so, let us compare $H_{\rm e}$ and ${\Gamma}_x$, which are listed below
\begin{eqnarray}\label{eq:DiracH-1}
 && H_{\rm e}=c \vec{\alpha}\cdot\vec{p}+\beta\,m c^2,\nonumber\\
 && {\Gamma}_x =  \dfrac{\kappa_1\vec{\alpha}\cdot\hat{n}+\kappa_2\,\beta}{
                  \sqrt{\kappa_1^2+\kappa_2^2}}.
\end{eqnarray}
By direct observation, if we let
\begin{eqnarray}\label{eq:DiracH-2}
 && \hat{n}=\hat{p}=\frac{\vec{p}}{|\vec{p}|},\;\;\;\;\;\; \kappa_1= pc, \;\;\;\;\;\;  \kappa_2=mc^2,
\end{eqnarray}
then we have
\begin{eqnarray}\label{eq:DiracH-3}
  && {\Gamma}_x =  \dfrac{H_{\rm e}}{\sqrt{\vec{p}^{\,2} c^2+m^2c^4}}.
\end{eqnarray}
The right-hand side of Eq. (\ref{eq:DiracH-3}) means that $H_{\rm e}$ is divided by its normalized factor, thus the physical meaning of ${\Gamma}_x$ is nothing but \emph{the normalized Dirac Hamiltonian operator} of a Dirac's electron.

However, this immediately leads to a direct question: If ${\Gamma}_x $ corresponds to the Dirac Hamiltonian of a Dirac's electron, then what will ${\Gamma}_y$ and ${\Gamma}_z$ correspond to? After substituting Eq. (\ref{eq:DiracH-2}) into Eq. (\ref{eq:s-3e}), we obtain
\begin{eqnarray}\label{eq:s-3g}
          && {\Gamma}_y = \dfrac{H_{\rm b}^{\rm I}}{\sqrt{\vec{p}^{\,2} c^2+m^2c^4}},\nonumber\\
          && {\Gamma}_z = \dfrac{H_{\rm b}^{\rm II}}{\sqrt{\vec{p}^{\,2} c^2+m^2c^4}},
      \end{eqnarray}
with
\begin{eqnarray}\label{eq:s-3h}
          && H_{\rm b}^{\rm I} = -mc^2\;\vec{\alpha}\cdot\hat{p}+\beta\,pc,
       \end{eqnarray}
and
\begin{eqnarray}\label{eq:s-3hh}
          && H_{\rm b}^{\rm II} = {\sqrt{\vec{p}^{\,2} c^2+m^2c^4}}\, {\rm i}\, \beta(\vec{\alpha}\cdot \hat{p}).
      \end{eqnarray}
Since ${\Gamma}_y$ and ${\Gamma}_x$ can form a \emph{braid} relation, we would like to call $H_{\rm b}^{\rm I}$ as \emph{the Dirac Hamiltonian of type-I Dirac's braidon}. Similarly,  $H_{\rm b}^{\rm II}$ is called \emph{the Dirac Hamiltonian of type-II Dirac's braidon}. Here, we have coined a term ``braidon'' (i.e., ``braid''$+$``on''), which denotes a certain particle connecting to a Dirac's electron via the braid relation.

\begin{figure}[t]
   \centering
   \includegraphics[width=90mm]{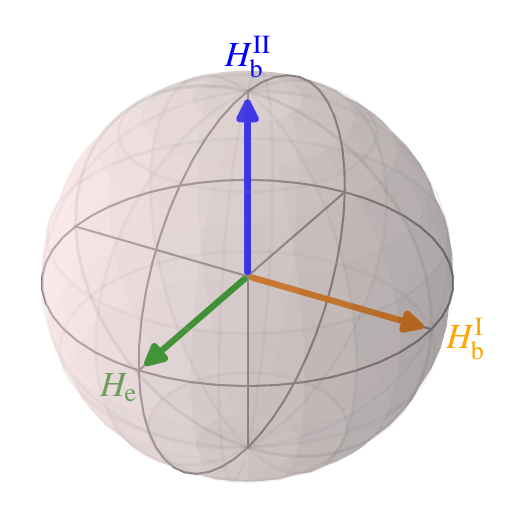}
   \caption{Illustration of Dirac's electron and Dirac's braidons in the angular momentum space. From the viewpoint of the generalized spin-$1/2$ angular momentum operator $\vec{\tau}=(\tau_x, \tau_y, \tau_z)$, the Hamiltonian $H_{\rm e}$ of a Dirac's electron corresponds to the $x$-component of spin operator $\vec{\tau}$, while the Hamiltonians $H_{\rm b}^{\rm I}$ and $H_{\rm b}^{\rm II}$ of Braidons correspond to the $y$-component and $z$-component of $\vec{\tau}$, respectively.}\label{fig:spin}
\end{figure}

In Fig. \ref{fig:spin}, we have illustrated Dirac's electron and Dirac's braidons in the spin-$1/2$ angular momentum space. Dirac's electron $H_{\rm e}$ locates on the $x$-axis, which corresponds to the $x$-component of the generalized spin-$1/2$ angular momentum operator $\vec{\tau}=(\tau_x, \tau_y, \tau_z)$, while the type-I braidon $H_{\rm b}^{\rm I}$ and the type-II braidon $H_{\rm b}^{\rm II}$ locate on the $y$-axis and $z$-axis, respectively.

\begin{remark}
Under the condition of Eq. (\ref{eq:DiracH-2}), from Eq. (\ref{eq:s-3f}) we can write down the explicit expressions for the spin-$1/2$ angular momentum operator as
\begin{eqnarray}\label{eq:s-3gg}
          &&  {\tau}_x = \frac{\hbar}{2}\, \dfrac{(c p\, \vec{\alpha}\cdot\hat{p}+\beta\,m c^2)}{
                  \sqrt{\vec{p}^{\,2} c^2+m^2c^4}},\nonumber\\
          && {\tau}_y =\frac{\hbar}{2}\, \dfrac{(-mc^2\;\vec{\alpha}\cdot\hat{p}+\beta\,pc)}{
                  \sqrt{\vec{p}^{\,2} c^2+m^2c^4}},\nonumber\\
          && {\tau}_z = \frac{\hbar}{2}\, {\rm i}\, \beta(\vec{\alpha}\cdot \hat{p}),
      \end{eqnarray}
or in the form of generalized Pauli's matrices as
\begin{eqnarray}
          &&  {\Gamma}_x =  \dfrac{c p \,\vec{\alpha}\cdot\hat{p}+\beta\,m c^2}{
                  \sqrt{\vec{p}^{\,2} c^2+m^2c^4}},\nonumber\\
          && {\Gamma}_y =\dfrac{-mc^2\;\vec{\alpha}\cdot\hat{p}+\beta\,pc}{
                  \sqrt{\vec{p}^{\,2} c^2+m^2c^4}},\nonumber\\
          && {\Gamma}_z =  {\rm i}\, \beta(\vec{\alpha}\cdot \hat{p}).
      \end{eqnarray}
One may notice that $\tau_x$ and $\tau_y$ (or $\Gamma_x$ and $\Gamma_y$) depend on the parameters $m$, $c$, and $p=|\vec{p}|$, while $\tau_z$ (or $\Gamma_z$) does not. $\blacksquare$
\end{remark}

\begin{remark}
According to the theory of angular momentum, one can transform $\tau_x$ to $\tau_y$ by a rotation around the $z$-axis through an angle $\pi/2$. Mathematically, it is expressed by
\begin{eqnarray}
 \tau_y &=& {\rm e}^{-\frac{\mathrm{i}}{\hbar} \frac{\pi}{2} \tau_z} \tau_x {\rm e}^{\frac{\mathrm{i}}{\hbar} \frac{\pi}{2} \tau_z},
\end{eqnarray}
or
\begin{eqnarray}
 \Gamma_y &=& {\rm e}^{-{\rm i} \frac{\pi}{4} \Gamma_z} \Gamma_x {\rm e}^{{\rm i} \frac{\pi}{4} \Gamma_z},
\end{eqnarray}
which is equivalent to
\begin{eqnarray}\label{eq:hh}
 &&  H_{\rm b}^{\rm I}= \mathcal{C} H_{\rm e} \mathcal{C}^\dagger, \;\;\; \mathcal{C} = {\rm e}^{-\mathrm{i} \frac{\pi}{4} \Gamma_z}.
\end{eqnarray}
The physical meaning of Eq. (\ref{eq:hh}) is that the Hamiltonian $H_{\rm e}$ of a Dirac's electron can be transformed to the Hamiltonian $H_{\rm b}^{\rm I}$ of the type-I Dirac's braidon via the unitary matrix $\mathcal{C}$. Similarly, one has
\begin{eqnarray}\label{eq:hhh}
  && H_{\rm b}^{\rm II}= \mathcal{D} H_{\rm e} \mathcal{D}^\dagger, \;\;\; \mathcal{D}= {\rm e}^{\mathrm{i} \frac{\pi}{4} \Gamma_y}.
\end{eqnarray}
$\blacksquare$
\end{remark}

\begin{table}[h]
	\centering
\caption{Dirac' electron versus Dirac's braidons.}
\begin{tabular}{lcc}
\hline\hline
 & Dirac's Electron $H_{\rm e}$ \;\;\;\;\;\;\;& Dirac's Braidons $H_{\rm b}$ \\
  \hline
mass & $m$ & $m$ \\
 \hline
magnitude of linear momentum &	$p$ & $p$ \\
  \hline
relativistic energy-momentum relation & \;\;$H_{\rm e}^2=\vec{p}^{\,2}c^2+m^2c^4$\;\; &  $H_{\rm b}^2=\vec{p}^{\,2}c^2+m^2c^4$\\
  \hline
conservation of total angular momentum operator \;\;\;\;\;\; & $[H_{\rm e}, \vec{J}]=0$ & $[H_{\rm b}, \vec{J}]=0$\\
  \hline
wave-function & $|\Psi\rangle_{\rm e}$ &  $|\Psi\rangle_{\rm b}=\mathcal{C}\;({\rm or}\; \mathcal{D})|\Psi\rangle_{\rm e}$\\
 \hline\hline
\end{tabular}\label{tab:compare}
\end{table}

\begin{remark}
In Table \ref{tab:compare}, we have compared some properties of Dirac' electron and Dirac's braidons.
They have the same $m$, the same magnitude of  linear momentum $p=|\vec{p}|$, and the same REM relation (\ref{eq:EP2}). If one considers the total angular momentum operator
\begin{eqnarray}\label{eq:EP12}
     && \vec{J}=\vec{\ell}+\vec{S},
\end{eqnarray}
where $\vec{\ell}$ is the orbital angular momentum operator, and
\begin{eqnarray}
\vec{S}&=&\frac{\hbar}{2}\,\vec{\Sigma}
\end{eqnarray}
is the spin-$1/2$ angular momentum operator of a Dirac's electron in the $4\times 4$ matrix-representation, with the operator
\begin{eqnarray}
\vec{\Sigma} &=& \openone\otimes\vec{\sigma}=
\left(\begin{array}{cc}
                   \vec{\sigma} & 0 \\
                   0 & \vec{\sigma}
                  \end{array}
                \right).
\end{eqnarray}
Then it is easy to check that
\begin{eqnarray}
&& [\vec{J}, \beta]=0, \;\;[\vec{J}, \vec{\alpha}\cdot\hat{p}]=0,
\end{eqnarray}
which leads to
\begin{eqnarray}
&& [H_{\rm e}, \vec{J}]=[H_{\rm b}^{\rm I}, \vec{J}]=[H_{\rm b}^{\rm II}, \vec{J}]=0.
\end{eqnarray}
This fact means that $\vec{J}$ is a conservation quantity for both Dirac's electron and Dirac's braidons. The above analysis shows that one cannot distinguish Dirac's electron from Dirac's braidons merely based on the mass, the magnitude of linear momentum, the REM relation, and the conservation of total angular momentum. Suppose $|\Psi\rangle_{\rm e}$ is the wave-function of Dirac's electron, then based on Eq. (\ref{eq:hh}) and Eq. (\ref{eq:hhh}) one knows that
\begin{eqnarray}
&& |\Psi\rangle_{\rm b}^{\rm I}=\mathcal{C}|\Psi\rangle_{\rm e}
\end{eqnarray}
is the wavefunction of the type-I Dirac's braidon, and
\begin{eqnarray}
&& |\Psi\rangle_{\rm b}^{\rm II}=\mathcal{D}|\Psi\rangle_{\rm e}
\end{eqnarray}
 is the wave-function of the type-II Dirac's braidon. However, the behaviors of Dirac's electron and Dirac's braidons are not always the same. For example, their behaviors can be different in the phenomenon of ``position Zitterbewegung''  \cite{1931Zitter}, which we shall study in the next Part. $\blacksquare$
\end{remark}

\newpage

\part{The Position Zitterbewegung of Dirac's Electron and Dirac's Braidons}

\section{Position Zitterbewegung for $H_{\rm e}$}

``Position Zitterbewegung'' is a kind of relativistic effect, which was first predicted by Schr{\" o}dinger \cite{1931Zitter}. He pointed out that a Dirac's electron would oscillate rapidly if it was in a superposition state of positive-energy and negative-energy. Here we make a brief review.

For a Dirac's electron, its eigen-equation is given by
\begin{equation}\label{eq:eigen}
H_{\rm e}\,|\Psi\rangle=E\,|\Psi\rangle,
\end{equation}
where $E$ is the eigenvalue, or the energy of a Dirac's electron. Since $H_{\rm e}$ is a $4\times 4$ traceless matrix, it has four eigenvalues; two of them have positive energies
\begin{equation}
E_+=+E_{\rm p}=\sqrt{p^2c^2+m^2c^4},
\end{equation}
and the other two have negative energies
\begin{equation}
E_-=-E_{\rm p}=-\sqrt{p^2c^2+m^2c^4}.
\end{equation}

For convenience, let us study the eigenstates in the momentum space.
Let us introduce the \emph{helicity} operator
\begin{equation}
\hat{\Lambda}=\vec{S} \cdot \hat{p}=\dfrac{\hbar}{2} \vec{\Sigma} \cdot \hat{p},
\end{equation}
because
\begin{equation}
\bigl[\vec{\Sigma}\cdot\hat{p},\ \vec{\alpha}\cdot\vec{p}\bigr]=0, \;\;\;\;\;\bigl[\vec{\Sigma},\ \beta\bigr]=0,
\end{equation}
then it is easy to prove that  helicity operator commutes with Hamiltonian $ H_{\rm e}$, i.e.,
\begin{equation}
[H_{\rm e}, \hat{\Lambda}]=0.
\end{equation}
Thereby, the set $\{H_{\rm e}, \hat{\Lambda}\}$ constitutes a complete set of commuting observables for the eigen-problem shown in Eq. (\ref{eq:eigen}). Let us denote $\Ket{{\Psi}_{\epsilon,\Lambda}}$ as the  simultaneous eigenstates of $\{H_{\rm e}, \hat{\Lambda}\}$, i.e.,
\begin{equation}
H_{\rm e}\,\Ket{{\Psi}_{\epsilon,\Lambda}}=E\,\Ket{{\Psi}_{\epsilon,\Lambda}},\quad \epsilon=\frac{E}{E_{\rm p}}=\pm 1,
\end{equation}
\begin{equation}
\hat{\Lambda}\Ket{{\Psi}_{\epsilon,\Lambda}}
=\Lambda\,\Ket{{\Psi}_{\epsilon,\Lambda}},\quad \Lambda=\pm\dfrac{\hbar}{2}.
\end{equation}
Then explicitly the four eigenstates are as follows
\begin{equation}\label{eq:mom}
\begin{split}
 & \epsilon=+1,\ \begin{cases}
                                          & |{\Psi}_1\rangle
                                          =\dfrac{1}{2\sqrt{E_{\rm p}\,p(p+p_z)}} \left[
                                                u_+ \Bigl(p+p_z ,\ p_+\Bigr)
                                                ,\dfrac{c\,p}{u_+}\Bigl(p+p_z,\ p_+ \Bigr)\right]^{\rm T},
                                          \quad \;\;\;\;\;\;\Lambda=+\dfrac{\hbar}{2}, \\
                                          & |{\Psi}_2\rangle
                                          =\dfrac{1}{2\sqrt{E_{\rm p}\,p(p+p_z)}}\left[
                                                u_+ \Bigl(
                                                      {-p_-}, p+p_z\Bigr)
                                                ,{-}\dfrac{c\,p}{u_+}\Bigl(
                                                      {-p_-}, p+p_z \Bigr)\right]^{\rm T},
                                          \quad \Lambda=-\dfrac{\hbar}{2},
                                    \end{cases} \\
                                    & \epsilon=-1,\ \begin{cases}
                                          & |{\Psi}_3\rangle
                                          =\dfrac{1}{2\sqrt{E_{\rm p}\,p(p+p_z)}} \left[
                                                u_- \Bigl(p+p_z ,\ p_+\Bigr)
                                                ,{-}\dfrac{c\,p}{u_-}\Bigl(
                                                      p+p_z,\ p_+ \Bigr)\right]^{\rm T},
                                          \quad \;\;\;\Lambda=+\dfrac{\hbar}{2}, \\
                                          & |{\Psi}_4\rangle
                                          =\dfrac{1}{2\sqrt{E_{\rm p}\,p(p+p_z)}}\left[
                                                u_- \Bigl(-p_-,p+p_z\Bigr)
                                                ,\dfrac{c\,p}{u_-}\Bigl(-p_-, p+p_z \Bigr)\right]^{\rm T},
                                          \quad \;\;\;\Lambda=-\dfrac{\hbar}{2},
                                    \end{cases}
\end{split}
\end{equation}
with
\begin{eqnarray}
u_{\pm} = \sqrt{E_{\rm p} \pm m\,c^2}, \;\;\;\; p_\pm =p_x \pm {\rm i} p_y.
\end{eqnarray}

From the Heisenberg equation
            \begin{eqnarray}\label{eq:D-2}
                  \frac{{\rm d}\,\hat{A}}{{\rm d}\,t}=\frac{1}{{\rm i}\hbar} \left[\hat{A}, H\right],
            \end{eqnarray}
            one has
            \begin{eqnarray}\label{eq:D-3}
                  &&\frac{{\rm d}\,\vec{p}}{{\rm d}\,t}=\frac{1}{{\rm i}\hbar}\left[\vec{p}, H_{\rm e}\right]=0,\nonumber\\
                  &&\frac{{\rm d}\,H_{\rm e}}{{\rm d}\,t}=\frac{1}{{\rm i}\hbar}\left[H_{\rm e}, H_{\rm e}\right]=0,
            \end{eqnarray}
            which means that $\vec{p}$ and $H_{\rm e}$ are conservation quantities. For position operator $\vec{r}$ and operator $\vec{\alpha}$, we have
            \begin{eqnarray}\label{eq:D-4}
                  \frac{{\rm d}\,\vec{r}}{{\rm d}\,t}=\frac{1}{{\rm i}\hbar}[\vec{r}, H_{\rm e}]=\frac{1}{{\rm i}\hbar} \big[\vec{r},\ c(\vec{\alpha}\cdot\vec{p})\big]=c\vec{\alpha},
            \end{eqnarray}
            and
            \begin{eqnarray}\label{eq:D-5}
                  \frac{{\rm d}\,\vec{\alpha}}{{\rm d}\,t}&=&\frac{1}{{\rm i}\hbar}[\vec{\alpha}, H_{\rm e}]=\frac{1}{{\rm i}\hbar}\left(\{\vec{\alpha}, H_{\rm e}\}-2\,H_{\rm e}\,\vec{\alpha}\right)\nonumber\\
                 &=&\frac{1}{{\rm i}\hbar}\left(2c\vec{p}-2\,H_{\rm e}\,\vec{\alpha}\right)=-\frac{2\,H_{\rm e}}{{\rm i}\hbar}\left(\vec{\alpha}-c H^{-1}_{\rm e}\vec{p}\right).
            \end{eqnarray}
            Then we have
            \begin{eqnarray}\label{eq:D-5a}
                  \frac{{\rm d}\,\left(\vec{\alpha}-c H^{-1}_{\rm e}\vec{p}\right)}{{\rm d}\,t}=-\frac{2\,H_{\rm e}}{{\rm i}\hbar}\left(\vec{\alpha}-c H^{-1}_{\rm e}\vec{p}\right),
            \end{eqnarray}
            i.e.,
            \begin{eqnarray}\label{eq:D-6}
                  \vec{\alpha}(t)-c H^{-1}_{\rm e}\vec{p}={{\rm e}^{\frac{{\rm i}\,2\,H_{\rm e}t}{\hbar}}\; \left[\vec{\alpha}(0)-c H^{-1}_{\rm e}\vec{p}\right]}.
            \end{eqnarray}
            One can check that
\begin{eqnarray}\label{eq:D-8a}
                  \{H_{\rm e}, \left[\vec{\alpha}(0)-c H^{-1}_{\rm e}\vec{p}\right]\}&=&[H_{\rm e} \vec{\alpha}(0)+\vec{\alpha}(0)H_{\rm e} ]-2c\vec{p}=[H_{\rm e} \vec{\alpha}+\vec{\alpha} H_{\rm e} ]-2c\vec{p}\nonumber\\
                  &=& c[(\vec{\alpha}\cdot\vec{p})\vec{\alpha}+\vec{\alpha}(\vec{\alpha}\cdot\vec{p})]-2c\vec{p}=2c\vec{p} -2c\vec{p}\nonumber\\
                  &=&0,
            \end{eqnarray}
            i.e.,
            \begin{eqnarray}\label{eq:D-8b}
                  H_{\rm e} \left[\vec{\alpha}(0)-c H^{-1}_{\rm e}\vec{p}\right]=-\left[\vec{\alpha}(0)-c H^{-1}_{\rm e}\vec{p}\right]H_{\rm e},
            \end{eqnarray}
            which leads to
            \begin{eqnarray}\label{eq:D-8c}
                  {{\rm e}^{\frac{{\rm i}\,2H_{\rm e}t}{\hbar}}\left[\vec{\alpha}(0)-c H^{-1}_{\rm e}\vec{p}\right]}=
                  {\left[\vec{\alpha}(0)-c H^{-1}_{\rm e}\vec{p}\right] {\rm e}^{\frac{-{\rm i}\,2H_{\rm e}t}{\hbar}}}.
            \end{eqnarray}
            From Eq. (\ref{eq:D-4}), we have
            \begin{eqnarray}\label{eq:D-9}
                  \frac{{\rm d}\,\vec{r}}{{\rm d}\,t}=c^2 H^{-1}_{\rm e}\vec{p}+\left[c\vec{\alpha}(0)-c^2 H^{-1}_{\rm e}\vec{p}\right]{\rm e}^{\frac{-{\rm i}\,2\,H_{\rm e}t}{\hbar}},
            \end{eqnarray}
            which results in
            \begin{eqnarray}\label{eq:D-10}
                  \vec{r}(t)&=&\vec{r}(0)+c^2 H^{-1}_{\rm e} \vec{p}\,t  +\dfrac{{\rm i}\hbar c}{2}\left[\vec{\alpha}(0)
                         -c\,H^{-1}_{\rm e} \vec{p}\right] H^{-1}_{\rm e} \left(
                                    {\rm e}^{\frac{-{\rm i}\,2H_{\rm e}t}{\hbar}} -1\right).
            \end{eqnarray}

            We now calculate the ``position Zitterbewegung''.  For the position operator in Eq. (\ref{eq:D-10}), the third term is an oscillation term, which is related to the ``position Zitterbewegung''. We need to calculate the expectation value of the ``position Zitterbewegung'' operator
            \begin{eqnarray}\label{eq:D-12a}
                  \hat{\mathcal{Z}}_{\rm e} &=& \frac{{\rm i}\hbar c}{2}\left[\vec{\alpha}(0)-c H^{-1}_{\rm e}\vec{p}\right] H^{-1}_{\rm e} \left(
                        {\rm e}^{\frac{-{\rm i}\,2\,H_{\rm e} t}{\hbar}} {-1}\right),
            \end{eqnarray}
            defined as
            \begin{eqnarray}\label{eq:D-12b}
                    \mathcal{Z}_{\rm e}&=&\langle \Psi| \hat{\mathcal{Z}}_{\rm e} |\Psi\rangle.
            \end{eqnarray}
            Here $|\Psi\rangle$ is the quantum state of Dirac's electron, and it is easy to prove that $\hat{\mathcal{Z}}_{\rm e}$ is Hermitian, i.e., $\hat{\mathcal{Z}}_{\rm e}^\dagger=\hat{\mathcal{Z}}_{\rm e}$.

            Schr{\" o}dinger has pointed out that, only when a Dirac's electron was in a superposition state of different energies, then $\mathcal{Z}_{\rm e}\neq 0$, thus appearing the relativistic effect of ``position Zitterbewegung''.
            In the following we provide a simple proof.

            \begin{proof}
            (i) Firstly, we prove that $\mathcal{Z}_{\rm e}= 0$ if Dirac's electron is in a superposition state of same energies.

            Let us introduce the following projection operators
            \begin{eqnarray}\label{eq:D-14}
                  &&\Pi_\pm=\frac{1}{2}\left(\mathbb{I}\pm \frac{H_{\rm e}}{\sqrt{p^2c^2+m^2c^4}}\right),\nonumber\\
                  && \Pi_\pm^2=\Pi_\pm,
            \end{eqnarray}
            we then obtain
            \begin{eqnarray}\label{eq:D-14a}
            && \Pi_+ |\Psi_1\rangle = |\Psi_1\rangle, \;\; \Pi_+ |\Psi_2\rangle = |\Psi_2\rangle,\nonumber\\
            && \Pi_+ |\Psi_3\rangle = 0, \;\; \Pi_+ |\Psi_4\rangle = 0,
            \end{eqnarray}
            and similarly
            \begin{eqnarray}\label{eq:D-14b}
            && \Pi_- |\Psi_1\rangle = 0, \;\; \Pi_- |\Psi_2\rangle = 0, \nonumber\\
            && \Pi_- |\Psi_3\rangle = |\Psi_3\rangle, \;\; \Pi_- |\Psi_4\rangle = |\Psi_4\rangle.
            \end{eqnarray}
            Based on Eq. (\ref{eq:D-8b}) we have
            \begin{eqnarray}\label{eq:D-17}
            &&    \Pi_+ \left[\vec{\alpha}(0)-c H^{-1}_{\rm e}\vec{p}\right]\Pi_+ =
                  \frac{1}{4} \left(\mathbb{I}+\frac{H_{\rm e}}{\sqrt{p^2c^2+m^2c^4}}\right) \left[\vec{\alpha}(0)-c H^{-1}_{\rm e}\vec{p}\right]
                  \left(\mathbb{I} + \frac{H_{\rm e}}{\sqrt{p^2c^2+m^2c^4}}\right)\nonumber\\
            &=& \frac{1}{4} \left[\left[\vec{\alpha}(0)-c H^{-1}_{\rm e}\vec{p}\right]+ \frac{1}{\sqrt{p^2c^2+m^2c^4}} \{H_{\rm e}, \left[\vec{\alpha}(0)-c H^{-1}_{\rm e}\vec{p}\right]\} +
                  \frac{1}{p^2c^2+m^2c^4} H_{\rm e} \left[\vec{\alpha}(0)-c H^{-1}_{\rm e}\vec{p} \right]H_{\rm e} \right]\nonumber\\
            &=& \frac{1}{4} \left[\left[\vec{\alpha}(0)-c H^{-1}_{\rm e}\vec{p}\right]-
                  \frac{1}{p^2c^2+m^2c^4} H^2_{\rm e} \left[\vec{\alpha}(0)-c H^{-1}_{\rm e}\vec{p} \right] \right]\nonumber\\
                  &=& \frac{1}{4} \left[\left[\vec{\alpha}(0)-c H^{-1}_{\rm e}\vec{p}\right]-
                  \left[\vec{\alpha}(0)-c H^{-1}_{\rm e}\vec{p} \right] \right]\nonumber\\
                  &=&0.
            \end{eqnarray}
            Similarly, we have
            \begin{eqnarray}\label{eq:D-18}
            &&    \Pi_- \left[\vec{\alpha}(0)-c H^{-1}_{\rm e}\vec{p}\right]\Pi_- =0.
            \end{eqnarray}
            The above results lead to
            \begin{eqnarray}\label{eq:D-19}
            &&    \Pi_+ \hat{\mathcal{Z}}_{\rm e} \Pi_+ =0,\;\;\;\;\; \Pi_- \hat{\mathcal{Z}}_{\rm e} \Pi_- =0.
            \end{eqnarray}

            Suppose the quantum state $|\Psi\rangle$ is the superposition of the positive-energy states, i.e.,
            \begin{eqnarray}\label{eq:D-20}
            &&  |\Psi\rangle\equiv|\Psi_+\rangle = c_1 |\Psi_1\rangle+ c_2 |\Psi_2\rangle,
            \end{eqnarray}
            then we have
            \begin{eqnarray}\label{eq:D-20a}
            &&  |\Psi_+\rangle=\Pi_+ \;|\Psi_+\rangle, \;\;\; \langle \Psi_+ | \; \Pi_+ = \langle \Psi_+ |,
            \end{eqnarray}
            therefore
            \begin{eqnarray}\label{eq:D-20b}
            &&  \mathcal{Z}_{\rm e}=\langle\Psi_+ | \hat{\mathcal{Z}}_{\rm e} |\Psi_+\rangle= \langle\Psi_+ | \left(\Pi_+ \; \hat{\mathcal{Z}}_{\rm e}\;\Pi_+ \right)|\Psi_+\rangle=0.
            \end{eqnarray}
            Similarly, if the quantum state $|\Psi\rangle$ is the superposition of the negative-energy states, i.e.,
            \begin{eqnarray}\label{eq:D-20c}
            &&  |\Psi\rangle\equiv|\Psi_-\rangle = c_3 |\Psi_3\rangle+ c_4 |\Psi_4\rangle,
            \end{eqnarray}
            then one has
            \begin{eqnarray}\label{eq:D-20d}
            && \mathcal{Z}_{\rm e}= \langle\Psi_- | \hat{\mathcal{Z}}_{\rm e} |\Psi_-\rangle= \langle\Psi_- | \left(\Pi_- \; \hat{\mathcal{Z}}_{\rm e}\;\Pi_- \right)|\Psi_-\rangle=0.
            \end{eqnarray}

            (ii) Secondly, we consider the case that Dirac's electron is in a superposition of different energies, for example,
            \begin{eqnarray}\label{eq:D-20e}
            &&  |\Psi\rangle = \cos\eta |\Psi_1\rangle+ \sin\eta |\Psi_3\rangle.
            \end{eqnarray}
            In this case, we have
            \begin{eqnarray}\label{eq:D-21-mb}
            \mathcal{Z}_{\rm e} &=& \langle \Psi| \hat{\mathcal{Z}}_{\rm e} |\Psi\rangle= \sin(2\eta)  {\rm Re}\left[\langle \Psi_1|\hat{\mathcal{Z}}_{\rm e}|\Psi_3 \rangle \right].
            \end{eqnarray}
            One can calculate that
            \begin{eqnarray}\label{eq:D-22-a}
            \langle\Psi_1|\hat{\mathcal{Z}}_{\rm e}|\Psi_3 \rangle
            &=&\langle\Psi_1|\left[ \frac{{\rm i}\hbar c}{2}\left[\vec{\alpha}(0)-c H^{-1}_{\rm e}\vec{p}\right] H^{-1}_{\rm e} \left(
                        {\rm e}^{\frac{-{\rm i}\,2\,H_{\rm e} t}{\hbar}} {-1}\right) \right]|\Psi_3 \rangle \nonumber\\
                        &=& \langle\Psi_1|\left[ \frac{{\rm i}\hbar c}{2}\left[\vec{\alpha}(0)-c \frac{1}{-E_{\rm p}}\vec{p}\right] \frac{1}{-E_{\rm p}} \left(
                        {\rm e}^{\frac{{\rm i}\,2\,E_{\rm p} t}{\hbar}} {-1}\right) \right]|\Psi_3 \rangle \nonumber\\
                        &=& \frac{-{\rm i}\hbar c}{2} \langle\Psi_1|\left[\vec{\alpha}(0)+c \frac{1}{E_{\rm p}}\vec{p}\right]|\Psi_3 \rangle
                              \frac{1}{E_{\rm p}} \left({\rm e}^{
                                    \frac{{\rm i}\,2\,E_{\rm p} t}{\hbar}} -1\right)
                              \nonumber\\
                        &=& \frac{-{\rm i}\hbar c}{2 E_{\rm p}} \left({\rm e}^{\frac{{\rm i}\,2\,E_{\rm p} t}{\hbar}} {-1} \right)\langle\Psi_1|\vec{\alpha}(0)|\Psi_3 \rangle,
            \end{eqnarray}
            where in the last step the orthogonal relation $\langle\Psi_1|\Psi_3 \rangle=0$ has been used.

            In the next, we need to calculate $\langle\Psi_1|\vec{\alpha}(0)|\Psi_3 \rangle$. We have
            \begin{eqnarray}\label{eq:D-28-qa0}
            && \langle\Psi_1|\vec{\alpha}(0)|\Psi_3 \rangle\nonumber\\
            &&= \left(\dfrac{1}{2\sqrt{E_{\rm p}\,p(p+p_z)}}\right)^2 \left[
                                                u_+ \Bigl(p+p_z ,\ (p_+)^\dagger\Bigr)
                                                ,\dfrac{c\,p}{u_+}\Bigl(p+p_z,\ (p_+)^\dagger \Bigr)\right]\vec{\alpha}(0)
                                                \left[
                                                u_- \Bigl(p+p_z ,\ p_+\Bigr)
                                                ,{-}\dfrac{c\,p}{u_-}\Bigl(
                                                      p+p_z,\ p_+ \Bigr)\right]^{\rm T}\nonumber\\
            &&= \dfrac{1}{4\,E_{\rm p}\,p(p+p_z)} \left[
                                                u_+ \Bigl(p+p_z ,\ p_-\Bigr)
                                                ,\dfrac{c\,p}{u_+}\Bigl(p+p_z,\ p_- \Bigr)\right]\vec{\alpha}(0)
                                                \left[
                                                u_- \Bigl(p+p_z ,\ p_+\Bigr)
                                                ,{-}\dfrac{c\,p}{u_-}\Bigl(
                                                      p+p_z,\ p_+ \Bigr)\right]^{\rm T}\nonumber\\
            &&= \dfrac{1}{4\,E_{\rm p}\,p(p+p_z)} \left[
                                                u_+ \Bigl(p+p_z ,\ p_-\Bigr)
                                                ,\dfrac{c\,p}{u_+}\Bigl(p+p_z,\ p_- \Bigr)\right]
            \left(\begin{array}{cc}
                    0 & \vec{\sigma} \\
                  \vec{\sigma} & 0 \\
                  \end{array}
                \right)     \left[
                                                u_- \Bigl(p+p_z ,\ p_+\Bigr)
                                                ,{-}\dfrac{c\,p}{u_-}\Bigl(
                                                      p+p_z,\ p_+ \Bigr)\right]^{\rm T}\nonumber\\
             &&= \dfrac{1}{4\,E_{\rm p}\,p(p+p_z)} \left[\dfrac{c\,p}{u_+}\Bigl(p+p_z,\ p_-\Bigr)\vec{\sigma},
                                                u_+ \Bigl(p+p_z ,\ p_-\Bigr)\vec{\sigma}
                                                 \right]
            \left[
                                                u_- \Bigl(p+p_z ,\ p_+\Bigr)
                                                ,{-}\dfrac{c\,p}{u_-}\Bigl(
                                                      p+p_z,\ p_+ \Bigr)\right]^{\rm T}\nonumber\\
             &&= \dfrac{1}{4\,E_{\rm p}\,p(p+p_z)} \left[\dfrac{c\,p}{u_+}\Bigl(p+p_z,\ p_-\Bigr)\vec{\sigma}u_- \Bigl(p+p_z ,\ p_+\Bigr)^{\rm T}+
                                                u_+ \Bigl(p+p_z ,\ p_-\Bigr)\vec{\sigma}\left[{-}\dfrac{c\,p}{u_-}\Bigl(
                                                      p+p_z,\ p_+ \Bigr)\right]^{\rm T}
                                                 \right]\nonumber\\
             &&= \dfrac{cp}{4\,E_{\rm p}\,p(p+p_z)} \left[\dfrac{u_-}{u_+}\Bigl(p+p_z,\ p_-\Bigr)\vec{\sigma} \Bigl(p+p_z ,\ p_+\Bigr)^{\rm T}-
                                                \dfrac{u_+}{u_-} \Bigl(p+p_z ,\ p_-\Bigr)\vec{\sigma}\Bigl(
                                                      p+p_z,\ p_+ \Bigr)^{\rm T}
                                                 \right]\nonumber\\
             &&= \dfrac{cp}{4\,E_{\rm p}\,p(p+p_z)} \left[\dfrac{u_-}{u_+}- \dfrac{u_+}{u_-} \right]\Bigl(p+p_z,\ p_-\Bigr)\vec{\sigma} \Bigl(p+p_z ,\ p_+\Bigr)^{\rm T}\nonumber\\
              &&= \dfrac{cp}{4\,E_{\rm p}\,p(p+p_z)} \left[\dfrac{u_-^2-u_+^2}{u_+u_-}\right]\Bigl(p+p_z,\ p_-\Bigr)\vec{\sigma} \Bigl(p+p_z ,\ p_+\Bigr)^{\rm T}\nonumber\\
               &&= \dfrac{cp}{4\,E_{\rm p}\,p(p+p_z)} \left[\dfrac{(\sqrt{E_{\rm p} - m\,c^2})^2-(\sqrt{E_{\rm p} + m\,c^2})^2}{\sqrt{E_{\rm p} + m\,c^2}\sqrt{E_{\rm p} -m\,c^2}}\right]\Bigl(p+p_z,\ p_-\Bigr)\vec{\sigma} \Bigl(p+p_z ,\ p_+\Bigr)^{\rm T}\nonumber\\
            &&= \dfrac{cp}{4\,E_{\rm p}\,p(p+p_z)}
            \left[\dfrac{-2 m\,c^2}{cp}\right]\Bigl(p+p_z,\ p_-\Bigr)\vec{\sigma} \Bigl(p+p_z ,\ p_+\Bigr)^{\rm T}\nonumber\\
            &&= \dfrac{-mc^2}{2E_{\rm p}p}\dfrac{1}{p+p_z}
            \Bigl(p+p_z,\ p_-\Bigr)\vec{\sigma} \Bigl(p+p_z ,\ p_+\Bigr)^{\rm T}.
            \end{eqnarray}
            Because
             \begin{eqnarray}
             \frac{1}{p+p_z}\Bigl(p+p_z,\ p_-\Bigr)\sigma_x \Bigl(p+p_z ,\ p_+\Bigr)^{\rm T}&=&\frac{1}{p+p_z}\Bigl(p_-, p+p_z\Bigr)\Bigl(p+p_z ,\ p_+\Bigr)^{\rm T}\nonumber\\
            &=&\frac{1}{p+p_z}\left[p_- (p+p_z)+(p+p_z)p_+\right]=2p_x, \nonumber\\
             \frac{1}{p+p_z}\Bigl(p+p_z,\ p_-\Bigr)\sigma_y \Bigl(p+p_z ,\ p_+\Bigr)^{\rm T}&=&\frac{1}{p+p_z}\Bigl({\rm i}p_-, -{\rm i}(p+p_z)\Bigr)\Bigl(p+p_z ,\ p_+\Bigr)^{\rm T}\nonumber\\
            &=&\frac{1}{p+p_z}\left[{\rm i}p_- (p+p_z)-{\rm i}(p+p_z)p_+\right]=2p_y, \nonumber\\
             \frac{1}{p+p_z}\Bigl(p+p_z,\ p_-\Bigr)\sigma_z\Bigl(p+p_z ,\ p_+\Bigr)^{\rm T}&=&\frac{1}{p+p_z}\Bigl(p+p_z,-p_-\Bigr)\Bigl(p+p_z ,\ p_+\Bigr)^{\rm T}\nonumber\\
            &=&\frac{1}{p+p_z}\left[(p+p_z)^2-p_-p_+\right]=2p_z,
            \end{eqnarray}
            then we have
            \begin{eqnarray}\label{eq:D-28-qa}
             \langle\Psi_1|\vec{\alpha}(0)|\Psi_3 \rangle &=& \dfrac{-mc^2}{2E_{\rm p}p}\dfrac{1}{p+p_z}
            \Bigl(p+p_z,\ p_-\Bigr)\vec{\sigma} \Bigl(p+p_z ,\ p_+\Bigr)^{\rm T}\nonumber\\
            &=& \dfrac{-mc^2}{2E_{\rm p}\,p}\; 2\vec{p}= -\dfrac{mc^2}{E_{\rm p}}\; \hat{p},
            \end{eqnarray}
            which yields
              \begin{eqnarray}\label{eq:ZZZ-a}
                  \langle\Psi_1|\hat{\mathcal{Z}}_{\rm e}|\Psi_3 \rangle
                  &=& \frac{-{\rm i}\hbar c}{2 E_{\rm p}} \left({\rm e}^{
                        \frac{{\rm i}\,2\,E_{\rm p} t}{\hbar}} {-1}
                        \right)\langle\Psi_1|\vec{\alpha}(0)|\Psi_3 \rangle
                  =\hat{p}\; \dfrac{{\rm i}\hbar c}{2}
                        \dfrac{mc^2}{E^2_{\rm p}}\left({\rm e}^{
                              \frac{{\rm i}\,2E_{\rm p}\,t}{\hbar}}-1\right).
            \end{eqnarray}
            Then we have
            \begin{eqnarray}\label{eq:ze1}
                        \mathcal{Z}_{\rm e} &=& \sin(2\eta){\rm Re}\left[
                              \langle \Psi_1|\hat{\mathcal{Z}}_{\rm e}|\Psi_3 \rangle \right] \nonumber \\
                        &=& \sin(2\eta){\rm Re}\left[
                              \frac{{\rm i}\,\hbar\,c}{2\,E_{\rm p}} \left(
                                    {\rm e}^{\frac{{\rm i}\,2\,E_{\rm p} t}{\hbar}} -1 \right)
                                    \dfrac{mc^2}{E_{\rm p}} \hat{p}\right] \nonumber \\
                        &=& \hat{p}\;\sin(2\eta)\dfrac{\hbar c}{2 E_{\rm p}} \dfrac{mc^2}{E_{\rm p}}
                               {\rm Re}\left[
                                    {\rm i} \left({\rm e}^{\frac{{\rm i}\,2\,E_{\rm p} t}{\hbar}} -1 \right)\right] \nonumber \\
                        &=& \hat{p}\;\sin(2\eta)\dfrac{\hbar c \times mc^2}{2 E^2_{\rm p}}
                              \left[-\sin\left(\frac{2\,E_{\rm p} t}{\hbar}\right)\right]\nonumber\\
                        &=& -\hat{p}\;\mathbb{A} \sin(\omega t).
                  \end{eqnarray}
            In this case one has $\mathcal{Z}_{\rm e}$.
            \end{proof}

                  Based on Eq. (\ref{eq:ze1}), we have the amplitude of the position oscillation as
            \begin{eqnarray}\label{eq:D-21-e1}
                        \mathbb{A} &=& \sin(2\eta) \frac{\hbar c}{2 E_{\rm p}} \dfrac{mc^2}{{E_{\rm p}}},
            \end{eqnarray}
            and the frequency as
            \begin{eqnarray}\label{eq:D-21-e2}
                        \omega &=& \frac{2\,E_{\rm p}}{\hbar}.
            \end{eqnarray}
            One may estimate the amplitude as
            \begin{eqnarray}\label{eq:D-21-f}
                        \mathbb{A} &=& \sin(2\eta) \frac{\hbar c}{2 E_{\rm p}} \dfrac{mc^2}{{E_{\rm p}}} \nonumber\\
                        &\leq &  \frac{\hbar c}{2 E_{\rm p}} \dfrac{mc^2}{{E_{\rm p}}}=\dfrac{\hbar c \times mc^2}{2E^2_{\rm p}}
                        =\dfrac{\hbar c \times mc^2}{2 (m^2c^4+p^2c^2)} \nonumber\\
                        &\leq &  \dfrac{\hbar c \times mc^2}{2 (m^2c^4)}=\frac{1}{4\pi} \dfrac{h}{mc}=\frac{1}{4\pi} \lambda_{\rm e},
            \end{eqnarray}
            where
            \begin{eqnarray}\label{eq:D-21-g}
                        \lambda_{\rm e} =\dfrac{h}{mc}\approx 2.42631\times 10^{-12} \;{\rm m},
            \end{eqnarray}
            is the Compton wavelength of a Dirac's electron. This means that the amplitude $\mathbb{A}$ of position oscillation is less than $ \lambda_{\rm e}/(4\pi)\approx 1.9308 \times 10^{-13} \;{\rm m}$. Here some constants are
            \begin{eqnarray}\label{eq:D-21-h}
                        h &=& 6.62607015\times 10^{-34} \; {\rm J}\cdot {\rm s}, \nonumber\\
                        c &=& 2.99792458 \times 10^8 \;{\rm m/s}, \nonumber\\
                        m &=& 9.10938188\times 10^{-31} \; {\rm kg}\nonumber\\
                        \pi &=& 3.141592654.
            \end{eqnarray}
            By using the data of Eq. (\ref{eq:D-21-h}), one may obtain Eq. (\ref{eq:D-21-g}).

            Similarly, one may estimate the frequency as
            \begin{eqnarray}\label{eq:D-21-i}
                        \omega &=& \frac{2\,E_{\rm p}}{\hbar}=\frac{2\,\sqrt{m^2c^4+p^2c^2}}{\hbar}\nonumber\\
                        &\geq& \frac{2\,mc^2}{\hbar}
                        =\frac{4\pi c}{\lambda_{\rm e}} \nonumber\\
                        &\approx& {1.55269 \times 10^{21}\; {\rm s}^{-1}}.
            \end{eqnarray}

\subsection{More General Results}

\begin{remark}
In the computation above, we have pondered only the case of $\langle\Psi_1|\vec{\alpha}(0)|\Psi_3\rangle$. In this section, let us consider more general cases. To do so, we need to calculate the other three terms, i.e., $\langle\Psi_2|\vec{\alpha}(0)|\Psi_4\rangle$, $\langle\Psi_1|\vec{\alpha}(0)|\Psi_4\rangle$, and $\langle\Psi_2|\vec{\alpha}(0)|\Psi_3\rangle$. $\blacksquare$
\end{remark}

\begin{remark}We can have
      \begin{eqnarray}
            && \langle\Psi_2|\vec{\alpha}(0)|\Psi_4\rangle \nonumber\\
            &=& \left(\dfrac{1}{2\sqrt{E_{\rm p}\,p(p+p_z)}}\right)^2 \left[
                  u_+ \Bigl(-(p_-)^\dagger ,p+p_z\Bigr),
                  -\dfrac{c\,p}{u_+}\Bigl(-(p_-)^\dagger ,p+p_z\Bigr)
                  \right]\vec{\alpha}(0)\left[u_- \Bigl(-p_- ,p+p_z\Bigr),
                        \dfrac{c\,p}{u_-}\Bigl(-p_- ,p+p_z\Bigr)\right]^{\rm T}
                  \nonumber \\
            &=& \dfrac{1}{4\,E_{\rm p}\,p(p+p_z)} \left[
                  u_+ \Bigl(-p_+ ,p+p_z\Bigr),-\dfrac{c\,p}{u_+}\Bigl(-p_+ ,p+p_z\Bigr)
                  \right]\vec{\alpha}(0)\left[u_- \Bigl(-p_- ,p+p_z\Bigr),
                        \dfrac{c\,p}{u_-}\Bigl(-p_- ,p+p_z\Bigr)\right]^{\rm T}
                  \nonumber\\
            &=& \dfrac{1}{4\,E_{\rm p}\,p(p+p_z)} \left[
                  u_+ \Bigl(-p_+ ,p+p_z\Bigr),-\dfrac{c\,p}{u_+}\Bigl(-p_+ ,p+p_z\Bigr)
                  \right]\left(\begin{array}{cc}
                              0 & \vec{\sigma} \\
                              \vec{\sigma} & 0 \\
                        \end{array}\right)\left[u_- \Bigl(-p_- ,p+p_z\Bigr),
                              \dfrac{c\,p}{u_-}\Bigl(-p_- ,p+p_z\Bigr)\right]^{\rm T}
                  \nonumber\\
            &=& \dfrac{1}{4\,E_{\rm p}\,p(p+p_z)} \left[
                  -\dfrac{c\,p}{u_+}\Bigl(-p_+ ,p+p_z\Bigr)\vec{\sigma},
                  u_+ \Bigl(-p_+ ,p+p_z\Bigr)\vec{\sigma}
                  \right]\left[u_- \Bigl(-p_- ,p+p_z\Bigr),
                        \dfrac{c\,p}{u_-}\Bigl(-p_- ,p+p_z\Bigr)\right]^{\rm T}
                  \nonumber\\
            &=& \dfrac{1}{4\,E_{\rm p}\,p(p+p_z)} \left\{
                  -\dfrac{c\,p}{u_+}\Bigl(-p_+ ,p+p_z\Bigr)\vec{\sigma}\,u_- \Bigl(
                        -p_- ,p+p_z\Bigr)^{\rm T}
                  +u_+ \Bigl(-p_+ ,p+p_z\Bigr)\vec{\sigma}\dfrac{c\,p}{u_-} \Bigl(
                        -p_- ,p+p_z\Bigr)^{\rm T}\right\} \nonumber\\
            &=& \dfrac{c\,p}{4\,E_{\rm p}\,p(p+p_z)} \left[
                  -\dfrac{u_-}{u_+}\Bigl(-p_+ ,p+p_z\Bigr)\vec{\sigma}\Bigl(
                        -p_- ,p+p_z\Bigr)^{\rm T}
                  +\dfrac{u_+}{u_-} \Bigl(-p_+ ,p+p_z\Bigr)\vec{\sigma}\Bigl(
                        -p_- ,p+p_z\Bigr)^{\rm T}
                                                 \right]\nonumber\\
            &=& \dfrac{c\,p}{4\,E_{\rm p}\,p(p+p_z)} \left(-\dfrac{u_-}{u_+}
                  +\dfrac{u_+}{u_-}\right)\Bigl(-p_+ ,p+p_z\Bigr)\vec{\sigma}\Bigl(
                        -p_- ,p+p_z\Bigr)^{\rm T} \nonumber\\
            &=& \dfrac{c\,p}{4\,E_{\rm p}\,p(p+p_z)} \left(
                  \dfrac{u_+^2 -u_-^2}{u_+ u_-}\right)\Bigl(-p_+ ,p+p_z\Bigr)
                        \vec{\sigma}\Bigl(-p_- ,p+p_z\Bigr)^{\rm T} \nonumber\\
            &=& \dfrac{c\,p}{4\,E_{\rm p}\,p(p+p_z)} \left[
                  \dfrac{(\sqrt{E_{\rm p} + m\,c^2})^2 -(\sqrt{E_{\rm p} -m\,c^2})^2}{
                        \sqrt{E_{\rm p} +m\,c^2}\sqrt{E_{\rm p} -m\,c^2}}\right]\Bigl(
                              -p_+ ,p+p_z\Bigr)\vec{\sigma}\Bigl(-p_- ,p+p_z
                              \Bigr)^{\rm T} \nonumber\\
            &=& \dfrac{c\,p}{4\,E_{\rm p}\,p(p+p_z)} \dfrac{2\,m\,c^2}{cp} \Bigl(
                  -p_+ ,p+p_z\Bigr)\vec{\sigma}\Bigl(-p_- ,p+p_z\Bigr)^{\rm T}
                  \nonumber\\
            &=& \dfrac{m\,c^2}{2\,E_{\rm p} p} \dfrac{1}{(p+p_z)} \Bigl(
                  -p_+ ,p+p_z\Bigr)\vec{\sigma}\Bigl(-p_- ,p+p_z\Bigr)^{\rm T}.
      \end{eqnarray}
      Because
      \begin{eqnarray}
             \dfrac{1}{(p+p_z)} \Bigl(-p_+ ,p+p_z\Bigr)\sigma_x \Bigl(
                        -p_- ,p+p_z\Bigr)^{\rm T}
                  &=&\frac{1}{p+p_z} \Bigl(p+p_z ,-p_+\Bigr)\Bigl(-p_- ,p+p_z
                        \Bigr)^{\rm T} \nonumber\\
                 & = &-\frac{1}{p+p_z} (p+p_z)(p_- +p_+) \notag
                  = -2\,p_x, \nonumber \\
            \dfrac{1}{(p+p_z)} \Bigl(-p_+ ,p+p_z\Bigr)\sigma_y \Bigl(
                        -p_- ,p+p_z\Bigr)^{\rm T}
                  &=&\frac{1}{p+p_z} \Bigl[{\rm i}(p+p_z),{\rm i}\,p_+\Bigr]\Bigl(
                        -p_- ,p+p_z\Bigr)^{\rm T} \nonumber\\
                 & =&\frac{1}{p+p_z} \left[-{\rm i}\,p_- (p+p_z)+{\rm i}\,p_+(p+p_z)
                        \right] \notag
                  = -2\,p_y, \nonumber\\
             \dfrac{1}{(p+p_z)} \Bigl(-p_+ ,p+p_z\Bigr)\sigma_z \Bigl(
                        -p_- ,p+p_z\Bigr)^{\rm T}
                  &=&\frac{1}{p+p_z} \Bigl[-p_+ ,-(p+p_z)\Bigr]\Bigl(
                        -p_- ,p+p_z\Bigr)^{\rm T} \nonumber\\
                  &=&\frac{1}{p+p_z} \left[p_- p_+ -(p+p_z)^2\right]
                  = -2\,p_z,
      \end{eqnarray}
      then we have
      \begin{eqnarray}
            \langle\Psi_2|\vec{\alpha}(0)|\Psi_4\rangle
            =\dfrac{m\,c^2}{2\,E_{\rm p} p} \dfrac{1}{(p+p_z)} \Bigl(
                  -p_+ ,p+p_z\Bigr)\vec{\sigma}\Bigl(-p_- ,p+p_z\Bigr)^{\rm T}
            =\dfrac{m\,c^2}{2\,E_{\rm p} p} (-2\,\vec{p})
            =-\dfrac{m\,c^2}{E_{\rm p}} \hat{p},
      \end{eqnarray}
      namely,
      \begin{eqnarray}
            \langle\Psi_2|\vec{\alpha}(0)|\Psi_4\rangle
            =\langle\Psi_1|\vec{\alpha}(0)|\Psi_3\rangle.
      \end{eqnarray}
      $\blacksquare$
      \end{remark}

      \begin{remark}Likewise, we have
      \begin{eqnarray}
            && \langle\Psi_1|\vec{\alpha}(0)|\Psi_4\rangle \nonumber\\
            &=& \left(\dfrac{1}{2\sqrt{E_{\rm p}\,p(p+p_z)}}\right)^2 \left[
                  u_+ \Bigl(p+p_z ,(p_+)^\dagger\Bigr),
                  \dfrac{c\,p}{u_+} \Bigl(p+p_z ,(p_+)^\dagger\Bigr)
                  \right]\vec{\alpha}(0)\left[u_- \Bigl(-p_- ,p+p_z\Bigr),
                        \dfrac{c\,p}{u_-}\Bigl(-p_- ,p+p_z\Bigr)\right]^{\rm T}
                  \nonumber \\
            &=& \dfrac{1}{4\,E_{\rm p}\,p(p+p_z)} \left[u_+ \Bigl(p+p_z ,\ p_-\Bigr),
                  \dfrac{c\,p}{u_+} \Bigl(p+p_z ,p_-\Bigr)\right]\vec{\alpha}(0)\left[
                        u_- \Bigl(-p_- ,p+p_z\Bigr),
                        \dfrac{c\,p}{u_-} \Bigl(-p_- ,p+p_z\Bigr)\right]^{\rm T}
                  \nonumber\\
            &=& \dfrac{1}{4\,E_{\rm p}\,p(p+p_z)} \left[u_+ \Bigl(p+p_z ,\ p_-\Bigr),
                  \dfrac{c\,p}{u_+} \Bigl(p+p_z ,p_-\Bigr)\right]\left(
                        \begin{array}{cc}
                              0 & \vec{\sigma} \\
                              \vec{\sigma} & 0 \\
                        \end{array}\right)\left[u_- \Bigl(-p_- ,p+p_z\Bigr),
                              \dfrac{c\,p}{u_-} \Bigl(-p_- ,p+p_z\Bigr)\right]^{\rm T}
                  \nonumber\\
            &=& \dfrac{1}{4\,E_{\rm p}\,p(p+p_z)} \left[
                  \dfrac{c\,p}{u_+} \Bigl(p+p_z ,p_-\Bigr)\vec{\sigma},
                  u_+ \Bigl(p+p_z ,\ p_-\Bigr)\vec{\sigma}\right]\left[
                        u_- \Bigl(-p_- ,p+p_z\Bigr),
                        \dfrac{c\,p}{u_-}\Bigl(-p_- ,p+p_z\Bigr)\right]^{\rm T}
                  \nonumber\\
            &=& \dfrac{1}{4\,E_{\rm p}\,p(p+p_z)} \left\{
                  \dfrac{c\,p}{u_+} \Bigl(p+p_z ,p_-\Bigr)\vec{\sigma}\,u_- \Bigl(
                        -p_- ,p+p_z\Bigr)^{\rm T}
                  +u_+ \Bigl(p+p_z ,\ p_-\Bigr)\vec{\sigma}\dfrac{c\,p}{u_-}\Bigl(
                        -p_- ,p+p_z\Bigr)^{\rm T}\right\} \nonumber\\
            &=& \dfrac{c\,p}{4\,E_{\rm p}\,p(p+p_z)} \left[
                  \dfrac{u_-}{u_+} \Bigl(p+p_z ,p_-\Bigr)\vec{\sigma}\Bigl(
                        -p_- ,p+p_z\Bigr)^{\rm T}
                  +\dfrac{u_+}{u_-} \Bigl(p+p_z ,p_-\Bigr)\vec{\sigma}\Bigl(
                        -p_- ,p+p_z\Bigr)^{\rm T}\right]\nonumber\\
            &=& \dfrac{c\,p}{4\,E_{\rm p}\,p(p+p_z)} \left(\dfrac{u_-}{u_+}
                  +\dfrac{u_+}{u_-}\right)\Bigl(p+p_z ,p_-\Bigr)\vec{\sigma}\Bigl(
                        -p_- ,p+p_z\Bigr)^{\rm T} \nonumber\\
            &=& \dfrac{c\,p}{4\,E_{\rm p}\,p(p+p_z)} \left(
                  \dfrac{u_+^2 +u_-^2}{u_+ u_-}\right)\Bigl(p+p_z ,p_-\Bigr)
                        \vec{\sigma}\Bigl(-p_- ,p+p_z\Bigr)^{\rm T} \nonumber\\
            &=& \dfrac{c\,p}{4\,E_{\rm p}\,p(p+p_z)} \left[
                  \dfrac{(\sqrt{E_{\rm p} + m\,c^2})^2 +(\sqrt{E_{\rm p} -m\,c^2})^2}{
                        \sqrt{E_{\rm p} +m\,c^2}\sqrt{E_{\rm p} -m\,c^2}}\right]\Bigl(
                              p+p_z ,p_-\Bigr)\vec{\sigma}\Bigl(-p_- ,p+p_z
                                    \Bigr)^{\rm T} \nonumber\\
            &=& \dfrac{c\,p}{4\,E_{\rm p}\,p(p+p_z)} \dfrac{2\,E_{\rm p}}{c\,p} \Bigl(
                  p+p_z ,p_-\Bigr)\vec{\sigma}\Bigl(-p_- ,p+p_z\Bigr)^{\rm T}
                  \nonumber\\
            &=& \dfrac{1}{2\,p} \dfrac{1}{(p+p_z)} \Bigl(p+p_z ,p_-\Bigr)\vec{\sigma}
                  \Bigl(-p_- ,p+p_z\Bigr)^{\rm T}.
      \end{eqnarray}
      Because
      \begin{eqnarray}
             \dfrac{1}{(p+p_z)} \Bigl(p+p_z ,p_-\Bigr)\sigma_x \Bigl(
                        -p_- ,p+p_z\Bigr)^{\rm T}
                  &=&\frac{1}{p+p_z} \Bigl(p_- ,p+p_z\Bigr)\Bigl(-p_- ,p+p_z
                        \Bigr)^{\rm T} \nonumber\\
                  &=&\frac{1}{p+p_z} \big[(p+p_z)^2 -p_-^2\big] \notag \\
                  &=& \frac{1}{p+p_z} \big[(p+p_z)^2 -(p_x -{\rm i}\,p_y)^2\big]\nonumber\\
                  &=&\frac{1}{p+p_z} \big[(p^2 +p_z^2 +2\,p\,p_z)-(
                        p_x^2 -p_y^2 -{\rm i}\,2\,p_x p_y)\big] \notag \\
                  &=& \frac{1}{p+p_z} \big[(p^2 -p_x^2)+(p_y^2 +p_z^2)+2(p\,p_z
                         +{\rm i}\,p_x p_y)\big], \nonumber \\
                  &=& \frac{2}{p+p_z} \big[(p^2 -p_x^2)+(p\,p_z +{\rm i}\,p_x p_y)
                        \big], \nonumber \\
             \dfrac{1}{(p+p_z)} \Bigl(p+p_z ,p_-\Bigr)\sigma_y \Bigl(
                        -p_- ,p+p_z\Bigr)^{\rm T}
                 & =&\frac{1}{p+p_z} \bigl[{\rm i}\,p_- ,-{\rm i}(p+p_z)\bigr]\bigl(
                        -p_- ,p+p_z\bigr)^{\rm T}\nonumber\\
                  &=&-\frac{\rm i}{p+p_z} \big[(p+p_z)^2 +p_-^2\big] \notag \\
                  &=& -\frac{\rm i}{p+p_z} \big[(p+p_z)^2 +(p_x -{\rm i}\,p_y)^2\big]\nonumber\\
                  &=&-\frac{\rm i}{p+p_z} \big[(p^2 +p_z^2 +2\,p\,p_z)+(
                        p_x^2 -p_y^2 -{\rm i}\,2\,p_x p_y)\big] \notag \\
                  &=& -\frac{\rm i}{p+p_z} \big[(p^2 -p_y^2)+(p_z^2 +p_x^2)+2(p\,p_z
                         -{\rm i}\,p_x p_y)\big], \nonumber \\
                  &=& -{\rm i}\frac{2}{p+p_z} \big[(p^2 -p_y^2)
                        +(p\,p_z -{\rm i}\,p_x p_y)\big], \nonumber\\
             \dfrac{1}{(p+p_z)} \bigl(p+p_z ,p_-\bigr)\sigma_z \bigl(
                        -p_- ,p+p_z\bigr)^{\rm T}
                 & =&\frac{1}{p+p_z} \bigl(p+p_z ,-p_-\bigr)\bigl(-p_- ,p+p_z
                        \bigr)^{\rm T}\nonumber\\
                  &=& -\frac{2}{p+p_z} (p+p_z)p_- \nonumber\\
                  &=& -2\,p_- \notag \\
                  &=& -2(p_x -{\rm i}\,p_y),
      \end{eqnarray}
      then we have
      \begin{eqnarray}
            && \langle\Psi_1|\vec{\alpha}(0)|\Psi_4\rangle
            =\dfrac{1}{2\,p} \dfrac{1}{(p+p_z)} \Bigl(p+p_z ,p_-\Bigr)\vec{\sigma}
                  \Bigl(-p_- ,p+p_z\Bigr)^{\rm T} \notag \\
            &=& \dfrac{1}{2\,p} \bigg\{\frac{2}{p+p_z} \big[(p^2 -p_x^2)
                        +(p\,p_z +{\rm i}\,p_x p_y)\big]\hat{e}_x
                  -{\rm i}\frac{2}{p+p_z} \big[(p^2 -p_y^2)+(p\,p_z -{\rm i}\,p_x p_y)
                        \big]\hat{e}_y
                  -2(p_x -{\rm i}\,p_y)\hat{e}_z\bigg\} \notag \\
            &=& \frac{1}{p(p+p_z)} \big[(p^2 -p_x^2)
                        +(p\,p_z +{\rm i}\,p_x p_y)\big]\hat{e}_x
                  -\frac{\rm i}{p(p+p_z)} \big[(p^2 -p_y^2)+(p\,p_z -{\rm i}\,p_x p_y)
                        \big]\hat{e}_y
                  -\frac{1}{p} (p_x -{\rm i}\,p_y)\hat{e}_z.
      \end{eqnarray}

      For convenience, we let
      \begin{equation}
            \langle\Psi_1|\vec{\alpha}(0)|\Psi_4\rangle=\vec{F}_1 +{\rm i}\,\vec{F}_2,
      \end{equation}
      where
      \begin{eqnarray}
           & & \vec{F}_1 =\frac{1}{p(p+p_z)} \big[(p^2 -p_x^2)+p\,p_z\big]
                        \hat{e}_x
                  -\frac{p_x p_y}{p(p+p_z)} \hat{e}_y
                  -\frac{p_x}{p} \hat{e}_z, \notag \\
            && \vec{F}_2 =\frac{p_x p_y}{p(p+p_z)} \hat{e}_x
                  -\frac{1}{p(p+p_z)} \big[(p^2 -p_y^2)+p\,p_z\big]\hat{e}_y
                  +\frac{p_y}{p} \hat{e}_z,
      \end{eqnarray}
      with
       \begin{eqnarray}
           & & \vec{F}_1^* =  \vec{F}_1, \;\;\; \vec{F}_2^* =  \vec{F}_2.
      \end{eqnarray}
      After that, we obtain
       \begin{eqnarray}
           & & |\vec{F}_1| = 1, \;\;\; |\vec{F}_2| = 1,
      \end{eqnarray}
      and
      \begin{eqnarray}
             \vec{F}_1 \cdot\vec{p}
            &=&\biggl\{\frac{1}{p(p+p_z)} \big[(p^2 -p_x^2)+p\,p_z\big]
                        \hat{e}_x
                  -\frac{p_x p_y}{p(p+p_z)} \hat{e}_y
                  -\frac{p_x}{p} \hat{e}_z\biggr\}\cdot\vec{p} \notag \\
           &=& \frac{p_x}{p(p+p_z)} \big[(p^2 -p_x^2)+p\,p_z\big]
                  -\frac{p_x p_y^2}{p(p+p_z)}
                  -\frac{p_x p_z}{p}
            =\dfrac{p_x}{p} \bigg\{\frac{\big[(p^2 -p_x^2)+p\,p_z\big]-p_y^2}{
                  (p+p_z)} -p_z\bigg\} \notag \\
           & =& \dfrac{p_x}{p} \bigg[\frac{p_z^2 +p\,p_z}{(p+p_z)} -p_z\bigg]
                  \notag \\
            &=& 0, \nonumber\\
      \vec{F}_2 \cdot\vec{p}
           & =&\biggl\{\frac{p_x p_y}{p(p+p_z)} \hat{e}_x
                  -\frac{1}{p(p+p_z)} \big[(p^2 -p_y^2)+p\,p_z\big]\hat{e}_y
                  +\frac{p_y}{p} \hat{e}_z\biggr\}\cdot\vec{p} \notag \\
            &=& \frac{p_x^2 p_y}{p(p+p_z)} -\frac{p_y}{p(p+p_z)} \big[
                        (p^2 -p_y^2)+p\,p_z\big]
                  +\frac{p_y p_z}{p} \notag \\
           & =& \dfrac{p_y}{p} \biggl\{\frac{p_x^2 -\big[
                        (p^2 -p_y^2)+p\,p_z\big]}{(p+p_z)} +p_z\biggr\}
            =\dfrac{p_y}{p} \biggl[-\frac{(p_z^2 +p\,p_z)}{(p+p_z)} +p_z\biggr]
                  \notag \\
            &=& 0,
      \end{eqnarray}
      \begin{eqnarray}
            \vec{F}_1 \cdot\vec{F}_2
            &=& \biggl\{\frac{1}{p(p+p_z)} \big[(p^2 -p_x^2)+p\,p_z\big]
                        \hat{e}_x
                  -\frac{p_x p_y}{p(p+p_z)} \hat{e}_y
                  -\frac{p_x}{p} \hat{e}_z
                  \biggr\}\nonumber\\
            && \;\;\;\;\cdot\biggl\{\frac{p_x p_y}{p(p+p_z)} \hat{e}_x
                        -\frac{1}{p(p+p_z)} \big[(p^2 -p_y^2)+p\,p_z
                              \big]\hat{e}_y
                        +\frac{p_y}{p} \hat{e}_z\biggr\} \notag \\
            &=& \frac{1}{p(p+p_z)} \big[(p^2 -p_x^2)+p\,p_z\big]
                        \frac{p_x p_y}{p(p+p_z)}
                  +\frac{p_x p_y}{p(p+p_z)} \frac{1}{p(p+p_z)} \big[
                        (p^2 -p_y^2)+p\,p_z\big]
                  -\frac{p_x p_y}{p^2} \notag \\
            &=& \frac{p_x p_y}{p^2}\biggl\{\frac{\big[(p^2 -p_x^2)+p\,p_z\big]}{
                        (p+p_z)^2}
                  +\frac{\big[(p^2 -p_y^2)+p\,p_z\big]}{(p+p_z)^2}
                  -1\biggr\}\nonumber\\
            &=& \frac{p_x p_y}{p^2}\biggl[\frac{p^2 +p_z^2 +2\,p\,p_z}{(p+p_z)^2}
                  -1\biggr]
            =\frac{p_x p_y}{p^2} (1-1) \notag \\
            &=& 0.
      \end{eqnarray}
      Thus, three vectors $\{\vec{F}_1, \vec{F}_2, \hat{p}\}$ are mutually perpendicular, i.e.,
      \begin{eqnarray}
            && \vec{F}_1 \perp \vec{F}_2,  \;\;\;\;\;  \vec{F}_1 \perp \hat{p},  \;\;\;\;\; \vec{F}_2 \perp \hat{p},
      \end{eqnarray}
      and
      \begin{eqnarray}
            && \vec{F}_2\times \vec{F}_1=\hat{p}, \;\;\;\;\;  \vec{F}_1\times \hat{p}=\vec{F}_2,  \;\;\;\;\; \hat{p} \times \vec{F}_2=\vec{F}_1.
      \end{eqnarray}
      $\blacksquare$
      \end{remark}

      \begin{remark}
      By the same token, we have
            \begin{eqnarray}
            && \langle\Psi_2|\vec{\alpha}(0)|\Psi_3\rangle \nonumber\\
            &=& \left(\dfrac{1}{2\sqrt{E_{\rm p}\,p(p+p_z)}}\right)^2 \left[
                  u_+ \Bigl(-(p_-)^\dagger ,p+p_z\Bigr),
                  -\dfrac{c\,p}{u_+}\Bigl(-(p_-)^\dagger ,p+p_z\Bigr)
                  \right]\vec{\alpha}(0)\left[u_- \Bigl(p+p_z ,\ p_+\Bigr),
                        -\dfrac{c\,p}{u_-} \Bigl(p+p_z ,p_+\Bigr)\right]^{\rm T}
                  \nonumber \\
            &=& \dfrac{1}{4\,E_{\rm p}\,p(p+p_z)} \left[
                  u_+ \Bigl(-p_+ ,p+p_z\Bigr),-\dfrac{c\,p}{u_+}\Bigl(-p_+ ,p+p_z\Bigr)
                  \right]\vec{\alpha}(0)\left[u_- \bigl(p+p_z ,\ p_+\bigr),
                        -\dfrac{c\,p}{u_-} \bigl(p+p_z ,p_+\bigr)\right]^{\rm T}
                  \nonumber\\
            &=& \dfrac{1}{4\,E_{\rm p}\,p(p+p_z)} \left[
                  u_+ \bigl(-p_+ ,p+p_z\bigr),-\dfrac{c\,p}{u_+}\bigl(-p_+ ,p+p_z\bigr)
                  \right]\left(\begin{array}{cc}
                              0 & \vec{\sigma} \\
                              \vec{\sigma} & 0 \\
                        \end{array}\right)\left[u_- \bigl(p+p_z ,\ p_+\bigr),
                              -\dfrac{c\,p}{u_-} \bigl(p+p_z ,p_+\bigr)\right]^{\rm T}
                  \nonumber\\
            &=& \dfrac{1}{4\,E_{\rm p}\,p(p+p_z)} \left[
                  -\dfrac{c\,p}{u_+}\bigl(-p_+ ,p+p_z\bigr)\vec{\sigma},
                  u_+ \bigl(-p_+ ,p+p_z\bigr)\vec{\sigma}
                  \right]\left[u_- \bigl(p+p_z ,\ p_+\bigr),
                        -\dfrac{c\,p}{u_-} \bigl(p+p_z ,p_+\bigr)\right]^{\rm T}
                  \nonumber\\
            &=& \dfrac{1}{4\,E_{\rm p}\,p(p+p_z)} \left\{
                  -\dfrac{c\,p}{u_+} \bigl(-p_+ ,p+p_z\bigr)\vec{\sigma}\,u_- \bigl(
                        p+p_z ,\ p_+\bigr)^{\rm T}
                  -u_+ \bigl(-p_+ ,p+p_z\bigr)\vec{\sigma}\dfrac{c\,p}{u_-} \bigl(
                        p+p_z ,p_+\bigr)^{\rm T}\right\} \nonumber\\
            &=& -\dfrac{c\,p}{4\,E_{\rm p}\,p(p+p_z)} \left[
                  \dfrac{u_-}{u_+} \bigl(-p_+ ,p+p_z\bigr)\vec{\sigma}\bigl(
                        p+p_z ,\ p_+\bigr)^{\rm T}
                  +\dfrac{u_+}{u_-} \bigl(-p_+ ,p+p_z\bigr)\vec{\sigma}\bigl(
                        p+p_z ,\ p_+\bigr)^{\rm T}\right]\nonumber\\
            &=& -\dfrac{c\,p}{4\,E_{\rm p}\,p(p+p_z)} \left(\dfrac{u_-}{u_+}
                  +\dfrac{u_+}{u_-}\right)\bigl(-p_+ ,p+p_z\bigr)\vec{\sigma}\bigl(
                        p+p_z ,\ p_+\bigr)^{\rm T} \nonumber\\
            &=& -\dfrac{c\,p}{4\,E_{\rm p}\,p(p+p_z)} \left(
                  \dfrac{u_+^2 +u_-^2}{u_+ u_-}\right)\bigl(-p_+ ,p+p_z\bigr)
                        \vec{\sigma}\bigl(p+p_z ,\ p_+\bigr)^{\rm T} \nonumber\\
            &=& -\dfrac{c\,p}{4\,E_{\rm p}\,p(p+p_z)} \left[
                  \dfrac{(\sqrt{E_{\rm p} + m\,c^2})^2 +(\sqrt{E_{\rm p} -m\,c^2})^2}{
                        \sqrt{E_{\rm p} +m\,c^2}\sqrt{E_{\rm p} -m\,c^2}}\right]\bigl(
                              -p_+ ,p+p_z\bigr)\vec{\sigma}\bigl(p+p_z ,p_+
                              \bigr)^{\rm T} \nonumber\\
            &=& -\dfrac{c\,p}{4\,E_{\rm p}\,p(p+p_z)} \dfrac{2\,E_{\rm p}}{c\,p} \bigl(
                  -p_+ ,p+p_z\bigr)\vec{\sigma}\bigl(p+p_z ,p_+\bigr)^{\rm T}
                  \nonumber\\
            &=& -\dfrac{1}{2\,p} \dfrac{1}{(p+p_z)} \bigl(-p_+ ,p+p_z\bigr)\vec{\sigma}
                  \bigl(p+p_z ,p_+\bigr)^{\rm T}.
      \end{eqnarray}
      Because
      \begin{eqnarray}
             \dfrac{1}{(p+p_z)} \bigl(-p_+ ,p+p_z\bigr)\sigma_x \bigl(
                        p+p_z ,p_+\bigr)^{\rm T}
                 & =&\frac{1}{p+p_z} \bigl(p+p_z ,-p_+\bigr)\bigl(p+p_z ,p_+
                        \bigr)^{\rm T}
                  =\frac{1}{p+p_z} \big[(p+p_z)^2 -p_+^2\big] \notag \\
                  &=& \frac{1}{p+p_z} \big[(p+p_z)^2 -(p_x +{\rm i}\,p_y)^2\big]\nonumber\\
                  &=&\frac{1}{p+p_z} \big[(p^2 +p_z^2 +2\,p\,p_z)-(
                        p_x^2 -p_y^2 +{\rm i}\,2\,p_x p_y)\big] \notag \\
                  &=& \frac{1}{p+p_z} \big[(p^2 -p_x^2)+(p_y^2 +p_z^2)+2(p\,p_z
                         -{\rm i}\,p_x p_y)\big], \nonumber \\
                  &=& \frac{2}{p+p_z} \big[(p^2 -p_x^2)+(p\,p_z -{\rm i}\,p_x p_y)
                        \big], \nonumber \\
             \dfrac{1}{(p+p_z)} \bigl(-p_+ ,p+p_z\bigr)\sigma_y \bigl(
                        p+p_z ,p_+\bigr)^{\rm T}
                 & =&\frac{1}{p+p_z} \bigl[{\rm i}(p+p_z),{\rm i}\,p_+\bigr]\bigl(
                        p+p_z ,p_+\bigr)^{\rm T}
                  =\frac{\rm i}{p+p_z} \left[(p+p_z)^2 +p_+^2\right] \notag \\
                  &=& \frac{\rm i}{p+p_z} \big[(p+p_z)^2 +(p_x +{\rm i}\,p_y)^2\big]\nonumber\\
                  &=&\frac{\rm i}{p+p_z} \big[(p^2 +p_z^2 +2\,p\,p_z)+(
                        p_x^2 -p_y^2 +{\rm i}\,2\,p_x p_y)\big] \notag \\
                  &=& {\rm i}\frac{2}{p+p_z} \big[(p^2 -p_y^2)+(p\,p_z
                        +{\rm i}\,p_x p_y)\big], \nonumber \\
             \dfrac{1}{(p+p_z)} \bigl(-p_+ ,p+p_z\bigr)\sigma_z \bigl(
                        p+p_z ,p_+\bigr)^{\rm T}
                 & =& \frac{1}{p+p_z} \bigl[-p_+ ,-(p+p_z)\bigr]\bigl(
                        p+p_z ,p_+\bigr)^{\rm T}
                  =-\frac{2}{p+p_z} p_+ (p+p_z)=-2\,p_+ \notag \\
                  &=& -2(p_x +{\rm i}\,p_y),
      \end{eqnarray}
      then we have
      \begin{eqnarray}
            && \langle\Psi_2|\vec{\alpha}(0)|\Psi_3\rangle\nonumber\\
            &=&-\dfrac{1}{2\,p} \dfrac{1}{(p+p_z)} \bigl(-p_+ ,p+p_z\bigr)\vec{\sigma}
                  \bigl(p+p_z ,p_+\bigr)^{\rm T} \notag \\
            &=& -\dfrac{1}{2\,p} \bigg\{\frac{2}{p+p_z} \big[(p^2 -p_x^2)+(
                        p\,p_z -{\rm i}\,p_x p_y)\big]\hat{e}_x
                  +{\rm i}\frac{2}{p+p_z} \big[(p^2 -p_y^2)+(p\,p_z +{\rm i}\,p_x p_y)
                        \big]\hat{e}_y
                  -2(p_x +{\rm i}\,p_y)\hat{e}_z\bigg\} \notag \\
            &=& -\frac{1}{p(p+p_z)} \big[(p^2 -p_x^2)+(
                        p\,p_z -{\rm i}\,p_x p_y)\big]\hat{e}_x
                  -\frac{\rm i}{p(p+p_z)} \big[(p^2 -p_y^2)+(p\,p_z
                        +{\rm i}\,p_x p_y)\big]\hat{e}_y
                  +\dfrac{1}{p} (p_x +{\rm i}\,p_y)\hat{e}_z \notag \\
            &=& -\left[\langle\Psi_1|\vec{\alpha}(0)|\Psi_4\rangle\right]^*.
      \end{eqnarray}
      i.e.,
      \begin{eqnarray}
            && \langle\Psi_2|\vec{\alpha}(0)|\Psi_3\rangle=-\left( \vec{F}_1 +{\rm i}\,\vec{F}_2\right)^*
            .
      \end{eqnarray}
      $\blacksquare$
      \end{remark}

\begin{remark}We can summarize the above results as the following Table \ref{tab:pz0} and Table \ref{tab:pz1}:
\begin{table}[h]
	\centering
\caption{The value of $\langle\Psi_j|\vec{\alpha}(0)|\Psi_k \rangle$, where $|\Psi_i\rangle$'s  ($i=1, 2, 3, 4$) are four eigenstates of Dirac's electron, and $|\Psi_j\rangle$ and $|\Psi_k\rangle$ correspond to different energies.}
\begin{tabular}{lllll}
\hline\hline
 & $|\Psi_1\rangle$ &  $|\Psi_2\rangle$& $|\Psi_3\rangle$ & $|\Psi_4\rangle$ \\
  \hline
$\langle \Psi_1| \vec{\alpha}(0)$ \;\;\;\quad& ---& --- & $\dfrac{-m\,c^2}{E_{\rm p}}\, \hat{p}$  \;\;\;\quad& $(\vec{F}_1 +{\rm i}\,\vec{F}_2)$  \\
 \hline
$\langle \Psi_2| \vec{\alpha}(0)$\;\;\;\quad &--- &--- & $-(\vec{F}_1 +{\rm i}\,\vec{F}_2)^* $ \quad\quad& $\dfrac{-m\,c^2}{E_{\rm p}}\, \hat{p}$ \\
 \hline
 $\langle \Psi_3| \vec{\alpha}(0)$\;\;\;\quad & $\dfrac{-m\,c^2}{E_{\rm p}}\,\hat{p} \quad$ \;\;\;& $-(\vec{F}_1 +{\rm i}\,\vec{F}_2) \quad\quad$ & --- & --- \\
 \hline
 $\langle \Psi_4| \vec{\alpha}(0)$ \;\;\;\quad& $(\vec{F}_1 +{\rm i}\,\vec{F}_2)^* \quad$ \quad&$\dfrac{-m\,c^2}{E_{\rm p}}\, \hat{p}$ \;\;\;& --- & --- \\
 \hline\hline
\end{tabular}\label{tab:pz0}
\end{table}

  \begin{table}[h]
	\centering
\caption{The results of ``position Zitterbewegung'' of Dirac's electron. In this case, the ``position Zitterbewegung'' operator reads $\hat{\mathcal{Z}}_{\rm e}^r \equiv \hat{\mathcal{Z}}_{\rm e}= \frac{{\rm i}\hbar c}{2}\left[\vec{\alpha}(0)-c H^{-1}_{\rm e}\vec{p}\right] H^{-1}_{\rm e} \left(
                        {\rm e}^{\frac{-{\rm i}\,2\,H_{\rm e} t}{\hbar}} {-1}\right)$,  and $\langle\Psi_j|\hat{\mathcal{Z}}_{\rm e}^r|\Psi_j \rangle
            =0$ $(j=1,2,3,4)$,
            $\langle\Psi_k|\hat{\mathcal{Z}}_{\rm e}^r|\Psi_l \rangle
            =\dfrac{-{\rm i}\hbar c}{2 E_{\rm p}} \left({\rm e}^{\frac{{\rm i}\,2\,E_{\rm p} t}{\hbar}} {-1} \right)\langle\Psi_k|\vec{\alpha}(0)|\Psi_l \rangle=\Delta_1\, \langle\Psi_k|\vec{\alpha}(0)|\Psi_l \rangle$ for $k\in\{1,2\}$, $l\in\{3,4\}$, and $\Delta_1=\dfrac{-{\rm i}\hbar c}{2 E_{\rm p}} \left({\rm e}^{\frac{{\rm i}\,2\,E_{\rm p} t}{\hbar}} {-1} \right)$ [Note: In this work, we use $\hat{\mathcal{Z}}_{\rm e}^r$ and $\hat{\mathcal{Z}}_{\rm e}^{\rm s}$ to denote the ``position Zitterbewegung'' operator and the ``spin Zitterbewegung'' operator, respectively].}
\begin{tabular}{lllll}
\hline\hline
 & $|\Psi_1\rangle$ &  $|\Psi_2\rangle$& $|\Psi_3\rangle$ & $|\Psi_4\rangle$ \\
  \hline
$\langle \Psi_1| \hat{\mathcal{Z}}_{\rm e}^r$ \;\;\;\quad& 0&0 & $\Delta_1\,\dfrac{-m\,c^2}{E_{\rm p}}\, \hat{p}$  \;\;\;\quad& $\Delta_1\,(\vec{F}_1 +{\rm i}\,\vec{F}_2)$  \\
 \hline
$\langle \Psi_2| \hat{\mathcal{Z}}_{\rm e}^r$\;\;\;\quad &0 &0 & $-\Delta_1\,(\vec{F}_1 +{\rm i}\,\vec{F}_2)^* $ \quad\quad& $\Delta_1\,\dfrac{-m\,c^2}{E_{\rm p}}\, \hat{p}$ \\
 \hline
 $\langle \Psi_3| \,\hat{\mathcal{Z}}_{\rm e}^r$\;\;\;\quad & $\Delta_1^*\,\dfrac{-m\,c^2}{E_{\rm p}}\,\hat{p} \quad$ \;\;\;& $-\Delta_1^*\,(\vec{F}_1 +{\rm i}\,\vec{F}_2) \quad\quad$ & 0 & 0 \\
 \hline
 $\langle \Psi_4| \hat{\mathcal{Z}}_{\rm e}^r$ \;\;\;\quad& $\Delta_1^*\,(\vec{F}_1 +{\rm i}\,\vec{F}_2)^* \quad$ \quad&$\Delta_1^*\,\dfrac{-m\,c^2}{E_{\rm p}}\, \hat{p}$ \;\;\;& 0 & 0 \\
 \hline\hline
\end{tabular}\label{tab:pz1}
\end{table}

By the way, based on $\langle\Psi_j|\hat{\mathcal{Z}}_{\rm e}^r|\Psi_j \rangle=0$, one can have $\langle\Psi_j|\left[\vec{\alpha}(0)-c H^{-1}_{\rm e}\vec{p}\right] |\Psi_j \rangle=0$, thus
 \begin{eqnarray}
     &&  \langle\Psi_j|\vec{\alpha}(0) |\Psi_j \rangle=\langle\Psi_j|\left[-c H^{-1}_{\rm e}\vec{p}\right] |\Psi_j \rangle, \;\;\;\; (j=1, 2, 3, 4).
 \end{eqnarray}
Explicitly, one has
 \begin{eqnarray}\label{eq:alpha-jk-1}
     &&  \langle\Psi_1|\vec{\alpha}(0) |\Psi_1 \rangle=\langle\Psi_2|\vec{\alpha}(0) |\Psi_2 \rangle=-\frac{cp}{E_{\rm p}}\, \hat{p}, \;\;\;  \langle\Psi_3|\vec{\alpha}(0) |\Psi_3 \rangle=\langle\Psi_4|\vec{\alpha}(0) |\Psi_4 \rangle=\frac{cp}{E_{\rm p}}\, \hat{p}.
 \end{eqnarray}
Due to $\Pi_+ \left[\vec{\alpha}(0)-c H^{-1}_{\rm e}\vec{p}\right]\Pi_+=0$ and $\Pi_- \left[\vec{\alpha}(0)-c H^{-1}_{\rm e}\vec{p}\right]\Pi_- =0$, one can similarly have
 \begin{eqnarray}\label{eq:alpha-jk-2}
     &&  \langle\Psi_1|\vec{\alpha}(0) |\Psi_2 \rangle=\langle\Psi_2|\vec{\alpha}(0) |\Psi_1 \rangle=0, \;\;\;  \langle\Psi_3|\vec{\alpha}(0) |\Psi_4 \rangle=\langle\Psi_4|\vec{\alpha}(0) |\Psi_3 \rangle=0.
 \end{eqnarray}
$\blacksquare$
\end{remark}

      \begin{remark}
      In general, we have the quantum state of Dirac's electron as
      \begin{eqnarray}
            |\Psi\rangle=c_1 |\Psi_1\rangle+c_2 |\Psi_2\rangle+c_3 |\Psi_3\rangle
                  +c_4 |\Psi_4\rangle,
      \end{eqnarray}
      where $c_j$'s $(j=1, 2, 3, 4)$ are some coefficients that satisfy the normalization condition
      \begin{eqnarray}
      |c_1|^2 +|c_2|^2 +|c_3|^2 +|c_4|^2 =1.
      \end{eqnarray}
      From Table \ref{tab:pz1} we have
      \begin{eqnarray}
            && \mathcal{Z}_{\rm e} =\bigl(c_1^* \Bra{\Psi_1}+c_2^* \Bra{\Psi_2}
                  +c_3^* \Bra{\Psi_3}+c_4^* \Bra{\Psi_4}
                  \bigr)\hat{\mathcal{Z}}_{\rm e} \bigl(c_1 |\Psi_1\rangle
                        +c_2 |\Psi_2\rangle+c_3 |\Psi_3\rangle
                        +c_4 |\Psi_4\rangle\bigr) \notag \\
            &=& |c_1|^2 \Bra{\Psi_1}\hat{\mathcal{Z}}_{\rm e}\Ket{\Psi_1}
                  +|c_2|^2 \Bra{\Psi_2}\hat{\mathcal{Z}}_{\rm e}\Ket{\Psi_2}
                  +|c_3|^2 \Bra{\Psi_3}\hat{\mathcal{Z}}_{\rm e}\Ket{\Psi_3}
                  +|c_4|^2 \Bra{\Psi_4}\hat{\mathcal{Z}}_{\rm e}\Ket{\Psi_4} \notag \\
                  && +c_1^* c_2 \Bra{\Psi_1}\hat{\mathcal{Z}}_{\rm e}\Ket{\Psi_2}
                  +c_2^* c_1 \Bra{\Psi_2}\hat{\mathcal{Z}}_{\rm e}\Ket{\Psi_1}
                  +c_3^* c_4 \Bra{\Psi_3}\hat{\mathcal{Z}}_{\rm e}\Ket{\Psi_4}
                  +c_4^* c_3 \Bra{\Psi_4}\hat{\mathcal{Z}}_{\rm e}\Ket{\Psi_3} \notag
                        \\
                  && +c_1^* c_3 \Bra{\Psi_1}\hat{\mathcal{Z}}_{\rm e}\Ket{\Psi_3}
                        +c_1^* c_4 \Bra{\Psi_1}\hat{\mathcal{Z}}_{\rm e}\Ket{\Psi_4}
                        +c_2^* c_3 \Bra{\Psi_2}\hat{\mathcal{Z}}_{\rm e}\Ket{\Psi_3}
                        +c_2^* c_4 \Bra{\Psi_2}\hat{\mathcal{Z}}_{\rm e}\Ket{\Psi_4}
                              \notag \\
                        && +c_3^* c_1 \Bra{\Psi_3}\hat{\mathcal{Z}}_{\rm e}\Ket{\Psi_1}
                        +c_3^* c_2 \Bra{\Psi_3}\hat{\mathcal{Z}}_{\rm e}\Ket{\Psi_2}
                        +c_4^* c_1 \Bra{\Psi_4}\hat{\mathcal{Z}}_{\rm e}\Ket{\Psi_1}
                        +c_4^* c_2 \Bra{\Psi_4}\hat{\mathcal{Z}}_{\rm e}\Ket{\Psi_2}
                  \notag \\
            &=& c_1^* c_3 \Bra{\Psi_1}\hat{\mathcal{Z}}_{\rm e}\Ket{\Psi_3}
                  +c_1^* c_4 \Bra{\Psi_1}\hat{\mathcal{Z}}_{\rm e}\Ket{\Psi_4}
                  +c_2^* c_3 \Bra{\Psi_2}\hat{\mathcal{Z}}_{\rm e}\Ket{\Psi_3}
                  +c_2^* c_4 \Bra{\Psi_2}\hat{\mathcal{Z}}_{\rm e}\Ket{\Psi_4}
                  +{\rm c.c.} \notag \\
            &=& \frac{-{\rm i}\hbar c}{2 E_{\rm p}} \left({\rm e}^{
                        \frac{{\rm i}\,2\,E_{\rm p} t}{\hbar}} -1
                        \right)\big[c_1^* c_3 \Bra{\Psi_1}\vec{\alpha}(0)\Ket{\Psi_3}
                              +c_2^* c_4 \Bra{\Psi_2}\vec{\alpha}(0)\Ket{\Psi_4}
                              +c_1^* c_4 \Bra{\Psi_1}\vec{\alpha}(0)\Ket{\Psi_4}
                              +c_2^* c_3 \Bra{\Psi_2}\vec{\alpha}(0)\Ket{\Psi_3}\big]
                  +{\rm c.c.} \notag \\
            &=& \frac{-{\rm i}\hbar c}{2 E_{\rm p}} \left({\rm e}^{
                  \frac{{\rm i}\,2\,E_{\rm p} t}{\hbar}} -1
                  \right)\Big\{(c_1^* c_3 +c_2^* c_4)\Bra{\Psi_1}\vec{\alpha}(0)
                              \Ket{\Psi_3}
                        +c_1^* c_4 \Bra{\Psi_1}\vec{\alpha}(0)\Ket{\Psi_4}
                        -c_2^* c_3 \big[\Bra{\Psi_1}\vec{\alpha}(0)\Ket{\Psi_4}\big]^*
                        \Big\}+{\rm c.c.} \notag \\
            &=& \frac{-{\rm i}\hbar c}{2 E_{\rm p}} \left({\rm e}^{
                  \frac{{\rm i}\,2\,E_{\rm p} t}{\hbar}} -1
                  \right)\Big\{-(c_1^* c_3 +c_2^* c_4)\dfrac{m\,c^2}{E_{\rm p}} \hat{p}
                        +c_1^* c_4 \Bra{\Psi_1}\vec{\alpha}(0)\Ket{\Psi_4}
                        -c_2^* c_3 \big[\Bra{\Psi_1}\vec{\alpha}(0)\Ket{\Psi_4}\big]^*
                        \Big\}+{\rm c.c.} \notag \\
            &=& 2\, {\rm Re} \left[\frac{-{\rm i}\hbar c}{2 E_{\rm p}} \left({\rm e}^{
                  \frac{{\rm i}\,2\,E_{\rm p} t}{\hbar}} -1
                  \right)\Big\{-(c_1^* c_3 +c_2^* c_4)\dfrac{m\,c^2}{E_{\rm p}} \hat{p}
                        +c_1^* c_4 \left( \vec{F}_1 +{\rm i}\,\vec{F}_2\right)
                        -c_2^* c_3 \left( \vec{F}_1 -{\rm i}\,\vec{F}_2\right)
                        \Big\}\right].
      \end{eqnarray}
      $\blacksquare$
      \end{remark}

\section{Position Zitterbewegung for $H_{\rm b}^{\rm I}$}

            The Hamiltonian of the type-I Dirac's braidon reads
            \begin{eqnarray}\label{eq:E-1}
                  H_{\rm b}^{\rm I} =-m\,c^2\;\vec{\alpha}\cdot\hat{p}+\beta\,p\,c.
            \end{eqnarray}
            One has
            \begin{eqnarray}\label{eq:E-2}
                  &&\dfrac{{\rm d}\,\vec{p}}{{\rm d}\,t} =\dfrac{1}{{\rm i}\hbar}
                  \left[\vec{p},\ H_{\rm b}^{\rm I}\right]=0,\nonumber\\
                  &&\dfrac{{\rm d}\,H_{\rm b}^{\rm I}}{{\rm d}\,t} =\dfrac{1}{{\rm i}\hbar}
                  \left[H_{\rm b}^{\rm I},\ H_{\rm b}^{\rm I}\right]=0,
            \end{eqnarray}
            which means that $\vec{p}$ and $H_{\rm b}^{\rm I}$ are conservation quantities.

            For the position operator $\vec{r}$, we have
            \begin{eqnarray}\label{eq:E-4}
            &&  \frac{1}{{\rm i}\hbar}\left[\vec{r}, \vec{\alpha}\cdot\hat{p}\right]=  \frac{1}{p} \vec{\alpha} -  \left(\vec{\alpha}\cdot\vec{p}\right) \dfrac{\vec{p}}{p^3},\nonumber\\
            &&  \frac{1}{{\rm i}\hbar}\left[\vec{r}, \beta p\right]=  \beta\, \dfrac{\vec{p}}{p},
            \end{eqnarray}
            which leads to
            \begin{eqnarray}\label{eq:E-7}
                        \dfrac{{\rm d}\,\vec{r}}{{\rm d}\,t}& =& \dfrac{1}{{\rm i}\hbar}
                        \left[\vec{r}, H_{\rm b}^{\rm I}\right]\nonumber\\
                         &=& -m c^2\;\left(\frac{1}{p} \vec{\alpha} -  \left(\vec{\alpha}\cdot\vec{p}\right) \dfrac{\vec{p}}{p^3}\right)+
                        c \beta \, \dfrac{\vec{p}}{p}.
            \end{eqnarray}
            For the the operator $\vec{\alpha}$, we have
            \begin{eqnarray}
                        \dfrac{{\rm d}\,\vec{\alpha}}{{\rm d}\,t}
                       & =&\dfrac{1}{{\rm i}\hbar} \left[\vec{\alpha},\ H_{\rm b}^{\rm I}\right]\nonumber\\
                        &=&\dfrac{1}{{\rm i}\hbar} \Bigl(\{\vec{\alpha},\ H_{\rm b}^{\rm I}\}
                              -2\,H_{\rm b}^{\rm I}\,\vec{\alpha}\Bigr)\nonumber\\
                       &=&-\dfrac{2}{{\rm i}\hbar} \left(m\,c^2\,\hat{p}
                              +H_{\rm b}^{\rm I}\,\vec{\alpha}\right)\nonumber\\
                        &=&-\dfrac{2}{{\rm i}\hbar}\,H_{\rm b}^{\rm I} \left[
                        m\,c^2 (H_{\rm b}^{\rm I})^{-1}\,\hat{p}+\vec{\alpha}\right],
            \end{eqnarray}
            then we have
            \begin{eqnarray}
                  \frac{{\rm d}\,\left[\vec{\alpha}
                        +m\,c^2 (H_{\rm b}^{\rm I})^{-1}\,\hat{p}\right]}{{\rm d}\,t}
                  =-\frac{2\,H_{\rm b}^{\rm I}}{{\rm i}\hbar}\left[\vec{\alpha}
                        +m\,c^2 (H_{\rm b}^{\rm I})^{-1}\,\hat{p}\right],
            \end{eqnarray}
            i.e.,
            \begin{eqnarray}
                  \vec{\alpha}(t)+m\,c^2 (H_{\rm b}^{\rm I})^{-1}\,\hat{p}
                  ={{\rm e}^{\frac{{\rm i}\,2\,H_{\rm b}^{\rm I}\,t}{\hbar}}
                  \left[\vec{\alpha}(0)+m\,c^2 (H_{\rm b}^{\rm I})^{-1}\,\hat{p}\right]},
            \end{eqnarray}
            i.e.,
            \begin{eqnarray}
                  \vec{\alpha}(t)={
                  {\rm e}^{\frac{{\rm i}\,2H_{\rm b}^{\rm I}\,t}{\hbar}}
                        \left[\vec{\alpha}(0)+m\,c^2 (H_{\rm b}^{\rm I})^{-1}\,\hat{p}\right]
                  -m\,c^2 (H_{\rm b}^{\rm I})^{-1}\,\hat{p}}.
            \end{eqnarray}
            One can check that
            \begin{eqnarray}
                 \Bigl\{H_{\rm b}^{\rm I},\ \left[
                  \vec{\alpha}(0)+m\,c^2 (H_{\rm b}^{\rm I})^{-1}\,\hat{p}\right]\Bigr\}
                                   &=& [H_{\rm b}^{\rm I} \vec{\alpha}(0)+\vec{\alpha}(0)H_{\rm b}^{\rm I}]
                  +2\,m\,c^2 \hat{p} \nonumber \\
                  &=& [H_{\rm b}^{\rm I} \vec{\alpha}+\vec{\alpha} H_{\rm b}^{\rm I}]
                  +2\,m\,c^2 \hat{p} \nonumber \\
                  &=& -m\,c^2 [(\vec{\alpha}\cdot\hat{p})\vec{\alpha}
                        +\vec{\alpha}(\vec{\alpha}\cdot\hat{p})]
                  +2\,m\,c^2 \hat{p} \nonumber \\
                  &=& -2\,m\,c^2 \hat{p}+2\,m\,c^2 \hat{p} \nonumber \\
                &=& 0,
            \end{eqnarray}
            thus
            \begin{eqnarray}
                  H_{\rm b}^{\rm I} \left[\vec{\alpha}(0)
                  +m\,c^2 (H_{\rm b}^{\rm I})^{-1}\,\hat{p}\right]
                  =-\left[\vec{\alpha}(0)+m\,c^2 (H_{\rm b}^{\rm I})^{-1}\,\hat{p}\right]
                  H_{\rm b}^{\rm I},
            \end{eqnarray}
            which leads to
            \begin{eqnarray}
                &&  {{\rm e}^{\frac{{\rm i}\,2\,H_{\rm b}^{\rm I}\,t}{\hbar}}
                  \left[\vec{\alpha}(0)+m\,c^2 (H_{\rm b}^{\rm I})^{-1}\,\hat{p}\right]}= {\left[\vec{\alpha}(0)
                        +m\,c^2 (H_{\rm b}^{\rm I})^{-1}\,\hat{p}\right]
                  {\rm e}^{\frac{-{\rm i}\,2\,H_{\rm b}^{\rm I}\,t}{\hbar}}}.
            \end{eqnarray}
            Further, we obtain
            \begin{eqnarray}\label{eq:E-8}
                  \vec{\alpha}(t)=\left[\vec{\alpha}(0)
                        +m\,c^2 (H_{\rm b}^{\rm I})^{-1}\,\hat{p}\right]
                        {\rm e}^{\frac{-{\rm i}\,2\,H_{\rm b}^{\rm I}\,t}{\hbar}}
                  -m\,c^2 (H_{\rm b}^{\rm I})^{-1}\,\hat{p},
            \end{eqnarray}
            and
            \begin{eqnarray}
                  \vec{\alpha}(t)\cdot\vec{p}=\left[\vec{\alpha}(0)\cdot\vec{p}+m\,c^2 (H_{\rm b}^{\rm I})^{-1}\,p\right]
                        {\rm e}^{\frac{-{\rm i}\,2\,H_{\rm b}^{\rm I}\,t}{\hbar}}
                  -m\,c^2 (H_{\rm b}^{\rm I})^{-1}\,p.
            \end{eqnarray}

            Likewise, for the the operator $\beta$, we have
            \begin{eqnarray}
                  \dfrac{{\rm d}\,\beta}{{\rm d}\,t}
                  &=&\dfrac{1}{{\rm i}\hbar} [\beta,\ H_{\rm b}^{\rm I}]
                  =\dfrac{1}{{\rm i}\hbar} \Bigl(\{\beta,\ H_{\rm b}^{\rm I}\}
                  -2\,H_{\rm b}^{\rm I}\,\beta\Bigr)\nonumber\\
                  &&=\dfrac{2}{{\rm i}\hbar} \left(c\,p-H_{\rm b}^{\rm I}\,\beta\right)
                  =-\dfrac{2}{{\rm i}\hbar} H_{\rm b}^{\rm I} \left(\beta
                  -c\,(H_{\rm b}^{\rm I})^{-1}\,p\right).
            \end{eqnarray}
            Then we have
            \begin{eqnarray}
                  \beta(t)-c\,(H_{\rm b}^{\rm I})^{-1}\,p
                  ={\rm e}^{\frac{{\rm i}\,2H_{\rm b}^{\rm I}\,t}{\hbar}}
                  \left[\beta(0)-c\,(H_{\rm b}^{\rm I})^{-1}\,p\right],
            \end{eqnarray}
            viz.,
            \begin{eqnarray}
                  \beta(t)=c\,(H_{\rm b}^{\rm I})^{-1}\,p
                  +{\rm e}^{\frac{{\rm i}\,2H_{\rm b}^{\rm I}\,t}{\hbar}}
                        \left[\beta(0)-c\,(H_{\rm b}^{\rm I})^{-1}\,p\right].
            \end{eqnarray}
            One can check that
            \begin{eqnarray}
                  \Bigl\{H_{\rm b}^{\rm I},\
                        \left[\beta(0)-c\,(H_{\rm b}^{\rm I})^{-1}\,p\right]\Bigr\}
                  &=& [H_{\rm b}^{\rm I}\,\beta(0)+\beta(0)\,H_{\rm b}^{\rm I} ]
                        -2\,c\,p \nonumber\\
                  &=& [H_{\rm b}^{\rm I}\,\beta+\beta\,H_{\rm b}^{\rm I}]-2\,c\,p \nonumber\\
                  &=& [(c\,\beta\,p)\beta+\beta(c\,\beta\,p)]-2\,c\,p\nonumber\\
                  &=& 2\,c\,p -2\,c\,p\nonumber\\
                  &=& 0,
            \end{eqnarray}
            thus
            \begin{eqnarray}
                  H_{\rm b}^{\rm I} \left[\beta(0)-c\,(H_{\rm b}^{\rm I})^{-1}\,p\right]
                  =-\left[\beta(0)-c\,(H_{\rm b}^{\rm I})^{-1}\,p\right]H_{\rm b}^{\rm I},\nonumber\\
            \end{eqnarray}
            which leads to
            \begin{eqnarray}
                  {\rm e}^{\frac{{\rm i}\,2H_{\rm b}^{\rm I}\,t}{\hbar}}
                        \left[\beta(0)-c\,(H_{\rm b}^{\rm I})^{-1}\,p\right]
                  ={\left[\beta(0)-c\,(H_{\rm b}^{\rm I})^{-1}\,p\right]
                        {\rm e}^{-\frac{{\rm i}\,2H_{\rm b}^{\rm I}\,t}{\hbar}}},\nonumber\\
            \end{eqnarray}
            i.e.,
            \begin{eqnarray}\label{eq:E-8a}
                  \beta(t)
                  =c\,(H_{\rm b}^{\rm I})^{-1}\,p
                  +\left[\beta(0)-c\,(H_{\rm b}^{\rm I})^{-1}\,p\right]
                        {\rm e}^{-\frac{{\rm i}\,2H_{\rm b}^{\rm I}\,t}{\hbar}}.
            \end{eqnarray}

            Based on above results, we have
            \begin{eqnarray}
                  \dfrac{{\rm d}\,\vec{r}}{{\rm d}\,t}
                  &=&\ -m\,c^2 \biggl[\dfrac{\vec{\alpha}}{p}-(\vec{\alpha}
                        \cdot\vec{p})\dfrac{\vec{p}}{p^3}\biggr]+c\,\beta\,\hat{p} \nonumber\\
                 & =&\ -m\,c^2 \Biggl\{
                              \dfrac{\left[\vec{\alpha}(0)
                                    +m\,c^2 (H_{\rm b}^{\rm I})^{-1}\,\hat{p}\right]
                              {\rm e}^{\frac{-{\rm i}\,2\,H_{\rm b}^{\rm I}\,t}{\hbar}}
                                    -m\,c^2 (H_{\rm b}^{\rm I})^{-1}\,\hat{p}}{p} \nonumber\\
                        && \     -\Bigl[\left(\vec{\alpha}(0)\cdot\vec{p}
                                    +m\,c^2 (H_{\rm b}^{\rm I})^{-1}\,p\right)
                                    {\rm e}^{-\frac{-{\rm i}\,2\,H_{\rm b}^{\rm I}\,t}{\hbar}}
                              -m\,c^2 (H_{\rm b}^{\rm I})^{-1}\,p\Bigr]\dfrac{\vec{p}}{p^3}\Biggr\} \nonumber\\
                        &&\ +c\,\Bigl\{c\,(H_{\rm b}^{\rm I})^{-1}\,p
                              +\left[\beta(0)-c\,(H_{\rm b}^{\rm I})^{-1}\,p\right]
                              {\rm e}^{-\frac{{\rm i}\,2H_{\rm b}^{\rm I}\,t}{\hbar}}\Bigr\}\,\hat{p} \nonumber\\
                  &=&\ -m\,c^2 \Biggl\{\dfrac{1}{p}\left[\vec{\alpha}(0)
                                    +m\,c^2 (H_{\rm b}^{\rm I})^{-1}\,\hat{p}\right]
                              {\rm e}^{\frac{-{\rm i}\,2\,H_{\rm b}^{\rm I}\,t}{\hbar}}
                              -\left[\vec{\alpha}(0)\cdot\vec{p}
                                    +m\,c^2 (H_{\rm b}^{\rm I})^{-1}\,p\right]
                                    {\rm e}^{\frac{-{\rm i}\,2\,H_{\rm b}^{\rm I}\,t}{\hbar}}
                              \dfrac{\vec{p}}{p^3}\Biggr\} \nonumber\\
                        &&\ +c\,\Bigl\{c\,(H_{\rm b}^{\rm I})^{-1}\,p
                              +\left[\beta(0)-c\,(H_{\rm b}^{\rm I})^{-1}\,p\right]
                              {\rm e}^{-\frac{{\rm i}\,2H_{\rm b}^{\rm I}\,t}{\hbar}}\Bigr\}\,\hat{p} \nonumber\\
                 & =&\ -m\,c^2 \Biggl\{\dfrac{1}{p} \vec{\alpha}(0)
                              -\Bigl[\vec{\alpha}(0)\cdot\vec{p}\Bigr]
                              \dfrac{\vec{p}}{p^3}\Biggr\}
                        {\rm e}^{\frac{-{\rm i}\,2\,H_{\rm b}^{\rm I}\,t}{\hbar}}
                        +c^2 (H_{\rm b}^{\rm I})^{-1}\,\vec{p}+\left[c\,\beta(0)\,\hat{p}
                              -c^2\,(H_{\rm b}^{\rm I})^{-1}\,\vec{p}\right]
                        {\rm e}^{-\frac{{\rm i}\,2H_{\rm b}^{\rm I}\,t}{\hbar}},
            \end{eqnarray}
            which yields
            \begin{eqnarray}\label{eq:E-9}
                  \vec{r}(t) &=& \vec{r}(0)+c^2 (H_{\rm b}^{\rm I})^{-1}\vec{p}\;t
                        +\int\left[c\,\beta(0)\,\hat{p}
                              -c^2\,(H_{\rm b}^{\rm I})^{-1}\,\vec{p}\right]
                        {\rm e}^{-\frac{{\rm i}\,2H_{\rm b}^{\rm I}\,t}{\hbar}}\:{\rm d}\,t
                        -m\,c^2 \int\Biggl\{\dfrac{1}{p} \vec{\alpha}(0)
                              -\Bigl[\vec{\alpha}(0)\cdot\vec{p}\Bigr]
                              \dfrac{\vec{p}}{p^3}\Biggr\}
                        {\rm e}^{\frac{-{\rm i}\,2\,H_{\rm b}^{\rm I}\,t}{\hbar}}
                        {\rm d}\,t\nonumber \\
                  &=& \vec{r}(0)+c^2 H^{-1}_{\rm b}\vec{p}\;t
                        +\left[c\,\beta(0)\,\hat{p}
                              -c^2\,(H_{\rm b}^{\rm I})^{-1}\,\vec{p}\right]
                        \int{\rm e}^{-\frac{{\rm i}\,2H_{\rm b}^{\rm I}\,t}{\hbar}}\:
                        {\rm d}\,t
                        -m\,c^2 \Biggl\{\dfrac{1}{p} \vec{\alpha}(0)
                              -\Bigl[\vec{\alpha}(0)\cdot\vec{p}\Bigr]
                              \dfrac{\vec{p}}{p^3}\Biggr\}
                        \int{\rm e}^{\frac{-{\rm i}\,2\,H_{\rm b}^{\rm I}\,t}{\hbar}}
                        {\rm d}\,t \nonumber \\
                  &=& \vec{r}(0)+c^2 (H_{\rm b}^{\rm I})^{-1}\vec{p}\;t
                        -\left[c\,\beta(0)\,\hat{p}-c^2 (H_{\rm b}^{\rm I})^{-1}\,\vec{p}\right]
                        \frac{\hbar}{{\rm i}\,2H_{\rm b}^{\rm I}} {\left(
                              {\rm e}^{\frac{-{\rm i}\,2H_{\rm b}^{\rm I}\,t}{\hbar}}-1\right)} \nonumber \\
                        && +m\,c^2 \Biggl\{\dfrac{1}{p} \vec{\alpha}(0)
                              -\Bigl[\vec{\alpha}(0)\cdot\vec{p}\Bigr]
                              \dfrac{\vec{p}}{p^3}\Biggr\}
                        \frac{\hbar}{{\rm i}\,2H_{\rm b}^{\rm I}} {\left(
                              {\rm e}^{\frac{-{\rm i}\,2H_{\rm b}^{\rm I}\,t}{\hbar}}-1\right)} \nonumber \\
                  &=& \vec{r}(0)+c^2 (H_{\rm b}^{\rm I})^{-1}\vec{p}\;t
                        +\frac{{\rm i}\hbar}{2} \left[c\,\beta(0)\,\hat{p}
                        -c^2\,(H_{\rm b}^{\rm I})^{-1}\,\vec{p}\right]
                              (H_{\rm b}^{\rm I})^{-1} {\left(
                                    {\rm e}^{\frac{-{\rm i}\,2H_{\rm b}^{\rm I}\,t}{\hbar}}-1\right)}
                                    \nonumber \\
                        && -\frac{{\rm i}\hbar\,m\,c^2 }{2} \Biggl\{
                        \dfrac{1}{p} \vec{\alpha}(0)
                              -\Bigl[\vec{\alpha}(0)\cdot\vec{p}\Bigr]
                              \dfrac{\vec{p}}{p^3}\Biggr\} (H_{\rm b}^{\rm I})^{-1} {\left(
                                    {\rm e}^{\frac{-{\rm i}\,2H_{\rm b}^{\rm I}\,t}{\hbar}}-1\right)} \nonumber \\
                  &=& \vec{r}(0)+c^2 (H_{\rm b}^{\rm I})^{-1}\vec{p}\;t
                        +\frac{{\rm i}\hbar}{2} \Biggl\{\left[c\,\beta(0)\,\hat{p}
                        -c^2\,(H_{\rm b}^{\rm I})^{-1}\,\vec{p}\right]
                        -m\,c^2 \dfrac{1}{p} \vec{\alpha}(0)
                              +m\,c^2 \Bigl[\vec{\alpha}(0)\cdot\vec{p}\Bigr]
                              \dfrac{\vec{p}}{p^3}\Biggr\}(H_{\rm b}^{\rm I})^{-1} {\left(
                                    {\rm e}^{\frac{-{\rm i}\,2H_{\rm b}^{\rm I}\,t}{\hbar}}-1\right)} \nonumber \\
                  &=& \vec{r}(0)+c^2 (H_{\rm b}^{\rm I})^{-1}\vec{p}\;t
                        +\frac{{\rm i}\hbar}{2} \Biggl\{
                        \dfrac{c\,\beta(0)\,\vec{p}-m\,c^2 \vec{\alpha}(0)}{p}
                        +m\,c^2 \Bigl[\vec{\alpha}(0)\cdot\vec{p}\Bigr]
                              \dfrac{\vec{p}}{p^3}
                        -c^2\,(H_{\rm b}^{\rm I})^{-1}\,\vec{p}\Biggr\}(H_{\rm b}^{\rm I})^{-1}
                        {\left(
                              {\rm e}^{\frac{-{\rm i}\,2H_{\rm b}^{\rm I}\,t}{\hbar}}-1\right)}.
            \end{eqnarray}

            \begin{remark}
            From Eq. (\ref{eq:E-1}) one obtains
            \begin{eqnarray}\label{eq:E-1a}
            c \beta = \frac{1}{p}\left(H_{\rm b}^{\rm I} +m\,c^2\;\vec{\alpha}\cdot\hat{p}\right).
            \end{eqnarray}
            After substituting Eq. (\ref{eq:E-1a}) into Eq. (\ref{eq:E-7}), we have
            \begin{eqnarray}\label{eq:E-1b}
                        \dfrac{{\rm d}\,\vec{r}}{{\rm d}\,t} &=& -m c^2\;\left(\frac{1}{p} \vec{\alpha} -  \left(\vec{\alpha}\cdot\vec{p}\right) \dfrac{\vec{p}}{p^3}\right)+ c \beta \dfrac{\vec{p}}{p}\nonumber\\
            &=& -m c^2\;\left(\frac{1}{p} \vec{\alpha} -  \left(\vec{\alpha}\cdot\vec{p}\right) \dfrac{\vec{p}}{p^3}\right)+
            \left[\frac{1}{p}\left(H_{\rm b}^{\rm I} +m\,c^2\;\vec{\alpha}\cdot\hat{p}\right)\right] \dfrac{\vec{p}}{p}\nonumber\\
                  &=& -m c^2\;\left(\frac{1}{p} \vec{\alpha} -  \left(\vec{\alpha}\cdot\vec{p}\right) \dfrac{\vec{p}}{p^3}\right)+
                  H_{\rm b}^{\rm I} \dfrac{\vec{p}}{p^2} +m\,c^2\;(\vec{\alpha}\cdot\vec{p}) \dfrac{\vec{p}}{p^3}\nonumber\\
                  &=& -m c^2\;\frac{1}{p} \vec{\alpha} + 2 m\,c^2\;(\vec{\alpha}\cdot\vec{p}) \dfrac{\vec{p}}{p^3}+
                  H_{\rm b}^{\rm I} \dfrac{\vec{p}}{p^2}.
            \end{eqnarray}
            After substituting Eq. (\ref{eq:E-8}) into Eq. (\ref{eq:E-1b}), we have
            \begin{eqnarray}\label{eq:E-1c}
                        \dfrac{{\rm d}\,\vec{r}}{{\rm d}\,t} &=& -m c^2\;\frac{1}{p} \vec{\alpha} + 2 m\,c^2\;(\vec{\alpha}\cdot\vec{p}) \dfrac{\vec{p}}{p^3}+
                  H_{\rm b}^{\rm I} \dfrac{\vec{p}}{p^2}\nonumber\\
                  &=& -m c^2\;\frac{1}{p} \left\{\left[\vec{\alpha}(0)
                        +m\,c^2 (H_{\rm b}^{\rm I})^{-1}\,\hat{p}\right]
                        {\rm e}^{\frac{-{\rm i}\,2\,H_{\rm b}^{\rm I}\,t}{\hbar}}
                  -m\,c^2 (H_{\rm b}^{\rm I})^{-1}\,\hat{p}\right\} \nonumber\\
                  &&+ 2 m\,c^2\;\vec{p}\cdot \left\{\left[\vec{\alpha}(0)
                        +m\,c^2 (H_{\rm b}^{\rm I})^{-1}\,\hat{p}\right]
                        {\rm e}^{\frac{-{\rm i}\,2\,H_{\rm b}^{\rm I}\,t}{\hbar}}
                  -m\,c^2 (H_{\rm b}^{\rm I})^{-1}\,\hat{p}\right\}\dfrac{\vec{p}}{p^3}+
                  H_{\rm b}^{\rm I} \dfrac{\vec{p}}{p^2}\nonumber\\
                  &=& \left(-m^2c^4 (H_{\rm b}^{\rm I})^{-1}+ H_{\rm b}^{\rm I} \right) \dfrac{\vec{p}}{p^2}-m c^2\;\frac{1}{p} \left\{\left[\vec{\alpha}(0)
                        +m\,c^2 (H_{\rm b}^{\rm I})^{-1}\,\hat{p}\right] \right\}
                        {\rm e}^{\frac{-{\rm i}\,2\,H_{\rm b}^{\rm I}\,t}{\hbar}}
                  \nonumber\\
                  &&+ 2 m\,c^2\; \left\{\left[\vec{\alpha}(0)\cdot \vec{p}+m\,c^2 (H_{\rm b}^{\rm I})^{-1}\,p\right]
                  \dfrac{\vec{p}}{p^3}
                  \right\} {\rm e}^{\frac{-{\rm i}\,2\,H_{\rm b}^{\rm I}\,t}{\hbar}}\nonumber\\
                  &=& c^2 (H_{\rm b}^{\rm I})^{-1} \vec{p}+m c^2\;\left\{-\frac{1}{p} \left[\vec{\alpha}(0)
                        +m\,c^2 (H_{\rm b}^{\rm I})^{-1}\,\hat{p}\right]+2\left[\vec{\alpha}(0)\cdot \vec{p}+m\,c^2 (H_{\rm b}^{\rm I})^{-1}\,p\right]
                  \dfrac{\vec{p}}{p^3} \right\}
                        {\rm e}^{\frac{-{\rm i}\,2\,H_{\rm b}^{\rm I}\,t}{\hbar}}\nonumber\\
                        &=& c^2 (H_{\rm b}^{\rm I})^{-1} \vec{p}+m c^2\;\left\{-\frac{1}{p} \vec{\alpha}(0)+2 \left[\vec{\alpha}(0)\cdot \vec{p}\right] \dfrac{\vec{p}}{p^3} +m\,c^2 (H_{\rm b}^{\rm I})^{-1}\,\dfrac{\vec{p}}{p^2}\right\}{\rm e}^{\frac{-{\rm i}\,2\,H_{\rm b}^{\rm I}\,t}{\hbar}}.
                        \end{eqnarray}
            On then has
            \begin{eqnarray}\label{eq:E-9-a}
                  \vec{r}(t)
                  &=& \vec{r}(0)+c^2 (H_{\rm b}^{\rm I})^{-1}\vec{p}\;t
                        +\frac{{\rm i}\hbar}{2} m c^2\; \Biggl\{
                        -\frac{1}{p} \vec{\alpha}(0)+2 \left[\vec{\alpha}(0)\cdot \vec{p}\right] \dfrac{\vec{p}}{p^3} +m\,c^2 (H_{\rm b}^{\rm I})^{-1}\,\dfrac{\vec{p}}{p^2}\Biggr\}(H_{\rm b}^{\rm I})^{-1} {\left(
                              {\rm e}^{\frac{-{\rm i}\,2H_{\rm b}^{\rm I}\,t}{\hbar}}-1\right)}.
            \end{eqnarray}
            One may verify that Eq. (\ref{eq:E-9}) and Eq. (\ref{eq:E-9-a}) are the same. Actually, by substituting
            \begin{eqnarray}\label{eq:E-9-b}
            c \beta(0) = \frac{1}{p}\left(H_{\rm b}^{\rm I} +m\,c^2\;\vec{\alpha}(0)\cdot\hat{p}\right)
            \end{eqnarray}
            into Eq. (\ref{eq:E-9}) one immediately has Eq. (\ref{eq:E-9-a}).
            The merit of Eq. (\ref{eq:E-9-a}) is that it does not contain the matrix $\beta(0)$. $\blacksquare$
            \end{remark}

            \begin{remark}
                  For the position operator in Eq. (\ref{eq:E-9-a}), the third term is an oscillation term, which is related to the ``position Zitterbewegung''. We need to calculate the following expectation value for the ``position Zitterbewegung'' operator
                  \begin{eqnarray}
                        \hat{\mathcal{Z}}_{\rm b}^{\rm I} &=& \dfrac{{\rm i}\hbar}{2} m\,c^2 \Biggl\{
                                    -\dfrac{1}{p} \vec{\alpha}(0)+2\left[
                                          \vec{\alpha}(0)\cdot\vec{p}\right]
                                                \dfrac{\vec{p}}{p^3}
                                    +m\,c^2 (H_{\rm b}^{\rm I})^{-1} \dfrac{\vec{p}}{p^2}\Biggr\}
                              (H_{\rm b}^{\rm I})^{-1} \left(
                                    {\rm e}^{\frac{-{\rm i}\,2H_{\rm b}^{\rm I}\,t}{\hbar}}-1\right),
                  \end{eqnarray}
                  defined by
                  \begin{eqnarray}
                        \mathcal{Z}_{\rm b}^{\rm I} &=& \langle\Psi'|\hat{\mathcal{Z}}_{\rm b}^{\rm I} |\Psi'\rangle,
                  \end{eqnarray}
                  where $|\Psi'\rangle$ is the quantum state of the type-I Dirac's braidon. Similarly, one may prove that $\hat{\mathcal{Z}}_{\rm b}^{\rm I}$ is Hermitian.


 Now we study the quantum states $|\Psi'\rangle$ of the type-I Dirac's braidon. Based on Eq. (\ref{eq:hh}), the transformation from $H_{\rm e}$ to $H_{\rm b}^{\rm I}$ is given by
            \begin{equation}\label{eq:bra-2a}
            H_{\rm b}^{\rm I}= \mathcal{C} H_{\rm e} \mathcal{C}^\dagger.
            \end{equation}
For a Dirac's electron, we have known that the common eigenstates of the set $\{H_{\rm e}, \hat{\Lambda}\}$ are $\{|\Psi_1\rangle, |\Psi_2\rangle, |\Psi_3\rangle, |\Psi_4\rangle\}$, where $\{|\Psi_1\rangle, |\Psi_2\rangle\}$ correspond to positive energy, while $\{|\Psi_3\rangle, |\Psi_4\rangle\}$ correspond to negative energy. The spectrum-decomposition of $H_{\rm e}$ is given by
            \begin{eqnarray}
            H_{\rm e}&=& E_{\rm p} \left(|\Psi_1\rangle\langle \Psi_1|+ |\Psi_2\rangle\langle \Psi_2| \right)-E_{\rm p} \left(|\Psi_3\rangle\langle \Psi_3|+ |\Psi_4\rangle\langle \Psi_4| \right).
            \end{eqnarray}
Based on the unitary matrix $\mathcal{C}$, from $|\Psi'_j\rangle=\mathcal{C}|\Psi_j\rangle$ we can determine the four eigenstates of type-I Dirac's braidon as
            \begin{eqnarray}\label{eq:braidoneigen-0}
            && \{|\Psi'_1\rangle, |\Psi'_2\rangle, |\Psi'_3\rangle, |\Psi'_4\rangle\} =\{\mathcal{C}|\Psi_1\rangle, \mathcal{C}|\Psi_2\rangle, \mathcal{C}|\Psi_3\rangle, \mathcal{C}|\Psi_4\rangle\},
            \end{eqnarray}
            which are common eigenstates of the set $\{H_{\rm b}^{\rm I}=\mathcal{C} H_{\rm e} \mathcal{C}^\dagger, \hat{\Lambda}'=\mathcal{C} \hat{\Lambda} \mathcal{C}^\dagger\}$. Similarly, $\{|\Psi'_1\rangle, |\Psi'_2\rangle\}$ correspond to positive energy of $H_{\rm b}^{\rm I}$, while $\{|\Psi'_3\rangle, |\Psi'_4\rangle\}$ correspond to negative energy of $H_{\rm b}^{\rm I}$. And $\{|\Psi'_1\rangle, |\Psi'_3\rangle\}$ correspond to positive helicity of $\hat{\Lambda}'$, while $\{|\Psi'_2\rangle, |\Psi'_4\rangle\}$ correspond to negative helicity of $\hat{\Lambda}'$.

            Because
            \begin{eqnarray}
            [\hat{\Lambda}, \vec{\alpha}\cdot\hat{p}]=0, \;\;\;\;\;\; [\hat{\Lambda}, \beta]=0,
            \end{eqnarray}
            thus
            \begin{eqnarray}
            [\hat{\Lambda}, \mathcal{C}]=0,
            \end{eqnarray}
            then we have
            \begin{eqnarray}
            \hat{\Lambda}'=\mathcal{C}\,\hat{\Lambda}\, \mathcal{C}^\dagger=\hat{\Lambda}.
            \end{eqnarray}
            Therefore the four states in
            Eq. (\ref{eq:braidoneigen-0}) are common eigenstates of the set $\{H_{\rm b}^{\rm I}, \hat{\Lambda}\}$. By the way, the spectrum-decomposition of $H_{\rm b}^{\rm I}$ is given by
            \begin{eqnarray}
            H_{\rm b}^{\rm I}&=& E_{\rm p} \left(|\Psi'_1\rangle\langle \Psi'_1|+ |\Psi'_2\rangle\langle \Psi'_2| \right)-E_{\rm p} \left(|\Psi'_3\rangle\langle \Psi'_3|+ |\Psi'_4\rangle\langle \Psi'_4| \right)=\mathcal{C} H_{\rm e} \mathcal{C}^\dagger,
            \end{eqnarray}
            which is consistent with Eq. (\ref{eq:bra-2a}).

            Let us introduce the following projection operators
                  \begin{eqnarray}
                        &&\Pi'_\pm=\frac{1}{2}\left(\mathbb{I}\pm
                              \frac{H_{\rm b}^{\rm I}}{\sqrt{p^2c^2+m^2c^4}}\right),\nonumber\\
                        && \left(\Pi'_\pm\right)^2=\Pi'_\pm,
                  \end{eqnarray}
                  and we easily have
                  \begin{eqnarray}
                        && \Pi'_+ |\Psi_1'\rangle = |\Psi_1'\rangle, \;\; \Pi'_+ |\Psi_2'\rangle = |\Psi_2'\rangle, \;\; \Pi'_+ |\Psi_3'\rangle = 0, \;\; \Pi'_+ |\Psi_4'\rangle = 0, \nonumber\\
                        && \Pi'_- |\Psi_1'\rangle = 0, \;\; \Pi'_- |\Psi_2'\rangle = 0, \;\; \Pi'_- |\Psi_3'\rangle = |\Psi_3'\rangle, \;\; \Pi'_- |\Psi_4'\rangle = |\Psi_4'\rangle.
                  \end{eqnarray}
                  We then have
                  \begin{eqnarray}
                        && \Pi'_+ \Biggl\{-\dfrac{1}{p} \vec{\alpha}(0)+2\left[
                                    \vec{\alpha}(0)\cdot\vec{p}\right]\dfrac{\vec{p}}{p^3}
                              +m\,c^2 (H_{\rm b}^{\rm I})^{-1} \dfrac{\vec{p}}{p^2}\Biggr\}\Pi'_+ \nonumber \\
                        &=& \dfrac{1}{4} \left(\mathbb{I}
                              +\dfrac{H_{\rm b}^{\rm I}}{\sqrt{p^2 c^2 +m^2 c^4}}\right)\Biggl\{
                                    -\dfrac{1}{p} \vec{\alpha}(0)+2\left[
                                          \vec{\alpha}(0)\cdot\vec{p}\right]
                                                \dfrac{\vec{p}}{p^3}
                                    +m\,c^2 (H_{\rm b}^{\rm I})^{-1} \dfrac{\vec{p}}{p^2}\Biggr\}
                              \left(\mathbb{I}+\frac{H_{\rm b}^{\rm I}}{\sqrt{p^2c^2+m^2c^4}}\right)\nonumber\\
                        &=& \dfrac{1}{4} \Biggl\{-\dfrac{1}{p} \vec{\alpha}(0)+2\left[
                                          \vec{\alpha}(0)\cdot\vec{p}\right]
                                                \dfrac{\vec{p}}{p^3}
                                    +m\,c^2 (H_{\rm b}^{\rm I})^{-1} \dfrac{\vec{p}}{p^2}\Biggr\}
                              + \frac{1}{4\sqrt{p^2c^2+m^2c^4}} \left\{H_{\rm b}^{\rm I},\
                                    -\dfrac{1}{p} \vec{\alpha}(0)+2\left[
                                          \vec{\alpha}(0)\cdot\vec{p}\right]
                                                \dfrac{\vec{p}}{p^3}
                                    +m\,c^2 (H_{\rm b}^{\rm I})^{-1} \dfrac{\vec{p}}{p^2}\right\} \nonumber \\
                              &&\quad +\dfrac{1}{4(p^2c^2+m^2c^4)} H_{\rm b}^{\rm I} \Biggl\{
                                    -\dfrac{1}{p} \vec{\alpha}(0)+2\left[
                                          \vec{\alpha}(0)\cdot\vec{p}\right]
                                                \dfrac{\vec{p}}{p^3}
                                    +m\,c^2 (H_{\rm b}^{\rm I})^{-1} \dfrac{\vec{p}}{p^2}\Biggr\}
                                    H_{\rm b}^{\rm I} \nonumber \\
                        &=& \dfrac{1}{4} \Biggl\{-\dfrac{1}{p} \vec{\alpha}(0)+2\left[
                                          \vec{\alpha}(0)\cdot\vec{p}\right]
                                                \dfrac{\vec{p}}{p^3}
                                    +m\,c^2 (H_{\rm b}^{\rm I})^{-1} \dfrac{\vec{p}}{p^2}\Biggr\}
                              -\dfrac{1}{4(p^2c^2+m^2c^4)} H^2_{\rm b} \Biggl\{
                                    -\dfrac{1}{p} \vec{\alpha}(0)+2\left[
                                          \vec{\alpha}(0)\cdot\vec{p}\right]
                                                \dfrac{\vec{p}}{p^3}
                                    +m\,c^2 (H_{\rm b}^{\rm I})^{-1} \dfrac{\vec{p}}{p^2}\Biggr\}\nonumber\\
                              &=& \frac{1}{4} \Biggl\{-\dfrac{1}{p} \vec{\alpha}(0)+2\left[
                                                \vec{\alpha}(0)\cdot\vec{p}\right]
                                                      \dfrac{\vec{p}}{p^3}
                                          +m\,c^2 (H_{\rm b}^{\rm I})^{-1} \dfrac{\vec{p}}{p^2}
                                    -\biggl[-\dfrac{1}{p} \vec{\alpha}(0)+2\left[
                                          \vec{\alpha}(0)\cdot\vec{p}\right]
                                                \dfrac{\vec{p}}{p^3}
                                    +m\,c^2 (H_{\rm b}^{\rm I})^{-1} \dfrac{\vec{p}}{p^2}\biggr]\Biggr\}
                                    \nonumber\\
                              &=& 0.
                  \end{eqnarray}
                  Similarly, we have
                  \begin{eqnarray}
                        \Pi'_- \Biggl\{-\dfrac{1}{p} \vec{\alpha}(0)+2\left[
                                    \vec{\alpha}(0)\cdot\vec{p}\right]
                                          \dfrac{\vec{p}}{p^3}
                              +m\,c^2 (H_{\rm b}^{\rm I})^{-1} \dfrac{\vec{p}}{p^2}\Biggr\}\Pi'_- =0.
                  \end{eqnarray}
                  The above results lead to
                  \begin{eqnarray}
                        \Pi'_+ \hat{\mathcal{Z}}_{\rm b}^{\rm I} \Pi'_+ =0,\;\;\;\;\;
                        \Pi'_- \hat{\mathcal{Z}}_{\rm b}^{\rm I} \Pi'_- =0.
                  \end{eqnarray}
                  Thus, if the type-I Dirac's braidon in a superposition state of only positive-energy, i.e.,
                   \begin{eqnarray}
                  &&  |\Psi'\rangle\equiv|\Psi'_+\rangle = c_1 |\Psi'_1\rangle+ c_2 |\Psi'_2\rangle,
                  \end{eqnarray}
                  or a superposition state of only negative-energy, i.e.,
                   \begin{eqnarray}
                  &&  |\Psi'\rangle\equiv|\Psi'_-\rangle = c_3 |\Psi'_3\rangle+ c_4 |\Psi'_4\rangle,
                  \end{eqnarray}
                  then one has
                  $\mathcal{Z}_{\rm b}^{\rm I}=\langle\Psi'|\hat{\mathcal{Z}}_{\rm b}^{\rm I} |\Psi'\rangle=0$, i.e., there is no any phenomenon of ``position Zitterbewegung''. $\blacksquare$
                  \end{remark}

            \begin{remark}
            Next, let us consider the case that the type-I Dirac's braidon is in a superposition state of positive-energy and negative-energy, for example,
                  \begin{eqnarray}
            &&  |\Psi'\rangle = \cos\eta |\Psi'_1\rangle+ \sin\eta |\Psi'_3\rangle.
            \end{eqnarray}
            In this case, we have
            \begin{eqnarray}
            \mathcal{Z}_{\rm b}^{\rm I} &=& \langle \Psi'| \hat{\mathcal{Z}}_{\rm b}^{\rm I} |\Psi'\rangle= \sin(2\eta)  {\rm Re}\left[\langle \Psi'_1|\hat{\mathcal{Z}}_{\rm b}^{\rm I}|\Psi'_3 \rangle \right],
            \end{eqnarray}
            where
                  \begin{eqnarray}\label{Eq:ZZ-r}
                        \langle\Psi'_1|\hat{\mathcal{Z}}_{\rm b}^{\rm I}|\Psi'_3 \rangle
                        &=& \dfrac{{\rm i}\hbar}{2} m\,c^2 \langle\Psi'_1|\Biggl\{
                                    -\dfrac{1}{p} \vec{\alpha}(0)+2\left[
                                          \vec{\alpha}(0)\cdot\vec{p}\right]
                                                \dfrac{\vec{p}}{p^3}
                                    +m\,c^2 (H_{\rm b}^{\rm I})^{-1} \dfrac{\vec{p}}{p^2}\Biggr\}
                              (H_{\rm b}^{\rm I})^{-1} \left(
                                    {\rm e}^{\frac{-{\rm i}\,2H_{\rm b}^{\rm I}\,t}{\hbar}}-1\right)
                                    |\Psi'_3\rangle.
                              \end{eqnarray}
            One needs to compute Eq. (\ref{Eq:ZZ-r}), he obtains
            \begin{eqnarray}\label{Eq:ZZ-s}
                        \langle\Psi'_1|\hat{\mathcal{Z}}_{\rm b}^{\rm I}|\Psi'_3 \rangle
                        &=& \dfrac{{\rm i}\hbar}{2} m\,c^2 \langle\Psi'_1|\Biggl\{
                                    -\dfrac{1}{p} \vec{\alpha}(0)+2\left[
                                          \vec{\alpha}(0)\cdot\vec{p}\right]
                                                \dfrac{\vec{p}}{p^3}
                                    +m\,c^2 (H_{\rm b}^{\rm I})^{-1} \dfrac{\vec{p}}{p^2}\Biggr\}
                              (H_{\rm b}^{\rm I})^{-1} \left(
                                    {\rm e}^{\frac{-{\rm i}\,2H_{\rm b}^{\rm I}\,t}{\hbar}}-1\right)
                                    |\Psi'_3\rangle\nonumber\\
                        &=& \dfrac{{\rm i}\hbar}{2} m\,c^2 \langle\Psi'_1|\Biggl\{
                                    -\dfrac{1}{p} \vec{\alpha}(0)+2\left[
                                          \vec{\alpha}(0)\cdot\vec{p}\right]
                                                \dfrac{\vec{p}}{p^3}
                                    -m\,c^2 E_{\rm p}^{-1} \dfrac{\vec{p}}{p^2}\Biggr\}
                              (-E^{-1}_{\rm p}) \left(
                                    {\rm e}^{\frac{{\rm i}\,2E_{\rm p}\,t}{\hbar}}-1\right)
                                    |\Psi'_3\rangle\nonumber\\
                        &=& {-}\dfrac{{\rm i}\hbar}{2} m\,c^2 E^{-1}_{\rm p} \left(
                                    {\rm e}^{\frac{{\rm i}\,2E_{\rm p}\,t}{\hbar}}-1\right) \langle\Psi'_1|\Biggl\{
                                    -\dfrac{1}{p} \vec{\alpha}(0)+2\left[
                                          \vec{\alpha}(0)\cdot\vec{p}\right]
                                                \dfrac{\vec{p}}{p^3}
                                    -m\,c^2 E_{\rm p}^{-1} \dfrac{\vec{p}}{p^2}\Biggr\}
                                    |\Psi'_3\rangle\nonumber\\
                        &=& {-}\dfrac{{\rm i}\hbar}{2} m\,c^2 E^{-1}_{\rm p} \left(
                                    {\rm e}^{\frac{{\rm i}\,2E_{\rm p}\,t}{\hbar}}-1\right) \langle\Psi'_1|\Biggl\{
                                    -\dfrac{1}{p} \vec{\alpha}(0)+2\left[
                                          \vec{\alpha}(0)\cdot\vec{p}\right]
                                                \dfrac{\vec{p}}{p^3}\Biggr\}
                                    |\Psi'_3\rangle\nonumber\\
                        &=& {-}\dfrac{{\rm i}\hbar}{2} m\,c^2 E^{-1}_{\rm p} \left(
                                    {\rm e}^{\frac{{\rm i}\,2E_{\rm p}\,t}{\hbar}}-1\right) \langle\Psi_1|\mathcal{C}^\dagger \Biggl\{
                                    -\dfrac{1}{p} \vec{\alpha}(0)+2\left[
                                          \vec{\alpha}(0)\cdot\vec{p}\right]
                                                \dfrac{\vec{p}}{p^3}\Biggr\}
                                    \mathcal{C} |\Psi_3\rangle.
                              \end{eqnarray}

            Because
            \begin{eqnarray}
                  \mathcal{C}^\dagger \alpha_z  \mathcal{C}&=&\frac{1}{\sqrt{2}}\left[\mathbb{I}- \beta\vec{\alpha}\cdot\hat{p}\right]\; \alpha_z \;\frac{1}{\sqrt{2}}\left[\mathbb{I}+ \beta\vec{\alpha}\cdot\hat{p}\right]=\frac{1}{2}\left[\mathbb{I}- \beta\vec{\alpha}\cdot\hat{p}\right]\; \alpha_z \;\left[\mathbb{I}+ \beta\vec{\alpha}\cdot\hat{p}\right]\nonumber\\
                  &=&\frac{1}{2}\left[\alpha_z- \beta\vec{\alpha}\cdot\hat{p}\; \alpha_z\right]\left[\mathbb{I}+ \beta\vec{\alpha}\cdot\hat{p}\right]=\frac{1}{2}\left[\alpha_z+\alpha_z \beta\vec{\alpha}\cdot\hat{p}- \beta\vec{\alpha}\cdot\hat{p}\; \alpha_z -\beta\vec{\alpha}\cdot\hat{p}\; \alpha_z\beta\vec{\alpha}\cdot\hat{p}\right]\nonumber\\
                  &=& \frac{1}{2}\left[\alpha_z-\beta \alpha_z \vec{\alpha}\cdot\hat{p}- \beta\vec{\alpha}\cdot\hat{p}\; \alpha_z -\vec{\alpha}\cdot\hat{p}\; \alpha_z\vec{\alpha}\cdot\hat{p}\right]\nonumber\\
                  &=& \frac{1}{2}\left[\alpha_z-\beta( \alpha_z \vec{\alpha}\cdot\hat{p}+\vec{\alpha}\cdot\hat{p}\; \alpha_z) -\vec{\alpha}\cdot\hat{p}\; \alpha_z\vec{\alpha}\cdot\hat{p}\right]\nonumber\\
                  &=& \frac{1}{2}\left[\alpha_z-\beta(2 \alpha_z ^2\;\hat{p}_z) -\vec{\alpha}\cdot\hat{p}\; \alpha_z\vec{\alpha}\cdot\hat{p}\right]=\frac{1}{2}\left[\alpha_z-2\beta\;\hat{p}_z -\vec{\alpha}\cdot\hat{p}\; \alpha_z\vec{\alpha}\cdot\hat{p}\right]\nonumber\\
                  &=& \frac{1}{2}\left[\alpha_z-2\beta\;\hat{p}_z -(\alpha_x \hat{p}_x+\alpha_y \hat{p}_y+\alpha_z \hat{p}_z)\; \alpha_z\;(\alpha_x \hat{p}_x+\alpha_y \hat{p}_y+\alpha_z \hat{p}_z)\right]\nonumber\\
                  &=& \frac{1}{2}\left[\alpha_z-2\beta\;\hat{p}_z -(\alpha_x \hat{p}_x \alpha_z+\alpha_y \hat{p}_y\alpha_z+\alpha_z \hat{p}_z\alpha_z)\; (\alpha_x \hat{p}_x+\alpha_y \hat{p}_y+\alpha_z \hat{p}_z)\right]\nonumber\\
                  &=& \frac{1}{2}\left[\alpha_z-2\beta\;\hat{p}_z +\alpha_z(\alpha_x \hat{p}_x +\alpha_y \hat{p}_y -\alpha_z \hat{p}_z)\; (\alpha_x \hat{p}_x+\alpha_y \hat{p}_y+\alpha_z \hat{p}_z)\right]\nonumber\\
                  &=& \frac{1}{2}\left[\alpha_z-2\beta\;\hat{p}_z +\alpha_z(\alpha_x \hat{p}_x +\alpha_y \hat{p}_y +\alpha_z \hat{p}_z)\; (\alpha_x \hat{p}_x+\alpha_y \hat{p}_y+\alpha_z \hat{p}_z)-2 \alpha_z \alpha_z \hat{p}_z \; (\alpha_x \hat{p}_x+\alpha_y \hat{p}_y+\alpha_z \hat{p}_z)\right]\nonumber\\
                  &=& \frac{1}{2}\left[\alpha_z-2\beta\;\hat{p}_z +\alpha_z(\vec{\alpha}\cdot\hat{p})\; (\vec{\alpha}\cdot\hat{p})-2  \hat{p}_z \; (\vec{\alpha}\cdot\hat{p})\right]=\frac{1}{2}\left[\alpha_z-2\beta\;\hat{p}_z +\alpha_z-2  \hat{p}_z \; (\vec{\alpha}\cdot\hat{p})\right]\nonumber\\
                  &=&\frac{1}{2}\left[2\alpha_z-2\beta\;\hat{p}_z -2  (\vec{\alpha}\cdot\hat{p}) \hat{p}_z \right]=\alpha_z-\beta\;\hat{p}_z -  (\vec{\alpha}\cdot\hat{p}) \hat{p}_z,
            \end{eqnarray}
            one generally has
            \begin{eqnarray}
                  \mathcal{C}^\dagger \vec{\alpha}  \mathcal{C}&=&
                  \vec{\alpha}-\beta\;\hat{p} -  (\vec{\alpha}\cdot\hat{p}) \hat{p},
            \end{eqnarray}
            which leads to
            \begin{eqnarray}
                  \mathcal{C}^\dagger \left[
                                          \vec{\alpha}\cdot\vec{p}\right]
                                                \dfrac{\vec{p}}{p^2} \mathcal{C}&=&
            \left\{ \left[ \vec{\alpha}-\beta\;\hat{p} -  (\vec{\alpha}\cdot\hat{p}) \hat{p}\right]\cdot\vec{p} \right\}\dfrac{\vec{p}}{p^2}=
            \left[ \vec{\alpha} \cdot\vec{p}-p \beta  -  (\vec{\alpha}\cdot\hat{p}) p\right]\dfrac{\vec{p}}{p^2}=
            \left[ -p \beta \right]\dfrac{\vec{p}}{p^2}=-\beta\; \hat{p}.
            \end{eqnarray}
            Based on the above result, we have
            \begin{eqnarray}\label{Eq:ZZ-t}
                        \mathcal{C}^\dagger \Biggl\{
                                    -\dfrac{1}{p} \vec{\alpha}(0)+2\left[
                                          \vec{\alpha}(0)\cdot\vec{p}\right]
                                                \dfrac{\vec{p}}{p^3}\Biggr\}
                                    \mathcal{C} &=&  -\dfrac{1}{p} (\vec{\alpha}-\beta\;\hat{p} -  (\vec{\alpha}\cdot\hat{p}) \hat{p})+\frac{2}{p}(-\beta\; \hat{p})\nonumber\\
                                    &=&  -\dfrac{1}{p} \left[\vec{\alpha}+\beta\;\hat{p} {-} (\vec{\alpha}\cdot\hat{p}) \hat{p}\right].
                              \end{eqnarray}
            From the Hamiltonian of a Dirac's electron
            \begin{equation}
                  H_{\rm e} =c\,\vec{\alpha}\cdot\vec{p} +\beta\,m\,c^2,
                  \end{equation}
            we have
            \begin{equation}\label{eq:beta}
                  \beta = \frac{1}{mc^2} (H_{\rm e} -c\,\vec{\alpha}\cdot\vec{p}).
                  \end{equation}
            Then Eq. (\ref{Eq:ZZ-t}) becomes
            \begin{eqnarray}\label{Eq:ZZ-p}
                        \mathcal{C}^\dagger \Biggl\{
                                    -\dfrac{1}{p} \vec{\alpha}(0)+2\left[
                                          \vec{\alpha}(0)\cdot\vec{p}\right]
                                                \dfrac{\vec{p}}{p^3}\Biggr\}
                                    \mathcal{C}
                                    &=&  -\dfrac{1}{p} \left[\vec{\alpha}+\beta\;\hat{p} {-}(\vec{\alpha}\cdot\hat{p}) \hat{p}\right]\nonumber\\
                                    &=&  -\dfrac{1}{p} \left[\vec{\alpha}+\frac{1}{mc^2} (H_{\rm e} -c\,\vec{\alpha}\cdot\vec{p})\;\hat{p} {-}(\vec{\alpha}\cdot\hat{p}) \hat{p}\right]\nonumber\\
                                    &=&  -\dfrac{1}{p} \left[\vec{\alpha}+\frac{1}{mc^2} H_{\rm e}\;\hat{p} {-}\left(\frac{cp}{mc^2}{+}1\right)  (\vec{\alpha}\cdot\hat{p}) \hat{p}\right].
                              \end{eqnarray}
            After substituting Eq. (\ref{Eq:ZZ-p}) into Eq. (\ref{Eq:ZZ-s}) we have
            \begin{eqnarray}\label{Eq:ZZ-q}
                  \langle\Psi'_1|\hat{\mathcal{Z}}_{\rm b}^{\rm I}|\Psi'_3 \rangle
                  &=& {-}\dfrac{{\rm i}\hbar}{2} m\,c^2 E^{-1}_{\rm p} \left(
                        {\rm e}^{\frac{{\rm i}\,2E_{\rm p}\,t}{\hbar}}-1\right) \langle\Psi_1|\mathcal{C}^\dagger \Biggl\{
                        -\dfrac{1}{p} \vec{\alpha}(0)+2\left[
                              \vec{\alpha}(0)\cdot\vec{p}\right]
                                    \dfrac{\vec{p}}{p^3}\Biggr\}
                        \mathcal{C} |\Psi_3\rangle\nonumber\\
                  &=& {-}\dfrac{{\rm i}\hbar}{2} m\,c^2 E^{-1}_{\rm p} \left(
                        {\rm e}^{\frac{{\rm i}\,2E_{\rm p}\,t}{\hbar}}-1\right) \langle\Psi_1|
                        \Biggl\{-\dfrac{1}{p} \left[\vec{\alpha}(0)+\frac{1}{mc^2} H_{\rm e}\;\hat{p} {-}\left(\frac{cp}{mc^2} {+}1\right)  [\vec{\alpha}(0)\cdot\hat{p}] \hat{p}\right]\Biggr\}  |\Psi_3\rangle\nonumber\\
                  &=& \dfrac{{\rm i}\hbar}{2p} m\,c^2 E^{-1}_{\rm p} \left(
                        {\rm e}^{\frac{{\rm i}\,2E_{\rm p}\,t}{\hbar}}-1\right) \langle\Psi_1|
                        \left[\vec{\alpha}(0)+\frac{1}{mc^2} H_{\rm e}\;\hat{p} {-}\left(\frac{cp}{mc^2}{+}1\right)  [\vec{\alpha}(0)\cdot\hat{p}] \hat{p}\right]  |\Psi_3\rangle\nonumber\\
                  &=& \dfrac{{\rm i}\hbar}{2p} m\,c^2 E^{-1}_{\rm p} \left(
                        {\rm e}^{\frac{{\rm i}\,2E_{\rm p}\,t}{\hbar}}-1\right) \langle\Psi_1|
                        \left[\vec{\alpha}(0){-}\left(\frac{cp}{mc^2}{+}1\right)  [\vec{\alpha}(0)\cdot\hat{p}] \hat{p}\right]  |\Psi_3\rangle.
            \end{eqnarray}
            From Eq. (\ref{eq:D-28-qa}) or Table \ref{tab:pz0}, we have known that
            \begin{eqnarray}
            && \langle\Psi_1|\vec{\alpha}(0)|\Psi_3 \rangle= -\dfrac{mc^2}{E_{\rm p}} \hat{p},
            \end{eqnarray}
            thus
            \begin{eqnarray}\label{eq:D-28-r}
            && \langle\Psi_1|[\vec{\alpha}(0)\cdot\hat{p}] \hat{p}|\Psi_3 \rangle= \left(-\dfrac{mc^2}{E_{\rm p}}
            \hat{p}\right)\cdot\hat{p} \;\hat{p}=-\dfrac{mc^2}{E_{\rm p}}\hat{p}=\langle\Psi_1|\vec{\alpha}(0)|\Psi_3 \rangle.
            \end{eqnarray}
            From Eq. (\ref{Eq:ZZ-q}) we then have
            \begin{eqnarray}\label{Eq:ZZ-u}
                  \langle\Psi'_1|\hat{\mathcal{Z}}_{\rm b}^{\rm I}|\Psi'_3 \rangle
                  &=& \dfrac{{\rm i}\hbar}{2p} m\,c^2 E^{-1}_{\rm p} \left(
                              {\rm e}^{\frac{{\rm i}\,2E_{\rm p}\,t}{\hbar}}-1\right) \langle\Psi_1|
                              \left[\vec{\alpha}(0){-}\left(\frac{cp}{mc^2}{+}1\right)  [\vec{\alpha}(0)\cdot\hat{p}] \hat{p}\right]  |\Psi_3\rangle\nonumber\\
                  &=& {-}\dfrac{{\rm i}\hbar}{2p} m\,c^2 E^{-1}_{\rm p} \left(
                              {\rm e}^{\frac{{\rm i}\,2E_{\rm p}\,t}{\hbar}}-1\right)
                              \frac{cp}{mc^2} \langle\Psi_1|\vec{\alpha}(0)|\Psi_3 \rangle\nonumber\\
                  &=& {-}\frac{{\rm i}\hbar c}{2 E_{\rm p}} \left({\rm e}^{\frac{{\rm i}\,2\,E_{\rm p} t}{\hbar}} {-1} \right)\langle\Psi_1|\vec{\alpha}(0)|\Psi_3 \rangle \equiv \langle\Psi_1|\hat{\mathcal{Z}}_{\rm e}|\Psi_3 \rangle,
            \end{eqnarray}
            which is the same as the result in Eq. (\ref{eq:D-22-a}). $\blacksquare$
            \end{remark}

        \subsection{More General Results}

\begin{remark}
In the above, we have calculated only the case of $\langle\Psi'_1|\hat{\mathcal{Z}}_{\rm b}^{\rm I}|\Psi'_3 \rangle$, and we find that
 \begin{eqnarray}\label{Eq:ZZ-uu}
                  \langle\Psi'_1|\hat{\mathcal{Z}}_{\rm b}^{\rm I}|\Psi'_3 \rangle
                  &=& \langle\Psi_1|\hat{\mathcal{Z}}_{\rm e}|\Psi_3 \rangle.
            \end{eqnarray}
In this section, let us consider the most general results. To do so, we need to calculate the other three terms, i.e., $\langle\Psi'_2|\hat{\mathcal{Z}}_{\rm b}^{\rm I}|\Psi'_4 \rangle$, $\langle\Psi'_1|\hat{\mathcal{Z}}_{\rm b}^{\rm I}|\Psi'_4 \rangle$, and $\langle\Psi'_2|\hat{\mathcal{Z}}_{\rm b}^{\rm I}|\Psi'_3 \rangle$. Similar to Eq. (\ref{eq:D-22-a}), we generally have
\begin{eqnarray}\label{eq:D-22-bb-1}
            \langle\Psi_k|\hat{\mathcal{Z}}_{\rm e}|\Psi_l\rangle
                        &=& \frac{-{\rm i}\hbar c}{2 E_{\rm p}} \left({\rm e}^{\frac{{\rm i}\,2\,E_{\rm p} t}{\hbar}} {-1} \right)\langle\Psi_k|\vec{\alpha}(0)|\Psi_l\rangle,
            \end{eqnarray}
and similar to Eq. (\ref{Eq:ZZ-u}), we generally have
\begin{eqnarray}\label{Eq:ZZ-u-1}
                  \langle\Psi'_k|\hat{\mathcal{Z}}_{\rm b}^{\rm I}|\Psi'_l\rangle
                  &=& \dfrac{{\rm i}\hbar}{2p} \dfrac{m\,c^2}{E_{\rm p}} \left(
                              {\rm e}^{\frac{{\rm i}\,2E_{\rm p}\,t}{\hbar}}-1\right) \langle\Psi_k|
                              \left[\vec{\alpha}(0)-\left(\frac{cp}{mc^2}+1\right)  [\vec{\alpha}(0)\cdot\hat{p}] \hat{p}\right]  |\Psi_l\rangle.
            \end{eqnarray}
where $k\in\{1, 2\}$, and $l\in\{3,4\}$. We would like to compare whether Eq. (\ref{Eq:ZZ-u-1}) is equal to Eq. (\ref{eq:D-22-bb-1}). $\blacksquare$
\end{remark}

\begin{remark}
By taking $j=2, k=4$, from Eq. (\ref{Eq:ZZ-u-1}) one obtains
      \begin{eqnarray}\label{Eq:ZZ-u-1a}
                  \langle\Psi'_2|\hat{\mathcal{Z}}_{\rm b}^{\rm I}|\Psi'_4 \rangle
                  &=& \dfrac{{\rm i}\hbar}{2p} \dfrac{m\,c^2}{E_{\rm p}} \left(
                              {\rm e}^{\frac{{\rm i}\,2E_{\rm p}\,t}{\hbar}}-1\right) \langle\Psi_2|
                              \left[\vec{\alpha}(0)-\left(\frac{cp}{mc^2}+1\right)  [\vec{\alpha}(0)\cdot\hat{p}] \hat{p}\right]  |\Psi_4\rangle.      \end{eqnarray}
      From Table \ref{tab:pz0}, we have known that
       \begin{eqnarray}\label{eq:4terms-1a}
           \langle\Psi_2|\vec{\alpha}(0)|\Psi_4\rangle
            &=&\langle\Psi_1|\vec{\alpha}(0)|\Psi_3\rangle=-\dfrac{m\,c^2}{E_{\rm p}} \hat{p},
            \end{eqnarray}
      thus
        \begin{eqnarray}\label{eq:4terms-1aa}
           \langle\Psi_2|\left[\vec{\alpha}(0)\cdot\hat{p}] \hat{p}\right] |\Psi_4\rangle
            &=&-\dfrac{m\,c^2}{E_{\rm p}} \hat{p}.
            \end{eqnarray}
      We then have
       \begin{eqnarray}
                  \langle\Psi'_2|\hat{\mathcal{Z}}_{\rm b}^{\rm I}|\Psi'_4 \rangle
                  &=& \dfrac{{\rm i}\hbar}{2p} \dfrac{m\,c^2}{E_{\rm p}} \left(
                              {\rm e}^{\frac{{\rm i}\,2E_{\rm p}\,t}{\hbar}}-1\right) \langle\Psi_2|
                              \left[\vec{\alpha}(0)-\left(\frac{cp}{mc^2}+1\right)  [\vec{\alpha}(0)\cdot\hat{p}] \hat{p}\right]  |\Psi_4\rangle\nonumber\\
                  &=& \dfrac{{\rm i}\hbar}{2p} \dfrac{m\,c^2}{E_{\rm p}} \, \frac{-cp}{mc^2}\left(
                              {\rm e}^{\frac{{\rm i}\,2E_{\rm p}\,t}{\hbar}}-1\right) \langle\Psi_2|
                              \vec{\alpha}(0)|\Psi_4\rangle\nonumber\\
                 &=& \frac{-{\rm i}\hbar c}{2 E_{\rm p}} \left({\rm e}^{\frac{{\rm i}\,2\,E_{\rm p} t}{\hbar}} {-1} \right)\langle\Psi_2|\vec{\alpha}(0)|\Psi_4\rangle \equiv \langle\Psi_2|\hat{\mathcal{Z}}_{\rm e}|\Psi_4 \rangle,
       \end{eqnarray}
       namely,
        \begin{eqnarray}
                  \langle\Psi'_2|\hat{\mathcal{Z}}_{\rm b}^{\rm I}|\Psi'_4 \rangle
                  &=& \langle\Psi_2|\hat{\mathcal{Z}}_{\rm e}|\Psi_4 \rangle,
            \end{eqnarray}
       which means that Eq. (\ref{Eq:ZZ-u-1}) is equal to Eq. (\ref{eq:D-22-bb-1}) when $k=2, l=4$. $\blacksquare$
      \end{remark}

    \begin{remark}
    By taking $k=1, l=4$, from Eq. (\ref{Eq:ZZ-u-1}) one obtains
      \begin{eqnarray}
                  \langle\Psi'_1|\hat{\mathcal{Z}}_{\rm b}^{\rm I}|\Psi'_4 \rangle
                  &=& \dfrac{{\rm i}\hbar}{2p} \dfrac{m\,c^2}{E_{\rm p}} \left(
                              {\rm e}^{\frac{{\rm i}\,2E_{\rm p}\,t}{\hbar}}-1\right) \langle\Psi_1|
                              \left[\vec{\alpha}(0)-\left(\frac{cp}{mc^2}+1\right)  [\vec{\alpha}(0)\cdot\hat{p}] \hat{p}\right]  |\Psi_4\rangle.      \end{eqnarray}
      From  Table \ref{tab:pz0}, we have known that
      \begin{equation}
            \langle\Psi_1|\vec{\alpha}(0)|\Psi_4\rangle=\vec{F}_1 +{\rm i}\,\vec{F}_2,
      \end{equation}
      thus
      \begin{eqnarray}
            & \langle\Psi_1|\big[\vec{\alpha}(0)\cdot\hat{p}\big]|\Psi_4\rangle=\left(\vec{F}_1 +{\rm i}\,\vec{F}_2\right)\cdot\hat{p}
            =0,
      \end{eqnarray}
      namely
      \begin{align}
            & \langle\Psi_1|\vec{\alpha}(0)|\Psi_4\rangle-\left(
                  \frac{c\,p}{m\,c^2} +1\right)\langle\Psi_1|\big[
                        \vec{\alpha}(0)\cdot\hat{p}\big]\hat{p}|\Psi_4\rangle
            =\langle\Psi_1|\vec{\alpha}(0)|\Psi_4\rangle.
      \end{align}
      We then have
      \begin{eqnarray}
            \langle\Psi'_1|\hat{\mathcal{Z}}_{\rm b}^{\rm I}|\Psi'_4 \rangle
            &=& \dfrac{{\rm i}\hbar}{2p} \dfrac{m\,c^2}{E_{\rm p}} \left(
                        {\rm e}^{\frac{{\rm i}\,2E_{\rm p}\,t}{\hbar}}-1\right) \langle\Psi_1|
                        \left[\vec{\alpha}(0)-\left(\frac{cp}{mc^2}+1\right)  [\vec{\alpha}(0)\cdot\hat{p}] \hat{p}\right]  |\Psi_4\rangle\nonumber\\
            &=& \dfrac{{\rm i}\hbar}{2\,p} \left(
                  {\rm e}^{\frac{{\rm i}\,2\,E_{\rm p} t}{\hbar}}-1\right)
                        \dfrac{1}{E_{\rm p}} m\,c^2 \langle\Psi_1|\left[
                              \vec{\alpha}(0)-\left(\frac{cp}{mc^2}+1\right)  [\vec{\alpha}(0)\cdot\hat{p}] \hat{p}\right]|
                                    \Psi_4\rangle \nonumber\\
            &=& \dfrac{{\rm i}\hbar}{2\,p} \left(
                  {\rm e}^{\frac{{\rm i}\,2\,E_{\rm p} t}{\hbar}}-1\right)
                        \dfrac{1}{E_{\rm p}} m\,c^2 \langle\Psi_1|\vec{\alpha}(0)|
                              \Psi_4\rangle.
      \end{eqnarray}
      Recall
      \begin{eqnarray}
            && \langle\Psi_1|\hat{\mathcal{Z}}_{\rm e}|\Psi_4\rangle
            =\frac{-{\rm i}\hbar\,c}{2\,E_{\rm p}} \left({\rm e}^{
                  \frac{{\rm i}\,2\,E_{\rm p} t}{\hbar}} -1\right)\langle\Psi_1|
                        \vec{\alpha}(0)|\Psi_4\rangle
            =\dfrac{{\rm i}\hbar}{2\,p}\left({\rm e}^{
                  \frac{{\rm i}\,2\,E_{\rm p} t}{\hbar}} -1\right)\frac{1}{E_{\rm p}} (
                        -p\,c)\langle\Psi_1|\vec{\alpha}(0)|\Psi_4\rangle,
      \end{eqnarray}
      which indicates $\langle\Psi'_1|\hat{\mathcal{Z}}_{\rm b}^{\rm I}|\Psi'_4 \rangle\neq\langle\Psi_1|\hat{\mathcal{Z}}_{\rm e}|\Psi_4\rangle$ in general. However, they are proportional to each other, i.e.,
      \begin{eqnarray}
            \langle\Psi'_1|\hat{\mathcal{Z}}_{\rm b}^{\rm I}|\Psi'_4 \rangle
            &=&\dfrac{{\rm i}\hbar}{2\,p} \left(
                  {\rm e}^{\frac{{\rm i}\,2\,E_{\rm p} t}{\hbar}}-1\right)
                        \dfrac{1}{E_{\rm p}} m\,c^2 \langle\Psi_1|\vec{\alpha}(0)|
                              \Psi_4\rangle \nonumber\\
            &=&-\dfrac{m\,c^2}{p\,c} \dfrac{{\rm i}\hbar}{2\,p} \left(
                  {\rm e}^{\frac{{\rm i}\,2\,E_{\rm p} t}{\hbar}}-1\right)
                        \dfrac{1}{E_{\rm p}} (-p\,c)\langle\Psi_1|\vec{\alpha}(0)|
                              \Psi_4\rangle \notag \\
           & =& -\dfrac{m\,c^2}{p\,c} \langle\Psi_1|\hat{\mathcal{Z}}_{\rm e}|
                  \Psi_4\rangle \notag \\
            &=& -\dfrac{m\,c}{p} \langle\Psi_1|\hat{\mathcal{Z}}_{\rm e}|
                  \Psi_4\rangle.
      \end{eqnarray}
$\blacksquare$
\end{remark}
\begin{remark}
      Similarly, for the superposition of $\Ket{\Psi'_2}$ and $\Ket{\Psi'_3}$, i.e. $k=2$, and $l=3$. We have known that
      \begin{eqnarray}
            && \langle\Psi_2|\vec{\alpha}(0)|\Psi_3\rangle = -\left[\langle\Psi_1|\vec{\alpha}(0)|\Psi_4\rangle\right]^*=
            -\left(\vec{F}_1 +{\rm i}\,\vec{F}_2\right)^*.
      \end{eqnarray}
      Thus
      \begin{eqnarray}
            && \langle\Psi_2|\big[\vec{\alpha}(0)\cdot\hat{p}\big]|
                  \Psi_3\rangle
             =0,
      \end{eqnarray}
      which indicates
      \begin{eqnarray}
            && \langle\Psi_2|\vec{\alpha}(0)|\Psi_3\rangle-\left(
                  \frac{c\,p}{m\,c^2} +1\right)\langle\Psi_2|\big[
                        \vec{\alpha}(0)\cdot\hat{p}\big]\hat{p}|\Psi_3\rangle
            =\langle\Psi_2|\vec{\alpha}(0)|\Psi_3\rangle.
      \end{eqnarray}
      After that, we attain
      \begin{eqnarray}
            \langle\Psi'_2|\hat{\mathcal{Z}}_{\rm b}^{\rm I}|\Psi'_3 \rangle
            &=& \dfrac{{\rm i}\hbar}{2\,p} \left(
                  {\rm e}^{\frac{{\rm i}\,2\,E_{\rm p} t}{\hbar}}-1\right)
                        \dfrac{1}{E_{\rm p}} m\,c^2 \langle\Psi_2|\left[
                              \vec{\alpha}(0)-\left(\frac{cp}{mc^2}+1\right)  [\vec{\alpha}(0)\cdot\hat{p}] \hat{p}\right]|
                                    \Psi_3\rangle \nonumber\\
            &=& \dfrac{{\rm i}\hbar}{2\,p} \left(
                  {\rm e}^{\frac{{\rm i}\,2\,E_{\rm p} t}{\hbar}}-1\right)
                        \dfrac{1}{E_{\rm p}} m\,c^2 \langle\Psi_2|\vec{\alpha}(0)|
                              \Psi_3\rangle \nonumber\\
            &=& -\dfrac{{\rm i}\hbar}{2\,p} \left(
                  {\rm e}^{\frac{{\rm i}\,2\,E_{\rm p} t}{\hbar}}-1\right)
                        \dfrac{1}{E_{\rm p}} m\,c^2 \left[\langle\Psi_1|\vec{\alpha}(0)|
                              \Psi_4\rangle\right]^*.
      \end{eqnarray}
      Recall
      \begin{eqnarray}
             \langle\Psi_2|\hat{\mathcal{Z}}_{\rm e}|\Psi_3\rangle
            &=&\frac{-{\rm i}\hbar\,c}{2\,E_{\rm p}} \left({\rm e}^{
                  \frac{{\rm i}\,2\,E_{\rm p} t}{\hbar}} -1\right)\langle\Psi_2|
                        \vec{\alpha}(0)|\Psi_3\rangle
            =\dfrac{{\rm i}\hbar}{2\,p}\left({\rm e}^{
                  \frac{{\rm i}\,2\,E_{\rm p} t}{\hbar}} -1\right)\frac{1}{E_{\rm p}} (
                        -p\,c)\langle\Psi_2|\vec{\alpha}(0)|\Psi_3\rangle \notag \\
            &=& \dfrac{{\rm i}\hbar}{2\,p}\left({\rm e}^{
                  \frac{{\rm i}\,2\,E_{\rm p} t}{\hbar}} -1\right)\frac{1}{E_{\rm p}}
                        p\,c \left[\langle\Psi_1|\vec{\alpha}(0)|\Psi_4\rangle\right]^*,
      \end{eqnarray}
      which indicates $\langle\Psi'_2|\hat{\mathcal{Z}}_{\rm b}^{\rm I}|\Psi'_3 \rangle\neq\langle\Psi_2|\hat{\mathcal{Z}}_{\rm e}|\Psi_3\rangle$ in general. However, they are proportional to each other, i.e.,
      \begin{eqnarray}
             \langle\Psi'_2|\hat{\mathcal{Z}}_{\rm b}^{\rm I}|\Psi'_3\rangle
            &=&-\dfrac{{\rm i}\hbar}{2\,p} \left(
                  {\rm e}^{\frac{{\rm i}\,2\,E_{\rm p} t}{\hbar}}-1\right)
                        \dfrac{1}{E_{\rm p}} m\,c^2 \left[\langle\Psi_1|\vec{\alpha}(0)|
                              \Psi_4\rangle\right]^*\nonumber\\
           & =&-\dfrac{m\,c^2}{p\,c} \dfrac{{\rm i}\hbar}{2\,p}\left({\rm e}^{
                  \frac{{\rm i}\,2\,E_{\rm p} t}{\hbar}} -1\right)\frac{1}{E_{\rm p}}
                        p\,c \left[\langle\Psi_1|\vec{\alpha}(0)|\Psi_4\rangle\right]^* \notag \\
            &=& -\dfrac{m\,c^2}{p\,c} \langle\Psi_2|\hat{\mathcal{Z}}_{\rm e}|
                  \Psi_3\rangle \notag \\
            &=& -\dfrac{m\,c}{p} \langle\Psi_2|\hat{\mathcal{Z}}_{\rm e}|
                  \Psi_3\rangle.
      \end{eqnarray}
      $\blacksquare$
\end{remark}

\begin{remark}We can summarize the above results as the following Table \ref{tab:pz2}:
  \begin{table}[h]
	\centering
\caption{The results of ``position Zitterbewegung'' of the type-I Dirac's braidon. $|\Psi'_j\rangle$'s  ($j=1, 2, 3, 4$) are four eigenstates of the type-I Dirac's braidon. The ``position Zitterbewegung'' operator reads $\hat{\mathcal{Z}}_{\rm b}^{{\rm I}, r} = \dfrac{{\rm i}\hbar}{2} m\,c^2 \Biggl\{
                                    -\dfrac{1}{p} \vec{\alpha}(0)+2\left[
                                          \vec{\alpha}(0)\cdot\vec{p}\right]
                                                \dfrac{\vec{p}}{p^3}
                                    +m\,c^2 (H_{\rm b}^{\rm I})^{-1} \dfrac{\vec{p}}{p^2}\Biggr\}
                              (H_{\rm b}^{\rm I})^{-1} \left(
                                    {\rm e}^{\frac{-{\rm i}\,2H_{\rm b}^{\rm I}\,t}{\hbar}}-1\right)$. One has $\langle\Psi'_j|\hat{\mathcal{Z}}_{\rm b}^{{\rm I},r}|\Psi'_j \rangle=0$,
$\langle\Psi'_k|\hat{\mathcal{Z}}_{\rm b}^{{\rm I},r}|\Psi'_l\rangle= \dfrac{{\rm i}\hbar}{2p} \dfrac{m\,c^2}{E_{\rm p}} \left(
                              {\rm e}^{\frac{{\rm i}\,2E_{\rm p}\,t}{\hbar}}-1\right) \langle\Psi_k|
                              \left[\vec{\alpha}(0)-\left(\frac{cp}{mc^2}+1\right)  [\vec{\alpha}(0)\cdot\hat{p}] \hat{p}\right]  |\Psi_l\rangle
                        =\Delta_2\, \langle\Psi_k|
                              \left[\vec{\alpha}(0)-\left(\frac{cp}{mc^2}+1\right)  [\vec{\alpha}(0)\cdot\hat{p}] \hat{p}\right]  |\Psi_l\rangle$ for $k\in\{1,2\}$, $l\in\{3,4\}$, and $\Delta_2=\dfrac{{\rm i}\hbar}{2p} \dfrac{m\,c^2}{E_{\rm p}} \left(
                              {\rm e}^{\frac{{\rm i}\,2E_{\rm p}\,t}{\hbar}}-1\right)=\dfrac{-mc}{p}\Delta_1$, $\Delta_1=\dfrac{-{\rm i}\hbar c}{2 E_{\rm p}} \left({\rm e}^{\frac{{\rm i}\,2\,E_{\rm p} t}{\hbar}} {-1} \right)$.}
\begin{tabular}{lllll}
\hline\hline
 & $|\Psi_1'\rangle$ &  $|\Psi'_2\rangle$& $|\Psi'_3\rangle$ & $|\Psi'_4\rangle$ \\
  \hline
$\langle \Psi'_1| \hat{\mathcal{Z}}_{\rm b}^{{\rm I}, r}$ \;\;\;\quad& 0&0 & $\Delta_1\,\dfrac{-m\,c^2}{E_{\rm p}}\, \hat{p}$  \;\;\;\quad& $\Delta_2\,(\vec{F}_1 +{\rm i}\,\vec{F}_2)$  \\
 \hline
$\langle \Psi'_2| \hat{\mathcal{Z}}_{\rm b}^{{\rm I}, r}$\;\;\;\quad &0 &0 & $-\Delta_2\,(\vec{F}_1 +{\rm i}\,\vec{F}_2)^* $ \quad\quad& $\Delta_1\,\dfrac{-m\,c^2}{E_{\rm p}}\, \hat{p}$ \\
 \hline
 $\langle \Psi'_3| \hat{\mathcal{Z}}_{\rm b}^{{\rm I}, r}$\;\;\;\quad & $\Delta_1^*\,\dfrac{-m\,c^2}{E_{\rm p}}\, \hat{p}$ \;\;\;& $-\Delta_2^*\,(\vec{F}_1 +{\rm i}\,\vec{F}_2) \quad\quad$ & 0 & 0 \\
 \hline
 $\langle \Psi'_4| \hat{\mathcal{Z}}_{\rm b}^{{\rm I}, r}$ \;\;\;\quad& $\Delta_2^*\,(\vec{F}_1 +{\rm i}\,\vec{F}_2)^* \quad$ \quad&$\Delta_1^*\,\dfrac{-m\,c^2}{E_{\rm p}}\, \hat{p}$ \;\;\;& 0 & 0 \\
 \hline\hline
\end{tabular}\label{tab:pz2}
\end{table}\\
$\blacksquare$
\end{remark}

\begin{remark}
From the above calculation, we have known
\begin{eqnarray}
       \mathcal{Z}_{\rm e}
      &=&c_1^* c_3 \Bra{\Psi_1}\hat{\mathcal{Z}}_{\rm e}\Ket{\Psi_3}
            +c_1^* c_4 \Bra{\Psi_1}\hat{\mathcal{Z}}_{\rm e}\Ket{\Psi_4}
            +c_2^* c_3 \Bra{\Psi_2}\hat{\mathcal{Z}}_{\rm e}\Ket{\Psi_3}
            +c_2^* c_4 \Bra{\Psi_2}\hat{\mathcal{Z}}_{\rm e}\Ket{\Psi_4}
            +{\rm c.c.} \notag \\
      &=& 2\,{\rm Re}\left(c_1^* c_3 \Bra{\Psi_1}\hat{\mathcal{Z}}_{\rm e}\Ket{\Psi_3}
            +c_1^* c_4 \Bra{\Psi_1}\hat{\mathcal{Z}}_{\rm e}\Ket{\Psi_4}
            +c_2^* c_3 \Bra{\Psi_2}\hat{\mathcal{Z}}_{\rm e}\Ket{\Psi_3}
            +c_2^* c_4 \Bra{\Psi_2}\hat{\mathcal{Z}}_{\rm e}\Ket{\Psi_4}\right).
      \end{eqnarray}
Note that
\begin{align}
      & \langle\Psi'_1|\hat{\mathcal{Z}}_{\rm b}^{\rm I}|\Psi'_3\rangle
            =\langle\Psi_1|\hat{\mathcal{Z}}_{\rm e}|\Psi_3\rangle, \notag \\
      & \langle\Psi'_2|\hat{\mathcal{Z}}_{\rm b}^{\rm I}|\Psi'_4\rangle
            =\langle\Psi_2|\hat{\mathcal{Z}}_{\rm e}|\Psi_4\rangle, \notag \\
      & \langle\Psi'_1|\hat{\mathcal{Z}}_{\rm b}^{\rm I}|\Psi'_4 \rangle
            =-\dfrac{m\,c}{p} \langle\Psi_1|\hat{\mathcal{Z}}_{\rm e}|
                  \Psi_4\rangle, \notag \\
      & \langle\Psi'_2|\hat{\mathcal{Z}}_{\rm b}^{\rm I}|\Psi'_3\rangle
            =-\dfrac{m\,c}{p} \langle\Psi_2|\hat{\mathcal{Z}}_{\rm e}|
                  \Psi_3\rangle,
\end{align}
thus
\begin{eqnarray}
       \mathcal{Z}_{\rm b}^{\rm I}
      &=& c_1^* c_3 \Bra{\Psi'_1}\hat{\mathcal{Z}}_{\rm b}^{\rm I}\Ket{\Psi'_3}
            +c_1^* c_4 \Bra{\Psi'_1}\hat{\mathcal{Z}}_{\rm b}^{\rm I}\Ket{\Psi'_4}
            +c_2^* c_3 \Bra{\Psi'_2}\hat{\mathcal{Z}}_{\rm b}^{\rm I}\Ket{\Psi'_3}
            +c_2^* c_4 \Bra{\Psi'_2}\hat{\mathcal{Z}}_{\rm b}^{\rm I}\Ket{\Psi'_4}
            +{\rm c.c.} \notag \\
      &=& 2\,{\rm Re}\left(
            c_1^* c_3 \Bra{\Psi'_1}\hat{\mathcal{Z}}_{\rm b}^{\rm I}\Ket{\Psi'_3}
            +c_1^* c_4 \Bra{\Psi'_1}\hat{\mathcal{Z}}_{\rm b}^{\rm I}\Ket{\Psi'_4}
            +c_2^* c_3 \Bra{\Psi'_2}\hat{\mathcal{Z}}_{\rm b}^{\rm I}\Ket{\Psi'_3}
            +c_2^* c_4 \Bra{\Psi'_2}\hat{\mathcal{Z}}_{\rm b}^{\rm I}\Ket{\Psi'_4}
            \right) \notag \\
      &=& 2\,{\rm Re}\left[
            c_1^* c_3 \Bra{\Psi_1}\hat{\mathcal{Z}}_{\rm e}\Ket{\Psi_3}
            +c_1^* c_4 \Bra{\Psi_1}\hat{\mathcal{Z}}_{\rm e}\Ket{\Psi_4}
            -\dfrac{m\,c}{p} \left(c_2^* c_3 \Bra{\Psi_2}\hat{\mathcal{Z}}_{\rm e}
                  \Ket{\Psi_3}
            +c_2^* c_4 \Bra{\Psi_2}\hat{\mathcal{Z}}_{\rm e} \Ket{\Psi_4}\right)
            \right],
\end{eqnarray}
which indicates $\mathcal{Z}_{\rm b}^{\rm I} \neq\mathcal{Z}_{\rm e}$ in general.
 $\blacksquare$
\end{remark}

\section{Position Zitterbewegung for the $H_{\rm e}$-$H_{\rm b}^{\rm I}$ Mixing}

            In this section, let us consider the mixture of Dirac's electron $H_{\rm e}$ and the type-I Dirac's braidon $H_{\rm b}^{\rm I}$. From the viewpoint of ``spin'', $H_{\rm e}$ plays a role of $\vec{S}_x=(\hbar/2)\sigma_x$, and $H_{\rm b}^{\rm I}$ plays a role of $\vec{S}_y=(\hbar/2)\sigma_y$, and furthermore $H_{\rm b}^{\rm II}$ plays a role of $\vec{S}_z=(\hbar/2)\sigma_z$. In quantum mechanics, it is natural to consider the general spin operator in the $\hat{n}$-direction
            \begin{eqnarray}
                  \vec{S}\cdot\hat{n} = n_x S_x + n_y S_y+ n_z S_z= \sin\vartheta \cos\varphi \; S_x + \sin\vartheta \sin\varphi\; S_y+ \cos\vartheta \;S_z,
            \end{eqnarray}
            with $\vec{S}=(S_x, S_y, S_z)$ and
            \begin{eqnarray}
                  \hat{n} = (n_x, n_y, n_z)=(\sin\vartheta \cos\varphi, \sin\vartheta \sin\varphi, \cos\vartheta).
            \end{eqnarray}
             Hence, similarly we consider the general Dirac's Hamiltonian operator
            \begin{eqnarray}
                  \mathcal{H}=\vec{H}\cdot\hat{n} =  \sin\vartheta \cos\varphi \; H_{\rm e} + \sin\vartheta \sin\varphi\; H_{\rm b}^{\rm I}+ \cos\vartheta \;H_{\rm b}^{\rm II},
            \end{eqnarray}
            with
            \begin{eqnarray}
                  \vec{H}=(H_x, H_y, H_z)\equiv(H_{\rm e}, H_{\rm b}^{\rm I}, H_{\rm b}^{\rm II}).
            \end{eqnarray}

            For simplicity, here we consider a simple case, i.e., the Hamiltonian is obtained by mixing $H_{\rm e}$ and $H_{\rm b}^{\rm I}$. In this situation, the mixed Dirac's Hamiltonian operator is given by
            \begin{eqnarray}
                  \mathcal{H}_{\rm m}=\cos\vartheta\;H_{\rm e}
                        +\sin\vartheta\;H_{\rm b}^{\rm I}.
            \end{eqnarray}
            Let us consider the following $\vartheta$-dependent unitary transformation
            \begin{eqnarray}\label{eq:EP8-y}
            \mathcal{C}(\vartheta) &=& {\rm e}^{-{\rm i} \frac{\vartheta}{2} \Gamma_z}=\cos\frac{\vartheta}{2}\;\mathbb{I}+ \sin\frac{\vartheta}{2}\;\beta\vec{\alpha}\cdot\hat{p},
            \end{eqnarray}
            we then have
            \begin{eqnarray}
            \mathcal{C}(\vartheta) H_{\rm e} \mathcal{C}(\vartheta)^\dagger &=& \left[\cos\frac{\vartheta}{2}\;\mathbb{I}+ \sin\frac{\vartheta}{2}\beta\vec{\alpha}\cdot\hat{p}\right]\;H_{\rm e}\; \left[\cos\frac{\vartheta}{2}\;\mathbb{I}- \sin\frac{\vartheta}{2}\beta\vec{\alpha}\cdot\hat{p}\right]\nonumber\\
            &=& \cos^2\left(\frac{\vartheta}{2}\right)\;H_{\rm e} + \sin\frac{\vartheta}{2}\cos\frac{\vartheta}{2}(\beta\vec{\alpha}\cdot\hat{p}\;H_{\rm e}-
            H_{\rm e}\;\beta\vec{\alpha}\cdot\hat{p}) - \sin^2\left(\frac{\vartheta}{2}\right)\beta\vec{\alpha}\cdot\hat{p} \;H_{\rm e}\; \beta\vec{\alpha}\cdot\hat{p}.
            \end{eqnarray}
            Because
            \begin{eqnarray}
            \beta\vec{\alpha}\cdot\hat{p}\;H_{\rm e}- H_{\rm e}\;\beta\vec{\alpha}\cdot\hat{p} &=&
            \beta\vec{\alpha}\cdot\hat{p}\;H_{\rm e}+ H_{\rm e}\;\vec{\alpha}\cdot\hat{p}\beta\nonumber\\
            &=& \beta\vec{\alpha}\cdot\hat{p}\;[c\,\vec{\alpha}\cdot\vec{p} +\beta\,m\,c^2]+ [c\,\vec{\alpha}\cdot\vec{p} +\beta\,m\,c^2]\;\vec{\alpha}\cdot\hat{p}\beta\nonumber\\
            &=& \beta\;[cp +\vec{\alpha}\cdot\hat{p} \beta\,m\,c^2]+ [cp  +\beta \vec{\alpha}\cdot\hat{p}\,m\,c^2]\beta\nonumber\\
            &=& [cp \beta -\vec{\alpha}\cdot\hat{p} \,m\,c^2]+ [cp \beta  - \vec{\alpha}\cdot\hat{p}\,m\,c^2]\nonumber\\
            &=& 2 [cp \beta - \,mc^2 \;\vec{\alpha}\cdot\hat{p}]\nonumber\\
            &=& 2 [- \,mc^2 \;\vec{\alpha}\cdot\hat{p}+cp \beta ]\nonumber\\
            &=& 2 H_{\rm b}^{\rm I},\nonumber\\
            \beta\vec{\alpha}\cdot\hat{p}\;H_{\rm e}\;\beta\vec{\alpha}\cdot\hat{p}
            &=& \beta\vec{\alpha}\cdot\hat{p}\;[c\,\vec{\alpha}\cdot\vec{p} +\beta\,m\,c^2] \;\beta\vec{\alpha}\cdot\hat{p} \nonumber\\
            &=& \beta\;[cp +\vec{\alpha}\cdot\hat{p} \beta\,m\,c^2]  \;\beta\vec{\alpha}\cdot\hat{p}\nonumber\\
            &=& [cp \beta -\vec{\alpha}\cdot\hat{p} \,m\,c^2]  \;\beta\vec{\alpha}\cdot\hat{p}\nonumber\\
            &=& cp \vec{\alpha}\cdot\hat{p} -mc^2 \vec{\alpha}\cdot\hat{p} \;\beta\vec{\alpha}\cdot\hat{p} \nonumber\\
            &=& c \vec{\alpha}\cdot\vec{p} +\beta mc^2  \nonumber\\
            &=& H_{\rm e},
            \end{eqnarray}
           thus we have
            \begin{eqnarray}
            \mathcal{H}_{\rm m}= \mathcal{C}(\vartheta) H_{\rm e} \mathcal{C}(\vartheta)^\dagger   &=& \cos^2\left(\frac{\vartheta}{2}\right)\;H_{\rm e} + \sin\frac{\vartheta}{2}\cos\frac{\vartheta}{2}(\beta\vec{\alpha}\cdot\hat{p}\;H_{\rm e}-
            H_{\rm e}\;\beta\vec{\alpha}\cdot\hat{p}) - \sin^2\left(\frac{\vartheta}{2}\right)\beta\vec{\alpha}\cdot\hat{p} \;H_{\rm e}\; \beta\vec{\alpha}\cdot\hat{p}\nonumber\\
            &=& \cos^2\left(\frac{\vartheta}{2}\right)\;H_{\rm e} + 2 \sin\frac{\vartheta}{2}\cos\frac{\vartheta}{2} \;H_{\rm b} - \sin^2\left(\frac{\vartheta}{2}\right)\;H_{\rm e}\nonumber\\
            &=& \cos\vartheta \;H_{\rm e} +  \sin\vartheta \;H_{\rm b}^{\rm I},
            \end{eqnarray}
            i.e., the unitary matrix $\mathcal{C}(\vartheta)$ transforms $H_{\rm e}$ to the mixed Dirac's Hamiltonian $\mathcal{H}_{\rm m}$.

            For Dirac's electron, we have known that the common eigenstates of the set $\{H_{\rm e}, \hat{\Lambda}\}$ are $\{|\Psi_1\rangle, |\Psi_2\rangle, |\Psi_3\rangle, |\Psi_4\rangle\}$, where $\{|\Psi_1\rangle, |\Psi_2\rangle\}$ correspond to positive energy, while $\{|\Psi_3\rangle, |\Psi_4\rangle\}$ correspond to negative energy. Based on the unitary matrix $\mathcal{C}(\vartheta)$, from $|\Psi'_j\rangle=\mathcal{C}(\vartheta)|\Psi_j\rangle$ we can have four eigenstates of the mixed Hamiltonian operator $\mathcal{H}_{\rm m}$ as
            \begin{eqnarray}\label{eq:braidoneigen}
            \{|\Psi'_1\rangle, |\Psi'_2\rangle, |\Psi'_3\rangle, |\Psi'_4\rangle\}=\{\mathcal{C}(\vartheta)|\Psi_1\rangle, \mathcal{C}(\vartheta)|\Psi_2\rangle, \mathcal{C}(\vartheta)|\Psi_3\rangle, \mathcal{C}(\vartheta)|\Psi_4\rangle\}.
            \end{eqnarray}
            which are common eigenstates of the set $\{\mathcal{H}_{\rm m}, \hat{\Lambda}'=\mathcal{C}(\vartheta) \hat{\Lambda} \mathcal{C}(\vartheta)^\dagger\}\equiv\{\mathcal{H}_{\rm m},\hat{\Lambda}\}$.
           Here $\{|\Psi'_1\rangle, |\Psi'_2\rangle\}$ correspond to positive energy of $\mathcal{H}_{\rm m}$ , while $\{|\Psi'_3\rangle, |\Psi'_4\rangle\}$ correspond to negative energy of $\mathcal{H}_{\rm m}$.  Likewise, $\{|\Psi'_1\rangle, |\Psi'_3\rangle\}$ correspond to positive helicity, while $\{|\Psi'_2\rangle, |\Psi'_4\rangle\}$ correspond to negative helicity of the Hamiltonian system.

            Let us consider the Heisenberg equation
            \begin{eqnarray}
                        \dfrac{{\rm d}\,\vec{r}}{{\rm d}\,t}& =& \dfrac{1}{{\rm i}\hbar}
                        \left[\vec{r}, \mathcal{H}_{\rm m}\right]=\dfrac{1}{{\rm i}\hbar}
                        \left[\vec{r}, \cos\vartheta \;H_{\rm e} +  \sin\vartheta \;H_{\rm b}^{\rm I}\right]\nonumber\\
                        &=& \cos\vartheta \; \dfrac{1}{{\rm i}\hbar}
                        \left[\vec{r}, \;H_{\rm e}\right]+ \sin\vartheta \; \dfrac{1}{{\rm i}\hbar}
                        \left[\vec{r}, \;H_{\rm b}^{\rm I}\right] \nonumber\\
                        &=& \cos\vartheta \; (c\vec{\alpha})+ \sin\vartheta \; \left[-m c^2\;\left(\frac{1}{p} \vec{\alpha} -  \left(\vec{\alpha}\cdot\vec{p}\right) \dfrac{\vec{p}}{p^3}\right)+
                        c \beta \dfrac{\vec{p}}{p}\right].
            \end{eqnarray}
            Because
            \begin{eqnarray}
            \mathcal{H}_{\rm m} &=& \cos\vartheta \;H_{\rm e} +  \sin\vartheta \;H_{\rm b}^{\rm I}\nonumber\\
                        &=& \cos\vartheta \;(c\,\vec{\alpha}\cdot\vec{p} +\beta\,m\,c^2) +  \sin\vartheta \;(-m\,c^2\;\vec{\alpha}\cdot\hat{p}+\beta\,p\,c) \nonumber\\
                        &=&\vec{\alpha}\cdot\hat{p} \left(c p \cos\vartheta - m\,c^2\;  \sin\vartheta \right)+\beta \left(m\,c^2\; \cos\vartheta +cp  \sin\vartheta \right),
            \end{eqnarray}
            thus
            \begin{eqnarray}
            \beta= \frac{1}{\left(m\,c^2\; \cos\vartheta +cp  \sin\vartheta \right)} [ \mathcal{H}_{\rm m}-
                        \vec{\alpha}\cdot\hat{p} \left(c p \cos\vartheta - m\,c^2\;  \sin\vartheta \right)],
            \end{eqnarray}
            we then have
            \begin{eqnarray}\label{eq:alpha-1}
                        \dfrac{{\rm d}\,\vec{r}}{{\rm d}\,t}
                        &=& \cos\vartheta \; (c\vec{\alpha})+ \sin\vartheta \; \left[-m c^2\;\left(\frac{1}{p} \vec{\alpha} -  \left(\vec{\alpha}\cdot\vec{p}\right) \dfrac{\vec{p}}{p^3}\right)+
                        c \beta \dfrac{\vec{p}}{p}\right]\nonumber\\
                        &=& \cos\vartheta \; (c\vec{\alpha})+ \sin\vartheta \; \left[-m c^2\;\left(\frac{1}{p} \vec{\alpha} -  \left(\vec{\alpha}\cdot\vec{p}\right) \dfrac{\vec{p}}{p^3}\right)+
                        \frac{c}{\left(m\,c^2\; \cos\vartheta +cp  \sin\vartheta \right)} [ \mathcal{H}_{\rm m}-
                        \vec{\alpha}\cdot\hat{p} \left(c p \cos\vartheta - m\,c^2\;  \sin\vartheta \right)] \dfrac{\vec{p}}{p}\right].\nonumber\\
            \end{eqnarray}

            From previous sections we know that
            \begin{eqnarray}
                  &&\frac{1}{{\rm i}\hbar}[\vec{\alpha}, H_{\rm e}]
                  =\frac{1}{{\rm i}\hbar}\left(2c\vec{p}-2\,H_{\rm e}\,\vec{\alpha}\right)=-\frac{2\,H_{\rm e}}{{\rm i}\hbar}\left(\vec{\alpha}-c H^{-1}_{\rm e}\vec{p}\right), \nonumber\\
                  &&\dfrac{1}{{\rm i}\hbar} [\vec{\alpha},\ H_{\rm b}^{\rm I}]
                        =-\dfrac{2}{{\rm i}\hbar} \left(m\,c^2\,\hat{p}
                              +H_{\rm b}^{\rm I}\,\vec{\alpha}\right)
                        =-\dfrac{2}{{\rm i}\hbar}\,H_{\rm b}^{\rm I} \left(
                        m\,c^2 (H_{\rm b}^{\rm I})^{-1}\,\hat{p}+\vec{\alpha}\right),
            \end{eqnarray}
            we then have
            \begin{eqnarray}
                  \dfrac{{\rm d}\,\vec{\alpha}}{{\rm d}\,t}& =& \dfrac{1}{{\rm i}\hbar}
                        [\vec{\alpha}, \mathcal{H}_{\rm m}]=\dfrac{1}{{\rm i}\hbar}
                        [\vec{\alpha}, \cos\vartheta \;H_{\rm e} +  \sin\vartheta \;H_{\rm b}^{\rm I}]\nonumber\\
                        &=& \cos\vartheta \; \dfrac{1}{{\rm i}\hbar}
                        [\vec{\alpha}, \;H_{\rm e}]+ \sin\vartheta \; \dfrac{1}{{\rm i}\hbar}
                        [\vec{\alpha}, \;H_{\rm b}^{\rm I}] \nonumber\\
                        &=& \cos\vartheta \; \left[\frac{1}{{\rm i}\hbar}\left(2c\vec{p}-2\,H_{\rm e}\,\vec{\alpha}\right)\right]+ \sin\vartheta \; \left[-\dfrac{2}{{\rm i}\hbar} \left(m\,c^2\,\hat{p}
                              +H_{\rm b}^{\rm I}\,\vec{\alpha}\right)\right]\nonumber\\
                        &=& \frac{-2}{{\rm i}\hbar} \left[\cos\vartheta \;H_{\rm e} +  \sin\vartheta \;H_{\rm b}^{\rm I}\right]\vec{\alpha} + \frac{2}{{\rm i}\hbar} \left[c p \cos\vartheta - m\,c^2\;  \sin\vartheta\right]\hat{p}\nonumber\\
                        &=& \frac{-2}{{\rm i}\hbar} \mathcal{H}_{\rm m}\; \vec{\alpha} + \frac{2}{{\rm i}\hbar} \left[c p \cos\vartheta - m\,c^2\;  \sin\vartheta\right]\hat{p}\nonumber\\
                        &=& \frac{-2 \mathcal{H}_{\rm m}}{{\rm i}\hbar} \; \left[\vec{\alpha} - \mathcal{H}^{-1}_{\rm m}\left[c p \cos\vartheta - m\,c^2\;  \sin\vartheta\right]\hat{p}\right],
            \end{eqnarray}
            i.e.,
            \begin{eqnarray}
                        \dfrac{{\rm d}\,[\vec{\alpha}-\mathcal{H}^{-1}_{\rm m}\left[c p \cos\vartheta - m\,c^2\;  \sin\vartheta\right]\hat{p}]}{{\rm d}\,t}
                        &=& \frac{-2 \mathcal{H}_{\rm m}}{{\rm i}\hbar} \; \left[\vec{\alpha} - \mathcal{H}^{-1}_{\rm m}\left[c p \cos\vartheta - m\,c^2\;  \sin\vartheta\right]\hat{p}\right],
            \end{eqnarray}
            i.e.,
            \begin{eqnarray}
                  \vec{\alpha}(t)-\mathcal{H}^{-1}_{\rm m}\left[c p \cos\vartheta - m\,c^2\;  \sin\vartheta\right]\hat{p}={\rm e}^{\frac{{\rm i}\,2\,H_{\rm m}t}{\hbar}}\; \left[\vec{\alpha}(0)-\mathcal{H}^{-1}_{\rm m}\left[c p \cos\vartheta - m\,c^2\;  \sin\vartheta\right]\hat{p}\right],
            \end{eqnarray}
            i.e.,
            \begin{eqnarray}
                  \vec{\alpha}(t)=\mathcal{H}^{-1}_{\rm m}\left[c p \cos\vartheta - m\,c^2\;  \sin\vartheta\right]\hat{p}+{\rm e}^{\frac{{\rm i}\,2\,H_{\rm m}t}{\hbar}}\; \left[\vec{\alpha}(0)-\mathcal{H}^{-1}_{\rm m}\left[c p \cos\vartheta - m\,c^2\;  \sin\vartheta\right]\hat{p}\right].
            \end{eqnarray}

            One can check that
            \begin{eqnarray}
                  && \left\{\mathcal{H}_{\rm m}, \left(\vec{\alpha}(0)-\mathcal{H}^{-1}_{\rm m}\left[c p \cos\vartheta - m\,c^2\;  \sin\vartheta\right]\hat{p}\right)\right\}=[\mathcal{H}_{\rm m} \vec{\alpha}(0)+\vec{\alpha}(0)\mathcal{H}_{\rm m} ]-2\left[c p \cos\vartheta - m\,c^2\;  \sin\vartheta\right]\hat{p}\nonumber\\
                  &=& \cos\vartheta\, [H_{\rm e} \vec{\alpha}+\vec{\alpha} H_{\rm e} ]+\sin\vartheta\,[H_{\rm b}^{\rm I} \vec{\alpha}+\vec{\alpha} H_{\rm b}^{\rm I} ]-2\left[c p \cos\vartheta - m\,c^2\;  \sin\vartheta\right]\hat{p}\nonumber\\
                  &=& \cos\vartheta\; c[(\vec{\alpha}\cdot\vec{p})\vec{\alpha}+\vec{\alpha}(\vec{\alpha}\cdot\vec{p})]+
                  \sin\vartheta\,\biggl\{\frac{-mc^2}{p}[(\vec{\alpha}\cdot\vec{p})\vec{\alpha}+\vec{\alpha}(\vec{\alpha}\cdot\vec{p})]\biggr\}-2\left[c p \cos\vartheta - m\,c^2\;  \sin\vartheta\right]\hat{p}\nonumber\\
                  &=&\cos\vartheta\, 2c\vec{p} -\sin\vartheta\;\frac{mc^2}{p} 2\vec{p}  -2\left[c p \cos\vartheta - m\,c^2\;  \sin\vartheta\right]\hat{p}\nonumber\\
                  &=& 2c p \cos\vartheta\; \hat{p} -2 mc^2  \sin\vartheta\;\hat{p}  -2\left[c p \cos\vartheta - m\,c^2\;  \sin\vartheta\right]\hat{p}\nonumber\\
                  &=&0,
            \end{eqnarray}
            thus
            \begin{eqnarray}\label{eq:D-8bb}
                  \mathcal{H}_{\rm m} \left(\vec{\alpha}(0)-\mathcal{H}^{-1}_{\rm m}\left[c p \cos\vartheta - m\,c^2\;  \sin\vartheta\right]\hat{p}\right)=-\left(\vec{\alpha}(0)-\mathcal{H}^{-1}_{\rm m}\left[c p \cos\vartheta - m\,c^2\;  \sin\vartheta\right]\hat{p}\right) \mathcal{H}_{\rm m},
            \end{eqnarray}
            which leads to
            \begin{eqnarray}\label{eq:D-8cc}
                  {\rm e}^{\frac{{\rm i}\,2\,\mathcal{H}_{\rm m}t}{\hbar}}\left[\vec{\alpha}(0)-\mathcal{H}^{-1}_{\rm m}\left[c p \cos\vartheta - m\,c^2\;  \sin\vartheta\right]\hat{p}\right]=
                 \left[\vec{\alpha}(0)-\mathcal{H}^{-1}_{\rm m}\left[c p \cos\vartheta - m\,c^2\;  \sin\vartheta\right]\hat{p}\right] {\rm e}^{\frac{-{\rm i}\,2\,\mathcal{H}_{\rm m}t}{\hbar}}.
            \end{eqnarray}
            Therefore we have
            \begin{eqnarray}\label{eq:alpha-2}
                  \vec{\alpha}(t)=\mathcal{H}^{-1}_{\rm m}\left[c p \cos\vartheta - m\,c^2\;  \sin\vartheta\right]\hat{p}+ \left[\vec{\alpha}(0)-\mathcal{H}^{-1}_{\rm m}\left[c p \cos\vartheta - m\,c^2\;  \sin\vartheta\right]\hat{p}\right]{\rm e}^{\frac{-{\rm i}\,2\,H_{\rm m}t}{\hbar}}.
            \end{eqnarray}
            Based on Eq. (\ref{eq:alpha-2}), we have
            \begin{eqnarray}
                  &&\frac{1}{p} \vec{\alpha} -   \left(\vec{\alpha}\cdot\vec{p}\right) \dfrac{\vec{p}}{p^3}=\frac{1}{p} \vec{\alpha} -   \dfrac{1}{p}\left(\vec{\alpha}\cdot\hat{p}\right) \hat{p}\nonumber\\
                  &=& \frac{1}{p} \left\{\mathcal{H}^{-1}_{\rm m}\left[c p \cos\vartheta - m\,c^2\;  \sin\vartheta\right]\hat{p}+ \left[\vec{\alpha}(0)-\mathcal{H}^{-1}_{\rm m}\left[c p \cos\vartheta - m\,c^2\;  \sin\vartheta\right]\hat{p}\right]{\rm e}^{\frac{-{\rm i}\,2\,H_{\rm m}t}{\hbar}}\right\}\nonumber\\
                  &&-\frac{1}{p} \left\{\mathcal{H}^{-1}_{\rm m}\left[c p \cos\vartheta - m\,c^2\;  \sin\vartheta\right]\hat{p}+ \left[\vec{\alpha}(0)-\mathcal{H}^{-1}_{\rm m}\left[c p \cos\vartheta - m\,c^2\;  \sin\vartheta\right]\hat{p}\right]{\rm e}^{\frac{-{\rm i}\,2\,H_{\rm m}t}{\hbar}}\right\}\cdot\hat{p} \;\hat{p}\nonumber\\
                        &=& \frac{1}{p} \left\{ \left[\vec{\alpha}(0)-\mathcal{H}^{-1}_{\rm m}\left[c p \cos\vartheta - m\,c^2\;  \sin\vartheta\right]\hat{p}\right]{\rm e}^{\frac{-{\rm i}\,2\,H_{\rm m}t}{\hbar}}\right\}\nonumber\\
                  &&-\frac{1}{p} \left\{\left[\vec{\alpha}(0)-\mathcal{H}^{-1}_{\rm m}\left[c p \cos\vartheta - m\,c^2\;  \sin\vartheta\right]\hat{p}\right]{\rm e}^{\frac{-{\rm i}\,2\,H_{\rm m}t}{\hbar}}\right\}\cdot\hat{p} \;\hat{p}\nonumber\\
                  &=& \frac{1}{p} \left\{ \left[\vec{\alpha}(0)\right]{\rm e}^{\frac{-{\rm i}\,2\,H_{\rm m}t}{\hbar}}\right\}-\frac{1}{p} \left\{ \left[\vec{\alpha}(0)\right]{\rm e}^{\frac{-{\rm i}\,2\,H_{\rm m}t}{\hbar}}\right\}\cdot\hat{p} \;\hat{p}\nonumber\\
                  &=& \frac{1}{p}\,\left[\vec{\alpha}(0)-(\vec{\alpha}(0)\cdot\hat{p})\hat{p}\right]{\rm e}^{\frac{-{\rm i}\,2\,H_{\rm m}t}{\hbar}},
                  \end{eqnarray}
            and
            \begin{eqnarray}
            &&\left[\mathcal{H}_{\rm m}-
                        \vec{\alpha}\cdot\hat{p} \left(c p \cos\vartheta - m\,c^2\;  \sin\vartheta \right)\right] \dfrac{\vec{p}}{p}
                        =\left[\mathcal{H}_{\rm m}-
                        \vec{\alpha}\cdot\hat{p} \left(c p \cos\vartheta - m\,c^2\;  \sin\vartheta \right)\right]  \hat{p}\nonumber\\
            &=& \mathcal{H}_{\rm m} \hat{p}- \left\{\mathcal{H}^{-1}_{\rm m}\left[c p \cos\vartheta - m\,c^2\;  \sin\vartheta\right]\hat{p}+ \left[\vec{\alpha}(0)-\mathcal{H}^{-1}_{\rm m}\left[c p \cos\vartheta - m\,c^2\;  \sin\vartheta\right]\hat{p}\right]{\rm e}^{\frac{-{\rm i}\,2\,H_{\rm m}t}{\hbar}}\right\}\cdot\hat{p} \left(c p \cos\vartheta - m\,c^2\;  \sin\vartheta \right)  \hat{p}\nonumber\\
            &=& \mathcal{H}_{\rm m} \hat{p}-\left\{\mathcal{H}^{-1}_{\rm m}\left[c p \cos\vartheta - m\,c^2\;  \sin\vartheta\right]+ \left[\vec{\alpha}(0)\cdot\hat{p}-\mathcal{H}^{-1}_{\rm m}\left[c p \cos\vartheta - m\,c^2\;  \sin\vartheta\right]\right]{\rm e}^{\frac{-{\rm i}\,2\,H_{\rm m}t}{\hbar}}\right\}\left(c p \cos\vartheta - m\,c^2\;  \sin\vartheta \right)  \hat{p}\nonumber\\
            &=& \mathcal{H}_{\rm m} \hat{p}- \mathcal{H}^{-1}_{\rm m}\left[c p \cos\vartheta - m\,c^2\;  \sin\vartheta\right]^2 \hat{p} -\left\{\left[\vec{\alpha}(0)\cdot\hat{p}-\mathcal{H}^{-1}_{\rm m}\left[c p \cos\vartheta - m\,c^2\;  \sin\vartheta\right]\right]{\rm e}^{\frac{-{\rm i}\,2\,H_{\rm m}t}{\hbar}}\right\}\left(c p \cos\vartheta - m\,c^2\;  \sin\vartheta \right)  \hat{p}\nonumber\\
            &=& \left\{\frac{\mathcal{H}^2_{\rm m} - \left[c p \cos\vartheta - m\,c^2\;  \sin\vartheta\right]^2}{\mathcal{H}_{\rm m}} \right\}\hat{p} -\left\{\left[\vec{\alpha}(0)\cdot\hat{p}-\mathcal{H}^{-1}_{\rm m}\left[c p \cos\vartheta - m\,c^2\;  \sin\vartheta\right]\right]{\rm e}^{\frac{-{\rm i}\,2\,H_{\rm m}t}{\hbar}}\right\}\left(c p \cos\vartheta - m\,c^2\;  \sin\vartheta \right)  \hat{p}.\nonumber\\
            \end{eqnarray}
            After substituting above results into Eq. (\ref{eq:alpha-1}), we have
            \begin{eqnarray}\label{eq:alpha-3}
                       && \dfrac{{\rm d}\,\vec{r}}{{\rm d}\,t}\nonumber\\
                        &=& \cos\vartheta \; (c\vec{\alpha})+ \sin\vartheta \; \left[-m c^2\;\left(\frac{1}{p} \vec{\alpha} -  \left(\vec{\alpha}\cdot\vec{p}\right) \dfrac{\vec{p}}{p^3}\right)+
                        \frac{c}{\left(m\,c^2\; \cos\vartheta +cp  \sin\vartheta \right)} \left[ \mathcal{H}_{\rm m}-
                        \vec{\alpha}\cdot\hat{p} \left(c p \cos\vartheta - m\,c^2\;  \sin\vartheta \right)\right] \dfrac{\vec{p}}{p}\right]\nonumber\\
            &=& \cos\vartheta \; c\left(\mathcal{H}^{-1}_{\rm m}\left[c p \cos\vartheta - m\,c^2\;  \sin\vartheta\right]\hat{p}+ \left[\vec{\alpha}(0)-\mathcal{H}^{-1}_{\rm m}\left[c p \cos\vartheta - m\,c^2\;  \sin\vartheta\right]\hat{p}\right]{\rm e}^{\frac{-{\rm i}\,2\,H_{\rm m}t}{\hbar}} \right)\nonumber\\
            &&+ \sin\vartheta \; \left[-m c^2\;\left(\frac{1}{p}  \left[\vec{\alpha}(0)-(\vec{\alpha}(0)\cdot\hat{p})\hat{p}\right]{\rm e}^{\frac{-{\rm i}\,2\,H_{\rm m}t}{\hbar}}\right)\right]\nonumber\\
            &&+\frac{\sin\vartheta \;c}{\left(m\,c^2\; \cos\vartheta +cp  \sin\vartheta \right)}\times \nonumber\\
            &&\left[\frac{\mathcal{H}^2_{\rm m} - \left[c p \cos\vartheta - m\,c^2\;  \sin\vartheta\right]^2}{\mathcal{H}_{\rm m}} \hat{p} -\left\{ \left[\vec{\alpha}(0)\cdot\hat{p}-\mathcal{H}^{-1}_{\rm m}\left[c p \cos\vartheta - m\,c^2\;  \sin\vartheta\right]\right]{\rm e}^{\frac{-{\rm i}\,2\,H_{\rm m}t}{\hbar}}\right\}\left(c p \cos\vartheta - m\,c^2\;  \sin\vartheta \right)  \hat{p} \right].\nonumber\\
            \end{eqnarray}
            One can have
            \begin{eqnarray}
            &&\cos\vartheta \; c\left(\mathcal{H}^{-1}_{\rm m}\left[c p \cos\vartheta - m\,c^2\;  \sin\vartheta\right]\hat{p}\right)
            +\frac{\sin\vartheta \;c}{\left(m\,c^2\; \cos\vartheta +cp  \sin\vartheta \right)}\times \left[\frac{\mathcal{H}^2_{\rm m} - \left[c p \cos\vartheta - m\,c^2\;  \sin\vartheta\right]^2}{\mathcal{H}_{\rm m}} \hat{p}\right]\nonumber\\
            &=&c\hat{p} \; \frac{ \cos\vartheta \left[c p \cos\vartheta - m\,c^2\;  \sin\vartheta\right]\left(m\,c^2\; \cos\vartheta +cp  \sin\vartheta \right)+ \sin\vartheta \left[\mathcal{H}^2_{\rm m} - \left[c p \cos\vartheta - m\,c^2\;  \sin\vartheta\right]^2\right]}{\mathcal{H}_{\rm m}\; \left(m\,c^2\; \cos\vartheta +cp  \sin\vartheta \right)}\nonumber\\
            &=&c\hat{p} \; \frac{ c p ( m c^2 \cos\vartheta + cp \sin\vartheta)}{\mathcal{H}_{\rm m}\; \left(m\,c^2 \cos\vartheta +cp  \sin\vartheta \right)}\nonumber\\
            &=&c\hat{p} \; \frac{ c  p }{\mathcal{H}_{\rm m}}=c^2 \mathcal{H}^{-1}_{\rm m} \vec{p},
            \end{eqnarray}
            which does not depend on the parameter $\vartheta$; here we have used the REM relation $\mathcal{H}_{\rm m}^2= c^2p^2+m^2c^4$.

            Furthermore, one can have
            \begin{eqnarray}
            &&  \left[\cos\vartheta \; c - \frac{1}{p} mc^2 \sin\vartheta \right] {\vec{\alpha}(0)\;{\rm e}^{\frac{-{\rm i}\,2\,H_{\rm m}t}{\hbar}}}=
            \frac{1}{p}\left[cp\; \cos\vartheta- mc^2 \sin\vartheta \right] {\vec{\alpha}(0)\;{\rm e}^{\frac{-{\rm i}\,2\,H_{\rm m}t}{\hbar}}},
            \end{eqnarray}

            \begin{eqnarray}
            &&  \left[\frac{1}{p} mc^2 \sin\vartheta - \frac{\sin\vartheta \;c}{\left(m\,c^2\; \cos\vartheta +cp  \sin\vartheta \right)}\times \left(c p \cos\vartheta - m\,c^2\;  \sin\vartheta \right) \right] {\left[\vec{\alpha}(0)\cdot\hat{p}\right]\; \hat{p}\;{\rm e}^{\frac{-{\rm i}\,2\,H_{\rm m}t}{\hbar}}}\nonumber\\
            &=&
            \frac{1}{p\left(m\,c^2\; \cos\vartheta +cp  \sin\vartheta \right)}\left[ mc^2 \sin\vartheta \left(m\,c^2\; \cos\vartheta +cp  \sin\vartheta \right)-
            cp \sin\vartheta \;\left(c p \cos\vartheta - m\,c^2\;  \sin\vartheta \right) \right]{\left[\vec{\alpha}(0)\cdot\hat{p}\right]\; \hat{p}\;{\rm e}^{\frac{-{\rm i}\,2\,H_{\rm m}t}{\hbar}}}\nonumber\\
            &=&
            \frac{\sin\vartheta \cos\vartheta\left[ m^2c^4-c^2 p^2  \right]+2cp\; mc^2 \sin^2\vartheta}{p\left(m\,c^2\; \cos\vartheta +cp  \sin\vartheta \right)}
            {\left[\vec{\alpha}(0)\cdot\hat{p}\right]\; \hat{p}\;{\rm e}^{\frac{-{\rm i}\,2\,H_{\rm m}t}{\hbar}}}\nonumber\\
            &=&
            \frac{\sin\vartheta \cos\vartheta\left[ m^2c^4-c^2 p^2  \right]+2cp\; mc^2 \sin^2\vartheta}{\left(m\,c^2\; \cos\vartheta +cp  \sin\vartheta \right)}
            {\left[\vec{\alpha}(0)\cdot\vec{p}\right]\; \frac{\vec{p}}{p^3}\;{\rm e}^{\frac{-{\rm i}\,2\,H_{\rm m}t}{\hbar}}},
            \end{eqnarray}
            and
            \begin{eqnarray}
            &&\left[-\cos\vartheta \; c + \frac{\sin\vartheta \;c}{\left(m\,c^2\; \cos\vartheta +cp  \sin\vartheta \right)}\times \left(c p \cos\vartheta - m\,c^2\;  \sin\vartheta \right) \right] {\left[\mathcal{H}^{-1}_{\rm m}\left[c p \cos\vartheta - m\,c^2\;  \sin\vartheta\right]\hat{p}\right]{\rm e}^{\frac{-{\rm i}\,2\,H_{\rm m}t}{\hbar}}}\nonumber\\
            &=& \frac{c \; \left[-\cos\vartheta \left(m\,c^2\; \cos\vartheta +cp  \sin\vartheta \right) + \sin\vartheta \left(c p \cos\vartheta - m\,c^2\;  \sin\vartheta \right) \right]}{\left(m\,c^2\; \cos\vartheta +cp  \sin\vartheta \right)} {\left[\mathcal{H}^{-1}_{\rm m}\left[c p \cos\vartheta - m\,c^2\;  \sin\vartheta\right]\hat{p}\right]{\rm e}^{\frac{-{\rm i}\,2\,H_{\rm m}t}{\hbar}}}\nonumber\\
            &=& -\frac{c \; mc^2}{\left(m\,c^2\; \cos\vartheta +cp  \sin\vartheta \right)} {\left[\mathcal{H}^{-1}_{\rm m}\left[c p \cos\vartheta - m\,c^2\;  \sin\vartheta\right]\frac{\vec{p}}{p}\right]{\rm e}^{\frac{-{\rm i}\,2\,H_{\rm m}t}{\hbar}}}.
            \end{eqnarray}
            Then from Eq. (\ref{eq:alpha-3}) we obtain
            \begin{eqnarray}\label{eq:r-1}
            \dfrac{{\rm d}\,\vec{r}}{{\rm d}\,t}
            &=& c^2 \mathcal{H}^{-1}_{\rm m} \vec{p}+\frac{1}{p}\left[cp\; \cos\vartheta- mc^2 \sin\vartheta \right] {\vec{\alpha}(0)\;{\rm e}^{\frac{-{\rm i}\,2\,H_{\rm m}t}{\hbar}}}\nonumber\\
            &&+\frac{\sin\vartheta \cos\vartheta\left[ m^2c^4-c^2 p^2  \right]+2 cp\; mc^2 \sin^2\vartheta}{\left(m\,c^2\; \cos\vartheta +cp  \sin\vartheta \right)}
           {\left[\vec{\alpha}(0)\cdot\vec{p}\right]\; \frac{\vec{p}}{p^3}\;{\rm e}^{\frac{-{\rm i}\,2\,H_{\rm m}t}{\hbar}}}
            \nonumber\\
            && -\frac{c \; mc^2}{\left(m\,c^2\; \cos\vartheta +cp  \sin\vartheta \right)} {\left[\mathcal{H}^{-1}_{\rm m}\left[c p \cos\vartheta - m\,c^2\;  \sin\vartheta\right]\frac{\vec{p}}{p}\right]{\rm e}^{\frac{-{\rm i}\,2\,H_{\rm m}t}{\hbar}}}
            \nonumber\\
            &=& c^2 \mathcal{H}^{-1}_{\rm m} \vec{p}+\Biggr\{\frac{1}{p}\left[cp\; \cos\vartheta- mc^2 \sin\vartheta \right] {\vec{\alpha}(0)
            }+\frac{\sin\vartheta \cos\vartheta\left[ m^2c^4-c^2 p^2  \right]+2 cp\; mc^2 \sin^2\vartheta}{\left(m\,c^2\; \cos\vartheta +cp  \sin\vartheta \right)}{\left[\vec{\alpha}(0)\cdot\vec{p}\right]\; \frac{\vec{p}}{p^3}}\nonumber\\
            &&-
            \frac{c \; mc^2 \left[c p \cos\vartheta-m\,c^2\;  \sin\vartheta \right]}{ \left(m\,c^2\; \cos\vartheta +cp  \sin\vartheta \right)} {\left[\mathcal{H}^{-1}_{\rm m}\frac{\vec{p}}{p}\right]\Biggr\}{\rm e}^{\frac{-{\rm i}\,2\,H_{\rm m}t}{\hbar}}}.
            \end{eqnarray}

            \begin{remark}
            When $\vartheta=0$, Eq. (\ref{eq:r-1}) reduces to the case of Dirac's electron as shown in Eq. (\ref{eq:D-9}), i.e.,
            \begin{eqnarray}\label{eq:D-9-r}
                  \frac{{\rm d}\,\vec{r}}{{\rm d}\,t}=c^2 H^{-1}_{\rm e}\vec{p}+\left[c\vec{\alpha}(0)-c^2 H^{-1}_{\rm e}\vec{p}\right]{\rm e}^{\frac{-{\rm i}\,2\,H_{\rm e}t}{\hbar}},
            \end{eqnarray}
            and when $\vartheta=\pi/2$, Eq. (\ref{eq:r-1}) reduces to the case of type-I Dirac's braidon as shown in Eq. (\ref{eq:E-1c}), i.e.,
            \begin{eqnarray}\label{eq:E-1r}
            \dfrac{{\rm d}\,\vec{r}}{{\rm d}\,t}
            &=& c^2 H_{\rm b}^{-1} \vec{p}+m c^2\;\left\{-\frac{1}{p} \vec{\alpha}(0)+2 \left[\vec{\alpha}(0)\cdot \vec{p}\right] \dfrac{\vec{p}}{p^3} +m\,c^2 H_{\rm b}^{-1}\,\dfrac{\vec{p}}{p^2}\right\}{\rm e}^{\frac{-{\rm i}\,2\,H_{\rm b}\,t}{\hbar}}.
            \end{eqnarray}
            $\blacksquare$
            \end{remark}

            Then from Eq. (\ref{eq:r-1}) we obtain
            \begin{eqnarray}\label{eq:r-2a}
            \vec{r}(t)
            &=& \vec{r}(0)+c^2 \mathcal{H}^{-1}_{\rm m} \vec{p} \; t+\Biggr\{\frac{1}{p}\left[cp\; \cos\vartheta- mc^2 \sin\vartheta \right] {\vec{\alpha}(0)
            }+\frac{\sin\vartheta \cos\vartheta\left[ m^2c^4-c^2 p^2  \right]+2 cp\; mc^2 \sin^2\vartheta}{\left(m\,c^2\; \cos\vartheta +cp  \sin\vartheta \right)}{\left[\vec{\alpha}(0)\cdot\vec{p}\right]\; \frac{\vec{p}}{p^3}}\nonumber\\
            &&-
            \frac{c \; mc^2 \left[c p \cos\vartheta-m\,c^2\;  \sin\vartheta \right]}{ \left(m\,c^2\; \cos\vartheta +cp  \sin\vartheta \right)} {\left[\mathcal{H}^{-1}_{\rm m}\frac{\vec{p}}{p}\right]\Biggr\}\frac{\hbar}{-{\rm i}\,2\,\mathcal{H}_{\rm m}}\left({\rm e}^{\frac{-{\rm i}\,2\,\mathcal{H}_{\rm m}t}{\hbar}}-1\right)}.
            \end{eqnarray}
            Let us define the third term of Eq. (\ref{eq:r-2a}) as the ``position Zitterbewegung'' operator, i.e.,
            \begin{eqnarray}\label{eq:r-2}
            \hat{\mathcal{Z}}_{\rm m}
            &=& \Biggr\{\frac{1}{p}\left[cp\; \cos\vartheta- mc^2 \sin\vartheta \right] {\vec{\alpha}(0)
            }+\frac{\sin\vartheta \cos\vartheta\left[ m^2c^4-c^2 p^2  \right]+2 cp\; mc^2 \sin^2\vartheta}{\left(m\,c^2\; \cos\vartheta +cp  \sin\vartheta \right)}{\left[\vec{\alpha}(0)\cdot\vec{p}\right]\; \frac{\vec{p}}{p^3}}\nonumber\\
            &&-
            \frac{c \; mc^2 \left[c p \cos\vartheta-m\,c^2\;  \sin\vartheta \right]}{ \left(m\,c^2\; \cos\vartheta +cp  \sin\vartheta \right)} \left[\mathcal{H}^{-1}_{\rm m}\frac{\vec{p}}{p}\right]\Biggr\}\frac{\hbar}{-{\rm i}\,2\,\mathcal{H}_{\rm m}}\left({\rm e}^{\frac{-{\rm i}\,2\,\mathcal{H}_{\rm m}t}{\hbar}}-1\right).
            \end{eqnarray}
            Because $\vec{r}(t)=\left(\vec{r}(t)\right)^\dagger$, then $\hat{\mathcal{Z}}_{\rm m}=\hat{\mathcal{Z}}^\dagger_{\rm m}$ is a Hermitian operator, which implies that
            \begin{eqnarray}\label{eq:r-3}
            && \mathcal{H}_{\rm m}\; \Biggr[ \frac{1}{p}\left[cp\; \cos\vartheta- mc^2 \sin\vartheta \right] {\vec{\alpha}(0)
            }+\frac{\sin\vartheta \cos\vartheta\left[ m^2c^4-c^2 p^2  \right]+2 cp\; mc^2 \sin^2\vartheta}{\left(m\,c^2\; \cos\vartheta +cp  \sin\vartheta \right)}{\left[\vec{\alpha}(0)\cdot\vec{p}\right]\; \frac{\vec{p}}{p^3}}\nonumber\\
            &&-
            \frac{c \; mc^2 \left[c p \cos\vartheta-m\,c^2\;  \sin\vartheta \right]}{ \left(m\,c^2\; \cos\vartheta +cp  \sin\vartheta \right)} \left[\mathcal{H}^{-1}_{\rm m}\frac{\vec{p}}{p}\right]\Biggr]\nonumber\\
            &=& -\Biggr[ \frac{1}{p}\left[cp\; \cos\vartheta- mc^2 \sin\vartheta \right] {\vec{\alpha}(0)
            }+\frac{\sin\vartheta \cos\vartheta\left[ m^2c^4-c^2 p^2  \right]+2 cp\; mc^2 \sin^2\vartheta}{\left(m\,c^2\; \cos\vartheta +cp  \sin\vartheta \right)}{\left[\vec{\alpha}(0)\cdot\vec{p}\right]\; \frac{\vec{p}}{p^3}}\nonumber\\
            &&-
            \frac{c \; mc^2 \left[c p \cos\vartheta-m\,c^2\;  \sin\vartheta \right]}{ \left(m\,c^2\; \cos\vartheta +cp  \sin\vartheta \right)} \left[\mathcal{H}^{-1}_{\rm m}\frac{\vec{p}}{p}\right]\Biggr]\; \mathcal{H}_{\rm m}.
            \end{eqnarray}
            Based on which one can have
            \begin{eqnarray}
            && \mathcal{H}_{\rm m}\hat{\mathcal{Z}}_{\rm m}+\hat{\mathcal{Z}}_{\rm m} \mathcal{H}_{\rm m}=0.
            \end{eqnarray}

            Let us introduce the following projection operators
            \begin{eqnarray}
                  \Pi_\pm=\frac{1}{2}\left( \mathbb{I}\pm \frac{\mathcal{H}_{\rm m}}{\sqrt{p^2c^2+m^2c^4}}\right), \;\;\;\;\; \Pi_\pm^2=\Pi_\pm,
            \end{eqnarray}
            we easily have
            \begin{eqnarray}
            && \Pi_+ |\Psi'_1\rangle = |\Psi'_1\rangle, \;\; \Pi_+ |\Psi'_2\rangle = |\Psi'_2\rangle, \;\; \Pi_+ |\Psi'_3\rangle = 0, \;\; \Pi_+ |\Psi_4\rangle = 0, \nonumber\\
            && \Pi_- |\Psi'_1\rangle = 0, \;\; \Pi_- |\Psi'_2\rangle = 0, \;\; \Pi_- |\Psi'_3\rangle = |\Psi'_3\rangle, \;\; \Pi_- |\Psi'_4\rangle = |\Psi'_4\rangle.
            \end{eqnarray}
            We then have
            \begin{eqnarray}
               \Pi_+ \; \hat{\mathcal{Z}}_{\rm m} \;\Pi_+ &=&
                  \frac{1}{4} \left(\mathbb{I}+\frac{\mathcal{H}_{\rm m}}{\sqrt{p^2c^2+m^2c^4}}\right) \;\hat{\mathcal{Z}}_{\rm m}\;
                  \left(\mathbb{I} + \frac{\mathcal{H}_{\rm m}}{\sqrt{p^2c^2+m^2c^4}}\right)\nonumber\\
            &=& \frac{1}{4} \left[\hat{\mathcal{Z}}_{\rm m}+ \frac{\mathcal{H}_{\rm m}\hat{\mathcal{Z}}_{\rm m}+\hat{\mathcal{Z}}_{\rm m}\mathcal{H}_{\rm m} }{\sqrt{p^2c^2+m^2c^4}}
            +\frac{1}{p^2c^2+m^2c^4} \mathcal{H}_{\rm m} \hat{\mathcal{Z}}_{\rm m}\mathcal{H}_{\rm m} \right]\nonumber\\
            &=& \frac{1}{4} \left[\hat{\mathcal{Z}}_{\rm m}-\frac{\mathcal{H}^2_{\rm m}}{p^2c^2+m^2c^4} \hat{\mathcal{Z}}_{\rm m} \right]\nonumber\\
            &=&\frac{1}{4} \left[\hat{\mathcal{Z}}_{\rm m}- \hat{\mathcal{Z}}_{\rm m} \right]=0.
            \end{eqnarray}
            Similarly, we have
            \begin{eqnarray}
            &&    \Pi_- \; \hat{\mathcal{Z}}_{\rm m} \;\Pi_- =0.
            \end{eqnarray}
            Therefore we have
            \begin{eqnarray}
            &&    \langle\Psi_+| \; \hat{\mathcal{Z}}_{\rm m} \;|\Psi_+\rangle =0, \nonumber\\
            &&     \langle \Psi_-| \; \hat{\mathcal{Z}}_{\rm m} \;|\Psi_-\rangle=0,
            \end{eqnarray}
            which means that there exists no phenomenon of position Zitterbewegung if the mixed Hamiltonian system (i.e., $\mathcal{H}_{\rm m}=\cos\vartheta \;H_{\rm e} +  \sin\vartheta \;H_{\rm b}^{\rm I}$) is in a superposition state of only positive energy (i.e., $|\Psi_+\rangle=\cos\theta |\Psi'_1\rangle+ \sin\theta |\Psi'_2\rangle$), or in a in a superposition state of only negative energy (i.e., $|\Psi_-\rangle=\cos\theta |\Psi'_3\rangle+ \sin\theta |\Psi'_4\rangle$).

            Now suppose the mixed Dirac's Hamiltonian system is in a superposition state of $|\Psi'_1\rangle$ and $|\Psi'_3\rangle$, i.e.,
            \begin{eqnarray}
            &&   |\Psi'\rangle=\cos\eta |\Psi'_1\rangle+ \sin\eta |\Psi'_3\rangle.
            \end{eqnarray}
            To calculate the Zitterbewegung, we only need to consider the term $\langle\Psi'_1|\hat{\mathcal{Z}}_{\rm m}|\Psi'_3 \rangle$. We can have
            \begin{eqnarray}
            &&  {\mathcal{Z}}_{\rm m}= \sin(2\eta){\rm Re}\left(
                  \langle\Psi'_1|\hat{\mathcal{Z}}_{\rm m}|\Psi'_3 \rangle\right),
            \end{eqnarray}
            where
            \begin{eqnarray}
                 && \langle\Psi'_1|\hat{\mathcal{Z}}_{\rm m}|\Psi'_3 \rangle\nonumber\\
                  &=&\langle\Psi'_1| \Biggr\{\frac{1}{p}\left[cp\; \cos\vartheta
                        - mc^2 \sin\vartheta \right] {\vec{\alpha}(0)}
                        +\frac{\sin\vartheta \cos\vartheta(m^2c^4-c^2 p^2)
                              +2cp\;mc^2 \sin^2\vartheta}{\left(m\,c^2\;\cos\vartheta
                                    +cp\sin\vartheta\right)}\left[
                                          \vec{\alpha}(0)\cdot\vec{p}\right]\;\frac{\vec{p}}{p^3}\nonumber\\
                        &&-\frac{c \; mc^2 \left[c p \cos\vartheta-m\,c^2\;  \sin\vartheta \right]}{ \left(m\,c^2\; \cos\vartheta +cp  \sin\vartheta \right)} \left[\mathcal{H}^{-1}_{\rm m}\frac{\vec{p}}{p}\right]\Biggr\}\frac{\hbar}{-{\rm i}\,2\,\mathcal{H}_{\rm m}}\left({\rm e}^{\frac{-{\rm i}\,2\,\mathcal{H}_{\rm m}t}{\hbar}}-1\right)|\Psi'_3 \rangle\nonumber\\
                  &=& \langle\Psi'_1| \Biggr\{\frac{1}{p}\left[cp\; \cos\vartheta
                        - mc^2 \sin\vartheta \right] {\vec{\alpha}(0)}
                        +\frac{\sin\vartheta \cos\vartheta\left[ m^2c^4-c^2 p^2  \right]+2 cp\; mc^2 \sin^2\vartheta}{\left(m\,c^2\; \cos\vartheta +cp  \sin\vartheta \right)}{\left[\vec{\alpha}(0)\cdot\vec{p}\right]\; \frac{\vec{p}}{p^3}}\nonumber\\
                        &&-\frac{c \; mc^2 \left[c p \cos\vartheta-m\,c^2\;  \sin\vartheta \right]}{ \left(m\,c^2\; \cos\vartheta +cp  \sin\vartheta \right)} \left[-{E}^{-1}_{\rm m}\frac{\vec{p}}{p}\right]\Biggr\}\frac{\hbar}{{\rm i}\,2\,{E}_{\rm m}}\left({\rm e}^{\frac{{\rm i}\,2\,{E}_{\rm m}t}{\hbar}}-1\right)|\Psi'_3 \rangle\nonumber\\
                  &=& \frac{\hbar}{{\rm i}\,2\,{E}_{\rm m}}\left({\rm e}^{\frac{{\rm i}\,2\,{E}_{\rm m}t}{\hbar}}-1\right)\langle\Psi'_1| \Biggr\{ \notag
                        \\
                        &&\qquad\qquad \frac{1}{p}\left[cp\; \cos\vartheta
                        - mc^2 \sin\vartheta \right] {\vec{\alpha}(0)}
                        +\frac{\sin\vartheta \cos\vartheta\left[ m^2c^4-c^2 p^2  \right]+2 cp\; mc^2 \sin^2\vartheta}{\left(m\,c^2\; \cos\vartheta +cp  \sin\vartheta \right)}{\left[\vec{\alpha}(0)\cdot\vec{p}\right]\; \frac{\vec{p}}{p^3}}\Biggr\}|\Psi'_3 \rangle,
            \end{eqnarray}
            where in the last step we have used $\langle\Psi'_1|\Psi'_3 \rangle=0$. Because
            \begin{eqnarray}
                   \vec{\alpha}(\vec{\alpha}\cdot\hat{p})
                              +(\vec{\alpha}\cdot\hat{p})\vec{\alpha}
                  &=& 2\hat{p}, \notag \\
                   \vec{\alpha}\cdot\hat{p}\; \alpha_z\vec{\alpha}\cdot\hat{p}
                  &=&(\alpha_x \hat{p}_x+\alpha_y \hat{p}_y+\alpha_z \hat{p}_z)\; \alpha_z\;(\alpha_x \hat{p}_x+\alpha_y \hat{p}_y+\alpha_z \hat{p}_z)\nonumber\\
                  & =&-\alpha_z (\alpha_x \hat{p}_x +\alpha_y \hat{p}_y
                              +\alpha_z \hat{p}_z)\;(\alpha_x \hat{p}_x
                                    +\alpha_y \hat{p}_y +\alpha_z \hat{p}_z)
                        +2 \alpha_z \alpha_z \hat{p}_z \; (\alpha_x \hat{p}_x
                              +\alpha_y \hat{p}_y+\alpha_z \hat{p}_z) \nonumber\\
                  & =&-\alpha_z(\vec{\alpha}\cdot\hat{p})\; (\vec{\alpha}\cdot\hat{p})
                        +2 \,\hat{p}_z \; (\vec{\alpha}\cdot\hat{p})
                  =-\alpha_z +2 \,\hat{p}_z \; (\vec{\alpha}\cdot\hat{p}), \notag \\
                   (\vec{\alpha}\cdot\hat{p})\vec{\alpha}(\vec{\alpha}\cdot\hat{p})
                  &=&-\vec{\alpha}+2\,\hat{p}(\vec{\alpha}\cdot\hat{p}),
            \end{eqnarray}
            and
            \begin{eqnarray}
                  \beta=\frac{1}{m\,c^2} \big[H_{\rm e} -c(\vec{\alpha}\cdot\vec{p})
                        \big],
            \end{eqnarray}
            then we obtain
            \begin{eqnarray}
                  \mathcal{C}^\dagger(\vartheta)\;\vec{\alpha}\;\mathcal{C}(\vartheta)
                  &=&\Bigl(\cos\frac{\vartheta}{2}\;\mathbb{I}-\sin\frac{\vartheta}{2}\;
                        \beta\vec{\alpha}\cdot\hat{p}\Bigr)\vec{\alpha}\Bigl(
                              \cos\frac{\vartheta}{2}\;\mathbb{I}
                              +\sin\frac{\vartheta}{2}\;\beta\vec{\alpha}\cdot\hat{p}
                              \Bigr) \notag \\
                 & =& {\cos^2}\frac{\vartheta}{2}\,\vec{\alpha}
                        -{\sin^2}\frac{\vartheta}{2} (\beta\vec{\alpha}\cdot\hat{p})
                              \vec{\alpha}(\beta\vec{\alpha}\cdot\hat{p})
                        +\sin\frac{\vartheta}{2}\,\cos\frac{\vartheta}{2} \Bigl[
                              \vec{\alpha}(\beta\vec{\alpha}\cdot\hat{p})
                              -(\beta\vec{\alpha}\cdot\hat{p})\vec{\alpha}\Bigr] \notag
                        \\
                  &=& {\cos^2}\frac{\vartheta}{2}\,\vec{\alpha}
                        -{\sin^2}\frac{\vartheta}{2} (\vec{\alpha}\cdot\hat{p})
                              \vec{\alpha}(\vec{\alpha}\cdot\hat{p})
                        -\sin\frac{\vartheta}{2}\,\cos\frac{\vartheta}{2} \beta\Bigl[
                              \vec{\alpha}(\vec{\alpha}\cdot\hat{p})
                              +(\vec{\alpha}\cdot\hat{p})\vec{\alpha}\Bigr] \notag
                        \\
                  &=& {\cos^2}\frac{\vartheta}{2}\,\vec{\alpha}
                        -{\sin^2}\frac{\vartheta}{2} \big[-\vec{\alpha}
                              +2\,\hat{p}(\vec{\alpha}\cdot\hat{p})\big]
                        -2\,\sin\frac{\vartheta}{2}\,\cos\frac{\vartheta}{2} \beta\,
                              \hat{p} \notag \\
                 & =& \vec{\alpha}-2\,\Bigl({\sin^2}\frac{\vartheta}{2}\Bigr)(
                              \vec{\alpha}\cdot\hat{p})\hat{p}
                        -(\sin\vartheta)\beta\,\hat{p}\nonumber\\
                 &=&\vec{\alpha}-2\,\Bigl({\sin^2}\frac{\vartheta}{2}\Bigr)(
                              \vec{\alpha}\cdot\hat{p})\hat{p}
                        -(\sin\vartheta)\frac{1}{m\,c^2} \big[H_{\rm e}
                              -c(\vec{\alpha}\cdot\vec{p})\big]\hat{p} \notag \\
                  &=& \vec{\alpha}-\sin\vartheta\,\frac{1}{m\,c^2} H_{\rm e} \hat{p}
                        +\bigg(\dfrac{p}{m\,c} \sin\vartheta
                              -2\,{\sin^2}\frac{\vartheta}{2}\bigg)(
                                    \vec{\alpha}\cdot\hat{p})\hat{p}.
            \end{eqnarray}
            Moreover,
            \begin{eqnarray}
                   \mathcal{C}^\dagger (\vartheta) \left(\vec{\alpha}\cdot\vec{p}
                        \right)\frac{\vec{p}}{p^3}\mathcal{C}(\vartheta)
                  &=& \Bigl\{\big[\mathcal{C}^\dagger (\vartheta)\,\vec{\alpha}\,
                        \mathcal{C}(\vartheta)\big]\cdot\vec{p}\Bigr\}
                              \frac{\vec{p}}{p^3} \nonumber\\
                  &=& \Biggl\{\bigg[\vec{\alpha}
                        -\sin\vartheta\,\frac{1}{m\,c^2} H_{\rm e} \hat{p}
                        +\Big(\dfrac{p}{m\,c} \sin\vartheta
                              -2\,{\sin^2}\frac{\vartheta}{2}\Big)(
                                    \vec{\alpha}\cdot\hat{p})\hat{p}\bigg]\cdot\vec{p}
                        \Biggr\}\frac{\vec{p}}{p^3} \notag \\
                 & =& \bigg[(\vec{\alpha}\cdot\vec{p})
                        -\sin\vartheta\,\frac{1}{m\,c^2} H_{\rm e} p
                        +\Big(\dfrac{p}{m\,c} \sin\vartheta
                              -2\,{\sin^2}\frac{\vartheta}{2}\Big)(
                                    \vec{\alpha}\cdot\vec{p})
                        \bigg]\frac{\vec{p}}{p^3} \notag \\
                  &=& \Big(\dfrac{p}{m\,c} \sin\vartheta+\cos\vartheta\Big)(
                              \vec{\alpha}\cdot\vec{p})\frac{\vec{p}}{p^3}
                        -\sin\vartheta\,\frac{1}{m\,c^2} H_{\rm e} \frac{\vec{p}}{p^2}.
            \end{eqnarray}
            After that, we get
            \begin{eqnarray}
                  && \langle\Psi'_1| \Biggr\{\frac{1}{p}\left[cp\; \cos\vartheta
                        - mc^2 \sin\vartheta \right] {\vec{\alpha}(0)}
                        +\frac{\sin\vartheta \cos\vartheta\left[ m^2c^4-c^2 p^2  \right]+2 cp\; mc^2 \sin^2\vartheta}{\left(m\,c^2\; \cos\vartheta +cp  \sin\vartheta \right)}{\left[\vec{\alpha}(0)\cdot\vec{p}\right]\; \frac{\vec{p}}{p^3}} \Biggr\}|\Psi'_3 \rangle\nonumber\\
                  &=& \langle\Psi_1| \mathcal{C}^\dagger(\vartheta)\Biggr\{
                        \frac{1}{p}\left[cp\; \cos\vartheta- mc^2 \sin\vartheta \right] {\vec{\alpha}(0)}
                        +\frac{\sin\vartheta \cos\vartheta\left[ m^2c^4-c^2 p^2  \right]+2 cp\; mc^2 \sin^2\vartheta}{\left(m\,c^2\; \cos\vartheta +cp  \sin\vartheta \right)}{\left[\vec{\alpha}(0)\cdot\vec{p}\right]\; \frac{\vec{p}}{p^3}} \Biggr\} \mathcal{C}(\vartheta)|\Psi_3 \rangle.\nonumber\\
                  &=& \frac{1}{p} \left(cp\;\cos\vartheta- mc^2 \sin\vartheta\right) \langle\Psi_1|
                              \mathcal{C}^\dagger(\vartheta)\;\vec{\alpha}(0)\;\mathcal{C}
                                    (\vartheta)|\Psi_3 \rangle \nonumber\\
                  &&      +\frac{\sin\vartheta \cos\vartheta\left(m^2 c^4 -c^2 p^2\right)
                              +2 cp\; mc^2 \sin^2\vartheta}{\left(m\,c^2 \cos\vartheta
                                    +cp \sin\vartheta\right)}\langle\Psi_1|\mathcal{C}^\dagger
                                          (\vartheta) \left[\vec{\alpha}(0)\cdot\vec{p}
                                                \right]\frac{\vec{p}}{p^3}\mathcal{C}(\vartheta)
                                                      |\Psi_3 \rangle.
                  \nonumber\\
                 &=& \frac{1}{p} \left(cp\;\cos\vartheta- mc^2 \sin\vartheta\right) \langle\Psi_1|
                             \Biggl\{\vec{\alpha}(0)-\dfrac{\sin\vartheta}{m\,c^2} H_{\rm e} \hat{p}
                             +\left[\dfrac{p}{m\,c} \sin\vartheta -2\,\sin^2 \Bigl(
                                   \frac{\vartheta}{2}\Bigr)\right][\vec{\alpha}(0)\cdot\hat{p}]
                                         \hat{p}\Biggr\}|\Psi_3 \rangle \notag\\
                       &&+\frac{\sin\vartheta \cos\vartheta\left(m^2 c^4 -c^2 p^2\right)
                             +2 cp\; mc^2 \sin^2\vartheta}{\left(m\,c^2 \cos\vartheta
                                   +cp \sin\vartheta\right)}\langle\Psi_1|\Biggl[\left(
                                               \cos\vartheta+\dfrac{p}{m\,c} \sin\vartheta\right)[
                                               \vec{\alpha}(0)\cdot\vec{p}]\frac{\vec{p}}{p^3}
                                         -\dfrac{\sin\vartheta}{m\,c^2} H_{\rm e}
                                               \frac{\vec{p}}{p^2}\Biggr]|\Psi_3 \rangle
                                               \nonumber\\
                  &=& \frac{1}{p} \left(cp\;\cos\vartheta- mc^2 \sin\vartheta\right)
                       \Biggl\{\langle\Psi_1|\vec{\alpha}(0)|\Psi_3 \rangle
                             +\left[\dfrac{p}{m\,c} \sin\vartheta
                                   -2\,\sin^2 \Bigl(\frac{\vartheta}{2}\Bigr)\right]\langle\Psi_1|[
                                         \vec{\alpha}(0)\cdot\hat{p}]|\Psi_3 \rangle\hat{p}\Biggr\}
                                         \notag\\
                       &&+\frac{\sin\vartheta \cos\vartheta\left(m^2 c^4 -c^2 p^2\right)
                             +2 cp\; mc^2 \sin^2\vartheta}{\left(m\,c^2 \cos\vartheta
                                   +cp \sin\vartheta\right)}\Biggl[\left(
                                         \cos\vartheta+\dfrac{p}{m\,c} \sin\vartheta\right)
                                               \langle\Psi_1|[\vec{\alpha}(0)\cdot\vec{p}]|
                                                     \Psi_3 \rangle\frac{\vec{p}}{p^3}\Biggr] \nonumber\\
                  &=& \frac{1}{p} \left(cp\;\cos\vartheta- mc^2 \sin\vartheta\right)
                       \Biggl\{\langle\Psi_1|\vec{\alpha}(0)|\Psi_3 \rangle
                             +\left[\dfrac{p}{m\,c} \sin\vartheta
                                   -2\,\sin^2 \Bigl(\frac{\vartheta}{2}\Bigr)\right]\langle\Psi_1 |\vec{\alpha}(0)|\Psi_3 \rangle\Biggr\} \notag\\
                       &&+\frac{\sin\vartheta \cos\vartheta\left(m^2 c^4 -c^2 p^2\right)
                             +2 cp\; mc^2 \sin^2\vartheta}{\left(m\,c^2 \cos\vartheta
                                   +cp \sin\vartheta\right)}\Biggl[
                                          \left(\cos\vartheta
                                                +\dfrac{p}{m\,c} \sin\vartheta\right)
                                                      \dfrac{1}{p} \langle\Psi_1 |
                                                            \vec{\alpha}(0)|\Psi_3
                                                                  \rangle\Biggr] \nonumber\\
                  &=& \frac{1}{p} \left(cp\;\cos\vartheta- mc^2 \sin\vartheta\right)
                        \left(\cos\vartheta+\dfrac{p}{m\,c} \sin\vartheta\right)
                              \langle\Psi_1 |\vec{\alpha}(0)|\Psi_3 \rangle \notag\\
                       &&+\frac{\sin\vartheta \cos\vartheta\left(m^2 c^4 -c^2 p^2\right)
                             +2 cp\; mc^2 \sin^2\vartheta}{p\left(m\,c^2 \cos\vartheta
                                   +cp \sin\vartheta\right)} \left(\cos\vartheta
                                          +\dfrac{p}{m\,c} \sin\vartheta\right)
                                                \langle\Psi_1 |\vec{\alpha}(0)|\Psi_3
                                                      \rangle \nonumber\\
                  &=& \frac{1}{p} \Biggl\{
                        \left(cp\;\cos\vartheta- mc^2 \sin\vartheta\right)
                        +\frac{\sin\vartheta \cos\vartheta\left(m^2 c^4 -c^2 p^2\right)
                             +2 cp\; mc^2 \sin^2\vartheta}{\left(m\,c^2 \cos\vartheta
                                   +cp \sin\vartheta\right)}
                        \Biggr\}\left(\cos\vartheta+\dfrac{p}{m\,c} \sin\vartheta\right)
                              \langle\Psi_1 |\vec{\alpha}(0)|\Psi_3 \rangle \nonumber \\
                  &=& \frac{1}{p\,m\,c^2} \Bigl[
                        \left(cp\;\cos\vartheta- mc^2 \sin\vartheta\right)(
                              m\,c^2 \cos\vartheta+c\,p\,\sin\vartheta)
                        +\sin\vartheta \cos\vartheta\left(m^2 c^4 -c^2 p^2\right)
                             +2 cp\; mc^2 \sin^2\vartheta
                        \Bigr]\langle\Psi_1 |\vec{\alpha}(0)|\Psi_3 \rangle \nonumber\\
                  &=& \frac{1}{p\,m\,c^2} \Bigl[
                        \sin\vartheta \cos\vartheta\left(c^2 p^2 -m^2 c^4\right)
                             +c\,p\,m\,c^2 \cos(2\vartheta)
                        +\sin\vartheta \cos\vartheta\left(m^2 c^4 -c^2 p^2\right)
                             +2 cp\; mc^2 \sin^2\vartheta
                        \Bigr]\langle\Psi_1 |\vec{\alpha}(0)|\Psi_3 \rangle \nonumber\\
                  &=& \frac{c\,p\,m\,c^2}{p m\,c^2} \langle\Psi_1 |\vec{\alpha}(0)|
                        \Psi_3 \rangle \nonumber\\
                  &=& c\langle\Psi_1 |\vec{\alpha}_3(0)|\Psi_3\rangle,
            \end{eqnarray}
            where we have used \Eq{eq:D-28-r} to simplify the expression. Then we have
            \begin{eqnarray}
                  && \langle\Psi'_1|\hat{\mathcal{Z}}_{\rm m}|\Psi'_3 \rangle\nonumber\\
                  &=& \frac{\hbar}{{\rm i}2{E}_{\rm m}}\left({\rm e}^{
                              \frac{{\rm i}2{E}_{\rm m}t}{\hbar}}-1
                        \right)\langle\Psi'_1| \Biggr\{\frac{1}{p}\left[
                              cp \cos\vartheta-mc^2 \sin\vartheta \right]
                                    {\vec{\alpha}(0)}
                        +\frac{\sin\vartheta \cos\vartheta\left[ m^2c^4-c^2 p^2  \right]+2 cp mc^2 \sin^2\vartheta}{\left(mc^2 \cos\vartheta +cp  \sin\vartheta \right)}{\left[\vec{\alpha}(0)\cdot\vec{p}\right]\frac{\vec{p}}{p^3}}\Biggr\}|\Psi'_3 \rangle \notag \\
                  &=& \frac{\hbar}{{\rm i}\,2\,{E}_{\rm m}}\left({\rm e}^{
                              \frac{{\rm i}\,2\,{E}_{\rm m}t}{\hbar}}-1
                        \right)c\langle\Psi_1 |\vec{\alpha}(0)|\Psi_3 \rangle \notag \\
                  &=& \frac{-{\rm i}\hbar c}{2 E_{\rm p}} \left({\rm e}^{\frac{{\rm i}\,2\,E_{\rm p} t}{\hbar}} {-1} \right)\langle\Psi_1|\vec{\alpha}(0)|\Psi_3 \rangle.
            \end{eqnarray}
            Thus the expectation value is given by
            \begin{eqnarray}\label{eq:ZZ-1}
                   {\mathcal{Z}}_{\rm m} &=& \sin(2\eta){\rm Re}\left(
                        \langle\Psi'_1|\hat{\mathcal{Z}}_{\rm m}|\Psi'_3 \rangle\right)\nonumber\\
                        &=&\sin(2\eta){\rm Re}\left[\frac{-{\rm i}\hbar c}{2 E_{\rm p}} \left({\rm e}^{\frac{{\rm i}\,2\,E_{\rm p} t}{\hbar}} {-1} \right)\langle\Psi_1|\vec{\alpha}(0)|\Psi_3 \rangle
                        \right]\nonumber\\
                  &=&\sin(2\eta){\rm Re}\left[\frac{-{\rm i}\hbar c}{2 E_{\rm p}} \left({\rm e}^{\frac{{\rm i}\,2\,E_{\rm p} t}{\hbar}} {-1} \right)\left(-\dfrac{mc^2}{E_{\rm p}}\; \hat{p}\right)
                        \right] \nonumber\\
                  &=& \hat{p}\;\sin(2\eta)\:{\rm Re}\bigg[{\rm i}
                        \frac{\hbar\,m\,c^3}{2\,{E}_{\rm m}^2} \Big({\rm e}^{
                              \frac{{\rm i}\,2\,{E}_{\rm m}t}{\hbar}}-1\Big)
                        \bigg] \notag \\
                  &=& -\hat{p}\;\sin(2\eta)\:\frac{\hbar\,m\,c^3}
                        {2\,{E}_{\rm m}^2} \sin\Bigl(\frac{2\,{E}_{\rm m}t}{\hbar}
                        \Bigr)\nonumber\\
                  &=& -\hat{p}\;\sin(2\eta)\:\frac{\hbar\,m\,c^3}
                        {2\,{E}_{\rm p}^2} \sin\Bigl(\frac{2\,{E}_{\rm p}t}{\hbar}
                        \Bigr) \notag \\
                  &=&-\hat{p}\;A \sin(\omega\,t).
            \end{eqnarray}
            Based on which, we have the amplitude and frequency as
            \begin{eqnarray}\label{eq:ZZ-2}
                  &&A=\sin(2\eta) \frac{\hbar c}{2 E_{\rm p}} \dfrac{mc^2}{
                        {E_{\rm p}}},\qquad\nonumber\\
                 && \omega=\frac{2\,E_{\rm p}}{\hbar},
            \end{eqnarray}
            which is the same as that of Dirac's electron and the type-I Dirac's braidon.

            Notice that the results in Eqs. (\ref{eq:ZZ-1}) and (\ref{eq:ZZ-2}) do not depend on the mixing angle $\vartheta$, whence we have the following observation:

            \emph{Observation 1.}---For the mixed Hamiltonian system of Dirac's electron $H_{\rm e}$ and the type-I Dirac's braidon $H_{\rm b}^{\rm I}$, i.e., $\mathcal{H}_{\rm m}=\cos\vartheta\;H_{\rm e}+\sin\vartheta\;H_{\rm b}^{\rm I}$, if it is in a superposition state
            of $|\Psi'_1\rangle$ (positive energy and positive helicity) and $|\Psi'_3\rangle$ (negative energy and positive helicity), i.e., $|\Psi'\rangle=\cos\eta |\Psi'_1\rangle+ \sin\eta |\Psi'_3\rangle$, then it has the same behavior of ``position Zitterbewegung'' for any mixing angle $\vartheta$. In other words, one cannot distinguish Dirac's electron $H_{\rm e}$ and the type-I Dirac's braidon $H_{\rm b}^{\rm I}$ in such a situation.

      \subsection{More General Results}

      \begin{remark}
In the above, we have calculated only the case of $\langle\Psi'_1|\hat{\mathcal{Z}}_{\rm m}|\Psi'_3\rangle$. In this section, let us consider more general cases. To do so, we need to calculate the other three terms, i.e., $\langle\Psi'_2|\hat{\mathcal{Z}}_{\rm m}|\Psi'_4\rangle$, $\langle\Psi'_1|\hat{\mathcal{Z}}_{\rm m}|\Psi'_4\rangle$, and $\langle\Psi'_2|\hat{\mathcal{Z}}_{\rm m}|\Psi'_3\rangle$. $\blacksquare$
\end{remark}

To reach this purpose, let us firstly develop a uniform formula. From previous results, we have known that
            \begin{eqnarray}
                  \hat{\mathcal{Z}}_{\rm m}
                  &=& \Biggr\{\frac{1}{p}\left[cp\; \cos\vartheta- mc^2 \sin\vartheta \right] {\vec{\alpha}(0)
                  }+\frac{\sin\vartheta \cos\vartheta\left[ m^2c^4-c^2 p^2  \right]+2 cp\; mc^2 \sin^2\vartheta}{\left(m\,c^2\; \cos\vartheta +cp  \sin\vartheta \right)}{\left[\vec{\alpha}(0)\cdot\vec{p}\right]\; \frac{\vec{p}}{p^3}}\nonumber\\
                  &&-
                  \frac{c \; mc^2 \left[c p \cos\vartheta-m\,c^2\;  \sin\vartheta \right]}{ \left(m\,c^2\; \cos\vartheta +cp  \sin\vartheta \right)} \left[\mathcal{H}^{-1}_{\rm m}\frac{\vec{p}}{p}\right]\Biggr\}\frac{\hbar}{-{\rm i}\,2\,\mathcal{H}_{\rm m}}\left({\rm e}^{\frac{-{\rm i}\,2\,\mathcal{H}_{\rm m}t}{\hbar}}-1\right).
            \end{eqnarray}
            and then we have
            \begin{eqnarray}\label{eq:zz-1a}
                  && \langle\Psi'_1|\hat{\mathcal{Z}}_{\rm m}|\Psi'_3 \rangle\nonumber\\
                  &=& \frac{\hbar}{{\rm i}\,2\,{E}_{\rm m}} \left({\rm e}^{
                        \frac{{\rm i}\,2\,{E}_{\rm m}t}{\hbar}}-1
                        \right)\times\nonumber\\
                  &&\langle\Psi'_1|\Biggr\{\frac{1}{p} (
                              c\,p\,\cos\vartheta-m\,c^2 \sin\vartheta)
                                    \vec{\alpha}(0)
                        +\frac{\sin\vartheta\,\cos\vartheta\,(m^2c^4-c^2 p^2)
                                    +2\,c\,p\,m\,c^2 \sin^2\vartheta}{
                                          \left(m\,c^2 \cos\vartheta
                                                +c\,p\, \sin\vartheta\right)}
                              \left[\vec{\alpha}(0)\cdot\vec{p}\right]
                                    \frac{\vec{p}}{p^3}\Biggr\}|\Psi'_3\rangle.
            \end{eqnarray}
            Because of
            \begin{eqnarray}
                  && \langle\Psi'_1| \Biggr\{\frac{1}{p}\left[cp\; \cos\vartheta
                        - mc^2 \sin\vartheta \right] {\vec{\alpha}(0)}
                        +\frac{\sin\vartheta \cos\vartheta\left[ m^2c^4-c^2 p^2  \right]+2 cp\; mc^2 \sin^2\vartheta}{\left(m\,c^2\; \cos\vartheta +cp  \sin\vartheta \right)}{\left[\vec{\alpha}(0)\cdot\vec{p}\right]\; \frac{\vec{p}}{p^3}} \Biggr\}|\Psi'_3 \rangle\nonumber\\
                  &=& \frac{1}{p} \left(cp\;\cos\vartheta- mc^2 \sin\vartheta\right)
                  \Biggl\{\langle\Psi_1|\vec{\alpha}(0)|\Psi_3 \rangle
                        +\left[\dfrac{p}{m\,c} \sin\vartheta
                              -2\,\sin^2 \Bigl(\frac{\vartheta}{2}\Bigr)\right]\langle\Psi_1|[
                                    \vec{\alpha}(0)\cdot\hat{p}]|\Psi_3 \rangle\hat{p}\Biggr\}
                                    \notag\\
                  &&+\frac{\sin\vartheta \cos\vartheta\left(m^2 c^4 -c^2 p^2\right)
                        +2 cp\; mc^2 \sin^2\vartheta}{\left(m\,c^2 \cos\vartheta
                              +cp \sin\vartheta\right)}\Biggl[\left(
                                    \cos\vartheta+\dfrac{p}{m\,c} \sin\vartheta\right)
                                          \langle\Psi_1|[\vec{\alpha}(0)\cdot\vec{p}]|
                                                \Psi_3 \rangle\frac{\vec{p}}{p^3}\Biggr]\nonumber\\
            &=& \frac{1}{p} \left(cp\;\cos\vartheta
                                    -mc^2 \sin\vartheta\right)\langle\Psi_1|
                                          \vec{\alpha}(0)|\Psi_3\rangle
                              +\biggl\{\frac{1}{p} \left(cp\;\cos\vartheta
                                                -mc^2 \sin\vartheta
                                          \right)\left[\dfrac{p}{m\,c} \sin\vartheta
                                                +\cos\vartheta-1\right] \notag \\
                                    &&\qquad +\dfrac{1}{p\,m\,c^2} \big[
                                          \sin\vartheta\,\cos\vartheta\,(m^2 c^4
                                                      -c^2 p^2)
                                                +2 cp\; mc^2 \sin^2\vartheta\big]
                                    \biggr\}\langle\Psi_1|[\vec{\alpha}(0)
                                          \cdot\hat{p}]|\Psi_3\rangle\hat{p}
                              \notag\\
                  &=& \frac{1}{p} \left(c\,p\,\cos\vartheta
                                    -m\,c^2 \sin\vartheta\right)\langle\Psi_1|
                                          \vec{\alpha}(0)|\Psi_3\rangle
                              +\frac{1}{p} \left(c\,p-c\,p\,\cos\vartheta
                                    +m\,c^2 \sin\vartheta\right)\langle\Psi_1|[
                                          \vec{\alpha}(0)\cdot\hat{p}]|
                                                \Psi_3\rangle\hat{p},
            \end{eqnarray}
            then from Eq. (\ref{eq:zz-1a}) we have
             \begin{eqnarray}\label{eq:ZmPsiPJK-0}
                  && \langle\Psi'_1|\hat{\mathcal{Z}}_{\rm m}|\Psi'_3\rangle
                        \notag \\
                  &=& \frac{\hbar\,c}{{\rm i}\,2\,{E}_{\rm m} p} \left({\rm e}^{
                              \frac{{\rm i}\,2\,{E}_{\rm m}t}{\hbar}} -1\right)
                        \Bigl[\left(p\,\cos\vartheta-m\,c\,\sin\vartheta\right)
                                    \langle\Psi_1|\vec{\alpha}(0)|\Psi_3\rangle
                              +\left(p-p\,\cos\vartheta
                                    +m\,c\,\sin\vartheta\right)\langle\Psi_1|[
                                          \vec{\alpha}(0)\cdot\hat{p}]|
                                                \Psi_3\rangle\hat{p}\Bigr],
            \end{eqnarray}
            which yields the general result as
             \begin{eqnarray}\label{eq:ZmPsiPJK}
                  && \langle\Psi'_k|\hat{\mathcal{Z}}_{\rm m}|\Psi'_l\rangle
                        \notag \\
                  &=& \frac{\hbar\,c}{{\rm i}\,2\,{E}_{\rm m} p} \left({\rm e}^{
                              \frac{{\rm i}\,2\,{E}_{\rm m}t}{\hbar}} -1\right)
                        \Bigl[\left(p\,\cos\vartheta-m\,c\,\sin\vartheta\right)
                                    \langle\Psi_k|\vec{\alpha}(0)|\Psi_l\rangle
                              +\left(p-p\,\cos\vartheta
                                    +m\,c\,\sin\vartheta\right)\langle\Psi_k|[
                                          \vec{\alpha}(0)\cdot\hat{p}]|
                                                \Psi_l\rangle\hat{p}\Bigr],
            \end{eqnarray}
            where $k\in\{1,2\}$, and $l\in\{3,4\}$.

            \begin{remark}
                  When $\vartheta=0$, then Eq. (\ref{eq:ZmPsiPJK}) reduces to
                        \begin{eqnarray}
                              && \langle\Psi'_k|\hat{\mathcal{Z}}_{\rm m}|
                                    \Psi'_l\rangle
                              =\frac{\hbar\,c}{{\rm i}\,2\,{E}_{\rm m} p} \left(
                                    {\rm e}^{\frac{{\rm i}\,2\,{E}_{\rm m}t}{
                                          \hbar}} -1\right)p\langle\Psi_k|
                                                \vec{\alpha}(0)|\Psi_l\rangle
                              =\frac{\hbar\,c}{{\rm i}\,2\,{E}_{\rm m}} \left(
                                    {\rm e}^{\frac{{\rm i}\,2\,{E}_{\rm m}t}{
                                          \hbar}} -1\right)\langle\Psi_k|
                                                \vec{\alpha}(0)|\Psi_l\rangle,
                        \end{eqnarray}
                        which coincides with
                        \begin{eqnarray}\label{eq:ZePsiJK}
                  \langle\Psi_k|\hat{\mathcal{Z}}_{\rm e}|\Psi_l \rangle
                  &=& \frac{-{\rm i}\hbar c}{2 E_{\rm p}} \left({\rm e}^{\frac{{\rm i}\,2\,E_{\rm p} t}{\hbar}} {-1} \right)\langle\Psi_k|\vec{\alpha}(0)|\Psi_l\rangle,
            \end{eqnarray}
            i.e., Eq. (\ref{eq:D-22-bb-1}). When $\vartheta=\pi/2$, then Eq. (\ref{eq:ZmPsiPJK}) reduces to
                        \begin{eqnarray}
                              \langle\Psi'_k|\hat{\mathcal{Z}}_{\rm m}|
                                    \Psi'_l\rangle
                             &=&\frac{\hbar\,c}{{\rm i}\,2\,{E}_{\rm m} p} \left(
                                    {\rm e}^{\frac{{\rm i}\,2\,{E}_{\rm m}t}{
                                          \hbar}} -1
                                    \right)\Bigl[-m\,c\langle\Psi_k|\vec{\alpha}(0)|
                                                \Psi_l\rangle
                                          +\left(p+m\,c\right)\langle\Psi_k|[
                                                \vec{\alpha}(0)\cdot\hat{p}]|
                                                      \Psi_l\rangle\hat{p}\Bigr]
                                    \notag \\
                              &=& \dfrac{{\rm i}\hbar}{2\,p} \dfrac{m\,c^2}{
                                    E_{\rm p}} \left({\rm e}^{
                                          \frac{{\rm i}\,2E_{\rm p}\,t}{\hbar}}
                                          -1\right)\bigg[\langle\Psi_k|
                                                      \vec{\alpha}(0)|\Psi_l\rangle
                                                -\left(\dfrac{p}{m\,c}+1\right)
                                                      \langle\Psi_k|[\vec{\alpha}(0)
                                                            \cdot\hat{p}]|
                                                                  \Psi_l\rangle
                                                                        \hat{p}
                                                \bigg] \notag \\
                              &=& \dfrac{{\rm i}\hbar}{2p} \dfrac{m\,c^2}{
                                    E_{\rm p}} \left({\rm e}^{
                                          \frac{{\rm i}\,2E_{\rm p}\,t}{\hbar}}
                                          -1\right) \langle\Psi_k|\left[
                                                \vec{\alpha}(0)-\left(
                                                      \frac{cp}{mc^2}+1
                                                      \right)[\vec{\alpha}(0)
                                                            \cdot\hat{p}]\hat{p}
                                                \right]|\Psi_l\rangle,
                        \end{eqnarray}
                        which coincides with
                        \begin{eqnarray}\label{eq:ZbIPsiJK}
                  \langle\Psi'_k|\hat{\mathcal{Z}}_{\rm b}^{\rm I}|\Psi'_l
                        \rangle
                  &=& \dfrac{{\rm i}\hbar}{2p} \dfrac{m\,c^2}{E_{\rm p}} \left(
                              {\rm e}^{\frac{{\rm i}\,2E_{\rm p}\,t}{\hbar}}-1\right) \langle\Psi_k|
                              \left[\vec{\alpha}(0)-\left(\frac{cp}{mc^2}+1\right)  [\vec{\alpha}(0)\cdot\hat{p}] \hat{p}\right]|\Psi_l\rangle,
            \end{eqnarray}
            i.e., Eq. (\ref{Eq:ZZ-u-1}). $\blacksquare$
            \end{remark}
            \begin{remark}
            By taking $k=2, l=4$, from Eq. (\ref{eq:ZmPsiPJK}) one obtains
            \begin{eqnarray}
                  && \langle\Psi'_2|\hat{\mathcal{Z}}_{\rm m}|\Psi'_4
                        \rangle \notag \\
                  &=& \frac{\hbar\,c}{{\rm i}\,2\,{E}_{\rm m} p} \left({\rm e}^{
                              \frac{{\rm i}\,2\,{E}_{\rm m}t}{\hbar}} -1\right)
                        \Bigl[\left(p\,\cos\vartheta-m\,c\,\sin\vartheta\right)
                                    \langle\Psi_2|\vec{\alpha}(0)|\Psi_4\rangle
                              +\left(p-p\,\cos\vartheta
                                    +m\,c\,\sin\vartheta\right)\langle\Psi_2|[
                                          \vec{\alpha}(0)\cdot\hat{p}]|
                                                \Psi_4\rangle\hat{p}\Bigr]
                        \notag \\
                  &=& \frac{\hbar\,c}{{\rm i}\,2\,{E}_{\rm m} p} \left({\rm e}^{
                              \frac{{\rm i}\,2\,{E}_{\rm m}t}{\hbar}} -1\right)
                        \Bigl[\left(p\,\cos\vartheta-m\,c\,\sin\vartheta\right)
                                    \langle\Psi_1|\vec{\alpha}(0)|\Psi_3\rangle
                              +\left(p-p\,\cos\vartheta
                                    +m\,c\,\sin\vartheta\right)\langle\Psi_1|[
                                          \vec{\alpha}(0)\cdot\hat{p}]|
                                                \Psi_3\rangle\hat{p}\Bigr]
                        \notag \\
                  &=& \langle\Psi'_1|\hat{\mathcal{Z}}_{\rm m}|\Psi'_3
                        \rangle \notag \\
                  &=& \frac{-{\rm i}\hbar c}{2 E_{\rm p}} \left({\rm e}^{
                        \frac{{\rm i}\,2\,E_{\rm p} t}{\hbar}} {-1} \right)\langle\Psi_1|\vec{\alpha}(0)|\Psi_3 \rangle.
            \end{eqnarray}
            Thereby,
            \begin{eqnarray}
                  \langle\Psi'_2|\hat{\mathcal{Z}}_{\rm m}|\Psi'_4\rangle=\langle\Psi'_1|\hat{\mathcal{Z}}_{\rm m}|\Psi'_3\rangle=\langle\Psi'_2|\hat{\mathcal{Z}}_{\rm b}^{\rm I}|\Psi'_4
                        \rangle
                  =\langle\Psi'_1|\hat{\mathcal{Z}}_{\rm b}^{\rm I}|\Psi'_3
                        \rangle
                  =\langle\Psi_1|\hat{\mathcal{Z}}_{\rm e}|\Psi_3\rangle
                  =\langle\Psi_2|\hat{\mathcal{Z}}_{\rm e}|\Psi_4\rangle.
            \end{eqnarray}
            $\blacksquare$
            \end{remark}
            \begin{remark}
                  By taking $k=1, l=4$, from Eq. (\ref{eq:ZmPsiPJK}) one obtains
                  \begin{eqnarray}\label{eq:ob-1a}
                        && \langle\Psi'_1|\hat{\mathcal{Z}}_{\rm m}|
                              \Psi'_4\rangle \notag \\
                        &=& \frac{\hbar\,c}{{\rm i}\,2\,{E}_{\rm m} p} \left(
                              {\rm e}^{\frac{{\rm i}\,2\,{E}_{\rm m}t}{\hbar}}
                              -1\right)\Bigl[\left(p\,\cos\vartheta
                                          -m\,c\,\sin\vartheta\right)
                                                \langle\Psi_1|\vec{\alpha}(0)|
                                                      \Psi_4\rangle
                                    +\left(p-p\,\cos\vartheta
                                          +m\,c\,\sin\vartheta\right)\langle\Psi_1|[
                                                \vec{\alpha}(0)\cdot\hat{p}]|
                                                      \Psi_4\rangle\hat{p}\Bigr]
                              \notag \\
                        &=& \frac{\hbar\,c}{{\rm i}\,2\,{E}_{\rm m} p} \left(
                              {\rm e}^{\frac{{\rm i}\,2\,{E}_{\rm m}t}{\hbar}}
                              -1\right)\left(p\,\cos\vartheta
                                    -m\,c\,\sin\vartheta\right)\langle\Psi_1|
                                          \vec{\alpha}(0)|\Psi_4\rangle
                        =\frac{-{\rm i}\,\hbar\,c}{2\,{E}_{\rm m}} \left(
                              {\rm e}^{\frac{{\rm i}\,2\,{E}_{\rm m}t}{\hbar}}
                              -1\right)\left(\cos\vartheta
                                    -\dfrac{m\,c}{p} \sin\vartheta
                                    \right)\langle\Psi_1|\vec{\alpha}(0)|
                                          \Psi_4\rangle \notag \\
                        &=& \left(\cos\vartheta-\dfrac{m\,c}{p} \sin\vartheta
                              \right)\langle\Psi_1|\hat{\mathcal{Z}}_{\rm e}|
                                    \Psi_4\rangle,
                  \end{eqnarray}
                  where we have used the following relations
                  \begin{align}
                        & \langle\Psi_1|\big[\vec{\alpha}(0)\cdot\hat{p}\big]|
                              \Psi_4\rangle=0,
                  \end{align}
                  and
                  \begin{eqnarray}
                        \langle\Psi_1|\hat{\mathcal{Z}}_{\rm e}|\Psi_4\rangle
                        &=& \frac{-{\rm i}\hbar c}{2 E_{\rm p}} \left({\rm e}^{
                              \frac{{\rm i}\,2\,E_{\rm p} t}{\hbar}} -1
                              \right)\langle\Psi_1|\vec{\alpha}(0)|\Psi_4\rangle.
                  \end{eqnarray}
                  $\blacksquare$
            \end{remark}
            \begin{remark}
                  Similarly, for the superposition of $\Ket{\Psi'_2}$ and $\Ket{\Psi'_3}$, i.e. $k=2$, and $l=3$.
                  \begin{align}
                        & \langle\Psi_2|\big[\vec{\alpha}(0)\cdot\hat{p}\big]\hat{p}|
                              \Psi_3\rangle
                        =\Bigl\{\big[\langle\Psi_2|\vec{\alpha}(0)|\Psi_3\rangle\big]
                              \cdot\hat{p}\Bigr\}\hat{p}
                        =\Bigl\{\big[-\langle\Psi_1|\vec{\alpha}(0)|\Psi_4\rangle^*\big]
                              \cdot\hat{p}\Bigr\}\hat{p}
                        =0,
                  \end{align}
                  which indicates
                  \begin{eqnarray}\label{eq:ob-1b}
                        && \langle\Psi'_2|\hat{\mathcal{Z}}_{\rm m}|
                              \Psi'_3\rangle \notag \\
                        &=& \frac{\hbar\,c}{{\rm i}\,2\,{E}_{\rm m} p} \left(
                              {\rm e}^{\frac{{\rm i}\,2\,{E}_{\rm m}t}{\hbar}}
                              -1\right)\Bigl[\left(p\,\cos\vartheta
                                          -m\,c\,\sin\vartheta\right)
                                                \langle\Psi_2|\vec{\alpha}|
                                                      \Psi_3\rangle
                                    +\left(p-p\,\cos\vartheta
                                          +m\,c\,\sin\vartheta\right)\langle\Psi_2|(
                                                \vec{\alpha}\cdot\hat{p})|
                                                      \Psi_3\rangle\hat{p}\Bigr]
                              \notag \\
                        &=& \frac{\hbar\,c}{{\rm i}\,2\,{E}_{\rm m} p} \left(
                              {\rm e}^{\frac{{\rm i}\,2\,{E}_{\rm m}t}{\hbar}}
                              -1\right)\left(p\,\cos\vartheta
                                    -m\,c\,\sin\vartheta\right)\langle\Psi_2|
                                          \vec{\alpha}|\Psi_3\rangle
                        =\frac{-{\rm i}\,\hbar\,c}{2\,{E}_{\rm m}} \left(
                              {\rm e}^{\frac{{\rm i}\,2\,{E}_{\rm m}t}{\hbar}}
                              -1\right)\left(\cos\vartheta
                                    -\dfrac{m\,c}{p} \sin\vartheta
                                    \right)\langle\Psi_2|\vec{\alpha}|
                                          \Psi_3\rangle \notag \\
                        &=& \left(\cos\vartheta-\dfrac{m\,c}{p} \sin\vartheta
                              \right)\langle\Psi_2|\hat{\mathcal{Z}}_{\rm e}|
                                    \Psi_3\rangle,
                  \end{eqnarray}
                  for
                  \begin{eqnarray}
                        && \langle\Psi_2|\hat{\mathcal{Z}}_{\rm e}|\Psi_3\rangle
                        =\frac{-{\rm i}\hbar\,c}{2\,E_{\rm p}} \left({\rm e}^{
                              \frac{{\rm i}\,2\,E_{\rm p} t}{\hbar}} -1\right)\langle\Psi_2|\vec{\alpha}(0)|\Psi_3\rangle.
                  \end{eqnarray}
            $\blacksquare$
            \end{remark}

            \begin{remark}
            It can be noticed that the results in Eqs. (\ref{eq:ob-1a}) and (\ref{eq:ob-1b}) depend on the mixing angle $\vartheta$, thus we have the following observation:

            \emph{Observation 2.}---For the mixed Hamiltonian system of Dirac's electron $H_{\rm e}$ and the type-I Dirac's braidon $H_{\rm b}^{\rm I}$, i.e., $\mathcal{H}_{\rm m}=\cos\vartheta\;H_{\rm e}+\sin\vartheta\;H_{\rm b}^{\rm I}$, if it is in a superposition state
            $|\Psi'\rangle=\cos\eta |\Psi'_1\rangle+ \sin\eta |\Psi'_4\rangle$, or $|\Psi'\rangle=\cos\eta |\Psi'_2\rangle+ \sin\eta |\Psi'_3\rangle$, then the amplitude of ``position Zitterbewegung'' is tuned by a factor
            \begin{eqnarray}
                        && \gamma(\vartheta)=\cos\vartheta-\dfrac{m\,c}{p} \sin\vartheta
             \end{eqnarray}
            in comparison to Dirac's electron $H_{\rm e}$. There are two special cases for the mixing angle $\vartheta$.

            (i) The factor is equal to 0, i.e.,
             \begin{eqnarray}
                        && \gamma(\vartheta)= \cos\vartheta-\dfrac{m\,c}{p} \sin\vartheta=0,
             \end{eqnarray}
             which leads to
             \begin{eqnarray}
                        && \tan\vartheta=\dfrac{p}{m\,c},
             \end{eqnarray}
            in such a case the phenomenon of ``position Zitterbewegung'' vanishes.

            (ii) The factor takes its maximum value, i.e.,
             \begin{eqnarray}
                        && \gamma(\vartheta)=\max\left[\cos\vartheta-\dfrac{m\,c}{p} \sin\vartheta\right]=\sqrt{1+\dfrac{m^2\,c^2}{p^2}}=\dfrac{E_{\rm p}}{pc}.
             \end{eqnarray}
             in such a case the amplitude of ``position Zitterbewegung'' is maximal. In this situation, one has
              \begin{eqnarray}
                        && \cos\vartheta=\dfrac{pc}{E_{\rm p}}, \;\;\;\; \sin\vartheta=-\dfrac{m c^2}{E_{\rm p}}, \;\;\;\;
                            \tan\vartheta=-\dfrac{m c}{p}.
             \end{eqnarray}
            For this case, because $p<mc$, thus $|\tan\vartheta|\ge 1$. $\blacksquare$
            \end{remark}

            \begin{remark}We can summarize the above results as the following Table \ref{tab:pz3}:
  \begin{table}[h]
	\centering
\caption{The results of ``position Zitterbewegung'' of the $H_{\rm e}$-$H_{\rm b}^{\rm I}$ mixing. $|\Psi'_j\rangle$'s  ($j=1, 2, 3, 4$) are four eigenstates of the mixed Hamiltonian $\mathcal{H}_{\rm m}=\cos\vartheta\;H_{\rm e}
                        +\sin\vartheta\;H_{\rm b}^{\rm I}$, The ``position Zitterbewegung'' operator reads $\hat{\mathcal{Z}}_{\rm m}^r
                  = \Biggr\{\frac{1}{p}\left[cp\; \cos\vartheta- mc^2 \sin\vartheta \right] {\vec{\alpha}(0)
                  }+\frac{\sin\vartheta \cos\vartheta\left[ m^2c^4-c^2 p^2  \right]+2 cp\; mc^2 \sin^2\vartheta}{\left(m\,c^2\; \cos\vartheta +cp  \sin\vartheta \right)}{\left[\vec{\alpha}(0)\cdot\vec{p}\right]\; \frac{\vec{p}}{p^3}}
                  -                  \frac{c \; mc^2 \left[c p \cos\vartheta-m\,c^2\;  \sin\vartheta \right]}{ \left(m\,c^2\; \cos\vartheta +cp  \sin\vartheta \right)} \left[\mathcal{H}^{-1}_{\rm m}\frac{\vec{p}}{p}\right]\Biggr\}\frac{\hbar}{-{\rm i}\,2\,\mathcal{H}_{\rm m}}\left({\rm e}^{\frac{-{\rm i}\,2\,\mathcal{H}_{\rm m}t}{\hbar}}-1\right)$.
One has $\langle\Psi'_j|\hat{\mathcal{Z}}_{\rm m}^{r}|\Psi'_j\rangle=0$, $\langle\Psi'_k|\hat{\mathcal{Z}}_{\rm m}^{r}|\Psi'_l\rangle
                   = \frac{\hbar\,c}{{\rm i}\,2\,{E}_{\rm m} p} \left({\rm e}^{
                              \frac{{\rm i}\,2\,{E}_{\rm m}t}{\hbar}} -1\right)
                        \Bigl[\left(p\,\cos\vartheta-m\,c\,\sin\vartheta\right)
                                    \langle\Psi_k|\vec{\alpha}(0)|\Psi_l\rangle
                              +\left(p-p\,\cos\vartheta
                                    +m\,c\,\sin\vartheta\right)\langle\Psi_k|[
                                          \vec{\alpha}(0)\cdot\hat{p}]|
                                                \Psi_l\rangle\hat{p}\Bigr]=\frac{1}{p} \Delta_1\,\Bigl[\left(p\,\cos\vartheta-m\,c\,\sin\vartheta\right)
                                    \langle\Psi_k|\vec{\alpha}(0)|\Psi_l\rangle
                              +\left(p-p\,\cos\vartheta
                                    +m\,c\,\sin\vartheta\right)\langle\Psi_k|[
                                          \vec{\alpha}(0)\cdot\hat{p}]|
                                                \Psi_l\rangle\hat{p}\Bigr]$ for $k\in\{1,2\}$, $l\in\{3,4\}$, and $\gamma(\vartheta)=\cos\vartheta-\frac{mc}{p} \sin\vartheta$. }
\begin{tabular}{lllll}
\hline\hline
 & $|\Psi_1'\rangle$ &  $|\Psi'_2\rangle$& $|\Psi'_3\rangle$ & $|\Psi'_4\rangle$ \\
  \hline
$\langle \Psi'_1| \hat{\mathcal{Z}}_{\rm m}^r$ \;\;\;\quad& 0&0 & $\Delta_1\,\dfrac{-m\,c^2}{E_{\rm p}}\, \hat{p}$  \;\;\;\quad& $\gamma(\vartheta)\Delta_1\,(\vec{F}_1 +{\rm i}\,\vec{F}_2)$  \\
 \hline
$\langle \Psi'_2| \hat{\mathcal{Z}}_{\rm m}^r$\;\;\;\quad &0 &0 & $-\gamma(\vartheta)\Delta_1\,(\vec{F}_1 +{\rm i}\,\vec{F}_2)^* $ \quad\quad& $\Delta_1\,\dfrac{-m\,c^2}{E_{\rm p}}\, \hat{p}$ \\
 \hline
 $\langle \Psi'_3| \hat{\mathcal{Z}}_{\rm m}^r$\;\;\;\quad & $\Delta_1^*\,\dfrac{-m\,c^2}{E_{\rm p}}\, \hat{p}$ \;\;\;& $-\gamma(\vartheta)\Delta_1^*\,(\vec{F}_1 +{\rm i}\,\vec{F}_2) \quad\quad$ & 0 & 0 \\
 \hline
 $\langle \Psi'_4| \hat{\mathcal{Z}}_{\rm m}^r$ \;\;\;\quad& $\gamma(\vartheta)\Delta_1^*\,(\vec{F}_1 +{\rm i}\,\vec{F}_2)^* \quad$ \quad&$\Delta_1^*\,\dfrac{-m\,c^2}{E_{\rm p}}\, \hat{p}$ \;\;\;& 0 & 0 \\
 \hline\hline
\end{tabular}\label{tab:pz3}
\end{table}

            \end{remark}

\section{Position Zitterbewegung for $H_{\rm b}^{\rm II}$}

            The Hamiltonian of type-II Dirac's braidon is given by
            \begin{eqnarray}\label{eq:F-1}
                  H_{\rm b}^{\rm II} = \sqrt{p^2c^2+m^2c^4}\;({\rm i}\beta\vec{\alpha}\cdot\hat{p})=\dfrac{\sqrt{p^2c^2+m^2c^4}}{p}\;({\rm i}\beta\vec{\alpha}\cdot\vec{p}),
            \end{eqnarray}
            one has
            \begin{eqnarray}\label{eq:F-2}
                 && \dfrac{{\rm d}\,\vec{p}}{{\rm d}\,t} =\dfrac{1}{{\rm i}\hbar}
                  \left[\vec{p},\ H_{\rm b}^{\rm II}\right]=0,\nonumber\\
                &&  \dfrac{{\rm d}\,H_{\rm b}^{\rm II}}{{\rm d}\,t} =\dfrac{1}{{\rm i}\hbar}
                  \left[H_{\rm b}^{\rm II},\ H_{\rm b}^{\rm II}\right]=0,
            \end{eqnarray}
            which means that $\vec{p}$ and $H_{\rm b}^{\rm II}$ are conservation quantities. For the position operator $\vec{r}=(x, y, z)$, we have
            \begin{eqnarray}\label{eq:F-3}
            &&  \left[x, \vec{\alpha}\cdot\hat{p}\right]= {\rm i}\hbar \frac{\partial \left(\vec{\alpha}\cdot\hat{p}\right)}{\partial p_x}= {\rm i}\hbar \frac{1}{p} \alpha_x - {\rm i}\hbar \left(\vec{\alpha}\cdot\vec{p}\right) \dfrac{p_x}{p^3},
            \end{eqnarray}
            moreover, we have
                  \begin{eqnarray}
            &&  \left[x, \sqrt{p^2c^2+m^2c^4}\right]= {\rm i}\hbar \frac{\partial \sqrt{p^2c^2+m^2c^4}}{\partial p_x}={\rm i}\hbar \frac{c^2 p_x}{\sqrt{p^2c^2+m^2c^4}},
            \end{eqnarray}
            which yield
            \begin{eqnarray}
            \left[x,  H_{\rm b}^{\rm II}\right] &=& \left[x, \sqrt{p^2c^2+m^2c^4}\;({\rm i}\beta\vec{\alpha}\cdot\hat{p})\right]= {\rm i}\beta \left\{ [x, \sqrt{p^2c^2+m^2c^4}] \vec{\alpha}\cdot\hat{p}+ \sqrt{p^2c^2+m^2c^4} [x, \vec{\alpha}\cdot\hat{p}]\right\}\nonumber\\
            &=& {\rm i}\beta \left\{ \left({\rm i}\hbar \frac{c^2p_x}{\sqrt{p^2c^2+m^2c^4}} \right) \vec{\alpha}\cdot\hat{p}+ \sqrt{p^2c^2+m^2c^4} \left( {\rm i}\hbar \frac{1}{p} \alpha_x - {\rm i}\hbar \left(\vec{\alpha}\cdot\vec{p}\right) \dfrac{p_x}{p^3} \right)\right\}\nonumber\\
            &=& {\rm i} \hbar({\rm i}\beta) \left\{ \left( \frac{c^2}{\sqrt{p^2c^2+m^2c^4}} \right) (\vec{\alpha}\cdot\vec{p})\;\frac{p_x}{p}+ \sqrt{p^2c^2+m^2c^4} \left(  \frac{1}{p} \alpha_x -  \left(\vec{\alpha}\cdot\vec{p}\right) \dfrac{p_x}{p^3} \right)\right\}\nonumber\\
                  &=& {\rm i}\hbar({\rm i}\beta) \left\{  \frac{\sqrt{p^2c^2+m^2c^4}}{p} \alpha_x + \left( \frac{c^2p^2}{\sqrt{p^2c^2+m^2c^4}} -\sqrt{p^2c^2+m^2c^4}\right) (\vec{\alpha}\cdot\vec{p})\;\frac{p_x}{p^3}\right\}\nonumber\\
                  &=& {\rm i}\hbar({\rm i}\beta) \left\{  \frac{\sqrt{p^2c^2+m^2c^4}}{p} \alpha_x -  \frac{m^2c^4}{\sqrt{p^2c^2+m^2c^4}} (\vec{\alpha}\cdot\vec{p})\;\frac{p_x}{p^3}\right\}.
            \end{eqnarray}
            Based on the above results, one obtains
            \begin{eqnarray}\label{eq:F-5}
                  \dfrac{{\rm d}\,\vec{r}}{{\rm d}\,t}
                  =\dfrac{1}{{\rm i}\hbar} \left[\vec{r},  H_{\rm b}^{\rm II}\right]
                  ={\rm i}\beta \left\{ \frac{\sqrt{p^2c^2+m^2c^4}}{p} \vec{\alpha}
                        -\frac{m^2c^4}{\sqrt{p^2c^2+m^2c^4}} (\vec{\alpha}\cdot\vec{p})\;\frac{\vec{p}}{p^3} \right\}.
            \end{eqnarray}

            Let us introduce the operator
            \begin{eqnarray}
                  \vec{\zeta}={\rm i}\:\beta\,\vec{\alpha},
            \end{eqnarray}
           which means
            \begin{eqnarray}
                 H_{\rm b}^{\rm II}=\sqrt{p^2 c^2 +m^2 c^4}\bigl(\vec{\zeta}\cdot\hat{p}\bigr).
            \end{eqnarray}
           Further we obtain
                  \begin{eqnarray}
                              \bigl\{\vec{\zeta},\ \vec{\zeta}\cdot\hat{p}\bigr\}_x
                              &=&-\bigl\{\beta\,\alpha_x,\ \beta(\alpha_x \hat{p}_x
                                    +\alpha_y \hat{p}_y +\alpha_z \hat{p}_z)\bigr\}
                              =-\{\beta\,\alpha_x,\ \beta\,\alpha_x \hat{p}_x\}-\bigl\{\beta\,
                                    \alpha_x,\ \beta(\alpha_y \hat{p}_y +\alpha_z \hat{p}_z)\bigr\}
                                    \nonumber\\
                              &=& -2\:\beta\,\alpha_x\,\beta\,\alpha_x \hat{p}_x
                                    -\beta\,\alpha_x \beta(\alpha_y \hat{p}_y +\alpha_z \hat{p}_z)
                                    -\beta(\alpha_y \hat{p}_y +\alpha_z \hat{p}_z)\beta\,\alpha_x \nonumber\\
                              &=& 2\:\beta^2 \alpha_x^2\,\hat{p}_x
                                    +\beta^2 \alpha_x (\alpha_y \hat{p}_y +\alpha_z \hat{p}_z)
                                    +\beta^2 (\alpha_y \hat{p}_y +\alpha_z \hat{p}_z)\alpha_x\nonumber\\
                              &=&2\,\mathbb{I}\,\hat{p}_x +\{\alpha_x,\ \alpha_y\} \hat{p}_y
                                    +\{\alpha_x,\ \alpha_z\} \hat{p}_z
                              =2\,\mathbb{I}\,\hat{p}_x,
                  \end{eqnarray}
                  and therefore
                  \begin{equation}
                        \bigl\{\vec{\zeta},\ \vec{\zeta}\cdot\hat{p}\bigr\}=2\,\mathbb{I}\,\hat{p},
                  \end{equation}
                  which alludes to
                  \begin{equation}
                        \bigl\{\vec{\zeta},\ H_{\rm b}^{\rm II}\bigr\}=\sqrt{p^2 c^2 +m^2 c^4}\bigl\{
                              \vec{\zeta},\ \vec{\zeta}\cdot\hat{p}\bigr\}
                        =2\sqrt{p^2 c^2 +m^2 c^4}\,\mathbb{I}\,\hat{p}.
                  \end{equation}
            After that, we have
            \begin{eqnarray}\label{eq:DTauDT}
                        \dfrac{{\rm d}\,\vec{\zeta}}{{\rm d}\,t}&=&\dfrac{1}{{\rm i}\hbar} \left[
                              \vec{\zeta},\  H_{\rm b}^{\rm II}\right]
                        =\dfrac{1}{{\rm i}\hbar} \Bigl(\{\vec{\zeta},\  H_{\rm b}^{\rm II}\}
                              -2\, H_{\rm b}^{\rm II}\,\vec{\zeta}\Bigr) =\dfrac{1}{{\rm i}\hbar} \Bigl( 2 \sqrt{p^2c^2+m^2c^4}\,\mathbb{I}\,\hat{p}
                              -2\, H_{\rm b}^{\rm II}\,\vec{\zeta}\Bigr)\nonumber\\
                        &=&-\dfrac{2}{{\rm i}\hbar} \Bigl(  H_{\rm b}^{\rm II}\,\vec{\zeta}
                              -\sqrt{p^2c^2+m^2c^4}\,\mathbb{I}\,\hat{p}\Bigr)=
                              -\dfrac{2}{{\rm i}\hbar}  H_{\rm b}^{\rm II}\Bigl(\vec{\zeta}-\sqrt{p^2c^2+m^2c^4}\;  (H_{\rm b}^{\rm II})^{-1}\hat{p}\Bigr),
            \end{eqnarray}
            and then
            \begin{eqnarray}
                  \frac{{\rm d}\,\left(\vec{\zeta}-\sqrt{p^2c^2+m^2c^4}\; (H_{\rm b}^{\rm II})^{-1}\hat{p}\right)}{{\rm d}\,t}
                  =-\dfrac{2}{{\rm i}\hbar} H_{\rm b}^{\rm II}\Bigl(\vec{\zeta}-\sqrt{p^2c^2+m^2c^4}\; (H_{\rm b}^{\rm II})^{-1}\hat{p}\Bigr),
            \end{eqnarray}
            i.e.,
            \begin{eqnarray}
                  \vec{\zeta}(t)-\sqrt{p^2c^2+m^2c^4}\; (H_{\rm b}^{\rm II})^{-1}\hat{p}
                  ={\rm e}^{\frac{{\rm i}\;2 H_{\rm b}^{\rm II} t}{\hbar}}\Bigl(\vec{\zeta}(0)-\sqrt{p^2c^2+m^2c^4}\; (H_{\rm b}^{\rm II})^{-1}\hat{p}\Bigr).
            \end{eqnarray}
            One can check that
            \begin{eqnarray}
                  \Bigl\{H_{\rm b}^{\rm II}, \left[
                  \vec{\zeta}(0)-\sqrt{p^2c^2+m^2c^4}\; (H_{\rm b}^{\rm II})^{-1}\hat{p}\right]\Bigr\}
                  &=& \left[ H_{\rm b}^{\rm II}\vec{\zeta}(0)+\vec{\zeta}(0)H_{\rm b}^{\rm II}\right]
                  -2 \sqrt{p^2c^2+m^2c^4}\; \hat{p}\nonumber\\
                  &=& 2\sqrt{p^2c^2+m^2c^4}\;\hat{p}-2\sqrt{p^2c^2+m^2c^4}\;\hat{p}=0,
            \end{eqnarray}
            which leads to
            \begin{eqnarray}
                  H_{\rm b}^{\rm II} \left[\vec{\zeta}(0)-\sqrt{p^2c^2+m^2c^4}\; (H_{\rm b}^{\rm II})^{-1}\hat{p}\right]
                  =-\left[\vec{\zeta}(0)-\sqrt{p^2c^2+m^2c^4}\; (H_{\rm b}^{\rm II})^{-1}\hat{p}\right]H_{\rm b}^{\rm II},
            \end{eqnarray}
            and then
                  \begin{eqnarray}\label{eq:F-5a}
                  \vec{\zeta}(t)&=&\sqrt{p^2c^2+m^2c^4}\; (H_{\rm b}^{\rm II})^{-1}\hat{p}
                  +\Bigl(\vec{\zeta}(0)-\sqrt{p^2c^2+m^2c^4}\; (H_{\rm b}^{\rm II})^{-1}\hat{p}\Bigr){\rm e}^{-\frac{{\rm i}\;2 H_{\rm b}^{\rm II}t}{\hbar}}\nonumber\\
                  &=&\frac{\sqrt{p^2c^2+m^2c^4}}{p}\; (H_{\rm b}^{\rm II})^{-1}\vec{p}
                  +\Bigl(\vec{\zeta}(0)-\sqrt{p^2c^2+m^2c^4}\; (H_{\rm b}^{\rm II})^{-1}\hat{p}\Bigr){\rm e}^{-\frac{{\rm i}\;2 H_{\rm b}^{\rm II}t}{\hbar}}.
            \end{eqnarray}
            Eq. (\ref{eq:F-5}) can be rewritten as
            \begin{eqnarray}\label{eq:F-5b}
                        \dfrac{{\rm d}\,\vec{r}}{{\rm d}\,t}& =& \dfrac{1}{{\rm i}\hbar}
                        [\vec{r}, H_{\rm b}^{\rm II}]={\rm i}\beta \left\{ \frac{\sqrt{p^2c^2+m^2c^4}}{p} \vec{\alpha} -  \frac{m^2c^4}{\sqrt{p^2c^2+m^2c^4}} (\vec{\alpha}\cdot\vec{p})\;\frac{\vec{p}}{p^3} \right\}\nonumber\\
            & =&   \frac{\sqrt{p^2c^2+m^2c^4}}{p} \vec{\zeta} -  \frac{m^2 c^4}{\sqrt{p^2c^2+m^2c^4}} (\vec{\zeta}\cdot\vec{p})\;\frac{\vec{p}}{p^3}.
            \end{eqnarray}
            Due to Eq. (\ref{eq:F-5a}), from Eq. (\ref{eq:F-5b}) we have
            \begin{eqnarray}\label{eq:F-5c}
                        \dfrac{{\rm d}\,\vec{r}}{{\rm d}\,t}& =&  \Biggr\{ \frac{\sqrt{p^2c^2+m^2c^4}}{p} \left[\frac{\sqrt{p^2c^2+m^2c^4}}{p}\; (H_{\rm b}^{\rm II})^{-1}\vec{p}
                  +\Bigl(\vec{\zeta}(0)-\sqrt{p^2c^2+m^2c^4}\; (H_{\rm b}^{\rm II})^{-1}\hat{p}\Bigr){\rm e}^{-\frac{{\rm i}\;2 H_{\rm b}^{\rm II}t}{\hbar}}\right]\nonumber\\
                  &&-  \frac{m^2c^4}{\sqrt{p^2c^2+m^2c^4}} \left[\frac{\sqrt{p^2c^2+m^2c^4}}{p}\; (H_{\rm b}^{\rm II})^{-1}\vec{p}
                  +\Bigl(\vec{\zeta}(0)-\sqrt{p^2c^2+m^2c^4}\; (H_{\rm b}^{\rm II})^{-1}\hat{p}\Bigr){\rm e}^{-\frac{{\rm i}\;2 H_{\rm b}^{\rm II}t}{\hbar}}\right]\cdot\vec{p}\;\frac{\vec{p}}{p^3} \Biggr\}\nonumber\\
                  &=& c^2 (H_{\rm b}^{\rm II})^{-1}\vec{p} +\Biggr\{ \frac{\sqrt{p^2c^2+m^2c^4}}{p} \Bigl(\vec{\zeta}(0)-\sqrt{p^2c^2+m^2c^4}\; (H_{\rm b}^{\rm II})^{-1}\hat{p}\Bigr)\nonumber\\
                  &&-  \frac{m^2c^4}{\sqrt{p^2c^2+m^2c^4}}\left[\left[\vec{\zeta}(0)\cdot \vec{p}\right]\frac{\vec{p}}{p^3}-\sqrt{p^2c^2+m^2c^4}\; (H_{\rm b}^{\rm II})^{-1}\frac{\vec{p}}{p^2}\right]  \Biggr\}{\rm e}^{-\frac{{\rm i}\;2 H_{\rm b}^{\rm II}t}{\hbar}}\nonumber\\
                  &=& c^2 (H_{\rm b}^{\rm II})^{-1}\vec{p} +\Biggr\{ \frac{\sqrt{p^2c^2+m^2c^4}}{p} \vec{\zeta}(0)-  \frac{m^2c^4}{\sqrt{p^2c^2+m^2c^4}}\left[\vec{\zeta}(0)\cdot \vec{p}\right]\frac{\vec{p}}{p^3}- c^2 (H_{\rm b}^{\rm II})^{-1} \vec{p}  \Biggr\}{\rm e}^{-\frac{{\rm i}\;2 H_{\rm b}^{\rm II}t}{\hbar}},
            \end{eqnarray}
            and thus
            \begin{eqnarray}\label{eq:RtHb2}
                  \vec{r}(t)&=&\vec{r}(0)+c^2 (H_{\rm b}^{\rm II})^{-1}\vec{p}\;t \nonumber\\
                     &&   +\frac{{\rm i}\hbar}{2} \; \Biggl\{
                        \frac{\sqrt{p^2c^2+m^2c^4}}{p} \vec{\zeta}(0)-  \frac{m^2c^4}{\sqrt{p^2c^2+m^2c^4}}\left[\vec{\zeta}(0)\cdot \vec{p}\right]\frac{\vec{p}}{p^3}- c^2 (H_{\rm b}^{\rm II})^{-1} \vec{p} \Biggr\} (H_{\rm b}^{\rm II})^{-1}
                        \left({\rm e}^{\frac{-{\rm i}\,2\, H_{\rm b}^{\rm II}\,t}{\hbar}}-1\right).
            \end{eqnarray}
            Because
            \begin{eqnarray}\label{eq:F-5d}
                  \frac{m^2c^4}{\sqrt{p^2c^2+m^2c^4}}\left[\vec{\zeta}(0)\cdot \vec{p}\right]\frac{\vec{p}}{p^3}+ c^2 (H_{\rm b}^{\rm II})^{-1} \vec{p} &=&
                  \frac{m^2c^4}{p^2c^2+m^2c^4}\left[\sqrt{p^2c^2+m^2c^4} [\vec{\zeta}(0)\cdot \vec{p}]\right]\frac{\vec{p}}{p^3}
                  + c^2 (H_{\rm b}^{\rm II})^{-1} \vec{p} \nonumber\\
                  & =&  \frac{m^2c^4}{p^2c^2+m^2c^4} H_{\rm b}^{\rm II} \frac{\vec{p}}{p^2} + c^2 (H_{\rm b}^{\rm II})^{-1} \vec{p}=  \frac{m^2c^4}{H_{\rm b}^{\rm II}} \frac{\vec{p}}{p^2} + c^2 (H_{\rm b}^{\rm II})^{-1} \vec{p}\nonumber\\
                  &=& \frac{m^2c^4}{H_{\rm b}^{\rm II}} \frac{\vec{p}}{p^2} + \frac{p^2c^2}{H_{\rm b}^{\rm II}} \frac{\vec{p}}{p^2}=H_{\rm b}^{\rm II}\; \frac{\vec{p}}{p^2},
            \end{eqnarray}
            then from Eq. (\ref{eq:F-5d}) we have
            \begin{eqnarray}\label{eq:F-5e}
                  \vec{r}(t)
                  &=& \vec{r}(0)+c^2 (H_{\rm b}^{\rm II})^{-1}\vec{p}\;t\nonumber\\
                  &&+\frac{{\rm i}\hbar}{2} \; \Biggl\{
                              \frac{\sqrt{p^2c^2+m^2c^4}}{p} \vec{\zeta}(0)-  \frac{m^2c^4}{\sqrt{p^2c^2+m^2c^4}}\left[\vec{\zeta}(0)\cdot \vec{p}\right]\frac{\vec{p}}{p^3}- c^2 (H_{\rm b}^{\rm II})^{-1} \vec{p} \Biggr\} (H_{\rm b}^{\rm II})^{-1}
                        \left({\rm e}^{\frac{-{\rm i}\,2\,H_{\rm b}^{\rm II}\,t}{\hbar}}-1\right)\nonumber\\
                  &=& \vec{r}(0)+c^2 (H_{\rm b}^{\rm II})^{-1}\vec{p}\;t
                        +\frac{{\rm i}\hbar}{2} \; \Biggl\{
                        \frac{\sqrt{p^2c^2+m^2c^4}}{p} \vec{\zeta}(0)-  H_{\rm b}^{\rm II}\; \frac{\vec{p}}{p^2}\Biggr\} (H_{\rm b}^{\rm II})^{-1}
                        \left({\rm e}^{\frac{-{\rm i}\,2\,H_{\rm b}^{\rm II}\,t}{\hbar}}-1\right).
            \end{eqnarray}

                 For the position operator in Eq. (\ref{eq:F-5e}), the third term is an oscillation term, which is related to the ``position Zitterbewegung''. We need to calculate the following expectation value for the ``position Zitterbewegung'' operator
                  \begin{eqnarray}
                        \hat{\mathcal{Z}}_{\rm b}^{\rm II} &=& \dfrac{{\rm i}\hbar}{2} \Biggl\{
                              \dfrac{\sqrt{p^2c^2+m^2c^4}}{p} \vec{\zeta}(0)
                              -H_{\rm b}^{\rm II} \dfrac{\vec{p}}{p^2}\Biggr\} (H_{\rm b}^{\rm II})^{-1} \left(
                                    {\rm e}^{\frac{-{\rm i}\,2\,H_{\rm b}^{\rm II}\,t}{\hbar}}-1\right),
                  \end{eqnarray}
                  denoted as
                  \begin{eqnarray}
                  {\mathcal{Z}}_{\rm b}^{\rm II} &=& \langle\Psi''|\hat{\mathcal{Z}}_{\rm b}^{\rm II} |\Psi''\rangle,
                  \end{eqnarray}
                  where $|\Psi''\rangle$ is the quantum state of $H_{\rm b}^{\rm II}$, and it is easy to prove that $\hat{\mathcal{Z}}_{\rm b}^{\rm II}$ is a Hermitian operator,
                  i.e., $(\hat{\mathcal{Z}}_{\rm b}^{\rm II})^\dagger=\hat{\mathcal{Z}}_{\rm b}^{\rm II}$.
                  \begin{proof}
                        First we prove
                        \begin{equation}
                              \Biggl\{H_{\rm b}^{\rm II},\ \dfrac{\sqrt{p^2c^2+m^2c^4}}{p} \vec{\zeta}(0)
                                    -H_{\rm b}^{\rm II} \dfrac{\vec{p}}{p^2}\Biggr\}=0.
                        \end{equation}
                        Since
                        \begin{equation}
                              \{\vec{\zeta},\ H_{\rm b}^{\rm II}\}=2\sqrt{p^2c^2+m^2c^4}\,\mathbb{I}\,\hat{p},
                        \end{equation}
                        then
                        \begin{eqnarray}
                              \Biggl\{H_{\rm b}^{\rm II},\ \dfrac{\sqrt{p^2c^2+m^2c^4}}{p}
                                          \vec{\zeta}(0)-H_{\rm b}^{\rm II} \dfrac{\vec{p}}{p^2}\Biggr\}
                                    &=&\Biggl\{H_{\rm b}^{\rm II},\ \dfrac{\sqrt{p^2c^2+m^2c^4}}{p}
                                          \vec{\zeta}-H_{\rm b}^{\rm II} \dfrac{\vec{p}}{p^2}\Biggr\} \nonumber\\
                                    &=& \dfrac{\sqrt{p^2c^2+m^2c^4}}{p} \{H_{\rm b}^{\rm II},\ \vec{\zeta}\}
                                          -2\,(H_{\rm b}^{\rm II})^2 \dfrac{\vec{p}}{p^2}\nonumber\\
                                    &=&2\dfrac{\sqrt{p^2c^2+m^2c^4}}{p}
                                                \sqrt{p^2c^2+m^2c^4}\,\mathbb{I}\,\hat{p}
                                          -2\,(p^2c^2+m^2c^4)\,\mathbb{I}\,\dfrac{\vec{p}}{p^2} \nonumber\\
                                    &=& 0,
                        \end{eqnarray}
                        which means
                        \begin{eqnarray}
                              H_{\rm b}^{\rm II} \left[\dfrac{\sqrt{p^2c^2+m^2c^4}}{p} \vec{\zeta}(0)
                                    -H_{\rm b}^{\rm II} \dfrac{\vec{p}}{p^2}\right]
                              =-\left[\dfrac{\sqrt{p^2c^2+m^2c^4}}{p} \vec{\zeta}(0)
                                    -H_{\rm b}^{\rm II} \dfrac{\vec{p}}{p^2}\right]H_{\rm b}^{\rm II},
                        \end{eqnarray}
                        viz.,
                        \begin{eqnarray}
                              {\rm e}^{\frac{{\rm i}\,2H_{\rm b}^{\rm II}\,t}{\hbar}} \left[
                                    \dfrac{\sqrt{p^2c^2+m^2c^4}}{p} \vec{\zeta}(0)
                                    -H_{\rm b}^{\rm II} \dfrac{\vec{p}}{p^2}\right]
                              =\left[\dfrac{\sqrt{p^2c^2+m^2c^4}}{p} \vec{\zeta}(0)
                                    -H_{\rm b}^{\rm II} \dfrac{\vec{p}}{p^2}\right]
                                    {\rm e}^{-\frac{{\rm i}\,2H_{\rm b}^{\rm II}\,t}{\hbar}}.
                        \end{eqnarray}
                        After that,
                        \begin{eqnarray}
                                    (\hat{\mathcal{Z}}_{\rm b}^{\rm II})^\dagger &=&
                                          -\dfrac{{\rm i}\hbar}{2} \left({\rm e}^{
                                                \frac{{\rm i}\,2 H_{\rm b}^{\rm II}\,t}{\hbar}}-1
                                          \right) (H_{\rm b}^{\rm II})^{-1} \Biggl\{
                                                \dfrac{\sqrt{p^2c^2+m^2c^4}}{p} \vec{\zeta}(0)
                                                      - H_{\rm b}^{\rm II} \dfrac{\vec{p}}{p^2}\Biggr\}\nonumber\\
                                    &=& -\dfrac{{\rm i}\hbar}{2} (H_{\rm b}^{\rm II})^{-1} \Biggl\{
                                                \dfrac{\sqrt{p^2c^2+m^2c^4}}{p} \vec{\zeta}(0)
                                                      -H_{\rm b}^{\rm II} \dfrac{\vec{p}}{p^2}
                                          \Biggr\}\left({\rm e}^{
                                                \frac{-{\rm i}\,2 H_{\rm b}^{\rm II}\,t}{\hbar}}-1\right) \nonumber\\
                                    &=& -\dfrac{{\rm i}\hbar}{2}
                                          \dfrac{H_{\rm b}^{\rm II}}{(p^2 c^2 +m^2 c^4)} \Biggl\{
                                                \dfrac{\sqrt{p^2c^2+m^2c^4}}{p} \vec{\zeta}(0)
                                                      -H_{\rm b}^{\rm II} \dfrac{\vec{p}}{p^2}
                                          \Biggr\}\left({\rm e}^{
                                                \frac{-{\rm i}\,2 H_{\rm b}^{\rm II}\,t}{\hbar}}-1\right) \nonumber\\
                                    &=& \dfrac{{\rm i}\hbar}{2} \Biggl\{
                                          \dfrac{\sqrt{p^2c^2+m^2c^4}}{p} \vec{\zeta}(0)
                                                -H_{\rm b}^{\rm II} \dfrac{\vec{p}}{p^2}
                                          \Biggr\}\dfrac{H_{\rm b}^{\rm II}}{(p^2 c^2 +m^2 c^4)} \left(
                                                {\rm e}^{\frac{-{\rm i}\,2 H_{\rm b}^{\rm II}\,t}{\hbar}}-1
                                                \right) \nonumber\\
                                    &=& \dfrac{{\rm i}\hbar}{2} \Biggl\{
                                          \dfrac{\sqrt{p^2c^2+m^2c^4}}{p} \vec{\zeta}(0)
                                                -H_{\rm b}^{\rm II} \dfrac{\vec{p}}{p^2}
                                          \Biggr\} (H_{\rm b}^{\rm II})^{-1} \left({\rm e}^{
                                                \frac{-{\rm i}\,2 H_{\rm b}^{\rm II}\,t}{\hbar}}-1\right) \nonumber\\
                                    &=& \hat{\mathcal{Z}}_{\rm b}^{\rm II}.
                        \end{eqnarray}
                  This ends the proof.
                  \end{proof}

                  \begin{remark}
                  Before studying the ``position Zitterbewegung'' of the type-II Dirac's braidon, let us study its wave-functions $|\Psi''\rangle$. Based on Eq. (\ref{eq:hhh}), the transformation from $H_{\rm e}$ to $H_{\rm b}^{\rm II}$ is given by
            \begin{equation}\label{eq:bra-2aa}
            H_{\rm b}^{\rm II}= \mathcal{D} H_{\rm e} \mathcal{D}^\dagger, \;\;\; \mathcal{D}= {\rm e}^{\mathrm{i} \frac{\pi}{4} \Gamma_y}.
            \end{equation}
For a Dirac's electron, we have known that the common eigenstates of the set $\{H_{\rm e}, \hat{\Lambda}\}$ are $\{|\Psi_1\rangle, |\Psi_2\rangle, |\Psi_3\rangle, |\Psi_4\rangle\}$, where $\{|\Psi_1\rangle, |\Psi_2\rangle\}$ correspond to positive energy, while $\{|\Psi_3\rangle, |\Psi_4\rangle\}$ correspond to negative energy. The spectrum-decomposition of $H_{\rm e}$ is given by
            \begin{eqnarray}
            H_{\rm e}&=& E_{\rm p} \left[|\Psi_1\rangle\langle \Psi_1|+ |\Psi_2\rangle\langle \Psi_2| \right)-E_{\rm p} \left(|\Psi_3\rangle\langle \Psi_3|+ |\Psi_4\rangle\langle \Psi_4| \right).
            \end{eqnarray}
Based on the unitary matrix $\mathcal{D}$, from $|\Psi''_j\rangle=\mathcal{D}|\Psi_j\rangle$ we can determine the four eigenstates of type-II Dirac's braidon as
            \begin{eqnarray}\label{eq:braidoneigen-00}
            && \{|\Psi''_1\rangle, |\Psi''_2\rangle, |\Psi''_3\rangle, |\Psi''_4\rangle\} =\{\mathcal{D}|\Psi_1\rangle, \mathcal{D}|\Psi_2\rangle, \mathcal{D}|\Psi_3\rangle, \mathcal{D}|\Psi_4\rangle\},
            \end{eqnarray}
            which are common eigenstates of the set $\{H_{\rm b}^{\rm II}=\mathcal{D} H_{\rm e} \mathcal{D}^\dagger, \hat{\Lambda}''=\mathcal{D} \hat{\Lambda} \mathcal{D}^\dagger\}$. Similarly, $\{|\Psi''_1\rangle, |\Psi''_2\rangle\}$ correspond to positive energy of $H_{\rm b}^{\rm II}$, while $\{|\Psi''_3\rangle, |\Psi''_4\rangle\}$ correspond to negative energy of $H_{\rm b}^{\rm II}$. And $\{|\Psi''_1\rangle, |\Psi''_3\rangle\}$ correspond to positive helicity of $\hat{\Lambda}''$, while $\{|\Psi''_2\rangle, |\Psi''_4\rangle\}$ correspond to negative helicity of $\hat{\Lambda}''$.

            Because
            \begin{eqnarray}
            [\hat{\Lambda}, \vec{\alpha}\cdot\hat{p}]=0, \;\;\;\;\;\; [\hat{\Lambda}, \beta]=0,
            \end{eqnarray}
            thus
            \begin{eqnarray}
            [\hat{\Lambda}, \mathcal{D}]=0,
            \end{eqnarray}
            and then
            \begin{eqnarray}
            \hat{\Lambda}''=\mathcal{D}\,\hat{\Lambda}\, \mathcal{D}^\dagger=\hat{\Lambda}.
            \end{eqnarray}
            Therefore the four states in
            Eq. (\ref{eq:braidoneigen-00}) are common eigenstates of the set $\{H_{\rm b}^{\rm II}, \hat{\Lambda}\}$. By the way, the spectrum-decomposition of $H_{\rm b}^{\rm II}$ is given by
            \begin{eqnarray}
            H_{\rm b}^{\rm II}&=& E_{\rm p} \left(|\Psi''_1\rangle\langle \Psi''_1|+ |\Psi''_2\rangle\langle \Psi''_2| \right)-E_{\rm p} \left(|\Psi''_3\rangle\langle \Psi''_3|+ |\Psi''_4\rangle\langle \Psi''_4| \right)=\mathcal{D} H_{\rm e} \mathcal{D}^\dagger,
            \end{eqnarray}
            which is consistent with Eq. (\ref{eq:bra-2aa}). $\blacksquare$
                  \end{remark}

                  \begin{remark}
                  Let us introduce the following projection operators
                  \begin{eqnarray}
                        \Pi_\pm=\frac{1}{2}\left(\mathbb{I}\pm
                              \frac{H_{\rm b}^{\rm II}}{\sqrt{p^2c^2+m^2c^4}}\right), \;\;\;\;\; \Pi_\pm^2=\Pi_\pm,
                  \end{eqnarray}
                  which lead to
                  \begin{eqnarray}
                        && \Pi_+ |\Psi_1''\rangle = |\Psi_1''\rangle, \;\; \Pi_+ |\Psi_2''\rangle = |\Psi_2''\rangle, \;\; \Pi_+ |\Psi_3''\rangle = 0, \;\; \Pi_+ |\Psi_4''\rangle = 0, \nonumber\\
                        && \Pi_- |\Psi_1''\rangle = 0, \;\; \Pi_- |\Psi_2''\rangle = 0, \;\; \Pi_- |\Psi_3''\rangle = |\Psi_3''\rangle, \;\; \Pi_- |\Psi_4''\rangle = |\Psi_4''\rangle.
                  \end{eqnarray}
                  We then have
                  \begin{eqnarray}
                        && \Pi_+ \Biggl\{\dfrac{\sqrt{p^2c^2+m^2c^4}}{p} \vec{\zeta}(0)
                                    -H_{\rm b}^{\rm II} \dfrac{\vec{p}}{p^2}\Biggr\}\Pi_+ \nonumber \\
                        &=& \dfrac{1}{4} \left(\mathbb{I}
                              +\dfrac{H_{\rm b}^{\rm II}}{\sqrt{p^2 c^2 +m^2 c^4}}\right)\Biggl\{
                                    \dfrac{\sqrt{p^2c^2+m^2c^4}}{p} \vec{\zeta}(0)
                              -H_{\rm b}^{\rm II} \dfrac{\vec{p}}{p^2}\Biggr\}
                              \left(\mathbb{I}+\frac{H_{\rm b}^{\rm II}}{\sqrt{p^2c^2+m^2c^4}}\right)\nonumber\\
                        &=& \dfrac{1}{4} \Biggl\{\dfrac{\sqrt{p^2c^2+m^2c^4}}{p} \vec{\zeta}(0)
                              -H_{\rm b}^{\rm II} \dfrac{\vec{p}}{p^2}\Biggr\}
                              + \frac{1}{4\sqrt{p^2c^2+m^2c^4}} \left\{H_{\rm b}^{\rm II},\
                                    \dfrac{\sqrt{p^2c^2+m^2c^4}}{p} \vec{\zeta}(0)
                              -H_{\rm b}^{\rm II} \dfrac{\vec{p}}{p^2}\right\} \nonumber \\
                              &&\quad +\dfrac{1}{4(p^2c^2+m^2c^4)} H_{\rm b}^{\rm II} \Biggl\{
                                    \dfrac{\sqrt{p^2c^2+m^2c^4}}{p} \vec{\zeta}(0)
                              -H_{\rm b}^{\rm II} \dfrac{\vec{p}}{p^2}\Biggr\}
                                    H_{\rm b}^{\rm II} \nonumber \\
                        &=& \dfrac{1}{4} \Biggl\{\dfrac{\sqrt{p^2c^2+m^2c^4}}{p} \vec{\zeta}(0)
                              -H_{\rm b}^{\rm II} \dfrac{\vec{p}}{p^2}\Biggr\}
                              -\dfrac{1}{4(p^2c^2+m^2c^4)} (H_{\rm b}^{\rm II})^2 \Biggl\{
                                    \dfrac{\sqrt{p^2c^2+m^2c^4}}{p} \vec{\zeta}(0)
                              -H_{\rm b}^{\rm II} \dfrac{\vec{p}}{p^2}\Biggr\}\nonumber\\
                        &=& \frac{1}{4} \Biggl\{\dfrac{\sqrt{p^2c^2+m^2c^4}}{p} \vec{\zeta}(0)
                                          -H_{\rm b}^{\rm II} \dfrac{\vec{p}}{p^2}
                              -\biggl[\dfrac{\sqrt{p^2c^2+m^2c^4}}{p} \vec{\zeta}(0)
                                    -H_{\rm b}^{\rm II} \dfrac{\vec{p}}{p^2}\biggr]\Biggr\} \nonumber \\
                        &=& 0;
                  \end{eqnarray}
                  similarly, we have
                  \begin{eqnarray}
                        \Pi_- \Biggl\{\dfrac{\sqrt{p^2c^2+m^2c^4}}{p} \vec{\zeta}(0)
                              -H_{\rm b}^{\rm II} \dfrac{\vec{p}}{p^2}\Biggr\}\Pi_- =0.
                  \end{eqnarray}
                  The above results lead to
                  \begin{eqnarray}
                        \Pi_+ \hat{\mathcal{Z}}_{\rm b}^{\rm II} \Pi_+ =0,\;\;\;\;\;
                        \Pi_- \hat{\mathcal{Z}}_{\rm b}^{\rm II} \Pi_- =0.
                  \end{eqnarray}

            Suppose the quantum state $|\Psi''\rangle$ is the superposition of the positive-energy states, i.e.,
            \begin{eqnarray}
            &&  |\Psi''\rangle\equiv|\Psi''_+\rangle = c_1 |\Psi''_1\rangle+ c_2 |\Psi''_2\rangle,
            \end{eqnarray}
            then we have
            \begin{eqnarray}
            &&  |\Psi''_+\rangle=\Pi_+ \;|\Psi''_+\rangle, \;\;\; \langle \Psi''_+ | \; \Pi_+ = \langle \Psi''_+ |,
            \end{eqnarray}
            therefore
            \begin{eqnarray}
            &&  \mathcal{Z}_{\rm b}^{\rm II}=\langle\Psi''_+ | \hat{\mathcal{Z}}_{\rm b}^{\rm II} |\Psi''_+\rangle= \langle\Psi_+ | \left(\Pi''_+ \; \hat{\mathcal{Z}}_{\rm b}^{\rm II}\;\Pi_+ \right)|\Psi''_+\rangle=0.
            \end{eqnarray}
            Similarly, if the quantum state $|\Psi''\rangle$ is the superposition of the negative-energy states, i.e.,
            \begin{eqnarray}
            &&  |\Psi''\rangle\equiv|\Psi''_-\rangle = c_3 |\Psi''_3\rangle+ c_4 |\Psi''_4\rangle,
            \end{eqnarray}
            then one has
            \begin{eqnarray}
            && \mathcal{Z}_{\rm b}^{\rm II}= \langle\Psi''_- | \hat{\mathcal{Z}}_{\rm b}^{\rm II} |\Psi''_-\rangle= \langle\Psi_- | \left(\Pi''_- \; \hat{\mathcal{Z}}_{\rm b}^{\rm II}\;\Pi_- \right)|\Psi''_-\rangle=0.
            \end{eqnarray}
            Thereby, if the type-II Dirac's braidon  is in a superposition state of only positive-energy (or negative-energy), then $\mathcal{Z}_{\rm b}^{\rm II} =0$, i.e, there is no phenomenon of ``position Zitterbewegung''. $\blacksquare$
            \end{remark}

            \begin{remark}
                  Let us introduce the projective operators
                  \begin{eqnarray}
                        \Pi^{\rm s}_\pm =\frac{1}{2}\left(\mathbb{I} \pm \dfrac{2}{\hbar} \hat{\Lambda}
                              \right),\qquad
                        (\Pi^{\rm s}_\pm)^2 =\Pi^{\rm s}_\pm,
                  \end{eqnarray}
                  with the helicity operator
                  \begin{eqnarray}
                  \hat{\Lambda}=\vec{S} \cdot \hat{p}=\dfrac{\hbar}{2} \vec{\Sigma} \cdot \hat{p}.
                  \end{eqnarray}
                 We easily have
                  \begin{eqnarray}
                        && \Pi^{\rm s}_+ |\Psi''_1\rangle = |\Psi''_1\rangle, \;\;
                        \Pi^{\rm s}_+ |\Psi''_2\rangle = 0, \;\;
                        \Pi^{\rm s}_+ |\Psi''_3\rangle = |\Psi''_3\rangle, \;\;
                        \Pi^{\rm s}_+ |\Psi''_4\rangle = 0, \nonumber\\
                        && \Pi^{\rm s}_- |\Psi''_1\rangle = 0, \;\;
                        \Pi^{\rm s}_- |\Psi''_2\rangle = |\Psi''_2\rangle, \;\;
                        \Pi^{\rm s}_- |\Psi''_3\rangle = 0, \;\;
                        \Pi^{\rm s}_- |\Psi''_4\rangle = |\Psi''_4\rangle.
                  \end{eqnarray}
                  After that,
                  \begin{eqnarray}
                      && \Bigg\{\hat{\Lambda},\ \biggl[
                                    \dfrac{\sqrt{p^2c^2+m^2c^4}}{p} \vec{\zeta}(0)
                                    -H_{\rm b}^{\rm II} \dfrac{\vec{p}}{p^2}\biggr]\Bigg\}
                              =\bigg\{\dfrac{\hbar}{2} \vec{\Sigma}\cdot\hat{p},\
                                    \dfrac{E_{\rm p}}{p} \bigl[\vec{\zeta}(0)
                                          -[\vec{\zeta}(0)\cdot\hat{p}]\hat{p}\bigr]\bigg\} \nonumber\\
                        &      =& \dfrac{\hbar}{2} \dfrac{E_{\rm p}}{p} \bigg[
                                    \Bigl\{\vec{\Sigma}\cdot\hat{p},\
                                          \vec{\zeta}(0)\Bigr\}
                                    -\Bigl\{\vec{\Sigma}\cdot\hat{p},\
                                          [\vec{\zeta}(0)\cdot\hat{p}]\hat{p}\Bigr\}\bigg]
                              ={\rm i}\dfrac{\hbar}{2} \dfrac{E_{\rm p}}{p} \bigg[
                                    \Bigl\{\vec{\Sigma}\cdot\hat{p},\
                                          \beta\,\vec{\alpha}\Bigr\}
                                    -\Bigl\{\vec{\Sigma}\cdot\hat{p},\
                                          \beta(\vec{\alpha}\cdot\hat{p})\hat{p}\Bigr\}
                                          \bigg] \nonumber\\
                          &    =& {\rm i}\dfrac{\hbar}{2} \dfrac{E_{\rm p}}{p} \Biggl(
                                          \begin{bmatrix}
                                                      \vec{\sigma}\cdot\hat{p} & 0 \\
                                                      0 & \vec{\sigma}\cdot\hat{p}
                                                \end{bmatrix}\begin{bmatrix}
                                                      0 & \vec{\sigma} \\
                                                      -\vec{\sigma} & 0
                                                \end{bmatrix}
                                          +\begin{bmatrix}
                                                      0 & \vec{\sigma} \\
                                                      -\vec{\sigma} & 0
                                                \end{bmatrix}\begin{bmatrix}
                                                      \vec{\sigma}\cdot\hat{p} & 0 \\
                                                      0 & \vec{\sigma}\cdot\hat{p}
                                                \end{bmatrix}\Biggr)
                                    -{\rm i}\hbar\dfrac{E_{\rm p}}{p} (
                                          \vec{\Sigma}\cdot\hat{p})\beta(\vec{\alpha}
                                                \cdot\hat{p})\hat{p} \nonumber\\
                           &   =& {\rm i}\dfrac{\hbar}{2\,p} E_{\rm p} \Biggl(
                                          \begin{bmatrix}
                                                      \vec{\sigma}\cdot\hat{p} & 0 \\
                                                      0 & \vec{\sigma}\cdot\hat{p}
                                                \end{bmatrix}\begin{bmatrix}
                                                      0 & \vec{\sigma} \\
                                                      -\vec{\sigma} & 0
                                                \end{bmatrix}
                                          +\begin{bmatrix}
                                                      0 & \vec{\sigma} \\
                                                      -\vec{\sigma} & 0
                                                \end{bmatrix}\begin{bmatrix}
                                                      \vec{\sigma}\cdot\hat{p} & 0 \\
                                                      0 & \vec{\sigma}\cdot\hat{p}
                                                \end{bmatrix}\Biggr)
                                    -{\rm i}\hbar\dfrac{E_{\rm p}}{p} \begin{bmatrix}
                                                      \vec{\sigma}\cdot\hat{p} & 0 \\
                                                      0 & \vec{\sigma}\cdot\hat{p}
                                                \end{bmatrix}\begin{bmatrix}
                                                      0 & \vec{\sigma}\cdot\hat{p} \\
                                                      -\vec{\sigma}\cdot\hat{p} & 0
                                                \end{bmatrix}\hat{p} \nonumber\\
                            &  =& {\rm i}\dfrac{\hbar}{2\,p} E_{\rm p} \Biggl\{
                                          \begin{bmatrix}
                                                      0 & (\vec{\sigma}\cdot\hat{p})
                                                            \vec{\sigma} \\
                                                      -(\vec{\sigma}\cdot\hat{p})
                                                            \vec{\sigma} & 0
                                                \end{bmatrix}
                                          +\begin{bmatrix}
                                                      0 & \vec{\sigma}(
                                                            \vec{\sigma}\cdot\hat{p}) \\
                                                      -\vec{\sigma}(\vec{\sigma}\cdot\hat{p})
                                                            & 0
                                                \end{bmatrix}\Biggr\}
                                    -{\rm i}\hbar\dfrac{E_{\rm p}}{p} \begin{bmatrix}
                                                0 & (\vec{\sigma}\cdot\hat{p})(
                                                      \vec{\sigma}\cdot\hat{p}) \\
                                                -(\vec{\sigma}\cdot\hat{p})(
                                                      \vec{\sigma}\cdot\hat{p}) & 0
                                          \end{bmatrix}\hat{p} \nonumber\\
                             & =& {\rm i}\dfrac{\hbar}{2\,p} E_{\rm p} \begin{bmatrix}
                                                0 & (\vec{\sigma}\cdot\hat{p})
                                                            \vec{\sigma}
                                                      +\vec{\sigma}(\vec{\sigma}
                                                            \cdot\hat{p}) \\
                                                -(\vec{\sigma}\cdot\hat{p})
                                                            \vec{\sigma}
                                                      -\vec{\sigma}(\vec{\sigma}\cdot
                                                            \hat{p}) & 0
                                          \end{bmatrix}
                                    -{\rm i}\hbar\dfrac{E_{\rm p}}{p} \begin{bmatrix}
                                                0 & \openone \\
                                                -\openone & 0
                                          \end{bmatrix}\hat{p} \nonumber\\
                            &  =& {\rm i}\dfrac{\hbar}{2\,p} E_{\rm p} \begin{bmatrix}
                                                0 &  2\,\hat{p} \,\openone  \\
                                                -2\,\hat{p}\,\openone & 0
                                          \end{bmatrix}
                                    -{\rm i}\hbar\dfrac{E_{\rm p}}{p} \begin{bmatrix}
                                                0 & \openone \\
                                                -\openone & 0
                                          \end{bmatrix}\hat{p}
                           =0,
                  \end{eqnarray}
                  and then
                  \begin{eqnarray}
                        && \Pi^{\rm s}_+ \biggl[\dfrac{\sqrt{p^2c^2+m^2c^4}}{p} \vec{\zeta}(0)
                              - H_{\rm b}^{\rm II} \dfrac{\vec{p}}{p^2}\biggr]\Pi^{\rm s}_+ \nonumber\\
                        &=&\frac{1}{4} \left(\mathbb{I}+\dfrac{2}{\hbar} \hat{\Lambda}\right)
                              \biggl[\dfrac{\sqrt{p^2c^2+m^2c^4}}{p} \vec{\zeta}(0)
                                          -H_{\rm b}^{\rm II} \dfrac{\vec{p}}{p^2}\biggr]\left(\mathbb{I}
                                    +\dfrac{2}{\hbar} \hat{\Lambda}\right) \notag\\
                        &=& \frac{1}{4} \biggr(\biggl[\dfrac{\sqrt{p^2c^2+m^2c^4}}{p}
                                    \vec{\zeta}(0)-H_{\rm b}^{\rm II} \dfrac{\vec{p}}{p^2}\biggr]
                              +\frac{2}{\hbar} \biggl\{\hat{\Lambda},\ \Bigl[
                                    \dfrac{\sqrt{p^2c^2+m^2c^4}}{p} \vec{\tau}(0)
                                    -H_{\rm b}^{\rm II} \dfrac{\vec{p}}{p^2}\Bigr]\biggr\} \nonumber\\
                        &&     +\frac{4}{\hbar^4} \hat{\Lambda} \biggl[
                                    \dfrac{\sqrt{p^2c^2+m^2c^4}}{p} \vec{\zeta}(0)
                                    -H_{\rm b}^{\rm II} \dfrac{\vec{p}}{p^2}\biggr]\hat{\Lambda}\biggr) \nonumber\\
                        &=& \frac{1}{4} \left\{\biggl[\dfrac{\sqrt{p^2c^2+m^2c^4}}{p}
                                    \vec{\zeta}(0)-H_{\rm b}^{\rm II} \dfrac{\vec{p}}{p^2}\biggr]
                              -\frac{4}{\hbar^2} \hat{\Lambda}^2 \biggl[
                                    \dfrac{\sqrt{p^2c^2+m^2c^4}}{p} \vec{\zeta}(0)
                                    -H_{\rm b}^{\rm II} \dfrac{\vec{p}}{p^2}\biggr]\right\} \notag\\
                        &=& \frac{1}{4} \Biggl\{\biggl[\dfrac{\sqrt{p^2c^2+m^2c^4}}{p}
                                    \vec{\zeta}(0)-H_{\rm b}^{\rm II} \dfrac{\vec{p}}{p^2}\biggr]
                              -\biggl[\dfrac{\sqrt{p^2c^2+m^2c^4}}{p} \vec{\zeta}(0)
                                    -H_{\rm b}^{\rm II} \dfrac{\vec{p}}{p^2}\biggr]\Biggr\}\nonumber\\
                       & =&0.
                  \end{eqnarray}
                  Similarly, we have
                  \begin{eqnarray}
                        \Pi^{\rm s}_- \biggl[\dfrac{\sqrt{p^2c^2+m^2c^4}}{p} \vec{\zeta}(0)
                              - H_{\rm b}^{\rm II} \dfrac{\vec{p}}{p^2}\biggr]\Pi^{\rm s}_- =0.
                  \end{eqnarray}
                  The above results lead to
                  \begin{eqnarray}
                        \Pi^{\rm s}_+ \hat{\mathcal{Z}}_{\rm b}^{\rm II} \Pi^{\rm s}_+ =0,\;\;\;\;\;
                        \Pi^{\rm s}_- \hat{\mathcal{Z}}_{\rm b}^{\rm II} \Pi^{\rm s}_- =0.
                  \end{eqnarray}
                  Thus, if the quantum state $|\Psi''\rangle$ is the superposition of the positive-helicity states, i.e.,
                  \begin{eqnarray}
                  &&  |\Psi''\rangle\equiv|\Psi''_+\rangle = c_1 |\Psi''_1\rangle+ c_3 |\Psi''_3\rangle,
                  \end{eqnarray}
                  or the superposition of the negative-helicity states, i.e.,
                  \begin{eqnarray}
                  &&  |\Psi''\rangle\equiv|\Psi''_-\rangle = c_2 |\Psi''_2\rangle+ c_4 |\Psi''_4\rangle,
                  \end{eqnarray}
                  then one obtains
                  \begin{eqnarray}\label{eq:II-1a}
                     && \mathcal{Z}_{\rm b}^{\rm II}=\langle \Psi''_+|\hat{\mathcal{Z}}_{\rm b}^{\rm II}|\Psi''_+\rangle=  \langle \Psi''_+|\Pi^{\rm s}_+ \hat{\mathcal{Z}}_{\rm b}^{\rm II} \Pi^{\rm s}_+|\Psi''_+\rangle =0,
                  \end{eqnarray}
                   \begin{eqnarray}\label{eq:II-1ab}
                     && \mathcal{Z}_{\rm b}^{\rm II}=\langle \Psi''_-|\hat{\mathcal{Z}}_{\rm b}^{\rm II}|\Psi''_-\rangle=  \langle \Psi''_-|\Pi^{\rm s}_- \hat{\mathcal{Z}}_{\rm b}^{\rm II} \Pi^{\rm s}_-|\Psi''_-\rangle =0.
                  \end{eqnarray}
                  Thereby, if the type-II Dirac's braidon  is in a superposition state of the same helicity, then there is no phenomenon of ``position Zitterbewegung''. Thus the behavior of the type-II Dirac's braidon is different from those of Dirac's electron and the type-I Dirac's braidon. $\blacksquare$
            \end{remark}

           \begin{remark}Now we would like to doubly check Eq. (\ref{eq:II-1a}) from the viewpoint of unitary transformation $\mathcal{D}$. Let
                  \begin{equation}\label{eq:EigenH3P3}
                        |\Psi''_j\rangle=\mathcal{D}\,|{\Psi}_j\rangle,\quad
                              j\in\{1,2,3,4\},
                  \end{equation}
                 where the unitary transformation is given by
                  \begin{eqnarray}
                        \mathcal{D}&=& {\rm e}^{\mathrm{i} \frac{\pi}{4} \Gamma_y}
                        =\dfrac{1}{\sqrt{2}} \left(\mathbb{I}+{\rm i}\,\Gamma_y\right)\nonumber\\
                       & =&\dfrac{1}{\sqrt{2}} \left[\mathbb{I}+{\rm i}\;
                              \dfrac{(-m\,c^2 \vec{\alpha}\cdot\hat{p}+\beta\,p\,c)}{
                                    \sqrt{p^2 c^2 +m^2 c^4}}\right],
                  \end{eqnarray}
                  which transforms Dirac's electron $H_{\rm e}$ to the type-II Dirac's braidon $H_{\rm b}^{\rm II}$ in the following way
                  \begin{equation}
                        H_{\rm b}^{\rm II} =\mathcal{D} H_{\rm e} \mathcal{D}^\dagger.
                  \end{equation}
                  Let us consider the quantum state
                  \begin{eqnarray}
                        |\Psi''\rangle=\cos\eta |\Psi''_1\rangle
                              +\sin\eta |\Psi''_3\rangle,
                  \end{eqnarray}
                  we need to calculate the expectation value
                  \begin{equation}
                        \mathcal{Z}_{\rm b}^{\rm II}  =\langle\Psi''|\hat{\mathcal{Z}}_{\rm b}^{\rm II} |\Psi''\rangle
                        =\sin(2\eta)\,{\rm Re}\left(
                              \langle\Psi''_1|\hat{\mathcal{Z}}_{\rm b}^{\rm II}
                                    |\Psi''_3 \rangle\right),
                  \end{equation}
                  with
                  \begin{eqnarray}
                        \langle\Psi''_1|\hat{\mathcal{Z}}_{\rm b}^{\rm II}|\Psi''_3 \rangle
                        &=& \dfrac{{\rm i}\hbar}{2} \langle\Psi''_1|\;\Biggl\{
                              \dfrac{\sqrt{p^2c^2+m^2c^4}}{p} \vec{\zeta}(0)
                              -H_{\rm b}^{\rm II}\; \frac{\vec{p}}{p^2}\Biggr\} (H_{\rm b}^{\rm II})^{-1} \left({\rm e}^{
                                    \frac{-{\rm i}\,2\,H_{\rm b}^{\rm II}\,t}{\hbar}}-1\right)|\Psi''_3\rangle.
                  \end{eqnarray}
                  Precisely, we have
                  \begin{eqnarray}
                       && \dfrac{{\rm i}\hbar}{2} \langle\Psi''_1|\Biggl\{
                                    \dfrac{\sqrt{p^2c^2+m^2c^4}}{p} \vec{\zeta}(0)
                                    -H_{\rm b}^{\rm II} \dfrac{\vec{p}}{p^2}\Biggr\} (H_{\rm b}^{\rm II})^{-1} \left(
                                          {\rm e}^{\frac{-{\rm i}\,2\,H_{\rm b}^{\rm II}\,t}{\hbar}}-1
                                          \right)|\Psi''_3\rangle \nonumber\\
                       & =& -\dfrac{{\rm i}\hbar}{2} \dfrac{1}{E_{\rm p}} \left(
                                    {\rm e}^{\frac{{\rm i}\,2\,E_{\rm p}\,t}{\hbar}}
                                    -1\right)\langle\Psi''_1|\Biggl(
                                          \dfrac{E_{\rm p}}{p} \vec{\zeta}(0)
                                          -H_{\rm b}^{\rm II} \dfrac{\vec{p}}{p^2}\Biggr)
                                                |\Psi''_3\rangle  \nonumber\\
                         &     =& -{\rm i}\dfrac{\hbar}{2} \dfrac{1}{E_{\rm p}} \left(
                                    {\rm e}^{\frac{{\rm i}\,2\,E_{\rm p}\,t}{\hbar}}
                                    -1\right)\langle\Psi_1|\mathcal{D}^\dagger \Biggl(
                                          \dfrac{E_{\rm p}}{p} \vec{\zeta}(0)
                                          -H_{\rm b}^{\rm II} \dfrac{\vec{p}}{p^2}\Biggr)\mathcal{D}
                                                |\Psi_3\rangle   \nonumber\\
                          &    =& -{\rm i}\dfrac{\hbar}{2} \dfrac{1}{E_{\rm p}} \left(
                                    {\rm e}^{\frac{{\rm i}\,2\,E_{\rm p}\,t}{\hbar}}
                                    -1\right)\langle\Psi_1|\Biggl[\dfrac{E_{\rm p}}{p}
                                          \bigl(\mathcal{D}^\dagger \vec{\zeta}(0)\,
                                                \mathcal{D}\bigr)
                                          -\bigl(\mathcal{D}^\dagger H_{\rm b}^{\rm II} \mathcal{D}
                                          \bigr)\dfrac{\vec{p}}{p^2}\Biggr]
                                                |\Psi_3\rangle    \nonumber\\
                           &   =& -{\rm i}\dfrac{\hbar}{2} \dfrac{1}{E_{\rm p}} \left(
                                    {\rm e}^{\frac{{\rm i}\,2\,E_{\rm p}\,t}{\hbar}}
                                    -1\right)\langle\Psi_1|\Biggl[\dfrac{E_{\rm p}}{p}
                                          \bigl(\mathcal{D}^\dagger \vec{\zeta}(0)\,
                                                \mathcal{D}\bigr)
                                          -H_{\rm e} \dfrac{\vec{p}}{p^2}\Biggr]
                                                |\Psi_3\rangle\nonumber\\
                              &=&-{\rm i}\dfrac{\hbar}{2} \dfrac{1}{E_{\rm p}} \left(
                                    {\rm e}^{\frac{{\rm i}\,2\,E_{\rm p}\,t}{\hbar}}
                                    -1\right)\langle\Psi_1|\dfrac{E_{\rm p}}{p}
                                          \bigl(\mathcal{D}^\dagger \vec{\zeta}(0)\,
                                                \mathcal{D}\bigr)|\Psi_3\rangle   \nonumber\\
                            &  =& -{\rm i}\dfrac{\hbar}{2} \dfrac{1}{p} \left(
                                    {\rm e}^{\frac{{\rm i}\,2\,E_{\rm p}\,t}{\hbar}}
                                    -1\right)\langle\Psi_1|\bigl(\mathcal{D}^\dagger
                                          \vec{\zeta}(0)\,\mathcal{D}\bigr)|\Psi_3\rangle.
                  \end{eqnarray}
                  Because
                  \begin{eqnarray}
                    && \mathcal{D}^\dagger \vec{\zeta}(0)\,\mathcal{D}
                              =\dfrac{1}{\sqrt{2}} \left[\mathbb{I}+{\rm i}\dfrac{
                                    (-m\,c^2 \vec{\alpha}\cdot\hat{p}+\beta\,p\,c)}{
                                    \sqrt{p^2 c^2 +m^2 c^4}}\right]^\dagger {\rm i}\,
                                          \beta\,\vec{\alpha}\dfrac{1}{\sqrt{2}} \left[
                                                \mathbb{I}+{\rm i}\dfrac{(
                                                      -m\,c^2 \vec{\alpha}\cdot\hat{p}
                                                      +\beta\,p\,c)}{
                                                            \sqrt{p^2 c^2 +m^2 c^4}}
                                                            \right] \nonumber\\
                            &  =& \dfrac{{\rm i}}{2} \left[\mathbb{I}
                                    -\dfrac{{\rm i}}{E_{\rm p}} (
                                          -m\,c^2 \vec{\alpha}\cdot\hat{p}+\beta\,p\,c)
                                          \right]\beta\,\vec{\alpha}\left[
                                                \mathbb{I}+\dfrac{{\rm i}}{E_{\rm p}} (
                                                      -m\,c^2 \vec{\alpha}\cdot\hat{p}
                                                      +\beta\,p\,c)\right] \nonumber\\
                             & =& \dfrac{{\rm i}}{2} \left[\mathbb{I}
                                    -\dfrac{{\rm i}}{E_{\rm p}} (
                                          -m\,c^2 \vec{\alpha}\cdot\hat{p}+\beta\,p\,c)
                                          \right]\left\{\beta\,\vec{\alpha}
                                                -\dfrac{{\rm i}}{E_{\rm p}} \bigl[
                                                      m\,c^2 \beta\,\vec{\alpha}
                                                            (\vec{\alpha}\cdot\hat{p})
                                                      +p\,c\,\vec{\alpha}\bigr]\right\} \nonumber\\
                             & =& \dfrac{{\rm i}}{2} \left[\mathbb{I}
                                    -\dfrac{{\rm i}}{E_{\rm p}} (
                                          -m\,c^2 \vec{\alpha}\cdot\hat{p}+\beta\,p\,c)
                                          \right]\left\{\beta\,\vec{\alpha}
                                                -\dfrac{{\rm i}}{E_{\rm p}} \Bigl\{
                                                      m\,c^2 \beta\,\bigl[2\,\mathbb{I}\,\hat{p}-
                                                            (\vec{\alpha}\cdot\hat{p})\vec{\alpha}\bigr]
                                                      +p\,c\,\vec{\alpha}\Bigr\}\right\} \nonumber\\
                              &=& \dfrac{{\rm i}}{2} \left[\mathbb{I}
                                    -\dfrac{{\rm i}}{E_{\rm p}} (
                                          -m\,c^2 \vec{\alpha}\cdot\hat{p}+\beta\,p\,c)
                                          \right]\left\{\beta\,\vec{\alpha}
                                                -\dfrac{{\rm i}}{E_{\rm p}} \Bigl[
                                                      2\,m\,c^2 \beta\,\hat{p}
                                                      -m\,c^2 \beta(\vec{\alpha}\cdot
                                                            \hat{p})\vec{\alpha}
                                                      +p\,c\,\vec{\alpha}\Bigr]\right\} \nonumber\\
                              &=& \dfrac{1}{2} \left[\mathbb{I}
                                    +\dfrac{{\rm i}}{E_{\rm p}} (
                                          m\,c^2 \vec{\alpha}\cdot\hat{p}-\beta\,p\,c)
                                          \right]\left\{{\rm i}\,\beta\,\vec{\alpha}
                                                +\dfrac{1}{E_{\rm p}} \Bigl[
                                                      2\,m\,c^2 \beta\,\hat{p}
                                                      -m\,c^2 \beta(\vec{\alpha}\cdot
                                                            \hat{p})\vec{\alpha}
                                                      +p\,c\,\vec{\alpha}\Bigr]\right\} \nonumber\\
                              &=& \dfrac{1}{2} \Biggl\{{\rm i}\,\beta\,\vec{\alpha}
                                    +\dfrac{1}{E_{\rm p}} \Bigl[2\,m\,c^2 \beta\,\hat{p}
                                          -m\,c^2 \beta(\vec{\alpha}\cdot\hat{p})
                                                \vec{\alpha}
                                          +p\,c\,\vec{\alpha}\Bigr]
                                    +\dfrac{{\rm i}}{E_{\rm p}} (
                                          m\,c^2 \vec{\alpha}\cdot\hat{p}-\beta\,p\,c)
                                          {\rm i}\,\beta\,\vec{\alpha} \nonumber\\
                               &     &\quad\ +\dfrac{{\rm i}}{E_{\rm p}} (
                                          m\,c^2 \vec{\alpha}\cdot\hat{p}-\beta\,p\,c)
                                          \dfrac{1}{E_{\rm p}} \Bigl[2\,m\,c^2 \beta\,
                                                      \hat{p}
                                                -m\,c^2 \beta(\vec{\alpha}\cdot
                                                      \hat{p})\vec{\alpha}
                                                +p\,c\,\vec{\alpha}\Bigr]
                                    \Biggr\} \nonumber\\
                              &=& \dfrac{1}{2} \Biggl\{{\rm i}\,\beta\,\vec{\alpha}
                                    +\dfrac{1}{E_{\rm p}} \Bigl[2\,m\,c^2 \beta\,\hat{p}
                                          -m\,c^2 \beta(\vec{\alpha}\cdot\hat{p})
                                                \vec{\alpha}+p\,c\,\vec{\alpha}\Bigr]
                                    -\dfrac{1}{E_{\rm p}}\bigl[
                                          m\,c^2 (\vec{\alpha}\cdot\hat{p})\beta\,\vec{\alpha}-p\,c\,\vec{\alpha}\bigr] \nonumber\\
                               &     &\quad\ +\dfrac{{\rm i}}{E_{\rm p}^2} (
                                          m\,c^2 \vec{\alpha}\cdot\hat{p}-\beta\,p\,c)
                                          \Bigl[2\,m\,c^2 \beta\,\hat{p}
                                                -m\,c^2 \beta(\vec{\alpha}\cdot\hat{p})
                                                      \vec{\alpha}+p\,c\,\vec{\alpha}\Bigr]
                                    \Biggr\} \nonumber\\
                             & =& \dfrac{1}{2} \Biggl\{{\rm i}\,\beta\,\vec{\alpha}
                                    +\dfrac{1}{E_{\rm p}} \Bigl[2\,m\,c^2 \beta\,\hat{p}
                                          -m\,c^2 \beta(\vec{\alpha}\cdot\hat{p})
                                                \vec{\alpha}+p\,c\,\vec{\alpha}\Bigr]
                                    +\dfrac{1}{E_{\rm p}} \bigl[
                                          m\,c^2 \beta(\vec{\alpha}\cdot\hat{p})\vec{\alpha}+p\,c\,\vec{\alpha}\bigr] \nonumber\\
                              &      &\quad\ +\dfrac{{\rm i}}{E_{\rm p}^2} (
                                          m\,c^2 \vec{\alpha}\cdot\hat{p}-\beta\,p\,c)
                                          \Bigl[2\,m\,c^2 \beta\,\hat{p}
                                                -m\,c^2 \beta(\vec{\alpha}\cdot\hat{p})
                                                      \vec{\alpha}+p\,c\,\vec{\alpha}\Bigr]
                                    \Biggr\} \nonumber\\
                              &=& \dfrac{1}{2} \Biggl\{{\rm i}\,\beta\,\vec{\alpha}
                                    +\dfrac{2}{E_{\rm p}} \Bigl(m\,c^2 \beta\,\hat{p}
                                          +p\,c\,\vec{\alpha}\Bigr)
                                    +\dfrac{{\rm i}}{E_{\rm p}^2} (
                                          m\,c^2 \vec{\alpha}\cdot\hat{p}-\beta\,p\,c)
                                          \Bigl[2\,m\,c^2 \beta\,\hat{p}
                                                -m\,c^2 \beta(\vec{\alpha}\cdot\hat{p})
                                                      \vec{\alpha}+p\,c\,\vec{\alpha}\Bigr]
                                    \Biggr\} \nonumber\\
                              &=& \dfrac{1}{2} \Biggl\{{\rm i}\,\beta\,\vec{\alpha}
                                    +\dfrac{2}{E_{\rm p}} \Bigl(m\,c^2 \beta\,\hat{p}
                                          +p\,c\,\vec{\alpha}\Bigr) \nonumber\\
                               &     &\quad\ +\dfrac{{\rm i}}{E_{\rm p}^2}\Bigl[
                                          2\,m^2 c^4 (\vec{\alpha}\cdot\hat{p})\beta\,
                                                \hat{p}
                                          -m^2 c^4 (\vec{\alpha}\cdot\hat{p})\beta(
                                                \vec{\alpha}\cdot\hat{p})\vec{\alpha}
                                          +m\,p\,c^3 (\vec{\alpha}\cdot\hat{p})
                                                \vec{\alpha}
                                          -2\,m\,p\,c^3 \mathbb{I}\,\hat{p}
                                          +m\,p\,c^3 (\vec{\alpha}\cdot\hat{p})
                                                \vec{\alpha}
                                          -p^2 c^2 \beta\,\vec{\alpha}\Bigr]
                                    \Biggr\} \nonumber\\
                              &=& \dfrac{1}{2} \Biggl\{{\rm i}\,\beta\,\vec{\alpha}
                                    +\dfrac{2}{E_{\rm p}} \Bigl(m\,c^2 \beta\,\hat{p}
                                          +p\,c\,\vec{\alpha}\Bigr) \nonumber\\
                               &     &\quad\ +\dfrac{{\rm i}}{E_{\rm p}^2}\Bigl[
                                          -2\,m^2 c^4 \beta(\vec{\alpha}\cdot\hat{p})
                                                \hat{p}
                                          +m^2 c^4 \beta\,\vec{\alpha}
                                          +m\,p\,c^3 (\vec{\alpha}\cdot\hat{p})
                                                \vec{\alpha}
                                          -2\,m\,p\,c^3 \mathbb{I}\,\hat{p}
                                          +m\,p\,c^3 (\vec{\alpha}\cdot\hat{p})
                                                \vec{\alpha}
                                          -p^2 c^2 \beta\,\vec{\alpha}\Bigr]
                                    \Biggr\} \nonumber\\
                              &=& \dfrac{1}{2} \Biggl\{{\rm i}
                                    \dfrac{2\,m^2 c^4}{E_{\rm p}^2}\beta\,\vec{\alpha}
                                    +\dfrac{2}{E_{\rm p}} \Bigl(m\,c^2 \beta\,\hat{p}
                                          +p\,c\,\vec{\alpha}\Bigr)
                                    +\dfrac{{\rm i}}{E_{\rm p}^2} \Bigl[
                                          -2\,m^2 c^4 \beta(\vec{\alpha}\cdot\hat{p})
                                                \hat{p}
                                          -2\,m\,p\,c^3 \mathbb{I}\,\hat{p}
                                          +2\,m\,p\,c^3 (\vec{\alpha}\cdot\hat{p})
                                                \vec{\alpha}\Bigr]
                                    \Biggr\} \nonumber\\
                              &=& {\rm i}\dfrac{m^2 c^4}{E_{\rm p}^2}\beta\,
                                          \vec{\alpha}
                                    +\dfrac{1}{E_{\rm p}} \Bigl(m\,c^2 \beta\,\hat{p}
                                          +p\,c\,\vec{\alpha}\Bigr)
                                    +\dfrac{{\rm i}}{E_{\rm p}^2} \Bigl[
                                          -m^2 c^4 \beta(\vec{\alpha}\cdot\hat{p})
                                                \hat{p}
                                          -m\,p\,c^3 \mathbb{I}\,\hat{p}
                                          +m\,p\,c^3 (\vec{\alpha}\cdot\hat{p})
                                                \vec{\alpha}\Bigr].
                  \end{eqnarray}
                  Because
                  \begin{equation}
                        \vec{\alpha}\cdot\hat{p}=\dfrac{1}{cp} \left(H_{\rm e}
                              -m c^2\; \beta\right),
                  \end{equation}
                  and
                  \begin{equation}
                        \beta=\dfrac{1}{m\,c^2} \left(H_{\rm e}
                              -c\,p\,\vec{\alpha}\cdot\hat{p}\right),
                  \end{equation}
                  which imply
                  \begin{eqnarray}
                        \mathcal{D}^\dagger \vec{\zeta}(0)\,\mathcal{D}
                        &=& {\rm i}\dfrac{m^2 c^4}{E_{\rm p}^2}\beta\,\vec{\alpha}
                              +\dfrac{1}{E_{\rm p}} \Bigl(m\,c^2 \beta\,\hat{p}
                                    +p\,c\,\vec{\alpha}\Bigr)
                              +\dfrac{{\rm i}}{E_{\rm p}^2} \Bigl[
                                    -m^2 c^4 \beta(\vec{\alpha}\cdot\hat{p})
                                          \hat{p}
                                    -m\,p\,c^3 \mathbb{I}\,\hat{p}
                                    +m\,p\,c^3 \left[\dfrac{1}{cp} \left(H_{\rm e}
                              -m c^2\; \beta\right)\right]\vec{\alpha}\Bigr]
                              \nonumber\\
                        &=& {\rm i}\dfrac{m^2 c^4}{E_{\rm p}^2}\beta\,
                                          \vec{\alpha}
                                    +\dfrac{1}{E_{\rm p}} \Bigl(m\,c^2 \beta\,\hat{p}
                                          +p\,c\,\vec{\alpha}\Bigr)
                                    +\dfrac{{\rm i}}{E_{\rm p}^2} \Bigl[
                                          -m^2 c^4 \beta(\vec{\alpha}\cdot\hat{p})
                                                \hat{p}
                                          -m\,p\,c^3 \mathbb{I}\,\hat{p}
                                          +m c^2 H_{\rm e} \vec{\alpha}- m^2 c^4 \beta
                                                \vec{\alpha}\Bigr]\nonumber\\
                        &=& {\rm i}\dfrac{m^2 c^4}{E_{\rm p}^2}\beta\,
                                          \vec{\alpha}
                                    +\dfrac{1}{E_{\rm p}} \Bigl(m\,c^2 \beta\,\hat{p}
                                          +p\,c\,\vec{\alpha}\Bigr)
                                    +\dfrac{{\rm i}}{E_{\rm p}^2} \Bigl[
                                          -m^2 c^4 \beta(\vec{\alpha}\cdot\hat{p})
                                                \hat{p}
                                          -m\,p\,c^3 \mathbb{I}\,\hat{p}
                                          +m c^2 H_{\rm e} \vec{\alpha}\Bigr]-{\rm i}\dfrac{m^2 c^4}{E_{\rm p}^2}\beta\,
                                          \vec{\alpha}\nonumber\\
                        &=& \dfrac{1}{E_{\rm p}} \Bigl(m\,c^2 \beta\,\hat{p}
                                          +p\,c\,\vec{\alpha}\Bigr)
                                    +\dfrac{{\rm i}}{E_{\rm p}^2} \Bigl[
                                          -m^2 c^4 \beta(\vec{\alpha}\cdot\hat{p})
                                                \hat{p}
                                          -m\,p\,c^3 \mathbb{I}\,\hat{p}
                                          +m c^2 H_{\rm e} \vec{\alpha}\Bigr]\nonumber\\
                        &=& \dfrac{1}{E_{\rm p}} \Bigl(m\,c^2 \beta\,\hat{p}
                                          +p\,c\,\vec{\alpha}\Bigr)
                                    +\dfrac{{\rm i}}{E_{\rm p}^2} \Bigl[
                                          -m^2 c^4 \left[\dfrac{1}{m\,c^2} \left(H_{\rm e}
                              -c\,p\,\vec{\alpha}\cdot\hat{p}\right)\right](\vec{\alpha}\cdot\hat{p})
                                                \hat{p}
                                          -m\,p\,c^3 \mathbb{I}\,\hat{p}
                                          +m c^2 H_{\rm e} \vec{\alpha}\Bigr]\nonumber\\
                        &=& \dfrac{1}{E_{\rm p}} \Bigl(m\,c^2 \beta\,\hat{p}
                                          +p\,c\,\vec{\alpha}\Bigr)
                                    +\dfrac{{\rm i}}{E_{\rm p}^2} \Bigl[
                                          -m c^2 H_{\rm e}(\vec{\alpha}\cdot\hat{p})
                                                \hat{p}+ mc^2 cp \mathbb{I}\,
                                                \hat{p}
                                          -m\,p\,c^3 \mathbb{I}\,\hat{p}
                                          +m c^2 H_{\rm e} \vec{\alpha}\Bigr]\nonumber\\
                        &=& \dfrac{1}{E_{\rm p}} \Bigl(m\,c^2 \beta\,\hat{p}
                                          +p\,c\,\vec{\alpha}\Bigr)
                                    +\dfrac{{\rm i}}{E_{\rm p}^2} \Bigl[
                                          -m c^2 H_{\rm e}(\vec{\alpha}\cdot\hat{p})
                                                \hat{p}
                                          +m c^2 H_{\rm e} \vec{\alpha}\Bigr]\nonumber\\
                        &=& \dfrac{1}{E_{\rm p}} \Bigl[\left(H_{\rm e}
                              -c\,p\,\vec{\alpha}\cdot\hat{p}\right)\,\hat{p}
                                          +p\,c\,\vec{\alpha}\Bigr]
                                    +\dfrac{{\rm i}}{E_{\rm p}^2} \Bigl[
                                          -m c^2 H_{\rm e}(\vec{\alpha}\cdot\hat{p})
                                                \hat{p}
                                          +m c^2 H_{\rm e} \vec{\alpha}\Bigr]\nonumber\\
                        & =& \dfrac{1}{E_{\rm p}} H_{\rm e} \hat{p}
                                    +\dfrac{p\,c}{E_{\rm p}} \Bigl[\vec{\alpha}
                                          -(\vec{\alpha}\cdot\hat{p})\hat{p}\Bigr]
                                    +{\rm i}\dfrac{m\,c^2}{E_{\rm p}^2} H_{\rm e} \Bigl[
                                          \vec{\alpha}-(\vec{\alpha}\cdot\hat{p})\hat{p}
                                          \Bigr] \nonumber\\
                         & =& \dfrac{1}{E_{\rm p}} H_{\rm e} \hat{p}
                                    +\dfrac{1}{E_{\rm p}} \left(p\,c\,\mathbb{I}
                                          +{\rm i}\dfrac{m\,c^2}{E_{\rm p}}
                                                H_{\rm e}\right)\Bigl[\vec{\alpha}
                                                      -(\vec{\alpha}\cdot\hat{p})
                                                            \hat{p}\Bigr].
                  \end{eqnarray}
                  Notice
                  \begin{eqnarray}
                        \mathcal{D}^\dagger [\vec{\zeta}(0)\cdot\hat{p}]\mathcal{D}
                       & =&\bigl(\mathcal{D}^\dagger \vec{\zeta}(0)\,\mathcal{D}\bigr)
                              \cdot\hat{p}
                        =\dfrac{1}{E_{\rm p}} H_{\rm e}
                              +\dfrac{p\,c}{E_{\rm p}} \Bigl[\vec{\alpha}\cdot\hat{p}
                                    -(\vec{\alpha}\cdot\hat{p})\Bigr]
                              +{\rm i}\dfrac{m\,c^2}{E_{\rm p}^2} H_{\rm e} \Bigl[
                                    \vec{\alpha}\cdot\hat{p}-(\vec{\alpha}\cdot\hat{p})
                                    \Bigr]
                        =\dfrac{1}{E_{\rm p}} H_{\rm e},
                  \end{eqnarray}
                  i.e., which is equivalent to $H_{\rm e} =\mathcal{D}^\dagger H_{\rm b}^{\rm II} \mathcal{D}$.

                  From Eq. (\ref{eq:D-28-qa}), we have known that
                 \begin{eqnarray}\label{eq:D-28-qaa}
                 \langle\Psi_1|\vec{\alpha}(0)|\Psi_3 \rangle &=& -\dfrac{mc^2}{E_{\rm p}}\; \hat{p},
                 \end{eqnarray}
                  which leads to
                  \begin{eqnarray}
                            \langle\Psi_1|\bigl[\vec{\alpha}
                                    -(\vec{\alpha}\cdot\hat{p})\hat{p}\bigr]|
                                          \Psi_3\rangle
                             & =&\langle\Psi_1|\vec{\alpha}|\Psi_3\rangle
                                    -\langle\Psi_1|(\vec{\alpha}\cdot\hat{p})\hat{p}|
                                          \Psi_3\rangle
                              =\langle\Psi_1|\vec{\alpha}|\Psi_3\rangle
                                    -\bigl[(\langle\Psi_1|\vec{\alpha}|\Psi_3\rangle)
                                          \cdot\hat{p}\bigr]\hat{p} \nonumber\\
                              &=& -\dfrac{m\,c^2}{E_{\rm p}} \hat{p}
                                    -\left(-\dfrac{m\,c^2}{E_{\rm p}} \hat{p}
                                          \cdot\hat{p}\right)\hat{p}
                              =-\dfrac{m\,c^2}{E_{\rm p}} \hat{p}
                                    +\dfrac{m\,c^2}{E_{\rm p}} \hat{p}
                              =0.
                  \end{eqnarray}
                  Therefore
                  \begin{eqnarray}\label{eq:13}
                             \langle\Psi''_1|\hat{\mathcal{Z}}_{\rm b}^{\rm II} |\Psi''_3 \rangle
                              &=&\dfrac{{\rm i}\hbar}{2} \langle\Psi''_1|\Biggl\{
                                    \dfrac{\sqrt{p^2c^2+m^2c^4}}{p} \vec{\zeta}(0)
                                    -H_{\rm b}^{\rm II} \dfrac{\vec{p}}{p^2}\Biggr\} (H_{\rm b}^{\rm II})^{-1} \left(
                                          {\rm e}^{\frac{-{\rm i}\,2\,H_{\rm b}^{\rm II}\,t}{\hbar}}-1
                                          \right)|\Psi''_3\rangle \nonumber\\
                              &=&-{\rm i}\dfrac{\hbar}{2} \dfrac{1}{p} \left(
                                    {\rm e}^{\frac{{\rm i}\,2\,E_{\rm p}\,t}{\hbar}}
                                    -1\right)\langle\Psi_1|\bigl(\mathcal{D}^\dagger
                                          \vec{\zeta}(0)\,\mathcal{D}\bigr)|\Psi_3\rangle  \nonumber\\
                              &=& -{\rm i}\dfrac{\hbar}{2} \dfrac{1}{p} \left({\rm e}^{
                                    \frac{{\rm i}\,2\,E_{\rm p}\,t}{\hbar}}-1\right)
                                          \langle\Psi_1|\biggl\{\dfrac{1}{E_{\rm p}}
                                                      H_{\rm e} \hat{p}
                                                +\dfrac{p\,c}{E_{\rm p}} \Bigl[
                                                      \vec{\alpha}-(\vec{\alpha}
                                                            \cdot\hat{p})\hat{p}\Bigr]
                                                +{\rm i}\dfrac{m\,c^2}{E_{\rm p}^2}
                                                      H_{\rm e} \Bigl[\vec{\alpha}
                                                            -(\vec{\alpha}\cdot\hat{p})
                                                                  \hat{p}\Bigr]\biggr\}|
                                                                        \Psi_3\rangle \nonumber\\
                             & =& -{\rm i}\dfrac{\hbar}{2} \dfrac{1}{p} \left({\rm e}^{
                                    \frac{{\rm i}\,2\,E_{\rm p}\,t}{\hbar}}-1\right)
                                          \biggl\{\dfrac{p\,c}{E_{\rm p}} \langle\Psi_1|
                                                \Bigl[\vec{\alpha}-(\vec{\alpha}
                                                      \cdot\hat{p})\hat{p}\Bigr]|
                                                            \Psi_3\rangle
                                                +{\rm i}\dfrac{m\,c^2}{E_{\rm p}^2}
                                                      \langle\Psi_1|H_{\rm e} \Bigl[
                                                            \vec{\alpha}-(\vec{\alpha}
                                                                  \cdot\hat{p})\hat{p}
                                                                  \Bigr]|\Psi_3\rangle
                                                                  \biggr\} \nonumber\\
                              &=& -{\rm i}\dfrac{\hbar}{2} \dfrac{1}{p} \left({\rm e}^{
                                    \frac{{\rm i}\,2\,E_{\rm p}\,t}{\hbar}}-1\right)
                                          \biggl\{{\rm i}\dfrac{m\,c^2}{E_{\rm p}}
                                                \langle\Psi_1|\Bigl[\vec{\alpha}
                                                      -(\vec{\alpha}\cdot\hat{p})\hat{p}
                                                      \Bigr]|\Psi_3\rangle\biggr\}
                              =0.
                  \end{eqnarray}
                  This leads to
                  \begin{equation}
                        \mathcal{Z}_{\rm b}^{\rm II}  =\langle\Psi''|\hat{\mathcal{Z}}_{\rm b}^{\rm II} |\Psi''\rangle
                        =\sin(2\eta)\,{\rm Re}\left(
                              \langle\Psi''_1|\hat{\mathcal{Z}}_{\rm b}^{\rm II}
                                    |\Psi''_3 \rangle\right)=0,
                  \end{equation}
                 which coincides with Eq. (\ref{eq:II-1a}). $\blacksquare$
            \end{remark}

        \subsection{More General Results}

         We would like to calculate more general results. Generally, for $k\in\{1,2\}$, and $l\in\{3,4\}$, we have
            \begin{eqnarray}\label{eq:ZbIIPsiJK}
                  \langle\Psi''_k|\hat{\mathcal{Z}}_{\rm b}^{\rm II} |\Psi''_l
                        \rangle
                 & =&-\dfrac{{\rm i}\hbar}{2\,p} \left({\rm e}^{
                        \frac{{\rm i}\,2\,E_{\rm p}\,t}{\hbar}}-1
                        \right)\biggl\{\dfrac{p\,c}{E_{\rm p}} \langle\Psi_k|
                                    \Bigl[\vec{\alpha}-(\vec{\alpha}\cdot\hat{p})
                                          \hat{p}\Bigr]|\Psi_l\rangle
                              +{\rm i}\dfrac{m\,c^2}{E_{\rm p}^2} \langle\Psi_k|
                                    H_{\rm e} \Bigl[\vec{\alpha}-(\vec{\alpha}
                                          \cdot\hat{p})\hat{p}\Bigr]|\Psi_l\rangle
                              \biggr\} \notag \\
                  &=& -\dfrac{{\rm i}\hbar\,c}{2\,E_{\rm p} p} \left({\rm e}^{
                        \frac{{\rm i}\,2\,E_{\rm p}\,t}{\hbar}}-1
                        \right)\biggl\{p\langle\Psi_k|\Bigl[\vec{\alpha}
                                    -(\vec{\alpha}\cdot\hat{p})\hat{p}\Bigr]|
                                          \Psi_l\rangle
                              +{\rm i}\,m\,c\langle\Psi_k|\Bigl[\vec{\alpha}
                                    -(\vec{\alpha}\cdot\hat{p})\hat{p}\Bigr]|
                                          \Psi_l\rangle\biggr\} \notag \\
                  &=& -\dfrac{{\rm i}\hbar\,c}{2\,E_{\rm p} p} \left({\rm e}^{
                        \frac{{\rm i}\,2\,E_{\rm p}\,t}{\hbar}}-1
                        \right)(p+{\rm i}\,m\,c)\langle\Psi_k|\bigl[\vec{\alpha}
                              -(\vec{\alpha}\cdot\hat{p})\hat{p}\bigr]|\Psi_l\rangle,
            \end{eqnarray}
           which is just Eq. (\ref{eq:13}) when $k=1, l=3$.
            \begin{remark}
                  By taking $k=2, l=4$, from Eq. (\ref{eq:ZbIIPsiJK}) one obtains
                  \begin{eqnarray}
                        \langle\Psi''_2|\hat{\mathcal{Z}}_{\rm b}^{\rm II}|\Psi''_4
                              \rangle
                        &=& -\dfrac{{\rm i}\hbar\,c}{2\,E_{\rm p} p} \left({\rm e}^{
                              \frac{{\rm i}\,2\,E_{\rm p}\,t}{\hbar}}-1
                              \right)(p+{\rm i}\,m\,c)\langle\Psi_2|\bigl[
                                    \vec{\alpha}-(\vec{\alpha}\cdot\hat{p})\hat{p}
                                    \bigr]|\Psi_4\rangle.
                  \end{eqnarray}
                  From Table \ref{tab:pz0}, we have known that
                  \begin{eqnarray}
                  \langle\Psi_2|\vec{\alpha}(0)|\Psi_4\rangle
                        &=&\langle\Psi_1|\vec{\alpha}(0)|\Psi_3\rangle=-\dfrac{m\,c^2}{E_{\rm p}} \hat{p},
                        \end{eqnarray}
                  thus
                  \begin{eqnarray}
                        \langle\Psi_2|\big[\vec{\alpha}(0)\cdot\hat{p}\big]\hat{p}|
                              \Psi_4\rangle
                        &=& \langle\Psi_1|\big[\vec{\alpha}(0)\cdot\hat{p}\big]
                        \hat{p}|\Psi_3\rangle=-\dfrac{m\,c^2}{E_{\rm p}} \hat{p}=\langle\Psi_1|\vec{\alpha}(0)|\Psi_3\rangle=-\dfrac{m\,c^2}{E_{\rm p}} \hat{p}.
                  \end{eqnarray}
                  We then have
                  \begin{eqnarray}
                        \langle\Psi''_2|\hat{\mathcal{Z}}_{\rm b}^{\rm II}|\Psi''_4
                              \rangle
                        &=& -\dfrac{{\rm i}\hbar\,c}{2\,E_{\rm p} p} \left({\rm e}^{
                              \frac{{\rm i}\,2\,E_{\rm p}\,t}{\hbar}}-1
                              \right)(p+{\rm i}\,m\,c)\langle\Psi_2|\bigl[
                                    \vec{\alpha}-(\vec{\alpha}\cdot\hat{p})\hat{p}
                                    \bigr]|\Psi_4\rangle=0.
                  \end{eqnarray}
                  $\blacksquare$
            \end{remark}

            \begin{remark}
                  By taking $k=1, l=4$, from Eq. (\ref{eq:ZbIIPsiJK}) one obtains
                  \begin{eqnarray}
                        \langle\Psi''_1|\hat{\mathcal{Z}}_{\rm b}^{\rm II}|\Psi''_4
                              \rangle
                        &=& -\dfrac{{\rm i}\hbar\,c}{2\,E_{\rm p} p} \left({\rm e}^{
                              \frac{{\rm i}\,2\,E_{\rm p}\,t}{\hbar}}-1
                              \right)(p+{\rm i}\,m\,c)\langle\Psi_1|\bigl[
                                    \vec{\alpha}-(\vec{\alpha}\cdot\hat{p})\hat{p}
                                    \bigr]|\Psi_4\rangle.
                  \end{eqnarray}
                  Note that
                  \begin{eqnarray}
                        & \langle\Psi_1|\big[\vec{\alpha}(0)\cdot\hat{p}\big]|
                              \Psi_4\rangle=\left(\vec{F}_1+{\rm i}\vec{F}_2\right)\cdot \hat{p}=0,
                  \end{eqnarray}
                  namely
                  \begin{eqnarray}
                         \langle\Psi''_1|\hat{\mathcal{Z}}_{\rm b}^{\rm II}|\Psi''_4
                              \rangle
                        &=&-\dfrac{{\rm i}\hbar\,c}{2\,E_{\rm p} p} \left({\rm e}^{
                              \frac{{\rm i}\,2\,E_{\rm p}\,t}{\hbar}}-1
                              \right)(p+{\rm i}\,m\,c)\langle\Psi_1|\bigl[
                                    \vec{\alpha}-(\vec{\alpha}\cdot\hat{p})\hat{p}
                                    \bigr]|\Psi_4\rangle \notag \\
                        &=& \dfrac{{\rm i}\hbar}{2\,E_{\rm p} p} \left({\rm e}^{
                              \frac{{\rm i}\,2\,E_{\rm p}\,t}{\hbar}}-1
                              \right)(-p\,c-{\rm i}\,m\,c^2)\langle\Psi_1|\vec{\alpha}|
                                    \Psi_4\rangle.
                  \end{eqnarray}
                  Recall
                  \begin{eqnarray}
                        && \langle\Psi_1|\hat{\mathcal{Z}}_{\rm e}|\Psi_4\rangle
                        =\frac{-{\rm i}\hbar\,c}{2\,E_{\rm p}} \left({\rm e}^{
                              \frac{{\rm i}\,2\,E_{\rm p} t}{\hbar}} -1\right)\langle\Psi_1|
                                    \vec{\alpha}(0)|\Psi_4\rangle
                        =\dfrac{{\rm i}\hbar}{2\,p}\left({\rm e}^{
                              \frac{{\rm i}\,2\,E_{\rm p} t}{\hbar}} -1\right)\frac{1}{E_{\rm p}} (
                                    -p\,c)\langle\Psi_1|\vec{\alpha}(0)|\Psi_4\rangle,
                  \end{eqnarray}
                  which indicates $\langle\Psi''_1|\hat{\mathcal{Z}}_{\rm b}^{\rm II}|\Psi''_4 \rangle\neq\langle\Psi_1|\hat{\mathcal{Z}}_{\rm e}|\Psi_4\rangle$ in general, but
                  \begin{eqnarray}
                         \langle\Psi''_1|\hat{\mathcal{Z}}_{\rm b}^{\rm II}|\Psi''_4
                              \rangle
                        &=& \dfrac{{\rm i}\hbar}{2\,E_{\rm p} p} \left({\rm e}^{
                              \frac{{\rm i}\,2\,E_{\rm p}\,t}{\hbar}}-1
                              \right)(-p\,c-{\rm i}\,m\,c^2)\langle\Psi_1|\vec{\alpha}|
                                    \Psi_4\rangle \notag \\
                        &=& \dfrac{{\rm i}\hbar}{2\,p} \left(
                                    {\rm e}^{\frac{{\rm i}\,2\,E_{\rm p} t}{\hbar}}
                                    -1\right)\dfrac{1}{E_{\rm p}} (-p\,c)\langle\Psi_1|
                                          \vec{\alpha}(0)|\Psi_4\rangle
                              +\dfrac{{\rm i}\,m\,c^2}{p\,c} \dfrac{{\rm i}\hbar}{
                                    2\,E_{\rm p} p} \left({\rm e}^{
                                          \frac{{\rm i}\,2\,E_{\rm p}\,t}{\hbar}}-1
                                    \right)(-p\,c)\langle\Psi_1|\vec{\alpha}|
                                          \Psi_4\rangle \notag \\
                        &=& \left(1+{\rm i}\dfrac{m\,c}{p}\right)\langle\Psi_1|
                              \hat{\mathcal{Z}}_{\rm e}|\Psi_4\rangle \propto \langle\Psi_1|
                              \hat{\mathcal{Z}}_{\rm e}|\Psi_4\rangle.
                  \end{eqnarray}
            $\blacksquare$
            \end{remark}
      \begin{remark}
            Similarly, for $k=2$, and $l=3$,
            \begin{eqnarray}
                   \langle\Psi''_2|\hat{\mathcal{Z}}_{\rm b}^{\rm II}|\Psi''_3
                        \rangle
                  &=& -\dfrac{{\rm i}\hbar\,c}{2\,E_{\rm p} p} \left({\rm e}^{
                        \frac{{\rm i}\,2\,E_{\rm p}\,t}{\hbar}}-1
                        \right)(p+{\rm i}\,m\,c)\langle\Psi_2|\bigl[
                              \vec{\alpha}-(\vec{\alpha}\cdot\hat{p})\hat{p}
                              \bigr]|\Psi_3\rangle \notag \\
                  &=& \dfrac{{\rm i}\hbar}{2\,E_{\rm p} p} \left({\rm e}^{
                        \frac{{\rm i}\,2\,E_{\rm p}\,t}{\hbar}}-1
                        \right)(-p\,c-{\rm i}\,m\,c^2)\langle\Psi_2|\vec{\alpha}|
                              \Psi_3\rangle \notag \\
                  &=& \dfrac{{\rm i}\hbar}{2\,E_{\rm p} p} \left({\rm e}^{
                        \frac{{\rm i}\,2\,E_{\rm p}\,t}{\hbar}}-1
                        \right)(p\,c+{\rm i}\,m\,c^2)\, \left[\langle\Psi_1|\vec{\alpha}|
                              \Psi_4\rangle\right]^*,
            \end{eqnarray}
            since
            \begin{eqnarray}
                  && \langle\Psi_2|\vec{\alpha}(0)|\Psi_3\rangle = -\langle\Psi_1|\vec{\alpha}(0)|\Psi_4\rangle^*,
            \end{eqnarray}
            and
            \begin{eqnarray}
                  && \langle\Psi_2|\big[\vec{\alpha}(0)\cdot\hat{p}\big]\hat{p}|
                        \Psi_3\rangle
                  =\Bigl\{\big[\langle\Psi_2|\vec{\alpha}(0)|\Psi_3\rangle\big]
                        \cdot\hat{p}\Bigr\}\hat{p}
                  =\Bigl\{-\big[\langle\Psi_1|\vec{\alpha}(0)|\Psi_4\rangle\big]^*
                        \cdot\hat{p}\Bigr\}\hat{p}
                  =0.
            \end{eqnarray}
            Recall
            \begin{eqnarray}
                   \langle\Psi_2|\hat{\mathcal{Z}}_{\rm e}|\Psi_3\rangle
                  &=& \frac{-{\rm i}\hbar\,c}{2\,E_{\rm p}} \left({\rm e}^{
                        \frac{{\rm i}\,2\,E_{\rm p} t}{\hbar}} -1\right)\langle\Psi_2|
                              \vec{\alpha}(0)|\Psi_3\rangle
                  =\dfrac{{\rm i}\hbar}{2\,p}\left({\rm e}^{
                        \frac{{\rm i}\,2\,E_{\rm p} t}{\hbar}} -1\right)\frac{1}{E_{\rm p}} (
                              -p\,c)\langle\Psi_2|\vec{\alpha}(0)|\Psi_3\rangle \notag \\
                  &=& \dfrac{{\rm i}\hbar}{2\,p}\left({\rm e}^{
                        \frac{{\rm i}\,2\,E_{\rm p} t}{\hbar}} -1\right)\frac{1}{E_{\rm p}}
                              p\,c\, \left[\langle\Psi_1|\vec{\alpha}(0)|\Psi_4\rangle\right]^*,
            \end{eqnarray}
            which indicates $\langle\Psi'_2|\hat{\mathcal{Z}}_{\rm b}^{\rm I}|\Psi'_3 \rangle\neq\langle\Psi_2|\hat{\mathcal{Z}}_{\rm e}|\Psi_3\rangle$ in general, but
            \begin{eqnarray}
                   \langle\Psi''_2|\hat{\mathcal{Z}}_{\rm b}^{\rm II}|\Psi''_3\rangle
                  &=&\dfrac{{\rm i}\hbar}{2\,E_{\rm p} p} \left({\rm e}^{
                        \frac{{\rm i}\,2\,E_{\rm p}\,t}{\hbar}}-1
                        \right)(p\,c+{\rm i}\,m\,c^2) \, \left[\langle\Psi_1|\vec{\alpha}|
                              \Psi_4\rangle\right]^* \notag \\
                  &=& \dfrac{{\rm i}\hbar}{2\,p}\left({\rm e}^{
                              \frac{{\rm i}\,2\,E_{\rm p} t}{\hbar}} -1\right)
                                    \frac{1}{E_{\rm p}} p\,c \, \left[\langle\Psi_1|
                                          \vec{\alpha}(0)|\Psi_4\rangle\right]^*
                        +\dfrac{{\rm i}\,m\,c^2}{p\,c} \dfrac{{\rm i}\hbar}{
                              2\,E_{\rm p} p} \left({\rm e}^{
                                    \frac{{\rm i}\,2\,E_{\rm p}\,t}{\hbar}}-1
                              \right)p\,c\, \left[\langle\Psi_1|\vec{\alpha}|
                                    \Psi_4\rangle\right]^* \notag \\
                  &=& \left(1+{\rm i}\dfrac{m\,c}{p}\right)\langle\Psi_2|
                        \hat{\mathcal{Z}}_{\rm e}|\Psi_3\rangle \propto \langle\Psi_2|
                        \hat{\mathcal{Z}}_{\rm e}|\Psi_3\rangle.
            \end{eqnarray}
      \end{remark}
      From the previous calculation, we have known
      \begin{eqnarray}
             \mathcal{Z}_{\rm e}
            &=& c_1^* c_3 \Bra{\Psi_1}\hat{\mathcal{Z}}_{\rm e}\Ket{\Psi_3}
                  +c_1^* c_4 \Bra{\Psi_1}\hat{\mathcal{Z}}_{\rm e}\Ket{\Psi_4}
                  +c_2^* c_3 \Bra{\Psi_2}\hat{\mathcal{Z}}_{\rm e}\Ket{\Psi_3}
                  +c_2^* c_4 \Bra{\Psi_2}\hat{\mathcal{Z}}_{\rm e}\Ket{\Psi_4}
                  +{\rm c.c.} \notag \\
            &=& 2\,{\rm Re}\left(c_1^* c_3 \Bra{\Psi_1}\hat{\mathcal{Z}}_{\rm e}
                        \Ket{\Psi_3}
                  +c_1^* c_4 \Bra{\Psi_1}\hat{\mathcal{Z}}_{\rm e}\Ket{\Psi_4}
                  +c_2^* c_3 \Bra{\Psi_2}\hat{\mathcal{Z}}_{\rm e}\Ket{\Psi_3}
                  +c_2^* c_4 \Bra{\Psi_2}\hat{\mathcal{Z}}_{\rm e}\Ket{\Psi_4}\right).
      \end{eqnarray}
      Note
      \begin{align}
            & \langle\Psi''_1|\hat{\mathcal{Z}}_{\rm b}^{\rm II}|\Psi''_3\rangle
                  =\langle\Psi''_2|\hat{\mathcal{Z}}_{\rm b}^{\rm II}|\Psi''_4\rangle =0, \notag \\
            & \langle\Psi''_1|\hat{\mathcal{Z}}_{\rm b}^{\rm II}|\Psi''_4 \rangle
                  =\left(1+{\rm i}\dfrac{m\,c}{p}\right)\langle\Psi_1|
                        \hat{\mathcal{Z}}_{\rm e}|\Psi_4\rangle, \notag \\
            & \langle\Psi''_2|\hat{\mathcal{Z}}_{\rm b}^{\rm II}|\Psi''_3\rangle
                  =\left(1+{\rm i}\dfrac{m\,c}{p}\right)\langle\Psi_2|
                        \hat{\mathcal{Z}}_{\rm e}|\Psi_3\rangle,
      \end{align}
      thus
      \begin{eqnarray}
             \mathcal{Z}_{\rm b}^{\rm II}
           &=& c_1^* c_3 \Bra{\Psi''_1}\hat{\mathcal{Z}}_{\rm b}^{\rm II}\Ket{\Psi''_3}
                  +c_1^* c_4 \Bra{\Psi''_1}\hat{\mathcal{Z}}_{\rm b}^{\rm II}\Ket{\Psi''_4}
                  +c_2^* c_3 \Bra{\Psi''_2}\hat{\mathcal{Z}}_{\rm b}^{\rm II}\Ket{\Psi''_3}
                  +c_2^* c_4 \Bra{\Psi''_2}\hat{\mathcal{Z}}_{\rm b}^{\rm II}\Ket{\Psi''_4}
                  +{\rm c.c.} \notag \\
            &=& 2\,{\rm Re}\left(
                  c_1^* c_3 \Bra{\Psi''_1}\hat{\mathcal{Z}}_{\rm b}^{\rm II}\Ket{\Psi''_3}
                  +c_1^* c_4 \Bra{\Psi''_1}\hat{\mathcal{Z}}_{\rm b}^{\rm II}\Ket{\Psi''_4}
                  +c_2^* c_3 \Bra{\Psi''_2}\hat{\mathcal{Z}}_{\rm b}^{\rm II}\Ket{\Psi''_3}
                  +c_2^* c_4 \Bra{\Psi''_2}\hat{\mathcal{Z}}_{\rm b}^{\rm II}\Ket{\Psi''_4}
                  \right) \notag \\
            &=& 2\,{\rm Re}\left[\left(1+{\rm i}\dfrac{m\,c}{p}\right)\left(
                  c_2^* c_3 \Bra{\Psi_2}\hat{\mathcal{Z}}_{\rm e} \Ket{\Psi_3}
                  +c_1^* c_4 \Bra{\Psi_1}\hat{\mathcal{Z}}_{\rm e} \Ket{\Psi_4}\right)
                  \right],
      \end{eqnarray}
      which indicates $\mathcal{Z}_{\rm b}^{\rm II} \neq\mathcal{Z}_{\rm e}$ in general.

      \begin{remark}We can summarize the above results as the following Table \ref{tab:pz4}:

        \begin{table}[h]
	\centering
\caption{The results of ``position Zitterbewegung'' of the type-II Dirac' braidon. $|\Psi''_j\rangle$'s  ($j=1, 2, 3, 4$) are four eigenstates of the type-II Dirac's braidon. In this case, the ``position Zitterbewegung'' operator reads $\hat{\mathcal{Z}}_{\rm b}^{\rm II} \equiv \hat{\mathcal{Z}}_{\rm b}^{{\rm II},r} = \dfrac{{\rm i}\hbar}{2} \Biggl\{ \dfrac{\sqrt{p^2c^2+m^2c^4}}{p} \vec{\zeta}(0)
                              -H_{\rm b}^{\rm II} \dfrac{\vec{p}}{p^2}\Biggr\} (H_{\rm b}^{\rm II})^{-1} \left(
                                    {\rm e}^{\frac{-{\rm i}\,2\,H_{\rm b}^{\rm II}\,t}{\hbar}}-1\right)$. One has $\langle\Psi''_j|\hat{\mathcal{Z}}_{\rm b}^{{\rm II},r}|\Psi''_j\rangle=0$, $\langle\Psi''_k|\hat{\mathcal{Z}}_{\rm b}^{{\rm II},r}|\Psi''_l\rangle=-\dfrac{{\rm i}\hbar\,c}{2\,E_{\rm p} p} \left({\rm e}^{
                              \frac{{\rm i}\,2\,E_{\rm p}\,t}{\hbar}}-1
                              \right)(p+{\rm i}\,m\,c)\langle\Psi_k|\bigl[
                                    \vec{\alpha}-(\vec{\alpha}\cdot\hat{p})\hat{p}
                                    \bigr]|\Psi_l\rangle$ for $k\in\{1,2\}$, $l\in\{3,4\}$, and $\Delta_3=-\dfrac{{\rm i}\hbar\,c}{2\,E_{\rm p} p} \left({\rm e}^{
                              \frac{{\rm i}\,2\,E_{\rm p}\,t}{\hbar}}-1
                              \right)(p+{\rm i}\,m\,c)$.}
\begin{tabular}{lllll}
\hline\hline
 & $|\Psi''_1\rangle$ &  $|\Psi''_2\rangle$& $|\Psi''_3\rangle$ & $|\Psi''_4\rangle$ \\
  \hline
$\langle \Psi''_1| \hat{\mathcal{Z}}_{\rm b}^{{\rm II},r}$ \;\;\;\quad& 0&0 & $0$  \;\;\;\quad& $\Delta_3\,(\vec{F}_1 +{\rm i}\,\vec{F}_2)$  \\
 \hline
$\langle \Psi''_2| \hat{\mathcal{Z}}_{\rm b}^{{\rm II},r}$\;\;\;\quad &0 &0 & $-\Delta_3\,(\vec{F}_1 +{\rm i}\,\vec{F}_2)^* $ \quad\quad& $0$ \\
 \hline
 $\langle \Psi''_3| \hat{\mathcal{Z}}_{\rm b}^{{\rm II},r}$\;\;\;\quad & $0$ \;\;\;& $-\Delta_3^*\,(\vec{F}_1 +{\rm i}\,\vec{F}_2) \quad\quad$ & 0 & 0 \\
 \hline
 $\langle \Psi''_4| \hat{\mathcal{Z}}_{\rm b}^{{\rm II},r}$ \;\;\;\quad& $\Delta_3^*\,(\vec{F}_1 +{\rm i}\,\vec{F}_2)^* \quad$ \quad&$0$ \;\;\;& 0 & 0 \\
 \hline\hline
\end{tabular}\label{tab:pz4}
\end{table}
$\blacksquare$
 \end{remark}

\section{Position Zitterbewegung for the $H_{\rm e}$-$H_{\rm b}^{\rm II}$ Mixing}

            In this section, let us consider the Hamiltonian mixed by
            $H_{\rm e}$ and $H_{\rm b}^{\rm II}$, whose Hamiltonian operator is given by
            \begin{eqnarray}
                  \mathcal{H}_{\rm m}'=\cos\vartheta\,H_{\rm e} +\sin\vartheta\,H_{\rm b}^{\rm II},
            \end{eqnarray}
            i.e.,
            \begin{eqnarray}
                  \mathcal{H}_{\rm m}'&=&\cos\vartheta\,H_{\rm e} +\sin\vartheta\,H_{\rm b}^{\rm II}\nonumber\\
                  &=& \cos\vartheta\,\left[c \vec{\alpha}\cdot\vec{p}+\beta\,m c^2\right] +\sin\vartheta\,\left[\dfrac{\sqrt{p^2c^2+m^2c^4}}{p}\;({\rm i}\beta\vec{\alpha}\cdot\vec{p})\right].
            \end{eqnarray}

            Let us consider the following $\vartheta$-dependent unitary transformation
            \begin{eqnarray}
                  \mathcal{D}(\vartheta) &=& {\rm e}^{{\rm i} \frac{\vartheta}{2} \Gamma_y} \nonumber\\
                  &=&\cos\Bigl(\frac{\vartheta}{2}\Bigr)\;\mathbb{I}+{\rm i}\,\sin\Bigl(
                        \frac{\vartheta}{2}\Bigr)\,\dfrac{(-m\,c^2 \vec{\alpha}
                              \cdot\hat{p}+\beta\,p\,c)}{\sqrt{p^2 c^2 +m^2 c^4}},
            \end{eqnarray}
            we then have
            \begin{eqnarray}
                   \mathcal{D}(\vartheta)\,H_{\rm e} \mathcal{D}(\vartheta)^\dagger
                  &=&\left[\cos\Bigl(\frac{\vartheta}{2}\Bigr)\;\mathbb{I}
                        +{\rm i}\,\sin\Bigl(\frac{\vartheta}{2}\Bigr)\,
                              \dfrac{(-m\,c^2 \vec{\alpha}\cdot\hat{p}+\beta\,p\,c)}{
                                    \sqrt{p^2 c^2 +m^2 c^4}}
                        \right]\,H_{\rm e} \left[
                              \cos\Bigl(\frac{\vartheta}{2}\Bigr)\;\mathbb{I}
                              -{\rm i}\,\sin\Bigl(\frac{\vartheta}{2}\Bigr)\,
                                    \dfrac{(-m\,c^2 \vec{\alpha}\cdot\hat{p}
                                          +\beta\,p\,c)}{\sqrt{p^2 c^2 +m^2 c^4}}
                              \right] \nonumber \\
                  &=& \cos^2 \left(\frac{\vartheta}{2}\right)\;H_{\rm e}
                        +{\rm i}\,\sin\frac{\vartheta}{2}\,\cos\frac{\vartheta}{2} (
                              \Gamma_y H_{\rm e} -H_{\rm e} \Gamma_y)
                        +\sin^2 \left(\frac{\vartheta}{2}\right) \Gamma_y H_{\rm e} \Gamma_y
                        \notag \\
                  &=& \cos^2 \left(\frac{\vartheta}{2}\right)\;H_{\rm e}
                        -{\rm i}\sin\frac{\vartheta}{2} \cos\frac{\vartheta}{2}\,\sqrt{
                              p^2 c^2 +m^2 c^4}[\Gamma_x,\ \Gamma_y]
                        +\sin^2 \left(\frac{\vartheta}{2}\right)\sqrt{p^2 c^2 +m^2 c^4}
                              \,\Gamma_y \Gamma_x \Gamma_y \notag \\
                  &=& \cos^2 \left(\frac{\vartheta}{2}\right)\;H_{\rm e}
                        +2\,\sin\frac{\vartheta}{2} \cos\frac{\vartheta}{2}\,\sqrt{
                              p^2 c^2 +m^2 c^4}\,\Gamma_z
                        +{\rm i}\,\sin^2 \left(\frac{\vartheta}{2}\right)\sqrt{
                              p^2 c^2 +m^2 c^4}\,\Gamma_y \Gamma_z \notag \\
                  &=& \cos^2 \left(\frac{\vartheta}{2}\right)\;H_{\rm e}
                        +2\,\sin\frac{\vartheta}{2} \cos\frac{\vartheta}{2}\,\sqrt{
                              p^2 c^2 +m^2 c^4}\,\Gamma_z
                        -\sin^2 \left(\frac{\vartheta}{2}\right)\sqrt{p^2 c^2 +m^2 c^4}
                              \,\Gamma_x \nonumber\\
                  &=&\cos\vartheta\,H_{\rm e} +\sin\vartheta\,H_{\rm b}^{\rm II} \notag \\
                  &=& \mathcal{H}_{\rm m}',
            \end{eqnarray}
            i.e., the unitary matrix transforms the Hamiltonian $H_{\rm e}$ of Dirac's electron to the mixed Hamiltonian $\mathcal{H}_{\rm m}'$.

            For $H_{\rm e}$, we have known that the common eigenstates of the set
            $\{H_{\rm e}, \hat{\Lambda}\}$ are $\{|\Psi_1\rangle,|\Psi_2\rangle,
            |\Psi_3\rangle, |\Psi_4\rangle\}$, where $\{|\Psi_1\rangle, |\Psi_2\rangle\}$ correspond to positive energy, while $\{|\Psi_3\rangle, |\Psi_4\rangle\}$ correspond to negative energy. Based on the unitary matrix $\mathcal{D}(\vartheta)$, from $|\Psi'''_j\rangle
            =\mathcal{D}(\vartheta)\,|\Psi_j\rangle$ we can have the four eigenstates of the $\mathcal{H}_m'$ as
            \begin{eqnarray}
                  \{|\Psi'''_1\rangle, |\Psi'''_2\rangle, |\Psi'''_3\rangle,
                        |\Psi'''_4\rangle\}
                  =\{\mathcal{D}(\vartheta)|\Psi_1\rangle,
                        \mathcal{D}(\vartheta)|\Psi_2\rangle,
                        \mathcal{D}(\vartheta)|\Psi_3\rangle,
                        \mathcal{D}(\vartheta)|\Psi_4\rangle\}.
            \end{eqnarray}
            which are common eigenstates of the set $\{\mathcal{H}_{\rm m}',
            \hat{\Lambda}'=\mathcal{D}(\vartheta) \hat{\Lambda} \mathcal{D}(\vartheta)^\dagger \equiv\hat{\Lambda}\}$. Likewise, $\{|\Psi'''_1\rangle,            |\Psi'''_2\rangle\}$ correspond to positive energy, while
            $\{|\Psi'''_3\rangle, |\Psi'''_4\rangle\}$ correspond to negative energy. And $\{|\Psi'''_1\rangle,
            |\Psi'''_3\rangle\}$ correspond to positive helicity, while
            $\{|\Psi'''_2\rangle, |\Psi'''_4\rangle\}$ correspond to negative helicity.

            Recall
            \begin{eqnarray}
                  \dfrac{1}{{\rm i}\hbar} [\vec{r}, H_{\rm b}^{\rm II}]
                  &=& \frac{\sqrt{p^2c^2+m^2c^4}}{p} \vec{\zeta}
                        -\frac{m^2 c^4}{\sqrt{p^2c^2+m^2c^4}} (\vec{\zeta}\cdot\vec{p})
                              \;\frac{\vec{p}}{p^3},
            \end{eqnarray}
            then
            \begin{eqnarray}\label{eq:DRDT}
                  \dfrac{{\rm d}\,\vec{r}}{{\rm d}\,t} &=& \dfrac{1}{{\rm i}\hbar}
                        [\vec{r}, \mathcal{H}_{\rm m}']
                  =\dfrac{1}{{\rm i}\hbar} [\vec{r}, \cos\vartheta \;H_{\rm e}
                        + \sin\vartheta \;H_{\rm b}^{\rm II}] \nonumber \\
                  &=& \cos\vartheta \; \dfrac{1}{{\rm i}\hbar} [\vec{r}, \;H_{\rm e}]
                        +\sin\vartheta \; \dfrac{1}{{\rm i}\hbar} [\vec{r}, \;H_{\rm b}^{\rm II}]
                        \nonumber \\
                  &=& \cos\vartheta\,(c\,\vec{\alpha})+\sin\vartheta\,\left[
                        \frac{\sqrt{p^2c^2+m^2c^4}}{p} \vec{\zeta}
                        -\frac{m^2c^4}{\sqrt{p^2c^2+m^2c^4}} (\vec{\zeta}
                              \cdot\vec{p})\;\frac{\vec{p}}{p^3}\right].
            \end{eqnarray}

            \begin{remark}
            It is not easy to calculate the Zitterbewegung for the ``position'' operator (i.e., $\vec{r}$) itself, when one considers the $H_{\rm e}$-$H_{\rm b}^{\rm II}$ mixing. However, it is more convenient to calculate the Zitterbewegung for the operator $(\vec{r}\cdot\hat{p})$, i.e., the projection of $\vec{r}$ on $\hat{p}$, here $\hat{p}$ is conserved in the Hamiltonian system $\mathcal{H}_{\rm m}'$. $\blacksquare$
            \end{remark}

            In the following, let us firstly calculate the Zitterbewegung for the operator $(\vec{r}\cdot\hat{p})$. From previous section we have known that
            \begin{eqnarray}
                  \frac{1}{{\rm i}\hbar}[\vec{\alpha}, H_{\rm e}]
                  =\frac{1}{{\rm i}\hbar}\left(2c\vec{p}-2\,H_{\rm e}\,\vec{\alpha}\right)
                  =-\frac{2\,H_{\rm e}}{{\rm i}\hbar}\left(\vec{\alpha}-c H^{-1}_{\rm e}\vec{p}\right).
            \end{eqnarray}
           Besides,
          \begin{eqnarray}
                  \{B\,C,\ A\}=\{A,\ B\,C\}=[A,\ B]C+B\{A,\ C\}=\{A,\ B\}C-B[A,\ C],
            \end{eqnarray}
            \begin{eqnarray}
                  && \bigl\{\vec{\alpha},\ (\vec{\alpha}\cdot\hat{p})\bigr\}
                  =2\,\hat{p},
            \end{eqnarray}
            and then
            \begin{eqnarray}
                   \{\vec{\alpha},\ H_{\rm b}^{\rm II}\}
                   &=&{\rm i}\sqrt{p^2 c^2 +m^2 c^4}\,
                        \big\{\vec{\alpha},\ \beta(\vec{\alpha}\cdot\hat{p})\big\}
                 = {\rm i}\sqrt{p^2 c^2 +m^2 c^4}\,\Bigl(\big[\vec{\alpha},\ \beta\big]
                              (\vec{\alpha}\cdot\hat{p})
                        +\beta\big\{\vec{\alpha},\ (\vec{\alpha}\cdot\hat{p})\big\}
                        \Bigr) \notag \\
                  &=& {\rm i}\sqrt{p^2 c^2 +m^2 c^4}\,\Big[\bigl(\{\vec{\alpha},\
                              \beta\big\}-2\,\beta\,\vec{\alpha}\bigr)(\vec{\alpha}
                                    \cdot\hat{p})
                        +\beta\big\{\vec{\alpha},\ (\vec{\alpha}\cdot\hat{p})\big\}
                        \Big] \notag \\
                  &=& {\rm i}\sqrt{p^2 c^2 +m^2 c^4}\,\big[-2\,\beta\,\vec{\alpha}(
                              \vec{\alpha}\cdot\hat{p})
                        +2\,\beta\,\hat{p}\big]
                  ={\rm i}\sqrt{p^2 c^2 +m^2 c^4}\,\Bigl\{-2\,\beta\big[
                        2\,\hat{p}-(\vec{\alpha}\cdot\hat{p})\vec{\alpha}\big]
                        +2\,\beta\,\hat{p}\Bigr\} \notag \\
                  &=& {\rm i}\sqrt{p^2 c^2 +m^2 c^4}\,\big[2\,\beta(\vec{\alpha}
                              \cdot\hat{p})\vec{\alpha}
                        -2\,\beta\,\hat{p}\big]
                  =2\, H_{\rm b}^{\rm II} \vec{\alpha}-{\rm i}\,2\sqrt{p^2 c^2 +m^2 c^4}\,\beta\,
                        \hat{p}, \notag \\
                  &=& 2\, H_{\rm b}^{\rm II} \vec{\alpha}-2\, H_{\rm b}^{\rm II} (\vec{\alpha}\cdot\hat{p})\hat{p},
            \end{eqnarray}
            which implies
            \begin{eqnarray}
                   \dfrac{1}{{\rm i}\hbar} [\vec{\alpha},\  H_{\rm b}^{\rm II}]
                  &=& \dfrac{1}{{\rm i}\hbar} \Bigl(\{\vec{\alpha},\  H_{\rm b}^{\rm II}\}
                        -2\, H_{\rm b}^{\rm II}\,\vec{\alpha}\Bigr)\nonumber\\
                  &=&\dfrac{1}{{\rm i}\hbar} \Bigl(2\, H_{\rm b}^{\rm II} \vec{\alpha}
                        -2\, H_{\rm b}^{\rm II} (\vec{\alpha}\cdot\hat{p})\hat{p}
                        -2\, H_{\rm b}^{\rm II}\,\vec{\alpha}\Bigr)\nonumber\\
                  &=&-\dfrac{2}{{\rm i}\hbar}  H_{\rm b}^{\rm II} (\vec{\alpha}\cdot\hat{p})\hat{p}.
            \end{eqnarray}
            Then we have
            \begin{eqnarray}
                  \dfrac{{\rm d}\,\vec{\alpha}}{{\rm d}\,t} &=& \dfrac{1}{{\rm i}\hbar}
                        [\vec{\alpha}, \mathcal{H}_{\rm m}']
                  =\dfrac{1}{{\rm i}\hbar} [\vec{\alpha}, \cos\vartheta \;H_{\rm e}
                        + \sin\vartheta \;H_{\rm b}^{\rm II}] \nonumber\\
                  &=& \cos\vartheta \; \dfrac{1}{{\rm i}\hbar} [
                              \vec{\alpha}, \;H_{\rm e}]
                        + \sin\vartheta \; \dfrac{1}{{\rm i}\hbar} [\vec{\alpha},
                              \;H_{\rm b}^{\rm II}] \nonumber\\
                  &=& \cos\vartheta \; \left[\frac{1}{{\rm i}\hbar} \left(2c\vec{p}
                              -2\,H_{\rm e}\,\vec{\alpha}\right)\right]
                        + \sin\vartheta \; \left[-\dfrac{2}{{\rm i}\hbar} H_{\rm b}^{\rm II} (
                              \vec{\alpha}\cdot\hat{p})\hat{p}\right],
            \end{eqnarray}
            viz.,
            \begin{eqnarray}\label{eq:DAlPDT}
                   \dfrac{{\rm d}(\vec{\alpha}\cdot\hat{p})}{{\rm d}\,t}
                  &=&\Bigl(\dfrac{{\rm d}\,\vec{\alpha}}{{\rm d}\,t}\Bigr)\cdot\hat{p}
                        +\vec{\alpha}\cdot\Bigl(\dfrac{{\rm d}\hat{p}}{{\rm d}\,t}
                              \Bigr)
                  =\Bigl(\dfrac{{\rm d}\,\vec{\alpha}}{{\rm d}\,t}\Bigr)\cdot\hat{p}
                  =\dfrac{1}{{\rm i}\hbar} [\vec{\alpha},\ \mathcal{H}_{\rm m}']
                        \cdot\hat{p}
                 \nonumber\\
                  &=&
                  \cos\vartheta \; \left[\frac{1}{{\rm i}\hbar} \left(2c\vec{p}
                              -2\,H_{\rm e}\,\vec{\alpha}\right)\right]\cdot\hat{p}
                        + \sin\vartheta \; \left[-\dfrac{2}{{\rm i}\hbar} H_{\rm b}^{\rm II} (
                              \vec{\alpha}\cdot\hat{p})\hat{p}\right]\cdot\hat{p}
                              \notag \\
                  &=& \frac{2\,c\,p}{{\rm i}\hbar} \cos\vartheta
                        -\frac{2}{{\rm i}\hbar} (\cos\vartheta\,H_{\rm e}
                              +\sin\vartheta\,H_{\rm b}^{\rm II})(\vec{\alpha}\cdot\hat{p})\nonumber\\
                  &=&\frac{2}{{\rm i}\hbar} \big[c\,p\,\cos\vartheta
                        -\mathcal{H}_{\rm m}'(\vec{\alpha}\cdot\hat{p})\big] \notag \\
                  &=& -\frac{2}{{\rm i}\hbar} \mathcal{H}_{\rm m}' \big[
                        (\vec{\alpha}\cdot\hat{p})-(\mathcal{H}_{\rm m}')^{-1} c\,p\,
                              \cos\vartheta\big], \notag \\
            \end{eqnarray}
            which indicates
            \begin{eqnarray}
                  \dfrac{{\rm d}}{{\rm d}\,t} \bigl[(\vec{\alpha}\cdot\hat{p})
                        -(\mathcal{H}_{\rm m}')^{-1} c\,p\,\cos\vartheta\bigr]
                  =-\frac{2}{{\rm i}\hbar} \mathcal{H}_{\rm m}' \big[
                        (\vec{\alpha}\cdot\hat{p})-(\mathcal{H}_{\rm m}')^{-1} c\,p\,
                              \cos\vartheta\big],
            \end{eqnarray}
            i.e.
            \begin{eqnarray}
                  \bigl[\vec{\alpha}(t)\cdot\hat{p}\bigr]-(
                        \mathcal{H}_{\rm m}')^{-1} c\,p\,\cos\vartheta
                  ={\rm e}^{{\rm i}\frac{2}{\hbar} \mathcal{H}_{\rm m}' t}\Bigl\{
                        \big[\vec{\alpha}(0)\cdot\hat{p}\big]-(
                        \mathcal{H}_{\rm m}')^{-1} c\,p\,\cos\vartheta\Bigr\},
            \end{eqnarray}
            i.e.
            \begin{eqnarray}\label{eq:AlP}
                  \bigl[\vec{\alpha}(t)\cdot\hat{p}\bigr]={\rm e}^{
                        {\rm i}\frac{2}{\hbar} \mathcal{H}_{\rm m}' t} \Bigl\{
                                    \big[\vec{\alpha}(0)\cdot\hat{p}\big]-(
                                    \mathcal{H}_{\rm m}')^{-1} c\,p\,\cos\vartheta
                                    \Bigr\}
                              +(\mathcal{H}_{\rm m}')^{-1} c\,p\,\cos\vartheta.
            \end{eqnarray}

            One can check that
            \begin{eqnarray}
                   \Big\{\mathcal{H}_{\rm m}',\ \big[\vec{\alpha}(0)\cdot\hat{p}
                        -(\mathcal{H}_{\rm m}')^{-1} \cos\vartheta\,c\,p\,\mathbb{I}\big]
                        \Big\}
                  &=&\Big\{\mathcal{H}_{\rm m}',\ \bigl[\vec{\alpha}(0)\cdot\hat{p}
                        \bigr]\Big\}-2\,c\,p\,\cos\vartheta\,\mathbb{I} \nonumber\\
                  &=& \cos\vartheta\,(H_{\rm e} \vec{\alpha}+\vec{\alpha} H_{\rm e})
                              \cdot\hat{p}
                        +\sin\vartheta\,(H_{\rm b}^{\rm II} \vec{\alpha}+\vec{\alpha} H_{\rm b}^{\rm II})\cdot\hat{p}
                        -2\,c\,p\,\cos\vartheta\,\mathbb{I} \nonumber \\
                  &=& \cos\vartheta\,2\,c\,p+\sin\vartheta\,\bigl[2\,H_{\rm b}^{\rm II} \vec{\alpha}
                              -2\,H_{\rm b}^{\rm II} (\vec{\alpha}\cdot\hat{p})\hat{p}\bigr]
                                    \cdot\hat{p}
                        -\cos\vartheta\,2\,c\,p\,\mathbb{I} \nonumber \\
                  &=& 0,
            \end{eqnarray}
            and thus
            \begin{eqnarray}
                  \mathcal{H}_{\rm m}' \big[\vec{\alpha}(0)\cdot\hat{p}
                        -(\mathcal{H}_{\rm m}')^{-1} \cos\vartheta\,c\,p\big]
                  =-\big[\vec{\alpha}(0)\cdot\hat{p}
                        -(\mathcal{H}_{\rm m}')^{-1} \cos\vartheta\,c\,p\big]
                              \mathcal{H}_{\rm m}',
            \end{eqnarray}
            which leads to
            \begin{eqnarray}
                  {\rm e}^{{\rm i}\frac{2}{\hbar} \mathcal{H}_{\rm m}' t} \Bigl\{
                        \big[\vec{\alpha}(0)\cdot\hat{p}\big]
                        -(\mathcal{H}_{\rm m}')^{-1} c\,p\,\cos\vartheta\Bigr\}
                  =\Bigl\{\big[\vec{\alpha}(0)\cdot\hat{p}\big]
                        -(\mathcal{H}_{\rm m}')^{-1} c\,p\,\cos\vartheta\Bigr\}{\rm e}^
                              {-{\rm i}\frac{2}{\hbar} \mathcal{H}_{\rm m}' t}.
            \end{eqnarray}
            Therefore we have
            \begin{eqnarray}
                  \bigl[\vec{\alpha}(t)\cdot\hat{p}\bigr]=\Bigl\{
                              \big[\vec{\alpha}(0)\cdot\hat{p}\big]
                              -(\mathcal{H}_{\rm m}')^{-1} c\,p\,\cos\vartheta\Bigr\}
                                    {\rm e}^{-{\rm i}\frac{2}{\hbar}
                                          \mathcal{H}_{\rm m}' t}
                        +(\mathcal{H}_{\rm m}')^{-1} c\,p\,\cos\vartheta.
            \end{eqnarray}

            Similarly, because of
            \begin{eqnarray}
                   \big\{\beta\,\vec{\alpha},\ (\vec{\alpha}\cdot\vec{p})\big\}
                  &=&\big[(\vec{\alpha}\cdot\vec{p}),\ \beta\big]\vec{\alpha}
                        +\beta\big\{(\vec{\alpha}\cdot\vec{p}),\ \vec{\alpha}\big\}
                  =\Bigl(\big\{(\vec{\alpha}\cdot\vec{p}),\ \beta\big\}
                              -2\,\beta(\vec{\alpha}\cdot\vec{p})\Bigr)\vec{\alpha}
                        +\beta\big\{(\vec{\alpha}\cdot\vec{p}),\ \vec{\alpha}\big\}
                        \notag \\
                  &=& -2\,p\,\beta(\vec{\alpha}\cdot\hat{p})\vec{\alpha}
                        +2\,p\,\beta\,\hat{p} = 2\,p\,\beta\bigl[\hat{p}-(\vec{\alpha}\cdot\hat{p})\vec{\alpha} \bigr], \nonumber\\
                      \{\vec{\zeta},\ H_{\rm e}\} &= & {\rm i}\{\beta\,\vec{\alpha},\
                        c\,\vec{\alpha}\cdot\vec{p}+m\,c^2 \beta\}
                  ={\rm i}\Bigl(c\big\{\beta\,\vec{\alpha},\ (
                              \vec{\alpha}\cdot\vec{p})\big\}
                        +m\,c^2 \{\beta\,\vec{\alpha},\ \beta\}\Bigr)
                  ={\rm i}\,c\big\{\beta\,\vec{\alpha},\ (\vec{\alpha}\cdot\vec{p})
                        \big\} \notag \\
                  &=& {\rm i}\,2\,c\,p\,\beta\bigl[\hat{p}-(\vec{\alpha}\cdot\hat{p})
                        \vec{\alpha}\bigr], \nonumber\\
                         \dfrac{1}{{\rm i}\hbar} [\vec{\zeta},\ H_{\rm e}]
                  &=&\dfrac{1}{{\rm i}\hbar} \Bigl(\{\vec{\zeta},\ H_{\rm e}\}
                        -2\,H_{\rm e}\,\vec{\zeta}\Bigr)
                  =\dfrac{2}{{\rm i}\hbar} \Bigl\{{\rm i}\,c\,p\,\beta\bigl[\hat{p}
                              -(\vec{\alpha}\cdot\hat{p})\vec{\alpha}\bigr]
                        -H_{\rm e}\, \vec{\zeta}\Bigr\},\nonumber\\
             \dfrac{1}{{\rm i}\hbar} \left[
                              \vec{\zeta},\  H_{\rm b}^{\rm II}\right]
                       & =& -\dfrac{2}{{\rm i}\hbar}  H_{\rm b}^{\rm II}\Bigl(\vec{\zeta}-\sqrt{p^2c^2+m^2c^4}\;  (H_{\rm b}^{\rm II})^{-1}\hat{p}\Bigr),
            \end{eqnarray}
            we then have
            \begin{eqnarray}
                   \dfrac{{\rm d}\,\vec{\zeta}}{{\rm d}\,t} &=& \dfrac{1}{{\rm i}\hbar}
                        [\vec{\zeta}, \mathcal{H}_{\rm m}']
                  =\dfrac{1}{{\rm i}\hbar} [\vec{\zeta}, \cos\vartheta \;H_{\rm e}
                        + \sin\vartheta \;H_{\rm b}^{\rm II}]\nonumber\\
                  &=&\cos\vartheta \; \dfrac{1}{{\rm i}\hbar} [
                              \vec{\zeta}, \;H_{\rm e}]
                        + \sin\vartheta \; \dfrac{1}{{\rm i}\hbar} [\vec{\zeta}, \;H_{\rm b}^{\rm II}]
                        \nonumber\\
                  &=& \cos\vartheta\,\dfrac{2}{{\rm i}\hbar} \Bigl\{{\rm i}\,c\,p\,\beta
                              \bigl[\hat{p}-(\vec{\alpha}\cdot\hat{p})\vec{\alpha}\bigr]
                              -H_{\rm e}\, \vec{\zeta}\Bigr\}
                        +\sin\vartheta\,\dfrac{2}{{\rm i}\hbar} \Bigl(
                              \sqrt{p^2 c^2 +m^2 c^4}\,\mathbb{I}\,\hat{p}
                              -H_{\rm b}^{\rm II}\,\vec{\zeta}\Bigr) \notag \\
                  &=& \sin\vartheta\,\dfrac{2}{{\rm i}\hbar} \sqrt{p^2 c^2 +m^2 c^4}\,
                              \mathbb{I}\,\hat{p}
                        -\dfrac{2}{{\rm i}\hbar} (\cos\vartheta\,H_{\rm e}
                              +\sin\vartheta\,H_{\rm b}^{\rm II})\vec{\zeta}
                        +\cos\vartheta\,\dfrac{2}{{\rm i}\hbar} {\rm i}\,c\,p\,\beta
                              \bigl[\hat{p}-(\vec{\alpha}\cdot\hat{p})\vec{\alpha}\bigr]
                        \notag \\
                  &=&\dfrac{2}{{\rm i}\hbar} (\sin\vartheta\,\sqrt{p^2 c^2 +m^2 c^4}\,
                                    \mathbb{I}\,\hat{p}
                              -\mathcal{H}_{\rm m}' \vec{\zeta})
                        +\cos\vartheta\,\dfrac{2}{{\rm i}\hbar} {\rm i}\,c\,p\,\beta
                              \bigl[\hat{p}-(\vec{\alpha}\cdot\hat{p})\vec{\alpha}
                              \bigr],
            \end{eqnarray}
            which leads to
            \begin{eqnarray}
                   \dfrac{{\rm d}(\vec{\zeta}\cdot\hat{p})}{{\rm d}\,t}
                  &=&\Bigl(\dfrac{{\rm d}\,\vec{\zeta}}{{\rm d}\,t}\Bigr)\cdot\hat{p}
                        +\vec{\zeta}\cdot\Bigl(\dfrac{{\rm d}\hat{p}}{{\rm d}\,t}\Bigr)
                  =\Bigl(\dfrac{{\rm d}\,\vec{\zeta}}{{\rm d}\,t}\Bigr)\cdot\hat{p}
                        \nonumber \\
                  &=& \dfrac{2}{{\rm i}\hbar} (\sin\vartheta\,\sqrt{p^2 c^2 +m^2 c^4}\,
                                    \mathbb{I}\,\hat{p}
                              -\mathcal{H}_{\rm m}' \vec{\zeta})\cdot\hat{p}
                        +\cos\vartheta\,\dfrac{2}{{\rm i}\hbar} {\rm i}\,c\,p\,\beta
                              \bigl[\hat{p}-(\vec{\alpha}\cdot\hat{p})\vec{\alpha}
                              \bigr]\cdot\hat{p}, \notag \\
                  &=& \dfrac{2}{{\rm i}\hbar} \bigl[\sin\vartheta\,
                              \sqrt{p^2 c^2 +m^2 c^4}\,\mathbb{I}
                        -\mathcal{H}_{\rm m}' (\vec{\zeta}\cdot\hat{p})\bigr]\nonumber\\
                 & =&\dfrac{2}{{\rm i}\hbar} \mathcal{H}_{\rm m}' \bigl[
                        (\mathcal{H}_{\rm m}')^{-1} \sin\vartheta\,\sqrt{
                              p^2 c^2 +m^2 c^4}
                        -(\vec{\zeta}\cdot\hat{p})\bigr] \notag \\
                  &=& {\rm i}\dfrac{2}{\hbar} \mathcal{H}_{\rm m}' \bigl[
                        (\vec{\zeta}\cdot\hat{p})-(\mathcal{H}_{\rm m}')^{-1}
                              \sin\vartheta\,\sqrt{p^2 c^2 +m^2 c^4}\bigr].
            \end{eqnarray}
            Moreover, we have
            \begin{eqnarray}
                  \dfrac{{\rm d}}{{\rm d}\,t} \bigl[(\vec{\zeta}\cdot\hat{p})-(
                        \mathcal{H}_{\rm m}')^{-1} \sin\vartheta\,\sqrt{
                              p^2 c^2 +m^2 c^4}\bigr]
                  ={\rm i}\dfrac{2}{\hbar} \mathcal{H}_{\rm m}' \bigl[
                        (\vec{\zeta}\cdot\hat{p})-(\mathcal{H}_{\rm m}')^{-1}
                              \sin\vartheta\,\sqrt{p^2 c^2 +m^2 c^4}\bigr],
            \end{eqnarray}
            i.e.,
            \begin{eqnarray}
                  \bigl[\vec{\zeta}(t)\cdot\hat{p}\bigr]-(
                        \mathcal{H}_{\rm m}')^{-1} \sin\vartheta\,\sqrt{
                              p^2 c^2 +m^2 c^4}
                  ={\rm e}^{{\rm i}\frac{2}{\hbar} \mathcal{H}_{\rm m}' t} \bigl[
                        \vec{\zeta}(0)\cdot\hat{p}-(\mathcal{H}_{\rm m}')^{-1}
                              \sin\vartheta\,\sqrt{p^2 c^2 +m^2 c^4}\bigr],
            \end{eqnarray}
            i.e.,
            \begin{eqnarray}
                  \bigl[\vec{\zeta}(t)\cdot\hat{p}\bigr]
                  ={\rm e}^{{\rm i}\frac{2}{\hbar} \mathcal{H}_{\rm m}' t} \bigl[
                              \vec{\zeta}(0)\cdot\hat{p}-(\mathcal{H}_{\rm m}')^{-1}
                                    \sin\vartheta\,\sqrt{p^2 c^2 +m^2 c^4}
                              \bigr]
                        +(\mathcal{H}_{\rm m}')^{-1} \sin\vartheta\,\sqrt{
                              p^2 c^2 +m^2 c^4}.
            \end{eqnarray}
            One can check that
            \begin{eqnarray}
                  && \Big\{\mathcal{H}_{\rm m}',\ \bigl[\vec{\zeta}(0)\cdot\hat{p}
                        -(\mathcal{H}_{\rm m}')^{-1} \sin\vartheta\,\sqrt{
                              p^2 c^2 +m^2 c^4}\bigr]\Big\}
                  =\Big\{\mathcal{H}_{\rm m}',\ \bigl[\vec{\zeta}(0)\cdot\hat{p}\bigr]
                        \Big\}-2\,\sin\vartheta\,\sqrt{p^2 c^2 +m^2 c^4}\,\mathbb{I}
                        \nonumber \\
                  &=& \cos\vartheta\,(H_{\rm e} \vec{\zeta}+\vec{\zeta} H_{\rm e})
                              \cdot\hat{p}
                        +\sin\vartheta\,(H_{\rm b}^{\rm II} \vec{\zeta}+\vec{\zeta} H_{\rm b}^{\rm II})\cdot\hat{p}
                        -2\,\sin\vartheta\,\sqrt{p^2 c^2 +m^2 c^4}\,\mathbb{I} \nonumber \\
                  &=& \cos\vartheta\,{\rm i}\,2\,c\,p\,\beta\bigl[\hat{p}
                              -(\vec{\alpha}\cdot\hat{p})\vec{\alpha}\bigr]\cdot\hat{p}
                        +\sin\vartheta\,(2\sqrt{p^2c^2+m^2c^4}\,\mathbb{I}\,\hat{p}
                              \bigr)\cdot\hat{p}
                        -2\,\sin\vartheta\,\sqrt{p^2 c^2 +m^2 c^4}\,\mathbb{I} \nonumber \\
                  &=& 0,
            \end{eqnarray}
            and thus
            \begin{eqnarray}
                  \mathcal{H}_{\rm m}'  \bigl[\vec{\zeta}(0)\cdot\hat{p}
                        -(\mathcal{H}_{\rm m}')^{-1} \sin\vartheta\,\sqrt{
                              p^2 c^2 +m^2 c^4}\bigr]
                  =-\bigl[\vec{\zeta}(0)\cdot\hat{p}-(\mathcal{H}_{\rm m}')^{-1}
                        \sin\vartheta\,\sqrt{p^2 c^2 +m^2 c^4}\bigr]
                              \mathcal{H}_{\rm m}',
            \end{eqnarray}
            which leads to
            \begin{eqnarray}
                  {\rm e}^{{\rm i}\frac{2}{\hbar} \mathcal{H}_{\rm m}' t} \bigl[
                        \vec{\zeta}(0)\cdot\hat{p}-(\mathcal{H}_{\rm m}')^{-1}
                              \sin\vartheta\,\sqrt{p^2 c^2 +m^2 c^4}\bigr]
                  =\bigl[\vec{\zeta}(0)\cdot\hat{p}-(\mathcal{H}_{\rm m}')^{-1}
                        \sin\vartheta\,\sqrt{p^2 c^2 +m^2 c^4}\bigr]{\rm e}^{
                              -{\rm i}\frac{2}{\hbar} \mathcal{H}_{\rm m}' t}.
            \end{eqnarray}
            Therefore we have
            \begin{eqnarray}
                  \bigl[\vec{\zeta}(t)\cdot\hat{p}\bigr]
                  =\bigl[\vec{\zeta}(0)\cdot\hat{p}-(\mathcal{H}_{\rm m}')^{-1}
                              \sin\vartheta\,\sqrt{p^2 c^2 +m^2 c^4}\bigr]
                              {\rm e}^{-{\rm i}\frac{2}{\hbar} \mathcal{H}_{\rm m}' t}
                        +(\mathcal{H}_{\rm m}')^{-1} \sin\vartheta\,\sqrt{
                              p^2 c^2 +m^2 c^4}.
            \end{eqnarray}
            After that, together with \Eq{eq:DRDT}, we attain
            \begin{eqnarray}\label{eq:DRPDT}
                   \dfrac{{\rm d}(\vec{r}\cdot\hat{p})}{{\rm d}\,t}
                  &=&\Bigl(\dfrac{{\rm d}\,\vec{r}}{{\rm d}\,t}\Bigr)\cdot\hat{p}
                        +\vec{r}\cdot\Bigl(\dfrac{{\rm d}\hat{p}}{{\rm d}\,t}\Bigr)
                  =\Bigl(\dfrac{{\rm d}\,\vec{r}}{{\rm d}\,t}\Bigr)\cdot\hat{p}
                        \notag \\
                  &=& \cos\vartheta\,c\,\bigl[\vec{\alpha}(t)\cdot\hat{p}\bigr]
                        +\sin\vartheta\,\left\{\frac{\sqrt{p^2c^2+m^2c^4}}{p} \bigl[
                                    \vec{\zeta}(t)\cdot\hat{p}\bigr]
                              -\frac{m^2c^4}{\sqrt{p^2c^2+m^2c^4}} \bigl[\vec{\zeta}(t)
                                    \cdot\vec{p}\bigr]\frac{(\vec{p}\cdot\hat{p})}{p^3}
                              \right\} \notag \\
                  &=& \cos\vartheta\,c\,\bigl[\vec{\alpha}(t)\cdot\hat{p}\bigr]
                        +\sin\vartheta\,\left(\frac{\sqrt{p^2c^2+m^2c^4}}{p}
                              -\frac{m^2 c^4}{p\sqrt{p^2c^2+m^2c^4}}\right)\bigl[
                                    \vec{\zeta}(t)\cdot\hat{p}\bigr] \notag \\
                  &=& c\Big\{\cos\vartheta\,\bigl[\vec{\alpha}(t)\cdot\hat{p}\bigr]
                        +\frac{p\,c}{\sqrt{p^2 c^2 +m^2 c^4}} \sin\vartheta\,\bigl[
                              \vec{\zeta}(t)\cdot\hat{p}\bigr]\Big\} \notag \\
                  &=& c\,\cos\vartheta\,\biggl(\Bigl\{\big[\vec{\alpha}(0)\cdot\hat{p}
                                    \big]-(\mathcal{H}_{\rm m}')^{-1} c\,p\,
                                          \cos\vartheta\Bigr\}{\rm e}^{
                                                -{\rm i}\frac{2}{\hbar}
                                                      \mathcal{H}_{\rm m}' t}
                              +(\mathcal{H}_{\rm m}')^{-1} c\,p\,\cos\vartheta\biggr)
                              \notag \\
                        && +c\frac{p\,c\,\sin\vartheta}{\sqrt{p^2 c^2 +m^2 c^4}} \Bigl\{
                              \bigl[\vec{\zeta}(0)\cdot\hat{p}
                                    -(\mathcal{H}_{\rm m}')^{-1} \sin\vartheta\,\sqrt{
                                          p^2 c^2 +m^2 c^4}\bigr]{\rm e}^{
                                                -{\rm i}\frac{2}{\hbar}
                                                      \mathcal{H}_{\rm m}' t}
                              +(\mathcal{H}_{\rm m}')^{-1} \sin\vartheta\,\sqrt{
                                    p^2 c^2 +m^2 c^4}\Bigr\} \notag \\
                  &=& c\biggl(\Bigl\{\cos\vartheta\,\big[\vec{\alpha}(0)\cdot\hat{p}
                                    \big]
                              +\frac{p\,c\,\sin\vartheta}{\sqrt{p^2 c^2 +m^2 c^4}}
                                    \bigl[\vec{\zeta}(0)\cdot\hat{p}\bigr]
                              -(\mathcal{H}_{\rm m}')^{-1} c\,p\Bigr\}{\rm e}^{
                                    -{\rm i}\frac{2}{\hbar} \mathcal{H}_{\rm m}' t}
                        +(\mathcal{H}_{\rm m}')^{-1} c\,p\biggr) \notag \\
                  &=& c^2 (\mathcal{H}_{\rm m}')^{-1} p +c\Bigl\{
                        \cos\vartheta\,\big[\vec{\alpha}(0)\cdot\hat{p}\big]
                        +\frac{p\,c\,\sin\vartheta}{\sqrt{p^2 c^2 +m^2 c^4}} \bigl[
                              \vec{\zeta}(0)\cdot\hat{p}\bigr]
                        -(\mathcal{H}_{\rm m}')^{-1} c\,p\Bigr\}{\rm e}^{
                              -{\rm i}\frac{2}{\hbar} \mathcal{H}_{\rm m}' t}.
            \end{eqnarray}
            \begin{remark}
                  When $\vartheta=0$, $\mathcal{H}_{\rm m}' =H_{\rm e}$, \Eq{eq:DRPDT} reduces to
                  \begin{eqnarray}
                        \dfrac{{\rm d}(\vec{r}\cdot\hat{p})}{{\rm d}\,t}
                        =c^2 H_{\rm e}^{-1} p +\Bigl\{c\big[\vec{\alpha}(0)\cdot\hat{p}
                              \big]-c^2 H_{\rm e}^{-1} p\Bigr\}{\rm e}^{
                                    -{\rm i}\frac{2}{\hbar} H_{\rm e} t},
                  \end{eqnarray}
                  which is in line with Eq. (\ref{eq:D-9}), i.e.,
                  \begin{eqnarray}
                        \frac{{\rm d}\,\vec{r}}{{\rm d}\,t}=c^2 H^{-1}_{\rm e}\vec{p}
                              +\left[c\vec{\alpha}(0)-c^2 H^{-1}_{\rm e}\vec{p}\right]
                                    {\rm e}^{\frac{-{\rm i}\,2\,H_{\rm e}t}{\hbar}},
                  \end{eqnarray}
                  after making dot product with $\hat{p}$.

                  While $\vartheta=\pi/2$, $\mathcal{H}_{\rm m}' =H_{\rm b}^{\rm II}$, and \Eq{eq:DRPDT} reduces to
                  \begin{eqnarray}
                        \dfrac{{\rm d}(\vec{r}\cdot\hat{p})}{{\rm d}\,t}
                        =c^2 (H_{\rm b}^{\rm II})^{-1} p +\Bigl\{
                              \frac{p\,c^2}{\sqrt{p^2 c^2 +m^2 c^4}} \bigl[
                                    \vec{\tau}(0)\cdot\hat{p}\bigr]
                              -c^2 ( H_{\rm b}^{\rm II})^{-1} p\Bigr\}{\rm e}^{
                                    -{\rm i}\frac{2}{\hbar}  H_{\rm b}^{\rm II} t},
                  \end{eqnarray}
            which coincides with the case shown in Eq. (\ref{eq:F-5c}), i.e.,
                  \begin{eqnarray}
                        \dfrac{{\rm d}\,\vec{r}}{{\rm d}\,t}
                        =c^2 (H_{\rm b}^{\rm II})^{-1}\vec{p} +\Biggr\{
                              \frac{\sqrt{p^2c^2+m^2c^4}}{p} \vec{\zeta}(0)
                              - \frac{m^2c^4}{\sqrt{p^2c^2+m^2c^4}} \left[
                                    \vec{\zeta}(0)\cdot \vec{p}\right]\frac{\vec{p}}{p^3}
                              -c^2 (H_{\rm b}^{\rm II})^{-1} \vec{p}\Biggr\}{\rm e}^{
                                    -\frac{{\rm i}\,2 H_{\rm b}^{\rm II} t}{\hbar}},
            \end{eqnarray}
            after making dot product with $\hat{p}$. $\blacksquare$
            \end{remark}
            Consequently, based on \Eq{eq:DRPDT} we have
            \begin{eqnarray}\label{eq:RtDotP}
                  \bigl[\vec{r}(t)\cdot\hat{p}\bigr]
                  &=& \vec{r}(0)\cdot\hat{p}+c^2 (\mathcal{H}_{\rm m}')^{-1} p\,t
                        \notag \\
                        && +{\rm i}\frac{c\,\hbar}{2} \Bigl\{
                              \cos\vartheta\,\big[\vec{\alpha}(0)\cdot\hat{p}\big]
                              +\frac{p\,c\,\sin\vartheta}{\sqrt{p^2 c^2 +m^2 c^4}}
                                    \bigl[\vec{\zeta}(0)\cdot\hat{p}\bigr]
                              -(\mathcal{H}_{\rm m}')^{-1} c\,p\Bigr\}
                                    \dfrac{1}{\mathcal{H}_{\rm m}'} \Bigl({\rm e}^{
                                          -{\rm i}\frac{2}{\hbar}
                                                \mathcal{H}_{\rm m}' t} -1\Bigr).
            \end{eqnarray}
            Let us define the third term of Eq. (\ref{eq:RtDotP}) as the ``Zitterbewegung'' operator, i.e.,
            \begin{eqnarray}
                  \hat{\mathcal{Z}}_{\rm m} ={\rm i}\frac{c\,\hbar}{2} \Bigl\{
                        \cos\vartheta\,\big[\vec{\alpha}(0)\cdot\hat{p}\big]
                        +\frac{p\,c\,\sin\vartheta}{\sqrt{p^2 c^2 +m^2 c^4}} \bigl[
                              \vec{\zeta}(0)\cdot\hat{p}\bigr]
                        -(\mathcal{H}_{\rm m}')^{-1} c\,p\Bigr\} \dfrac{1}{
                              \mathcal{H}_{\rm m}'} \Bigl({\rm e}^{-{\rm i}\frac{2}{
                                    \hbar} \mathcal{H}_{\rm m}' t} -1\Bigr).
            \end{eqnarray}
            Because $\bigl[\vec{r}(t)\cdot\hat{p}\bigr]$ is Hermitian, so does $\mathcal{Z}_{\rm m}$, which implies that
            \begin{eqnarray}\label{eq:herm-1a}
                  \Bigl\{\mathcal{H}_{\rm m}',\ \cos\vartheta\,\big[\vec{\alpha}(0)
                              \cdot\hat{p}\big]
                        +\frac{p\,c\,\sin\vartheta}{\sqrt{p^2 c^2 +m^2 c^4}} \bigl[
                              \vec{\zeta}(0)\cdot\hat{p}\bigr]
                        -(\mathcal{H}_{\rm m}')^{-1} c\,p\Bigr\}=0.
            \end{eqnarray}
            Based on which one can have
            \begin{eqnarray}
            && \mathcal{H}_{\rm m}'\; \hat{\mathcal{Z}}_{\rm m}+\hat{\mathcal{Z}}_{\rm m} \mathcal{H}_{\rm m}'=0.
            \end{eqnarray}

            Let us introduce the following projection operators
            \begin{eqnarray}
                  \Pi_\pm=\frac{1}{2}\left(\mathbb{I} \pm \frac{\mathcal{H}_{\rm m}'}{\sqrt{p^2c^2+m^2c^4}}\right), \;\;\;\;\; \Pi_\pm^2=\Pi_\pm,
            \end{eqnarray}
            we easily have
            \begin{eqnarray}
                  && \Pi_+ |\Psi'''_1\rangle = |\Psi'''_1\rangle, \;\; \Pi_+ |\Psi'''_2\rangle = |\Psi'''_2\rangle, \;\; \Pi_+ |\Psi'''_3\rangle = 0, \;\; \Pi_+ |\Psi_4\rangle = 0, \nonumber\\
                  && \Pi_- |\Psi'''_1\rangle = 0, \;\; \Pi_- |\Psi'''_2\rangle = 0, \;\; \Pi_- |\Psi'''_3\rangle = |\Psi'''_3\rangle, \;\; \Pi_- |\Psi'''_4\rangle = |\Psi'''_4\rangle.
            \end{eqnarray}
            We then have
            \begin{eqnarray}
                   \Pi_+ \; \hat{\mathcal{Z}}_{\rm m} \;\Pi_+ &=&
                        \frac{1}{4} \left(\mathbb{I}+\frac{\mathcal{H}_{\rm m}'}{\sqrt{p^2c^2+m^2c^4}}\right) \;\hat{\mathcal{Z}}_{\rm m}\;
                        \left(\mathbb{I} + \frac{\mathcal{H}_{\rm m}'}{\sqrt{p^2c^2+m^2c^4}}\right)\nonumber\\
                  &=& \frac{1}{4} \left(\hat{\mathcal{Z}}_{\rm m}
                        + \frac{\mathcal{H}_{\rm m}'\hat{\mathcal{Z}}_{\rm m}+\mathcal{Z}_{\rm m}\mathcal{H}_{\rm m}' }{\sqrt{p^2c^2+m^2c^4}}
                        +\frac{1}{p^2c^2+m^2c^4} \mathcal{H}_{\rm m}'\hat{\mathcal{Z}}_{\rm m}\mathcal{H}_{\rm m}' \right)\nonumber\\
                  &=& \frac{1}{4} \left(\hat{\mathcal{Z}}_{\rm m}
                        -\frac{\mathcal{H}^2_{\rm m}}{p^2c^2+m^2c^4} \hat{\mathcal{Z}}_{\rm m} \right) \nonumber \\
                  &=& \frac{1}{4} \left(\hat{\mathcal{Z}}_{\rm m}- \hat{\mathcal{Z}}_{\rm m}\right)
                  =0.
            \end{eqnarray}
            Similarly, we have
            \begin{eqnarray}
                  &&    \Pi_- \; \hat{\mathcal{Z}}_{\rm m} \;\Pi_- =0.
            \end{eqnarray}
            Then we have
            \begin{eqnarray}
                  &&    \langle\Psi_+| \; \hat{\mathcal{Z}}_{\rm m} \;|\Psi_+\rangle =0, \;\;\;\;\;\; \langle \Psi_-| \; \hat{\mathcal{Z}}_{\rm m} \;|\Psi_-\rangle=0.
            \end{eqnarray}
            Thereby, if the mixed Hamiltonian system is in a superposition state of only positive-energy (or negative-energy), then $\mathcal{Z}_{\rm m}=0$, i.e, there is no phenomenon of ``Zitterbewegung''.

            Now suppose the mixed Hamiltonian system is in a superposition state of $|\Psi'''_1\rangle$ and $|\Psi'''_3\rangle$, i.e.,
            \begin{eqnarray}
                  &&   |\Psi'''\rangle=\cos\eta |\Psi'''_1\rangle+ \sin\eta |\Psi'''_3\rangle.
            \end{eqnarray}
            To calculate the Zitterbewegung, we only need to consider the term $\langle\Psi'''_1|\hat{\mathcal{Z}}_{\rm m}|\Psi'''_3 \rangle$. We can have
            \begin{eqnarray}
                  &&  {\mathcal{Z}}_{\rm m}= \sin(2\eta){\rm Re}\left(
                  \langle\Psi'''_1|\hat{\mathcal{Z}}_{\rm m}|\Psi'''_3 \rangle\right).
            \end{eqnarray}
            Since
            \begin{eqnarray}
                  \mathcal{D}(\vartheta) &=& {\rm e}^{{\rm i} \frac{\vartheta}{2} \Gamma_y}=\cos\Bigl(\frac{\vartheta}{2}\Bigr)\;\mathbb{I}+{\rm i}\,\sin\Bigl(
                        \frac{\vartheta}{2}\Bigr)\,\dfrac{(-m\,c^2 \vec{\alpha}
                              \cdot\hat{p}+\beta\,p\,c)}{\sqrt{p^2 c^2 +m^2 c^4}},\nonumber\\
                    \mathcal{D}^\dagger (\vartheta) &=& {\rm e}^{
                        -{\rm i} \frac{\vartheta}{2}  \Gamma_y}
                  =\cos\Bigl(\frac{\vartheta}{2}\Bigr)\;\mathbb{I}-{\rm i}\,\sin\Bigl(
                        \frac{\vartheta}{2}\Bigr)\,\dfrac{(-m\,c^2 \vec{\alpha}
                              \cdot\hat{p}+\beta\,p\,c)}{\sqrt{p^2 c^2 +m^2 c^4}},
            \end{eqnarray}

            \begin{eqnarray}\label{eq:AlphaP}
                   \big[\vec{\alpha}(0)\cdot\hat{p},\ (
                        -m\,c^2 \vec{\alpha}\cdot\hat{p}+\beta\,p\,c)\big]
                  &=&\big[\vec{\alpha}\cdot\hat{p},\ (
                        -m\,c^2 \vec{\alpha}\cdot\hat{p}+\beta\,p\,c)\big]
                  =p\,c\big[(\vec{\alpha}\cdot\hat{p}),\ \beta\big] \notag \\
                  &=& p\,c\Bigl(\big\{(\vec{\alpha}\cdot\hat{p}),\ \beta\big\}
                        -2\,\beta(\vec{\alpha}\cdot\hat{p})\Bigr)
                  =-2\,p\,c\,\beta(\vec{\alpha}\cdot\hat{p})
                  ={\rm i}\,2\,p\,c\bigl[{\rm i}\,\beta(\vec{\alpha}\cdot\hat{p})
                        \bigr] \notag \\
                  &=& {\rm i}\,2\dfrac{p\,c}{\sqrt{p^2 c^2 +m^2 c^4}} H_{\rm b}^{\rm II}, \notag \\
                    (-m\,c^2 \vec{\alpha}\cdot\hat{p}+\beta\,p\,c)\bigl[\vec{\alpha}(0)
                        \cdot\hat{p}\bigr](-m\,c^2 \vec{\alpha}\cdot\hat{p}+\beta\,p\,c)
                  &=&(-m\,c^2 \vec{\alpha}\cdot\hat{p}+\beta\,p\,c)(\vec{\alpha}
                        \cdot\hat{p})(-m\,c^2 \vec{\alpha}\cdot\hat{p}+\beta\,p\,c)
                        \notag \\
                  &=& m^2 c^4 (\vec{\alpha}\cdot\hat{p})-2\,p\,c\,m\,c^2 \beta
                        +p^2 c^2 \beta(\vec{\alpha}\cdot\hat{p})\beta\nonumber\\
                  &=&m^2 c^4 (\vec{\alpha}\cdot\hat{p})-2\,p\,c\,m\,c^2 \beta
                        -p^2 c^2 (\vec{\alpha}\cdot\hat{p}) \notag \\
                  &=& m^2 c^4 (\vec{\alpha}\cdot\hat{p})
                              +p^2 c^2 (\vec{\alpha}\cdot\hat{p})
                        -2\,p\,c\bigl[m\,c^2 \beta+p\,c(\vec{\alpha}\cdot\hat{p})\bigr]
                        \notag \\
                  &=& (p^2 c^2 +m^2 c^4)(\vec{\alpha}\cdot\hat{p})
                        -2\,p\,c\,H_{\rm e},
            \end{eqnarray}
            we then have
            \begin{eqnarray}\label{eq:DAlphaPD}
                  && \mathcal{D}^\dagger (\vartheta)\big[\vec{\alpha}(0)\cdot\hat{p}
                        \big]\mathcal{D}(\vartheta) \notag \\
                  &=& \bigg[\cos\Bigl(\frac{\vartheta}{2}\Bigr)\;\mathbb{I}
                        -{\rm i}\,\sin\Bigl(\frac{\vartheta}{2}\Bigr)
                              \dfrac{(-m\,c^2 \vec{\alpha}\cdot\hat{p}+\beta\,p\,c)}{
                                    \sqrt{p^2 c^2 +m^2 c^4}}
                        \bigg]\big[\vec{\alpha}(0)\cdot\hat{p}\big]\bigg[
                              \cos\Bigl(\frac{\vartheta}{2}\Bigr)\;\mathbb{I}
                              +{\rm i}\,\sin\Bigl(\frac{\vartheta}{2}\Bigr)
                                    \dfrac{(-m\,c^2 \vec{\alpha}\cdot\hat{p}
                                          +\beta\,p\,c)}{\sqrt{p^2 c^2 +m^2 c^4}}\bigg]
                        \notag \\
                  &=& \cos^2 \Bigl(\frac{\vartheta}{2}\Bigr)\big[\vec{\alpha}(0)
                              \cdot\hat{p}\big]
                        +{\rm i}\,\sin\Bigl(\frac{\vartheta}{2}\Bigr)\cos\Bigl(
                              \frac{\vartheta}{2}\Bigr)\dfrac{1}{
                                    \sqrt{p^2 c^2 +m^2 c^4}} \big[
                                          \vec{\alpha}(0)\cdot\hat{p},\ (
                                                -m\,c^2 \vec{\alpha}\cdot\hat{p}
                                                +\beta\,p\,c)\big] \notag \\
                        && +\sin^2 \Bigl(\frac{\vartheta}{2}\Bigr)\dfrac{1}{
                              p^2 c^2 +m^2 c^4} (-m\,c^2 \vec{\alpha}\cdot\hat{p}
                                    +\beta\,p\,c)\bigl[\vec{\alpha}(0)\cdot\hat{p}\bigr]
                                          (-m\,c^2 \vec{\alpha}\cdot\hat{p}+\beta\,p\,c)
                        \notag \\
                  &=& \cos^2 \Bigl(\frac{\vartheta}{2}\Bigr)\big[\vec{\alpha}(0)
                              \cdot\hat{p}\big]
                        +{\rm i}\,\sin\Bigl(\frac{\vartheta}{2}\Bigr)\cos\Bigl(
                              \frac{\vartheta}{2}\Bigr)\dfrac{1}{
                                    \sqrt{p^2 c^2 +m^2 c^4}} {\rm i}\,2\dfrac{p\,c}{
                                          \sqrt{p^2 c^2 +m^2 c^4}} H_{\rm b}^{\rm II} \notag \\
                        && +\sin^2 \Bigl(\frac{\vartheta}{2}\Bigr)\dfrac{1}{
                              p^2 c^2 +m^2 c^4} \bigl[(p^2 c^2 +m^2 c^4)(
                                    \vec{\alpha}\cdot\hat{p})-2\,p\,c\,H_{\rm e}\bigr]
                        \notag \\
                  &=& \vec{\alpha}\cdot\hat{p}
                        -2\,\sin\Bigl(\frac{\vartheta}{2}\Bigr) \cos\Bigl(
                              \frac{\vartheta}{2}\Bigr)\,\dfrac{p\,c}{p^2 c^2 +m^2 c^4}
                                    H_{\rm b}^{\rm II}
                        -2\,\sin^2 \Bigl(\frac{\vartheta}{2}\Bigr)\dfrac{p\,c}{
                              p^2 c^2 +m^2 c^4} H_{\rm e} \notag \\
                  &=& \vec{\alpha}\cdot\hat{p}
                        -\dfrac{p\,c}{p^2 c^2 +m^2 c^4} \sin\vartheta\,H_{\rm b}^{\rm II}
                        +\dfrac{p\,c}{p^2 c^2 +m^2 c^4} (\cos\vartheta -1)H_{\rm e}.
            \end{eqnarray}
            Because
            \begin{eqnarray}
                   \big[\vec{\zeta}\cdot\hat{p},\ (
                        -m\,c^2 \vec{\alpha}\cdot\hat{p}+\beta\,p\,c)\big]
                  &=&[\Gamma_z,\ \sqrt{p^2 c^2 +m^2 c^4}\,\Gamma_y]
                  =-{\rm i}\,2\,H_{\rm e},\nonumber\\
                 (-m\,c^2 \vec{\alpha}\cdot\hat{p}+\beta\,p\,c)(\vec{\zeta}
                        \cdot\hat{p})(-m\,c^2 \vec{\alpha}\cdot\hat{p}+\beta\,p\,c)
                  &=&(p^2 c^2 +m^2 c^4) \Gamma_y \Gamma_z \Gamma_y ={\rm i}(p^2 c^2 +m^2 c^4) \Gamma_x \Gamma_y
                        \notag \\
                  &=& {\rm i}\,{\rm i}(p^2 c^2 +m^2 c^4) \Gamma_z \notag \\
                  &=& -\sqrt{p^2 c^2 +m^2 c^4}\,H_{\rm b}^{\rm II},
            \end{eqnarray}
            we then have
            \begin{eqnarray}\label{eq:DD-1a}
                  && \mathcal{D}^\dagger (\vartheta)\bigl[\vec{\zeta}(0)\cdot\hat{p}
                        \bigr]\mathcal{D}(\vartheta)
                  \notag \\
                  &=& \bigg[\cos\Bigl(\frac{\vartheta}{2}\Bigr)\;\mathbb{I}
                        -{\rm i}\,\sin\Bigl(\frac{\vartheta}{2}\Bigr)
                              \dfrac{(-m\,c^2 \vec{\alpha}\cdot\hat{p}+\beta\,p\,c)}{
                                    \sqrt{p^2 c^2 +m^2 c^4}}
                        \bigg](\vec{\zeta}\cdot\hat{p})\bigg[
                              \cos\Bigl(\frac{\vartheta}{2}\Bigr)\;\mathbb{I}
                              +{\rm i}\,\sin\Bigl(\frac{\vartheta}{2}\Bigr)
                                    \dfrac{(-m\,c^2 \vec{\alpha}\cdot\hat{p}
                                          +\beta\,p\,c)}{\sqrt{p^2 c^2 +m^2 c^4}}\bigg]
                        \notag \\
                  &=& \cos^2 \Bigl(\frac{\vartheta}{2}\Bigr)(\vec{\zeta}\cdot\hat{p})
                        +{\rm i}\,\sin\Bigl(\frac{\vartheta}{2}\Bigr)\cos\Bigl(
                              \frac{\vartheta}{2}\Bigr)\dfrac{1}{
                                    \sqrt{p^2 c^2 +m^2 c^4}} \big[
                                          (\vec{\zeta}\cdot\hat{p}),\ (
                                                -m\,c^2 \vec{\alpha}\cdot\hat{p}
                                                +\beta\,p\,c)\big] \notag \\
                        && +\sin^2 \Bigl(\frac{\vartheta}{2}\Bigr)\dfrac{1}{
                              p^2 c^2 +m^2 c^4} (-m\,c^2 \vec{\alpha}\cdot\hat{p}
                                    +\beta\,p\,c)(\vec{\zeta}\cdot\hat{p})(
                                          -m\,c^2 \vec{\alpha}\cdot\hat{p}+\beta\,p\,c)
                        \notag \\
                  &=& \cos^2 \Bigl(\frac{\vartheta}{2}\Bigr)(\vec{\zeta}\cdot\hat{p})
                        -{\rm i}\,\sin\Bigl(\frac{\vartheta}{2}\Bigr)\cos\Bigl(
                              \frac{\vartheta}{2}\Bigr)\dfrac{1}{
                                    \sqrt{p^2 c^2 +m^2 c^4}} {\rm i}\,2\,H_{\rm e}
                        -\sin^2 \Bigl(\frac{\vartheta}{2}\Bigr)\dfrac{1}{
                              p^2 c^2 +m^2 c^4} \sqrt{p^2 c^2 +m^2 c^4}\,H_{\rm b}^{\rm II}
                        \notag \\
                  &=& \cos^2 \Bigl(\frac{\vartheta}{2}\Bigr)\dfrac{H_{\rm b}^{\rm II}}{
                              \sqrt{p^2 c^2 +m^2 c^4}}
                        +2\,\sin\Bigl(\frac{\vartheta}{2}\Bigr) \cos\Bigl(
                              \frac{\vartheta}{2}\Bigr)\,\dfrac{H_{\rm e}}{
                                    \sqrt{p^2 c^2 +m^2 c^4}}
                        -\sin^2 \Bigl(\frac{\vartheta}{2}\Bigr)\dfrac{H_{\rm b}^{\rm II}}{
                              \sqrt{p^2 c^2 +m^2 c^4}} \notag \\
                  &=& \cos\vartheta \dfrac{H_{\rm b}^{\rm II}}{\sqrt{p^2 c^2 +m^2 c^4}}
                        +\sin\vartheta \dfrac{H_{\rm e}}{\sqrt{p^2 c^2 +m^2 c^4}},
                        \notag \\
            \end{eqnarray}
            therefore
            \begin{eqnarray}
                  && \langle\Psi'''_1|\Bigl\{\cos\vartheta\,\big[\vec{\alpha}(0)
                              \cdot\hat{p}\big]
                        +\frac{p\,c\,\sin\vartheta}{\sqrt{p^2 c^2 +m^2 c^4}} \bigl[
                              \vec{\zeta}(0)\cdot\hat{p}\bigr]
                        -(\mathcal{H}_{\rm m}')^{-1} c\,p\Bigr\}|\Psi'''_3 \rangle
                        \nonumber \\
                  &=& \langle\Psi_1|\mathcal{D}^\dagger (\vartheta)\Bigl\{
                        \cos\vartheta\,\vec{\alpha}(0)\cdot\hat{p}
                        +\frac{p\,c\,\sin\vartheta}{\sqrt{p^2 c^2 +m^2 c^4}} \bigl[
                              \vec{\zeta}(0)\cdot\hat{p}\bigr]
                        -(\mathcal{H}_{\rm m}')^{-1} c\,p\Bigr\}\mathcal{D}(\vartheta)|
                              \Psi_3 \rangle \nonumber \\
                  &=& \langle\Psi_1|\Bigl\{
                        \cos\vartheta\,\mathcal{D}^\dagger (\vartheta)\big[
                              \vec{\alpha}(0)\cdot\hat{p}\big]\mathcal{D}(\vartheta)
                        +\frac{p\,c\,\sin\vartheta}{\sqrt{p^2 c^2 +m^2 c^4}}
                              \mathcal{D}^\dagger (\vartheta)\bigl[\vec{\zeta}(0)
                                    \cdot\hat{p}\bigr]\mathcal{D}(\vartheta)
                        -\mathcal{D}^\dagger (\vartheta)\,(\mathcal{H}_{\rm m}')^{-1}
                              \mathcal{D}^\dagger (\vartheta) c\,p\Bigr\}|
                                    \Psi_3 \rangle \nonumber \\
                  &=& \langle\Psi_1|\bigg\{\cos\vartheta\,\Bigl[
                              \vec{\alpha}\cdot\hat{p}
                              -\dfrac{p\,c}{p^2 c^2 +m^2 c^4} \sin\vartheta\,H_{\rm b}^{\rm II}
                              +\dfrac{p\,c}{p^2 c^2 +m^2 c^4} (\cos\vartheta -1)
                                    H_{\rm e}\Bigr] \notag \\
                        &&\qquad\quad +\frac{p\,c\,\sin\vartheta}{\sqrt{
                                    p^2 c^2 +m^2 c^4}}
                              \Big(\cos\vartheta \dfrac{H_{\rm b}^{\rm II}}{\sqrt{p^2 c^2 +m^2 c^4}}
                                    +\sin\vartheta \dfrac{H_{\rm e}}{\sqrt{
                                          p^2 c^2 +m^2 c^4}}\Big)
                        -H_{\rm e}^{-1} c\,p\bigg\}|\Psi_3 \rangle \nonumber \\
                  &=& \langle\Psi_1|\Big[\cos\vartheta (\vec{\alpha}\cdot\hat{p})
                        +\dfrac{p\,c}{p^2 c^2 +m^2 c^4} (1-\cos\vartheta)H_{\rm e}
                        -H_{\rm e}^{-1} c\,p\Big]|\Psi_3 \rangle
                  =\cos\vartheta \langle\Psi_1|(\vec{\alpha}\cdot\hat{p})|\Psi_3\rangle
                        \notag \\
                  &=& \cos\vartheta \langle\Psi_1|\vec{\alpha}|\Psi_3\rangle
                        \cdot\hat{p}
                  =-\cos\vartheta \dfrac{m\,c^2}{E_{\rm p}} \hat{p}\cdot\hat{p}
                        \nonumber \\
                  &=& -\dfrac{m\,c^2}{E_{\rm m}} \cos\vartheta.
            \end{eqnarray}
            Then we get
            \begin{eqnarray}
                  &&\langle\Psi'''_1|\hat{\mathcal{Z}}_{\rm m}|\Psi'''_3 \rangle
                  =\langle\Psi'''_1| \biggr[{\rm i}\frac{c\,\hbar}{2} \Bigl\{
                        \cos\vartheta\,\big[\vec{\alpha}(0)\cdot\hat{p}\big]
                        +\frac{p\,c\,\sin\vartheta}{\sqrt{p^2 c^2 +m^2 c^4}} \bigl[
                              \vec{\zeta}(0)\cdot\hat{p}\bigr]
                        -(\mathcal{H}_{\rm m}')^{-1} c\,p\Bigr\} \dfrac{1}{
                              \mathcal{H}_{\rm m}'} \Bigl({\rm e}^{-{\rm i}\frac{2}{
                                    \hbar} \mathcal{H}_{\rm m}' t} -1\Bigr)
                        \biggr]|\Psi'''_3 \rangle \nonumber \\
                  &=& -\langle\Psi'''_1| \biggr[{\rm i}\frac{c\,\hbar}{2} \Bigl\{
                        \cos\vartheta\,\big[\vec{\alpha}(0)\cdot\hat{p}\big]
                        +\frac{p\,c\,\sin\vartheta}{\sqrt{p^2 c^2 +m^2 c^4}} \bigl[
                              \vec{\zeta}(0)\cdot\hat{p}\bigr]
                        -(\mathcal{H}_{\rm m}')^{-1} c\,p\Bigr\}\biggr]|\Psi'''_3\rangle
                        \frac{1}{E_{\rm m}} \left({\rm e}^{
                              \frac{{\rm i}\,2\,{E}_{\rm m}t}{\hbar}}-1\right) \notag \\
                  &=& \cos\vartheta\bigg[{\rm i}\frac{\hbar\,m\,c^3}{2\,{E}_{\rm m}^2}
                        \Big({\rm e}^{\frac{{\rm i}\,2\,{E}_{\rm m}t}{\hbar}}-1\Big)
                        \bigg].
            \end{eqnarray}
            Therefore the expectation value is
            \begin{eqnarray}
                   {\mathcal{Z}}_{\rm m} &=&\sin(2\eta){\rm Re}\left(
                        \langle\Psi'''_1|\hat{\mathcal{Z}}_{\rm m}|\Psi'''_3 \rangle\right)
                  =\cos\vartheta \sin(2\,\eta) {\rm Re}\bigg[{\rm i}
                        \frac{\hbar\,m\,c^3}{2\,{E}_{\rm m}^2} \Big({\rm e}^{
                              \frac{{\rm i}\,2\,{E}_{\rm m}t}{\hbar}}-1\Big)
                        \bigg] \notag \\
                  &=& \cos\vartheta \bigg[-\sin(2\,\eta)\:\frac{\hbar\,m\,c^3}
                        {2\,{E}_{\rm m}^2} \sin\Bigl(\frac{2\,{E}_{\rm m}t}{\hbar}
                        \Bigr)\bigg], \notag \\
                  &=& -A\,\sin(\omega\,t).
            \end{eqnarray}
            Based on which, we have the amplitude and frequency as
            \begin{eqnarray}
                  A&=& \cos\vartheta\,\sin(2\eta) \frac{\hbar c}{2 E_{\rm p}} \dfrac{mc^2}{
                        {E_{\rm p}}},\nonumber\\
                  \omega &=&\frac{2\,E_{\rm p}}{\hbar}.
            \end{eqnarray}
           One may notice that the amplitude of ``Zitterbewegung'' is tuned by a factor $\cos\vartheta$. When the mixing angle $\vartheta=0$ and $\vartheta=\pi/2$, then the above results reduce to those of Dirac's electron and the type-II Dirac's braidon, respectively.

\begin{remark}In the above, we have obtained the following differential equation for $\vec{r}(t)$, i.e.,
  \begin{eqnarray}\label{eq:DRDT-0}
                  \dfrac{{\rm d}\,\vec{r}}{{\rm d}\,t}
                  &=& \cos\vartheta\,(c\,\vec{\alpha})+\sin\vartheta\,\left[
                        \frac{\sqrt{p^2c^2+m^2c^4}}{p} \vec{\zeta}
                        -\frac{m^2c^4}{\sqrt{p^2c^2+m^2c^4}} (\vec{\zeta}
                              \cdot\vec{p})\;\frac{\vec{p}}{p^3}\right].
            \end{eqnarray}
However, we have not yet obtained directly the result for $\vec{r}(t)$ itself. Indeed, we have had the result of
the operator $\vec{r}(t)\cdot\hat{p}$, which is given in Eq. (\ref{eq:RtDotP}), i.e.,
 \begin{eqnarray}\label{eq:RtDotP-1}
                  \vec{r}(t)\cdot\hat{p}
                  &=& \vec{r}(0)\cdot\hat{p}+c^2 (\mathcal{H}_{\rm m}')^{-1} p\,t
                        \notag \\
                        && +{\rm i}\frac{c\,\hbar}{2} \Bigl\{
                              \cos\vartheta\,\big[\vec{\alpha}(0)\cdot\hat{p}\big]
                              +\frac{p\,c\,\sin\vartheta}{\sqrt{p^2 c^2 +m^2 c^4}}
                                    \bigl[\vec{\zeta}(0)\cdot\hat{p}\bigr]
                              -(\mathcal{H}_{\rm m}')^{-1} c\,p\Bigr\}
                                    \dfrac{1}{\mathcal{H}_{\rm m}'} \Bigl({\rm e}^{
                                          -{\rm i}\frac{2}{\hbar}
                                                \mathcal{H}_{\rm m}' t} -1\Bigr).
            \end{eqnarray}
Here, based on Eq. (\ref{eq:RtDotP-1}) we try to guess the result of $\vec{r}(t)$. We may recast Eq. (\ref{eq:RtDotP-1}) to the following form
\begin{eqnarray}\label{eq:RtDotP-1b}
                  \vec{r}(t)\cdot\hat{p}
                  &=& \biggr[\vec{r}(0)+c^2 (\mathcal{H}_{\rm m}')^{-1} \vec{p}\,t
                        \notag \\
                        && +{\rm i}\frac{c\,\hbar}{2} \Bigl\{
                              \cos\vartheta\,\vec{\alpha}(0)
                              +\frac{p\,c\,\sin\vartheta}{\sqrt{p^2 c^2 +m^2 c^4}}
                                    \vec{\zeta}(0)
                              -(\mathcal{H}_{\rm m}')^{-1} c\,\vec{p}\Bigr\}
                                    \dfrac{1}{\mathcal{H}_{\rm m}'} \Bigl({\rm e}^{
                                          -{\rm i}\frac{2}{\hbar}
                                                \mathcal{H}_{\rm m}' t} -1\Bigr)\biggr]\cdot\hat{p},
            \end{eqnarray}
then we may set $\vec{r}(t)$ as the following form
\begin{eqnarray}\label{eq:RtDotP-1c}
\vec{r}(t) &=& \vec{r}_{\parallel}(t)+\vec{r}_{\perp}(t),
            \end{eqnarray}
with
\begin{eqnarray}\label{eq:RtDotP-1d}
\vec{r}_{\parallel}(t)
                  &=& \biggr[\vec{r}(0)+c^2 (\mathcal{H}_{\rm m}')^{-1} \vec{p}\,t
                         +{\rm i}\frac{c\,\hbar}{2} \Bigl\{
                              \cos\vartheta\,\vec{\alpha}(0)
                              +\frac{p\,c\,\sin\vartheta}{\sqrt{p^2 c^2 +m^2 c^4}}
                                    \vec{\zeta}(0)
                              -(\mathcal{H}_{\rm m}')^{-1} c\,\vec{p}\Bigr\}
                                    \dfrac{1}{\mathcal{H}_{\rm m}'} \Bigl({\rm e}^{
                                          -{\rm i}\frac{2}{\hbar}
                                                \mathcal{H}_{\rm m}' t} -1\Bigr)\biggr],
\end{eqnarray}
and
\begin{eqnarray}\label{eq:RtDotP-1e}
\vec{r}_{\perp}(t)\cdot\hat{p} &=& 0.
            \end{eqnarray}

From previous results, we have known that two extreme cases
\begin{subequations}
 \begin{eqnarray}
 && \vec{r}(t)|_{\vartheta=0}=\vec{r}(0)+c^2 H^{-1}_{\rm e} \vec{p}\,t  +\dfrac{{\rm i}\hbar c}{2}\left[\vec{\alpha}(0)
                         -c\,H^{-1}_{\rm e} \vec{p}\right] H^{-1}_{\rm e} \left(
                                    {\rm e}^{\frac{-{\rm i}\,2H_{\rm e}t}{\hbar}} -1\right), \label{eq:RtDotP-2a-a}\\
 && \vec{r}(t)|_{\vartheta=\pi/2}  \notag \\
 &=&\vec{r}(0)+c^2 (H_{\rm b}^{\rm II})^{-1}\vec{p}\;t \nonumber\\
                     &&+\frac{{\rm i}\hbar}{2} \; \Biggl\{
                        \frac{\sqrt{p^2c^2+m^2c^4}}{p} \vec{\zeta}(0)-  \frac{m^2c^4}{\sqrt{p^2c^2+m^2c^4}}\left[\vec{\zeta}(0)\cdot \vec{p}\right]\frac{\vec{p}}{p^3}- c^2 (H_{\rm b}^{\rm II})^{-1} \vec{p} \Biggr\} (H_{\rm b}^{\rm II})^{-1}
                        \left({\rm e}^{\frac{-{\rm i}\,2\, H_{\rm b}^{\rm II}\,t}{\hbar}}-1\right), \label{eq:RtDotP-2b}
             \end{eqnarray}
\end{subequations}
and then
\begin{subequations}
 \begin{eqnarray}
 \vec{r}_{\perp}(t)|_{\vartheta=0}&=& \vec{r}(t)|_{\vartheta=0}-\vec{r}_{\parallel}(t)|_{\vartheta=0}=0, \\
 \vec{r}(t)_{\perp}|_{\vartheta=\pi/2}  &=& \vec{r}(t)|_{\vartheta=\pi/2}-\vec{r}_{\parallel}(t)|_{\vartheta=\pi/2}\nonumber\\
  &=&\frac{{\rm i}\hbar}{2} \; \Biggl\{
                        \frac{\sqrt{p^2c^2+m^2c^4}}{p} \vec{\zeta}(0)-  \frac{m^2c^4}{\sqrt{p^2c^2+m^2c^4}}\left[\vec{\zeta}(0)\cdot \vec{p}\right]\frac{\vec{p}}{p^3}- c^2 (H_{\rm b}^{\rm II})^{-1} \vec{p} \Biggr\} (H_{\rm b}^{\rm II})^{-1}
                        \left({\rm e}^{\frac{-{\rm i}\,2\, H_{\rm b}^{\rm II}\,t}{\hbar}}-1\right)\nonumber\\
                        && -{\rm i}\frac{c\,\hbar}{2} \Bigl\{
                              \frac{p\,c}{\sqrt{p^2 c^2 +m^2 c^4}}
                                    \vec{\zeta}(0)
                              -(H_{\rm b}^{\rm II})^{-1} c\,\vec{p}\Bigr\}
                                    \dfrac{1}{H_{\rm b}^{\rm II}} \Bigl({\rm e}^{
                                          -{\rm i}\frac{2}{\hbar}
                                                H_{\rm b}^{\rm II} t} -1\Bigr) \nonumber\\
  &=&\frac{{\rm i}\hbar c}{2} \; \Biggl\{
                        \left[\frac{\sqrt{p^2c^2+m^2c^4}}{pc}-\frac{p\,c}{\sqrt{p^2 c^2 +m^2 c^4}}\right] \vec{\zeta}(0)-  \frac{m^2c^4}{\sqrt{p^2c^2+m^2c^4}}\left[\vec{\zeta}(0)\cdot \hat{p}\right]\frac{\hat{p}}{p\,c} \Biggr\} (H_{\rm b}^{\rm II})^{-1}
                        \left({\rm e}^{\frac{-{\rm i}\,2\, H_{\rm b}^{\rm II}\,t}{\hbar}}-1\right) \nonumber\\
  &=&\frac{{\rm i}\hbar c}{2} \; \Biggl\{
                        \frac{m^2c^4}{\sqrt{p^2c^2+m^2c^4}}\frac{1}{p\,c} \vec{\zeta}(0)-  \frac{m^2c^4}{\sqrt{p^2c^2+m^2c^4}} \frac{1}{p\,c}\left[\vec{\zeta}(0)\cdot \hat{p}\right]\hat{p} \Biggr\} (H_{\rm b}^{\rm II})^{-1}
                        \left({\rm e}^{\frac{-{\rm i}\,2\, H_{\rm b}^{\rm II}\,t}{\hbar}}-1\right)\nonumber\\
  &=&\frac{{\rm i}\hbar c }{2} \frac{m^2c^4}{\sqrt{p^2c^2+m^2c^4}}\frac{1}{p\,c}\; \Biggl\{
                         \vec{\zeta}(0)-  \left[\vec{\zeta}(0)\cdot \hat{p}\right]\hat{p} \Biggr\} (H_{\rm b}^{\rm II})^{-1}
                        \left({\rm e}^{\frac{-{\rm i}\,2\, H_{\rm b}^{\rm II}\,t}{\hbar}}-1\right).
             \end{eqnarray}
\end{subequations}
It is easy to check that
\begin{eqnarray}\label{eq:RtDotP-1f}
&&\left[\vec{r}_{\perp}(t)|_{\vartheta=0}\right]\cdot\hat{p} = 0,\nonumber\\
&&\left[\vec{r}_{\perp}(t)|_{\vartheta=\pi/2}\right]\cdot\hat{p} = 0.
\end{eqnarray}
Hence,  a very possible solution for $\vec{r}(t)_{\perp}$ reads
\begin{eqnarray}\label{eq:RtDotP-1g}
 \vec{r}(t)_{\perp} &=& \sin\vartheta\, \frac{{\rm i}\hbar c}{2} \frac{m^2c^4}{\sqrt{p^2c^2+m^2c^4}}\frac{1}{p\, c}\; \Biggl\{
                         \vec{\zeta}(0)-  \left[\vec{\zeta}(0)\cdot \hat{p}\right]\hat{p} \Biggr\} \dfrac{1}{\mathcal{H}_{\rm m}'} \Bigl({\rm e}^{
                                          -{\rm i}\frac{2}{\hbar}
                                                \mathcal{H}_{\rm m}' t} -1\Bigr).
             \end{eqnarray}
Then we have
\begin{eqnarray}\label{eq:RtDotP-1h}
\vec{r}(t) &=& \vec{r}_{\parallel}(t)+\vec{r}_{\perp}(t)\nonumber\\
&=& \biggr[\vec{r}(0)+c^2 (\mathcal{H}_{\rm m}')^{-1} \vec{p}\,t
                         +{\rm i}\frac{c\,\hbar}{2} \Bigl\{
                              \cos\vartheta\,\vec{\alpha}(0)
                              +\frac{p\,c\,\sin\vartheta}{\sqrt{p^2 c^2 +m^2 c^4}}
                                    \vec{\zeta}(0)
                              -(\mathcal{H}_{\rm m}')^{-1} c\,\vec{p}\Bigr\}
                                    \dfrac{1}{\mathcal{H}_{\rm m}'} \Bigl({\rm e}^{
                                          -{\rm i}\frac{2}{\hbar}
                                                \mathcal{H}_{\rm m}' t} -1\Bigr)\biggr]\nonumber\\
&&+ \sin\vartheta\, \frac{{\rm i}\hbar c }{2} \frac{m^2c^4}{\sqrt{p^2c^2+m^2c^4}}\frac{1}{p\, c}\; \Biggl\{
                         \vec{\zeta}(0)-  \left[\vec{\zeta}(0)\cdot \hat{p}\right]\hat{p} \Biggr\} \dfrac{1}{\mathcal{H}_{\rm m}'} \Bigl({\rm e}^{
                                          -{\rm i}\frac{2}{\hbar}
                                                \mathcal{H}_{\rm m}' t} -1\Bigr)\nonumber\\
&=& \vec{r}(0)+c^2 (\mathcal{H}_{\rm m}')^{-1} \vec{p}\,t
+\frac{{\rm i}\hbar c }{2} \biggr\{\cos\vartheta\,\vec{\alpha}(0)+\sin\vartheta\, \frac{\sqrt{p^2c^2+m^2c^4}}{p\,c} \vec{\zeta}(0) \nonumber\\
&&- \sin\vartheta\, \frac{m^2c^4}{\sqrt{p^2c^2+m^2c^4}}\frac{1}{p\, c}\;\left[\vec{\zeta}(0)\cdot \hat{p}\right]\hat{p} -(\mathcal{H}_{\rm m}')^{-1} c\,\vec{p}\biggr\} \dfrac{1}{\mathcal{H}_{\rm m}'} \Bigl({\rm e}^{
                                          -{\rm i}\frac{2}{\hbar}
                                                \mathcal{H}_{\rm m}' t} -1\Bigr).
            \end{eqnarray}
One then defines the third term of $\vec{r}$ as the ``position Zitterbewegung'' operator, which is given by
\begin{eqnarray}\label{eq:RtDotP-2a}
\hat{\mathcal{Z}}_{\rm m}
&=&
\frac{{\rm i}\hbar c }{2} \biggr\{\cos\vartheta\,\vec{\alpha}(0)+\sin\vartheta\, \frac{\sqrt{p^2c^2+m^2c^4}}{p\,c} \vec{\zeta}(0) \nonumber\\
&&- \sin\vartheta\, \frac{m^2c^4}{\sqrt{p^2c^2+m^2c^4}}\frac{1}{p\, c}\;\left[\vec{\zeta}(0)\cdot \hat{p}\right]\hat{p} -(\mathcal{H}_{\rm m}')^{-1} c\,\vec{p}\biggr\} \dfrac{1}{\mathcal{H}_{\rm m}'} \Bigl({\rm e}^{
                                          -{\rm i}\frac{2}{\hbar}
                                                \mathcal{H}_{\rm m}' t} -1\Bigr).
            \end{eqnarray}

$\blacksquare$
\end{remark}

\begin{remark} To ensure $\vec{r}(t)$ is an hermitian operator, one needs to prove
            \begin{eqnarray}
            && \mathcal{H}_{\rm m}'\; \hat{\mathcal{Z}}_{\rm m}+\hat{\mathcal{Z}}_{\rm m} \mathcal{H}_{\rm m}'=0,
            \end{eqnarray}
  i.e., to prove
            \begin{eqnarray}\label{eq:G-0}
                  \Bigl\{\mathcal{H}_{\rm m}',\ \cos\vartheta\,\vec{\alpha}(0)+\sin\vartheta\, \frac{\sqrt{p^2c^2+m^2c^4}}{p\,c} \vec{\zeta}(0) - \sin\vartheta\, \frac{m^2c^4}{\sqrt{p^2c^2+m^2c^4}}\frac{1}{p\, c}\;\left[\vec{\zeta}(0)\cdot \hat{p}\right]\hat{p} -(\mathcal{H}_{\rm m}')^{-1} c\,\vec{p}\Bigr\}=0.
            \end{eqnarray}
\begin{proof}
From previous results, we have known that
\begin{subequations}
\begin{eqnarray}
&&\mathcal{H}_{\rm m}'=\cos\vartheta\,H_{\rm e} +\sin\vartheta\,H_{\rm b}^{\rm II},\label{eq:G-1a}\\
&&\Big\{\mathcal{H}_{\rm m}',\ \bigl[\vec{\zeta}(0)\cdot\hat{p}
                        -(\mathcal{H}_{\rm m}')^{-1} \sin\vartheta\,\sqrt{
                              p^2 c^2 +m^2 c^4}\bigr]\Big\}
                                    = 0,\label{eq:G-1b}\\
&&  \Big\{\mathcal{H}_{\rm m}',\ \big[\vec{\alpha}(0)\cdot\hat{p}
                        -(\mathcal{H}_{\rm m}')^{-1} \cos\vartheta\,c\,p \big]
                        \Big\}  = 0, \label{eq:G-1c}\\
&&     \{H_{\rm e}, \left[\vec{\alpha}(0)-c H^{-1}_{\rm e}\vec{p}\right]\} =0, \label{eq:G-1d}\\
&&      \Biggl\{H_{\rm b}^{\rm II},\ \dfrac{\sqrt{p^2c^2+m^2c^4}}{p} \vec{\zeta}(0)
                                    -H_{\rm b}^{\rm II} \dfrac{\vec{p}}{p^2}\Biggr\}=0. \label{eq:G-1e}
\end{eqnarray}
\end{subequations}
Based on Eq. (\ref{eq:G-1b}), we have
\begin{eqnarray}\label{eq:G-2a}
&&\Big\{\mathcal{H}_{\rm m}',\ \vec{\zeta}(0)\cdot\hat{p}\Big\}=
                        \Big\{\mathcal{H}_{\rm m}',(\mathcal{H}_{\rm m}')^{-1} \sin\vartheta\,\sqrt{
                              p^2 c^2 +m^2 c^4}\Big\},
\end{eqnarray}
then Eq. (\ref{eq:G-0}) becomes
\begin{eqnarray}
               &&   \Bigl\{\mathcal{H}_{\rm m}',\ \cos\vartheta\,\vec{\alpha}(0)+\sin\vartheta\, \frac{\sqrt{p^2c^2+m^2c^4}}{p\,c} \vec{\zeta}(0) \nonumber\\
               &&- \sin\vartheta\, \frac{m^2c^4}{\sqrt{p^2c^2+m^2c^4}}\frac{1}{p\, c}\;\left[(\mathcal{H}_{\rm m}')^{-1} \sin\vartheta\,\sqrt{
                              p^2 c^2 +m^2 c^4}\right]\hat{p} -(\mathcal{H}_{\rm m}')^{-1} c\,\vec{p}\Bigr\}=0,
            \end{eqnarray}
i.e.,
\begin{eqnarray}
               &&   \Bigl\{\mathcal{H}_{\rm m}',\ \cos\vartheta\,\vec{\alpha}(0)+\sin\vartheta\, \frac{\sqrt{p^2c^2+m^2c^4}}{p\,c} \vec{\zeta}(0) - \sin^2\vartheta\, \frac{m^2c^4}{p\, c}\;(\mathcal{H}_{\rm m}')^{-1} \hat{p} -(\mathcal{H}_{\rm m}')^{-1} c\,\vec{p}\Bigr\}=0,
            \end{eqnarray}
i.e.,
\begin{eqnarray}
               &&   \Bigl\{\mathcal{H}_{\rm m}',\ \cos\vartheta\,\vec{\alpha}(0)+\sin\vartheta\, \frac{\sqrt{p^2c^2+m^2c^4}}{p\,c} \vec{\zeta}(0) - \sin^2\vartheta\, \frac{m^2c^4}{p^2\, c^2}\;(\mathcal{H}_{\rm m}')^{-1} c\, \vec{p} -(\mathcal{H}_{\rm m}')^{-1} c\,\vec{p}\Bigr\}=0,
\end{eqnarray}
i.e.,
\begin{eqnarray}
               &&   \Bigl\{\mathcal{H}_{\rm m}',\ \cos\vartheta\,\vec{\alpha}(0)+\sin\vartheta\, \frac{\sqrt{p^2c^2+m^2c^4}}{p\,c} \vec{\zeta}(0) - \left[\sin^2\vartheta\, \frac{m^2c^4}{p^2\, c^2}+1\right]\;(\mathcal{H}_{\rm m}')^{-1} c\, \vec{p}\Bigr\}=0,
\end{eqnarray}
i.e.,
\begin{eqnarray}
               &&   \Bigl\{\mathcal{H}_{\rm m}',\ \cos\vartheta\,\vec{\alpha}(0)+\sin\vartheta\, \frac{\sqrt{p^2c^2+m^2c^4}}{p\,c} \vec{\zeta}(0)\Bigr\} - 2\, \left[\sin^2\vartheta\, \frac{m^2c^4}{p^2\, c^2}+1\right]\; c\, \vec{p}=0,
\end{eqnarray}
i.e.,
\begin{eqnarray}\label{eq:G-2b}
               &&   \Bigl\{\cos\vartheta\,H_{\rm e} +\sin\vartheta\,H_{\rm b}^{\rm II},\ \cos\vartheta\,\vec{\alpha}(0)+\sin\vartheta\, \frac{\sqrt{p^2c^2+m^2c^4}}{p\,c} \vec{\zeta}(0)\Bigr\} - 2\, \left[\sin^2\vartheta\, \frac{m^2c^4}{p^2\, c^2}+1\right]\; c\, \vec{p}=0.
\end{eqnarray}

One can have
\begin{eqnarray}\label{eq:G-3a}
               &&   \Bigl\{\cos\vartheta\,H_{\rm e},\ \cos\vartheta\,\vec{\alpha}(0)+\sin\vartheta\, \frac{\sqrt{p^2c^2+m^2c^4}}{p\,c} \vec{\zeta}(0)\Bigr\}=  \cos^2\vartheta\, \Bigl\{H_{\rm e}, \vec{\alpha}(0)\Bigr\}+\sin\vartheta\cos\vartheta\, \frac{\sqrt{p^2c^2+m^2c^4}}{p\,c}\Bigl\{H_{\rm e}, \vec{\zeta}(0)\Bigr\}\nonumber\\
&=& \cos^2\vartheta\, \Bigl\{H_{\rm e}, c H^{-1}_{\rm e}\vec{p}\Bigr\} +\sin\vartheta\cos\vartheta\, \frac{\sqrt{p^2c^2+m^2c^4}}{p\,c}\Bigl\{H_{\rm e}, \vec{\zeta}(0)\Bigr\}\nonumber\\
&=& 2 \cos^2\vartheta\,c \vec{p}+ \sin\vartheta\cos\vartheta\, \frac{\sqrt{p^2c^2+m^2c^4}}{p\,c}\Bigl\{H_{\rm e}, \vec{\zeta}(0)\Bigr\},
\end{eqnarray}
and
\begin{eqnarray}\label{eq:G-3b}
               &&   \Bigl\{\sin\vartheta\,H_{\rm b}^{\rm II},\ \cos\vartheta\,\vec{\alpha}(0)+\sin\vartheta\, \frac{\sqrt{p^2c^2+m^2c^4}}{p\,c} \vec{\zeta}(0)\Bigr\}= \sin^2\vartheta\,\frac{1}{c} \Bigl\{H_{\rm b}^{\rm II},\frac{\sqrt{p^2c^2+m^2c^4}}{p}\vec{\zeta}(0)\Bigr\}+
               \sin\vartheta\cos\vartheta \Bigl\{H_{\rm b}^{\rm II},\ \vec{\alpha}(0)\Bigr\}\nonumber\\
&=&\sin^2\vartheta\,\frac{1}{c} \Bigl\{H_{\rm b}^{\rm II},H_{\rm b}^{\rm II} \dfrac{\vec{p}}{p^2}\Bigr\}+
               \sin\vartheta\cos\vartheta \Bigl\{H_{\rm b}^{\rm II},\ \vec{\alpha}(0)\Bigr\}=
               2 \sin^2\vartheta\,\frac{1}{p^2 c^2} (H_{\rm b}^{\rm II})^2 c \vec{p}+
               \sin\vartheta\cos\vartheta \Bigl\{H_{\rm b}^{\rm II},\ \vec{\alpha}(0)\Bigr\}\nonumber\\
&=& 2 \sin^2\vartheta\,\frac{p^2c^2+m^2c^4}{p^2 c^2}  c \vec{p}+
               \sin\vartheta\cos\vartheta \Bigl\{H_{\rm b}^{\rm II},\ \vec{\alpha}(0)\Bigr\}= 2 \sin^2\vartheta\,\left[1+\frac{m^2c^4}{p^2 c^2} \right] c \vec{p}+
               \sin\vartheta\cos\vartheta \Bigl\{H_{\rm b}^{\rm II},\ \vec{\alpha}(0)\Bigr\}.
\end{eqnarray}
After substituting Eq. (\ref{eq:G-3a}) and Eq. (\ref{eq:G-3b}) into Eq. (\ref{eq:G-2b}), we obtain
\begin{eqnarray}\label{eq:G-4}
               &&   \sin\vartheta\cos\vartheta\, \left(\frac{\sqrt{p^2c^2+m^2c^4}}{p\,c}\Bigl\{H_{\rm e}, \vec{\zeta}(0)\Bigr\}+\Bigl\{H_{\rm b}^{\rm II},\ \vec{\alpha}(0)\Bigr\}\right)=0.
\end{eqnarray}
Since we need to prove Eq. (\ref{eq:G-4}) is valid for any $\vartheta$, thus we need to prove
\begin{eqnarray}\label{eq:G-5}
               &&   \frac{\sqrt{p^2c^2+m^2c^4}}{p\,c}\Bigl\{H_{\rm e}, \vec{\zeta}(0)\Bigr\}+\Bigl\{H_{\rm b}^{\rm II},\ \vec{\alpha}(0)\Bigr\}=0,
\end{eqnarray}
i.e.,
\begin{eqnarray}
               &&   \frac{\sqrt{p^2c^2+m^2c^4}}{p\,c}\Bigl\{c \vec{\alpha}(0)\cdot\vec{p}+\beta(0)\,m c^2, {\rm i}\beta(0)\vec{\alpha}(0)\Bigr\}+\Bigl\{\dfrac{\sqrt{p^2c^2+m^2c^4}}{p}\;[{\rm i}\beta(0)\vec{\alpha}(0)\cdot\vec{p}],\ \vec{\alpha}(0)\Bigr\}=0,
\end{eqnarray}
i.e.,
\begin{eqnarray}
               &&   \Bigl\{ \vec{\alpha}(0)\cdot\vec{p}+\beta(0)\,m c, \beta(0)\vec{\alpha}(0)\Bigr\}+\Bigl\{\beta(0)[\vec{\alpha}(0)\cdot\vec{p}],\ \vec{\alpha}(0)\Bigr\}=0,
\end{eqnarray}
i.e.,
\begin{eqnarray}
               &&   \Bigl\{ \vec{\alpha}(0)\cdot\vec{p}, \beta(0)\vec{\alpha}(0)\Bigr\}+\Bigl\{\beta(0)[\vec{\alpha}(0)\cdot\vec{p}],\ \vec{\alpha}(0)\Bigr\}=0,
\end{eqnarray}
i.e.,
\begin{eqnarray}
               &&    [\vec{\alpha}(0)\cdot\vec{p}]\, \beta(0)\vec{\alpha}(0) + \beta(0)\vec{\alpha}(0)\,[\vec{\alpha}(0)\cdot\vec{p}]
               +\beta(0)[\vec{\alpha}(0)\cdot\vec{p}]\, \vec{\alpha}(0)+ \vec{\alpha}(0)\, \beta(0)[\vec{\alpha}(0)\cdot\vec{p}]=0
\end{eqnarray}
i.e.,
\begin{eqnarray}\label{eq:G-6}
               &&    -\beta(0)\, [\vec{\alpha}(0)\cdot\vec{p}]\, \vec{\alpha}(0) + \beta(0)\vec{\alpha}(0)\,[\vec{\alpha}(0)\cdot\vec{p}]
               +\beta(0)[\vec{\alpha}(0)\cdot\vec{p}]\, \vec{\alpha}(0)- \beta(0)\,\vec{\alpha}(0)\, [\vec{\alpha}(0)\cdot\vec{p}]=0.
\end{eqnarray}
Obviously, Eq. (\ref{eq:G-6}) is valid, thus Eq. (\ref{eq:G-5}) is valid. Thus, we have proved that the operator $\vec{r}$ is hermitian. This implies that the results in Eq. (\ref{eq:RtDotP-1h}) and Eq. (\ref{eq:RtDotP-2a}) are correct. This ends the proof.
\end{proof}
$\blacksquare$
\end{remark}

\begin{remark}Based on above results, we have
                  \begin{eqnarray}\label{eq:RtDotP-1hh}
                        \vec{r}(t) &=& \vec{r}(0)+c^2 (\mathcal{H}_{\rm m}')^{-1} \vec{p}\,t
                              +\frac{{\rm i}\,\hbar\,c}{2} \biggr\{\cos\vartheta\,\vec{\alpha}(0)
                                    +\sin\vartheta\,\frac{\sqrt{p^2 c^2 +m^2 c^4}}{p\,c}
                                          \vec{\zeta}(0) \nonumber\\
                                    && -\sin\vartheta\,\frac{m^2c^4}{\sqrt{p^2c^2+m^2c^4}}
                                          \frac{1}{p\,c} \;\left[\vec{\zeta}(0)\cdot\hat{p}
                                                \right]\hat{p}
                                    -(\mathcal{H}_{\rm m}')^{-1} c\,\vec{p}
                                    \biggr\}\dfrac{1}{\mathcal{H}_{\rm m}'} \Bigl({\rm e}^{
                                          -{\rm i}\frac{2}{\hbar} \mathcal{H}_{\rm m}'\,t} -1\Bigr).
                  \end{eqnarray}

                  \begin{eqnarray}
            \hat{\mathcal{Z}}_{\rm m}
            &=&
            \frac{{\rm i}\hbar c }{2} \biggr\{\cos\vartheta\,\vec{\alpha}(0)+\sin\vartheta\, \frac{\sqrt{p^2c^2+m^2c^4}}{p\,c} \vec{\zeta}(0) \nonumber\\
            &&- \sin\vartheta\, \frac{m^2c^4}{\sqrt{p^2c^2+m^2c^4}}\frac{1}{p\, c}\;\left[\vec{\zeta}(0)\cdot \hat{p}\right]\hat{p} -(\mathcal{H}_{\rm m}')^{-1} c\,\vec{p}\biggr\} \dfrac{1}{\mathcal{H}_{\rm m}'} \Bigl({\rm e}^{
                                                      -{\rm i}\frac{2}{\hbar}
                                                            \mathcal{H}_{\rm m}' t} -1\Bigr).
                  \end{eqnarray}
            When $\vartheta=0$, Eq. (\ref{eq:RtDotP-1hh}) reduces to the case of Dirac's electron as shown in Eq. (\ref{eq:D-10}), i.e.,
                  \begin{eqnarray}
                        \vec{r}(t)&=&\vec{r}(0)+c^2 H^{-1}_{\rm e} \vec{p}\,t  +\dfrac{{\rm i}\hbar c}{2}\left[\vec{\alpha}(0)
                              -c\,H^{-1}_{\rm e} \vec{p}\right] H^{-1}_{\rm e} \left(
                                          {\rm e}^{\frac{-{\rm i}\,2H_{\rm e}t}{\hbar}} -1\right).
                  \end{eqnarray}
                  and when $\vartheta=\pi/2$, Eq. (\ref{eq:RtDotP-1hh}) reduces to the case of type-II Dirac's braidon as shown in Eq. (\ref{eq:RtHb2}), i.e.,
                  \begin{eqnarray}
                        \vec{r}(t) &=& \vec{r}(0)+c^2 (H_{\rm b}^{\rm II})^{-1}\vec{p}\;t \nonumber\\
                        &&   +\frac{{\rm i}\hbar}{2} \; \Biggl\{
                              \frac{\sqrt{p^2c^2+m^2c^4}}{p} \vec{\zeta}(0)-  \frac{m^2c^4}{\sqrt{p^2c^2+m^2c^4}}\left[\vec{\zeta}(0)\cdot \vec{p}\right]\frac{\vec{p}}{p^3}- c^2 (H_{\rm b}^{\rm II})^{-1} \vec{p} \Biggr\} (H_{\rm b}^{\rm II})^{-1}
                              \left({\rm e}^{\frac{-{\rm i}\,2\, H_{\rm b}^{\rm II}\,t}{\hbar}}-1\right).
                  \end{eqnarray}
                  Because $\left\{\mathcal{H}_{\rm m}', \hat{\mathcal{Z}}_{\rm m}\right\}=\mathcal{H}_{\rm m}'\; \hat{\mathcal{Z}}_{\rm m}+\hat{\mathcal{Z}}_{\rm m} \mathcal{H}_{\rm m}'=0$, similarly one can prove that $\langle \Psi'''| \hat{\mathcal{Z}}_{\rm m}|\Psi'''\rangle=0$ if $|\Psi'''\rangle$ is a superposition of only positive energy (or only negative energy) state.
                  $\blacksquare$
            \end{remark}

             \begin{remark}
            Now suppose the mixed Dirac's Hamiltonian system is in a superposition state of $|\Psi'''_1\rangle$ and $|\Psi'''_3\rangle$, i.e.,
            \begin{eqnarray}
            &&   |\Psi'''\rangle=\cos\eta |\Psi'''_1\rangle+ \sin\eta |\Psi'''_3\rangle.
            \end{eqnarray}
            To calculate the Zitterbewegung, we only need to consider the term $\langle\Psi'''_1|\hat{\mathcal{Z}}_{\rm m}|\Psi'''_3 \rangle$. We can have
            \begin{eqnarray}
            &&  {\mathcal{Z}}_{\rm m}= \sin(2\eta){\rm Re}\left(
                  \langle\Psi'''_1|\hat{\mathcal{Z}}_{\rm m}|\Psi'''_3 \rangle\right),
            \end{eqnarray}
            where
            \begin{eqnarray}\label{eq:ZZZ-1a}
                   \langle\Psi'''_1|\hat{\mathcal{Z}}_{\rm m}|\Psi'''_3 \rangle
                  &=& \frac{{\rm i}\,\hbar\,c}{2} \biggr\{
                        \cos\vartheta\,\langle\Psi'''_1|\vec{\alpha}(0)
                        +\sin\vartheta\,\frac{\sqrt{p^2 c^2 +m^2 c^4}}{p\,c}
                              \langle\Psi'''_1|\vec{\zeta}(0) \nonumber \\
                        && -\sin\vartheta\,\frac{m^2 c^4}{\sqrt{p^2 c^2 +m^2 c^4}}
                              \frac{1}{p\, c} \langle\Psi'''_1|\left[
                                    \vec{\zeta}(0)\cdot\hat{p}\right]\hat{p}
                        -(\mathcal{H}_{\rm m}')^{-1} c\langle\Psi'''_1|\vec{p}
                        \biggr\}\dfrac{1}{\mathcal{H}_{\rm m}'} \Bigl({\rm e}^{
                              -{\rm i}\frac{2}{\hbar} \mathcal{H}_{\rm m}' t} -1
                              \Bigr)|\Psi'''_3\rangle \notag \\
                  &=& -\frac{{\rm i}\,\hbar\,c}{2} \biggr\{
                        \cos\vartheta\,\langle\Psi'''_1|\vec{\alpha}(0)|\Psi'''_3\rangle
                        +\sin\vartheta\,\frac{\sqrt{p^2 c^2 +m^2 c^4}}{p\,c}
                              \langle\Psi'''_1|\vec{\zeta}(0)|\Psi'''_3\rangle\nonumber\\
                      &&  -\sin\vartheta\,\frac{m^2 c^4}{\sqrt{p^2 c^2 +m^2 c^4}}
                              \frac{1}{p\, c} \langle\Psi'''_1|\left[
                                    \vec{\zeta}(0)\cdot\hat{p}
                                    \right]|\Psi'''_3\rangle\hat{p}  -E_{\rm m}^{-1} c\langle\Psi'''_1|\vec{p}|
                              \Psi'''_3\rangle\biggr\}\dfrac{1}{E_{\rm m}} \Bigl(
                                    {\rm e}^{{\rm i}\frac{2}{\hbar} E_{\rm m} t}
                                          -1\Bigr) \notag \\
                  &=& -\frac{{\rm i}\,\hbar\,c}{2} \biggr\{
                        \cos\vartheta\,\langle\Psi'''_1|\vec{\alpha}(0)|\Psi'''_3\rangle
                        +\sin\vartheta\,\frac{\sqrt{p^2 c^2 +m^2 c^4}}{p\,c}
                              \langle\Psi'''_1|\vec{\zeta}(0)|\Psi'''_3\rangle
                              \nonumber \\
                        &&\qquad\qquad -\sin\vartheta\,\frac{m^2 c^4}{
                              \sqrt{p^2 c^2 +m^2 c^4}} \frac{1}{p\, c} \langle\Psi'''_1|
                                    \left[\vec{\zeta}(0)\cdot\hat{p}
                                          \right]|\Psi'''_3\rangle\hat{p}
                        \biggr\}\dfrac{1}{E_{\rm m}} \Bigl({\rm e}^{
                              {\rm i}\frac{2}{\hbar} E_{\rm m} t} -1\Bigr),
            \end{eqnarray}
            where in the last step we have used $\langle\Psi'''_1|\Psi'''_3 \rangle=0$.
            One can have
            \begin{eqnarray}
                  \mathcal{D}(\vartheta) &=& {\rm e}^{{\rm i} \frac{\vartheta}{2} \Gamma_y}=\cos\Bigl(\frac{\vartheta}{2}\Bigr)\;\mathbb{I}+{\rm i}\,\sin\Bigl(
                        \frac{\vartheta}{2}\Bigr)\,\dfrac{(-m\,c^2 \vec{\alpha}
                              \cdot\hat{p}+\beta\,p\,c)}{\sqrt{p^2 c^2 +m^2 c^4}},\nonumber\\
                    \mathcal{D}^\dagger (\vartheta) &=& {\rm e}^{
                        -{\rm i} \frac{\vartheta}{2}  \Gamma_y}
                  =\cos\Bigl(\frac{\vartheta}{2}\Bigr)\;\mathbb{I}-{\rm i}\,\sin\Bigl(
                        \frac{\vartheta}{2}\Bigr)\,\dfrac{(-m\,c^2 \vec{\alpha}
                              \cdot\hat{p}+\beta\,p\,c)}{\sqrt{p^2 c^2 +m^2 c^4}},
            \end{eqnarray}
            \begin{eqnarray}
                   \big[\vec{\alpha},\ (\vec{\alpha}\cdot\hat{p})\big]
                  &=&\big\{\vec{\alpha},\ (\vec{\alpha}\cdot\hat{p})\big\}
                        -2(\vec{\alpha}\cdot\hat{p})\vec{\alpha}
                  =\bigg\{\Bigl(\sum_u \vec{e}_u \alpha_u\Bigr),\ \Bigl(\sum_v
                              \alpha_v \hat{p}_v\Bigr)\bigg\}
                        -2(\vec{\alpha}\cdot\hat{p})\vec{\alpha} \notag \\
                  &=& \sum_{u,v} \vec{e}_u \hat{p}_v \{\alpha_u,\ \alpha_v\}
                        -2(\vec{\alpha}\cdot\hat{p})\vec{\alpha}
                  =2\sum_{u,v} \vec{e}_u \hat{p}_v \delta_{uv}
                        -2(\vec{\alpha}\cdot\hat{p})\vec{\alpha} \notag \\
                  &=& 2\,\hat{p}-2(\vec{\alpha}\cdot\hat{p})\vec{\alpha}.
            \end{eqnarray}

            \begin{eqnarray}
                  \big[\vec{\alpha}(0),\ (-m\,c^2 \vec{\alpha}\cdot\hat{p}+\beta\,p\,c)
                        \big]
                  &=& -m\,c^2 \big[\vec{\alpha},\ (\vec{\alpha}\cdot\hat{p})\big]
                        +p\,c[\vec{\alpha},\ \beta] \notag \\
                  &=& -m\,c^2 \big[2\,\hat{p}-2(\vec{\alpha}\cdot\hat{p})\vec{\alpha}
                              \big]
                        +p\,c\big[\{\vec{\alpha},\ \beta\}-2\,\beta\,\vec{\alpha}\big]
                        \notag \\
                  &=& -m\,c^2 \big[2\,\hat{p}-2(\vec{\alpha}\cdot\hat{p})\vec{\alpha}
                        \big]-2\,p\,c\,\beta\,\vec{\alpha} \notag \\
                  &=& 2\Bigl\{\big[m\,c^2 (\vec{\alpha}\cdot\hat{p})
                        -p\,c\,\beta\big]\vec{\alpha}-m\,c^2 \hat{p}\Bigr\} \notag \\
                  &=& -2(H_{\rm b}^{\rm I} \vec{\alpha}+m\,c^2 \hat{p}).
            \end{eqnarray}
            Note that
            \begin{eqnarray}
                   \big[\vec{\alpha}(0)\cdot\hat{p},\ (
                        -m\,c^2 \vec{\alpha}\cdot\hat{p}+\beta\,p\,c)\big]
                  &=&\big[\vec{\alpha}(0),\ (-m\,c^2 \vec{\alpha}\cdot\hat{p}+\beta\,p\,c)
                        \big]\cdot\hat{p}
                  =2\Bigl\{\big[m\,c^2 (\vec{\alpha}\cdot\hat{p})
                        -p\,c\,\beta\big]\vec{\alpha}-m\,c^2 \hat{p}\Bigr\}\cdot\hat{p}
                        \notag \\
                 & =& 2\Bigl\{\big[m\,c^2 (\vec{\alpha}\cdot\hat{p})
                        -p\,c\,\beta\big](\vec{\alpha}\cdot\hat{p})-m\,c^2\Bigr\}
                  =-2\,p\,c\,\beta(\vec{\alpha}\cdot\hat{p}) \notag \\
                  &=& {\rm i}\,2\dfrac{p\,c}{\sqrt{p^2 c^2 +m^2 c^4}} H_{\rm b}^{\rm II},
            \end{eqnarray}
            which coincides with the first equation in \Eq{eq:AlphaP}. Moreover,
            \begin{eqnarray}
                  && (-m\,c^2 \vec{\alpha}\cdot\hat{p}+\beta\,p\,c)\vec{\alpha}(0)\,(
                        -m\,c^2 \vec{\alpha}\cdot\hat{p}+\beta\,p\,c)
                  =(-m\,c^2 \vec{\alpha}\cdot\hat{p}+\beta\,p\,c)\vec{\alpha}(
                        -m\,c^2 \vec{\alpha}\cdot\hat{p}+\beta\,p\,c) \notag \\
                  &=& m^2 c^4 (\vec{\alpha}\cdot\hat{p})\vec{\alpha}(
                              \vec{\alpha}\cdot\hat{p})
                        -m\,c^2 p\,c\big[(\vec{\alpha}\cdot\hat{p})\vec{\alpha}\,\beta
                              +\beta\,\vec{\alpha}(\vec{\alpha}\cdot\hat{p})\big]
                        +p^2 c^2 (\beta\,\vec{\alpha}\,\beta) \nonumber \\
                  &=& m^2 c^4 (\vec{\alpha}\cdot\hat{p})\Big(
                              \big\{\vec{\alpha},\ (\vec{\alpha}\cdot\hat{p})\big\}
                              -(\vec{\alpha}\cdot\hat{p})\vec{\alpha}\Big)
                        -m\,c^2 p\,c\,\beta\big\{\vec{\alpha},\ (
                              \vec{\alpha}\cdot\hat{p})\big\}-p^2 c^2 \vec{\alpha}
                        \nonumber \\
                  &=& m^2 c^4 (\vec{\alpha}\cdot\hat{p})\big[2\,\hat{p}
                              -(\vec{\alpha}\cdot\hat{p})\vec{\alpha}\big]
                        -2\,m\,c^2 p\,c\,\beta\,\hat{p}-p^2 c^2 \vec{\alpha}
                  =-2\,m\,c^2 \big[-m\,c^2 (\vec{\alpha}\cdot\hat{p})+p\,c\,\beta
                        \big]\hat{p}-(p^2 c^2 +m^2 c^4)\vec{\alpha} \nonumber \\
                  &=& -2\,m\,c^2 \big[-m\,c^2 (\vec{\alpha}\cdot\hat{p})+p\,c\,\beta
                        \big]\hat{p}-(p^2 c^2 +m^2 c^4)\vec{\alpha} \nonumber \\
                  &=& -2\,m\,c^2 H_{\rm b}^{\rm I}\,\hat{p}-E_{\rm m}^2 \vec{\alpha},
            \end{eqnarray}
            then we have
            \begin{eqnarray}
                  && (-m\,c^2 \vec{\alpha}\cdot\hat{p}+\beta\,p\,c)\bigl[\vec{\alpha}(0)
                        \cdot\hat{p}\bigr](-m\,c^2 \vec{\alpha}\cdot\hat{p}+\beta\,p\,c)
                  =\big[(-m\,c^2 \vec{\alpha}\cdot\hat{p}+\beta\,p\,c)\vec{\alpha}\,(
                        -m\,c^2 \vec{\alpha}\cdot\hat{p}+\beta\,p\,c)\big]\cdot\hat{p}
                        \notag \\
                  &=& -2\,m\,c^2 \big[-m\,c^2 (\vec{\alpha}\cdot\hat{p})+p\,c\,\beta
                        \big]-(p^2 c^2 +m^2 c^4)(\vec{\alpha}\cdot\hat{p})
                  =m^2 c^4 (\vec{\alpha}\cdot\hat{p})-2\,m\,c^2 p\,c\,\beta
                        -p^2 c^2 (\vec{\alpha}\cdot\hat{p}) \notag \\
                  &=& (m^2 c^4 +p^2 c^2)(\vec{\alpha}\cdot\hat{p})-2\,m\,c^2 p\,c\,\beta
                        -2\,p^2 c^2 (\vec{\alpha}\cdot\hat{p}) = (m^2 c^4 +p^2 c^2)(\vec{\alpha}\cdot\hat{p})
                        -2\,p\,c\bigl[m\,c^2 \beta+p\,c(\vec{\alpha}\cdot\hat{p})\bigr]
                        \notag \\
                  &=& (p^2 c^2 +m^2 c^4)(\vec{\alpha}\cdot\hat{p})-2\,p\,c\,H_{\rm e},
            \end{eqnarray}
            which coincides with the second equation in \Eq{eq:AlphaP}.

            We then have
            \begin{eqnarray}\label{eq:DAlphaPD-aa}
                  && \mathcal{D}^\dagger (\vartheta)\,\vec{\alpha}(0)\,\mathcal{D}(
                        \vartheta) \notag \\
                  &=& \bigg[\cos\Bigl(\frac{\vartheta}{2}\Bigr)\;\mathbb{I}
                        -{\rm i}\,\sin\Bigl(\frac{\vartheta}{2}\Bigr)
                              \dfrac{(-m\,c^2 \vec{\alpha}\cdot\hat{p}+\beta\,p\,c)}{
                                    \sqrt{p^2 c^2 +m^2 c^4}}
                        \bigg]\vec{\alpha}(0)\,\bigg[
                              \cos\Bigl(\frac{\vartheta}{2}\Bigr)\;\mathbb{I}
                              +{\rm i}\,\sin\Bigl(\frac{\vartheta}{2}\Bigr)
                                    \dfrac{(-m\,c^2 \vec{\alpha}\cdot\hat{p}
                                          +\beta\,p\,c)}{\sqrt{p^2 c^2 +m^2 c^4}}\bigg]
                        \notag \\
                  &=& \cos^2 \Bigl(\frac{\vartheta}{2}\Bigr)\,\vec{\alpha}(0)
                        +{\rm i}\,\sin\Bigl(\frac{\vartheta}{2}\Bigr)\cos\Bigl(
                              \frac{\vartheta}{2}\Bigr)\dfrac{1}{
                                    \sqrt{p^2 c^2 +m^2 c^4}} \big[\vec{\alpha}(0),\ (
                                          -m\,c^2 \vec{\alpha}\cdot\hat{p}+\beta\,p\,c)
                                          \big] \notag \\
                        && +\sin^2 \Bigl(\frac{\vartheta}{2}\Bigr)\dfrac{1}{
                              (p^2 c^2 +m^2 c^4)} (-m\,c^2 \vec{\alpha}\cdot\hat{p}
                                    +\beta\,p\,c)\,\vec{\alpha}(0)\,(
                                          -m\,c^2 \vec{\alpha}\cdot\hat{p}+\beta\,p\,c)
                        \notag \\
                  &=& \cos^2 \Bigl(\frac{\vartheta}{2}\Bigr)\,\vec{\alpha}
                        +{\rm i}\,\sin\Bigl(\frac{\vartheta}{2}\Bigr)\cos\Bigl(
                              \frac{\vartheta}{2}\Bigr)\dfrac{1}{
                                    \sqrt{p^2 c^2 +m^2 c^4}} 2\Bigl\{\big[
                                          m\,c^2 (\vec{\alpha}\cdot\hat{p})
                                          -p\,c\,\beta\big]\vec{\alpha}-m\,c^2 \hat{p}
                                          \Bigr\} \notag \\
                        && +\sin^2 \Bigl(\frac{\vartheta}{2}\Bigr)\dfrac{1}{
                              (p^2 c^2 +m^2 c^4)} \Bigl\{-2\,m\,c^2 \big[
                                    -m\,c^2 (\vec{\alpha}\cdot\hat{p})+p\,c\,\beta
                                    \big]\hat{p}-(p^2 c^2 +m^2 c^4)\vec{\alpha}\Bigr\}
                        \notag \\
                  &=& \biggl\{\cos\vartheta+{\rm i}\,\sin\vartheta\,\dfrac{\big[
                              m\,c^2 (\vec{\alpha}\cdot\hat{p})-p\,c\,\beta\big]}{
                                    \sqrt{p^2 c^2 +m^2 c^4}}\biggr\}\vec{\alpha}
                              \notag \\
                        && -\dfrac{2\,m\,c^2}{\sqrt{p^2 c^2 +m^2 c^4}} \sin\Bigl(
                              \frac{\vartheta}{2}
                              \Bigr)\bigg\{\sin\Bigl(\frac{\vartheta}{2}
                                    \Bigr)\dfrac{\big[
                                          -m\,c^2 (\vec{\alpha}\cdot\hat{p})
                                          +p\,c\,\beta\big]}{\sqrt{p^2 c^2 +m^2 c^4}}
                              +{\rm i}\,\cos\Bigl(\frac{\vartheta}{2}\Bigr)
                              \bigg\}\hat{p}.
            \end{eqnarray}
            One may check that
            \begin{eqnarray}
                  && \big[\mathcal{D}^\dagger (\vartheta)\,\vec{\alpha}(0)\,
                              \mathcal{D}(\vartheta)\big]\cdot\hat{p}
                        -\mathcal{D}^\dagger (\vartheta)\big[
                              \vec{\alpha}(0)\cdot\hat{p}\big]\mathcal{D}(\vartheta)=0,
            \end{eqnarray}
            i.e., based on Eq. (\ref{eq:DAlphaPD}) and Eq. (\ref{eq:DAlphaPD-aa}) one indeed has
            \begin{eqnarray}
                  && \big[\mathcal{D}^\dagger (\vartheta)\,\vec{\alpha}(0)\,
                              \mathcal{D}(\vartheta)\big]\cdot\hat{p}
                        -\mathcal{D}^\dagger (\vartheta)\big[
                              \vec{\alpha}(0)\cdot\hat{p}\big]\mathcal{D}(\vartheta)
                        \notag \\
                  &=& \biggl\{\cos\vartheta+{\rm i}\,\sin\vartheta\,\dfrac{\big[
                                    m\,c^2 (\vec{\alpha}\cdot\hat{p})-p\,c\,\beta\big]}{
                                          \sqrt{p^2 c^2 +m^2 c^4}}
                                    \biggr\}(\vec{\alpha}\cdot\hat{p})\nonumber\\
                         &&     -\dfrac{2\,m\,c^2}{\sqrt{p^2 c^2 +m^2 c^4}} \sin\Bigl(
                                    \frac{\vartheta}{2}
                                    \Bigr)\bigg\{\sin\Bigl(\frac{\vartheta}{2}
                                          \Bigr)\dfrac{\big[
                                                -m\,c^2 (\vec{\alpha}\cdot\hat{p})
                                                +p\,c\,\beta\big]}{\sqrt{p^2 c^2 +m^2 c^4}}
                                    +{\rm i}\,\cos\Bigl(\frac{\vartheta}{2}\Bigr)
                                    \bigg\} \notag \\
                        && -\bigg[(\vec{\alpha}\cdot\hat{p})
                              -\dfrac{p\,c}{p^2 c^2 +m^2 c^4} \sin\vartheta\,
                                    H_{\rm b}^{\rm II}
                              +\dfrac{p\,c}{p^2 c^2 +m^2 c^4} (\cos\vartheta-1)H_{\rm e}
                              \bigg] \notag \\
                  &=& \biggl\{\cos\vartheta+{\rm i}\,\sin\vartheta\,\dfrac{\big[
                                    m\,c^2 (\vec{\alpha}\cdot\hat{p})-p\,c\,\beta\big]}{
                                          \sqrt{p^2 c^2 +m^2 c^4}}
                                    \biggr\}(\vec{\alpha}\cdot\hat{p})
                              -\dfrac{m\,c^2}{\sqrt{p^2 c^2 +m^2 c^4}} \bigg\{
                                    2\,{\sin^2}\Bigl(\frac{\vartheta}{2}
                                          \Bigr)\dfrac{\big[
                                                -m\,c^2 (\vec{\alpha}\cdot\hat{p})
                                                +p\,c\,\beta\big]}{\sqrt{p^2 c^2 +m^2 c^4}}
                                    +{\rm i}\,\sin\vartheta\Bigr)
                                    \bigg\} \notag \\
                        && -\bigg\{(\vec{\alpha}\cdot\hat{p})
                              -\dfrac{p\,c}{p^2 c^2 +m^2 c^4} \sin\vartheta\,
                                    \sqrt{\vec{p}^{\,2} c^2+m^2c^4}\,{\rm i}\,\beta(
                                          \vec{\alpha}\cdot\hat{p})
                              -2\dfrac{p\,c}{p^2 c^2 +m^2 c^4} {\sin^2}\Bigl(
                                    \frac{\vartheta}{2}
                                    \Bigr)\big[c\,p(\vec{\alpha}\cdot\hat{p})
                                          +\beta\,m\,c^2\big]\bigg\} \notag \\
                  &=& \biggl[\cos\vartheta-{\rm i}\,\sin\vartheta\,
                                    \dfrac{p\,c\,\beta}{\sqrt{p^2 c^2 +m^2 c^4}}
                              +2\,{\sin^2}\Bigl(\frac{\vartheta}{2}
                                    \Bigr)\dfrac{m^2 c^4 }{(p^2 c^2 +m^2 c^4)}
                              \biggr](\vec{\alpha}\cdot\hat{p})
                        +{\rm i}\,\sin\vartheta\,\dfrac{m\,c^2}{
                              \sqrt{p^2 c^2 +m^2 c^4}} \notag \\
                        && -2\,m\,c^2 {\sin^2}\Bigl(\frac{\vartheta}{2}
                              \Bigr)\dfrac{p\,c\,\beta}{(p^2 c^2 +m^2 c^4)}
                        -{\rm i}\,\sin\vartheta\,\dfrac{m\,c^2}{
                              \sqrt{p^2 c^2 +m^2 c^4}} -\bigg\{(\vec{\alpha}\cdot\hat{p})\notag \\
                        &&                               -\dfrac{p\,c}{p^2 c^2 +m^2 c^4} \sin\vartheta\,
                                    \sqrt{\vec{p}^{\,2} c^2+m^2c^4}\,{\rm i}\,\beta(
                                          \vec{\alpha}\cdot\hat{p})
                              -2\dfrac{p^2 c^2}{p^2 c^2 +m^2 c^4} {\sin^2}\Bigl(
                                    \frac{\vartheta}{2}\Bigr)(\vec{\alpha}\cdot\hat{p})
                              -2\dfrac{m\,c^2}{p^2 c^2 +m^2 c^4} {\sin^2}\Bigl(
                                    \frac{\vartheta}{2}\Bigr)p\,c\,\beta\bigg\}
                              \notag \\
                  &=& \biggl[\cos\vartheta+2\,{\sin^2}\Bigl(\frac{\vartheta}{2}\Bigr)
                              \dfrac{(p^2 c^2 +m^2 c^4)}{(p^2 c^2 +m^2 c^4)}
                        -1\biggr](\vec{\alpha}\cdot\hat{p}) \notag \\
                  &=& 0.
            \end{eqnarray}
            Further, because of
            \begin{eqnarray}
                  \beta=\frac{1}{m\,c^2} \big[H_{\rm e} -c(\vec{\alpha}\cdot\vec{p})
                        \big],
            \end{eqnarray}
            which leads to
            \begin{eqnarray}
                   m\,c^2 (\vec{\alpha}\cdot\hat{p})-p\,c\,\beta
                  &=& m\,c^2 (\vec{\alpha}\cdot\hat{p})-p\,c\,\frac{1}{m\,c^2} \big[
                        H_{\rm e} -c(\vec{\alpha}\cdot\vec{p})\big]\nonumber\\
                  &=& m\,c^2 (\vec{\alpha}\cdot\hat{p})-\,\frac{p\,c}{m\,c^2} H_{\rm e}
                        +\frac{p^2 c^2}{m\,c^2} (\vec{\alpha}\cdot\hat{p}) \notag \\
                  &=& \frac{(p^2 c^2 +m^2 c^4)}{m\,c^2} (\vec{\alpha}\cdot\hat{p})
                        -\,\frac{p\,c}{m\,c^2} H_{\rm e},
            \end{eqnarray}
            then we finally have
            \begin{eqnarray}
                  && \mathcal{D}^\dagger (\vartheta)\,\vec{\alpha}(0)\,\mathcal{D}(
                        \vartheta) \notag \\
                  &=& \biggl\{\cos\vartheta+{\rm i}\,\sin\vartheta\,\dfrac{\big[
                              m\,c^2 (\vec{\alpha}\cdot\hat{p})-p\,c\,\beta\big]}{
                                    \sqrt{p^2 c^2 +m^2 c^4}}\biggr\}\vec{\alpha}
                              \notag \\
                        && -\dfrac{2\,m\,c^2}{\sqrt{p^2 c^2 +m^2 c^4}} \sin\Bigl(
                              \frac{\vartheta}{2}
                              \Bigr)\bigg\{\sin\Bigl(\frac{\vartheta}{2}
                                    \Bigr)\dfrac{\big[
                                          -m\,c^2 (\vec{\alpha}\cdot\hat{p})
                                          +p\,c\,\beta\big]}{\sqrt{p^2 c^2 +m^2 c^4}}
                              +{\rm i}\,\cos\Bigl(\frac{\vartheta}{2}\Bigr)
                              \bigg\}\hat{p} \notag \\
                  &=& \biggl\{\cos\vartheta+{\rm i}\,\sin\vartheta\,\dfrac{\big[
                              m\,c^2 (\vec{\alpha}\cdot\hat{p})-p\,c\,\beta\big]}{
                                    \sqrt{p^2 c^2 +m^2 c^4}}\biggr\}\vec{\alpha}
                        +\dfrac{2\,m\,c^2}{(p^2 c^2 +m^2 c^4)} {\sin^2}\Bigl(
                              \frac{\vartheta}{2}\Bigr)\,\big[
                                          m\,c^2 (\vec{\alpha}\cdot\hat{p})
                                          -p\,c\,\beta\big]\hat{p}
                        -{\rm i}\dfrac{m\,c^2 \sin\vartheta}{\sqrt{p^2 c^2 +m^2 c^4}}
                              \hat{p} \notag \\
                  &=& \Biggl\{\cos\vartheta+\dfrac{{\rm i}\,\sin\vartheta}{
                                    \sqrt{p^2 c^2 +m^2 c^4}} \bigg[
                                          \frac{(p^2 c^2 +m^2 c^4)}{m\,c^2} (
                                                \vec{\alpha}\cdot\hat{p})
                              -\,\frac{p\,c}{m\,c^2} H_{\rm e}\bigg]
                              \Biggr\}\vec{\alpha} \notag \\
                        && +\dfrac{2\,m\,c^2}{(p^2 c^2 +m^2 c^4)} {\sin^2}\Bigl(
                              \frac{\vartheta}{2}\Bigr)\,\bigg[
                                    \frac{(p^2 c^2 +m^2 c^4)}{m\,c^2} (
                                          \vec{\alpha}\cdot\hat{p})
                              -\,\frac{p\,c}{m\,c^2} H_{\rm e}\bigg]\hat{p}
                        -{\rm i}\dfrac{m\,c^2 \sin\vartheta}{\sqrt{p^2 c^2 +m^2 c^4}}
                              \hat{p}.
            \end{eqnarray}

            Besides,
            \begin{eqnarray}
                  \big[\vec{\alpha}(0),\ (-m\,c^2 \vec{\alpha}\cdot\hat{p}+\beta\,p\,c)
                        \big]
                  =2\Bigl\{\big[m\,c^2 (\vec{\alpha}\cdot\hat{p})
                        -p\,c\,\beta\big]\vec{\alpha}-m\,c^2 \hat{p}\Bigr\}.
            \end{eqnarray}
            \begin{eqnarray}
                   \big[\vec{\zeta},\ (-m\,c^2 \vec{\alpha}\cdot\hat{p}+\beta\,p\,c)
                        \big]
                  &=&\big[{\rm i}\,\beta\,\vec{\alpha},\ (
                        -m\,c^2 \vec{\alpha}\cdot\hat{p}+\beta\,p\,c)\big]\nonumber\\
                  &=&{\rm i}\,\beta\big[\vec{\alpha},\ (
                              -m\,c^2 \vec{\alpha}\cdot\hat{p}+\beta\,p\,c)\big]
                        +{\rm i}\big[\beta,\ (
                              -m\,c^2 \vec{\alpha}\cdot\hat{p}+\beta\,p\,c)
                              \big]\vec{\alpha} \notag \\
                  &=& {\rm i}\,\beta\big[\vec{\alpha},\ (
                              -m\,c^2 \vec{\alpha}\cdot\hat{p}+\beta\,p\,c)\big]
                        -{\rm i}\,m\,c^2 \big[\beta,\ (\vec{\alpha}\cdot\hat{p})
                              \big]\vec{\alpha}\nonumber\\
                  &=&{\rm i}\,\beta\big[\vec{\alpha},\ (
                              -m\,c^2 \vec{\alpha}\cdot\hat{p}+\beta\,p\,c)\big]
                        -{\rm i}\,m\,c^2 \Bigl(\big\{\beta,\ (\vec{\alpha}\cdot\hat{p})
                              \big\}-2(\vec{\alpha}\cdot\hat{p})\beta
                              \Bigr)\vec{\alpha} \notag \\
                  &=& {\rm i}\,2\,\beta\Bigl\{\big[m\,c^2 (\vec{\alpha}\cdot\hat{p})
                        -p\,c\,\beta\big]\vec{\alpha}-m\,c^2 \hat{p}\Bigr\}
                        +{\rm i}\,2\,m\,c^2 (\vec{\alpha}\cdot\hat{p})\beta\,
                              \vec{\alpha}\nonumber\\
                  &=& {\rm i}\,2\,m\,c^2 \bigl\{\beta,\ (\vec{\alpha}\cdot\hat{p})
                              \bigr\}\vec{\alpha}
                        -{\rm i}\,2(p\,c\,\vec{\alpha}+m\,c^2 \beta\,\hat{p}) \notag \\
                  &=& -{\rm i}\,2(p\,c\,\vec{\alpha}+m\,c^2 \beta\,\hat{p}).
            \end{eqnarray}
            Apparently,
            \begin{equation}
                  \big[\vec{\zeta},\ (-m\,c^2 \vec{\alpha}\cdot\hat{p}+\beta\,p\,c)
                        \big]\cdot\hat{p}
                  =-{\rm i}\,2(p\,c\,\vec{\alpha}+m\,c^2 \beta\,\hat{p})\cdot\hat{p}
                  =-{\rm i}\,2\big[p\,c(\vec{\alpha}\cdot\hat{p})+m\,c^2 \beta\big]
                  =-{\rm i}\,2\,H_{\rm e}.
            \end{equation}
            Additionally,
            \begin{eqnarray}
                  && (-m\,c^2 \vec{\alpha}\cdot\hat{p}+\beta\,p\,c)\vec{\zeta}(
                        -m\,c^2 \vec{\alpha}\cdot\hat{p}+\beta\,p\,c)\nonumber\\
                  &=&(-m\,c^2 \vec{\alpha}\cdot\hat{p}+\beta\,p\,c){\rm i}\,\beta\,
                        \vec{\alpha}(-m\,c^2 \vec{\alpha}\cdot\hat{p}+\beta\,p\,c)
                        \notag \\
                  &=& m^2 c^4 (\vec{\alpha}\cdot\hat{p}){\rm i}\,\beta\,\vec{\alpha}(
                              \vec{\alpha}\cdot\hat{p})
                        -m\,c^2 p\,c\big[
                              (\vec{\alpha}\cdot\hat{p})({\rm i}\,\beta\,\vec{\alpha})
                                    \beta
                              +\beta({\rm i}\,\beta\,\vec{\alpha})(
                                    \vec{\alpha}\cdot\hat{p})\big]
                        +p^2 c^2 \beta({\rm i}\,\beta\,\vec{\alpha})\beta \nonumber \\
                  &=& -m^2 c^4 {\rm i}\,\beta(\vec{\alpha}\cdot\hat{p})\vec{\alpha}(
                              \vec{\alpha}\cdot\hat{p})
                        -{\rm i}\,m\,c^2 p\,c\big[
                              -(\vec{\alpha}\cdot\hat{p})\vec{\alpha}
                              +\vec{\alpha}(\vec{\alpha}\cdot\hat{p})\big]
                        -p^2 c^2 ({\rm i}\,\beta\,\vec{\alpha}) \nonumber \\
                  &=& -m^2 c^4 {\rm i}\,\beta(\vec{\alpha}\cdot\hat{p})\Bigl\{\bigl[
                              \vec{\alpha},\ (\vec{\alpha}\cdot\hat{p})
                              \bigr]+(\vec{\alpha}\cdot\hat{p})\vec{\alpha}\Bigr\}
                        -{\rm i}\,m\,c^2 p\,c\big[\vec{\alpha},\ (
                              \vec{\alpha}\cdot\hat{p})\big]
                        -p^2 c^2 ({\rm i}\,\beta\,\vec{\alpha}) \nonumber \\
                  &=& -m^2 c^4 {\rm i}\,\beta(\vec{\alpha}\cdot\hat{p})\Bigl\{
                              2\,\hat{p}-2(\vec{\alpha}\cdot\hat{p})\vec{\alpha}
                              +(\vec{\alpha}\cdot\hat{p})\vec{\alpha}\Bigr\}
                        -{\rm i}\,m\,c^2 p\,c\big[2\,\hat{p}
                              -2(\vec{\alpha}\cdot\hat{p})\vec{\alpha}\big]
                        -p^2 c^2 ({\rm i}\,\beta\,\vec{\alpha}) \nonumber \\
                  &=& -m^2 c^4 {\rm i}\,2\,\beta(\vec{\alpha}\cdot\hat{p})\hat{p}
                        +m^2 c^4 {\rm i}\,\beta\,\vec{\alpha}
                        -{\rm i}\,2\,m\,c^2 p\,c\,\hat{p}
                        +{\rm i}\,2\,m\,c^2 p\,c(\vec{\alpha}\cdot\hat{p})\vec{\alpha}
                        -p^2 c^2 ({\rm i}\,\beta\,\vec{\alpha}) \nonumber \\
                  &=& -m^2 c^4 {\rm i}\,2\,\beta(\vec{\alpha}\cdot\hat{p})\hat{p}
                        +(m^2 c^4 -p^2 c^2){\rm i}\,\beta\,\vec{\alpha}
                        +{\rm i}\,2\,m\,c^2 p\,c(\vec{\alpha}\cdot\hat{p})\vec{\alpha}
                        -{\rm i}\,2\,m\,c^2 p\,c\,\hat{p},
            \end{eqnarray}
            which implies
            \begin{eqnarray}
                  && (-m\,c^2 \vec{\alpha}\cdot\hat{p}+\beta\,p\,c)(\vec{\zeta}
                        \cdot\hat{p})(-m\,c^2 \vec{\alpha}\cdot\hat{p}+\beta\,p\,c)
                  =\bigl[(-m\,c^2 \vec{\alpha}\cdot\hat{p}+\beta\,p\,c)\vec{\zeta}
                        (-m\,c^2 \vec{\alpha}\cdot\hat{p}+\beta\,p\,c)
                        \bigr]\cdot\hat{p} \notag \\
                  &=& -m^2 c^4 {\rm i}\,2\,\beta(\vec{\alpha}\cdot\hat{p})
                        +(m^2 c^4 -p^2 c^2){\rm i}\,\beta(\vec{\alpha}\cdot\hat{p})
                        +{\rm i}\,2\,m\,c^2 p\,c-{\rm i}\,2\,m\,c^2 p\,c
                  =-(m^2 c^4 +p^2 c^2){\rm i}\,\beta(\vec{\alpha}\cdot\hat{p})
                        \notag \\
                  &=& -\sqrt{p^2 c^2 +m^2 c^4}\,H_{\rm b}^{\rm II}.
            \end{eqnarray}
            We can have
            \begin{eqnarray}\label{eq:DD-1b}
                  && \mathcal{D}^\dagger (\vartheta)\,\vec{\zeta}\,\mathcal{D}(
                        \vartheta)
                  =\bigg[\cos\Bigl(\frac{\vartheta}{2}\Bigr)\;\mathbb{I}
                        -{\rm i}\,\sin\Bigl(\frac{\vartheta}{2}\Bigr)
                              \dfrac{(-m\,c^2 \vec{\alpha}\cdot\hat{p}+\beta\,p\,c)}{
                                    \sqrt{p^2 c^2 +m^2 c^4}}
                        \bigg]\vec{\zeta}\bigg[
                              \cos\Bigl(\frac{\vartheta}{2}\Bigr)\mathbb{I}
                              +{\rm i}\,\sin\Bigl(\frac{\vartheta}{2}\Bigr)
                                    \dfrac{(-m\,c^2 \vec{\alpha}\cdot\hat{p}
                                          +\beta\,p\,c)}{\sqrt{p^2 c^2 +m^2 c^4}}\bigg]
                        \notag \\
                  &=& \cos^2 \Bigl(\frac{\vartheta}{2}\Bigr)\,\vec{\zeta}
                        +{\rm i}\,\sin\Bigl(\frac{\vartheta}{2}\Bigr)\cos\Bigl(
                              \frac{\vartheta}{2}\Bigr)\dfrac{1}{
                                    \sqrt{p^2 c^2 +m^2 c^4}} \big[\vec{\zeta},\ (
                                          -m\,c^2 \vec{\alpha}\cdot\hat{p}+\beta\,p\,c)
                                          \big] \notag \\
                        && +\sin^2 \Bigl(\frac{\vartheta}{2}\Bigr)\dfrac{1}{
                              p^2 c^2 +m^2 c^4} \big[-m\,c^2 (\vec{\alpha}\cdot\hat{p})
                                    +\beta\,p\,c\big]\vec{\zeta}\big[
                                          -m\,c^2 (\vec{\alpha}\cdot\hat{p})
                                          +\beta\,p\,c\big] \notag \\
                  &=& \cos^2 \Bigl(\frac{\vartheta}{2}\Bigr)\,\vec{\zeta}
                        -{\rm i}\,\sin\Bigl(\frac{\vartheta}{2}\Bigr)\cos\Bigl(
                              \frac{\vartheta}{2}\Bigr)\dfrac{{\rm i}\,2(
                                    p\,c\,\vec{\alpha}+m\,c^2 \beta\,\hat{p})
                                    }{\sqrt{p^2 c^2 +m^2 c^4}} \notag \\
                        && +\sin^2 \Bigl(\frac{\vartheta}{2}\Bigr)\dfrac{1}{
                              p^2 c^2 +m^2 c^4} \big[-m^2 c^4 {\rm i}\,2\,\beta(
                                          \vec{\alpha}\cdot\hat{p})\hat{p}
                                    +(m^2 c^4 -p^2 c^2){\rm i}\,\beta\,\vec{\alpha}
                                    +{\rm i}\,2\,m\,c^2 p\,c(\vec{\alpha}\cdot\hat{p})
                                          \vec{\alpha}
                                    -{\rm i}\,2\,m\,c^2 p\,c\,\hat{p}\big] \notag \\
                  &=& \bigg[\cos^2 \Bigl(\frac{\vartheta}{2}\Bigr)
                              +\sin^2 \Bigl(\frac{\vartheta}{2}\Bigr)\,
                                    \dfrac{(m^2 c^4 -p^2 c^2)}{p^2 c^2 +m^2 c^4}
                              \bigg]{\rm i}\,\beta\,\vec{\alpha}\nonumber\\
                       && +\sin^2 \Bigl(\frac{\vartheta}{2}\Bigr)\dfrac{1}{
                              p^2 c^2 +m^2 c^4} \big[
                                    {\rm i}\,2\,m\,c^2 p\,c(\vec{\alpha}\cdot\hat{p})
                                          \vec{\alpha}
                                    -{\rm i}\,2\,m\,c^2 p\,c\,\hat{p}
                                    -m^2 c^4 {\rm i}\,2\,\beta(
                                          \vec{\alpha}\cdot\hat{p})\hat{p}\big] \nonumber\\
                                    && -{\rm i}\,\sin\Bigl(\frac{\vartheta}{2}\Bigr)\cos\Bigl(
                              \frac{\vartheta}{2}\Bigr)\dfrac{{\rm i}\,2(
                                    p\,c\,\vec{\alpha}+m\,c^2 \beta\,\hat{p})
                                    }{\sqrt{p^2 c^2 +m^2 c^4}}.
            \end{eqnarray}
            From Eq. (\ref{eq:DD-1a}) we have known that
            \begin{eqnarray}\label{eq:DD-1c}
                   \mathcal{D}^\dagger (\vartheta)\bigl[\vec{\zeta}(0)\cdot\hat{p}
                        \bigr]\mathcal{D}(\vartheta)
                        &=& \cos\vartheta \dfrac{H_{\rm b}^{\rm II}}{\sqrt{p^2 c^2 +m^2 c^4}}
                        +\sin\vartheta \dfrac{H_{\rm e}}{\sqrt{p^2 c^2 +m^2 c^4}}
                        \notag \\
                  &=& \cos\vartheta\,{\rm i}\,\beta(\vec{\alpha}\cdot\hat{p})
                        +\sin\vartheta \dfrac{H_{\rm e}}{\sqrt{p^2 c^2 +m^2 c^4}},
            \end{eqnarray}
            thus based on Eq. (\ref{eq:DD-1b}) and Eq. (\ref{eq:DD-1c}) we can verify that
            \begin{eqnarray}\label{eq:DD-1d}
                  \left[\mathcal{D}^\dagger (\vartheta)\vec{\zeta}(0)
                        \mathcal{D}(\vartheta)\right]\cdot\hat{p}- \mathcal{D}^\dagger (\vartheta)\bigl[\vec{\zeta}(0)\cdot\hat{p}
                        \bigr]\mathcal{D}(\vartheta) &=& 0.
            \end{eqnarray}
            $\blacksquare$
            \end{remark}

            \begin{remark}After making these preparations, we now come to calculate Eq. (\ref{eq:ZZZ-1a}), i.e.,
             \begin{eqnarray}\label{eq:ZZZ-1b}
                   \langle\Psi'''_1|\hat{\mathcal{Z}}_{\rm m}|\Psi'''_3 \rangle
                  &=& -\frac{{\rm i}\,\hbar\,c}{2} \biggr\{
                        \cos\vartheta\,\langle\Psi'''_1|\vec{\alpha}(0)|\Psi'''_3\rangle
                        +\sin\vartheta\,\frac{\sqrt{p^2 c^2 +m^2 c^4}}{p\,c}
                              \langle\Psi'''_1|\vec{\zeta}(0)|\Psi'''_3\rangle
                              \nonumber \\
                        &&\qquad\qquad -\sin\vartheta\,\frac{m^2 c^4}{
                              \sqrt{p^2 c^2 +m^2 c^4}} \frac{1}{p\, c} \langle\Psi'''_1|
                                    \left[\vec{\zeta}(0)\cdot\hat{p}
                                          \right]|\Psi'''_3\rangle\hat{p}
                        \biggr\}\dfrac{1}{E_{\rm m}} \Bigl({\rm e}^{
                              {\rm i}\frac{2}{\hbar} E_{\rm m} t} -1\Bigr)\nonumber\\
                  &=& -\frac{{\rm i}\,\hbar\,c}{2} \biggr\{
                        \cos\vartheta\,\langle\Psi'''_1|\vec{\alpha}(0)|\Psi'''_3\rangle
                        +\sin\vartheta\,\frac{\sqrt{p^2 c^2 +m^2 c^4}}{p\,c}
                              \langle\Psi'''_1|\vec{\zeta}(0)|\Psi'''_3\rangle
                              \nonumber \\
                        &&\qquad\qquad -\sin\vartheta\,\frac{m^2 c^4}{
                              \sqrt{p^2 c^2 +m^2 c^4}} \frac{1}{p\, c} \langle\Psi'''_1|
                                    \left[\vec{\zeta}(0)\cdot\hat{p}
                                          \right]|\Psi'''_3\rangle\hat{p}
                        \biggr\}\dfrac{1}{E_{\rm p}} \Bigl({\rm e}^{
                              {\rm i}\frac{2}{\hbar} E_{\rm p} t} -1\Bigr).
            \end{eqnarray}
            We need to handle the following three terms:
             \begin{eqnarray}
             \langle\Psi'''_1|\vec{\alpha}(0)|\Psi'''_3\rangle,\qquad
                           \langle\Psi'''_1|\vec{\zeta}(0)|\Psi'''_3\rangle,\qquad
                              \langle\Psi'''_1| \left[\vec{\zeta}(0)\cdot\hat{p}
                                          \right]|\Psi'''_3\rangle\hat{p}.
            \end{eqnarray}

           (i) The first term. One can have
             \begin{eqnarray}
            && \langle\Psi'''_1|\vec{\alpha}(0)|\Psi'''_3\rangle =  \langle\Psi_1|\left[\mathcal{D}^\dagger (\vartheta)\,\vec{\alpha}\,
                              \mathcal{D}(\vartheta)\right]|\Psi_3\rangle\nonumber\\
            &=&\langle\Psi_1|\biggr[ \Biggl\{\cos\vartheta+\dfrac{{\rm i}\,\sin\vartheta}{
                                    \sqrt{p^2 c^2 +m^2 c^4}} \bigg[
                                          \frac{(p^2 c^2 +m^2 c^4)}{m\,c^2} (
                                                \vec{\alpha}\cdot\hat{p})
                              -\,\frac{p\,c}{m\,c^2} H_{\rm e}\bigg]
                              \Biggr\}\vec{\alpha} \notag \\
                        && +\dfrac{2\,m\,c^2}{(p^2 c^2 +m^2 c^4)} {\sin^2}\Bigl(
                              \frac{\vartheta}{2}\Bigr)\,\bigg[
                                    \frac{(p^2 c^2 +m^2 c^4)}{m\,c^2} (
                                          \vec{\alpha}\cdot\hat{p})
                              -\,\frac{p\,c}{m\,c^2} H_{\rm e}\bigg]\hat{p}
                        -{\rm i}\dfrac{m\,c^2 \sin\vartheta}{\sqrt{p^2 c^2 +m^2 c^4}}
                              \hat{p}\biggr]|\Psi_3\rangle\nonumber\\
             &=&\langle\Psi_1|\biggr[ \Biggl\{\cos\vartheta+\dfrac{{\rm i}\,\sin\vartheta}{
                                    \sqrt{p^2 c^2 +m^2 c^4}} \bigg[
                                          \frac{(p^2 c^2 +m^2 c^4)}{m\,c^2} (
                                                \vec{\alpha}\cdot\hat{p})
                              -\,\frac{p\,c}{m\,c^2} E_{\rm p}\bigg]
                              \Biggr\}\vec{\alpha} \notag \\
                        && +\dfrac{2\,m\,c^2}{(p^2 c^2 +m^2 c^4)} {\sin^2}\Bigl(
                              \frac{\vartheta}{2}\Bigr)\,\bigg[
                                    \frac{(p^2 c^2 +m^2 c^4)}{m\,c^2} (
                                          \vec{\alpha}\cdot\hat{p})
                            \bigg]\hat{p} \biggr]|\Psi_3\rangle\nonumber\\
              &=&\langle\Psi_1|\biggr[ \Biggl\{\cos\vartheta-{\rm i}\,\sin\vartheta \frac{p}{m\,c}
              +{\rm i}\,\sin\vartheta\bigg[
                                          \frac{\sqrt{p^2 c^2 +m^2 c^4}}{m\,c^2} (
                                                \vec{\alpha}\cdot\hat{p})
                             \bigg]
                              \Biggr\}\vec{\alpha} +2 {\sin^2}\Bigl(
                              \frac{\vartheta}{2}\Bigr)\, (
                                          \vec{\alpha}\cdot\hat{p})
                            \hat{p} \biggr]|\Psi_3\rangle\nonumber\\
             &=& \left(\cos\vartheta-{\rm i}\,\sin\vartheta \frac{p}{m\,c}\right)\langle\Psi_1|\vec{\alpha}|\Psi_3\rangle+2 {\sin^2}\Bigl(
                              \frac{\vartheta}{2}\Bigr)\, \left[\left(\langle\Psi_1|
                                          \vec{\alpha}|\Psi_3\rangle)\cdot\hat{p}\right)\right]
                            \hat{p}+
             {\rm i}\,\sin\vartheta \frac{\sqrt{p^2 c^2 +m^2 c^4}}{m\,c^2}\langle\Psi_1|(
                                                \vec{\alpha}\cdot\hat{p})
                                                          \vec{\alpha}|\Psi_3\rangle.\nonumber\\
            \end{eqnarray}

           Based on Table \ref{tab:pz0}, we have
           \begin{eqnarray}
     &&  \langle\Psi_1|\vec{\alpha}(0) |\Psi_3 \rangle=\dfrac{-m\,c^2}{E_{\rm p}}\, \hat{p},
      \end{eqnarray}
      thus
             \begin{eqnarray}
     &&  \left[\left(\langle\Psi_1| \vec{\alpha}(0)|\Psi_3\rangle)\cdot\hat{p}\right)\right]
                            \hat{p}=\dfrac{-m\,c^2}{E_{\rm p}}\, \hat{p}=\langle\Psi_1|\vec{\alpha}(0) |\Psi_3 \rangle,
      \end{eqnarray}
     then one has
                \begin{eqnarray}
       \left(\cos\vartheta-{\rm i}\,\sin\vartheta \frac{p}{m\,c}\right)\langle\Psi_1|\vec{\alpha}|\Psi_3\rangle+2 {\sin^2}\Bigl(
                              \frac{\vartheta}{2}\Bigr)\, \left[\left(\langle\Psi_1|
                                          \vec{\alpha}|\Psi_3\rangle)\cdot\hat{p}\right)\right]
                            \hat{p}&=&\left(1-{\rm i}\,\sin\vartheta \frac{p}{m\,c}\right)\dfrac{-m\,c^2}{E_{\rm p}}\, \hat{p}\nonumber\\
    & =& \left(1-{\rm i}\,\sin\vartheta \frac{p}{m\,c}\right)\langle\Psi_1|\vec{\alpha}(0) |\Psi_3 \rangle.
      \end{eqnarray}

For the term $\langle\Psi_1|(\vec{\alpha}\cdot\hat{p})\vec{\alpha}|\Psi_3\rangle$, one can calculate it in the following way
 \begin{eqnarray}\label{eq:alp-13-a}
      \langle\Psi_1|(\vec{\alpha}\cdot\hat{p})\vec{\alpha}|\Psi_3\rangle&=&\langle\Psi_1|(\vec{\alpha}\cdot\hat{p})\,\mathbb{I}\, \vec{\alpha}|\Psi_3\rangle= \langle\Psi_1|(\vec{\alpha}\cdot\hat{p})\,\left[\sum_{j=1}^4 |\Psi_j\rangle\langle\Psi_j|\right]\, \vec{\alpha}|\Psi_3\rangle\nonumber\\
      &=& \sum_{j=1}^4 \left[\left( \langle\Psi_1|(\vec{\alpha}\cdot\hat{p})\, |\Psi_j\rangle\right) \left(\langle\Psi_j| \vec{\alpha}|\Psi_3\rangle\right)\right].
\end{eqnarray}

To reach this purpose, one needs to complete Table \ref{tab:pz0}. From Eq. (\ref{eq:alpha-jk-1}) and Eq. (\ref{eq:alpha-jk-2}), we have the complete table as shown in Table \ref{tab:pz0-a}.
\begin{table}[t]
                              \centering
                              \caption{The value of $\langle\Psi_j|\vec{\alpha}(0)|\Psi_k \rangle$, where $j,k=1, 2, 3, 4$ are four eigenstates of Dirac's electron.}
                              \begin{tabular}{lllll}
                                    \hline\hline
                                    & $|\Psi_1\rangle$ &  $|\Psi_2\rangle$& $|\Psi_3\rangle$ & $|\Psi_4\rangle$ \\
                                    \hline
                                    $\langle \Psi_1| \vec{\alpha}(0)$ \;\;\;\quad& $-\frac{cp}{E_{\rm p}}\, \hat{p}$& 0 & $\dfrac{-m\,c^2}{E_{\rm p}}\, \hat{p}$  \;\;\;\quad& $(\vec{F}_1 +{\rm i}\,\vec{F}_2)$  \\
                                    \hline
                                    $\langle \Psi_2| \vec{\alpha}(0)$\;\;\;\quad &0 & $-\frac{cp}{E_{\rm p}}\, \hat{p}$ & $-(\vec{F}_1 +{\rm i}\,\vec{F}_2)^* $ \quad\quad& $\dfrac{-m\,c^2}{E_{\rm p}}\, \hat{p}$ \\
                                    \hline
                                    $\langle \Psi_3| \vec{\alpha}(0)$\;\;\;\quad & $\dfrac{-m\,c^2}{E_{\rm p}}\,\hat{p} \quad$ \;\;\;& $-(\vec{F}_1 +{\rm i}\,\vec{F}_2) \quad\quad$ & $\frac{cp}{E_{\rm p}}\, \hat{p}$ & 0 \\
                                    \hline
                                    $\langle \Psi_4| \vec{\alpha}(0)$ \;\;\;\quad& $(\vec{F}_1 +{\rm i}\,\vec{F}_2)^* \quad$ \quad&$\dfrac{-m\,c^2}{E_{\rm p}}\, \hat{p}$ \;\;\;& 0 & $\frac{cp}{E_{\rm p}}\, \hat{p}$ \\
                                    \hline\hline
                              \end{tabular}\label{tab:pz0-a}
      \end{table}
Then from Eq. (\ref{eq:alp-13-a}) we have
 \begin{eqnarray}\label{eq:alp-13-b}
      \langle\Psi_1|(\vec{\alpha}\cdot\hat{p})\vec{\alpha}|\Psi_3\rangle
      &=& \sum_{j=1}^4 \left[\left( \langle\Psi_1|(\vec{\alpha}\cdot\hat{p})\, |\Psi_j\rangle\right) \left(\langle\Psi_j| \vec{\alpha}|\Psi_3\rangle\right)\right]\nonumber\\
      &=& \left[\left( \langle\Psi_1|(\vec{\alpha}\cdot\hat{p})\, |\Psi_1\rangle\right) \left(\langle\Psi_1| \vec{\alpha}|\Psi_3\rangle\right)\right]+\left[\left( \langle\Psi_1|(\vec{\alpha}\cdot\hat{p})\, |\Psi_3\rangle\right) \left(\langle\Psi_3| \vec{\alpha}|\Psi_3\rangle\right)\right]\nonumber\\
      &=& \frac{-cp}{E_{\rm p}}\, \dfrac{-m\,c^2}{E_{\rm p}}\, \hat{p}+\dfrac{-m\,c^2}{E_{\rm p}}\, \frac{cp}{E_{\rm p}}\, \hat{p}=0.
\end{eqnarray}
Finally we have
  \begin{eqnarray}
             \langle\Psi'''_1|\vec{\alpha}(0)|\Psi'''_3\rangle = \left(1-{\rm i}\,\sin\vartheta \frac{p}{m\,c}\right)\langle\Psi_1|\vec{\alpha}(0) |\Psi_3 \rangle.
  \end{eqnarray}

    (ii) The second term. One can have
     \begin{eqnarray}\label{eq:alp-13-c}
           &&  \langle\Psi'''_1| \vec{\zeta}(0) |\Psi'''_3\rangle =  \langle\Psi_1| \mathcal{D}^\dagger (\vartheta)\,\vec{\zeta}(0)\,
                                    \mathcal{D}(\vartheta) |\Psi_3\rangle \nonumber\\
           &=& \langle\Psi_1| \biggr[ \bigg[\cos^2 \Bigl(\frac{\vartheta}{2}\Bigr)
                              +\sin^2 \Bigl(\frac{\vartheta}{2}\Bigr)\,
                                    \dfrac{(m^2 c^4 -p^2 c^2)}{p^2 c^2 +m^2 c^4}
                              \bigg]{\rm i}\,\beta(0)\,\vec{\alpha}(0)\nonumber\\
                       && +\sin^2 \Bigl(\frac{\vartheta}{2}\Bigr)\dfrac{1}{
                              p^2 c^2 +m^2 c^4} \big[
                                    {\rm i}\,2\,m\,c^2 p\,c(\vec{\alpha}(0)\cdot\hat{p})
                                          \vec{\alpha}(0)
                                    -{\rm i}\,2\,m\,c^2 p\,c\,\hat{p}
                                    -m^2 c^4 {\rm i}\,2\,\beta(0)(
                                          \vec{\alpha}(0)\cdot\hat{p})\hat{p}\big] \nonumber\\
                                    && -{\rm i}\,\sin\Bigl(\frac{\vartheta}{2}\Bigr)\cos\Bigl(
                              \frac{\vartheta}{2}\Bigr)\dfrac{{\rm i}\,2(
                                    p\,c\,\vec{\alpha}(0)+m\,c^2 \beta(0)\,\hat{p})
                                    }{\sqrt{p^2 c^2 +m^2 c^4}}\biggr] |\Psi_3\rangle \nonumber\\
      &=& \langle\Psi_1| \biggr[ \bigg[\cos^2 \Bigl(\frac{\vartheta}{2}\Bigr)
                              +\sin^2 \Bigl(\frac{\vartheta}{2}\Bigr)\,
                                    \dfrac{(m^2 c^4 -p^2 c^2)}{p^2 c^2 +m^2 c^4}
                              \bigg]{\rm i}\,\frac{1}{m\,c^2} \big[H_{\rm e} -c(\vec{\alpha}(0)\cdot\vec{p})
                        \big]\,\vec{\alpha}(0)\nonumber\\
                       && +\sin^2 \Bigl(\frac{\vartheta}{2}\Bigr)\dfrac{1}{
                              p^2 c^2 +m^2 c^4} \big[
                                    {\rm i}\,2\,m\,c^2 p\,c(\vec{\alpha}(0)\cdot\hat{p})
                                          \vec{\alpha}(0)
                                    -m^2 c^4 {\rm i}\,2\,\frac{1}{m\,c^2} \big[H_{\rm e} -c(\vec{\alpha}(0)\cdot\vec{p})
                        \big](
                                          \vec{\alpha}(0)\cdot\hat{p})\hat{p}\big] \nonumber\\
                                    && -{\rm i}\,\sin\Bigl(\frac{\vartheta}{2}\Bigr)\cos\Bigl(
                              \frac{\vartheta}{2}\Bigr)\dfrac{1
                                    }{\sqrt{p^2 c^2 +m^2 c^4}} {\rm i}\,2 \left(
                                    p\,c\,\vec{\alpha}(0)+m\,c^2 \frac{1}{m\,c^2} \big[H_{\rm e} -c(\vec{\alpha}(0)\cdot\vec{p})
                        \big]\,\hat{p}\right)\biggr] |\Psi_3\rangle.
  \end{eqnarray}
  Due to
  \begin{eqnarray}
  && \langle\Psi_1|[\vec{\alpha}(0)\cdot\hat{p}]\vec{\alpha}(0)|\Psi_3\rangle=0, \nonumber\\
     &&  \left[\left(\langle\Psi_1| \vec{\alpha}(0)|\Psi_3\rangle)\cdot\hat{p}\right)\right]
                            \hat{p}=\langle\Psi_1|\vec{\alpha}(0) |\Psi_3 \rangle,
      \end{eqnarray}
 Eq. (\ref{eq:alp-13-c}) becomes
  \begin{eqnarray}
             \langle\Psi'''_1| \vec{\zeta}(0) |\Psi'''_3\rangle
      &=& \langle\Psi_1| \biggr\{ \bigg[\cos^2 \Bigl(\frac{\vartheta}{2}\Bigr)
                              +\sin^2 \Bigl(\frac{\vartheta}{2}\Bigr)\,
                                    \dfrac{(m^2 c^4 -p^2 c^2)}{p^2 c^2 +m^2 c^4}
                              \bigg]{\rm i}\,\frac{1}{m\,c^2} H_{\rm e}
                       \,\vec{\alpha}\nonumber\\
                       && +\sin^2 \Bigl(\frac{\vartheta}{2}\Bigr)\dfrac{1}{
                              p^2 c^2 +m^2 c^4} \big[  -m^2 c^4 {\rm i}\,2\,\frac{1}{m\,c^2} \big[H_{\rm e}(\vec{\alpha}\cdot\hat{p})\hat{p} -c\vec{p}
                        \big]\big] \nonumber\\
                                    && -{\rm i}\,\sin\Bigl(\frac{\vartheta}{2}\Bigr)\cos\Bigl(
                              \frac{\vartheta}{2}\Bigr)\dfrac{1
                                    }{\sqrt{p^2 c^2 +m^2 c^4}} {\rm i}\,2 \left[
                                    c\,p\vec{\alpha}-c(\vec{\alpha}\cdot\vec{p})\hat{p}\right]\biggr\} |\Psi_3\rangle\nonumber\\
      &=& \langle\Psi_1| \biggr\{ \bigg[\cos^2 \Bigl(\frac{\vartheta}{2}\Bigr)
                              +\sin^2 \Bigl(\frac{\vartheta}{2}\Bigr)\,
                                    \dfrac{(m^2 c^4 -p^2 c^2)}{p^2 c^2 +m^2 c^4}
                              \bigg]{\rm i}\,\frac{1}{m\,c^2} H_{\rm e}
                       \,\vec{\alpha}\nonumber\\
                       && +\sin^2 \Bigl(\frac{\vartheta}{2}\Bigr)\dfrac{1}{
                              p^2 c^2 +m^2 c^4} \big[  -m^2 c^4 {\rm i}\,2\,\frac{1}{m\,c^2} \big[H_{\rm e}(\vec{\alpha}\cdot\hat{p})\hat{p}
                        \big]\big]\biggr\} |\Psi_3\rangle \nonumber\\
       &=&   \bigg[\cos^2 \Bigl(\frac{\vartheta}{2}\Bigr)
                              +\sin^2 \Bigl(\frac{\vartheta}{2}\Bigr)\,
                                    \dfrac{(m^2 c^4 -p^2 c^2)}{p^2 c^2 +m^2 c^4}
                              \bigg]{\rm i}\,\frac{1}{m\,c^2} E_{\rm p}
                       \,\langle\Psi_1| \vec{\alpha}|\Psi_3\rangle \nonumber\\
                       && +\sin^2 \Bigl(\frac{\vartheta}{2}\Bigr)\dfrac{1}{
                              p^2 c^2 +m^2 c^4} \big[  -m^2 c^4 {\rm i}\,2\,\frac{1}{m\,c^2} E_{\rm p} \big] \langle\Psi_1|(\vec{\alpha}\cdot\hat{p})\hat{p}
                         |\Psi_3\rangle \nonumber\\
        &=&   \frac{{\rm i} E_{\rm e} }{m\,c^2}\bigg[\cos^2 \Bigl(\frac{\vartheta}{2}\Bigr)
                              +\sin^2 \Bigl(\frac{\vartheta}{2}\Bigr)\,
                                    \dfrac{(-m^2 c^4 -p^2 c^2)}{p^2 c^2 +m^2 c^4}
                              \bigg]
                       \,\langle\Psi_1| \vec{\alpha}|\Psi_3\rangle \nonumber\\
         &=&   \frac{{\rm i}  E_{\rm p}  }{m\,c^2} \cos\vartheta
                       \,\langle\Psi_1| \vec{\alpha}|\Psi_3\rangle=\frac{{\rm i}  E_{\rm p}  }{m\,c^2} \cos\vartheta \dfrac{-m\,c^2}{E_{\rm p}}\, \hat{p}=-{\rm i} \cos\vartheta \,\hat{p},
  \end{eqnarray}
  which yields
     \begin{eqnarray}
     \langle\Psi'''_1| \left[\vec{\zeta}(0)\cdot\hat{p} \right]|\Psi'''_3\rangle\hat{p}=-{\rm i} \cos\vartheta \,\hat{p}.
     \end{eqnarray}
 Then from Eq. (\ref{eq:ZZZ-1b}) we finally have
   \begin{eqnarray}\label{eq:ZZZ-1c}
                   \langle\Psi'''_1|\hat{\mathcal{Z}}_{\rm m}|\Psi'''_3 \rangle
                &=& -\frac{{\rm i}\,\hbar\,c}{2} \biggr\{
                        \cos\vartheta\,\langle\Psi'''_1|\vec{\alpha}(0)|\Psi'''_3\rangle
                        +\sin\vartheta\,\frac{\sqrt{p^2 c^2 +m^2 c^4}}{p\,c}
                              \langle\Psi'''_1|\vec{\zeta}(0)|\Psi'''_3\rangle
                              \nonumber \\
                        &&\qquad\qquad -\sin\vartheta\,\frac{m^2 c^4}{
                              \sqrt{p^2 c^2 +m^2 c^4}} \frac{1}{p\, c} \langle\Psi'''_1|
                                    \left[\vec{\zeta}(0)\cdot\hat{p}
                                          \right]|\Psi'''_3\rangle\hat{p}
                        \biggr\}\dfrac{1}{E_{\rm p}} \Bigl({\rm e}^{
                              {\rm i}\frac{2}{\hbar} E_{\rm p} t} -1\Bigr)\nonumber\\
     &=& -\frac{{\rm i}\,\hbar\,c}{2} \biggr\{
                        \cos\vartheta\,\left(1-{\rm i}\,\sin\vartheta \frac{p}{m\,c}\right)\langle\Psi_1|\vec{\alpha}(0) |\Psi_3 \rangle
                        +\sin\vartheta\,\frac{\sqrt{p^2 c^2 +m^2 c^4}}{p\,c}
                               \frac{{\rm i}  E_{\rm p}  }{m\,c^2} \cos\vartheta
                       \,\langle\Psi_1| \vec{\alpha}|\Psi_3\rangle
                              \nonumber \\
                        &&\qquad\qquad -\sin\vartheta\,\frac{m^2 c^4}{
                              \sqrt{p^2 c^2 +m^2 c^4}} \frac{1}{p\, c}  \frac{{\rm i}  E_{\rm p}  }{m\,c^2} \cos\vartheta
                       \,\langle\Psi_1| \vec{\alpha}|\Psi_3\rangle
                        \biggr\}\dfrac{1}{E_{\rm p}} \Bigl({\rm e}^{
                              {\rm i}\frac{2}{\hbar} E_{\rm p} t} -1\Bigr)\nonumber\\
   &=& -\frac{{\rm i}\,\hbar\,c}{2} \biggr\{
                        \cos\vartheta\,\left(1-{\rm i}\,\sin\vartheta \frac{p}{m\,c}\right)
                        +{\rm i}\sin\vartheta \cos\vartheta\,\frac{{p^2 c^2 +m^2 c^4}}{p\,c}
                               \frac{1 }{m\,c^2}
                                                     \nonumber \\
                        &&\qquad\qquad -{\rm i}\sin\vartheta \cos\vartheta\, \frac{1}{p\, c}  \frac{m^2 c^4 }{m\,c^2}
                                              \biggr\}\dfrac{1}{E_{\rm p}} \Bigl({\rm e}^{
                              {\rm i}\frac{2}{\hbar} E_{\rm p} t} -1\Bigr)\langle\Psi_1|\vec{\alpha}(0) |\Psi_3 \rangle\nonumber\\
   &=& \cos\vartheta\,\frac{-{\rm i}\,\hbar\,c}{2} \dfrac{1}{E_{\rm p}} \Bigl({\rm e}^{
                              {\rm i}\frac{2}{\hbar} E_{\rm p} t} -1\Bigr)\langle\Psi_1|\vec{\alpha}(0) |\Psi_3 \rangle \nonumber\\
   &=& \cos\vartheta\,  \langle\Psi_1|\hat{\mathcal{Z}}_{\rm e}|\Psi_3 \rangle.
            \end{eqnarray}
   When $\vartheta=0$, Eq. (\ref{eq:ZZZ-1c}) reduces to the result is Eq. (\ref{eq:D-22-a}), and when $\vartheta=\pi/2$, Eq. (\ref{eq:ZZZ-1c}) reduces to the result is Eq. (\ref{eq:13}). Thus the results are consistent.
   $\blacksquare$
   \end{remark}

      \subsection{More General Results}

      \begin{remark}
In the above, we have calculated only the case of $\langle\Psi'''_1|\hat{\mathcal{Z}}_{\rm m}|\Psi'''_3\rangle$. In this section, let us consider more general cases. To do so, we need to calculate the other three terms, i.e., $\langle\Psi'''_2|\hat{\mathcal{Z}}_{\rm m}|\Psi'''_4\rangle$, $\langle\Psi'''_1|\hat{\mathcal{Z}}_{\rm m}|\Psi'''_4\rangle$, and $\langle\Psi'''_2|\hat{\mathcal{Z}}_{\rm m}|\Psi'''_3\rangle$. $\blacksquare$
\end{remark}

            \begin{remark}
            By taking $j=2, k=4$, from Eq. (\ref{eq:ZZZ-1b})
            \begin{eqnarray}\label{eq:ZZZ-24-a}
                   \langle\Psi'''_2|\hat{\mathcal{Z}}_{\rm m}|\Psi'''_4 \rangle
                  &=& -\frac{{\rm i}\,\hbar\,c}{2} \biggr\{
                        \cos\vartheta\,\langle\Psi'''_2|\vec{\alpha}(0)|\Psi'''_4\rangle
                        +\sin\vartheta\,\frac{\sqrt{p^2 c^2 +m^2 c^4}}{p\,c}
                              \langle\Psi'''_2|\vec{\zeta}(0)|\Psi'''_4\rangle
                              \nonumber \\
                        &&\qquad\qquad -\sin\vartheta\,\frac{m^2 c^4}{
                              \sqrt{p^2 c^2 +m^2 c^4}} \frac{1}{p\, c} \langle\Psi'''_2|
                                    \left[\vec{\zeta}(0)\cdot\hat{p}
                                          \right]|\Psi'''_4\rangle\hat{p}
                        \biggr\}\dfrac{1}{E_{\rm p}} \Bigl({\rm e}^{
                              {\rm i}\frac{2}{\hbar} E_{\rm p} t} -1\Bigr).
            \end{eqnarray}
            Similarly, due to
             \begin{eqnarray}
        && \langle\Psi_2|\vec{\alpha}(0)|\Psi_4\rangle=\langle\Psi_1|\vec{\alpha}(0)|\Psi_3\rangle, \nonumber\\
  && \langle\Psi_2|[\vec{\alpha}(0)\cdot\hat{p}]\vec{\alpha}(0)|\Psi_4\rangle
  =\langle\Psi_1|[\vec{\alpha}(0)\cdot\hat{p}]\vec{\alpha}(0)|\Psi_3\rangle=0, \nonumber\\
     &&  \left[\left(\langle\Psi_2| \vec{\alpha}(0)|\Psi_4\rangle)\cdot\hat{p}\right)\right]
                            \hat{p}=\langle\Psi_2|\vec{\alpha}(0) |\Psi_4 \rangle,\nonumber\\
    && \langle\Psi'''_2| \vec{\zeta}(0)|\Psi'''_4\rangle\hat{p}=\langle\Psi'''_1| \vec{\zeta}(0)|\Psi'''_3\rangle\hat{p}=-{\rm i} \cos\vartheta \,\hat{p},\nonumber\\
   && \langle\Psi'''_2| \left[\vec{\zeta}(0)\cdot\hat{p} \right]|\Psi'''_4\rangle\hat{p}=\langle\Psi'''_1| \left[\vec{\zeta}(0)\cdot\hat{p} \right]|\Psi'''_3\rangle\hat{p}=-{\rm i} \cos\vartheta \,\hat{p}.
      \end{eqnarray}
           we finally have
   \begin{eqnarray}\label{eq:ZZZ-24-b}
                   \langle\Psi'''_2|\hat{\mathcal{Z}}_{\rm m}|\Psi'''_4 \rangle
                &=& \cos\vartheta\,  \langle\Psi_2|\hat{\mathcal{Z}}_{\rm e}|\Psi_4 \rangle.
            \end{eqnarray}
            $\blacksquare$
            \end{remark}

            \begin{remark}
                  By taking $j=1, k=4$, from Eq. (\ref{eq:ZZZ-1b})
            \begin{eqnarray}\label{eq:ZZZ-14-a}
                   \langle\Psi'''_1|\hat{\mathcal{Z}}_{\rm m}|\Psi'''_4 \rangle
                  &=& -\frac{{\rm i}\,\hbar\,c}{2} \biggr\{
                        \cos\vartheta\,\langle\Psi'''_1|\vec{\alpha}(0)|\Psi'''_4\rangle
                        +\sin\vartheta\,\frac{\sqrt{p^2 c^2 +m^2 c^4}}{p\,c}
                              \langle\Psi'''_1|\vec{\zeta}(0)|\Psi'''_4\rangle
                              \nonumber \\
                        &&\qquad\qquad -\sin\vartheta\,\frac{m^2 c^4}{
                              \sqrt{p^2 c^2 +m^2 c^4}} \frac{1}{p\, c} \langle\Psi'''_1|
                                    \left[\vec{\zeta}(0)\cdot\hat{p}
                                          \right]|\Psi'''_4\rangle\hat{p}
                        \biggr\}\dfrac{1}{E_{\rm p}} \Bigl({\rm e}^{
                              {\rm i}\frac{2}{\hbar} E_{\rm p} t} -1\Bigr).
            \end{eqnarray}
              We need to handle the following three terms:
             \begin{eqnarray}
             \langle\Psi'''_1|\vec{\alpha}(0)|\Psi'''_4\rangle,\qquad
                           \langle\Psi'''_1|\vec{\zeta}(0)|\Psi'''_4\rangle,\qquad
                              \langle\Psi'''_1| \left[\vec{\zeta}(0)\cdot\hat{p}
                                          \right]|\Psi'''_4\rangle\hat{p}.
            \end{eqnarray}

            (i) The first term. One can have
            \begin{eqnarray}
                  && \langle\Psi'''_1|\vec{\alpha}(0)|\Psi'''_4\rangle =  \langle\Psi_1|\left[\mathcal{D}^\dagger (\vartheta)\,\vec{\alpha}\,
                                    \mathcal{D}(\vartheta)\right]|\Psi_4\rangle\nonumber\\
                  &=&\langle\Psi_1|\biggr[ \Biggl\{\cos\vartheta+\dfrac{{\rm i}\,\sin\vartheta}{
                                          \sqrt{p^2 c^2 +m^2 c^4}} \bigg[
                                                \frac{(p^2 c^2 +m^2 c^4)}{m\,c^2} (
                                                      \vec{\alpha}\cdot\hat{p})
                                    -\,\frac{p\,c}{m\,c^2} H_{\rm e}\bigg]
                                    \Biggr\}\vec{\alpha} \notag \\
                              && +\dfrac{2\,m\,c^2}{(p^2 c^2 +m^2 c^4)} {\sin^2}\Bigl(
                                    \frac{\vartheta}{2}\Bigr)\,\bigg[
                                          \frac{(p^2 c^2 +m^2 c^4)}{m\,c^2} (
                                                \vec{\alpha}\cdot\hat{p})
                                    -\,\frac{p\,c}{m\,c^2} H_{\rm e}\bigg]\hat{p}
                              -{\rm i}\dfrac{m\,c^2 \sin\vartheta}{\sqrt{p^2 c^2 +m^2 c^4}}
                                    \hat{p}\biggr]|\Psi_4\rangle\nonumber\\
                  &=&\langle\Psi_1|\biggr[ \Biggl\{\cos\vartheta+\dfrac{{\rm i}\,\sin\vartheta}{
                                          \sqrt{p^2 c^2 +m^2 c^4}} \bigg[
                                                \frac{(p^2 c^2 +m^2 c^4)}{m\,c^2} (
                                                      \vec{\alpha}\cdot\hat{p})
                                    -\,\frac{p\,c}{m\,c^2} E_{\rm p}\bigg]
                                    \Biggr\}\vec{\alpha} \notag \\
                              && +\dfrac{2\,m\,c^2}{(p^2 c^2 +m^2 c^4)} {\sin^2}\Bigl(
                                    \frac{\vartheta}{2}\Bigr)\,\bigg[
                                          \frac{(p^2 c^2 +m^2 c^4)}{m\,c^2} (
                                                \vec{\alpha}\cdot\hat{p})
                              \bigg]\hat{p} \biggr]|\Psi_4\rangle\nonumber\\
                  &=&\langle\Psi_1|\biggr[ \Biggl\{\cos\vartheta-{\rm i}\,\sin\vartheta \frac{p}{m\,c}
                  +{\rm i}\,\sin\vartheta\bigg[
                                                \frac{\sqrt{p^2 c^2 +m^2 c^4}}{m\,c^2} (
                                                      \vec{\alpha}\cdot\hat{p})
                              \bigg]
                                    \Biggr\}\vec{\alpha} +2 {\sin^2}\Bigl(
                                    \frac{\vartheta}{2}\Bigr)\, (
                                                \vec{\alpha}\cdot\hat{p})
                              \hat{p} \biggr]|\Psi_4\rangle\nonumber\\
                  &=& \left(\cos\vartheta-{\rm i}\,\sin\vartheta \frac{p}{m\,c}\right)\langle\Psi_1|\vec{\alpha}|\Psi_4\rangle+2 {\sin^2}\Bigl(
                                    \frac{\vartheta}{2}\Bigr)\, \left[\left(\langle\Psi_1|
                                                \vec{\alpha}|\Psi_4\rangle)\cdot\hat{p}\right)\right]
                              \hat{p}+
                  {\rm i}\,\sin\vartheta \frac{\sqrt{p^2 c^2 +m^2 c^4}}{m\,c^2}\langle\Psi_1|(
                                                      \vec{\alpha}\cdot\hat{p})
                                                            \vec{\alpha}|\Psi_4\rangle \nonumber\\
                  &=& \left(\cos\vartheta-{\rm i}\,\sin\vartheta \frac{p}{m\,c}\right)
                              \langle\Psi_1|\vec{\alpha}|\Psi_4\rangle
                        -{\rm i}\,\sin\vartheta\,\frac{c\,p}{E_{\rm p}}
                              \frac{\sqrt{p^2 c^2 +m^2 c^4}}{m\,c^2}
                              \langle\Psi_1|\vec{\alpha}|\Psi_4\rangle \nonumber\\
                  &=& \left(\cos\vartheta-{\rm i}\,2\,\sin\vartheta \frac{p}{m\,c}
                        \right)\langle\Psi_1|\vec{\alpha}|\Psi_4\rangle.
            \end{eqnarray}
            since
            \begin{align}
                  \langle\Psi_1|\big[\vec{\alpha}(0)\cdot\hat{p}\big]|\Psi_4\rangle=0,
            \end{align}
            and
            \begin{eqnarray}
                  \langle\Psi_1|(\vec{\alpha}\cdot\hat{p})\vec{\alpha}|\Psi_4\rangle
                  &=& \sum_{j=1}^4 \left[\left( \langle\Psi_1|(\vec{\alpha}\cdot\hat{p})\, |\Psi_j\rangle\right) \left(\langle\Psi_j| \vec{\alpha}|\Psi_4\rangle\right)\right]
                  =\left[\left( \langle\Psi_1|(\vec{\alpha}\cdot\hat{p})\, |\Psi_1\rangle\right) \left(\langle\Psi_1| \vec{\alpha}|\Psi_4\rangle\right)\right]\nonumber\\
                 & =&\frac{-cp}{E_{\rm p}} \langle\Psi_1| \vec{\alpha}|\Psi_4\rangle.
            \end{eqnarray}
          Because
            \begin{eqnarray}
                         \langle\Psi'''_1| \vec{\zeta}(0) |\Psi'''_4\rangle
                        &=& \langle\Psi_1| \biggr\{ \bigg[\cos^2 \Bigl(\frac{\vartheta}{2}\Bigr)
                                                +\sin^2 \Bigl(\frac{\vartheta}{2}\Bigr)\,
                                                      \dfrac{(m^2 c^4 -p^2 c^2)}{p^2 c^2 +m^2 c^4}
                                                \bigg]{\rm i}\,\frac{1}{m\,c^2} H_{\rm e}
                                    \,\vec{\alpha}\nonumber\\
                                    && +\sin^2 \Bigl(\frac{\vartheta}{2}\Bigr)\dfrac{1}{
                                                p^2 c^2 +m^2 c^4} \big[  -m^2 c^4 {\rm i}\,2\,\frac{1}{m\,c^2} \big[H_{\rm e}(\vec{\alpha}\cdot\hat{p})\hat{p} -c\vec{p}
                                          \big]\big] \nonumber\\
                                                      && -{\rm i}\,\sin\Bigl(\frac{\vartheta}{2}\Bigr)\cos\Bigl(
                                                \frac{\vartheta}{2}\Bigr)\dfrac{1
                                                      }{\sqrt{p^2 c^2 +m^2 c^4}} {\rm i}\,2 \left[
                                                      c\,p\vec{\alpha}-c(\vec{\alpha}\cdot\vec{p})\hat{p}\right]\biggr\} |\Psi_4\rangle\nonumber\\
                        &=& \langle\Psi_1|\biggr\{
                              \bigg[\cos^2 \Bigl(\frac{\vartheta}{2}\Bigr)
                                    +\sin^2 \Bigl(\frac{\vartheta}{2}\Bigr)\,
                                          \dfrac{(m^2 c^4 -p^2 c^2)}{p^2 c^2 +m^2 c^4}
                                    \bigg]{\rm i}\,\frac{1}{m\,c^2} H_{\rm e}
                                          \vec{\alpha} \notag \\
                              &&\qquad\ +\sin^2 \Bigl(\frac{\vartheta}{2}
                                    \Bigr)\dfrac{1}{p^2 c^2 +m^2 c^4} \big[
                                          -m^2 c^4 {\rm i}\,2\,\frac{1}{m\,c^2} \big[
                                                H_{\rm e}(\vec{\alpha}\cdot\hat{p}
                                                      )\hat{p}\big]\big]
                              +\sin\vartheta\,\dfrac{c\,p}{\sqrt{p^2 c^2 +m^2 c^4}}
                                    \vec{\alpha}\biggr\}|\Psi_4\rangle \nonumber\\
                        &=&   \bigg[\cos^2 \Bigl(\frac{\vartheta}{2}\Bigr)
                                                +\sin^2 \Bigl(\frac{\vartheta}{2}\Bigr)\,
                                                      \dfrac{(m^2 c^4 -p^2 c^2)}{p^2 c^2 +m^2 c^4}
                                                \bigg]{\rm i}\,\frac{1}{m\,c^2} E_{\rm p}
                                    \,\langle\Psi_1| \vec{\alpha}|\Psi_4\rangle
                              +\sin\vartheta\,\dfrac{c\,p}{\sqrt{p^2 c^2 +m^2 c^4}}
                                    \langle\Psi_1| \vec{\alpha}|\Psi_4\rangle
                                    \nonumber\\
                                    && +\sin^2 \Bigl(\frac{\vartheta}{2}\Bigr)\dfrac{1}{
                                                p^2 c^2 +m^2 c^4} \big[  -m^2 c^4 {\rm i}\,2\,\frac{1}{m\,c^2} E_{\rm p} \big] \langle\Psi_1|(\vec{\alpha}\cdot\hat{p})\hat{p}
                                          |\Psi_4\rangle \nonumber\\
                        &=& {\rm i}\frac{E_{\rm p}}{m\,c^2} \bigg[
                                    \cos^2 \Bigl(\frac{\vartheta}{2}\Bigr)
                                    +\sin^2 \Bigl(\frac{\vartheta}{2}\Bigr)\,
                                          \dfrac{(m^2 c^4 -p^2 c^2)}{p^2 c^2 +m^2 c^4}
                                    \bigg]\langle\Psi_1|\vec{\alpha}|\Psi_4\rangle
                              +\sin\vartheta\,\dfrac{c\,p}{\sqrt{p^2 c^2 +m^2 c^4}}
                                    \langle\Psi_1| \vec{\alpha}|\Psi_4\rangle,
                  \end{eqnarray}
                  and
                  \begin{eqnarray}
                        && \langle\Psi'''_1|\left[\vec{\zeta}(0)\cdot\hat{p}\right]|
                              \Psi'''_4\rangle\hat{p}
                        =\left[\langle\Psi'''_1|\vec{\zeta}(0)|\Psi'''_4\rangle
                              \cdot\hat{p}\right]\hat{p} \notag \\
                       & =& {\rm i}\frac{E_{\rm p}}{m\,c^2} \bigg[
                                    \cos^2 \Bigl(\frac{\vartheta}{2}\Bigr)
                                    +\sin^2 \Bigl(\frac{\vartheta}{2}\Bigr)\,
                                          \dfrac{(m^2 c^4 -p^2 c^2)}{p^2 c^2 +m^2 c^4}
                                    \bigg]\langle\Psi_1|(\vec{\alpha}\cdot\hat{p}
                                          )|\Psi_4\rangle\hat{p}
                              -{\rm i}\frac{E_{\rm p}}{m\,c^2} {\sin^2}\Bigl(
                                    \frac{\vartheta}{2}
                                    \Bigr)\dfrac{2\,m^2 c^4 }{p^2 c^2 +m^2 c^4}
                                          \langle\Psi_1|(\vec{\alpha}\cdot\hat{p})|
                                                \Psi_4\rangle\hat{p} \notag \\
                       & =& 0,
                  \end{eqnarray}
                  then we obtain
                  \begin{eqnarray}\label{eq:ZZZ-14-b}
                        && \langle\Psi'''_1|\hat{\mathcal{Z}}_{\rm m}|\Psi'''_4
                              \rangle \notag \\
                        &=& -\frac{{\rm i}\,\hbar\,c}{2} \biggr\{
                              \cos\vartheta\,\langle\Psi'''_1|\vec{\alpha}(0)|\Psi'''_4\rangle
                              +\sin\vartheta\,\frac{\sqrt{p^2 c^2 +m^2 c^4}}{p\,c}
                                    \langle\Psi'''_1|\vec{\zeta}(0)|\Psi'''_4\rangle
                                    \nonumber \\
                              &&\qquad\quad\ -\sin\vartheta\,\frac{m^2 c^4}{
                                    \sqrt{p^2 c^2 +m^2 c^4}} \frac{1}{p\, c} \langle\Psi'''_1|
                                          \left[\vec{\zeta}(0)\cdot\hat{p}
                                                \right]|\Psi'''_4\rangle\hat{p}
                              \biggr\}\dfrac{1}{E_{\rm p}} \Bigl({\rm e}^{
                                    {\rm i}\frac{2}{\hbar} E_{\rm p} t} -1\Bigr)
                              \notag \\
                        &=& -\frac{{\rm i}\,\hbar\,c}{2} \biggr\{
                              \cos\vartheta\,\left(\cos\vartheta
                                    -{\rm i}\,2\,\sin\vartheta\,\frac{p}{m\,c}
                                    \right)\langle\Psi_1|\vec{\alpha}|\Psi_4\rangle
                              +{\sin^2}\vartheta\,\langle\Psi_1| \vec{\alpha}|
                                    \Psi_4\rangle \notag \\
                              &&\qquad\quad\ +{\rm i}\sin\vartheta\,
                                    \frac{\sqrt{p^2 c^2 +m^2 c^4}}{p\,c}
                                    \frac{E_{\rm p}}{m\,c^2} \bigg[
                                          \cos^2 \Bigl(\frac{\vartheta}{2}\Bigr)
                                          +\sin^2 \Bigl(\frac{\vartheta}{2}\Bigr)\,
                                                \dfrac{(m^2 c^4 -p^2 c^2)}{
                                                      p^2 c^2 +m^2 c^4}
                                          \bigg]\langle\Psi_1|\vec{\alpha}|
                                                \Psi_4\rangle
                              \biggr\}\dfrac{1}{E_{\rm p}} \Bigl({\rm e}^{
                                    {\rm i}\frac{2}{\hbar} E_{\rm p} t} -1\Bigr)
                              \notag \\
                        &=& \biggr\{1-{\rm i}\,\sin(2\vartheta)\,\frac{p}{m\,c}
                              +{\rm i}\sin\vartheta\,\frac{1}{
                                    p\,c} \frac{E_{\rm p}^2}{m\,c^2} \bigg[
                                          \cos^2 \Bigl(\frac{\vartheta}{2}\Bigr)
                                          +\sin^2 \Bigl(\frac{\vartheta}{2}\Bigr)\,
                                                \dfrac{(m^2 c^4 -p^2 c^2)}{
                                                      p^2 c^2 +m^2 c^4}\bigg]
                              \biggr\}\langle\Psi_1|\hat{\mathcal{Z}}_{\rm e}|
                                    \Psi_4\rangle\nonumber\\
                  &=& \biggr\{1-{\rm i}\,\sin(2\vartheta)\,\frac{p}{m\,c}
                              +{\rm i}\sin\vartheta\,\frac{1}{
                                    p\,c} \frac{1}{m\,c^2} \bigg[
                                          \cos^2 \Bigl(\frac{\vartheta}{2}\Bigr)\, ( p^2 c^2 +m^2 c^4)
                                          +\sin^2 \Bigl(\frac{\vartheta}{2}\Bigr)\,(m^2 c^4 -p^2 c^2)
                                                \bigg]
                              \biggr\}\langle\Psi_1|\hat{\mathcal{Z}}_{\rm e}|
                                    \Psi_4\rangle  \nonumber\\
                   &=& \biggr\{1-{\rm i}\,\sin(2\vartheta)\,\frac{p}{m\,c}
                              +{\rm i}\sin\vartheta\,\frac{1}{
                                    p\,c} \frac{1}{m\,c^2} \bigg[
                                          m^2 c^4 + p^2 c^2  \cos\vartheta \bigg]
                              \biggr\}\langle\Psi_1|\hat{\mathcal{Z}}_{\rm e}|
                                    \Psi_4\rangle\nonumber\\
                   &=& \biggr\{1-{\rm i}\,\sin(2\vartheta)\,\frac{p}{m\,c}
                              +{\rm i}\sin\vartheta\,\bigg[
                                         \frac{mc}{p} + \frac{p}{mc}  \cos\vartheta \bigg]
                              \biggr\}\langle\Psi_1|\hat{\mathcal{Z}}_{\rm e}|
                                    \Psi_4\rangle\nonumber\\
                   &=& \biggr[1 + {\rm i}\sin\vartheta \frac{mc}{p}-{\rm i}\,\sin(2\vartheta)\,\frac{p}{2 m\,c}
                                            \biggr]\langle\Psi_1|\hat{\mathcal{Z}}_{\rm e}|
                                    \Psi_4\rangle,
                  \end{eqnarray}
                  here we have used the following relation
                  \begin{eqnarray}
                        \langle\Psi_1|\hat{\mathcal{Z}}_{\rm e}|\Psi_4\rangle
                        &=& \frac{-{\rm i}\hbar c}{2 E_{\rm p}} \left({\rm e}^{
                              \frac{{\rm i}\,2\,E_{\rm p} t}{\hbar}} -1
                              \right)\langle\Psi_1|\vec{\alpha}(0)|\Psi_4\rangle.
                  \end{eqnarray}
                  Obviously, when $\vartheta=0$ and $\vartheta=\pi/2$, Eq. (\ref{eq:ZZZ-14-b}) reduces to the previous results.
                  $\blacksquare$
            \end{remark}

            \begin{remark}
                  Similarly, for the superposition of $\Ket{\Psi'''_2}$ and $\Ket{\Psi'''_3}$, i.e. $j=2$, and $k=3$.
                  \begin{align}
                        & \langle\Psi_2|\big[\vec{\alpha}(0)\cdot\hat{p}\big]\hat{p}|
                              \Psi_3\rangle
                        =\Bigl\{\big[\langle\Psi_2|\vec{\alpha}(0)|\Psi_3\rangle\big]
                              \cdot\hat{p}\Bigr\}\hat{p}
                        =\Bigl\{\big[-\langle\Psi_1|\vec{\alpha}(0)|\Psi_4\rangle^*\big]
                              \cdot\hat{p}\Bigr\}\hat{p}
                        =0,
                  \end{align}
                  and
                  \begin{eqnarray}
                        \langle\Psi_2|(\vec{\alpha}\cdot\hat{p})\vec{\alpha}|\Psi_3\rangle
                        &=& \sum_{j=1}^4 \left[\left( \langle\Psi_2|(\vec{\alpha}\cdot\hat{p})\, |\Psi_j\rangle\right) \left(\langle\Psi_j| \vec{\alpha}|\Psi_3\rangle\right)\right]
                        =\left[\left( \langle\Psi_2|(\vec{\alpha}\cdot\hat{p})\, |\Psi_2\rangle\right) \left(\langle\Psi_2| \vec{\alpha}|\Psi_3\rangle\right)\right]
                        =\frac{-cp}{E_{\rm p}} \langle\Psi_2| \vec{\alpha}|\Psi_3\rangle.
                  \end{eqnarray}
                  which indicates
                  \begin{eqnarray}
                        && \langle\Psi'''_2|\hat{\mathcal{Z}}_{\rm m}|\Psi'''_3
                              \rangle \notag \\
                        &=& -\frac{{\rm i}\,\hbar\,c}{2} \biggr\{
                              \cos\vartheta\,\langle\Psi'''_2|\vec{\alpha}(0)|\Psi'''_3\rangle
                              +\sin\vartheta\,\frac{\sqrt{p^2 c^2 +m^2 c^4}}{p\,c}
                                    \langle\Psi'''_2|\vec{\zeta}(0)|\Psi'''_3\rangle
                                    \nonumber \\
                              &&\qquad\quad\ -\sin\vartheta\,\frac{m^2 c^4}{
                                    \sqrt{p^2 c^2 +m^2 c^4}} \frac{1}{p\, c} \langle\Psi'''_2|
                                          \left[\vec{\zeta}(0)\cdot\hat{p}
                                                \right]|\Psi'''_3\rangle\hat{p}
                              \biggr\}\dfrac{1}{E_{\rm p}} \Bigl({\rm e}^{
                                    {\rm i}\frac{2}{\hbar} E_{\rm p} t} -1\Bigr)
                              \notag \\
                        &=& \biggr\{1-{\rm i}\,\sin(2\vartheta)\,\frac{p}{m\,c}
                              +{\rm i}\sin\vartheta\,\frac{1}{
                                    p\,c} \frac{E_{\rm p}^2}{m\,c^2} \bigg[
                                          \cos^2 \Bigl(\frac{\vartheta}{2}\Bigr)
                                          +\sin^2 \Bigl(\frac{\vartheta}{2}\Bigr)\,
                                                \dfrac{(m^2 c^4 -p^2 c^2)}{
                                                      p^2 c^2 +m^2 c^4}\bigg]
                              \biggr\}\langle\Psi_2|\hat{\mathcal{Z}}_{\rm e}|
                                    \Psi_3\rangle\nonumber\\
                  &=& \biggr[1 + {\rm i}\sin\vartheta \frac{mc}{p}-{\rm i}\,\sin(2\vartheta)\,\frac{p}{2 m\,c}
                                            \biggr]\langle\Psi_2|\hat{\mathcal{Z}}_{\rm e}|
                                    \Psi_3\rangle,
                  \end{eqnarray}
                  for
                  \begin{eqnarray}
                        && \langle\Psi_2|\hat{\mathcal{Z}}_{\rm e}|\Psi_3\rangle
                        =\frac{-{\rm i}\hbar\,c}{2\,E_{\rm p}} \left({\rm e}^{
                              \frac{{\rm i}\,2\,E_{\rm p} t}{\hbar}} -1\right)\langle\Psi_2|\vec{\alpha}(0)|\Psi_3\rangle.
                  \end{eqnarray}
            Analogously, we have the following observation:

            \emph{Observation 3.}---For the mixed Hamiltonian system of Dirac's electron $H_{\rm e}$ and the type-II Dirac's braidon $H_{\rm b}^{\rm II}$, i.e., $\mathcal{H}_{\rm m}=\cos\vartheta\;H_{\rm e}+\sin\vartheta\;H_{\rm b}^{\rm II}$, if it is in a superposition state
            $|\Psi'''\rangle=\cos\eta |\Psi'''_1\rangle+ \sin\eta |\Psi'''_4\rangle$, or $|\Psi'''\rangle=\cos\eta |\Psi'''_2\rangle+ \sin\eta |\Psi'''_3\rangle$, then the amplitude of ``position Zitterbewegung'' is tuned by a factor
            \begin{eqnarray}
                  && \gamma(\vartheta)=\biggr[1 + {\rm i}\sin\vartheta \frac{mc}{p}-{\rm i}\,\sin(2\vartheta)\,\frac{p}{2 m\,c}
                                            \biggr]
             \end{eqnarray}
            in comparison to Dirac's electron $H_{\rm e}$.
            $\blacksquare$
            \end{remark}

            \begin{remark}We can summarize the above results as the following Table \ref{tab:pz5}:

              \begin{table}[h]
	\centering
\caption{The results of ``position Zitterbewegung'' of the $H_{\rm e}$-$H_{\rm b}^{\rm II}$ mixing. $|\Psi'''_j\rangle$'s  ($j=1, 2, 3, 4$) are four eigenstates of the mixed Hamiltonian $\mathcal{H}_{\rm m}=\cos\vartheta\;H_{\rm e}
                        +\sin\vartheta\;H_{\rm b}^{\rm II}$, The ``position Zitterbewegung'' operator reads $\hat{\mathcal{Z}}_{\rm m}^r
                  =\frac{{\rm i}\hbar c }{2} \biggr\{\cos\vartheta\,\vec{\alpha}(0)+\sin\vartheta\, \frac{\sqrt{p^2c^2+m^2c^4}}{p\,c} \vec{\zeta}(0) - \sin\vartheta\, \frac{m^2c^4}{\sqrt{p^2c^2+m^2c^4}}\frac{1}{p\, c}\;\left[\vec{\zeta}(0)\cdot \hat{p}\right]\hat{p} -(\mathcal{H}_{\rm m}')^{-1} c\,\vec{p}\biggr\} \dfrac{1}{\mathcal{H}_{\rm m}'} \Bigl({\rm e}^{
                                                      -{\rm i}\frac{2}{\hbar}
                                                            \mathcal{H}_{\rm m}' t} -1\Bigr)$.
One has $\langle\Psi'''_j|\hat{\mathcal{Z}}_{\rm m}^{r}|\Psi'''_j\rangle=0$, $\langle\Psi'''_k|\hat{\mathcal{Z}}_{\rm m}^{r}|\Psi'''_l\rangle
                   =  -\frac{{\rm i}\,\hbar\,c}{2} \biggr\{
                        \cos\vartheta\,\langle\Psi'''_k|\vec{\alpha}(0)|\Psi'''_l\rangle
                        +\sin\vartheta\,\frac{\sqrt{p^2 c^2 +m^2 c^4}}{p\,c}
                              \langle\Psi'''_k|\vec{\zeta}(0)|\Psi'''_l\rangle
                               -\sin\vartheta\,\frac{m^2 c^4}{
                              \sqrt{p^2 c^2 +m^2 c^4}} \frac{1}{p\, c} \langle\Psi'''_k|
                                    \left[\vec{\zeta}(0)\cdot\hat{p}
                                          \right]|\Psi'''_l\rangle\hat{p}
                        \biggr\}\dfrac{1}{E_{\rm p}} \Bigl({\rm e}^{
                              {\rm i}\frac{2}{\hbar} E_{\rm p} t} -1\Bigr)$ for $k\in\{1,2\}$, $l\in\{3,4\}$, and $\gamma(\vartheta)=\biggr[1 + {\rm i}\sin\vartheta \frac{mc}{p}-{\rm i}\,\sin(2\vartheta)\,\frac{p}{2 m\,c}
                                            \biggr]$.}
\begin{tabular}{lllll}
\hline\hline
 & $|\Psi'''_1\rangle$ &  $|\Psi'''_2\rangle$& $|\Psi'''_3\rangle$ & $|\Psi'''_4\rangle$ \\
  \hline
$\langle \Psi'''_1| \hat{\mathcal{Z}}_{\rm m}^r$ \;\;\;\quad& 0&0 & $ \cos\vartheta\, \Delta_1\,\dfrac{-m\,c^2}{E_{\rm p}}\, \hat{p}$  \;\;\;\quad& $\gamma(\vartheta)\, \Delta_1\,(\vec{F}_1 +{\rm i}\,\vec{F}_2)$  \\
 \hline
$\langle \Psi'''_2| \hat{\mathcal{Z}}_{\rm m}^r$\;\;\;\quad &0 &0 & $-\gamma(\vartheta)\,\Delta_1\,(\vec{F}_1 +{\rm i}\,\vec{F}_2)^* $ \quad\quad& $\cos\vartheta\, \Delta_1\,\dfrac{-m\,c^2}{E_{\rm p}}\, \hat{p}$ \\
 \hline
 $\langle \Psi'''_3| \,\hat{\mathcal{Z}}_{\rm m}^r$\;\;\;\quad & $ \cos\vartheta\, \Delta_1^*\,\dfrac{-m\,c^2}{E_{\rm p}}\,\hat{p} \quad$ \;\;\;& $-\gamma^*(\vartheta)\,\Delta_1^*\,(\vec{F}_1 +{\rm i}\,\vec{F}_2) \quad\quad$ & 0 & 0 \\
 \hline
 $\langle \Psi'''_4| \hat{\mathcal{Z}}_{\rm m}^r$ \;\;\;\quad& $\gamma^*(\vartheta)\,\Delta_1^*\,(\vec{F}_1 +{\rm i}\,\vec{F}_2)^* \quad$ \quad&$\cos\vartheta\, \Delta_1^*\,\dfrac{-m\,c^2}{E_{\rm p}}\, \hat{p}$ \;\;\;& 0 & 0 \\
 \hline\hline
\end{tabular}\label{tab:pz5}
\end{table}
    $\blacksquare$
    \end{remark}

\newpage

\part{The Spin Zitterbewegung of Dirac's Electron and Dirac's Braidons}

\section{Spin Zitterbewegung for $H_{\rm e}$}

 Spin Zitterbewegung (SZ) is another kind of relativistic effect. It is easy to calculate the ``spin Zitterbewegung'', if one has known the corresponding ``position Zitterbewegung'' (PZ).

  \begin{remark}
   The spin operator has been introduced as
            \begin{equation}
                  \vec{S}=\dfrac{\hbar}{2} \vec{\Sigma}\equiv\dfrac{\hbar}{2}
                        \begin{bmatrix}
                              \vec{\sigma} & 0 \\
                              0 & \vec{\sigma}
                        \end{bmatrix}.
            \end{equation}
            Here we introduce a simple method derive the ``spin Zitterbewegung'' operator, which shows a direct connection with the ``position Zitterbewegung'' operator.

            Let us introduce the total angular momentum operator as
            \begin{equation}
            \vec{J}=\vec{\ell}+\vec{S},
            \end{equation}
            with
            \begin{equation}
            \vec{\ell}=\vec{r}\times \vec{p}
            \end{equation}
            being the orbital angular momentum operator. It is well-known that the operator $\vec{J}$ is conserved for a free Dirac's electron, i.e.,
            \begin{equation}
            [\vec{J}, H_{\rm e}]=0,
            \end{equation}
            which leads to
            \begin{equation}\label{eq:total-1}
            \vec{J}(t)=\vec{J}(0)
            \end{equation}
            for any time $t$. Based on Eq. (\ref{eq:total-1}) one obtains
            \begin{equation}\label{eq:total-2}
            \vec{\ell}(t)+\vec{S}(t)=\vec{\ell}(0)+\vec{S}(0),
            \end{equation}
            i.e., the time evolution of the spin operator $\vec{S}(t)$ is given by
            \begin{eqnarray}
                        \vec{S}(t)&=&\vec{L}(0)+\vec{S}(0)-\vec{L}(t)=\vec{r}(0)\times\vec{p}(0)+\vec{S}(0)-\vec{r}(t)\times\vec{p}(t)= \vec{r}(0)\times\vec{p}+\vec{S}(0)-\vec{r}(t)\times\vec{p}\nonumber\\
                        &=& \vec{r}(0)\times\vec{p}+\vec{S}(0)-\biggl\{
                              \vec{r}(0)+c^2 H^{-1}_{\rm e} \vec{p}\,t
                              +\dfrac{{\rm i}\hbar c}{2}\left[\vec{\alpha}(0)
                                    -c\,H^{-1}_{\rm e} \vec{p}\right] H^{-1}_{\rm e}
                                          \left({\rm e}^{
                                                \frac{-{\rm i}\,2\,H_{\rm e}t}{\hbar}}
                                                -1\right)\biggr\}\times\vec{p} \nonumber\\
                        &=& \vec{r}(0)\times\vec{p}+\vec{S}(0)-\biggl\{
                              \vec{r}(0)+c^2 H^{-1}_{\rm e} \vec{p}\,t
                              +\hat{\mathcal{Z}}_r\biggr\}\times\vec{p} = \vec{r}(0)\times\vec{p}+\vec{S}(0)-\biggl\{
                              \vec{r}(0)\times\vec{p}+\hat{\mathcal{Z}}^r \times\vec{p}\biggr\}\nonumber\\
                        &=& \vec{S}(0)-\hat{\mathcal{Z}}^r \times\vec{p}.
            \end{eqnarray}
            Based on $\vec{S}(t)$, we immediately extract out its second term as the SZ operator, i.e.,
            \begin{eqnarray}
            \hat{\mathcal{Z}}^s=-\hat{\mathcal{Z}}^r \times\vec{p},
            \end{eqnarray}
            or explicitly
            \begin{eqnarray}
            \hat{\mathcal{Z}}^s=-\hat{\mathcal{Z}}^r \times\vec{p}=-\dfrac{{\rm i}\hbar c}{2}\left[
                                    \vec{\alpha}(0)\times\vec{p}\right] H^{-1}_{\rm e}
                                          \left({\rm e}^{\frac{-{\rm i}\,2\,H_{\rm e}t}{
                                                \hbar}} -1\right).
            \end{eqnarray}
            When Dirac's electron is in a quantum state $|\Psi\rangle$, the expectation value of the SZ operator reads
            \begin{eqnarray}\label{eq:sr-1}
                  \mathcal{Z}^s=\langle \Psi| \hat{\mathcal{Z}}^s |\Psi\rangle= -\langle \Psi| \hat{\mathcal{Z}}^r |\Psi\rangle \times\vec{p}=-{\mathcal{Z}}^r\times\vec{p}.
            \end{eqnarray}
            If $\mathcal{Z}^s\neq 0$, we then say that there is a SZ phenomenon for Dirac's electron. Based on Eq. (\ref{eq:sr-1}), we easily know that if ${\mathcal{Z}}^r=0$, or the direction of ${\mathcal{Z}}^r$ is $\hat{p}$, then we must have $\mathcal{Z}^s = 0$. $\blacksquare$
             \end{remark}

  \begin{table}[t]
	\centering
\caption{The results of ``spin Zitterbewegung'' of Dirac' electron. Here the ``spin Zitterbewegung'' operator reads  $\hat{\mathcal{Z}}_{\rm e}^{\rm s}=-\hat{\mathcal{Z}}_{\rm e}^r \times\vec{p}$, and $\langle\Psi_j|\hat{\mathcal{Z}}_{\rm e}^{\rm s}|\Psi_k \rangle=-\langle\Psi_j|\hat{\mathcal{Z}}_{\rm e}^r|\Psi_k \rangle \times \vec{p}
            = \langle\Psi_j|\hat{\mathcal{Z}}_{\rm e}^r|\Psi_k \rangle \times (-\vec{p})$.}
\begin{tabular}{lllll}
\hline\hline
 & $|\Psi_1\rangle$ &  $|\Psi_2\rangle$& $|\Psi_3\rangle$ & $|\Psi_4\rangle$ \\
  \hline
$\langle \Psi_1| \hat{\mathcal{Z}}_{\rm e}^{\rm s}$ \;\;\;\quad& 0&0 & $0$  \;\;\;\quad& $\Delta_1\,(\vec{F}_1 +{\rm i}\,\vec{F}_2)\times (-\vec{p})$  \\
 \hline
$\langle \Psi_2| \hat{\mathcal{Z}}_{\rm e}^{\rm s}$\;\;\;\quad &0 &0 & $-\Delta_1\,(\vec{F}_1 +{\rm i}\,\vec{F}_2)^*\times (-\vec{p}) $ \quad\quad& $0$ \\
 \hline
 $\langle \Psi_3| \,\hat{\mathcal{Z}}_{\rm e}^{\rm s}$\;\;\;\quad & $0$ \;\;\;& $-\Delta_1^*\,(\vec{F}_1 +{\rm i}\,\vec{F}_2)\times (-\vec{p}) \quad\quad$ & 0 & 0 \\
 \hline
 $\langle \Psi_4| \hat{\mathcal{Z}}_{\rm e}^{\rm s}$ \;\;\;\quad& $\Delta_1^*\,(\vec{F}_1 +{\rm i}\,\vec{F}_2)^* \times (-\vec{p})\quad$ \quad &0 & 0 & 0 \\
 \hline\hline
\end{tabular}\label{tab:sz1}
\end{table}

We have listed the results of ``spin Zitterbewegung''  in Table \ref{tab:sz1}. From Table \ref{tab:sz1}, for the ``spin Zitterbewegung'' we have the following observation.

            \emph{Observation 3.---}Only when Dirac's electron is in a superposition state with \emph{opposite} energy and \emph{opposite} helicity, is there a ``spin Zitterbewegung'' phenomenon.

            \begin{proof} We need to calculate the expectation value of the SZ operator. From Eq. (\ref{eq:sr-1}) we have known $\mathcal{Z}^s= 0$
             if $\mathcal{Z}^r= 0$. Based on Table \ref{tab:pz1}, to obtain the non-zero expectation value of $\mathcal{Z}^s$, at least Dirac's electron is in a superposition state with opposite energy. In the following, we further show that $\mathcal{Z}^s= 0$ if Dirac's electron is in a superposition state with the same helicity. Let us introduce the projective operators
            \begin{eqnarray}
                  \Pi^s_\pm =\frac{1}{2}\left(\mathbb{I}\pm \dfrac{2}{\hbar} \hat{\Lambda}
                        \right),\qquad
                  (\Pi^s_\pm)^2 =\Pi^s_\pm,
            \end{eqnarray}
            we easily have
            \begin{eqnarray}
                  && \Pi^s_+ |\Psi_1\rangle = |\Psi_1\rangle, \;\;
                  \Pi^s_+ |\Psi_2\rangle = 0, \;\;
                  \Pi^s_+ |\Psi_3\rangle = |\Psi_3\rangle, \;\;
                  \Pi^s_+ |\Psi_4\rangle = 0, \nonumber\\
                  && \Pi^s_- |\Psi_1\rangle = 0, \;\;
                  \Pi^s_- |\Psi_2\rangle = |\Psi_2\rangle, \;\;
                  \Pi^s_- |\Psi_3\rangle = 0, \;\;
                  \Pi^s_- |\Psi_4\rangle = |\Psi_4\rangle.
            \end{eqnarray}
            Because
            \begin{eqnarray}
                    \bigl\{\hat{\Lambda},\ \left[\vec{\alpha}(0)\times\vec{p}
                              \right]\bigr\}
                        &=&\dfrac{\hbar}{2} \Bigl\{\vec{\Sigma}\cdot\hat{p},\
                              \vec{\alpha}\times\vec{p}\Bigr\}
                        =\dfrac{\hbar}{2} \Bigl\{\vec{\Sigma}\cdot\hat{p},\
                              \vec{\alpha}\Bigr\}\times\vec{p} =\dfrac{\hbar}{2} \Biggl(\begin{bmatrix}
                                          \vec{\sigma}\cdot\hat{p} & 0 \\
                                          0 & \vec{\sigma}\cdot\hat{p}
                                    \end{bmatrix}\begin{bmatrix}
                                          0 & \vec{\sigma} \\
                                          \vec{\sigma} & 0
                                    \end{bmatrix}
                              +\begin{bmatrix}
                                          0 & \vec{\sigma} \\
                                          \vec{\sigma} & 0
                                    \end{bmatrix}\begin{bmatrix}
                                          \vec{\sigma}\cdot\hat{p} & 0 \\
                                          0 & \vec{\sigma}\cdot\hat{p}
                                    \end{bmatrix}\Biggr)\times\vec{p} \nonumber\\
                       & =&\dfrac{\hbar}{2} \Biggl\{\begin{bmatrix}
                                          0 & (\vec{\sigma}\cdot\hat{p})
                                                \vec{\sigma} \\
                                          (\vec{\sigma}\cdot\hat{p})
                                                \vec{\sigma} & 0
                                    \end{bmatrix}
                              +\begin{bmatrix}
                                          0 & \vec{\sigma}(
                                                \vec{\sigma}\cdot\hat{p}) \\
                                          \vec{\sigma}(\vec{\sigma}\cdot\hat{p})
                                                & 0
                                    \end{bmatrix}\Biggr\}\times\vec{p} \nonumber\\
                       & =& \dfrac{\hbar}{2} \begin{bmatrix}
                                    0 & (\vec{\sigma}\cdot\hat{p})\vec{\sigma}
                                          +\vec{\sigma}(\vec{\sigma}\cdot\hat{p})
                                          \\
                                    (\vec{\sigma}\cdot\hat{p})\vec{\sigma}
                                          +\vec{\sigma}(\vec{\sigma}\cdot
                                                \hat{p}) & 0
                              \end{bmatrix}\times\vec{p} = \dfrac{\hbar}{2} \begin{bmatrix}
                                    0 & 2\,\hat{p} \\
                                    2\,\hat{p} & 0
                              \end{bmatrix}\times\vec{p}
                        =0,
            \end{eqnarray}
            we then have
            \begin{eqnarray}
                   \Pi^s_+ \left[\vec{\alpha}(0)\times\vec{p}\right]\Pi^s_+
                  &=&\frac{1}{4} \left(\openone+\dfrac{2}{\hbar} \hat{\Lambda}\right)\left[
                        \vec{\alpha}(0)\times\vec{p}\right]\left(\openone
                              +\dfrac{2}{\hbar} \hat{\Lambda}\right) \notag\\
                  &=& \frac{1}{4} \left(\left[\vec{\alpha}(0)\times\vec{p}\right]
                        +\frac{2}{\hbar} \bigl\{\hat{\Lambda},\
                              \left[\vec{\alpha}(0)\times\vec{p}\right]\bigr\}
                        +\frac{4}{\hbar^4} \hat{\Lambda} \left[
                              \vec{\alpha}(0)\times\vec{p}\right]\hat{\Lambda}\right)
                              \nonumber\\
                  &=& \frac{1}{4} \left[\left[\vec{\alpha}(0)\times\vec{p}\right]
                        -\frac{4}{\hbar^2} \hat{\Lambda}^2 \left[
                              \vec{\alpha}(0)\times\vec{p}\right] \right]= \frac{1}{4} \Bigl\{\left[\vec{\alpha}(0)\times\vec{p}\right]
                        -\left[\vec{\alpha}(0)\times\vec{p}\right]\Bigr\}
                  =0,
            \end{eqnarray}
            similarly, we have
            \begin{eqnarray}
                  \Pi^s_- \left[\vec{\alpha}(0)\times\vec{p}\right]\Pi^s_- =0.
            \end{eqnarray}
            The above results lead to
            \begin{eqnarray}
                  \Pi^s_+ \hat{\mathcal{Z}}_s \Pi^s_+ =0,\;\;\;\;\;
                  \Pi^s_- \hat{\mathcal{Z}}_s \Pi^s_- =0.
            \end{eqnarray}
            Thus, if $|\Psi\rangle$ is in a superposition state of the same helicity, one will have
            \begin{eqnarray}
                  \mathcal{Z}_s=\langle \Psi|\hat{\mathcal{Z}}_s|\Psi\rangle=  \langle \Psi|\Pi^s_+ \hat{\mathcal{Z}}_s \Pi^s_+|\Psi\rangle =0,\;\;\;{\rm or}\;\;\;
                  \mathcal{Z}_s=\langle \Psi|\hat{\mathcal{Z}}_s|\Psi\rangle=  \langle \Psi|\Pi^s_- \hat{\mathcal{Z}}_s \Pi^s_-|\Psi\rangle =0.
            \end{eqnarray}
            Therefore, there is a SZ phenomenon only when Dirac's electron is in a superposition state with opposite energy and opposite helicity. This ends the proof.
            \end{proof}

           \section{Spin Zitterbewegung for $H_{\rm b}^{\rm I}$}

            For the type-I Dirac's braidon, the total angular momentum operator $\vec{J}$ is also conserved, thus
            \begin{eqnarray}
            \hat{\mathcal{Z}}_{\rm b}^{{\rm I},s}=-\hat{\mathcal{Z}}_{\rm b}^{{\rm I},r}\times\vec{p}.
            \end{eqnarray}
             We have listed the results of ``spin Zitterbewegung''  in Table \ref{tab:sz2}.

  \begin{table}[h]
	\centering
\caption{The results of ``spin Zitterbewegung'' of the type-I Dirac's braidon. $|\Psi'_j\rangle$'s  ($j=1, 2, 3, 4$) are four eigenstates of the  type-I Dirac's braidon. Here $\langle\Psi'_j|\hat{\mathcal{Z}}_{\rm b}^{{\rm I},s}|\Psi'_k \rangle=-\langle\Psi'_j|\hat{\mathcal{Z}}_{\rm b}^{{\rm I},r}|\Psi'_k \rangle \times \vec{p} = \langle\Psi'_j|\hat{\mathcal{Z}}_{\rm b}^{{\rm I},r}|\Psi'_k \rangle \times (-\vec{p})$.}
\begin{tabular}{lllll}
\hline\hline
 & $|\Psi_1'\rangle$ &  $|\Psi'_2\rangle$& $|\Psi'_3\rangle$ & $|\Psi'_4\rangle$ \\
  \hline
$\langle \Psi'_1| \hat{\mathcal{Z}}_{\rm b}^{{\rm I}, s}$ \;\;\;\quad& 0&0 & $0$  \;\;\;\quad& $\Delta_2\,(\vec{F}_1 +{\rm i}\,\vec{F}_2) \times (-\vec{p})$  \\
 \hline
$\langle \Psi'_2| \hat{\mathcal{Z}}_{\rm b}^{{\rm I}, s}$\;\;\;\quad &0 &0 & $-\Delta_2\,(\vec{F}_1 +{\rm i}\,\vec{F}_2)^* \times (-\vec{p}) $ \quad\quad & 0 \\
 \hline
 $\langle \Psi'_3| \hat{\mathcal{Z}}_{\rm b}^{{\rm I}, s}$\;\;\;\quad & 0 & $-\Delta_2^*\,(\vec{F}_1 +{\rm i}\,\vec{F}_2) \times (-\vec{p})$  & 0 & 0 \\
 \hline
 $\langle \Psi'_4| \hat{\mathcal{Z}}_{\rm b}^{{\rm I}, s}$ \;\;\;\quad& $\Delta_2^*\,(\vec{F}_1 +{\rm i}\,\vec{F}_2)^*  \times (-\vec{p})\quad$ \quad&0 & 0 & 0 \\
 \hline\hline
\end{tabular}\label{tab:sz2}
\end{table}

\section{Spin Zitterbewegung for the $H_{\rm e}$-$H_{\rm b}^{\rm I}$ Mixing}

            In this section, let us consider the mixture of Dirac's electron $H_{\rm e}$ and the type-I Dirac's braidon $H_{\rm b}^{\rm I}$.
            For the mixed system, the total angular momentum operator $\vec{J}$ is also conserved, thus
            \begin{eqnarray}
            \hat{\mathcal{Z}}_{\rm m}^s=-\hat{\mathcal{Z}}_{\rm m}^r\times\vec{p}.
            \end{eqnarray}
            We listed the results of ``spin Zitterbewegung''  in Table \ref{tab:sz3}.

  \begin{table}[h]
	\centering
\caption{The results of ``spin Zitterbewegung'' of of the $H_{\rm e}$-$H_{\rm b}^{\rm I}$ mixing. $|\Psi'_j\rangle$'s  ($j=1, 2, 3, 4$) are four eigenstates of the mixed Hamiltonian $\mathcal{H}_{\rm m}=\cos\vartheta\;H_{\rm e} +\sin\vartheta\;H_{\rm b}^{\rm I}$, The ``spin Zitterbewegung'' operator reads $\hat{\mathcal{Z}}_{\rm m}^s=\hat{\mathcal{Z}}_{\rm m}^r \times (-\vec{p})$, and $\gamma(\vartheta)=\cos\vartheta-\frac{mc}{p} \sin\vartheta$.}
\begin{tabular}{lllll}
\hline\hline
 & $|\Psi_1'\rangle$ &  $|\Psi'_2\rangle$& $|\Psi'_3\rangle$ & $|\Psi'_4\rangle$ \\
  \hline
$\langle \Psi'_1| \hat{\mathcal{Z}}_{\rm m}^s$ \;\;\;\quad& 0&0 & 0  \;\;\;\quad& $\gamma(\vartheta)\Delta_1\,(\vec{F}_1 +{\rm i}\,\vec{F}_2)\times (-\vec{p})$  \\
 \hline
$\langle \Psi'_2| \hat{\mathcal{Z}}_{\rm m}^s$\;\;\;\quad &0 &0 & $-\gamma(\vartheta)\Delta_1\,(\vec{F}_1 +{\rm i}\,\vec{F}_2)^* \times (-\vec{p})$ & 0 \\
 \hline
 $\langle \Psi'_3| \hat{\mathcal{Z}}_{\rm m}^s$\;\;\;\quad & 0 \;\;\;& $-\gamma(\vartheta)\Delta_1\,(\vec{F}_1 +{\rm i}\,\vec{F}_2)\times (-\vec{p}) $ & 0 & 0 \\
 \hline
 $\langle \Psi'_4| \hat{\mathcal{Z}}_{\rm m}^s$ \;\;\;\quad& $\gamma(\vartheta)\Delta_1\,(\vec{F}_1 +{\rm i}\,\vec{F}_2)^* \times (-\vec{p})$ \quad&0 & 0 & 0 \\
 \hline\hline
\end{tabular}\label{tab:sz3}
\end{table}

\begin{remark}From Table \ref{tab:sz3}, one may notice an interesting thing: when the factor
    \begin{eqnarray}
                        && \gamma(\vartheta)= \cos\vartheta-\dfrac{m\,c}{p} \sin\vartheta=0,
             \end{eqnarray}
  i.e.,
             \begin{eqnarray}
                        && \tan\vartheta=\dfrac{p}{m\,c},
             \end{eqnarray}
            then for any superposition state of the $H_{\rm e}$-$H_{\rm b}^{\rm I}$ mixing system, the phenomenon of ``spin Zitterbewegung'' (together with the ``orbit-angular-momentum Zitterbewegung'' defined by $\hat{\mathcal{Z}}_{\rm m}^\ell=\hat{\mathcal{Z}}_{\rm m}^r \times \vec{p}$ ) vanishes. This is also one prediction of our work based on the electron-braidon mixing.
\end{remark}

\section{Spin Zitterbewegung for $H_{\rm b}^{\rm II}$}

 For the type-II Dirac's braidon, the total angular momentum operator $\vec{J}$ is also conserved, thus
            \begin{eqnarray}
            \hat{\mathcal{Z}}_{\rm b}^{{\rm II},s}=-\hat{\mathcal{Z}}_{\rm b}^{{\rm II},r}\times\vec{p}.
            \end{eqnarray}
             We have listed the results of ``spin Zitterbewegung''  in Table \ref{tab:sz4}.

        \begin{table}[h]
	\centering
\caption{The results of ``spin Zitterbewegung'' of the type-II Dirac' braidon. $|\Psi''_j\rangle$'s  ($j=1, 2, 3, 4$) are four eigenstates of the type-II Dirac's braidon.In this case, the ``spin Zitterbewegung'' operator reads $\hat{\mathcal{Z}}_{\rm b}^{{\rm II},s} = \hat{\mathcal{Z}}_{\rm b}^{{\rm II},r} \times (-\vec{p})$, and $\Delta_3=\left(1+{\rm i}\dfrac{m\,c}{p}\right)$.}
\begin{tabular}{lllll}
\hline\hline
 & $|\Psi''_1\rangle$ &  $|\Psi''_2\rangle$& $|\Psi''_3\rangle$ & $|\Psi''_4\rangle$ \\
  \hline
$\langle \Psi''_1| \hat{\mathcal{Z}}_{\rm b}^{{\rm II},r}$ \;\;\;\quad& 0&0 & $0$  \;\;\;\quad& $\Delta_3\,(\vec{F}_1 +{\rm i}\,\vec{F}_2)\times (-\vec{p})$  \\
 \hline
$\langle \Psi''_2| \hat{\mathcal{Z}}_{\rm b}^{{\rm II},r}$\;\;\;\quad &0 &0 & $-\Delta_3\,(\vec{F}_1 +{\rm i}\,\vec{F}_2)^*\times (-\vec{p}) $ \quad\quad& $0$ \\
 \hline
 $\langle \Psi''_3| \hat{\mathcal{Z}}_{\rm b}^{{\rm II},r}$\;\;\;\quad & $0$ \;\;\;& $-\Delta_3^*\,(\vec{F}_1 +{\rm i}\,\vec{F}_2) \times (-\vec{p})$ & 0 & 0 \\
 \hline
 $\langle \Psi''_4| \hat{\mathcal{Z}}_{\rm b}^{{\rm II},r}$ \;\;\;\quad& $\Delta_3^*\,(\vec{F}_1 +{\rm i}\,\vec{F}_2)^*\times (-\vec{p})$ \quad&$0$ \;\;\;& 0 & 0 \\
 \hline\hline
\end{tabular}\label{tab:sz4}
\end{table}

\section{Spin Zitterbewegung for the $H_{\rm e}$-$H_{\rm b}^{\rm II}$ Mixing}

            In this section, let us consider the mixture of Dirac's electron $H_{\rm e}$ and the type-I Dirac's braidon $H_{\rm b}^{\rm II}$.
            For the mixed system, the total angular momentum operator $\vec{J}$ is also conserved, thus
            \begin{eqnarray}
            \hat{\mathcal{Z}}_{\rm m}^s=-\hat{\mathcal{Z}}_{\rm m}^r\times\vec{p}.
            \end{eqnarray}
            We listed the results of ``spin Zitterbewegung''  in Table \ref{tab:sz5}.

  \begin{table}[h]
	\centering
\caption{The results of ``spin Zitterbewegung'' of of the $H_{\rm e}$-$H_{\rm b}^{\rm II}$ mixing. $|\Psi'''_j\rangle$'s  ($j=1, 2, 3, 4$) are four eigenstates of the mixed Hamiltonian $\mathcal{H}_{\rm m}=\cos\vartheta\;H_{\rm e} +\sin\vartheta\;H_{\rm b}^{\rm II}$, The ``spin Zitterbewegung'' operator reads $\hat{\mathcal{Z}}_{\rm m}^s=\hat{\mathcal{Z}}_{\rm m}^r \times (-\vec{p})$, and $\gamma(\vartheta)=\biggr[1 + {\rm i}\sin\vartheta \frac{mc}{p}-{\rm i}\,\sin(2\vartheta)\,\frac{p}{2 m\,c}
                                            \biggr]$.}
\begin{tabular}{lllll}
\hline\hline
 & $|\Psi'''_1\rangle$ &  $|\Psi'''_2\rangle$& $|\Psi'''_3\rangle$ & $|\Psi'''_4\rangle$ \\
  \hline
$\langle \Psi'''_1| \hat{\mathcal{Z}}_{\rm m}^s$ \;\;\;& 0&0 & $ 0$  \;\;\;\quad& $\gamma(\vartheta)\, \Delta_1\,(\vec{F}_1 +{\rm i}\,\vec{F}_2)\times (-\vec{p})$  \\
 \hline
$\langle \Psi'''_2| \hat{\mathcal{Z}}_{\rm m}^s$\;\;\; &0 &0 & $\gamma(\vartheta)\,\Delta_1\,(\vec{F}_1 +{\rm i}\,\vec{F}_2)^* \times \vec{p}$ & $0$ \\
 \hline
 $\langle \Psi'''_3| \,\hat{\mathcal{Z}}_{\rm m}^s$\;\;\;& $ 0$ \;\;\;& $\gamma^*(\vartheta)\,\Delta_1^*\,(\vec{F}_1 +{\rm i}\,\vec{F}_2)\times \vec{p}$ & 0 & 0 \\
 \hline
 $\langle \Psi'''_4| \hat{\mathcal{Z}}_{\rm m}^s$ \;\;\;& $\gamma^*(\vartheta)\,\Delta_1^*\,(\vec{F}_1 +{\rm i}\,\vec{F}_2)^* \times (-\vec{p})$ &$0$ \;\;\;& 0 & 0 \\
 \hline\hline
\end{tabular}\label{tab:sz5}
\end{table}

\newpage

\part{A Possible Mechanism to Alter Gyromagnetic Factor}

\section{The $g$ Factor of an Electron with Respect to $\vec{\ell}$}

      Refer to \cite{GreineQMIntroM}, we first recall the formula of currents in H atom (i.e., the hydrogen atom).

      \subsection{The Currents in H Atom}

           Retrospect the operator of current density $\vec{j}$ is
            \begin{equation}
                  \vec{j}=\dfrac{{\rm i}\hbar}{2\, m_{\rm e}} \left[\Psi( \vec{\nabla}\,\Psi^*)
                        -\Psi^* (\vec{\nabla}\,\Psi)\right],
            \end{equation}
            and the eigenstate of H atom reads
            \begin{equation}
                  \Psi\equiv\Psi_{nlm}(\vec{r}) =N_{nl} \dfrac{R_{nl} (r)}{r}
                        P^{\Abs{m}}_l (\theta)\:{\rm e}^{{\rm i}\,m\,\varphi},
            \end{equation}
            where $m_{\rm e}$ is the mass of an electron. Note in spherical coordinates,
            \begin{equation}
                  \vec{\nabla}=\vec{e}_r \p{}{r} +\vec{e}_\theta \dfrac{1}{r} \p{}{\theta}
                        +\vec{e}_\varphi \dfrac{1}{r\,\sin\theta} \p{}{\varphi},
            \end{equation}
            then we obtain
            \begin{align}
                  & j_r =\dfrac{{\rm i}\hbar}{2\,m_{\rm e}} \left(\Psi\p{\,\Psi^*}{r}
                        -\Psi^* \p{\,\Psi}{r}\right)
                  =\dfrac{{\rm i}\hbar}{2\,m_{\rm e}} \dfrac{1}{r} \left(
                        \Psi\p{\,\Psi^*}{\theta} -\Psi^* \p{\,\Psi}{\theta}
                        \right)=j_\theta
                  =0,
            \end{align}
            since $N_{nl} \bigl[R_{nl} (r)/r\bigr]P^{\Abs{m}}_l (\theta) \in\Rel$ for given $n,l,m$. Finally, one obtains
            \begin{eqnarray}
                   \vec{j} &=& \vec{e}_\varphi j_\varphi
                  =\vec{e}_\varphi \dfrac{{\rm i}\hbar}{2\,m_{\rm e}} \dfrac{1}{r\,\sin\theta}
                        \left(\Psi\p{\,\Psi^*}{\varphi} -\Psi^* \p{\,\Psi}{\varphi}\right)\nonumber\\
                  &=&\vec{e}_\varphi \dfrac{{\rm i}\hbar}{2\,m_{\rm e}} \dfrac{1}{r\,\sin\theta}
                        \Abs{\Psi}^2 (-{\rm i}\,m-{\rm i}\,m)
                  =\vec{e}_\varphi \dfrac{\hbar\,m}{m_{\rm e}} \dfrac{1}{r\,\sin\theta}
                        \Abs{\Psi}^2.
            \end{eqnarray}

      \subsection{The Magnetic Moment}

\begin{figure}[t]
                  \centering
                  \includegraphics[width=150mm]{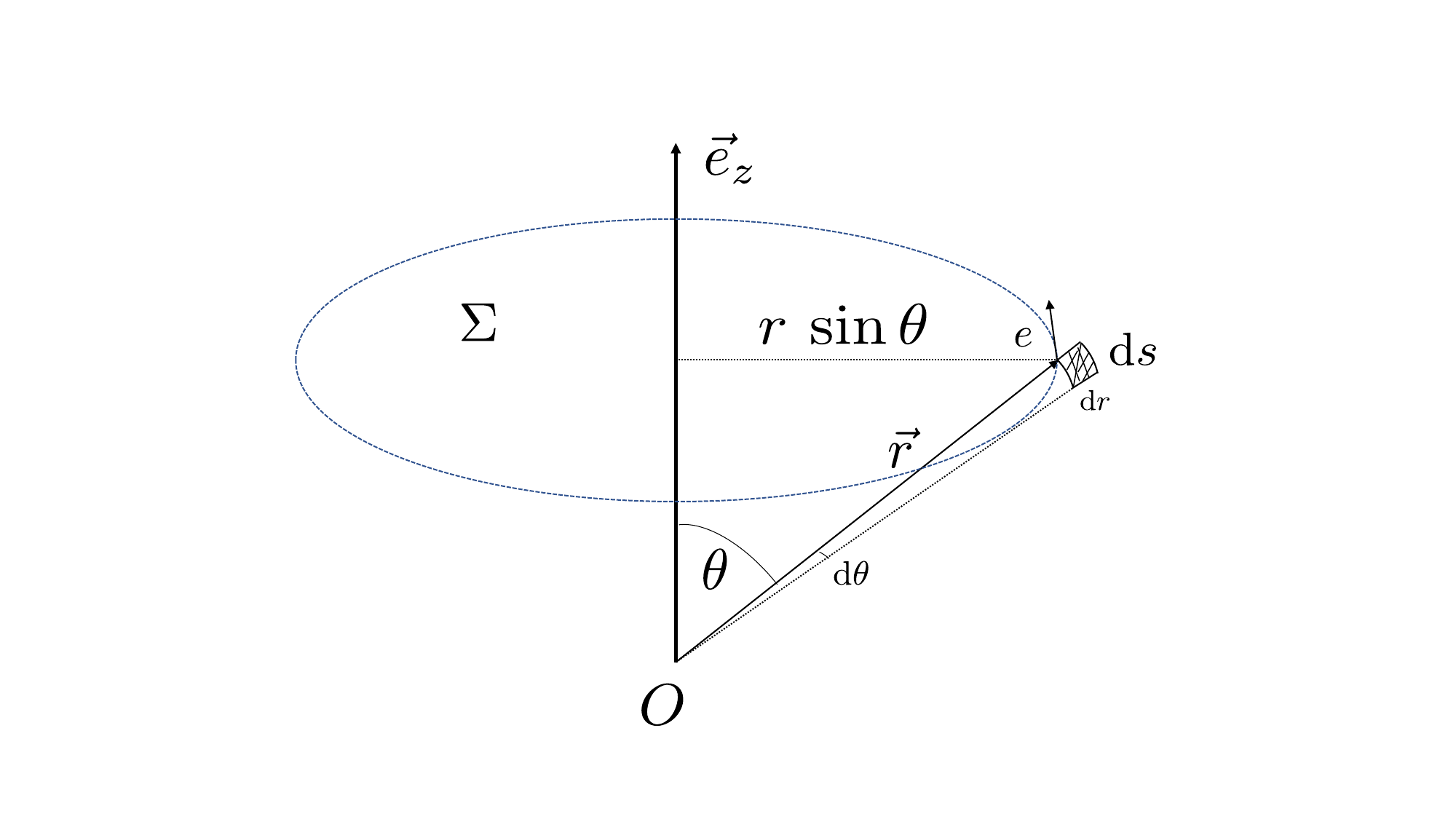}
                  \caption{Illustration of the magnetic moment in H atom.}\label{fig:MH}
            \end{figure}

            Because $\vec{j}=\vec{e}_\varphi j_\varphi$ according to our calculation about the current density in H atom above, the magnetic moment appears along the direction of $\vec{e}_z$. Explicitly, electrodynamics (see e.g. \cite{GreineCEDM}) tells us that the magnetic moment of a current element ${\rm d} \mathcal{I}$ takes the form of
            \begin{equation}
                  {\rm d} \mathcal{M}=\dfrac{\Sigma}{c}\, {\rm d} \mathcal{I},
            \end{equation}
            where $\Sigma$ indicates the area of the plane $\mathcal{I}$ circling around (i.e., $\Sigma=\pi\,r^2 {\sin^2}\theta$ in Fig.~\ref{fig:MH}). From Fig.~\ref{fig:MH}, we have
            \begin{eqnarray}
                   {\rm d}\mathcal{M} &=& {\rm d} \mathcal{M}_z =\dfrac{\Sigma}{c} \, {\rm d} \mathcal{I}
                  =-e\dfrac{\Sigma}{c} j_\varphi {\rm d}s \nonumber\\
                  &=&-e\dfrac{\pi\,r^2 {\sin^2}\theta}{c} \left(\dfrac{\hbar\,m}{m_{\rm e}}
                        \dfrac{1}{r\,\sin\theta} \Abs{\Psi}^2\right) (r\,{\rm d}\theta\,
                              {\rm d}r)
                  =-e\dfrac{\pi}{c} \left(\dfrac{\hbar\,m}{m_{\rm e}} \Abs{\Psi}^2\right) (
                        r^2 \sin\theta\,{\rm d}r\,{\rm d}\theta).
            \end{eqnarray}
            Note that the integral for the square norm $\Abs{\Psi}^2$ over the whole space reads
            \begin{equation}
                  \int\Abs{\Psi}^2 {\rm d}V=\int \Abs{\Psi}^2 r^2 \sin\theta\,{\rm d}r\,
                        {\rm d}\theta\,{\rm d}\varphi
                  =2\,\pi\int\Abs{\Psi}^2 r^2 \sin\theta\,{\rm d}r\,{\rm d}\theta
                  =1
            \end{equation}
            for the symmetry of $\varphi$, which means
            \begin{equation}
                  \int\Abs{\Psi}^2 r^2 \sin\theta\,{\rm d}r\,{\rm d}\theta
                  =\dfrac{1}{2\,\pi}.
            \end{equation}

            After that, we have
            \begin{eqnarray}
                   \mathcal{M}_z &=&\int{\rm d} \mathcal{M}=\int{\rm d} \mathcal{M}_z
                  =\int-e\dfrac{\pi}{c} \left(\dfrac{\hbar\,m}{m_{\rm e}} \Abs{\Psi}^2\right) (
                        r^2 \sin\theta\,{\rm d}r\,{\rm d}\theta)\nonumber\\
                 &=& -\dfrac{e}{c} \dfrac{\hbar\,m}{2\,m_{\rm e}}=-m \dfrac{e \hbar}{2 m_{\rm e} c}=-m\,\mu_{\rm B},
            \end{eqnarray}
            with
             \begin{equation}
               \mu_{\rm B} \equiv \dfrac{e \hbar}{2 m_{\rm e} c}
            \end{equation}
             being the Bohr magneton. Further the gyromagnetic factor (i.e., the $g$ factor) with respect to the $z$-components of the orbit angular momentum $\ell_z$ is defined as the following ratio
            \begin{equation}
                  g_{\rm orbit}=\dfrac{\Abs{\mathcal{M}_z}/\mu_{\rm B}}{\Abs{\ell_z}/\hbar} =1,
            \end{equation}
            for $\langle \ell_z\rangle =m \hbar$, (or $\ell_z \,{\rm e}^{{\rm i}\,m\,\varphi} =m\,\hbar \,{\rm e}^{{\rm i}\,m\,\varphi}$).

\section{The $g$ Factor of an Electron with Respect to $\vec{S}$}

      The relativistic Hamiltonian of a Dirac's electron in an electromagnetic field is given by
      \begin{equation}\label{eq:Hev-1}
            \mathcal{H}_{\rm e} =c\,\vec{\alpha}\cdot\vec{\Pi}
                  +\beta\,m_{\rm e}\,c^2 +q\,\phi
            \equiv c\,\vec{\alpha}\cdot\Bigl(\vec{p}-\dfrac{q}{c} \vec{A}\Bigr)
                  +\beta\,m_{\rm e}\,c^2 +q\,\phi,
      \end{equation}
      where $\phi$ depicts the scalar potential.
      \begin{remark}
                  For the convenience of describing other Dirac particles, we use $q$ to depict the charge of a Dirac particle. In the case of electron, $q=-e$.
            \end{remark}
      Then Dirac's equation reads
      \begin{equation}\label{eq:DiracEqE}
            {\rm i}\hbar\p{}{t} \Psi=\mathcal{H}_{\rm e} \Psi.
      \end{equation}
      Next we study the Dirac equation in the representation of
      \begin{equation}
            \Psi=\begin{bmatrix}
                  \tilde{\eta} \\ \tilde{\chi}
            \end{bmatrix},
      \end{equation}
      then \Eq{eq:DiracEqE} becomes
      \begin{equation}
            {\rm i}\hbar\p{}{t} \begin{bmatrix}
                        \tilde{\eta} \\ \tilde{\chi}
                  \end{bmatrix}
            =c\,\begin{bmatrix}
                              0 & \vec{\sigma} \\
                              \vec{\sigma} & 0
                        \end{bmatrix}\cdot\vec{\Pi}\begin{bmatrix}
                                    \tilde{\eta} \\ \tilde{\chi}
                              \end{bmatrix}
                  +\begin{bmatrix}
                              \openone & 0 \\
                              0 & -\openone
                        \end{bmatrix}\,m_{\rm e}\,c^2 \begin{bmatrix}
                              \tilde{\eta} \\ \tilde{\chi}
                        \end{bmatrix}
                  +q\,\phi\begin{bmatrix}
                              \tilde{\eta} \\ \tilde{\chi}
                        \end{bmatrix},
      \end{equation}
      i.e.,
      \begin{equation}\label{eq:DiracEExpand-0}
            {\rm i}\hbar\p{}{t} \begin{bmatrix}
                        \tilde{\eta} \\ \tilde{\chi}
                  \end{bmatrix}
            =c(\vec{\sigma}\cdot\vec{\Pi})\begin{bmatrix}
                              \tilde{\chi} \\ \tilde{\eta}
                        \end{bmatrix}
                  +m_{\rm e}\,c^2 \openone\begin{bmatrix}
                              \tilde{\eta} \\ -\tilde{\chi}
                        \end{bmatrix}
                  +q\,\phi\begin{bmatrix}
                              \tilde{\eta} \\ \tilde{\chi}
                        \end{bmatrix}.
      \end{equation}

      Now, we consider the non-relativistic approximation of Dirac's equation. In the case of low energy, the term involving $m_{\rm e}\,c^2$ is far larger than other ones. Then a part of time factor can be separated as follows.
      \begin{equation}
            \begin{bmatrix}
                  \tilde{\eta} \\ \tilde{\chi}
            \end{bmatrix}=\begin{bmatrix}
                  \eta \\ \chi
            \end{bmatrix}{\rm e}^{-{\rm i}\frac{m_{\rm e}\,c^2}{\hbar} t},
      \end{equation}
      thus \Eq{eq:DiracEExpand-0} is transformed as
      \begin{equation}
            {\rm i}\hbar\Biggl\{\p{}{t} \begin{bmatrix}
                        \eta \\ \chi
                  \end{bmatrix}\Biggr\}{\rm e}^{-{\rm i}\frac{m_{\rm e}\,c^2}{\hbar} t}
            +{\rm i}\begin{bmatrix}
                        \eta \\ \chi
                  \end{bmatrix}\hbar\Bigl(\p{}{t} {\rm e}^{-{\rm i}\frac{m_{\rm e}\,c^2}{
                        \hbar} t}\Bigr)
            =c(\vec{\sigma}\cdot\vec{\Pi})\begin{bmatrix}
                              \chi \\ \eta
                        \end{bmatrix}{\rm e}^{-{\rm i}\frac{m_{\rm e}\,c^2}{\hbar} t}
                  +m_{\rm e}\,c^2 \openone\begin{bmatrix}
                              \eta \\ -\chi
                        \end{bmatrix}{\rm e}^{-{\rm i}\frac{m_{\rm e}\,c^2}{\hbar} t}
                  +q\,\phi\begin{bmatrix}
                              \eta \\ \chi
                        \end{bmatrix}{\rm e}^{-{\rm i}\frac{m_{\rm e}\,c^2}{\hbar} t},
      \end{equation}
      i.e.,
      \begin{equation}
            {\rm i}\hbar\p{}{t} \begin{bmatrix}
                        \eta \\ \chi
                  \end{bmatrix}
            +m_{\rm e}\,c^2 \openone\begin{bmatrix}
                        \eta \\ \chi
                  \end{bmatrix}
            =c(\vec{\sigma}\cdot\vec{\Pi})\begin{bmatrix}
                              \chi \\ \eta
                        \end{bmatrix}
                  +m_{\rm e}\,c^2 \openone\begin{bmatrix}
                              \eta \\ -\chi
                        \end{bmatrix}
                  +q\,\phi\begin{bmatrix}
                              \eta \\ \chi
                        \end{bmatrix},
      \end{equation}
      i.e.,
      \begin{equation}\label{eq:DiracELowE-0}
            {\rm i}\hbar\p{}{t} \begin{bmatrix}
                        \eta \\ \chi
                  \end{bmatrix}
            =c(\vec{\sigma}\cdot\vec{\Pi})\begin{bmatrix}
                              \chi \\ \eta
                        \end{bmatrix}
                  +m_{\rm e}\,c^2 \openone\begin{bmatrix}
                              0 \\ -2\,\chi
                        \end{bmatrix}
                  +q\,\phi\begin{bmatrix}
                              \eta \\ \chi
                        \end{bmatrix}.
      \end{equation}
      For the 2nd row of \Eq{eq:DiracELowE-0}, we have
      \begin{equation}\label{eq:chi-1}
            {\rm i}\hbar\p{\chi}{t} =c(\vec{\sigma}\cdot\vec{\Pi})\eta
                  -2\,m_{\rm e}\,c^2 \openone\chi+q\,\phi\,\chi.
      \end{equation}
      Since
      \begin{eqnarray}
        && m_{\rm e}\,c^2 \,\chi\gg   {\rm i}\hbar\p{\chi}{t}, \nonumber\\
        && m_{\rm e}\,c^2 \,\chi\gg q\,\phi\,\chi,
      \end{eqnarray}
      then from Eq. (\ref{eq:chi-1}) we arrive at
      \begin{equation}
            c(\vec{\sigma}\cdot\vec{\Pi})\eta=2\,m_{\rm e}\,c^2 \openone\chi,
      \end{equation}
      which means
      \begin{equation}\label{eq:chieta}
            \chi=\dfrac{c(\vec{\sigma}\cdot\vec{\Pi})}{2\,m_{\rm e}\,c^2} \eta,
      \end{equation}
      i.e., the wavefunction $\eta$ can be connected to the wavefunction $\chi$ through the relation (\ref{eq:chieta}).

      After that, for the first row, we obtain
      \begin{equation}
            {\rm i}\hbar\p{\eta}{t} =c(\vec{\sigma}\cdot\vec{\Pi})\chi
                  +q\,\phi\,\eta,
      \end{equation}
      i.e.,
      \begin{equation}
            {\rm i}\hbar\p{\eta}{t} =\dfrac{1}{2\,m_{\rm e}} (\vec{\sigma}\cdot\vec{\Pi})(
                        \vec{\sigma}\cdot\vec{\Pi})\eta
                  +q\,\phi\,\eta.
      \end{equation}
      Due to
      \begin{eqnarray}\label{eq:SigmPi-1a}
             (\vec{\sigma}\cdot\vec{\Pi})(\vec{\sigma}\cdot\vec{\Pi})
            &=&\Bigl(\vec{p}-\dfrac{q}{c} \vec{A}\Bigr)^2 +{\rm i}\,\vec{\sigma}
                  \cdot\bigg[\Bigl(\vec{p}-\dfrac{q}{c} \vec{A}\Bigr)\times\Bigl(
                        \vec{p}-\dfrac{q}{c} \vec{A}\Bigr)\bigg] \notag \\
           & =& \Bigl(\vec{p}-\dfrac{q}{c} \vec{A}\Bigr)^2 +{\rm i}\,\vec{\sigma}
                  \cdot\bigg[\Bigl(-{\rm i}\hbar\,\vec{\nabla}-\dfrac{q}{c} \vec{A}\Bigr)
                        \times\Bigl(-{\rm i}\hbar\,\vec{\nabla}-\dfrac{q}{c} \vec{A}\Bigr)
                        \bigg]\nonumber\\
           & =&\Bigl(\vec{p}-\dfrac{q}{c} \vec{A}\Bigr)^2 +{\rm i}\,\vec{\sigma}
                  \cdot\bigg[{\rm i}\hbar\dfrac{q}{c}\Bigl(\vec{\nabla}\times\vec{A}
                        +\vec{A}\times\nabla\Bigr)\bigg] \notag \\
           & =& \Bigl(\vec{p}-\dfrac{q}{c} \vec{A}\Bigr)^2 -\dfrac{q\,\hbar}{c} \vec{
                  \sigma}\cdot\left[(\vec{\nabla}\times\vec{A})-\vec{A}\times\vec{\nabla}+\vec{A}\times\vec{\nabla}\right] \notag \\
           & =& \Bigl(\vec{p}-\dfrac{q}{c} \vec{A}\Bigr)^2 -\dfrac{q\,\hbar}{c} \vec{
                  \sigma}\cdot(\vec{\nabla}\times\vec{A}) \notag \\
           &=& \Bigl(\vec{p}-\dfrac{q}{c} \vec{A}\Bigr)^2 -\dfrac{q\,\hbar}{c} \vec{
                  \sigma}\cdot\vec{B},
      \end{eqnarray}
      we then have
      \begin{equation}
            {\rm i}\hbar\p{\eta}{t} = \bigg[
                        \dfrac{1}{2\,m_{\rm e}} \Bigl(\vec{p}-\dfrac{q}{c} \vec{A}\Bigr)^2
                        -\dfrac{q\,\hbar}{2\,m_{\rm e}\,c} \vec{\sigma}\cdot\vec{B}\bigg]\eta
                  +q\,\phi\,\eta,
      \end{equation}
      i.e.,
      \begin{equation}\label{eq:pauli-1}
            {\rm i}\hbar\p{\eta}{t} = \bigg[
                        \dfrac{1}{2\,m_{\rm e}} \Bigl(\vec{p}-\dfrac{q}{c} \vec{A}\Bigr)^2
                        -\dfrac{q\,\hbar}{2\,m_{\rm e}\,c} \vec{\sigma}\cdot\vec{B}+q\,\phi\bigg]\eta.
      \end{equation}

      \begin{remark}Under the non-relativistic approximation, Dirac's equation of an electron \Eq{eq:DiracEqE} reduces to Eq. (\ref{eq:pauli-1}), which is nothing but the well-known Pauli equation in non-relativistic quantum mechanics. From the Schr{\" o}dinger equation as shown in Eq. (\ref{eq:pauli-1}), one may extract a Hamiltonian (which can be called Pauli Hamiltonian) as
      \begin{equation}\label{eq:pauli-2}
            H_{\rm e}^{\rm appro} =\dfrac{1}{2\,m_{\rm e}} \Bigl(\vec{p}
                        -\dfrac{q}{c} \vec{A}\Bigr)^2
                  -\dfrac{q\,\hbar}{2\,m_{\rm e}\,c} \vec{\sigma}\cdot\vec{B}
                  +q\,\phi.
      \end{equation}
      As described by Greiner in Ref. \cite{GreineQMIntroM}: ``The second term in the equation of motion (\ref{eq:pauli-1}) is the interaction energy of
     the magnetic field $\vec{B}$ with the intrinsic magnetic moment of the particle
     \begin{eqnarray}
     && \vec{\mu}=\dfrac{q\,\hbar}{2\,m_{\rm e}\,c} \vec{\sigma},
      \end{eqnarray}
      or, because the spin operator of the particle is $\vec{S}=(1/2)\vec{\sigma}$,
      \begin{eqnarray}
     && \vec{\mu}=\dfrac{q\,\hbar}{2\,m_{\rm e}\,c} \vec{\sigma}=\dfrac{q\,\hbar}{\,m_{\rm e}\,c} \vec{S}=g_{\rm spin}\, \mu_{\rm B} \vec{S}=2 \, \mu_{\rm B} \vec{S}.
     \end{eqnarray}
     The factor $g_{\rm spin}$ is called the gyromagnetic ratio or gyromagnetic factor and turns out to be twice as large as that coming from the orbital motion. The ratio $g_{\rm spin}/g_{\rm orbit}$ is called the spin-Land{\' q} factor $g_{\rm s}$. For the particle in question, $g_{\rm s}$ is therefore 2. \emph{Thus a completely nonrelativistic linearized theory predicts the correct intrinsic magnetic moment of a spin-$\frac{1}{2}$ particle.}'' $\blacksquare$
     \end{remark}

     \begin{remark}
     As described by Greiner in Ref. \cite{GreinerRQM}: ``This can be demonstrated once again by turning on a weak, homogeneous magnetic field
     \begin{eqnarray}
     && \vec{B}=\vec{\nabla}\times \vec{A}, \;\;\;\;\;\; \vec{A}=\frac{1}{2} \vec{B}\times \vec{r},
     \end{eqnarray}
      where the quadratic terms of $\vec{A}$ have been neglected. With
     \begin{eqnarray}
     \vec{\Pi}^2=\Bigl(\vec{p}-\dfrac{q}{c} \vec{A}\Bigr)^2  &=& \Bigl(\vec{p}-\dfrac{q}{2c} \vec{B}\times \vec{r} \Bigr)^2
     \approx \vec{p}^{\,2}- \dfrac{q}{c} \left(\vec{B}\times \vec{r}\right)\cdot \vec{p}=
     \vec{p}^{\,2}- \dfrac{q}{c} \vec{B}\cdot \left(\vec{r}\times \vec{p}\right) = \vec{p}^{\,2}- \dfrac{q}{c}\vec{B}\cdot \vec{\ell},
     \end{eqnarray}
     where $\vec{\ell}=\vec{r}\times \vec{p}$ is the operator of angular momentum, and
      \begin{eqnarray}
      \vec{S} &=& \frac{1}{2} \hbar\, \vec{\sigma}
     \end{eqnarray}
     is the spin operator, it follows for the Pauli equation that
      \begin{equation}\label{eq:pauli-3}
            {\rm i}\hbar\p{\eta}{t} = \bigg[
                        \dfrac{\vec{p}^{\, 2}}{2\,m_{\rm e}}
                        -\dfrac{q}{2\,m_{\rm e}\,c} \left(\vec{\ell}+2 \vec{S}\right)\cdot\vec{B}+q\,\phi\bigg]\eta.
      \end{equation}
      This form shows explicitly the $g$ factor 2.'' $\blacksquare$
     \end{remark}


      \begin{remark}
            Set the scalar potential $\phi=0$, from Eq. (\ref{eq:Hev-1}) we then have
            \begin{equation}
                  \mathcal{H}_{\rm e} =c\,\vec{\alpha}\cdot\vec{\Pi}+\beta\,m_{\rm e}\,c^2
                  \equiv c\,\vec{\alpha}\cdot\Bigl(\vec{p}-\dfrac{q}{c} \vec{A}\Bigr)
                        +\beta\,m_{\rm e} \,c^2,
            \end{equation}
            then
            \begin{eqnarray}
                   \mathcal{H}_{\rm e}^2 &=&\bigg[c\,\vec{\alpha}\cdot\Bigl(
                        \vec{p}-\dfrac{q}{c} \vec{A}\Bigr)+\beta\,m_{\rm e}\,c^2\bigg]^2
                  =  \bigg[c(\sigma_x \otimes\vec{\sigma})\cdot\Bigl(
                              \vec{p}-\dfrac{q}{c} \vec{A}\Bigr)
                        +\beta m_{\rm e}\,c^2\bigg]^2 \notag \\
                  &=& c^2 \openone\otimes\bigg[\vec{\sigma}\cdot\Bigl(
                              \vec{p}-\dfrac{q}{c} \vec{A}\Bigr)\bigg]\bigg[
                                    \vec{\sigma}\cdot\Bigl(\vec{p}-\dfrac{q}{c} \vec{A}
                                    \Bigr)\bigg]
                        +\openone\otimes\openone\,m_{\rm e}^2 c^4 \nonumber\\
                  &=& c^2 \openone\otimes\bigg[\openone\Bigl(\vec{p}-\dfrac{q}{c} \vec{A}
                              \Bigr)^2 -\dfrac{q\,\hbar}{c} \vec{\sigma}\cdot\vec{B}
                              \bigg]
                        +\openone\otimes\openone\,m_{\rm e}^2 c^4 \notag \\
                  &=& \openone\otimes \bigg[c^2 \openone\Bigl(\vec{p}-\dfrac{q}{c} \vec{A}
                              \Bigr)^2 -q\,\hbar\,c\,\vec{\sigma}\cdot\vec{B}
                        +\openone\,m_{\rm e}^2 c^4\bigg],
            \end{eqnarray}
            which indicates
            \begin{equation}\label{eq:H-1b}
                  \dfrac{\mathcal{H}_{\rm e}^2}{2\, m_{\rm e}\,c^2} =\openone\otimes \bigg[
                        \dfrac{1}{2\,m_{\rm e}} \openone\Bigl(\vec{p}-\dfrac{q}{c} \vec{A}
                              \Bigr)^2 -\dfrac{q\,\hbar}{2\,m_{\rm e}\,c}\,\vec{\sigma}\cdot\vec{B}
                        +\dfrac{m_{\rm e}\,c^2}{2} \openone\bigg].
            \end{equation}
            One can note that the right-hand side of Eq. (\ref{eq:H-1b}) is almost the same as that of Eq. (\ref{eq:pauli-2}).
      \end{remark}

      \section{Calculating the $g$ Factor for the $H_{\rm e}$-${H}_{\rm b}^{\rm I}$ Mixing}

      In this section, we would like to calculate the $g$ factor for the case of $H_{\rm e}$-${H}_{\rm b}^{\rm I}$ mixing.
      The mixed Hamiltonian is given by
      \begin{eqnarray}\label{eq:mix-1aa}
            H_{\rm mix}=\cos\vartheta\,H_{\rm e} +\sin\vartheta\,H_{\rm b}^{\rm I}.
      \end{eqnarray}
      If without the vector potential $\vec{A}$ and the scalar potential $\phi$, then the Hamiltonian for Dirac's electron $H_{\rm e}$ and the Hamiltonian for the first-type Dirac's braidon ${H}_{\rm b}^{\rm I}$ are given as usual, i.e.,
      \begin{eqnarray}
           && H_{\rm e} = c\,\vec{\alpha}\cdot \vec{p}+\beta\,m_{\rm e}\,c^2,
      \end{eqnarray}
      \begin{eqnarray}\label{eq:Hev-1bb}
           && H_{\rm b}^{\rm I} =-m_{\rm e}\,c^2\;\vec{\alpha}\cdot\hat{p}+\beta\,p\,c.
      \end{eqnarray}
       However, if there is external magnetic field $\vec{B}$, or there is the vector potential $\vec{A}$, the Hamiltonian for Dirac's electron $H_{\rm e}$ then turns to Eq. (\ref{eq:Hev-1}), i.e.,
       \begin{eqnarray}\label{eq:Hev-1aa}
            \mathcal{H}_{\rm e} &=& c\,\vec{\alpha}\cdot\Bigl(\vec{p}-\dfrac{q}{c} \vec{A}\Bigr)+\beta\,m\,c^2=
            c\,\vec{\alpha}\cdot \vec{\Pi}+\beta\,m_{\rm e}\,c^2.
      \end{eqnarray}
      In other words, to achieve Eq. (\ref{eq:Hev-1aa}) from Eq. (\ref{eq:Hev-1}), one needs to perform the following replacement for the linear momentum:
       \begin{eqnarray}\label{eq:mom-1}
            \vec{p} & \mapsto & \vec{\Pi}=  \vec{p}-\dfrac{q}{c} \vec{A}.
      \end{eqnarray}

      For the Hamiltonian in Eq. (\ref{eq:Hev-1bb}), one may recast it to the following form
      \begin{eqnarray}\label{eq:Hev-1cc}
            H_{\rm b}^{\rm I} &= & -m_{\rm e}\,c^2\;\vec{\alpha}\cdot\hat{p}+\beta\,p\,c = -m_{\rm e}\,c^2\;\vec{\alpha}\cdot\frac{\vec{p}}{p}+\beta\,p\,c \nonumber\\
            &= & -m_{\rm e}\,c^2\;\vec{\alpha}\cdot\frac{\vec{p}}{\sqrt{\vec{p}^{\,2}}}+\beta\,\sqrt{\vec{p}^{\,2}}\,c \nonumber\\
             &= & -\dfrac{1}{2} m_{\rm e}\,c^2\;\left(\frac{1}{\sqrt{\vec{p}^{\,2}}}\;\vec{\alpha}\cdot \vec{p} + \vec{\alpha}\cdot \vec{p} \;\frac{1}{\sqrt{\vec{p}^{\,2}}} \right)+\beta\,\sqrt{\vec{p}^{\,2}}\,c.
      \end{eqnarray}
      Therefore, after performing the replacement in Eq. (\ref{eq:mom-1}), one have from Eq. (\ref{eq:Hev-1cc}) that
      \begin{eqnarray}\label{eq:Hev-1d}
            \mathcal{H}_{\rm b}^{\rm I}
           &=&-\dfrac{1}{2} m_{\rm e}\,c^2 \Big(
                  \dfrac{1}{\sqrt{\vec{\Pi}^2}} \vec{\alpha}\cdot\vec{\Pi}
                  +\vec{\alpha}\cdot\vec{\Pi}\dfrac{1}{\sqrt{\vec{\Pi}^2}}\Big)
                        +\beta\sqrt{\vec{\Pi}^2}\,c,
      \end{eqnarray}
      which is the Hamiltonian for the first-type Dirac's braidon in presence of an external magnetic field. One may note that the Hamiltonian is hermitian, i.e., $\left(\mathcal{H}_{\rm b}^{\rm I}\right)^\dagger= \mathcal{H}_{\rm b}^{\rm I}$. Note that we have recast  $H_{\rm b}^{\rm I}$ to the form of Eq. (\ref{eq:Hev-1cc}) in order to ensure $\mathcal{H}_{\rm b}^{\rm I}$ to be hermitian.

      After substituting Eq. (\ref{eq:Hev-1aa}) and Eq. (\ref{eq:Hev-1d}) into Eq. (\ref{eq:mix-1aa}), and considering an additional scalar potential $\phi$, then one obtains the mixed Hamiltonian as
      \begin{eqnarray}\label{eq:mix-1bb}
            \mathcal{H}_{\rm mix} &=& \cos\vartheta\,\mathcal{H}_{\rm e} +\sin\vartheta\,\mathcal{H}_{\rm b}^I +q\,\phi\nonumber\\
            &=&=\cos\vartheta\,\bigl(c\,\vec{\alpha}\cdot \vec{\Pi}
                        +\beta\,m_{\rm e}\,c^2\bigr)
                  +\sin\vartheta\,\bigg[-\dfrac{1}{2} m_{\rm e}\,c^2 \Big(
                        \dfrac{1}{\sqrt{\vec{\Pi}^2}} \vec{\alpha}\cdot\vec{\Pi}
                        +\vec{\alpha}\cdot\vec{\Pi}\dfrac{1}{\sqrt{\vec{\Pi}^2}}\Big)
                              +\beta\sqrt{\vec{\Pi}^2}\,c\bigg]
                  +q\,\phi.
      \end{eqnarray}

      \begin{remark}There is a clear physical picture of obtaining $\mathcal{H}_{\rm mix}$ from $H_{\rm e}$. Let us start from the Hamiltonian $H_{\rm e}$ of an Dirac's electron. Quantum mechanically we perform a unitary transformation $\mathcal{C}(\vartheta)$ on Dirac's electron, then it becomes a mixture of $H_{\rm e}$ and ${H}_{\rm b}^I$, i.e.,
      \begin{eqnarray}\label{eq:mix-1cc}
         H_{\rm e} \mapsto   H_{\rm mix} &= & \mathcal{C}(\vartheta) H_{\rm e} \mathcal{C}(\vartheta)^\dagger= \cos\vartheta\,H_{\rm e} +\sin\vartheta\,H_{\rm b}^I,
      \end{eqnarray}
      with the unitary transformation
      \begin{eqnarray}
            \mathcal{C}(\vartheta) &=& {\rm e}^{-{\rm i} \frac{\vartheta}{2} \Gamma_z}=\cos\frac{\vartheta}{2}\;\mathbb{I}+ \sin\frac{\vartheta}{2}\;\beta\vec{\alpha}\cdot\hat{p}.
            \end{eqnarray}
      We then put the physical system into an external electromagnetic fields $\{\vec{B}, \vec{q}\}$ (or represented by the vector potential $\vec{A}$ and the scalar potential $\phi$), thus we have the mixed Hamiltonian $\mathcal{H}_{\rm mix}$ as shown in Eq. (\ref{eq:mix-1bb}).
      The physical picture is very helpful for designing the experiment. In addition, the unitary transformation $\mathcal{C}(\vartheta)$ does not depend on mass $m_{\rm e}$, thus it is feasible to be realized in experiment. $\blacksquare$
      \end{remark}

      Now we come to study the Dirac equation
      \begin{equation}\label{eq:DiracHeHb1-1}
            {\rm i}\hbar\p{}{t} \Psi=\mathcal{H}_{\rm mix} \Psi.
      \end{equation}
      Similarly we consider the Dirac equation in the representation of
      \begin{equation}
            \Psi=\begin{bmatrix}
                  \tilde{\eta} \\ \tilde{\chi}
            \end{bmatrix},
      \end{equation}
      then \Eq{eq:DiracHeHb1-1} becomes

      \begin{eqnarray}
            {\rm i}\hbar\p{}{t} \begin{bmatrix}
                        \tilde{\eta} \\ \tilde{\chi}
                  \end{bmatrix}
            &=& \cos\vartheta\,c\,\begin{bmatrix}
                              0 & \vec{\sigma} \\
                              \vec{\sigma} & 0
                        \end{bmatrix}\cdot\vec{\Pi}\begin{bmatrix}
                                    \tilde{\eta} \\ \tilde{\chi}
                              \end{bmatrix}
                  -\sin\vartheta\,\dfrac{1}{2} m_{\rm e}\,c^2 \begin{bmatrix}
                              0 & \dfrac{1}{\sqrt{\vec{\Pi}^2}}\vec{\sigma}\cdot\vec{\Pi}+ \vec{\sigma}\cdot\vec{\Pi}\dfrac{1}{\sqrt{\vec{\Pi}^2}} \\
                             \dfrac{1}{\sqrt{\vec{\Pi}^2}}\vec{\sigma}\cdot\vec{\Pi}+ \vec{\sigma}\cdot\vec{\Pi}\dfrac{1}{\sqrt{\vec{\Pi}^2}} & 0
                        \end{bmatrix}\begin{bmatrix}
                                    \tilde{\eta} \\ \tilde{\chi}
                              \end{bmatrix}\notag \\
                  && +\cos\vartheta\,\begin{bmatrix}
                              \openone & 0 \\
                              0 & -\openone
                        \end{bmatrix}\,m_{\rm e}\,c^2 \begin{bmatrix}
                              \tilde{\eta} \\ \tilde{\chi}
                        \end{bmatrix}
                  +\sin\vartheta\,\begin{bmatrix}
                              \openone & 0 \\
                              0 & -\openone
                        \end{bmatrix}\sqrt{\vec{\Pi}^2}\,c \begin{bmatrix}
                              \tilde{\eta} \\ \tilde{\chi}
                        \end{bmatrix}
                  +q\,\phi\begin{bmatrix}
                              \tilde{\eta} \\ \tilde{\chi}
                        \end{bmatrix},
      \end{eqnarray}
      i.e.,
      \begin{eqnarray}\label{eq:DiracEExpand-1A}
            {\rm i}\hbar\p{}{t} \begin{bmatrix}
                        \tilde{\eta} \\ \tilde{\chi}
                  \end{bmatrix}
            &=&  \Biggl\{\cos\vartheta\,c(\vec{\sigma}\cdot\vec{\Pi})
                        -\sin\vartheta\,\dfrac{1}{2} m_{\rm e}\,c^2 \bigg[
                              \dfrac{1}{\sqrt{\vec{\Pi}^2}}\vec{\sigma}\cdot\vec{\Pi}
                              +\vec{\sigma}\cdot\vec{\Pi}\dfrac{1}{\sqrt{\vec{\Pi}^2}}\bigg]\Biggr\}\begin{bmatrix}
                                    \tilde{\chi} \\ \tilde{\eta}
                              \end{bmatrix} \notag \\
                  && +(\cos\vartheta\,m_{\rm e}\,c^2 +\sin\vartheta\,\sqrt{\vec{\Pi}^2}\,c)
                        \begin{bmatrix}
                              \tilde{\eta} \\ -\tilde{\chi}
                        \end{bmatrix}
                  +q\,\phi\begin{bmatrix}
                              \tilde{\eta} \\ \tilde{\chi}
                        \end{bmatrix}.
      \end{eqnarray}

      Let us investigate the nonrelativistic approximation. In the case of low energy, the term involving $m_{\rm e}\,c^2$ is far larger than other ones. Then a part of time factor can be separated as follows.
      \begin{equation}
            \begin{bmatrix}
                  \tilde{\eta} \\ \tilde{\chi}
            \end{bmatrix}=\begin{bmatrix}
                  \eta \\ \chi
            \end{bmatrix}{\rm e}^{-{\rm i}\,\cos\vartheta\,\frac{m_{\rm e}\,c^2}{\hbar} t},
      \end{equation}
      then \Eq{eq:DiracEExpand-1A} is transformed as
      \begin{eqnarray}
            && {\rm i}\hbar\Biggl\{\p{}{t} \begin{bmatrix}
                        \eta \\ \chi
                  \end{bmatrix}\Biggr\}{\rm e}^{-{\rm i}\,\cos\vartheta\,\frac{m_{\rm e}\,c^2}{\hbar} t}
            +{\rm i}\begin{bmatrix}
                        \eta \\ \chi
                  \end{bmatrix}\hbar\Bigl(\p{}{t} {\rm e}^{
                        -{\rm i}\,\cos\vartheta\,\frac{m_{\rm e}\,c^2}{\hbar} t}\Bigr) \notag \\
            &=& \Biggl\{\cos\vartheta\,c(\vec{\sigma}\cdot\vec{\Pi})
                        -\sin\vartheta\,\dfrac{1}{2} m_{\rm e}\,c^2 \bigg[
                              \dfrac{1}{\sqrt{\vec{\Pi}^2}}\vec{\sigma}\cdot\vec{\Pi}
                              +\vec{\sigma}\cdot\vec{\Pi}\dfrac{1}{\sqrt{\vec{\Pi}^2}}\bigg]\Biggr\}\begin{bmatrix}
                                    \chi \\ \eta
                              \end{bmatrix}{\rm e}^{-{\rm i}\,\cos\vartheta\,\frac{m_{\rm e}\,c^2}{\hbar} t}
                              \notag \\
                  && +(\cos\vartheta\,m_{\rm e}\,c^2 +\sin\vartheta\,\sqrt{\vec{\Pi}^2}\,c)
                        \begin{bmatrix}
                              \eta \\ -\chi
                        \end{bmatrix}{\rm e}^{-{\rm i}\,\cos\vartheta\,\frac{m_{\rm e}\,c^2}{\hbar} t}
                  +q\,\phi\begin{bmatrix}
                              \eta \\ \chi
                        \end{bmatrix}{\rm e}^{-{\rm i}\,\cos\vartheta\,\frac{m_{\rm e}\,c^2}{\hbar} t},
      \end{eqnarray}
      i.e.,
      \begin{eqnarray}
            {\rm i}\hbar\p{}{t} \begin{bmatrix}
                        \eta \\ \chi
                  \end{bmatrix}+\cos\vartheta\,m_{\rm e}\,c^2 \begin{bmatrix}
                        \eta \\ \chi
                  \end{bmatrix}
            &=& \Biggl\{\cos\vartheta\,c(\vec{\sigma}\cdot\vec{\Pi})
                        -\sin\vartheta\,\dfrac{1}{2} m_{\rm e}\,c^2 \bigg[
                              \dfrac{1}{\sqrt{\vec{\Pi}^2}}\vec{\sigma}\cdot\vec{\Pi}
                              +\vec{\sigma}\cdot\vec{\Pi}\dfrac{1}{\sqrt{\vec{\Pi}^2}}\bigg]\Biggr\}\begin{bmatrix}
                                    \chi \\ \eta
                              \end{bmatrix} \notag \\
                  && +(\cos\vartheta\,m_{\rm e}\,c^2 +\sin\vartheta\,\sqrt{\vec{\Pi}^2}\,c)
                        \begin{bmatrix}
                              \eta \\ -\chi
                        \end{bmatrix}
                  +q\,\phi\begin{bmatrix}
                              \eta \\ \chi
                        \end{bmatrix},
      \end{eqnarray}
      i.e.,
      \begin{eqnarray}\label{eq:mixeq-1a}
            {\rm i}\hbar\p{}{t} \begin{bmatrix}
                        \eta \\ \chi
                  \end{bmatrix}
            &=& \Biggl\{\cos\vartheta\,c(\vec{\sigma}\cdot\vec{\Pi})
                        -\sin\vartheta\,\dfrac{1}{2} m_{\rm e}\,c^2 \bigg[
                              \dfrac{1}{\sqrt{\vec{\Pi}^2}}\vec{\sigma}\cdot\vec{\Pi}
                              +\vec{\sigma}\cdot\vec{\Pi}\dfrac{1}{\sqrt{\vec{\Pi}^2}}\bigg]\Biggr\}\begin{bmatrix}
                                    \chi \\ \eta
                              \end{bmatrix} \notag \\
                  && +\begin{bmatrix}
                        \sin\vartheta\,\sqrt{\vec{\Pi}^2}\,c\:\eta \\ -\big(
                              2\,\cos\vartheta\,m_{\rm e}\,c^2
                              +\sin\vartheta\,\sqrt{\vec{\Pi}^2}\,c\big)\chi
                  \end{bmatrix}
                  +q\,\phi\begin{bmatrix}
                              \eta \\ \chi
                        \end{bmatrix}.
      \end{eqnarray}

      \subsection{Approximations: Weak magnetic field $B=|\vec{B}|$}

     Let us consider the following approximations: Weak magnetic field, i.e., $B=|\vec{B}|$ is small so that $B^2\approx 0$.

     In classical mechanics, the linear momentum $\vec{p}=m_{\rm e} \vec{v}$, where $\vec{v}$ is the velocity of the particle. Usually the speed $v=|\vec{v}|$ is much smaller than the speed of light $c$, i.e., $c \gg v$, thus we have $m_{\rm e}\,c^2 \gg p\,c$. From
      \begin{eqnarray}
       \vec{\Pi}^2  &\approx & \vec{p}^{\,2}- \dfrac{q}{c}\vec{B}\cdot \vec{\ell},
      \end{eqnarray}
      we have
        \begin{eqnarray}
      \sqrt{\vec{\Pi}^2} & \approx &  \sqrt{\vec{p}^{\,2}- \dfrac{q}{c}\vec{B}\cdot \vec{\ell}}=\sqrt{\vec{p}^{\,2}} \sqrt{1 -\dfrac{q}{c}\frac{1}{\vec{p}^{\,2}}\vec{B}\cdot \vec{\ell}}\nonumber\\
       & \approx &  \sqrt{\vec{p}^{\,2}} \left(1 -\frac{1}{2}\dfrac{q}{c}\frac{1}{\vec{p}^{\,2}}\vec{B}\cdot \vec{\ell}\right)
       = p  -\dfrac{q}{2 c}\frac{1}{p}\vec{B}\cdot \vec{\ell},
      \end{eqnarray}
     then for the weak value $B$ and the mixing angle
     \begin{eqnarray}
     \vartheta \in [-\frac{\pi}{4}, \frac{\pi}{4}],
      \end{eqnarray}
     we have
     \begin{eqnarray}
     \cos\vartheta\,m_{\rm e}\,c^2 \gg \left|\sin\vartheta\,\sqrt{\vec{\Pi}^2}\,c\right|.
      \end{eqnarray}

            Then for the 2nd row of \Eq{eq:mixeq-1a}, we have
            \begin{eqnarray}
                  \openone\,{\rm i}\hbar\p{\chi}{t} &= &\Biggl\{
                              \cos\vartheta\,c(\vec{\sigma}\cdot\vec{\Pi})
                              -\sin\vartheta\,\dfrac{1}{2} m_{\rm e}\,c^2 \bigg[
                                    \dfrac{1}{\sqrt{\vec{\Pi}^2}}\vec{\sigma}\cdot\vec{\Pi}
                                    +\vec{\sigma}\cdot\vec{\Pi}\dfrac{1}{\sqrt{\vec{\Pi}^2}}\bigg]\Biggr\}\eta \nonumber\\
                    &&    -\openone\,\left(2\,\cos\vartheta\,m_{\rm e}\,c^2+ \sin\vartheta\,\sqrt{\vec{\Pi}^2}\,c\right)\chi+\openone\,q\,\phi\,\chi.
            \end{eqnarray}
            Because
            \begin{eqnarray}
               && m_{\rm e}\,c^2 \chi\gg{\rm i}\hbar\,\dfrac{\partial\chi}{\partial t}, \nonumber\\
               && m_{\rm e}\,c^2 \chi\gg q\,\phi\,\chi, \nonumber\\
               && m_{\rm e}\,c^2 \chi \gg \sqrt{\vec{\Pi}^2}\,c\,\chi,
            \end{eqnarray}
            then we arrive at
            \begin{equation}
                  \Biggl\{\cos\vartheta\,c(\vec{\sigma}\cdot\vec{\Pi})
                        -\sin\vartheta\,\dfrac{1}{2} m_{\rm e}\,c^2 \bigg[
                              \dfrac{1}{\sqrt{\vec{\Pi}^2}}\vec{\sigma}\cdot\vec{\Pi}
                              +\vec{\sigma}\cdot\vec{\Pi}\dfrac{1}{\sqrt{\vec{\Pi}^2}}\bigg]\Biggr\}\eta
                  = \cos\vartheta\, 2\,m_{\rm e}\,c^2 \openone\chi,
            \end{equation}
            i.e.,
            \begin{equation}
                  \chi=\dfrac{1}{ \cos\vartheta\,2\,m_{\rm e}\,c^2} \Biggl\{\cos\vartheta\,c(
                              \vec{\sigma}\cdot\vec{\Pi})
                        -\sin\vartheta\,\dfrac{1}{2} m_{\rm e}\,c^2 \bigg[
                              \dfrac{1}{\sqrt{\vec{\Pi}^2}}\vec{\sigma}\cdot\vec{\Pi}
                              +\vec{\sigma}\cdot\vec{\Pi}\dfrac{1}{\sqrt{\vec{\Pi}^2}}\bigg]\Biggr\}\eta.
            \end{equation}
            After that, for the first row of \Eq{eq:mixeq-1a}, we obtain
            \begin{equation}
                  {\rm i}\hbar\p{\eta}{t} =\Biggl\{
                              \cos\vartheta\,c(\vec{\sigma}\cdot\vec{\Pi})
                              -\sin\vartheta\,\dfrac{1}{2} m_{\rm e}\,c^2 \bigg[
                                    \dfrac{1}{\sqrt{\vec{\Pi}^2}}\vec{\sigma}\cdot\vec{\Pi}
                                    +\vec{\sigma}\cdot\vec{\Pi}\dfrac{1}{\sqrt{\vec{\Pi}^2}}\bigg]\Biggr\}\chi
                     +\sin\vartheta\,\sqrt{\vec{\Pi}^2}\,c\:\eta   +q\,\phi\,\eta,
            \end{equation}
            i.e.,
            \begin{eqnarray}\label{eq:EtaHeHb1}
            {\rm i}\hbar\p{\eta}{t} &=& \dfrac{1}{\cos\vartheta\,2\,m_{\rm e}\,c^2} \Biggl\{
                        \cos\vartheta\,c(\vec{\sigma}\cdot\vec{\Pi})
                        -\sin\vartheta\,\dfrac{1}{2} m_{\rm e}\,c^2\bigg[
                              \dfrac{1}{\sqrt{\vec{\Pi}^2}} (
                                    \vec{\sigma}\cdot\vec{\Pi})
                              +(\vec{\sigma}\cdot\vec{\Pi})
                                    \dfrac{1}{\sqrt{\vec{\Pi}^2}}\bigg]
                        \Biggr\}^2 \eta  +\sin\vartheta\,\sqrt{\vec{\Pi}^2}\,c\:\eta+q\,\phi\,\eta\nonumber\\
            &=& \dfrac{1}{\cos\vartheta\,2\,m_{\rm e}} \Biggl\{
                        \cos\vartheta\,(\vec{\sigma}\cdot\vec{\Pi})
                        -\sin\vartheta\,\dfrac{1}{2} m_{\rm e}\,c\bigg[
                              \dfrac{1}{\sqrt{\vec{\Pi}^2}} (
                                    \vec{\sigma}\cdot\vec{\Pi})
                              +(\vec{\sigma}\cdot\vec{\Pi})
                                    \dfrac{1}{\sqrt{\vec{\Pi}^2}}\bigg]
                        \Biggr\}^2 \eta  +\sin\vartheta\,\sqrt{\vec{\Pi}^2}\,c\:\eta+q\,\phi\,\eta.
            \end{eqnarray}

            \begin{remark}We need to calculate the following term
            \begin{eqnarray}\label{eq:mix-1b}
                  && \Biggl\{\cos\vartheta\,(\vec{\sigma}\cdot\vec{\Pi})
                        -\sin\vartheta\,\dfrac{1}{2} m_{\rm e}\,c\bigg[
                              \dfrac{1}{\sqrt{\vec{\Pi}^2}} (
                                    \vec{\sigma}\cdot\vec{\Pi})
                              +(\vec{\sigma}\cdot\vec{\Pi})
                                    \dfrac{1}{\sqrt{\vec{\Pi}^2}}\bigg]\Biggr\}^2
                        \notag \\
                 & =& {\cos^2}\vartheta\,(\vec{\sigma}\cdot\vec{\Pi})(
                              \vec{\sigma}\cdot\vec{\Pi})
                        +\dfrac{{\sin^2}\vartheta}{4} m_{\rm e}^2 c^2 \bigg[
                              \dfrac{1}{\sqrt{\vec{\Pi}^2}} (
                                    \vec{\sigma}\cdot\vec{\Pi})
                              +(\vec{\sigma}\cdot\vec{\Pi})
                                    \dfrac{1}{\sqrt{\vec{\Pi}^2}}
                              \bigg]\bigg[\dfrac{1}{\sqrt{\vec{\Pi}^2}} (
                                          \vec{\sigma}\cdot\vec{\Pi})
                                    +(\vec{\sigma}\cdot\vec{\Pi})
                                          \dfrac{1}{\sqrt{\vec{\Pi}^2}}\bigg] \notag \\
                       & & -\sin\vartheta\,\cos\vartheta\,\dfrac{1}{2} m_{\rm e}\,c\Biggl\{(
                              \vec{\sigma}\cdot\vec{\Pi})\bigg[
                                    \dfrac{1}{\sqrt{\vec{\Pi}^2}} (
                                          \vec{\sigma}\cdot\vec{\Pi})
                              +(\vec{\sigma}\cdot\vec{\Pi})
                                    \dfrac{1}{\sqrt{\vec{\Pi}^2}}\bigg]
                              +\bigg[\dfrac{1}{\sqrt{\vec{\Pi}^2}} (
                                          \vec{\sigma}\cdot\vec{\Pi})
                                    +(\vec{\sigma}\cdot\vec{\Pi})
                                          \dfrac{1}{\sqrt{\vec{\Pi}^2}}
                                    \bigg](\vec{\sigma}\cdot\vec{\Pi})\Biggr\} \notag \\
                  &=& {\cos^2}\vartheta\,(\vec{\sigma}\cdot\vec{\Pi})(
                              \vec{\sigma}\cdot\vec{\Pi})
                        +\dfrac{{\sin^2}\vartheta}{4} m_{\rm e}^2 c^2 \bigg[
                              \dfrac{1}{\sqrt{\vec{\Pi}^2}} (
                                    \vec{\sigma}\cdot\vec{\Pi})
                              +(\vec{\sigma}\cdot\vec{\Pi})
                                    \dfrac{1}{\sqrt{\vec{\Pi}^2}}
                              \bigg]^2 \notag \\
                       & & -\sin\vartheta\,\cos\vartheta\,\dfrac{1}{2} m_{\rm e}\,c\Biggl(2(
                              \vec{\sigma}\cdot\vec{\Pi})
                                    \dfrac{1}{\sqrt{\vec{\Pi}^2}} (
                                          \vec{\sigma}\cdot\vec{\Pi})
                              +\bigg\{(\vec{\sigma}\cdot\vec{\Pi})(
                                    \vec{\sigma}\cdot\vec{\Pi}),\
                                    \dfrac{1}{\sqrt{\vec{\Pi}^2}}\bigg\}\Biggr).
            \end{eqnarray}
            $\blacksquare$
            \end{remark}

            \begin{remark}
            We study Eq. (\ref{eq:mix-1b}) with the approximation of weak $\vec{B}$. We need to know the following four terms.

            (i) From Eq. (\ref{eq:SigmPi-1a}) we have known that
            \begin{eqnarray}
             (\vec{\sigma}\cdot\vec{\Pi})^2 &=& (\vec{\sigma}\cdot\vec{\Pi})(
                  \vec{\sigma}\cdot\vec{\Pi})= \vec{\Pi}^2 \openone +{\rm i} \vec{\sigma}\cdot \left(\vec{\Pi}\times \vec{\Pi}\right)
            = \vec{\Pi}^2 \openone -\dfrac{q\,\hbar}{c} (\vec{\sigma}\cdot\vec{B}).
            \end{eqnarray}

            (ii) One can have
            \begin{eqnarray}
                  \Big(\dfrac{1}{\sqrt{\vec{\Pi}^2}} \vec{\sigma}\cdot\vec{\Pi}
                        +\vec{\sigma}\cdot\vec{\Pi}\dfrac{1}{\sqrt{\vec{\Pi}^2}}\Big)^2
                  \approx 4\, \left[\openone -\dfrac{q\,\hbar}{c} \dfrac{1}{\vec{p}^4} \left[(\vec{p}\cdot\vec{B})(\vec{\sigma}\cdot\vec{p})\right]\right].
            \end{eqnarray}
            \begin{proof}
            Because
            \begin{eqnarray}
            \dfrac{1}{\sqrt{\vec{\Pi}^2}}
                       & =& \dfrac{1}{\sqrt{\Bigl(\vec{p}-\dfrac{q}{c} \vec{A}\Bigr)^2}} \approx  \dfrac{1}{\sqrt{\vec{p}^2 \Big\{1-\dfrac{q}{c} \dfrac{1}{\vec{p}^2} (\vec{B}\cdot\vec{\ell})\Big\}}}  \approx
                              \dfrac{1}{\sqrt{\vec{p}^2}} \Big\{1
                                    +\dfrac{q}{2\,c} \dfrac{1}{\vec{p}^2} (\vec{B}\cdot\vec{\ell}) \Big\}, \nonumber\\
             \Big[\vec{\Pi},\ \dfrac{1}{\sqrt{\vec{\Pi}^2}}\Big]
                        & \approx& \left[\vec{p}-\dfrac{q}{2\,c} (\vec{B}\times\vec{r}),\
                              \dfrac{1}{\sqrt{\vec{p}^2}} \Big\{1
                                    +\dfrac{q}{2\,c} \dfrac{1}{\vec{p}^2} (\vec{B}\cdot\vec{\ell})
                                    \Big\}\right] \notag \\
                       & =& \dfrac{1}{\sqrt{\vec{p}^2}} \bigg[\vec{p},\ \Big\{1+\dfrac{q}{2\,c}
                                    \dfrac{1}{\vec{p}^2} (\vec{B}\cdot\vec{\ell})\Big\}\bigg]
                              -\dfrac{q}{2\,c} \vec{B}\times\bigg[\vec{r},\
                                    \dfrac{1}{\sqrt{\vec{p}^2}} \Big\{1+\dfrac{q}{2\,c}
                                          \dfrac{1}{\vec{p}^2} (\vec{B}\cdot\vec{\ell})\Big\}\bigg]
                              \notag \\
                       &=& \dfrac{q}{2\,c} \dfrac{1}{\sqrt{\vec{p}^6}} \big[\vec{p},\ (
                                    \vec{B}\cdot\vec{\ell})\big]
                              -\dfrac{q}{2\,c} \vec{B}\times\bigg[\vec{r},\
                                    \dfrac{1}{\sqrt{\vec{p}^2}}\bigg]\nonumber\\
                       &=& {\rm i}\hbar\dfrac{q}{2\,c} \dfrac{1}{\sqrt{\vec{p}^6}} (
                                    \vec{B}\times\vec{p})
                              +{\rm i}\hbar\dfrac{q}{2\,c} \vec{B}
                                    \times\dfrac{\vec{p}}{\sqrt{\vec{p}^6}} \notag \\
                        &\approx & {\rm i}\hbar\dfrac{q}{c} \dfrac{1}{\sqrt{\vec{p}^6}} (
                                    \vec{B}\times\vec{p})= {\rm i}\hbar\dfrac{q}{c} \dfrac{1}{p^3} (
                                    \vec{B}\times\vec{p}),
                  \end{eqnarray}

                  \begin{eqnarray}
                         \dfrac{1}{\vec{\Pi}^2}
                       & =& \dfrac{1}{\Bigl(\vec{p}-\dfrac{q}{c} \vec{A}\Bigr)^2} \approx  \dfrac{1}{{\vec{p}^2 \Big\{1-\dfrac{q}{c} \dfrac{1}{\vec{p}^2} (\vec{B}\cdot\vec{\ell})\Big\}}}  \approx
                              \dfrac{1}{{\vec{p}^2}} \Big\{1
                                    +\dfrac{q}{\,c} \dfrac{1}{\vec{p}^2} (\vec{B}\cdot\vec{\ell}) \Big\},\nonumber\\
                         \Big[\vec{\Pi},\ \dfrac{1}{{\vec{\Pi}^2}}\Big]
                        & \approx& \left[\vec{p}-\dfrac{q}{2\,c} (\vec{B}\times\vec{r}),\
                              \dfrac{1}{{\vec{p}^2}} \Big\{1
                                    +\dfrac{q}{c} \dfrac{1}{\vec{p}^2} (\vec{B}\cdot\vec{\ell})
                                    \Big\}\right] \notag \\
                       & =& \dfrac{1}{{\vec{p}^2}} \bigg[\vec{p},\ \Big\{1+\dfrac{q}{c}
                                    \dfrac{1}{\vec{p}^2} (\vec{B}\cdot\vec{\ell})\Big\}\bigg]
                              -\dfrac{q}{2\,c} \vec{B}\times\bigg[\vec{r},\
                                    \dfrac{1}{{\vec{p}^2}} \Big\{1+\dfrac{q}{c}
                                          \dfrac{1}{\vec{p}^2} (\vec{B}\cdot\vec{\ell})\Big\}\bigg]
                              \notag \\
                       &=& \dfrac{q}{c} \dfrac{1}{{\vec{p}^4}} \big[\vec{p},\ (
                                    \vec{B}\cdot\vec{\ell})\big]
                              -\dfrac{q}{2\,c} \vec{B}\times\bigg[\vec{r},\
                                    \dfrac{1}{{\vec{p}^2}}\bigg]\nonumber\\
                       &=& {\rm i}\hbar\dfrac{q}{c} \dfrac{1}{{\vec{p}^4}} (
                                    \vec{B}\times\vec{p})
                              +{\rm i}\hbar\dfrac{q}{c} \vec{B}
                                    \times\dfrac{\vec{p}}{{\vec{p}^4}} \notag \\
                        &\approx &  2\,{\rm i}\hbar\dfrac{q}{c} \dfrac{1}{\vec{p}^4} (
                  \vec{B}\times\vec{p})=2\,{\rm i}\hbar\dfrac{q}{c} \dfrac{1}{p^4} (
                  \vec{B}\times\vec{p}),
                  \end{eqnarray}

             \begin{eqnarray}
             \big[\vec{\sigma}\cdot(\vec{B}\times\vec{p})\big](\vec{\sigma}\cdot\vec{\Pi})
            &\approx& \big[\vec{\sigma}\cdot(\vec{B}\times\vec{p})\big](\vec{\sigma}\cdot\vec{p})={\rm i}\vec{\sigma}\cdot\left[(\vec{B}\times\vec{p})\times \vec{p}\right]\nonumber\\
            &=& {\rm i}\vec{\sigma}\cdot\left[ (\vec{B}\cdot\vec{p})\vec{p}-\vec{B}\vec{p}^{\, 2}\right]={\rm i}\left[  (\vec{B}\cdot\vec{p})(\vec{\sigma}\cdot\vec{p})-\vec{p}^{\, 2} (\vec{\sigma}\cdot\vec{B})\right],\nonumber\\
             (\vec{\sigma}\cdot\vec{\Pi})\big[\vec{\sigma}\cdot(\vec{B}\times\vec{p})\big]
            &\approx& (\vec{\sigma}\cdot\vec{p})\big[\vec{\sigma}\cdot(\vec{B}\times\vec{p})\big]=
            \left(\big[\vec{\sigma}\cdot(\vec{B}\times\vec{p})\big](\vec{\sigma}\cdot\vec{p})\right)^\dagger \nonumber\\
            &=& -{\rm i}\left[  (\vec{B}\cdot\vec{p})(\vec{\sigma}\cdot\vec{p})-\vec{p}^{\, 2} (\vec{\sigma}\cdot\vec{B})\right],
      \end{eqnarray}
      and
      \begin{eqnarray}
             (\vec{\sigma}\cdot\vec{\Pi})\dfrac{1}{\vec{\Pi}^2} (
                  \vec{\sigma}\cdot\vec{\Pi})
            &=&\vec{\sigma}\cdot\biggl\{\Big[\vec{\Pi},\ \dfrac{1}{\vec{\Pi}^2}\Big]
                  +\dfrac{1}{\vec{\Pi}^2} \vec{\Pi}\biggr\} (\vec{\sigma}\cdot\vec{\Pi})\nonumber\\
           & \approx &  2\,{\rm i}\hbar\dfrac{q}{c} \dfrac{1}{\vec{p}^4} \big[\vec{\sigma}
                        \cdot(\vec{B}\times\vec{p})\big](\vec{\sigma}\cdot\vec{\Pi})
                  +\dfrac{1}{\vec{\Pi}^2} (\vec{\sigma}\cdot\vec{\Pi})(\vec{\sigma}
                        \cdot\vec{\Pi}) \notag \\
           & \approx& 2\,{\rm i}\hbar\dfrac{q}{c} \dfrac{1}{\vec{p}^4}  {\rm i}\,
                  \big[-\vec{p}^2 (\vec{\sigma}\cdot\vec{B})
                        +(\vec{p}\cdot\vec{B})(\vec{\sigma}\cdot\vec{p})\big]
                  +\dfrac{1}{\vec{\Pi}^2} \Bigl[\openone\,\vec{\Pi}^2
                        -\dfrac{q\,\hbar}{c} (\vec{\sigma}\cdot\vec{B})\Bigr] \notag \\
          &  \approx& 2\,\hbar\dfrac{q}{c} \dfrac{1}{\vec{p}^4}\big[
                  \vec{p}^2 (\vec{\sigma}\cdot\vec{B})
                        -(\vec{p}\cdot\vec{B})(\vec{\sigma}\cdot\vec{p})\big]
                  +\Bigl[\openone-\dfrac{q\,\hbar}{c}
                        \dfrac{1}{\vec{p}^2}(\vec{\sigma}\cdot\vec{B})\Bigr] \notag \\
           & =& \openone {+} \dfrac{q\,\hbar}{c}
                        \dfrac{1}{\vec{p}^2}(\vec{\sigma}\cdot\vec{B})
                 {-2}\dfrac{q\,\hbar}{c} \dfrac{1}{\vec{p}^4}(\vec{p}\cdot\vec{B})(
                        \vec{\sigma}\cdot\vec{p}).
      \end{eqnarray}

                  then we have
                  \begin{eqnarray}
                        && \Big(\dfrac{1}{\sqrt{\vec{\Pi}^2}} \vec{\sigma}\cdot\vec{\Pi}
                              +\vec{\sigma}\cdot\vec{\Pi}\dfrac{1}{\sqrt{\vec{\Pi}^2}}\Big)^2
                              \notag \\
                        &=& \dfrac{1}{\sqrt{\vec{\Pi}^2}} (\vec{\sigma}\cdot\vec{\Pi})
                                    \dfrac{1}{\sqrt{\vec{\Pi}^2}} (\vec{\sigma}\cdot\vec{\Pi})
                              +(\vec{\sigma}\cdot\vec{\Pi})\dfrac{1}{\sqrt{\vec{\Pi}^2}}
                                    (\vec{\sigma}\cdot\vec{\Pi})\dfrac{1}{\sqrt{\vec{\Pi}^2}}
                              +\dfrac{1}{\sqrt{\vec{\Pi}^2}} (\vec{\sigma}\cdot\vec{\Pi})(
                                    \vec{\sigma}\cdot\vec{\Pi})\dfrac{1}{\sqrt{\vec{\Pi}^2}}
                              +(\vec{\sigma}\cdot\vec{\Pi})\dfrac{1}{\sqrt{\vec{\Pi}^2}}
                                    \dfrac{1}{\sqrt{\vec{\Pi}^2}} (\vec{\sigma}\cdot\vec{\Pi})
                              \notag \\
                        &=& \dfrac{1}{\sqrt{\vec{\Pi}^2}} \vec{\sigma}\cdot\Biggl\{\bigg[\vec{\Pi},\
                                          \dfrac{1}{\sqrt{\vec{\Pi}^2}}\bigg]
                                    +\dfrac{1}{\sqrt{\vec{\Pi}^2}} \vec{\Pi}\Biggr\} (\vec{\sigma}
                                          \cdot\vec{\Pi})
                              +(\vec{\sigma}\cdot\vec{\Pi})\vec{\sigma}\cdot\Biggl\{\vec{\Pi}
                                          \dfrac{1}{\sqrt{\vec{\Pi}^2}}
                                    -\bigg[\vec{\Pi},\ \dfrac{1}{\sqrt{\vec{\Pi}^2}}\bigg]
                                    \Biggr\}\dfrac{1}{\sqrt{\vec{\Pi}^2}} \notag \\
                             & & +\dfrac{1}{\sqrt{\vec{\Pi}^2}} (\vec{\sigma}\cdot\vec{\Pi})^2
                                    \dfrac{1}{\sqrt{\vec{\Pi}^2}}
                              +(\vec{\sigma}\cdot\vec{\Pi})\dfrac{1}{\sqrt{\vec{\Pi}^4}} (
                                    \vec{\sigma}\cdot\vec{\Pi}) \notag \\
                       & \approx& \Biggl\{\dfrac{1}{\sqrt{\vec{\Pi}^2}} \vec{\sigma}
                                          \cdot{\rm i}\hbar\dfrac{q}{c} \dfrac{1}{\sqrt{\vec{p}^6}}
                                                (\vec{B}\times\vec{p})
                                    +\dfrac{1}{\vec{\Pi}^2} (\vec{\sigma}\cdot\vec{\Pi})\Biggr\}(
                                          \vec{\sigma}\cdot\vec{\Pi})
                              +\Biggl\{(\vec{\sigma}\cdot\vec{\Pi})(\vec{\sigma}\cdot\vec{\Pi})
                                          \dfrac{1}{\sqrt{\vec{\Pi}^2}}
                                    -(\vec{\sigma}\cdot\vec{\Pi})\vec{\sigma}\cdot{\rm i}\hbar
                                          \dfrac{q}{c} \dfrac{1}{\sqrt{\vec{p}^6}} (
                                                \vec{B}\times\vec{p})
                                    \Biggr\}\dfrac{1}{\sqrt{\vec{\Pi}^2}} \notag \\
                             & & +\dfrac{1}{\sqrt{\vec{\Pi}^2}} (\vec{\sigma}\cdot\vec{\Pi})^2
                                    \dfrac{1}{\sqrt{\vec{\Pi}^2}}
                              +(\vec{\sigma}\cdot\vec{\Pi})\dfrac{1}{\vec{\Pi}^2} (
                                    \vec{\sigma}\cdot\vec{\Pi}) \notag \\
                        &\approx& \dfrac{1}{\vec{p}^4} {\rm i}\hbar\dfrac{q}{c} \big[\vec{\sigma}
                                    \cdot(\vec{B}\times\vec{p})\big](\vec{\sigma}\cdot\vec{\Pi})
                              -\dfrac{1}{\vec{p}^4} {\rm i}\hbar\dfrac{q}{c} (\vec{\sigma}
                                    \cdot\vec{\Pi})\big[\vec{\sigma}\cdot(\vec{B}\times\vec{p})
                                    \big]
                              +\dfrac{1}{\vec{\Pi}^2} (\vec{\sigma}\cdot\vec{\Pi})(
                                    \vec{\sigma}\cdot\vec{\Pi})
                              +(\vec{\sigma}\cdot\vec{\Pi})(\vec{\sigma}\cdot\vec{\Pi})
                                    \dfrac{1}{\vec{\Pi}^2}  \notag \\
                              && +\dfrac{1}{\sqrt{\vec{\Pi}^2}} (\vec{\sigma}\cdot\vec{\Pi})^2
                                    \dfrac{1}{\sqrt{\vec{\Pi}^2}}
                              +(\vec{\sigma}\cdot\vec{\Pi})\dfrac{1}{\vec{\Pi}^2} (
                                    \vec{\sigma}\cdot\vec{\Pi}) \notag \\
                        &=& \dfrac{1}{\vec{p}^4} {\rm i}\hbar\dfrac{q}{c} \big[\vec{\sigma}
                                    \cdot(\vec{B}\times\vec{p})\big](\vec{\sigma}\cdot\vec{\Pi})
                              -\dfrac{1}{\vec{p}^4} {\rm i}\hbar\dfrac{q}{c} (\vec{\sigma}
                                    \cdot\vec{\Pi})\big[\vec{\sigma}\cdot(\vec{B}\times\vec{p})
                                    \big]
                              +\dfrac{1}{\vec{\Pi}^2} (\vec{\sigma}\cdot\vec{\Pi})(
                                    \vec{\sigma}\cdot\vec{\Pi})
                              +(\vec{\sigma}\cdot\vec{\Pi})(\vec{\sigma}\cdot\vec{\Pi})
                                    \dfrac{1}{\vec{\Pi}^2}  \notag \\
                              && +\dfrac{1}{\sqrt{\vec{\Pi}^2}} (\vec{\sigma}\cdot\vec{\Pi})(
                                    \vec{\sigma}\cdot\vec{\Pi})\dfrac{1}{\sqrt{\vec{\Pi}^2}}
                              +(\vec{\sigma}\cdot\vec{\Pi})\dfrac{1}{\vec{\Pi}^2} (
                                    \vec{\sigma}\cdot\vec{\Pi}) \notag \\
                        &=& \dfrac{1}{\vec{p}^4} {\rm i}\hbar\dfrac{q}{c} {\rm i}\,\big[(\vec{p}\cdot\vec{B})(\vec{\sigma}
                                                \cdot\vec{p})
                                    -\vec{p}^2 (\vec{\sigma}\cdot\vec{B})\big]
                              -\dfrac{1}{\vec{p}^4} {\rm i}\hbar\dfrac{q}{c} {\rm i}\,\big[\vec{p}^2 (\vec{\sigma}\cdot\vec{B})-(
                                          \vec{p}\cdot\vec{B})(\vec{\sigma}\cdot\vec{p})\big]
                                    \notag \\
                             & & +\dfrac{1}{\vec{\Pi}^2} \Bigl[\openone\,\vec{\Pi}^2
                                    -\dfrac{q\,\hbar}{c} (\vec{\sigma}\cdot\vec{B})\Bigr]
                              +\Bigl[\openone\,\vec{\Pi}^2
                                    -\dfrac{q\,\hbar}{c} (\vec{\sigma}\cdot\vec{B})
                                    \Bigr]\dfrac{1}{\vec{\Pi}^2}
                              +\dfrac{1}{\sqrt{\vec{\Pi}^2}} \Bigl[
                                    \openone\,\vec{\Pi}^2 -\dfrac{q\,\hbar}{c} (\vec{\sigma}
                                          \cdot\vec{B})\Bigr]\dfrac{1}{\sqrt{\vec{\Pi}^2}} \notag \\
                             & & +\Bigl[\openone {+} \dfrac{q\,\hbar}{c}
                        \dfrac{1}{\vec{p}^2}(\vec{\sigma}\cdot\vec{B})
                 {-2}\dfrac{q\,\hbar}{c} \dfrac{1}{\vec{p}^4}(\vec{p}\cdot\vec{B})(
                        \vec{\sigma}\cdot\vec{p})\Bigr] \notag \\
                       & \approx& -2\dfrac{q\,\hbar}{c} \dfrac{1}{\vec{p}^4} \big[(\vec{p}\cdot\vec{B})(\vec{\sigma}
                                                \cdot\vec{p})
                                    -\vec{p}^2 (\vec{\sigma}\cdot\vec{B})\big]
                              +3\,\Bigl[\openone
                                    -\dfrac{q\,\hbar}{c} \dfrac{1}{\vec{p}^2} (\vec{\sigma}
                                          \cdot\vec{B})\Bigr] +\Bigl[\openone {+} \dfrac{q\,\hbar}{c}
                        \dfrac{1}{\vec{p}^2}(\vec{\sigma}\cdot\vec{B})
                 {-2}\dfrac{q\,\hbar}{c} \dfrac{1}{\vec{p}^4}(\vec{p}\cdot\vec{B})(
                        \vec{\sigma}\cdot\vec{p})\Bigr]\notag \\
                  & =& 4\, \openone +(2-3+1)\dfrac{q\,\hbar}{c}
                              \dfrac{1}{\vec{p}^2}(\vec{\sigma}\cdot\vec{B})+(-2-2)\dfrac{q\,\hbar}{c} \dfrac{1}{\vec{p}^4} \left[(\vec{p}\cdot\vec{B})(\vec{\sigma}\cdot\vec{p})\right]\nonumber\\
                  & =& 4\, \left[\openone -\dfrac{q\,\hbar}{c} \dfrac{1}{\vec{p}^4} \left[(\vec{p}\cdot\vec{B})(\vec{\sigma}\cdot\vec{p})\right]\right].
                  \end{eqnarray}
            \end{proof}

            (iii) One can have
            \begin{eqnarray}
                  & (\vec{\sigma}\cdot\vec{\Pi})\dfrac{1}{\sqrt{\vec{\Pi}^2}} (
                        \vec{\sigma}\cdot\vec{\Pi})
                  \approx \openone\sqrt{\vec{\Pi}^2}{-}\dfrac{q\,\hbar}{c}
                        \dfrac{1}{p^3}(\vec{p}\cdot\vec{B})(
                              \vec{\sigma}\cdot\vec{p}).
            \end{eqnarray}
            \begin{proof}
                  \begin{eqnarray}
                         (\vec{\sigma}\cdot\vec{\Pi})\dfrac{1}{\sqrt{\vec{\Pi}^2}} (
                              \vec{\sigma}\cdot\vec{\Pi})
                        &=&\vec{\sigma}\cdot\biggl\{\Big[\vec{\Pi},\ \dfrac{1}{\sqrt{\vec{\Pi}^2}}
                                    \Big]
                              +\dfrac{1}{\sqrt{\vec{\Pi}^2}} \vec{\Pi}\biggr\} (\vec{\sigma}
                                    \cdot\vec{\Pi}) \nonumber\\
                        &\approx & {\rm i}\hbar\dfrac{q}{c} \dfrac{1}{\sqrt{\vec{p}^6}} \big[
                              \vec{\sigma}\cdot(\vec{B}\times\vec{p})\big](\vec{\sigma}
                                    \cdot\vec{\Pi})
                              +\dfrac{1}{\sqrt{\vec{\Pi}^2}} (\vec{\sigma}\cdot\vec{\Pi})(
                                    \vec{\sigma}\cdot\vec{\Pi}) \notag \\
                        &\approx& {\rm i}\hbar\dfrac{q}{c} \dfrac{1}{\sqrt{\vec{p}^6}} {\rm i}\,
                  \big[-\vec{p}^2 (\vec{\sigma}\cdot\vec{B})
                        +(\vec{p}\cdot\vec{B})(\vec{\sigma}\cdot\vec{p})\big]
                              +\dfrac{1}{\sqrt{\vec{\Pi}^2}} \Bigl[
                                    \openone\,\vec{\Pi}^2 -\dfrac{q\,\hbar}{c} (
                                          \vec{\sigma}\cdot\vec{B})\Bigr] \notag \\
                  &\approx& \hbar\dfrac{q}{c} \dfrac{1}{\sqrt{\vec{p}^6}} \big[
                        \vec{p}^2 (\vec{\sigma}\cdot\vec{B})
                              -(\vec{p}\cdot\vec{B})(\vec{\sigma}\cdot\vec{p})\big]
                        +\Bigl[\openone\sqrt{\vec{\Pi}^2}-\dfrac{q\,\hbar}{c}
                                    \dfrac{1}{\sqrt{\vec{p}^2}}(\vec{\sigma}\cdot\vec{B})\Bigr]
                              \notag \\
                  & =& \openone\sqrt{\vec{\Pi}^2}-\dfrac{q\,\hbar}{c}
                        \dfrac{1}{\sqrt{\vec{p}^6}}(\vec{p}\cdot\vec{B})(
                              \vec{\sigma}\cdot\vec{p})\nonumber\\
                  &=& \openone\sqrt{\vec{\Pi}^2}{-}\dfrac{q\,\hbar}{c}
                        \dfrac{1}{p^3}(\vec{p}\cdot\vec{B})(
                              \vec{\sigma}\cdot\vec{p}).
                  \end{eqnarray}
            \end{proof}
            (iv) One can have
            \begin{eqnarray}
                  & \left\{(\vec{\sigma}\cdot\vec{\Pi})(\vec{\sigma}\cdot\vec{\Pi}),\
                        \dfrac{1}{\sqrt{\vec{\Pi}^2}}\right\}
                  = 2\,\openone\,\sqrt{\vec{\Pi}^2}
                        -2\dfrac{q\,\hbar}{c} \dfrac{1}{\sqrt{\vec{p}^2}} (
                              \vec{\sigma}\cdot\vec{B}),
            \end{eqnarray}
            \begin{proof}
                  \begin{eqnarray}
                         \left\{(\vec{\sigma}\cdot\vec{\Pi})(\vec{\sigma}\cdot\vec{\Pi}),\
                              \dfrac{1}{\sqrt{\vec{\Pi}^2}}\right\}
                       &=& \left[(\vec{\sigma}\cdot\vec{\Pi})(\vec{\sigma}\cdot\vec{\Pi})\right]
                                    \dfrac{1}{\sqrt{\vec{\Pi}^2}}
                              +\dfrac{1}{\sqrt{\vec{\Pi}^2}} \left[(\vec{\sigma}\cdot\vec{\Pi})(
                                    \vec{\sigma}\cdot\vec{\Pi})\right] \notag \\
                        &=& \left(\openone\,\vec{\Pi}^2
                                    -\hbar\dfrac{q}{c} \vec{\sigma}\cdot\vec{B}\right)
                                          \dfrac{1}{\sqrt{\vec{\Pi}^2}}+ \dfrac{1}{\sqrt{\vec{\Pi}^2}} \left(\openone\,\vec{\Pi}^2
                                    -\hbar\dfrac{q}{c} \vec{\sigma}\cdot\vec{B}\right)\nonumber\\
                        &\approx&  2\,\openone\,\sqrt{\vec{\Pi}^2}
                              -2\dfrac{q\,\hbar}{c} \dfrac{1}{\sqrt{\vec{p}^2}} (\vec{\sigma}\cdot\vec{B})
                  \end{eqnarray}
             \end{proof}
           $\blacksquare$
            \end{remark}

           Based on above results, from Eq. (\ref{eq:mix-1b}) we then have
            \begin{eqnarray}
                  && \Biggl\{\cos\vartheta\,(\vec{\sigma}\cdot\vec{\Pi})
                        -\sin\vartheta\,\dfrac{1}{2} m_{\rm e}\,c\bigg[
                              \dfrac{1}{\sqrt{\vec{\Pi}^2}} (
                                    \vec{\sigma}\cdot\vec{\Pi})
                              +(\vec{\sigma}\cdot\vec{\Pi})
                                    \dfrac{1}{\sqrt{\vec{\Pi}^2}}\bigg]\Biggr\}^2
                        \notag \\
                  &=& {\cos^2}\vartheta\,(\vec{\sigma}\cdot\vec{\Pi})(
                              \vec{\sigma}\cdot\vec{\Pi})
                        +\dfrac{{\sin^2}\vartheta}{4} m_{\rm e}^2 c^2 \bigg[
                              \dfrac{1}{\sqrt{\vec{\Pi}^2}} (
                                    \vec{\sigma}\cdot\vec{\Pi})
                              +(\vec{\sigma}\cdot\vec{\Pi})
                                    \dfrac{1}{\sqrt{\vec{\Pi}^2}}
                              \bigg]^2 \notag \\
                       & & -\sin\vartheta\,\cos\vartheta\,\dfrac{1}{2} m_{\rm e}\,c\Biggl(2(
                              \vec{\sigma}\cdot\vec{\Pi})
                                    \dfrac{1}{\sqrt{\vec{\Pi}^2}} (
                                          \vec{\sigma}\cdot\vec{\Pi})
                              +\bigg\{(\vec{\sigma}\cdot\vec{\Pi})(
                                    \vec{\sigma}\cdot\vec{\Pi}),\
                                    \dfrac{1}{\sqrt{\vec{\Pi}^2}}\bigg\}\Biggr)
                        \notag \\
                  &\approx& {\cos^2}\vartheta\,\Big[\vec{\Pi}^2 \,\openone
                              -\dfrac{q\,\hbar}{c} (\vec{\sigma}\cdot\vec{B})\Big]
                        +\dfrac{{\sin^2}\vartheta}{4} m_{\rm e}^2 c^2 \Bigl[4\, \left[\openone -\dfrac{q\,\hbar}{c} \dfrac{1}{\vec{p}^4} \left[(\vec{p}\cdot\vec{B})(\vec{\sigma}\cdot\vec{p})\right]\right]\Bigr] \notag \\
                        && -\sin\vartheta\,\cos\vartheta\,\dfrac{1}{2} m_{\rm e}\,c\Biggl\{2\bigg[\openone\sqrt{\vec{\Pi}^2}
                              -\dfrac{q\,\hbar}{c} \dfrac{1}{\sqrt{\vec{p}^6}}(
                                    \vec{p}\cdot\vec{B})(\vec{\sigma}\cdot\vec{p})
                              \bigg]+\left(2\,\openone\,\sqrt{\vec{\Pi}^2}
                        -2\dfrac{q\,\hbar}{c} \dfrac{1}{\sqrt{\vec{p}^2}} (
                              \vec{\sigma}\cdot\vec{B})\right)\Biggr\} \notag \\
                  &=& {\cos^2}\vartheta\,\Big[\vec{\Pi}^2 \,\openone
                              -\dfrac{q\,\hbar}{c} (\vec{\sigma}\cdot\vec{B})\Big]
                        +{\sin^2}\vartheta m_{\rm e}^2 c^2 \left[\openone -\dfrac{q\,\hbar}{c} \dfrac{1}{\vec{p}^4} \left[(\vec{p}\cdot\vec{B})(\vec{\sigma}\cdot\vec{p})\right]\right] \notag \\
                  && -\sin\vartheta\,\cos\vartheta\, m_{\rm e}\,c\Biggl\{
                        \bigg[\openone\sqrt{\vec{\Pi}^2}
                              -\dfrac{q\,\hbar}{c} \dfrac{1}{\sqrt{\vec{p}^6}}(
                                    \vec{p}\cdot\vec{B})(\vec{\sigma}\cdot\vec{p})
                              \bigg]+\left(\openone\,\sqrt{\vec{\Pi}^2}
                        -\dfrac{q\,\hbar}{c} \dfrac{1}{\sqrt{\vec{p}^2}} (
                              \vec{\sigma}\cdot\vec{B})\right)\Biggr\} \notag \\
                  &=& {\cos^2}\vartheta\,\Big[\vec{\Pi}^2 \,\openone
                              -\dfrac{q\,\hbar}{c} (\vec{\sigma}\cdot\vec{B})\Big]
                        +{\sin^2}\vartheta m_{\rm e}^2 c^2 \left[\openone -\dfrac{q\,\hbar}{c} \dfrac{1}{\vec{p}^4} \left[(\vec{p}\cdot\vec{B})(\vec{\sigma}\cdot\vec{p})\right]\right] \notag \\
                  && -\sin\vartheta\,\cos\vartheta\, m_{\rm e}\,c\Biggl[
                        2\,\openone\sqrt{\vec{\Pi}^2}
                        -\dfrac{q\,\hbar}{c} \dfrac{1}{\sqrt{\vec{p}^6}} (
                              \vec{p}\cdot\vec{B})(\vec{\sigma}\cdot\vec{p})
                        -\dfrac{q\,\hbar}{c} \dfrac{1}{\sqrt{\vec{p}^2}} (
                              \vec{\sigma}\cdot\vec{B})\Biggr] \notag \\
                  &=& {\cos^2}\vartheta\,\vec{\Pi}^2 \,\openone
                        -\left[{\cos^2}\vartheta-\sin\vartheta\,\cos\vartheta
                              \dfrac{m_{\rm e}\,c}{\sqrt{\vec{p}^2}}
                              \right]\dfrac{q\,\hbar}{c} (\vec{\sigma}\cdot\vec{B})
                        -\left[{\sin^2}\vartheta\,m_{\rm e}^2 c^2
                                    \frac{1}{\sqrt{\vec{p}^2}}
                              -\sin\vartheta\,\cos\vartheta\,m_{\rm e}\,c
                              \right]\dfrac{q\,\hbar}{c} \dfrac{1}{\sqrt{\vec{p}^6}}(
                                    \vec{p}\cdot\vec{B})(\vec{\sigma}\cdot\vec{p})
                        \nonumber\\
                  && -2 \sin\vartheta\,\cos\vartheta\, m_{\rm e}\,c
                             \,\sqrt{\vec{\Pi}^2}\, \openone +{\sin^2}\vartheta m_{\rm e}^2 c^2 \openone \nonumber\\
                  &=& {\cos^2}\vartheta\,\left[\vec{\Pi}^2\,\openone-\left(1
                              -\tan\vartheta\,\dfrac{m_{\rm e}\,c}{p}
                              \right)\dfrac{q\,\hbar}{c} (\vec{\sigma}\cdot\vec{B})
                              \right]
                        -\left[{\sin^2}\vartheta m_{\rm e}^2 c^2 \frac{1}{p}
                              -\sin\vartheta\,\cos\vartheta\, m_{\rm e}\,c
                              \right]\dfrac{q\,\hbar}{c} \dfrac{1}{p^3} (
                                    \vec{p}\cdot\vec{B})(\vec{\sigma}\cdot\vec{p})
                        \nonumber\\
                  && -2 \sin\vartheta\,\cos\vartheta\, m_{\rm e}\,c
                             \,\sqrt{\vec{\Pi}^2}\, \openone +{\sin^2}\vartheta m_{\rm e}^2 c^2 \openone \nonumber\\
                  \end{eqnarray}
            Thus \Eq{eq:EtaHeHb1} can be rewritten as
            \begin{eqnarray}
                  {\rm i}\hbar\p{\eta}{t} &=& \dfrac{1}{\cos\vartheta}\,2\,m_{\rm e} \Biggl\{
                  {\cos^2}\vartheta\,\left[\vec{\Pi}^2 \,\openone - \left(1 -\tan\vartheta\,\dfrac{m_{\rm e}\,c}{p} \right)\dfrac{q\,\hbar}{c} (\vec{\sigma}\cdot\vec{B})\right]\nonumber\\
                  &&-\left[ {\sin^2}\vartheta m_{\rm e}^2 c^2 \frac{1}{p}
                        -\sin\vartheta\,\cos\vartheta\, m_{\rm e}\,c
                        \right]\dfrac{q\,\hbar}{c} \dfrac{1}{p^3}(\vec{p}\cdot\vec{B})(
                              \vec{\sigma}\cdot\vec{p})
                   -2 \sin\vartheta\,\cos\vartheta\, m_{\rm e}\,c
                   \,\sqrt{\vec{\Pi}^2}\, \openone +{\sin^2}\vartheta m_{\rm e}^2 c^2 \openone \Biggr\}\eta  \nonumber\\
                   &&+\sin\vartheta\,\sqrt{\vec{\Pi}^2}\,c\:\eta+q\,\phi\,\eta,
            \end{eqnarray}
            i.e.,
            \begin{eqnarray}\label{eq:EtaHeHb1v1}
                  {\rm i}\hbar\p{\eta}{t} &=& \dfrac{1}{\cos\vartheta\,2\,m_{\rm e}} \Biggl\{
                  {\cos^2}\vartheta\,\left[\vec{\Pi}^2 \,\openone - \left(1 -\tan\vartheta\,\dfrac{m_{\rm e}\,c}{p} \right)\dfrac{q\,\hbar}{c} (\vec{\sigma}\cdot\vec{B})\right]\nonumber\\
                  &&-\left[ {\sin^2}\vartheta m_{\rm e}^2 c^2 \frac{1}{p}
                        -\sin\vartheta\,\cos\vartheta\, m_{\rm e}\,c\right]\dfrac{q\,\hbar}{c} \dfrac{1}{p^3}(\vec{p}\cdot\vec{B})(
                                    \vec{\sigma}\cdot\vec{p})
                   +{\sin^2}\vartheta m_{\rm e}^2 c^2 \openone \Biggr\}\eta +q\,\phi\,\eta,
            \end{eqnarray}
            Notice when $\vartheta=0$, \Eq{eq:EtaHeHb1v1} reduces into
            \begin{equation}
                  {\rm i}\hbar\p{\eta}{t} =\bigg[\dfrac{1}{2\,m} \Bigl(
                                    \vec{p}-\dfrac{q}{c} \vec{A}\Bigr)^2
                              -\dfrac{q\,\hbar}{2\,m\,c} \vec{\sigma}\cdot\vec{B}\bigg]
                                    \eta
                        +q\,\phi\,\eta,
            \end{equation}
            which is just Eq. (\ref{eq:pauli-1}).

            Due to
            \begin{eqnarray}
                  & \vec{\Pi}^2 =\vec{p}^2 -\dfrac{q}{c} \vec{B}\cdot\vec{\ell}.
            \end{eqnarray}
            from \Eq{eq:EtaHeHb1v1} we attain
            \begin{eqnarray}
                  {\rm i}\hbar\p{\eta}{t} &\approx& \cos\vartheta\,\bigg[
                              \dfrac{1}{2\,m_{\rm e}} \Bigl(\vec{p}^2
                                          -\dfrac{q}{c} \vec{B}\cdot\vec{\ell}\Bigr)
                                    -2\left(1 -\tan\vartheta\,\dfrac{m_{\rm e}\,c}{p} \right)\dfrac{q}{2\,m_{\rm e}\,c} (\vec{B}\cdot\vec{S})
                              \bigg]\eta \notag \\
                        && -\dfrac{1}{\cos\vartheta}\left[ {\sin^2}\vartheta m_{\rm e}^2 c^2 \frac{1}{p} -\sin\vartheta\,\cos\vartheta\, m_{\rm e}\,c\right]\dfrac{q\,\hbar}{2\,m_{\rm e}\,c}
                              \dfrac{1}{p^3} (\vec{p}\cdot\vec{B})(
                                    \vec{\sigma}\cdot\vec{p})\, \eta  +\dfrac{1}{2\,\cos\vartheta} {\sin^2}\vartheta\,m_{\rm e} c^2 \,\eta
                        +q\,\phi\,\eta,
            \end{eqnarray}
            i.e.,
            \begin{eqnarray}\label{eq:pauli-c1}
                  {\rm i}\hbar\p{\eta}{t} &=& {\cos}\vartheta\,\Bigg\{
                              \dfrac{\vec{p}^2}{2\,m_{\rm e}}-\dfrac{q}{2\,m_{\rm e}\,c} \bigg[
                                    \vec{\ell}+2\Big(1 -\tan\vartheta\,\dfrac{m_{\rm e}\,c}{p}
                                    \Big)\vec{S}\bigg]\cdot\vec{B}\Bigg\}\, \eta \notag \\
                  && -\dfrac{1}{\cos\vartheta}\left[ {\sin^2}\vartheta m_{\rm e}^2 c^2 \frac{1}{p} -\sin\vartheta\,\cos\vartheta\, m_{\rm e}\,c\right]\dfrac{q\,\hbar}{2\,m_{\rm e}\,c}
                              \dfrac{1}{p^3} (\vec{p}\cdot\vec{B})(
                                    \vec{\sigma}\cdot\vec{p})\, \eta  +\dfrac{1}{2\,\cos\vartheta} {\sin^2}\vartheta\,m_{\rm e} c^2 \,\eta
                        +q\,\phi\,\eta.
            \end{eqnarray}
            When $\vartheta=0$, Eq. (\ref{eq:pauli-c1}) reduces to
            \begin{equation}
                  {\rm i}\hbar\p{\eta}{t} =\Big[\dfrac{\vec{p}^2}{2\,m}
                              -\dfrac{q}{2\,m\,c} (\vec{\ell}+2\,\vec{S})\cdot\vec{B}
                              \Big]\eta
                        +q\,\phi\,\eta,
            \end{equation}
            which is just Eq. (\ref{eq:pauli-3}).

           Based on (\ref{eq:pauli-c1}), we have the spin-Land{\' q} factor $g_{\rm s}$ (the ratio $g_{\rm spin}/g_{\rm orbit}$) as \begin{eqnarray}\label{eq:pauli-c2}
                 g_{\rm s}=\dfrac{g_{\rm spin}}{g_{\rm orbit}}=2\left(1 -\tan\vartheta\,\dfrac{m_{\rm e}\,c}{p}\right).
            \end{eqnarray}
            If we view $g_{\rm orbit}$ as 1, then the $g$ factor for the $H_{\rm e}$-${H}_{\rm b}^{\rm I}$ mixing system is given by
            \begin{eqnarray}\label{eq:pauli-c3}
                 g=2\left(1 -\tan\vartheta\,\dfrac{m_{\rm e}\,c}{p}\right).
            \end{eqnarray}
             This means that the mixture of $H_{\rm e}$ and ${H}_{\rm b}^{\rm I}$ can indeed alter the $g$ factor.

            \begin{remark}We can find something interesting:

            (i) If there is no the ``spin Zitterbewegung'' (and the ``position Zitterbewegung''), i.e.,
            \begin{eqnarray}
                        && \tan\vartheta=\dfrac{p}{m_{\rm e}\,c},
             \end{eqnarray}
             then we have the $g$ factor as
            \begin{eqnarray}
                        &&  g=2\left(1 -\tan\vartheta\,\dfrac{m_{\rm e}\,c}{p}\right)=0.
             \end{eqnarray}
             This is also a prediction of our theory.

             \end{remark}

\section{Calculating the $g$ Factor for the $H_{\rm e}$-${H}_{\rm b}^{\rm II}$ Mixing}

      In this section, we would like to calculate the $g$ factor for the mixing system $H_{\rm e}$-${H}_{\rm b}^{\rm II}$.
      The mixed Hamiltonian is given by
      \begin{eqnarray}
           {H}_{\rm mix} =\cos\vartheta\,H_{\rm e} +\sin\vartheta\,{H}_{\rm b}^{\rm II}.
      \end{eqnarray}
      According to the Hamiltonian of an electron in an electromagnetic field, the Hamiltonians $H_{\rm e}$ and ${H}_{\rm b}^{\rm II}$ become
      \begin{eqnarray}
            \mathcal{H}_{\rm e} &=& c\,\vec{\alpha}\cdot\Bigl(\vec{p}-\dfrac{q}{c} \vec{A}\Bigr)+\beta\,m\,c^2=
            c\,\vec{\alpha}\cdot \vec{\Pi}+\beta\,m_{\rm e}\,c^2,
      \end{eqnarray}
      \begin{eqnarray}
            \mathcal{H}_{\rm b}^{\rm II} =\frac{c}{2}\left[\sqrt{1+\dfrac{m_{\rm e}^2 c^2}{\vec{\Pi}^2}}\,{\rm i}\,
                  \beta(\vec{\alpha}\cdot \vec{\Pi})+{\rm i}\,
                  \beta(\vec{\alpha}\cdot \vec{\Pi}) \sqrt{1+\dfrac{m_{\rm e}^2 c^2}{\vec{\Pi}^2}}\right].
      \end{eqnarray}
      Note that $\mathcal{H}_{\rm b}^{\rm II}=(\mathcal{H}_{\rm b}^{\rm II})^\dagger$ is hermitian, and it reduces to $\mathcal{H}_{\rm b}^{II}=\sqrt{p^2c^2+m^2c^4}\;({\rm i}\beta\vec{\alpha}\cdot\hat{p})$ if without the magnetic field.

      Then we have the mixed Hamiltonian in the presence of electromagnetic field as
      \begin{equation}
            \mathcal{H}_{\rm mix} =\cos\vartheta\,\bigl(c\,\vec{\alpha}\cdot \vec{\Pi}
                        +\beta\,m_{\rm e}\,c^2\bigr)
                  +\sin\vartheta\, \frac{c}{2}\left[
                        \sqrt{1+\dfrac{m_{\rm e}^2 c^2}{\vec{\Pi}^2}}\,{\rm i}\,\beta(
                              \vec{\alpha}\cdot \vec{\Pi})
                        +{\rm i}\,\beta(\vec{\alpha}\cdot \vec{\Pi})
                              \sqrt{1+\dfrac{m_{\rm e}^2 c^2}{\vec{\Pi}^2}}\right]+q\,\phi.
      \end{equation}
      where $\phi$ depicts the scalar potential. Then the Dirac equation reads
      \begin{equation}\label{eq:DiracHeHb2}
            {\rm i}\hbar\p{}{t} \Psi=\mathcal{H}_{\rm mix} \Psi.
      \end{equation}
      Next we study the Dirac equation in the representation of
      \begin{equation}
            \Psi=\begin{bmatrix}
                  \tilde{\eta} \\ \tilde{\chi}
            \end{bmatrix},
      \end{equation}
      then \Eq{eq:DiracHeHb2} becomes
      \begin{eqnarray}
            {\rm i}\hbar\p{}{t} \begin{bmatrix}
                        \tilde{\eta} \\ \tilde{\chi}
                  \end{bmatrix}
            &=& \cos\vartheta\,c\,\begin{bmatrix}
                              0 & \vec{\sigma} \\
                              \vec{\sigma} & 0
                        \end{bmatrix}\cdot\vec{\Pi}\begin{bmatrix}
                                    \tilde{\eta} \\ \tilde{\chi}
                              \end{bmatrix}
                  +\cos\vartheta\,\begin{bmatrix}
                              \openone & 0 \\
                              0 & -\openone
                        \end{bmatrix}\,m_{\rm e}\,c^2 \begin{bmatrix}
                              \tilde{\eta} \\ \tilde{\chi}
                        \end{bmatrix} \notag \\
                  && +\sin\vartheta\,\frac{c}{2} \begin{bmatrix}
                              0 & \sqrt{1+\dfrac{m_{\rm e}^2 c^2}{\vec{\Pi}^2}}\,{\rm i}\,(
                                          \vec{\sigma}\cdot\vec{\Pi})
                                    +{\rm i}\,(\vec{\sigma}\cdot\vec{\Pi})
                                          \sqrt{1+\dfrac{m_{\rm e}^2 c^2}{\vec{\Pi}^2}} \\
                              -\sqrt{1+\dfrac{m_{\rm e}^2 c^2}{\vec{\Pi}^2}}\,{\rm i}(
                                          \vec{\sigma}\cdot\vec{\Pi})
                                    -{\rm i}\,(\vec{\sigma}\cdot\vec{\Pi})
                                          \sqrt{1+\dfrac{m_{\rm e}^2 c^2}{\vec{\Pi}^2}} & 0
                        \end{bmatrix}\begin{bmatrix}
                                    \tilde{\eta} \\ \tilde{\chi}
                              \end{bmatrix}\nonumber\\
                 && +q\,\phi\begin{bmatrix}
                              \tilde{\eta} \\ \tilde{\chi}
                        \end{bmatrix},
      \end{eqnarray}
      i.e.,
      \begin{eqnarray}\label{eq:DiracEHb2Expand}
            {\rm i}\hbar\p{}{t} \begin{bmatrix}
                        \tilde{\eta} \\ \tilde{\chi}
                  \end{bmatrix}
            &=&  \cos\vartheta\,c(\vec{\sigma}\cdot\vec{\Pi})\begin{bmatrix}
                                    \tilde{\chi} \\ \tilde{\eta}
                              \end{bmatrix}
                        +\sin\vartheta\,\dfrac{c}{2} \left[
                              \sqrt{1+\dfrac{m^2 c^2}{\vec{\Pi}^2}}\,{\rm i}\,(
                                          \vec{\sigma}\cdot\vec{\Pi})
                                    +{\rm i}\,(\vec{\sigma}\cdot\vec{\Pi})
                                          \sqrt{1+\dfrac{m^2 c^2}{\vec{\Pi}^2}}
                              \right]\begin{bmatrix}
                                          \tilde{\chi} \\ -\tilde{\eta}
                                    \end{bmatrix} \notag \\
                  && +\cos\vartheta\,m_{\rm e}\,c^2 \begin{bmatrix}
                              \tilde{\eta} \\ -\tilde{\chi}
                        \end{bmatrix}
                  +q\,\phi\begin{bmatrix}
                              \tilde{\eta} \\ \tilde{\chi}
                        \end{bmatrix}.
      \end{eqnarray}

      In the case of low energy, the term involving $m_{\rm e}\,c^2$ is far larger than other ones. Then a part of time factor can be separated as follows:
      \begin{equation}
            \begin{bmatrix}
                  \tilde{\eta} \\ \tilde{\chi}
            \end{bmatrix}=\begin{bmatrix}
                  \eta \\ \chi
            \end{bmatrix}{\rm e}^{-{\rm i}\,\cos\vartheta\,\frac{m_{\rm e}\,c^2}{\hbar} t},
      \end{equation}
      then \Eq{eq:DiracEHb2Expand} is transformed as
      \begin{eqnarray}
            && {\rm i}\hbar\Biggl\{\p{}{t} \begin{bmatrix}
                        \eta \\ \chi
                  \end{bmatrix}\Biggr\}{\rm e}^{-{\rm i}\,\cos\vartheta\,\frac{m_{\rm e}\,c^2}{\hbar} t}
            +{\rm i}\begin{bmatrix}
                        \eta \\ \chi
                  \end{bmatrix}\hbar\Bigl(\p{}{t} {\rm e}^{
                        -{\rm i}\,\cos\vartheta\,\frac{m_{\rm e}\,c^2}{\hbar} t}\Bigr) \notag \\
            &=&  \cos\vartheta\,c(\vec{\sigma}\cdot\vec{\Pi})\begin{bmatrix}
                                    \chi \\ \eta
                              \end{bmatrix}{\rm e}^{
                                    -{\rm i}\,\cos\theta\,\frac{m_{\rm e}\,c^2}{\hbar} t}
                        +\sin\vartheta\,\dfrac{c}{2} \left[
                              \sqrt{1+\dfrac{m_{\rm e}^2 c^2}{\vec{\Pi}^2}}\,{\rm i}\,(
                                          \vec{\sigma}\cdot\vec{\Pi})
                                    +{\rm i}\,(\vec{\sigma}\cdot\vec{\Pi})
                                          \sqrt{1+\dfrac{m_{\rm e}^2 c^2}{\vec{\Pi}^2}}
                              \right]\begin{bmatrix}
                                          \chi \\ -\eta
                                    \end{bmatrix}{\rm e}^{
                                          -{\rm i}\,\cos\vartheta\,\frac{m_{\rm e}\,c^2}{\hbar} t}
                        \notag \\
                  && +\cos\vartheta\,m_{\rm e}\,c^2 \begin{bmatrix}
                              \eta \\ -\chi
                        \end{bmatrix}{\rm e}^{
                              -{\rm i}\,\cos\vartheta\,\frac{m_{\rm e}\,c^2}{\hbar} t}
                  +q\,\phi\begin{bmatrix}
                              \eta \\ \chi
                        \end{bmatrix}{\rm e}^{
                              -{\rm i}\,\cos\vartheta\,\frac{m_{\rm e}\,c^2}{\hbar} t},
      \end{eqnarray}
      i.e.,
      \begin{eqnarray}
            {\rm i}\hbar\p{}{t} \begin{bmatrix}
                        \eta \\ \chi
                  \end{bmatrix}+\cos\vartheta\,m_{\rm e}\,c^2 \begin{bmatrix}
                        \eta \\ \chi
                  \end{bmatrix}
            &=&  \cos\vartheta\,c(\vec{\sigma}\cdot\vec{\Pi})\begin{bmatrix}
                                    \chi \\ \eta
                              \end{bmatrix}
                        +\sin\vartheta\,\dfrac{c}{2} \left[
                              \sqrt{1+\dfrac{m_{\rm e}^2 c^2}{\vec{\Pi}^2}}\,{\rm i}\,(
                                          \vec{\sigma}\cdot\vec{\Pi})
                                    +{\rm i}\,(\vec{\sigma}\cdot\vec{\Pi})
                                          \sqrt{1+\dfrac{m_{\rm e}^2 c^2}{\vec{\Pi}^2}}
                              \right]\begin{bmatrix}
                                          \chi \\ -\eta
                                    \end{bmatrix} \notag \\
                  && +\cos\vartheta\,m_{\rm e}\,c^2 \begin{bmatrix}
                              \eta \\ -\chi
                        \end{bmatrix}
                  +q\,\phi\begin{bmatrix}
                              \eta \\ \chi
                        \end{bmatrix},
      \end{eqnarray}
      i.e.,
      \begin{eqnarray}\label{eq:DiracHeHb2Expand}
            {\rm i}\hbar\p{}{t} \begin{bmatrix}
                        \eta \\ \chi
                  \end{bmatrix}
            &=&  \cos\vartheta\,c(\vec{\sigma}\cdot\vec{\Pi})\begin{bmatrix}
                                    \chi \\ \eta
                              \end{bmatrix}
                        +\sin\vartheta\,\dfrac{c}{2} \left[
                              \sqrt{1+\dfrac{m_{\rm e}^2 c^2}{\vec{\Pi}^2}}\,{\rm i}\,(
                                          \vec{\sigma}\cdot\vec{\Pi})
                                    +{\rm i}\,(\vec{\sigma}\cdot\vec{\Pi})
                                          \sqrt{1+\dfrac{m_{\rm e}^2 c^2}{\vec{\Pi}^2}}
                              \right]\begin{bmatrix}
                                          \chi \\ -\eta
                                    \end{bmatrix} \notag \\
                  && +\cos\vartheta\,m_{\rm e}\,c^2 \begin{bmatrix}
                              0 \\ -2\,\chi
                        \end{bmatrix}
                  +q\,\phi\begin{bmatrix}
                              \eta \\ \chi
                        \end{bmatrix},
      \end{eqnarray}

      For the 2nd row of \Eq{eq:DiracHeHb2Expand}, we have
      \begin{equation}
            \openone\,{\rm i}\hbar\p{\chi}{t} =\left\{
                        \cos\vartheta\,c(\vec{\sigma}\cdot\vec{\Pi})
                  -\sin\vartheta\,\dfrac{c}{2} \Bigg[
                              \sqrt{1+\dfrac{m_{\rm e}^2 c^2}{\vec{\Pi}^2}}\,{\rm i}\,(
                                          \vec{\sigma}\cdot\vec{\Pi})
                                    +{\rm i}\,(\vec{\sigma}\cdot\vec{\Pi})
                                          \sqrt{1+\dfrac{m_{\rm e}^2 c^2}{\vec{\Pi}^2}}
                              \Bigg]\right\}\eta
                  -\openone\,2\,\cos\vartheta\,m_{\rm e}\,c^2 \chi+\openone\,q\,\phi\,\chi.
      \end{equation}
      Since the term containing $m_{\rm e}\,c^2 \openone\chi\gg{\rm i}\hbar\,\partial\chi/\partial t$, and $m_{\rm e}\,c^2 \openone\chi\gg q\,\phi\,\chi$, then we arrive at
      \begin{equation}
            \left\{\cos\vartheta\,c(\vec{\sigma}\cdot\vec{\Pi})-\sin\vartheta\,\dfrac{c}{2}
                  \Bigg[\sqrt{1+\dfrac{m_{\rm e}^2 c^2}{\vec{\Pi}^2}}\,{\rm i}\,(
                              \vec{\sigma}\cdot\vec{\Pi})
                        +{\rm i}\,(\vec{\sigma}\cdot\vec{\Pi})
                              \sqrt{1+\dfrac{m_{\rm e}^2 c^2}{\vec{\Pi}^2}}\Bigg]\right\}\eta
            = \cos\vartheta\,2\,m_{\rm e}\,c^2 \openone\chi,
      \end{equation}
      i.e.,
      \begin{equation}
            \chi=\dfrac{1}{2\,\cos\vartheta\,m_{\rm e}\,c^2} \left\{\cos\vartheta\,c(\vec{\sigma}\cdot\vec{\Pi})
                  -\sin\vartheta\,\dfrac{c}{2} \Bigg[
                        \sqrt{1+\dfrac{m_{\rm e}^2 c^2}{\vec{\Pi}^2}}\,{\rm i}\,(
                              \vec{\sigma}\cdot\vec{\Pi})
                        +{\rm i}\,(\vec{\sigma}\cdot\vec{\Pi})
                              \sqrt{1+\dfrac{m_{\rm e}^2 c^2}{\vec{\Pi}^2}}\Bigg]\right\}\eta.
      \end{equation}
      After that, for the first row, we obtain
      \begin{equation}
            \openone\,{\rm i}\hbar\p{\eta}{t} =\left\{
                        \cos\vartheta\,c(\vec{\sigma}\cdot\vec{\Pi})
                        +\sin\vartheta\,\dfrac{c}{2} \Bigg[
                              \sqrt{1+\dfrac{m_{\rm e}^2 c^2}{\vec{\Pi}^2}}\,{\rm i}\,(
                                          \vec{\sigma}\cdot\vec{\Pi})
                                    +{\rm i}\,(\vec{\sigma}\cdot\vec{\Pi})
                                          \sqrt{1+\dfrac{m_{\rm e}^2 c^2}{\vec{\Pi}^2}}
                              \Bigg]\right\}\chi
                  +q\,\phi\,\eta,
      \end{equation}
      i.e.,
      \begin{eqnarray}\label{eq:EtaHeHb2}
            \openone\,{\rm i}\hbar\p{\eta}{t} &=& \dfrac{1}{\cos\vartheta\,2\,m_{\rm e}} \left\{
                  \cos\vartheta\,(\vec{\sigma}\cdot\vec{\Pi})+\dfrac{\sin\vartheta}{2} \Bigg[
                              \sqrt{1+\dfrac{m_{\rm e}^2 c^2}{\vec{\Pi}^2}}\,{\rm i}\,(
                                          \vec{\sigma}\cdot\vec{\Pi})
                                    +{\rm i}\,(\vec{\sigma}\cdot\vec{\Pi})
                                          \sqrt{1+\dfrac{m_{\rm e}^2 c^2}{\vec{\Pi}^2}}\Bigg]
                        \right\} \notag \\
                 & &\qquad \left\{\cos\vartheta\,(\vec{\sigma}\cdot\vec{\Pi})
                              -\dfrac{\sin\vartheta}{2} \Bigg[
                                    \sqrt{1+\dfrac{m_{\rm e}^2 c^2}{\vec{\Pi}^2}}\,{\rm i}\,(
                                          \vec{\sigma}\cdot\vec{\Pi})
                                    +{\rm i}\,(\vec{\sigma}\cdot\vec{\Pi})
                                          \sqrt{1+\dfrac{m_{\rm e}^2 c^2}{\vec{\Pi}^2}}\Bigg]
                              \right\}\eta
                  +q\,\phi\,\eta.
      \end{eqnarray}

      \begin{remark}We need to calculate the following three terms.

      (i) From previous result, we have
      \begin{eqnarray}
            && (\vec{\sigma}\cdot\vec{\Pi})^2 =(\vec{\sigma}\cdot\vec{\Pi})(
                  \vec{\sigma}\cdot\vec{\Pi})
            \approx \vec{\Pi}^2 \, \openone-\dfrac{q\,\hbar}{c} (\vec{\sigma}\cdot\vec{B}).
      \end{eqnarray}

      (ii) We have
      \begin{eqnarray}\label{eq:cc-1}
            && \Bigg[\sqrt{1+\dfrac{m_{\rm e}^2 c^2}{\vec{\Pi}^2}}\,{\rm i}\,(
                  \vec{\sigma}\cdot\vec{\Pi})+{\rm i}\,(\vec{\sigma}\cdot\vec{\Pi})
                        \sqrt{1+\dfrac{m_{\rm e}^2 c^2}{\vec{\Pi}^2}}\Bigg]^2 \notag \\
            &=& -\left[\sqrt{1+\dfrac{m_{\rm e}^2 c^2}{\vec{\Pi}^2}}\,(
                        \vec{\sigma}\cdot\vec{\Pi})\,
                              \sqrt{1+\dfrac{m_{\rm e}^2 c^2}{\vec{\Pi}^2}}\,(
                                    \vec{\sigma}\cdot \vec{\Pi})\right]
                  -\left[\sqrt{1+\dfrac{m_{\rm e}^2 c^2}{\vec{\Pi}^2}}\,(
                        \vec{\sigma}\cdot\vec{\Pi})^2
                              \sqrt{1+\dfrac{m_{\rm e}^2 c^2}{\vec{\Pi}^2}}\right]
                        \nonumber \\
                  &&-\left[(\vec{\sigma}\cdot \vec{\Pi}) \left(
                        1+\dfrac{m_{\rm e}^2 c^2}{\vec{\Pi}^2}\right)(
                              \vec{\sigma}\cdot\vec{\Pi})\right]
                  -\left[(\vec{\sigma}\cdot\vec{\Pi})
                        \sqrt{1+\dfrac{m_{\rm e}^2 c^2}{\vec{\Pi}^2}}\,(
                              \vec{\sigma}\cdot\vec{\Pi})
                                    \sqrt{1+\dfrac{m_{\rm e}^2 c^2}{\vec{\Pi}^2}}\right] \notag
                  \\
             &\approx& -4\bigg[\openone\, \vec{\Pi}^2
                  -\dfrac{q\,\hbar}{c} (\vec{\sigma}\cdot\vec{B})
                  -\dfrac{q\,\hbar}{c} \dfrac{m_{\rm e}^2 c^2}{p^4} (
                        \vec{p}\cdot\vec{B})(\vec{\sigma}\cdot\vec{p})
                  +\openone\, m_{\rm e}^2 c^2\bigg].
      \end{eqnarray}
      \emph{Proof.---}We can have
            \begin{align}
                  & [\Pi,\ \vec{\Pi}^2]\approx\bigg[
                        \vec{p}-\dfrac{q}{2\,c} (\vec{B}\times\vec{r}),\
                        \vec{p}^2 -\dfrac{q}{c} (\vec{B}\cdot\vec{\ell})\bigg]
                  \approx-\dfrac{q}{c} \big[\vec{p},\ (\vec{B}\cdot\vec{\ell})\big]
                        -\dfrac{q}{2\,c} \big[(\vec{B}\times\vec{r}),\ \vec{p}^2\big]
                        \notag \\
                  =& -\dfrac{q}{c} \sum_{u,v} \vec{e}_u B_v [p_u,\ \ell_v]
                        -\dfrac{q}{2\,c} \vec{B}\times\sum_{u,v} \vec{e}_u [u,\ p_v^2]
                  =-{\rm i}\hbar\dfrac{q}{c} \sum_{u,v,w} \vec{e}_u B_v \epsilon_{uvw}
                              p_w
                        -{\rm i}\hbar\dfrac{q}{2\,c} \vec{B}\times\sum_{u,v} \vec{e}_u
                              \p{p_v^2}{p_u} \notag \\
                  =& -{\rm i}\hbar\dfrac{q}{c} \sum_{u,v,w} \epsilon_{uvw} \vec{e}_u
                              B_v p_w
                        -{\rm i}\hbar\dfrac{q}{2\,c} \vec{B}\times\sum_u \vec{e}_u 2\,
                              p_u
                  =-{\rm i}\hbar\dfrac{q}{c} (\vec{B}\times\vec{p})
                        -{\rm i}\hbar\dfrac{q}{c} (\vec{B}\times\vec{p}) \notag \\
                  =& -{\rm i}\hbar\dfrac{2\,q}{c} (\vec{B}\times\vec{p}).
            \end{align}
       \begin{eqnarray}
             \Bigg[\vec{\Pi},\ \sqrt{1+\dfrac{m_{\rm e}^2 c^2}{\vec{\Pi}^2}}\Bigg]_x
            &=&\Bigg[p_x -\dfrac{q}{2\,c} (B_y z-B_z y),\
                  \sqrt{1+\dfrac{m_{\rm e}^2 c^2}{\vec{\Pi}^2}}\Bigg]\nonumber\\
            &=&\Bigg[p_x,\ \sqrt{1+\dfrac{m_{\rm e}^2 c^2}{\vec{\Pi}^2}}\Bigg]
                  -\dfrac{q}{2\,c} \Bigg[(B_y z-B_z y),\ \sqrt{1+\dfrac{m_{\rm e}^2 c^2}{\vec{\Pi}^2}}\Bigg] \notag \\
            &=& \Bigg[p_x,\ \sqrt{1+\dfrac{m_{\rm e}^2 c^2}{\vec{\Pi}^2}}\Bigg]
                  -\dfrac{q}{2\,c} B_y \Bigg[z,\ \sqrt{1+\dfrac{m_{\rm e}^2 c^2}{\vec{\Pi}^2}}
                        \Bigg]
                  +\dfrac{q}{2\,c} B_z \Bigg[y,\ \sqrt{1+\dfrac{m_{\rm e}^2 c^2}{\vec{\Pi}^2}}
                        \Bigg] \notag \\
            &=& -{\rm i}\hbar\Biggl(\p{}{x} \sqrt{1+\dfrac{m_{\rm e}^2 c^2}{\vec{\Pi}^2}}\Biggr)
                  -\dfrac{q}{2\,c} B_y {\rm i}\hbar\Biggl(
                        \p{}{p_z} \sqrt{1+\dfrac{m_{\rm e}^2 c^2}{\vec{\Pi}^2}}\Biggr)
                  +\dfrac{q}{2\,c} B_z {\rm i}\hbar\Biggl(
                        \p{}{p_y} \sqrt{1+\dfrac{m_{\rm e}^2 c^2}{\vec{\Pi}^2}}\Biggr) \notag \\
            &=& {\rm i}\hbar\dfrac{m_{\rm e}^2 c^2}{2} \dfrac{1}{\vec{\Pi}^4} \Biggl(
                        \sqrt{\dfrac{\vec{\Pi}^2}{m_{\rm e}^2 c^2 +\vec{\Pi}^2}}\Biggr)
                              \p{\vec{\Pi}^2}{x}
                  +{\rm i}\hbar\dfrac{q}{2\,c} B_y \dfrac{m_{\rm e}^2 c^2}{2}
                        \dfrac{1}{\vec{\Pi}^4} \Biggl(\sqrt{\dfrac{\vec{\Pi}^2}{
                              m_{\rm e}^2 c^2 +\vec{\Pi}^2}}\Biggr)\p{\vec{\Pi}^2}{p_z} \nonumber\\
                &&  -{\rm i}\hbar\dfrac{q}{2\,c} B_z \dfrac{m_{\rm e}^2 c^2}{2}
                        \dfrac{1}{\vec{\Pi}^4} \Biggl(\sqrt{\dfrac{\vec{\Pi}^2}{
                              m_{\rm e}^2 c^2 +\vec{\Pi}^2}}\Biggr)\p{\vec{\Pi}^2}{p_y} \notag \\
            &=& -\dfrac{m_{\rm e}^2 c^2}{2} \dfrac{1}{\vec{\Pi}^4} \Biggl(
                        \sqrt{\dfrac{\vec{\Pi}^2}{m_{\rm e}^2 c^2 +\vec{\Pi}^2}}\Biggr)[p_x,\
                              \vec{\Pi}^2]
                  +\dfrac{q}{c} \dfrac{m_{\rm e}^2 c^2}{2} \dfrac{1}{\vec{\Pi}^4} \Biggl(
                        \sqrt{\dfrac{\vec{\Pi}^2}{m_{\rm e}^2 c^2 +\vec{\Pi}^2}}\Biggr)[A_x,\
                              \vec{\Pi}^2]\nonumber\\
            &=&-\dfrac{m_{\rm e}^2 c^2}{2} \dfrac{1}{\vec{\Pi}^4} \Biggl(
                  \sqrt{\dfrac{\vec{\Pi}^2}{m_{\rm e}^2 c^2 +\vec{\Pi}^2}}
                  \Biggr)[\Pi_x,\ \vec{\Pi}^2] \notag \\
           &\approx & -{\rm i}\hbar\dfrac{m_{\rm e}^2 c^2}{\vec{\Pi}^4} \Biggl(
                  \sqrt{\dfrac{\vec{\Pi}^2}{m_{\rm e}^2 c^2 +\vec{\Pi}^2}}
                  \Biggr)\dfrac{q}{c} (\vec{p}\times\vec{B})_x,
      \end{eqnarray}
      thus we have
       \begin{eqnarray}
            && \Bigg[\vec{\Pi},\ \sqrt{1+\dfrac{m_{\rm e}^2 c^2}{\vec{\Pi}^2}}\Bigg]
            \approx-{\rm i}\hbar\dfrac{m_{\rm e}^2 c^2}{\vec{\Pi}^4} \Biggl(
                  \sqrt{\dfrac{\vec{\Pi}^2}{m_{\rm e}^2 c^2 +\vec{\Pi}^2}}
                  \Biggr)\dfrac{q}{c} (\vec{p}\times\vec{B}).
      \end{eqnarray}
      Similarly, we have
      \begin{eqnarray}
             \bigg[\vec{\Pi},\ \Bigl(1+\dfrac{m_{\rm e}^2 c^2}{\vec{\Pi}^2}\Bigr)\bigg]
            &=&-{\rm i}\hbar\dfrac{m_{\rm e}^2 c^2}{\vec{\Pi}^4}\Bigl\{
                  2\dfrac{q}{c} (\vec{p}\times\vec{B})+\dfrac{q^2}{c^2} \bigl[
                        (\vec{B}\cdot\vec{r})\vec{B}-\vec{B}^2 \vec{r}\bigr]\Bigr\}\nonumber\\
            &\approx& - 2\,{\rm i}\hbar\dfrac{m_{\rm e}^2 c^2}{\vec{\Pi}^4}
                  \dfrac{q}{c} (\vec{p}\times\vec{B}),
      \end{eqnarray}
      which can also be obtained through
      \begin{eqnarray}
             \bigg[\vec{\Pi},\ \Bigl(1+\dfrac{m_{\rm e}^2 c^2}{\vec{\Pi}^2}\Bigr)\bigg]
            &=& \bigg[\vec{\Pi},\ \left(\sqrt{1+\dfrac{m_{\rm e}^2 c^2}{\vec{\Pi}^2}}\right)^2\bigg]\nonumber\\
            &=& \sqrt{1+\dfrac{m_{\rm e}^2 c^2}{\vec{\Pi}^2}}\,\bigg[\vec{\Pi},\ \sqrt{1+\dfrac{m_{\rm e}^2 c^2}{\vec{\Pi}^2}}\bigg]+
            \bigg[\vec{\Pi},\ \sqrt{1+\dfrac{m_{\rm e}^2 c^2}{\vec{\Pi}^2}}\bigg]\,\sqrt{1+\dfrac{m_{\rm e}^2 c^2}{\vec{\Pi}^2}}\nonumber\\
             &\approx& - 2\,{\rm i}\hbar\dfrac{m_{\rm e}^2 c^2}{\vec{\Pi}^4}
                  \dfrac{q}{c} (\vec{p}\times\vec{B}).
      \end{eqnarray}
      Then one obtains
      \begin{eqnarray}
            && \Bigg[\vec{\sigma}\cdot \vec{\Pi},\ \sqrt{1+\dfrac{m_{\rm e}^2 c^2}{\vec{\Pi}^2}}\Bigg]
            \approx-{\rm i}\hbar\dfrac{m_{\rm e}^2 c^2}{\vec{\Pi}^4} \Biggl(
                  \sqrt{\dfrac{\vec{\Pi}^2}{m_{\rm e}^2 c^2 +\vec{\Pi}^2}}
                  \Biggr)\dfrac{q}{c} \, \vec{\sigma} \cdot (\vec{p}\times\vec{B}).
      \end{eqnarray}
      \begin{eqnarray}
             \bigg[\vec{\sigma}\cdot \vec{\Pi},\ \Bigl(1+\dfrac{m_{\rm e}^2 c^2}{\vec{\Pi}^2}\Bigr)\bigg]
             &\approx& - 2\,{\rm i}\hbar\dfrac{m_{\rm e}^2 c^2}{\vec{\Pi}^4}
                  \dfrac{q}{c} \, \vec{\sigma}\cdot (\vec{p}\times\vec{B}).
            \end{eqnarray}

      (ii-1)  Then we have
       \begin{eqnarray}
            && \left[\sqrt{1+\dfrac{m_{\rm e}^2 c^2}{\vec{\Pi}^2}}\,(
                        \vec{\sigma}\cdot\vec{\Pi})^2
                              \sqrt{1+\dfrac{m_{\rm e}^2 c^2}{\vec{\Pi}^2}}\right]
             = \sqrt{1+\dfrac{m_{\rm e}^2 c^2}{\vec{\Pi}^2}}\,
                        \left(\vec{\Pi}^2 \, \openone-\dfrac{q\,\hbar}{c} (\vec{\sigma}\cdot\vec{B})\right)
                              \sqrt{1+\dfrac{m_{\rm e}^2 c^2}{\vec{\Pi}^2}}\nonumber\\
              &=& \left(1+\dfrac{m_{\rm e}^2 c^2}{\vec{\Pi}^2}\right)\,
                        \left(\vec{\Pi}^2 \, \openone -\dfrac{q\,\hbar}{c} (\vec{\sigma}\cdot\vec{B})\right)
              = \left(\vec{\Pi}^2 +m_{\rm e}^2 c^2\right)\, \openone -  \left(1+\dfrac{m_{\rm e}^2 c^2}{\vec{\Pi}^2}\right)\,
                        \dfrac{q\,\hbar}{c} \left(\vec{\sigma}\cdot\vec{B}\right) \nonumber\\
                        &\approx& \left(\vec{\Pi}^2 +m_{\rm e}^2 c^2\right)\, \openone -  \left(1+\dfrac{m_{\rm e}^2 c^2}{\vec{p}^2}\right)\,
                        \dfrac{q\,\hbar}{c} \left(\vec{\sigma}\cdot\vec{B}\right).
      \end{eqnarray}

  (ii-2)  We have
  \begin{eqnarray}
           && \left[(\vec{\sigma}\cdot \vec{\Pi}) \left(
                        1+\dfrac{m_{\rm e}^2 c^2}{\vec{\Pi}^2}\right)(
                              \vec{\sigma}\cdot\vec{\Pi})\right] = \left\{\left[\vec{\sigma}\cdot \vec{\Pi},
                        1+\dfrac{m_{\rm e}^2 c^2}{\vec{\Pi}^2}\right]+\left(1+\dfrac{m_{\rm e}^2 c^2}{\vec{\Pi}^2}\right)(
                              \vec{\sigma}\cdot\vec{\Pi})\right\}(\vec{\sigma}\cdot\vec{\Pi})\nonumber\\
             &\approx& - 2\,{\rm i}\hbar\dfrac{m_{\rm e}^2 c^2}{\vec{\Pi}^4}
                  \dfrac{q}{c} \, \vec{\sigma}\cdot (\vec{p}\times\vec{B})(\vec{\sigma}\cdot\vec{\Pi})+\left(1+\dfrac{m_{\rm e}^2 c^2}{\vec{\Pi}^2}\right)\left(\vec{\Pi}^2 \, \openone-\dfrac{q\,\hbar}{c} (\vec{\sigma}\cdot\vec{B})\right)\nonumber\\
             &\approx& - 2\,{\rm i}\hbar\dfrac{m_{\rm e}^2 c^2}{\vec{\Pi}^4}
                  \dfrac{q}{c} \, \vec{\sigma}\cdot (\vec{p}\times\vec{B})(\vec{\sigma}\cdot\vec{p})
                  + \vec{\Pi}^2 \, \openone+ m_{\rm e}^2 c^2 \, \openone-\left(1+\dfrac{m_{\rm e}^2 c^2}{\vec{p}^2}\right)\dfrac{q\,\hbar}{c} (\vec{\sigma}\cdot\vec{B})\nonumber\\
              &\approx& - 2\,{\rm i}\hbar\dfrac{m_{\rm e}^2 c^2}{\vec{\Pi}^4}
                  \dfrac{q}{c} \, {\rm i} \vec{\sigma}\cdot \left[(\vec{p}\times\vec{B})\times\vec{p}\right]
                  + \vec{\Pi}^2 \, \openone+ m_{\rm e}^2 c^2 \, \openone-\left(1+\dfrac{m_{\rm e}^2 c^2}{\vec{p}^2}\right)\dfrac{q\,\hbar}{c} (\vec{\sigma}\cdot\vec{B})\nonumber\\
              &\approx&  2\,\dfrac{m_{\rm e}^2 c^2}{\vec{p}^4}
                  \dfrac{q\,\hbar}{c} \, \vec{\sigma}\cdot \left[\vec{p}^{2}\vec{B}-\vec{p}(\vec{B}\cdot{\vec{p}})\right]
                  + \vec{\Pi}^2 \, \openone+ m_{\rm e}^2 c^2 \, \openone-\left(1+\dfrac{m_{\rm e}^2 c^2}{\vec{p}^2}\right)\dfrac{q\,\hbar}{c} (\vec{\sigma}\cdot\vec{B})\nonumber\\
              &\approx&  2\,\dfrac{m_{\rm e}^2 c^2}{\vec{p}^2} \dfrac{q\,\hbar}{c} (
                        \vec{\sigma}\cdot\vec{B})
                  -2\,\dfrac{m_{\rm e}^2 c^2}{\vec{p}^4} \dfrac{q\,\hbar}{c} (
                        \vec{B}\cdot{\vec{p}})(\vec{\sigma}\cdot\vec{p})
                  + \vec{\Pi}^2 \, \openone+ m_{\rm e}^2 c^2 \, \openone-\left(1+\dfrac{m_{\rm e}^2 c^2}{\vec{p}^2}\right)\dfrac{q\,\hbar}{c} (\vec{\sigma}\cdot\vec{B})\nonumber\\
               &\approx& \vec{\Pi}^2 \, \openone+ m_{\rm e}^2 c^2 \, \openone -\left(1-\dfrac{m_{\rm e}^2 c^2}{\vec{p}^2}\right)\dfrac{q\,\hbar}{c} (\vec{\sigma}\cdot\vec{B}) -2\,\dfrac{m_{\rm e}^2 c^2}{\vec{p}^4} \dfrac{q\,\hbar}{c}(\vec{B}\cdot{\vec{p}}) (
                  \vec{\sigma}\cdot\vec{p}).
      \end{eqnarray}
      or in the other way as
       \begin{eqnarray}
           && \left[(\vec{\sigma}\cdot \vec{\Pi}) \left(
                        1+\dfrac{m_{\rm e}^2 c^2}{\vec{\Pi}^2}\right)(
                              \vec{\sigma}\cdot\vec{\Pi})\right] = \left\{\left[\vec{\sigma}\cdot \vec{\Pi},
                        1+\dfrac{m_{\rm e}^2 c^2}{\vec{\Pi}^2}\right]+\left(1+\dfrac{m_{\rm e}^2 c^2}{\vec{\Pi}^2}\right)(
                              \vec{\sigma}\cdot\vec{\Pi})\right\}(\vec{\sigma}\cdot\vec{\Pi})\nonumber\\
             &=&  (\vec{\sigma}\cdot \vec{\Pi})^2+ m_{\rm e}^2 c^2 (\vec{\sigma}\cdot \vec{\Pi}) \dfrac{m_{\rm e}^2 c^2}{\vec{\Pi}^2}(\vec{\sigma}\cdot \vec{\Pi}) \nonumber\\
             &\approx& \left[ \vec{\Pi}^2 \, \openone-\dfrac{q\,\hbar}{c} (\vec{\sigma}\cdot\vec{B})\right]+ m_{\rm e}^2 c^2\, \left[\openone {+} \dfrac{q\,\hbar}{c}
                        \dfrac{1}{\vec{p}^2}(\vec{\sigma}\cdot\vec{B})
                 {-2}\dfrac{q\,\hbar}{c} \dfrac{1}{\vec{p}^4}(\vec{p}\cdot\vec{B})(
                        \vec{\sigma}\cdot\vec{p})\right] \nonumber\\
             &=& \vec{\Pi}^2 \, \openone+ m_{\rm e}^2 c^2 \, \openone {-}\left(1-\dfrac{m_{\rm e}^2 c^2}{\vec{p}^2}\right)\dfrac{q\,\hbar}{c} (\vec{\sigma}\cdot\vec{B}) -2\,\dfrac{m_{\rm e}^2 c^2}{\vec{p}^4} \dfrac{q\,\hbar}{c}(\vec{B}\cdot{\vec{p}}) ( \vec{\sigma}\cdot\vec{p}).
             \end{eqnarray}

    (ii-3)  We have

      \begin{eqnarray}
           && \left[\sqrt{1+\dfrac{m_{\rm e}^2 c^2}{\vec{\Pi}^2}}\,(
                        \vec{\sigma}\cdot\vec{\Pi})\,
                              \sqrt{1+\dfrac{m_{\rm e}^2 c^2}{\vec{\Pi}^2}}\,(
                                    \vec{\sigma}\cdot \vec{\Pi})\right] =
                                    \sqrt{1+\dfrac{m_{\rm e}^2 c^2}{\vec{\Pi}^2}}\,\left\{\left[
                        \vec{\sigma}\cdot\vec{\Pi},
                              \sqrt{1+\dfrac{m_{\rm e}^2 c^2}{\vec{\Pi}^2}}\right]+\sqrt{1+\dfrac{m_{\rm e}^2 c^2}{\vec{\Pi}^2}}(\vec{\sigma}\cdot\vec{\Pi})\right\}\,(
                                    \vec{\sigma}\cdot \vec{\Pi})\nonumber\\
             &\approx& \sqrt{1+\dfrac{m_{\rm e}^2 c^2}{\vec{\Pi}^2}}\,\left\{
             \left[-{\rm i}\hbar\dfrac{m_{\rm e}^2 c^2}{\vec{\Pi}^4} \Biggl(
                  \sqrt{\dfrac{\vec{\Pi}^2}{m_{\rm e}^2 c^2 +\vec{\Pi}^2}}
                  \Biggr)\dfrac{q}{c} \, \vec{\sigma} \cdot (\vec{p}\times\vec{B})\right]+\sqrt{1+\dfrac{m_{\rm e}^2 c^2}{\vec{\Pi}^2}}(\vec{\sigma}\cdot\vec{\Pi})\right\}\,(
                                    \vec{\sigma}\cdot \vec{\Pi})\nonumber\\
             &=&
             -{\rm i}\hbar\dfrac{m_{\rm e}^2 c^2}{\vec{\Pi}^4} \dfrac{q}{c} \, \vec{\sigma} \cdot (\vec{p}\times\vec{B})(
                                    \vec{\sigma}\cdot \vec{\Pi})+\left(\sqrt{1+\dfrac{m_{\rm e}^2 c^2}{\vec{\Pi}^2}}\right)^2(\vec{\sigma}\cdot\vec{\Pi})\,(
                                    \vec{\sigma}\cdot \vec{\Pi})\nonumber\\
              &\approx& -{\rm i}\hbar\dfrac{m_{\rm e}^2 c^2}{\vec{p}^4} \dfrac{q}{c} \, \vec{\sigma} \cdot (\vec{p}\times\vec{B})(
                                    \vec{\sigma}\cdot \vec{p})+\left({1+\dfrac{m_{\rm e}^2 c^2}{\vec{\Pi}^2}}\right)
                                    \left(\vec{\Pi}^2 \, \openone-\dfrac{q\,\hbar}{c} (\vec{\sigma}\cdot\vec{B})\right)\nonumber\\
              &\approx& -{\rm i}\hbar\dfrac{m_{\rm e}^2 c^2}{\vec{p}^4} \dfrac{q}{c} \, {\rm i} \vec{\sigma}\cdot \left[(\vec{p}\times\vec{B})\times\vec{p}\right]
              +\left(\vec{\Pi}^2+m_{\rm e}^2 c^2\right) \, \openone
                                   -\left({1+\dfrac{m_{\rm e}^2 c^2}{\vec{\Pi}^2}}\right)\dfrac{q\,\hbar}{c} (\vec{\sigma}\cdot\vec{B})\nonumber\\
              &\approx& \hbar\dfrac{m_{\rm e}^2 c^2}{\vec{p}^4} \dfrac{q}{c} \, \vec{\sigma}\cdot \left[\vec{p}^{2}\vec{B}-\vec{p}(\vec{B}\cdot{\vec{p}})\right]
              +\left(\vec{\Pi}^2+m_{\rm e}^2 c^2\right)\, \openone
                                   -\left({1+\dfrac{m_{\rm e}^2 c^2}{\vec{p}^2 -\dfrac{q}{c} \vec{B}\cdot\vec{\ell}}}\right)\dfrac{q\,\hbar}{c} (\vec{\sigma}\cdot\vec{B})\nonumber\\
              &\approx& \hbar\dfrac{m_{\rm e}^2 c^2}{\vec{p}^4} \dfrac{q}{c} \, \vec{\sigma}\cdot \left[\vec{p}^{2}\vec{B}-\vec{p}(\vec{B}\cdot{\vec{p}})\right]
              +\left(\vec{\Pi}^2+m_{\rm e}^2 c^2\right) \, \openone
                                   -\left({1+\dfrac{m_{\rm e}^2 c^2}{\vec{p}^2 }\frac{1}{1 -\dfrac{q}{c}\dfrac{1}{\vec{p}^2} \vec{B}\cdot\vec{\ell}}}\right)\dfrac{q\,\hbar}{c} (\vec{\sigma}\cdot\vec{B})\nonumber\\
                &\approx& \hbar\dfrac{m_{\rm e}^2 c^2}{\vec{p}^4} \dfrac{q}{c} \, \vec{\sigma}\cdot \left[\vec{p}^{2}\vec{B}-\vec{p}(\vec{B}\cdot{\vec{p}})\right]
              +\left(\vec{\Pi}^2+m_{\rm e}^2 c^2\right) \, \openone
                                   -\left[{1+\dfrac{m_{\rm e}^2 c^2}{\vec{p}^2 }\left({1+\dfrac{q}{c}\dfrac{1}{\vec{p}^2} \vec{B}\cdot\vec{\ell}}\right)}\right]\dfrac{q\,\hbar}{c} (\vec{\sigma}\cdot\vec{B})\nonumber\\
                        &\approx& \hbar\dfrac{m_{\rm e}^2 c^2}{\vec{p}^4} \dfrac{q}{c} \, \vec{\sigma}\cdot \left[\vec{p}^{2}\vec{B}-\vec{p}(\vec{B}\cdot{\vec{p}})\right]
              +\left(\vec{\Pi}^2+m_{\rm e}^2 c^2\right) \, \openone
                                   -\left({1+\dfrac{m_{\rm e}^2 c^2}{\vec{p}^2 }}\right)\dfrac{q\,\hbar}{c} (\vec{\sigma}\cdot\vec{B})\nonumber\\
              &=&\left(\vec{\Pi}^2+m_{\rm e}^2 c^2\right) \, \openone+ \dfrac{m_{\rm e}^2 c^2}{\vec{p}^2} \dfrac{q\,\hbar}{c} \, (\vec{\sigma}\cdot \vec{B)}-\dfrac{m_{\rm e}^2 c^2}{\vec{p}^4} \dfrac{q\,\hbar}{c} (\vec{B}\cdot{\vec{p}})(\vec{\sigma}\cdot\vec{p})                                   -\left({1+\dfrac{m_{\rm e}^2 c^2}{\vec{p}^2 }}\right)\dfrac{q\,\hbar}{c} (\vec{\sigma}\cdot\vec{B})\nonumber\\
              &=&\left(\vec{\Pi}^2+m_{\rm e}^2 c^2\right) \, \openone- \dfrac{q\,\hbar}{c} \, (\vec{\sigma}\cdot \vec{B)}-\dfrac{m_{\rm e}^2 c^2}{\vec{p}^4} \dfrac{q\,\hbar}{c} (\vec{B}\cdot{\vec{p}})(\vec{\sigma}\cdot\vec{p}).
          \end{eqnarray}

 (ii-4)  We have
      \begin{eqnarray}
           && \left[(\vec{\sigma}\cdot\vec{\Pi})
                        \sqrt{1+\dfrac{m_{\rm e}^2 c^2}{\vec{\Pi}^2}}\,(
                              \vec{\sigma}\cdot\vec{\Pi})
                                    \sqrt{1+\dfrac{m_{\rm e}^2 c^2}{\vec{\Pi}^2}}\right]=\left[\sqrt{1+\dfrac{m_{\rm e}^2 c^2}{\vec{\Pi}^2}}\,(
                        \vec{\sigma}\cdot\vec{\Pi})\,
                              \sqrt{1+\dfrac{m_{\rm e}^2 c^2}{\vec{\Pi}^2}}\,(
                                    \vec{\sigma}\cdot \vec{\Pi})\right]^\dagger \nonumber\\
             &=& \left[\left(\vec{\Pi}^2+m_{\rm e}^2 c^2\right) \, \openone- \dfrac{q\,\hbar}{c} \, (\vec{\sigma}\cdot \vec{B)}-\dfrac{m_{\rm e}^2 c^2}{\vec{p}^4} \dfrac{q\,\hbar}{c} (\vec{B}\cdot{\vec{p}})(\vec{\sigma}\cdot\vec{p})\right]^\dagger\nonumber\\
             &=&\left(\vec{\Pi}^2+m_{\rm e}^2 c^2\right) \, \openone- \dfrac{q\,\hbar}{c} \, (\vec{\sigma}\cdot \vec{B)}-\dfrac{m_{\rm e}^2 c^2}{\vec{p}^4} \dfrac{q\,\hbar}{c} (\vec{B}\cdot{\vec{p}})(\vec{\sigma}\cdot\vec{p}).                               \end{eqnarray}

      Therefore, from Eq. (\ref{eq:cc-1}) we obtain
      \begin{eqnarray}\label{eq:cc-2}
            && \Bigg[\sqrt{1+\dfrac{m_{\rm e}^2 c^2}{\vec{\Pi}^2}}\,{\rm i}\,(
                  \vec{\sigma}\cdot\vec{\Pi})+{\rm i}\,(\vec{\sigma}\cdot\vec{\Pi})
                        \sqrt{1+\dfrac{m_{\rm e}^2 c^2}{\vec{\Pi}^2}}\Bigg]^2 \notag \\
            &=& -\left[\sqrt{1+\dfrac{m_{\rm e}^2 c^2}{\vec{\Pi}^2}}\,(
                        \vec{\sigma}\cdot\vec{\Pi})\,
                              \sqrt{1+\dfrac{m_{\rm e}^2 c^2}{\vec{\Pi}^2}}\,(
                                    \vec{\sigma}\cdot \vec{\Pi})\right]
                  -\left[\sqrt{1+\dfrac{m_{\rm e}^2 c^2}{\vec{\Pi}^2}}\,(
                        \vec{\sigma}\cdot\vec{\Pi})^2
                              \sqrt{1+\dfrac{m_{\rm e}^2 c^2}{\vec{\Pi}^2}}\right]
                        \nonumber \\
                  &&-\left[(\vec{\sigma}\cdot \vec{\Pi}) \left(
                        1+\dfrac{m_{\rm e}^2 c^2}{\vec{\Pi}^2}\right)(
                              \vec{\sigma}\cdot\vec{\Pi})\right]
                  -\left[(\vec{\sigma}\cdot\vec{\Pi})
                        \sqrt{1+\dfrac{m_{\rm e}^2 c^2}{\vec{\Pi}^2}}\,(
                              \vec{\sigma}\cdot\vec{\Pi})
                                    \sqrt{1+\dfrac{m_{\rm e}^2 c^2}{\vec{\Pi}^2}}\right] \notag  \\
      &=& -2\left[\left(\vec{\Pi}^2+m_{\rm e}^2 c^2\right) \, \openone- \dfrac{q\,\hbar}{c} \, (\vec{\sigma}\cdot \vec{B)}-\dfrac{m_{\rm e}^2 c^2}{\vec{p}^4} \dfrac{q\,\hbar}{c} (\vec{B}\cdot{\vec{p}})(\vec{\sigma}\cdot\vec{p})\right]\nonumber\\
      && -\left[\left(\vec{\Pi}^2 +m_{\rm e}^2 c^2\right)\, \openone -  \left(1+\dfrac{m_{\rm e}^2 c^2}{\vec{p}^2}\right)\,
                        \dfrac{q\,\hbar}{c} \left(\vec{\sigma}\cdot\vec{B}\right)\right] \nonumber\\
      && -\left[\vec{\Pi}^2 \, \openone+ m_{\rm e}^2 c^2 \, \openone -\left(1-\dfrac{m_{\rm e}^2 c^2}{\vec{p}^2}\right)\dfrac{q\,\hbar}{c} (\vec{\sigma}\cdot\vec{B}) -2\,\dfrac{m_{\rm e}^2 c^2}{\vec{p}^4} \dfrac{q\,\hbar}{c}(\vec{B}\cdot{\vec{p}}) ( \vec{\sigma}\cdot\vec{p})\right]\nonumber\\
      &=& -4\left(\vec{\Pi}^2+m_{\rm e}^2 c^2\right) \, \openone+4 \dfrac{q\,\hbar}{c} \, (\vec{\sigma}\cdot \vec{B})+4 \dfrac{m_{\rm e}^2 c^2}{\vec{p}^4} \dfrac{q\,\hbar}{c} (\vec{B}\cdot{\vec{p}})(\vec{\sigma}\cdot\vec{p})\nonumber\\
       &\approx& -4\bigg[\openone\, \vec{\Pi}^2
                  -\dfrac{q\,\hbar}{c} (\vec{\sigma}\cdot\vec{B})
                  -\dfrac{q\,\hbar}{c} \dfrac{m_{\rm e}^2 c^2}{p^4} (
                        \vec{p}\cdot\vec{B})(\vec{\sigma}\cdot\vec{p})
                  +\openone\, m_{\rm e}^2 c^2\bigg].
     \end{eqnarray}

     (iii) We have
      \begin{eqnarray}
            && \left[(\vec{\sigma}\cdot\vec{\Pi}),\ \Bigg\{
                  \sqrt{1+\dfrac{m^2 c^2}{\vec{\Pi}^2}}\,{\rm i}\,(
                        \vec{\sigma}\cdot\vec{\Pi})
                  +{\rm i}\,(\vec{\sigma}\cdot\vec{\Pi})
                        \sqrt{1+\dfrac{m^2 c^2}{\vec{\Pi}^2}}\Bigg\}\right] \notag \\
            &=& {\rm i}\left[(\vec{\sigma}\cdot\vec{\Pi}),\
                        \sqrt{1+\dfrac{m^2 c^2}{\vec{\Pi}^2}}\right](
                              \vec{\sigma}\cdot\vec{\Pi})
                  +{\rm i}\,(\vec{\sigma}\cdot\vec{\Pi})\left[
                        (\vec{\sigma}\cdot\vec{\Pi}),\
                        \sqrt{1+\dfrac{m^2 c^2}{\vec{\Pi}^2}}\right] \notag \\
            &=& \dfrac{q\,\hbar}{c} \dfrac{m^2 c^2}{\vec{\Pi}^4} \Biggl(
                        \sqrt{\dfrac{\vec{\Pi}^2}{m^2 c^2 +\vec{\Pi}^2}}
                        \Biggr)\bigl[\vec{\sigma}\cdot(\vec{p}\times\vec{B})\bigr](
                              \vec{\sigma}\cdot\vec{\Pi})
                  +\dfrac{q\,\hbar}{c} (\vec{\sigma}\cdot\vec{\Pi})
                        \dfrac{m^2 c^2}{\vec{\Pi}^4} \Biggl(
                                    \sqrt{\dfrac{\vec{\Pi}^2}{m^2 c^2 +\vec{\Pi}^2}}
                              \Biggr)\bigl[\vec{\sigma}\cdot(\vec{p}\times\vec{B})
                              \bigr] \notag \\
             &\approx& \dfrac{q\,\hbar}{c} \dfrac{m^2 c^2}{\vec{p}^4} \Biggl(
                        \sqrt{\dfrac{\vec{p}^2}{m^2 c^2 +\vec{p}^2}}
                        \Biggr)\bigl[\vec{\sigma}\cdot(\vec{p}\times\vec{B})\bigr](
                              \vec{\sigma}\cdot\vec{p})
                  +\dfrac{q\,\hbar}{c} (\vec{\sigma}\cdot\vec{p})
                        \dfrac{m^2 c^2}{\vec{p}^4} \Biggl(
                                    \sqrt{\dfrac{\vec{p}^2}{m^2 c^2 +\vec{p}^2}}
                              \Biggr)\bigl[\vec{\sigma}\cdot(\vec{p}\times\vec{B})
                              \bigr] \notag \\
            &=& \dfrac{q\,\hbar}{c} \dfrac{m^2 c^2}{\vec{p}^4} \Biggl(
                        \sqrt{\dfrac{\vec{p}^2}{m^2 c^2 +\vec{p}^2}}
                        \Biggr)\left\{\bigl[\vec{\sigma}\cdot(\vec{p}\times\vec{B})\bigr](
                              \vec{\sigma}\cdot\vec{p})+ (
                              \vec{\sigma}\cdot\vec{p})\bigl[\vec{\sigma}\cdot(\vec{p}\times\vec{B})\bigr] \right\}
                   \notag \\
           & =& 2\dfrac{q\,\hbar}{c} \dfrac{m^2 c^2}{\vec{p}^4} \Biggl(
                        \sqrt{\dfrac{\vec{p}^2}{m^2 c^2 +\vec{p}^2}}
                        \Biggr)\bigl[\vec{p}\cdot(\vec{p}\times\vec{B})\bigr]\notag \\
           & =& 0.
           \end{eqnarray}
      $\blacksquare$

      \end{remark}

      Based on above results, we have
      \begin{eqnarray}
            && \left\{\cos\vartheta\,(\vec{\sigma}\cdot\vec{\Pi})+\dfrac{\sin\vartheta}{2}
                  \Bigg[\sqrt{1+\dfrac{m_{\rm e}^2 c^2}{\vec{\Pi}^2}}\,{\rm i}\,(
                                          \vec{\sigma}\cdot\vec{\Pi})
                                    +{\rm i}\,(\vec{\sigma}\cdot\vec{\Pi})
                                          \sqrt{1+\dfrac{m_{\rm e}^2 c^2}{\vec{\Pi}^2}}\Bigg]
                        \right\} \notag \\
                  && \left\{\cos\vartheta\,(\vec{\sigma}\cdot\vec{\Pi})
                              -\dfrac{\sin\vartheta}{2} \Bigg[
                                    \sqrt{1+\dfrac{m_{\rm e}^2 c^2}{\vec{\Pi}^2}}\,{\rm i}\,(
                                          \vec{\sigma}\cdot\vec{\Pi})
                                    +{\rm i}\,(\vec{\sigma}\cdot\vec{\Pi})
                                          \sqrt{1+\dfrac{m_{\rm e}^2 c^2}{\vec{\Pi}^2}}\Bigg]
                              \right\} \notag \\
            =&& {\cos^2}\vartheta\,(\vec{\sigma}\cdot\vec{\Pi})^2
                  -\dfrac{{\sin^2}\vartheta}{4} \Bigg[\sqrt{1+\dfrac{m_{\rm e}^2 c^2}{\vec{\Pi}^2}}
                              \,{\rm i}\,(\vec{\sigma}\cdot\vec{\Pi})
                        +{\rm i}\,(\vec{\sigma}\cdot\vec{\Pi})
                              \sqrt{1+\dfrac{m_{\rm e}^2 c^2}{\vec{\Pi}^2}}\Bigg]^2
                        \notag \\
                  && -\dfrac{\sin\vartheta\,\cos\theta}{2} \left[
                        (\vec{\sigma}\cdot\vec{\Pi}),\ \Bigg\{
                              \sqrt{1+\dfrac{m_{\rm e}^2 c^2}{\vec{\Pi}^2}}\,{\rm i}\,(
                                    \vec{\sigma}\cdot\vec{\Pi})
                              +{\rm i}\,(\vec{\sigma}\cdot\vec{\Pi})
                                    \sqrt{1+\dfrac{m_{\rm e}^2 c^2}{\vec{\Pi}^2}}\Bigg\}
                        \right] \notag \\
       =&& {\cos^2}\vartheta\,\left[\vec{\Pi}^2 \, \openone-\dfrac{q\,\hbar}{c} (\vec{\sigma}\cdot\vec{B})\right]         -\dfrac{{\sin^2}\vartheta}{4} \left[-4\bigg(\openone\, \vec{\Pi}^2
                  -\dfrac{q\,\hbar}{c} (\vec{\sigma}\cdot\vec{B})
                  -\dfrac{q\,\hbar}{c} \dfrac{m_{\rm e}^2 c^2}{p^4} (
                        \vec{p}\cdot\vec{B})(\vec{\sigma}\cdot\vec{p})
                  +\openone\, m_{\rm e}^2 c^2\bigg)\right]
                        \nonumber\\
       =&& {\cos^2}\vartheta\,\left[\vec{\Pi}^2 \, \openone-\dfrac{q\,\hbar}{c} (\vec{\sigma}\cdot\vec{B})\right]         +{\sin^2}\vartheta\left[\openone\, \vec{\Pi}^2
                  -\dfrac{q\,\hbar}{c} (\vec{\sigma}\cdot\vec{B})
                  -\dfrac{q\,\hbar}{c} \dfrac{m_{\rm e}^2 c^2}{p^4} (
                        \vec{p}\cdot\vec{B})(\vec{\sigma}\cdot\vec{p})
                  +\openone\, m_{\rm e}^2 c^2\right]\nonumber\\
      =&& \left[\vec{\Pi}^2 \, \openone-\dfrac{q\,\hbar}{c} (\vec{\sigma}\cdot\vec{B})\right]
      +{\sin^2}\vartheta\left[-\dfrac{q\,\hbar}{c} \dfrac{m_{\rm e}^2 c^2}{p^4} (
                        \vec{p}\cdot\vec{B})(\vec{\sigma}\cdot\vec{p})
                  +\openone\, m_{\rm e}^2 c^2\right].
      \end{eqnarray}
      After that, \Eq{eq:EtaHeHb2} can be rewritten as
      \begin{eqnarray}
            {\rm i}\hbar\p{\eta}{t} &=& \dfrac{1}{\cos\vartheta\,2\,m_{\rm e}} \Biggl[\left[\vec{\Pi}^2 \, \openone-\dfrac{q\,\hbar}{c} (\vec{\sigma}\cdot\vec{B})\right]
      +{\sin^2}\vartheta\left[-\dfrac{q\,\hbar}{c} \dfrac{m_{\rm e}^2 c^2}{p^4} (
                        \vec{p}\cdot\vec{B})(\vec{\sigma}\cdot\vec{p})
                  +\openone\, m_{\rm e}^2 c^2\right]\Biggr]\eta+q\,\phi\,\eta,
      \end{eqnarray}
      i.e.,
      \begin{eqnarray}\label{eq:EtaHeHb2v1}
            {\rm i}\hbar\p{\eta}{t} &=& \frac{1}{\cos\vartheta}\Bigg[\dfrac{1}{2\,m_{\rm e}} \vec{\Pi}^2
                  - \dfrac{q\,\hbar}{2\,m_{\rm e}\,c} (\vec{\sigma}\cdot\vec{B})
                        \Bigg]\eta
            +\frac{{\sin^2}\vartheta}{\cos\vartheta\,2\,m_{\rm e}}\left[-\dfrac{q\,\hbar}{c} \dfrac{m_{\rm e}^2 c^2}{p^4} (
                        \vec{p}\cdot\vec{B})(\vec{\sigma}\cdot\vec{p})
                  +\openone\, m_{\rm e}^2 c^2\right]\eta+q\,\phi\,\eta,
      \end{eqnarray}
      Notice when $\vartheta=0$, \Eq{eq:EtaHeHb2v1} degenerates into
      \begin{equation}
            {\rm i}\hbar\p{\eta}{t} =\bigg[\dfrac{1}{2\,m} \Bigl(
                              \vec{p}-\dfrac{q}{c} \vec{A}\Bigr)^2
                        -\dfrac{q\,\hbar}{2\,m\,c} \vec{\sigma}\cdot\vec{B}\bigg]
                              \eta
                  +q\,\phi\,\eta.
      \end{equation}
      From Eq. (\ref{eq:EtaHeHb2v1}), we have
       \begin{eqnarray}\label{eq:EtaHeHb2v2}
            {\rm i}\hbar\p{\eta}{t} &=& \frac{1}{\cos\vartheta} \Bigg[\dfrac{1}{2\,m_{\rm e}} \left(\vec{p}^2
                        -\dfrac{q}{c} \vec{B}\cdot\vec{\ell}\right)
                  - \dfrac{q\,\hbar}{2\,m_{\rm e}\,c} (\vec{\sigma}\cdot\vec{B})
                        \Bigg]\eta
            +\frac{{\sin^2}\vartheta}{\cos\vartheta\,2\,m_{\rm e}}\left[-\dfrac{q\,\hbar}{c} \dfrac{m_{\rm e}^2 c^2}{p^4} (
                        \vec{p}\cdot\vec{B})(\vec{\sigma}\cdot\vec{p})
                  +\openone\, m_{\rm e}^2 c^2\right]\eta+q\,\phi\,\eta,\nonumber\\
      \end{eqnarray}
      i.e.,
       \begin{eqnarray}\label{eq:EtaHeHb2v3}
            {\rm i}\hbar\p{\eta}{t} &=& \frac{1}{\cos\vartheta}\Bigg[\dfrac{\vec{p}^2}{2\,m}
                        -\dfrac{q}{2\,m\,c} (\vec{\ell}+2\,\vec{S})\cdot\vec{B}
                        \Bigg]\eta
            +\frac{{\sin^2}\vartheta}{\cos\vartheta\,2\,m_{\rm e}}\left[-\dfrac{q\,\hbar}{c} \dfrac{m_{\rm e}^2 c^2}{p^4} (
                        \vec{p}\cdot\vec{B})(\vec{\sigma}\cdot\vec{p})
                  +\openone\, m_{\rm e}^2 c^2\right]\eta+q\,\phi\,\eta,
      \end{eqnarray}
      and when $\vartheta=0$, we have
      \begin{equation}
            {\rm i}\hbar\p{\eta}{t} =\Big[\dfrac{\vec{p}^2}{2\,m}
                        -\dfrac{q}{2\,m\,c} (\vec{\ell}+2\,\vec{S})\cdot\vec{B}
                        \Big]\eta
                  +q\,\phi\,\eta.
      \end{equation}

        Based on (\ref{eq:EtaHeHb2v3}), we have the spin-Land{\' q} factor $g_{\rm s}$ (the ratio $g_{\rm spin}/g_{\rm orbit}$) as \begin{eqnarray}\label{eq:pauli-c4}
                 g_{\rm s}=\dfrac{g_{\rm spin}}{g_{\rm orbit}}=2.
            \end{eqnarray}
            If we view $g_{\rm orbit}$ as 1, then the $g$ factor for the $H_{\rm e}$-$\mathcal{H}_{\rm b}^{\rm II}$ mixing system is given by
            \begin{eqnarray}\label{eq:pauli-c5}
                 g=2.
            \end{eqnarray}
        This means that the mixture of $H_{\rm e}$ and $\mathcal{H}_{\rm b}^{\rm II}$ cannot alter the $g$ factor.

\section{Calculating the $g$ Factor for the $H_{\rm e}$-${H}_{\rm b}^{\rm I}$-${H}_{\rm b}^{\rm II}$ Mixing}

      In this section, we would like to calculate the $g$ factor for the mixed system $H_{\rm e}$-${H}_{\rm b}^{\rm I}$-${H}_{\rm b}^{\rm II}$, whose Hamiltonian is given by
      \begin{eqnarray}
            H_{\rm mix}=\cos\vartheta\,H_{\rm e} +\sin\vartheta\,(
                  \cos\varphi\,{H}_{\rm b}^I
                  +\sin\varphi\,{H}_{\rm b}^{II}).
      \end{eqnarray}

      From previous sections, in the presence of magnetic field we have known that
      \begin{eqnarray}
            \mathcal{H}_{\rm e} &=& c\,\vec{\alpha}\cdot\Bigl(\vec{p}-\dfrac{q}{c} \vec{A}\Bigr)+\beta\,m_{\rm e}\,c^2=
            c\,\vec{\alpha}\cdot \vec{\Pi}+\beta\,m_{\rm e}\,c^2,
      \end{eqnarray}
      \begin{equation}
            \mathcal{H}_{\rm b}^{\rm I}
            =-\dfrac{1}{2} m_{\rm e}\,c^2 \Big(
                  \dfrac{1}{\sqrt{\vec{\Pi}^2}} \vec{\alpha}\cdot\vec{\Pi}
                  +\vec{\alpha}\cdot\vec{\Pi}\dfrac{1}{\sqrt{\vec{\Pi}^2}}\Big)
                        +\beta\sqrt{\vec{\Pi}^2}\,c,
      \end{equation}
      \begin{eqnarray}
            \mathcal{H}_{\rm b}^{\rm II} =\frac{c}{2}\bigg[\sqrt{1+\dfrac{m_{\rm e}^2 c^2}{\vec{\Pi}^2}}\,{\rm i}\,
                  \beta(\vec{\alpha}\cdot \vec{\Pi})+{\rm i}\,
                  \beta(\vec{\alpha}\cdot \vec{\Pi}) \sqrt{1+\dfrac{m_{\rm e}^2 c^2}{\vec{\Pi}^2}}\bigg],
      \end{eqnarray}
      i.e.,
      \begin{eqnarray}
            \mathcal{H}_{\rm mix} &=& \cos\vartheta\,\bigl(c\,\vec{\alpha}\cdot \vec{\Pi}
                        +\beta\,m_{\rm e}\,c^2\bigr)
                  +\sin\vartheta\,\cos\varphi\,\bigg[-\dfrac{1}{2} m_{\rm e}\,c^2 \Big(
                        \dfrac{1}{\sqrt{\vec{\Pi}^2}} \vec{\alpha}\cdot\vec{\Pi}
                        +\vec{\alpha}\cdot\vec{\Pi}\dfrac{1}{\sqrt{\vec{\Pi}^2}}\Big)
                              +\beta\sqrt{\vec{\Pi}^2}\,c\bigg] \notag \\
                  && +\sin\vartheta\,\sin\varphi\,\frac{c}{2}\bigg[
                        \sqrt{1+\dfrac{m_{\rm e}^2 c^2}{\vec{\Pi}^2}}\,{\rm i}\,\beta(
                              \vec{\alpha}\cdot \vec{\Pi})
                        +{\rm i}\,\beta(\vec{\alpha}\cdot \vec{\Pi})
                              \sqrt{1+\dfrac{m_{\rm e}^2 c^2}{\vec{\Pi}^2}}\bigg]
                  +q\,\phi,
      \end{eqnarray}
      where $\phi$ depicts the scalar potential. Then the Dirac equation reads
      \begin{equation}\label{eq:DiracHeHb1}
            {\rm i}\hbar\p{}{t} \Psi=\mathcal{H}_{\rm mix} \Psi.
      \end{equation}

      Next we study the Dirac equation in the representation of
      \begin{equation}
            \Psi=\begin{bmatrix}
                  \tilde{\eta} \\ \tilde{\chi}
            \end{bmatrix},
      \end{equation}
      then we have
      \begin{eqnarray}
           && {\rm i}\hbar\p{}{t} \begin{bmatrix}
                        \tilde{\eta} \\ \tilde{\chi}
                  \end{bmatrix}
            = \cos\vartheta\,c\,\begin{bmatrix}
                              0 & \vec{\sigma} \\
                              \vec{\sigma} & 0
                        \end{bmatrix}\cdot\vec{\Pi}\begin{bmatrix}
                                    \tilde{\eta} \\ \tilde{\chi}
                              \end{bmatrix}
                  -\sin\vartheta\,\cos\varphi\,\dfrac{1}{2} m_{\rm e}\,c^2 \begin{bmatrix}
                              0 & \dfrac{1}{\sqrt{\vec{\Pi}^2}}\vec{\sigma}\cdot\vec{\Pi}+ \vec{\sigma}\cdot\vec{\Pi}\dfrac{1}{\sqrt{\vec{\Pi}^2}} \\
                             \dfrac{1}{\sqrt{\vec{\Pi}^2}}\vec{\sigma}\cdot\vec{\Pi}+ \vec{\sigma}\cdot\vec{\Pi}\dfrac{1}{\sqrt{\vec{\Pi}^2}} & 0
                        \end{bmatrix}\begin{bmatrix}
                                    \tilde{\eta} \\ \tilde{\chi}
                              \end{bmatrix} \notag \\
                  && +\sin\vartheta\,\sin\varphi\,\frac{c}{2} \begin{bmatrix}
                              0 & \sqrt{1+\dfrac{m_{\rm e}^2 c^2}{\vec{\Pi}^2}}\,{\rm i}\,(
                                          \vec{\sigma}\cdot\vec{\Pi})
                                    +{\rm i}\,(\vec{\sigma}\cdot\vec{\Pi})
                                          \sqrt{1+\dfrac{m_{\rm e}^2 c^2}{\vec{\Pi}^2}} \\
                              -\sqrt{1+\dfrac{m_{\rm e}^2 c^2}{\vec{\Pi}^2}}\,{\rm i}(
                                          \vec{\sigma}\cdot\vec{\Pi})
                                    -{\rm i}\,(\vec{\sigma}\cdot\vec{\Pi})
                                          \sqrt{1+\dfrac{m_{\rm e}^2 c^2}{\vec{\Pi}^2}} & 0
                        \end{bmatrix}\begin{bmatrix}
                                    \tilde{\eta} \\ \tilde{\chi}
                              \end{bmatrix} \notag \\
                  && +\cos\vartheta\,\begin{bmatrix}
                              \openone & 0 \\
                              0 & -\openone
                        \end{bmatrix}\,m_{\rm e}\,c^2 \begin{bmatrix}
                              \tilde{\eta} \\ \tilde{\chi}
                        \end{bmatrix}
                  +\sin\vartheta\,\cos\varphi\,\begin{bmatrix}
                              \openone & 0 \\
                              0 & -\openone
                        \end{bmatrix}\sqrt{\vec{\Pi}^2}\,c \begin{bmatrix}
                              \tilde{\eta} \\ \tilde{\chi}
                        \end{bmatrix}
                  +q\,\phi\begin{bmatrix}
                              \tilde{\eta} \\ \tilde{\chi}
                        \end{bmatrix},
      \end{eqnarray}
      i.e.,
      \begin{eqnarray}\label{eq:EHb12Expand}
            {\rm i}\hbar\p{}{t} \begin{bmatrix}
                        \tilde{\eta} \\ \tilde{\chi}
                  \end{bmatrix}
            &=&  \Biggl\{\cos\vartheta\,c(\vec{\sigma}\cdot\vec{\Pi})
                        -\sin\vartheta\,\cos\varphi\,\dfrac{1}{2} m_{\rm e}\,c^2 \bigg[
                              \dfrac{1}{\sqrt{\vec{\Pi}^2}}\vec{\sigma}\cdot\vec{\Pi}
                              +\vec{\sigma}\cdot\vec{\Pi}\dfrac{1}{\sqrt{\vec{\Pi}^2}}\bigg]\Biggr\}\begin{bmatrix}
                                    \tilde{\chi} \\ \tilde{\eta}
                              \end{bmatrix} \notag \\
                  && +\sin\vartheta\,\sin\varphi\,\dfrac{c}{2} \left[
                              \sqrt{1+\dfrac{m_{\rm e}^2 c^2}{\vec{\Pi}^2}}\,{\rm i}\,(
                                          \vec{\sigma}\cdot\vec{\Pi})
                                    +{\rm i}\,(\vec{\sigma}\cdot\vec{\Pi})
                                          \sqrt{1+\dfrac{m_{\rm e}^2 c^2}{\vec{\Pi}^2}}
                              \right]\begin{bmatrix}
                                          \tilde{\chi} \\ -\tilde{\eta}
                                    \end{bmatrix} \notag \\
                  && +(\cos\vartheta\,m_{\rm e}\,c^2 +\sin\theta\,\cos\varphi\,
                        \sqrt{\vec{\Pi}^2}\,c)\begin{bmatrix}
                                    \tilde{\eta} \\ -\tilde{\chi}
                              \end{bmatrix}
                  +q\,\phi\begin{bmatrix}
                              \tilde{\eta} \\ \tilde{\chi}
                        \end{bmatrix}.
      \end{eqnarray}

      In the case of low energy, the term involving $m_{\rm e}\,c^2$ is far larger than other ones. Then a part of time factor can be separated as follows.
      \begin{equation}
            \begin{bmatrix}
                  \tilde{\eta} \\ \tilde{\chi}
            \end{bmatrix}=\begin{bmatrix}
                  \eta \\ \chi
            \end{bmatrix}{\rm e}^{-{\rm i}\,\cos\vartheta\,\frac{m_{\rm e}\,c^2}{\hbar} t},
      \end{equation}
      then \Eq{eq:EHb12Expand} is transformed as
      \begin{eqnarray}
            && {\rm i}\hbar\Biggl\{\p{}{t} \begin{bmatrix}
                        \eta \\ \chi
                  \end{bmatrix}\Biggr\}{\rm e}^{-{\rm i}\,\cos\vartheta\,\frac{m_{\rm e}\,c^2}{\hbar} t}
            +{\rm i}\begin{bmatrix}
                        \eta \\ \chi
                  \end{bmatrix}\hbar\Bigl(\p{}{t} {\rm e}^{
                        -{\rm i}\,\cos\theta\,\frac{m_{\rm e}\,c^2}{\hbar} t}\Bigr) \notag \\
            &=& \Biggl\{\cos\vartheta\,c(\vec{\sigma}\cdot\vec{\Pi})
                        -\sin\vartheta\,\cos\varphi\,\dfrac{1}{2} m_{\rm e}\,c^2 \bigg[
                              \dfrac{1}{\sqrt{\vec{\Pi}^2}}\vec{\sigma}\cdot\vec{\Pi}
                              +\vec{\sigma}\cdot\vec{\Pi}\dfrac{1}{\sqrt{\vec{\Pi}^2}}\bigg]\Biggr\}\begin{bmatrix}
                                    \chi \\ \eta
                              \end{bmatrix}{\rm e}^{
                                    -{\rm i}\,\cos\vartheta\,\frac{m_{\rm e}\,c^2}{\hbar} t}
                        \notag \\
                  && +\sin\vartheta\,\sin\varphi\,\dfrac{c}{2} \left[
                              \sqrt{1+\dfrac{m_{\rm e}^2 c^2}{\vec{\Pi}^2}}\,{\rm i}\,(
                                          \vec{\sigma}\cdot\vec{\Pi})
                                    +{\rm i}\,(\vec{\sigma}\cdot\vec{\Pi})
                                          \sqrt{1+\dfrac{m_{\rm e}^2 c^2}{\vec{\Pi}^2}}
                              \right]\begin{bmatrix}
                                          \chi \\ -\eta
                                    \end{bmatrix}{\rm e}^{
                                          -{\rm i}\,\cos\vartheta\,\frac{m_{\rm e}\,c^2}{\hbar} t}
                        \notag \\
                  && +(\cos\vartheta\,m_{\rm e}\,c^2 +\sin\vartheta\,\cos\varphi\,
                        \sqrt{\vec{\Pi}^2}\,c)\begin{bmatrix}
                                    \eta \\ -\chi
                              \end{bmatrix}{\rm e}^{
                                    -{\rm i}\,\cos\vartheta\,\frac{m_{\rm e}\,c^2}{\hbar} t}
                  +q\,\phi\begin{bmatrix}
                              \eta \\ \chi
                        \end{bmatrix}{\rm e}^{
                              -{\rm i}\,\cos\vartheta\,\frac{m_{\rm e}\,c^2}{\hbar} t},
      \end{eqnarray}
      i.e.,
      \begin{eqnarray}
            {\rm i}\hbar\p{}{t} \begin{bmatrix}
                        \eta \\ \chi
                  \end{bmatrix}+\cos\vartheta\,m_{\rm e}\,c^2 \begin{bmatrix}
                        \eta \\ \chi
                  \end{bmatrix}
            &=& \Biggl\{\cos\vartheta\,c(\vec{\sigma}\cdot\vec{\Pi})
                        -\sin\vartheta\,\cos\varphi\,\dfrac{1}{2} m_{\rm e}\,c^2 \bigg[
                              \dfrac{1}{\sqrt{\vec{\Pi}^2}}\vec{\sigma}\cdot\vec{\Pi}
                              +\vec{\sigma}\cdot\vec{\Pi}\dfrac{1}{\sqrt{\vec{\Pi}^2}}\bigg]\Biggr\}\begin{bmatrix}
                                    \chi \\ \eta
                              \end{bmatrix} \notag \\
                  && +\sin\vartheta\,\sin\varphi\,\dfrac{c}{2} \left[
                              \sqrt{1+\dfrac{m_{\rm e}^2 c^2}{\vec{\Pi}^2}}\,{\rm i}\,(
                                          \vec{\sigma}\cdot\vec{\Pi})
                                    +{\rm i}\,(\vec{\sigma}\cdot\vec{\Pi})
                                          \sqrt{1+\dfrac{m_{\rm e}^2 c^2}{\vec{\Pi}^2}}
                              \right]\begin{bmatrix}
                                          \chi \\ -\eta
                                    \end{bmatrix} \notag \\
                  && +(\cos\vartheta\,m_{\rm e}\,c^2 +\sin\vartheta\,\cos\varphi\,
                        \sqrt{\vec{\Pi}^2}\,c)\begin{bmatrix}
                                    \eta \\ -\chi
                              \end{bmatrix}
                  +q\,\phi\begin{bmatrix}
                              \eta \\ \chi
                        \end{bmatrix},
      \end{eqnarray}
      i.e.,
      \begin{eqnarray}\label{eq:DiracHeHb12Aprox}
            {\rm i}\hbar\p{}{t} \begin{bmatrix}
                        \eta \\ \chi
                  \end{bmatrix}
            &=& \Biggl\{\cos\vartheta\,c(\vec{\sigma}\cdot\vec{\Pi})
                        -\sin\vartheta\,\cos\varphi\,\dfrac{1}{2} m_{\rm e}\,c^2 \bigg[
                              \dfrac{1}{\sqrt{\vec{\Pi}^2}}\vec{\sigma}\cdot\vec{\Pi}
                              +\vec{\sigma}\cdot\vec{\Pi}\dfrac{1}{\sqrt{\vec{\Pi}^2}}\bigg]\Biggr\}\begin{bmatrix}
                                    \chi \\ \eta
                              \end{bmatrix} \notag \\
                  && +\sin\vartheta\,\sin\varphi\,\dfrac{c}{2} \left[
                              \sqrt{1+\dfrac{m_{\rm e}^2 c^2}{\vec{\Pi}^2}}\,{\rm i}\,(
                                          \vec{\sigma}\cdot\vec{\Pi})
                                    +{\rm i}\,(\vec{\sigma}\cdot\vec{\Pi})
                                          \sqrt{1+\dfrac{m_{\rm e}^2 c^2}{\vec{\Pi}^2}}
                              \right]\begin{bmatrix}
                                          \chi \\ -\eta
                                    \end{bmatrix} \notag \\
                  && +\begin{bmatrix}
                        \sin\vartheta\,\cos\varphi\,\sqrt{\vec{\Pi}^2}\,c\:\eta \\ -\big(
                             \openone\, 2\,\cos\vartheta\,m_{\rm e}\,c^2
                              +\sin\vartheta\,\cos\varphi\,\sqrt{\vec{\Pi}^2}\,c\big)\chi
                  \end{bmatrix}
                  +q\,\phi\begin{bmatrix}
                              \eta \\ \chi
                        \end{bmatrix}.
      \end{eqnarray}

            From the second row of \Eq{eq:DiracHeHb12Aprox}, we have
            \begin{eqnarray}
                  \openone\,{\rm i}\hbar\p{\chi}{t} &=& \Biggl\{
                              \cos\vartheta\,c(\vec{\sigma}\cdot\vec{\Pi})
                              -\sin\vartheta\,\cos\varphi\,\dfrac{1}{2} m_{\rm e}\,c^2 \bigg[
                                    \dfrac{1}{\sqrt{\vec{\Pi}^2}}\vec{\sigma}\cdot\vec{\Pi}
                                    +\vec{\sigma}\cdot\vec{\Pi}\dfrac{1}{\sqrt{\vec{\Pi}^2}}\bigg] \notag \\
                              &&-\sin\vartheta\,\sin\varphi\,\dfrac{c}{2} \left[
                                    \sqrt{1+\dfrac{m_{\rm e}^2 c^2}{\vec{\Pi}^2}}\,{\rm i}\,(
                                                \vec{\sigma}\cdot\vec{\Pi})
                                          +{\rm i}\,(\vec{\sigma}\cdot\vec{\Pi})
                                                \sqrt{1+\dfrac{m_{\rm e}^2 c^2}{\vec{\Pi}^2}}
                                    \right]\Biggr\}\eta \notag \\
                        &&-\big(\openone\, 2\,\cos\vartheta\,m_{\rm e}\,c^2
                              +\sin\vartheta\,\cos\varphi\,\sqrt{\vec{\Pi}^2}\,c\big)\chi+\openone\,q\,\phi\,
                              \chi.
            \end{eqnarray}
            Since $m_{\rm e}\,c^2 \openone\chi\gg{\rm i}\hbar\,\partial\chi/\partial t$, $m_{\rm e}\,c^2 \openone\chi\gg q\,\phi\,\chi$, and
            $\cos\vartheta\,m_{\rm e}\,c^2 \chi\gg|\sin\vartheta\,\cos\varphi\,\sqrt{\vec{\Pi}^2}\,c \chi|$, $\vartheta\in [-\pi/4, \pi/4]$,
            then we arrive at
            \begin{eqnarray}
                  && \Biggl\{\cos\vartheta\,c(\vec{\sigma}\cdot\vec{\Pi})
                        -\sin\vartheta\,\cos\varphi\,\dfrac{1}{2} m_{\rm e}\,c^2 \bigg[
                              \dfrac{1}{\sqrt{\vec{\Pi}^2}}\vec{\sigma}\cdot\vec{\Pi}
                              +\vec{\sigma}\cdot\vec{\Pi}\dfrac{1}{\sqrt{\vec{\Pi}^2}}\bigg] \notag \\
                        &&\quad -\sin\vartheta\,\sin\varphi\,\dfrac{c}{2} \left[
                              \sqrt{1+\dfrac{m_{\rm e}^2 c^2}{\vec{\Pi}^2}}\,{\rm i}\,(
                                          \vec{\sigma}\cdot\vec{\Pi})
                                    +{\rm i}\,(\vec{\sigma}\cdot\vec{\Pi})
                                          \sqrt{1+\dfrac{m_{\rm e}^2 c^2}{\vec{\Pi}^2}}
                              \right]\Biggr\}\eta =\cos\vartheta\, 2\,m_{\rm e}\,c^2 \openone\chi,
            \end{eqnarray}
            i.e.,
            \begin{eqnarray}
                  \chi &=& \dfrac{1}{\cos\vartheta\,2\,m\,c^2} \Biggl\{
                        \cos\vartheta\,c(\vec{\sigma}\cdot\vec{\Pi})
                        -\sin\vartheta\,\cos\varphi\,\dfrac{1}{2} m_{\rm e}\,c^2 \bigg[
                              \dfrac{1}{\sqrt{\vec{\Pi}^2}}\vec{\sigma}\cdot\vec{\Pi}
                              +\vec{\sigma}\cdot\vec{\Pi}\dfrac{1}{\sqrt{\vec{\Pi}^2}}\bigg] \notag \\
                        &&\qquad\qquad -\sin\vartheta\,\sin\varphi\,\dfrac{c}{2} \left[
                              \sqrt{1+\dfrac{m_{\rm e}^2 c^2}{\vec{\Pi}^2}}\,{\rm i}\,(
                                          \vec{\sigma}\cdot\vec{\Pi})
                                    +{\rm i}\,(\vec{\sigma}\cdot\vec{\Pi})
                                          \sqrt{1+\dfrac{m_{\rm e}^2 c^2}{\vec{\Pi}^2}}
                              \right]\Biggr\}\eta.
            \end{eqnarray}

            After that, for the first row of \Eq{eq:DiracHeHb12Aprox}, we obtain
            \begin{eqnarray}
                  {\rm i}\hbar\p{\eta}{t} &=& \Biggl\{
                        \cos\vartheta\,c(\vec{\sigma}\cdot\vec{\Pi})
                        -\sin\vartheta\,\cos\varphi\,\dfrac{1}{2} m_{\rm e}\,c^2 \bigg[
                              \dfrac{1}{\sqrt{\vec{\Pi}^2}}\vec{\sigma}\cdot\vec{\Pi}
                              +\vec{\sigma}\cdot\vec{\Pi}\dfrac{1}{\sqrt{\vec{\Pi}^2}}\bigg] \notag \\
                        &&\quad +\sin\vartheta\,\sin\varphi\,\dfrac{c}{2} \left[
                              \sqrt{1+\dfrac{m_{\rm e}^2 c^2}{\vec{\Pi}^2}}\,{\rm i}\,(
                                          \vec{\sigma}\cdot\vec{\Pi})
                                    +{\rm i}\,(\vec{\sigma}\cdot\vec{\Pi})
                                          \sqrt{1+\dfrac{m_{\rm e}^2 c^2}{\vec{\Pi}^2}}
                              \right]\Biggr\}\chi
                              +\sin\vartheta\,\cos\varphi\,\sqrt{\vec{\Pi}^2}\,c\:\eta+q\,\phi\,\eta,
            \end{eqnarray}
            i.e.,
            \begin{eqnarray}
                  {\rm i}\hbar\p{\eta}{t} &=& \dfrac{1}{\cos\vartheta\,2\,m_{\rm e}\,c^2} \Biggl\{
                        \cos\vartheta\,c(\vec{\sigma}\cdot\vec{\Pi})
                        -\sin\vartheta\,\cos\varphi\,\dfrac{1}{2} m_{\rm e}\,c^2 \bigg[
                              \dfrac{1}{\sqrt{\vec{\Pi}^2}}\vec{\sigma}\cdot\vec{\Pi}
                              +\vec{\sigma}\cdot\vec{\Pi}\dfrac{1}{\sqrt{\vec{\Pi}^2}}\bigg] \notag \\
                        &&\qquad\qquad +\sin\vartheta\,\sin\varphi\,\dfrac{c}{2} \left[
                              \sqrt{1+\dfrac{m_{\rm e}^2 c^2}{\vec{\Pi}^2}}\,{\rm i}\,(
                                          \vec{\sigma}\cdot\vec{\Pi})
                                    +{\rm i}\,(\vec{\sigma}\cdot\vec{\Pi})
                                          \sqrt{1+\dfrac{m_{\rm e}^2 c^2}{\vec{\Pi}^2}}
                              \right]\Biggr\}\Biggl\{ \notag \\
                        &&\qquad\qquad \cos\vartheta\,c(\vec{\sigma}\cdot\vec{\Pi})
                        -\sin\vartheta\,\cos\varphi\,\dfrac{1}{2} m_{\rm e}\,c^2 \bigg[
                              \dfrac{1}{\sqrt{\vec{\Pi}^2}}\vec{\sigma}\cdot\vec{\Pi}
                              +\vec{\sigma}\cdot\vec{\Pi}\dfrac{1}{\sqrt{\vec{\Pi}^2}}\bigg] \notag \\
                        &&\qquad\qquad -\sin\vartheta\,\sin\varphi\,\dfrac{c}{2} \left[
                              \sqrt{1+\dfrac{m_{\rm e}^2 c^2}{\vec{\Pi}^2}}\,{\rm i}\,(
                                          \vec{\sigma}\cdot\vec{\Pi})
                                    +{\rm i}\,(\vec{\sigma}\cdot\vec{\Pi})
                                          \sqrt{1+\dfrac{m_{\rm e}^2 c^2}{\vec{\Pi}^2}}
                              \right]\Biggr\}\eta\nonumber\\
                       && +\sin\vartheta\,\cos\varphi\,\sqrt{\vec{\Pi}^2}\,c\:\eta+q\,\phi\,\eta,
            \end{eqnarray}
            i.e.,
            \begin{eqnarray}\label{eq:3mix-1a}
                  {\rm i}\hbar\p{\eta}{t} &=& \dfrac{1}{\cos\vartheta\,2\,m_{\rm e}\,c^2} \Biggl\{
                        \cos\vartheta\,c(\vec{\sigma}\cdot\vec{\Pi})
                        -\sin\vartheta\,\cos\varphi\,\dfrac{1}{2} m_{\rm e}\,c^2 \bigg[
                              \dfrac{1}{\sqrt{\vec{\Pi}^2}}\vec{\sigma}\cdot\vec{\Pi}
                              +\vec{\sigma}\cdot\vec{\Pi}\dfrac{1}{\sqrt{\vec{\Pi}^2}}\bigg]\Biggr\}^2 \eta \notag \\
                        && -\dfrac{1}{\cos\vartheta\,2\,m_{\rm e}\,c^2} \Bigg[
                                    \cos\vartheta\,c(\vec{\sigma}\cdot\vec{\Pi})
                                    -\sin\vartheta\,\cos\varphi\,\dfrac{1}{2} m_{\rm e}\,c^2 \bigg(
                                    \dfrac{1}{\sqrt{\vec{\Pi}^2}} \vec{\sigma}
                                          \cdot\vec{\Pi}
                                    +\vec{\sigma}\cdot\vec{\Pi}
                                          \dfrac{1}{\sqrt{\vec{\Pi}^2}}\bigg),\ \notag
                                    \\
                              &&\qquad\qquad \sin\vartheta\,\sin\varphi\,\dfrac{c}{2}
                                    \biggl(\sqrt{1+\dfrac{m_{\rm e}^2 c^2}{\vec{\Pi}^2}}\,
                                          {\rm i}\,(\vec{\sigma}\cdot\vec{\Pi})
                                          +{\rm i}\,(\vec{\sigma}\cdot\vec{\Pi})
                                                \sqrt{1+\dfrac{m_{\rm e}^2 c^2}{\vec{\Pi}^2}}
                                          \biggr)\Bigg]\eta \notag \\
                        && -\dfrac{1}{\cos\vartheta\,2\,m_{\rm e}\,c^2} {\sin^2}\vartheta\,
                              {\sin^2}\varphi\,\dfrac{c^2}{4} \left[
                                    \sqrt{1+\dfrac{m_{\rm e}^2 c^2}{\vec{\Pi}^2}}\,{\rm i}\,(
                                          \vec{\sigma}\cdot\vec{\Pi})
                                    +{\rm i}\,(\vec{\sigma}\cdot\vec{\Pi})
                                          \sqrt{1+\dfrac{m_{\rm e}^2 c^2}{\vec{\Pi}^2}}
                              \right]^2 \eta \nonumber\\
                        && +\sin\vartheta\,\cos\varphi\,\sqrt{\vec{\Pi}^2}\,c\:\eta+q\,\phi\,\eta.
            \end{eqnarray}

           \begin{remark}To calculate Eq. (\ref{eq:3mix-1a}), we need to know the following terms.

           (i) We can have
            \begin{eqnarray}\label{eq:3mix-1aaa}
                &&  \Biggl\{
                        \cos\vartheta\,c(\vec{\sigma}\cdot\vec{\Pi})
                        -\sin\vartheta\,\cos\varphi\,\dfrac{1}{2} m_{\rm e}\,c^2 \bigg[
                              \dfrac{1}{\sqrt{\vec{\Pi}^2}}\vec{\sigma}\cdot\vec{\Pi}
                              +\vec{\sigma}\cdot\vec{\Pi}\dfrac{1}{\sqrt{\vec{\Pi}^2}}\bigg]\Biggr\}^2 \nonumber\\
                &=& \cos^2\vartheta\,c^2\,\left[\vec{\Pi}^2 \openone -\dfrac{q\,\hbar}{c} (\vec{\sigma}\cdot\vec{B})\right]+
                \sin^2\vartheta\,\cos^2\varphi\,m^2_{\rm e}\,c^4 \left[\openone -\dfrac{q\,\hbar}{c} \dfrac{1}{p^4} \left[(\vec{p}\cdot\vec{B})(\vec{\sigma}\cdot\vec{p})\right]\right]\nonumber\\
                &&-\cos\vartheta \sin\vartheta\,\cos\varphi\, m_{\rm e}\,c^3
                \left[2 \openone\sqrt{\vec{\Pi}^2}-\dfrac{1}{p} \dfrac{q\,\hbar}{c} (\vec{\sigma}\cdot\vec{B})
                -  \dfrac{q\,\hbar}{c} \dfrac{1}{p^3}(\vec{p}\cdot\vec{B})(
                                    \vec{\sigma}\cdot\vec{p})\right].
            \end{eqnarray}
           \emph{Proof.---}From previous results, we have known that
           \begin{eqnarray}
            && (\vec{\sigma}\cdot\vec{\Pi})^2 = (\vec{\sigma}\cdot\vec{\Pi})(
                  \vec{\sigma}\cdot\vec{\Pi})= \vec{\Pi}^2 \openone +{\rm i} \vec{\sigma}\cdot \left(\vec{\Pi}\times \vec{\Pi}\right)
            = \vec{\Pi}^2 \openone -\dfrac{q\,\hbar}{c} (\vec{\sigma}\cdot\vec{B})\nonumber\\
           &&  \Big(\dfrac{1}{\sqrt{\vec{\Pi}^2}} \vec{\sigma}\cdot\vec{\Pi}
                        +\vec{\sigma}\cdot\vec{\Pi}\dfrac{1}{\sqrt{\vec{\Pi}^2}}\Big)^2
                  \approx 4\, \left[\openone -\dfrac{q\,\hbar}{c} \dfrac{1}{\vec{p}^4} \left[(\vec{p}\cdot\vec{B})(\vec{\sigma}\cdot\vec{p})\right]\right].
            \end{eqnarray}
          and
          \begin{eqnarray}
                 && (\vec{\sigma}\cdot\vec{\Pi}) \bigg[
                              \dfrac{1}{\sqrt{\vec{\Pi}^2}}\vec{\sigma}\cdot\vec{\Pi}
                              +\vec{\sigma}\cdot\vec{\Pi}\dfrac{1}{\sqrt{\vec{\Pi}^2}}\bigg]+\bigg[
                              \dfrac{1}{\sqrt{\vec{\Pi}^2}}\vec{\sigma}\cdot\vec{\Pi}
                              +\vec{\sigma}\cdot\vec{\Pi}\dfrac{1}{\sqrt{\vec{\Pi}^2}}\bigg] (\vec{\sigma}\cdot\vec{\Pi}) \nonumber\\
                 &=& 2(\vec{\sigma}\cdot\vec{\Pi})
                              \dfrac{1}{\sqrt{\vec{\Pi}^2}}(\vec{\sigma}\cdot\vec{\Pi})+(\vec{\sigma}\cdot\vec{\Pi})^2
                              \dfrac{1}{\sqrt{\vec{\Pi}^2}}+\dfrac{1}{\sqrt{\vec{\Pi}^2}}(\vec{\sigma}\cdot\vec{\Pi})^2 \nonumber\\
                 &  \approx& 2\, \left[\openone\sqrt{\vec{\Pi}^2}{-}\dfrac{q\,\hbar}{c}
                        \dfrac{1}{p^3}(\vec{p}\cdot\vec{B})(
                              \vec{\sigma}\cdot\vec{p})\right]\nonumber\\
                && + \left[\vec{\Pi}^2 \openone -\dfrac{q\,\hbar}{c} (\vec{\sigma}\cdot\vec{B})\right]\dfrac{1}{\sqrt{\vec{\Pi}^2}} + \dfrac{1}{\sqrt{\vec{\Pi}^2}}\left[\vec{\Pi}^2 \openone -\dfrac{q\,\hbar}{c} (\vec{\sigma}\cdot\vec{B})\right] \nonumber\\
                &  \approx& 2\, \left[\openone\sqrt{\vec{\Pi}^2}{-}\dfrac{q\,\hbar}{c}
                        \dfrac{1}{p^3}(\vec{p}\cdot\vec{B})(
                              \vec{\sigma}\cdot\vec{p})\right] + 2 \sqrt{\vec{\Pi}^2} \openone -2 \dfrac{1}{\sqrt{\vec{p}^2}} \dfrac{q\,\hbar}{c} (\vec{\sigma}\cdot\vec{B}) \nonumber\\
                &=& 4 \openone\sqrt{\vec{\Pi}^2}-2 \dfrac{1}{\sqrt{\vec{p}^2}} \dfrac{q\,\hbar}{c} (\vec{\sigma}\cdot\vec{B})
                - 2 \dfrac{q\,\hbar}{c} \dfrac{1}{\sqrt{\vec{p}^6}}(\vec{p}\cdot\vec{B})(
                                    \vec{\sigma}\cdot\vec{p}).
            \end{eqnarray}
            Therefore we have
             \begin{eqnarray}
                &&  \Biggl\{
                        \cos\vartheta\,c(\vec{\sigma}\cdot\vec{\Pi})
                        -\sin\vartheta\,\cos\varphi\,\dfrac{1}{2} m_{\rm e}\,c^2 \bigg[
                              \dfrac{1}{\sqrt{\vec{\Pi}^2}}\vec{\sigma}\cdot\vec{\Pi}
                              +\vec{\sigma}\cdot\vec{\Pi}\dfrac{1}{\sqrt{\vec{\Pi}^2}}\bigg]\Biggr\}^2 \nonumber\\
                &=& \cos^2\vartheta\,c^2\,(\vec{\sigma}\cdot\vec{\Pi})^2+
                \sin^2\vartheta\,\cos^2\varphi\,\dfrac{1}{4} m^2_{\rm e}\,c^4 \bigg[
                              \dfrac{1}{\sqrt{\vec{\Pi}^2}}\vec{\sigma}\cdot\vec{\Pi}
                              +\vec{\sigma}\cdot\vec{\Pi}\dfrac{1}{\sqrt{\vec{\Pi}^2}}\bigg]^2\nonumber\\
                &&-\cos\vartheta \sin\vartheta\,\cos\varphi\,\dfrac{1}{2} m_{\rm e}\,c^3
                \left((\vec{\sigma}\cdot\vec{\Pi}) \bigg[
                              \dfrac{1}{\sqrt{\vec{\Pi}^2}}\vec{\sigma}\cdot\vec{\Pi}
                              +\vec{\sigma}\cdot\vec{\Pi}\dfrac{1}{\sqrt{\vec{\Pi}^2}}\bigg]+\bigg[
                              \dfrac{1}{\sqrt{\vec{\Pi}^2}}\vec{\sigma}\cdot\vec{\Pi}
                              +\vec{\sigma}\cdot\vec{\Pi}\dfrac{1}{\sqrt{\vec{\Pi}^2}}\bigg] (\vec{\sigma}\cdot\vec{\Pi})\right)\nonumber\\
             &=& \cos^2\vartheta\,c^2\,\left[\vec{\Pi}^2 \openone -\dfrac{q\,\hbar}{c} (\vec{\sigma}\cdot\vec{B})\right]+
                \sin^2\vartheta\,\cos^2\varphi\,\dfrac{1}{4} m^2_{\rm e}\,c^4 \left[4\, \left[\openone -\dfrac{q\,\hbar}{c} \dfrac{1}{\vec{p}^4} \left[(\vec{p}\cdot\vec{B})(\vec{\sigma}\cdot\vec{p})\right]\right]\right]\nonumber\\
                &&-\cos\vartheta \sin\vartheta\,\cos\varphi\,\dfrac{1}{2} m_{\rm e}\,c^3
                \left[4 \openone\sqrt{\vec{\Pi}^2}-2 \dfrac{1}{\sqrt{\vec{p}^2}} \dfrac{q\,\hbar}{c} (\vec{\sigma}\cdot\vec{B})
                - 2 \dfrac{q\,\hbar}{c} \dfrac{1}{\sqrt{\vec{p}^6}}(\vec{p}\cdot\vec{B})(
                                    \vec{\sigma}\cdot\vec{p})\right]\nonumber\\
             &=& \cos^2\vartheta\,c^2\,\left[\vec{\Pi}^2 \openone -\dfrac{q\,\hbar}{c} (\vec{\sigma}\cdot\vec{B})\right]+
                \sin^2\vartheta\,\cos^2\varphi\,m^2_{\rm e}\,c^4 \left[\openone -\dfrac{q\,\hbar}{c} \dfrac{1}{p^4} \left[(\vec{p}\cdot\vec{B})(\vec{\sigma}\cdot\vec{p})\right]\right]\nonumber\\
                &&-\cos\vartheta \sin\vartheta\,\cos\varphi\, m_{\rm e}\,c^3
                \left[2 \openone\sqrt{\vec{\Pi}^2}- \dfrac{1}{\sqrt{\vec{p}^2}} \dfrac{q\,\hbar}{c} (\vec{\sigma}\cdot\vec{B})
                -  \dfrac{q\,\hbar}{c} \dfrac{1}{\sqrt{\vec{p}^6}}(\vec{p}\cdot\vec{B})(
                                    \vec{\sigma}\cdot\vec{p})\right].
            \end{eqnarray}

             (ii) We can have
            \begin{eqnarray}\label{eq:3mix-1b}
                &&\Bigg[\cos\vartheta\,c(\vec{\sigma}\cdot\vec{\Pi})
                                    -\sin\vartheta\,\cos\varphi\,\dfrac{1}{2} m_{\rm e}\,c^2 \bigg(
                                    \dfrac{1}{\sqrt{\vec{\Pi}^2}} \vec{\sigma}
                                          \cdot\vec{\Pi}
                                    +\vec{\sigma}\cdot\vec{\Pi}
                                          \dfrac{1}{\sqrt{\vec{\Pi}^2}}\bigg), \nonumber\\
                &&\, \sin\vartheta\,\sin\varphi\,\dfrac{c}{2}
                                    \biggl(\sqrt{1+\dfrac{m_{\rm e}^2 c^2}{\vec{\Pi}^2}}\,
                                          {\rm i}\,(\vec{\sigma}\cdot\vec{\Pi})
                                          +{\rm i}\,(\vec{\sigma}\cdot\vec{\Pi})
                                                \sqrt{1+\dfrac{m_{\rm e}^2 c^2}{\vec{\Pi}^2}}
                                          \biggr)\Bigg]=0.
            \end{eqnarray}
           \emph{Proof.---}From previous results, we have known that

           \begin{eqnarray}\label{eq:3mix-1b1}
            && \left[(\vec{\sigma}\cdot\vec{\Pi}),\
                  \sqrt{1+\dfrac{m^2 c^2}{\vec{\Pi}^2}}\,{\rm i}\,(
                        \vec{\sigma}\cdot\vec{\Pi})
                  +{\rm i}\,(\vec{\sigma}\cdot\vec{\Pi})
                        \sqrt{1+\dfrac{m^2 c^2}{\vec{\Pi}^2}}\right]=0,
           \end{eqnarray}
           thus
            \begin{eqnarray}\label{eq:3mix-1b2}
                &&\Bigg[\dfrac{1}{\sqrt{\vec{\Pi}^2}} \vec{\sigma} \cdot\vec{\Pi}+\vec{\sigma}\cdot\vec{\Pi}
                                          \dfrac{1}{\sqrt{\vec{\Pi}^2}},  \sqrt{1+\dfrac{m_{\rm e}^2 c^2}{\vec{\Pi}^2}}\,{\rm i}\,(\vec{\sigma}\cdot\vec{\Pi})+{\rm i}\,(\vec{\sigma}\cdot\vec{\Pi})\sqrt{1+\dfrac{m_{\rm e}^2 c^2}{\vec{\Pi}^2}}\Bigg]\nonumber\\
                &=& \Bigg[\dfrac{1}{\sqrt{\vec{\Pi}^2}},  \sqrt{1+\dfrac{m_{\rm e}^2 c^2}{\vec{\Pi}^2}}\,{\rm i}\,(\vec{\sigma}\cdot\vec{\Pi})+{\rm i}\,(\vec{\sigma}\cdot\vec{\Pi})\sqrt{1+\dfrac{m_{\rm e}^2 c^2}{\vec{\Pi}^2}}\Bigg] (\vec{\sigma} \cdot\vec{\Pi})\nonumber\\
                &&+ (\vec{\sigma} \cdot\vec{\Pi}) \Bigg[\dfrac{1}{\sqrt{\vec{\Pi}^2}},  \sqrt{1+\dfrac{m_{\rm e}^2 c^2}{\vec{\Pi}^2}}\,{\rm i}\,(\vec{\sigma}\cdot\vec{\Pi})+{\rm i}\,(\vec{\sigma}\cdot\vec{\Pi})\sqrt{1+\dfrac{m_{\rm e}^2 c^2}{\vec{\Pi}^2}}\Bigg] \nonumber\\
                &=& \left\{{\rm i}\, \sqrt{1+\dfrac{m_{\rm e}^2 c^2}{\vec{\Pi}^2}}\, \left[\dfrac{1}{\sqrt{\vec{\Pi}^2}}, (\vec{\sigma}\cdot\vec{\Pi})\right]+ {\rm i}\, \left[\dfrac{1}{\sqrt{\vec{\Pi}^2}}, (\vec{\sigma}\cdot\vec{\Pi})\right]\sqrt{1+\dfrac{m_{\rm e}^2 c^2}{\vec{\Pi}^2}}\right\}\,(\vec{\sigma}\cdot\vec{\Pi})\nonumber\\
                &&+(\vec{\sigma}\cdot\vec{\Pi})\, \left\{{\rm i}\, \sqrt{1+\dfrac{m_{\rm e}^2 c^2}{\vec{\Pi}^2}}\, \left[\dfrac{1}{\sqrt{\vec{\Pi}^2}}, (\vec{\sigma}\cdot\vec{\Pi})\right]+ {\rm i}\, \left[\dfrac{1}{\sqrt{\vec{\Pi}^2}}, (\vec{\sigma}\cdot\vec{\Pi})\right]\sqrt{1+\dfrac{m_{\rm e}^2 c^2}{\vec{\Pi}^2}}\right\}\nonumber\\
                &=& \left\{ \sqrt{1+\dfrac{m_{\rm e}^2 c^2}{\vec{\Pi}^2}}\, \hbar\dfrac{q}{c} \dfrac{1}{p^3} \vec{\sigma}\cdot (
                                    \vec{B}\times\vec{p})+  \hbar\dfrac{q}{c} \dfrac{1}{p^3} \vec{\sigma}\cdot (
                                    \vec{B}\times\vec{p})\sqrt{1+\dfrac{m_{\rm e}^2 c^2}{\vec{\Pi}^2}}\right\}\,(\vec{\sigma}\cdot\vec{\Pi})\nonumber\\
                &&+(\vec{\sigma}\cdot\vec{\Pi})\, \left\{ \sqrt{1+\dfrac{m_{\rm e}^2 c^2}{\vec{\Pi}^2}}\, \hbar\dfrac{q}{c} \dfrac{1}{p^3} \vec{\sigma}\cdot (
                                    \vec{B}\times\vec{p})+  \hbar\dfrac{q}{c} \dfrac{1}{p^3} \vec{\sigma}\cdot (
                                    \vec{B}\times\vec{p})\sqrt{1+\dfrac{m_{\rm e}^2 c^2}{\vec{\Pi}^2}}\right\}\nonumber\\
                    &=& \left\{ \sqrt{1+\dfrac{m_{\rm e}^2 c^2}{\vec{p}^2}}\, \hbar\dfrac{q}{c} \dfrac{1}{p^3} \vec{\sigma}\cdot (
                                    \vec{B}\times\vec{p})+  \hbar\dfrac{q}{c} \dfrac{1}{p^3} \vec{\sigma}\cdot (
                                    \vec{B}\times\vec{p})\sqrt{1+\dfrac{m_{\rm e}^2 c^2}{\vec{p}^2}}\right\}\,(\vec{\sigma}\cdot\vec{\Pi})\nonumber\\
                &&+(\vec{\sigma}\cdot\vec{\Pi})\, \left\{ \sqrt{1+\dfrac{m_{\rm e}^2 c^2}{\vec{p}^2}}\, \hbar\dfrac{q}{c} \dfrac{1}{p^3} \vec{\sigma}\cdot (
                                    \vec{B}\times\vec{p})+  \hbar\dfrac{q}{c} \dfrac{1}{p^3} \vec{\sigma}\cdot (
                                    \vec{B}\times\vec{p})\sqrt{1+\dfrac{m_{\rm e}^2 c^2}{\vec{p}^2}}\right\}\nonumber\\
                        &=& \left\{ \sqrt{1+\dfrac{m_{\rm e}^2 c^2}{\vec{p}^2}}\, \hbar\dfrac{q}{c} \dfrac{1}{p^3} \vec{\sigma}\cdot (
                                    \vec{B}\times\vec{p})+  \hbar\dfrac{q}{c} \dfrac{1}{p^3} \vec{\sigma}\cdot (
                                    \vec{B}\times\vec{p})\sqrt{1+\dfrac{m_{\rm e}^2 c^2}{\vec{p}^2}}\right\}\,(\vec{\sigma}\cdot\vec{p})\nonumber\\
                &&+(\vec{\sigma}\cdot\vec{p})\, \left\{ \sqrt{1+\dfrac{m_{\rm e}^2 c^2}{\vec{p}^2}}\, \hbar\dfrac{q}{c} \dfrac{1}{p^3} \vec{\sigma}\cdot (
                                    \vec{B}\times\vec{p})+  \hbar\dfrac{q}{c} \dfrac{1}{p^3} \vec{\sigma}\cdot (
                                    \vec{B}\times\vec{p})\sqrt{1+\dfrac{m_{\rm e}^2 c^2}{\vec{p}^2}}\right\}\nonumber\\
            &=& 2 \sqrt{1+\dfrac{m_{\rm e}^2 c^2}{\vec{p}^2}}\, \dfrac{q \hbar}{c} \dfrac{1}{p^3} \left[\vec{\sigma}\cdot (
                                    \vec{B}\times\vec{p})(\vec{\sigma}\cdot\vec{p})+(\vec{\sigma}\cdot\vec{p})\vec{\sigma}\cdot (
                                    \vec{B}\times\vec{p})\right] \nonumber\\
             &=& 4 \sqrt{1+\dfrac{m_{\rm e}^2 c^2}{\vec{p}^2}}\, \dfrac{q \hbar}{c} \dfrac{1}{p^3} \left[(
                                    \vec{B}\times\vec{p})\cdot\vec{p}\right] =0.
            \end{eqnarray}
           where we have used
            \begin{eqnarray}
            \Big[\vec{\sigma}\cdot \vec{\Pi},\ \dfrac{1}{\sqrt{\vec{\Pi}^2}}\Big]
            \approx  {\rm i}\hbar\dfrac{q}{c} \dfrac{1}{p^3} \vec{\sigma}\cdot (
                                    \vec{B}\times\vec{p}),\;\;\;\; \Big[\dfrac{1}{\sqrt{\vec{\Pi}^2}}, \, \vec{\sigma}\cdot \vec{\Pi}\Big]
            \approx  -{\rm i}\hbar\dfrac{q}{c} \dfrac{1}{p^3} \vec{\sigma}\cdot (
                                    \vec{B}\times\vec{p}).
             \end{eqnarray}
             Thus based on Eq. (\ref{eq:3mix-1b1}) and Eq. (\ref{eq:3mix-1b2}) one can directly prove Eq. (\ref{eq:3mix-1b}).

           (iii) From Eq. (\ref{eq:cc-2}), we have known that
            \begin{eqnarray}\label{eq:3mix-1c}
                &&  \left[
                                    \sqrt{1+\dfrac{m_{\rm e}^2 c^2}{\vec{\Pi}^2}}\,{\rm i}\,(
                                          \vec{\sigma}\cdot\vec{\Pi})
                                    +{\rm i}\,(\vec{\sigma}\cdot\vec{\Pi})
                                          \sqrt{1+\dfrac{m_{\rm e}^2 c^2}{\vec{\Pi}^2}}
                              \right]^2 \nonumber\\
               & \approx& -4\bigg[\openone\, \vec{\Pi}^2
                  -\dfrac{q\,\hbar}{c} (\vec{\sigma}\cdot\vec{B})
                  -\dfrac{q\,\hbar}{c} \dfrac{m_{\rm e}^2 c^2}{p^4} (
                        \vec{p}\cdot\vec{B})(\vec{\sigma}\cdot\vec{p})
                  +\openone\, m_{\rm e}^2 c^2\bigg].
            \end{eqnarray}
           $\blacksquare$
           \end{remark}

           Then Eq. (\ref{eq:3mix-1a}) becomes

           \begin{eqnarray}\label{eq:3mix-2a}
                  {\rm i}\hbar\p{\eta}{t} &=& \dfrac{1}{\cos\vartheta\,2\,m_{\rm e}\,c^2} \Biggl\{
                        \cos\vartheta\,c(\vec{\sigma}\cdot\vec{\Pi})
                        -\sin\vartheta\,\cos\varphi\,\dfrac{1}{2} m_{\rm e}\,c^2 \bigg[
                              \dfrac{1}{\sqrt{\vec{\Pi}^2}}\vec{\sigma}\cdot\vec{\Pi}
                              +\vec{\sigma}\cdot\vec{\Pi}\dfrac{1}{\sqrt{\vec{\Pi}^2}}\bigg]\Biggr\}^2 \eta \notag \\
                        && -\dfrac{1}{\cos\vartheta\,2\,m_{\rm e}\,c^2} \Bigg[
                                    \cos\vartheta\,c(\vec{\sigma}\cdot\vec{\Pi})
                                    -\sin\vartheta\,\cos\varphi\,\dfrac{1}{2} m_{\rm e}\,c^2 \bigg(
                                    \dfrac{1}{\sqrt{\vec{\Pi}^2}} \vec{\sigma}
                                          \cdot\vec{\Pi}
                                    +\vec{\sigma}\cdot\vec{\Pi}
                                          \dfrac{1}{\sqrt{\vec{\Pi}^2}}\bigg),\ \notag
                                    \\
                              &&\qquad\qquad \sin\vartheta\,\sin\varphi\,\dfrac{c}{2}
                                    \biggl(\sqrt{1+\dfrac{m_{\rm e}^2 c^2}{\vec{\Pi}^2}}\,
                                          {\rm i}\,(\vec{\sigma}\cdot\vec{\Pi})
                                          +{\rm i}\,(\vec{\sigma}\cdot\vec{\Pi})
                                                \sqrt{1+\dfrac{m_{\rm e}^2 c^2}{\vec{\Pi}^2}}
                                          \biggr)\Bigg]\eta \notag \\
                        && -\dfrac{1}{\cos\vartheta\,2\,m_{\rm e}\,c^2} {\sin^2}\vartheta\,
                              {\sin^2}\varphi\,\dfrac{c^2}{4} \left[
                                    \sqrt{1+\dfrac{m_{\rm e}^2 c^2}{\vec{\Pi}^2}}\,{\rm i}\,(
                                          \vec{\sigma}\cdot\vec{\Pi})
                                    +{\rm i}\,(\vec{\sigma}\cdot\vec{\Pi})
                                          \sqrt{1+\dfrac{m_{\rm e}^2 c^2}{\vec{\Pi}^2}}
                              \right]^2 \eta \nonumber\\
                        && +\sin\vartheta\,\cos\varphi\,\sqrt{\vec{\Pi}^2}\,c\:\eta+q\,\phi\,\eta\nonumber\\
           &=& \dfrac{1}{\cos\vartheta\,2\,m_{\rm e}\,c^2} \Biggl\{\cos^2\vartheta\,c^2\,\left[\vec{\Pi}^2 \openone -\dfrac{q\,\hbar}{c} (\vec{\sigma}\cdot\vec{B})\right]+
                \sin^2\vartheta\,\cos^2\varphi\,m^2_{\rm e}\,c^4 \left[\openone -\dfrac{q\,\hbar}{c} \dfrac{1}{p^4} \left[(\vec{p}\cdot\vec{B})(\vec{\sigma}\cdot\vec{p})\right]\right]\nonumber\\
                &&-\cos\vartheta \sin\vartheta\,\cos\varphi\, m_{\rm e}\,c^3
                \left[2 \openone\sqrt{\vec{\Pi}^2}- \dfrac{1}{\sqrt{\vec{p}^2}} \dfrac{q\,\hbar}{c} (\vec{\sigma}\cdot\vec{B})
                -  \dfrac{q\,\hbar}{c} \dfrac{1}{\sqrt{\vec{p}^6}}(\vec{p}\cdot\vec{B})(
                                    \vec{\sigma}\cdot\vec{p})\right]
                       \Biggr\} \eta \notag \\
           &&-\dfrac{1}{\cos\vartheta\,2\,m_{\rm e}\,c^2}\times 0\, \eta\nonumber\\
           && -\dfrac{1}{\cos\vartheta\,2\,m_{\rm e}\,c^2} {\sin^2}\vartheta\,
                              {\sin^2}\varphi\,\dfrac{c^2}{4} \left\{
                                                                 -4\bigg[\openone\, \vec{\Pi}^2
                  -\dfrac{q\,\hbar}{c} (\vec{\sigma}\cdot\vec{B})
                  -\dfrac{q\,\hbar}{c} \dfrac{m_{\rm e}^2 c^2}{p^4} (
                        \vec{p}\cdot\vec{B})(\vec{\sigma}\cdot\vec{p})
                  +\openone\, m_{\rm e}^2 c^2\bigg] \right\} \eta \nonumber\\
                   && +\sin\vartheta\,\cos\varphi\,\sqrt{\vec{\Pi}^2}\,c\:\eta+q\,\phi\,\eta\nonumber\\
            &=& \dfrac{1}{\cos\vartheta\,2\,m_{\rm e}} \Biggl\{\cos^2\vartheta\,\left[\vec{\Pi}^2 \openone -\dfrac{q\,\hbar}{c} (\vec{\sigma}\cdot\vec{B})\right]+
                \sin^2\vartheta\,\cos^2\varphi\,m^2_{\rm e}\,c^2 \left[\openone -\dfrac{q\,\hbar}{c} \dfrac{1}{p^4} \left[(\vec{p}\cdot\vec{B})(\vec{\sigma}\cdot\vec{p})\right]\right]\nonumber\\
           &&-\cos\vartheta \sin\vartheta\,\cos\varphi\, m_{\rm e}c
                \left[- \dfrac{1}{\sqrt{\vec{p}^2}} \dfrac{q\,\hbar}{c} (\vec{\sigma}\cdot\vec{B})
                -  \dfrac{q\,\hbar}{c} \dfrac{1}{\sqrt{\vec{p}^6}}(\vec{p}\cdot\vec{B})(
                                    \vec{\sigma}\cdot\vec{p})\right]
                       \Biggr\} \eta \notag \\
            && +\dfrac{1}{\cos\vartheta\,2\,m_{\rm e}} {\sin^2}\vartheta\,
                              {\sin^2}\varphi \bigg[\openone\, \vec{\Pi}^2
                  -\dfrac{q\,\hbar}{c} (\vec{\sigma}\cdot\vec{B})
                  -\dfrac{q\,\hbar}{c} \dfrac{m_{\rm e}^2 c^2}{p^4} (
                        \vec{p}\cdot\vec{B})(\vec{\sigma}\cdot\vec{p})
                  +\openone\, m_{\rm e}^2 c^2\bigg] \eta  +q\,\phi\,\eta\nonumber\\
           &=& \dfrac{1}{\cos\vartheta\,2\,m_{\rm e}} \left\{  \left(\cos^2\vartheta+{\sin^2}\vartheta\,
                              {\sin^2}\varphi \right)\,\vec{\Pi}^2 \openone -\left(\cos^2\vartheta+{\sin^2}\vartheta\,
                              {\sin^2}\varphi- \dfrac{m_{\rm e}c}{p}\cos\vartheta \sin\vartheta\,\cos\varphi\right)\dfrac{q\,\hbar}{c} (\vec{\sigma}\cdot\vec{B})\right\}\eta\nonumber\\
           &&+ \dfrac{1}{\cos\vartheta\,2\,m_{\rm e}} \left\{
           -{\sin^2}\vartheta\,{\cos^2}\varphi \dfrac{m_{\rm e}^2c^2}{p^4} +\cos\vartheta \sin\vartheta\,\cos\varphi\,\dfrac{m_{\rm e}c}{p^3} -{\sin^2}\vartheta\,{\sin^2}\varphi \dfrac{m_{\rm e}^2c^2}{p^4}\right\}\dfrac{q\,\hbar}{c} (\vec{p}\cdot\vec{B})(
                                    \vec{\sigma}\cdot\vec{p})\eta\nonumber\\
           &&+ \dfrac{1}{\cos\vartheta\,2\,m_{\rm e}} \left\{
           {\sin^2}\vartheta\,{\cos^2}\varphi +{\sin^2}\vartheta\,{\sin^2}\varphi\right\}m_{\rm e}^2 c^2 \openone\,\eta +q\,\phi\,\eta\nonumber\\
           &=& \dfrac{1}{\cos\vartheta\,2\,m_{\rm e}} \left\{  \left(\cos^2\vartheta+{\sin^2}\vartheta\,
                              {\sin^2}\varphi \right)\,\vec{\Pi}^2 \openone -\left(\cos^2\vartheta+{\sin^2}\vartheta\,
                              {\sin^2}\varphi- \dfrac{m_{\rm e}c}{p}\,\cos\vartheta \sin\vartheta\,\cos\varphi\right)\dfrac{q\,\hbar}{c} (\vec{\sigma}\cdot\vec{B}) \right\}\eta\nonumber\\
           &&+\dfrac{1}{\cos\vartheta\,2\,m_{\rm e}} \left\{
           -\dfrac{m_{\rm e}^2c^2}{p^4} {\sin^2}\vartheta +\dfrac{m_{\rm e}c}{p^3}\cos\vartheta \sin\vartheta\,\cos\varphi \right\}\dfrac{q\,\hbar}{c} (\vec{p}\cdot\vec{B})(
                                    \vec{\sigma}\cdot\vec{p})\eta\nonumber\\
           &&+ \dfrac{1}{\cos\vartheta\,2\,m_{\rm e}}
           {\sin^2}\vartheta m_{\rm e}^2 c^2 \openone\,\eta +q\,\phi\,\eta,
            \end{eqnarray}
           i.e.,
           \begin{eqnarray}\label{eq:3mix-2b}
            {\rm i}\hbar\p{\eta}{t}
            &=& \dfrac{1}{\cos\vartheta} \left\{  \left(\cos^2\vartheta+{\sin^2}\vartheta\,
                              {\sin^2}\varphi \right)\,\dfrac{\vec{\Pi}^2}{2\,m_{\rm e}} -\left(\cos^2\vartheta+{\sin^2}\vartheta\,
                              {\sin^2}\varphi- \dfrac{m_{\rm e}c}{p}\cos\vartheta \sin\vartheta\,\cos\varphi\right)\dfrac{q\,\hbar}{2\,m_{\rm e}c} (\vec{\sigma}\cdot\vec{B}) \right\}\eta\nonumber\\
           &&+\dfrac{1}{\cos\vartheta} \left\{
           -\dfrac{m_{\rm e}^2c^2}{p^4} {\sin^2}\vartheta +\dfrac{m_{\rm e}c}{p^3}\cos\vartheta \sin\vartheta\,\cos\varphi \right\}\dfrac{q\,\hbar}{2\,m_{\rm e}c} (\vec{p}\cdot\vec{B})(
                                    \vec{\sigma}\cdot\vec{p})\eta+ \dfrac{1}{\cos\vartheta}
           {\sin^2}\vartheta \frac{m_{\rm e} c^2}{2} \openone\,\eta +q\,\phi\,\eta.
           \end{eqnarray}

           \begin{remark}Let us study Eq. (\ref{eq:3mix-2b}).

           (i) When $\vartheta=0$, Eq. (\ref{eq:3mix-2b}) reduces to
           \begin{eqnarray}\label{eq:3mix-2b-1}
            {\rm i}\hbar\p{\eta}{t}
            &=& \left[ \dfrac{\vec{\Pi}^2}{2\,m_{\rm e}} -\dfrac{q\,\hbar}{2\,m_{\rm e}c} (\vec{\sigma}\cdot\vec{B}) \right] \eta+q\,\phi\,\eta,
            \end{eqnarray}
            which is the Pauli equation Eq. (\ref{eq:pauli-1}) for Dirac's electron.

           (ii) When $\vartheta\neq 0$, $\varphi=0$, Eq. (\ref{eq:3mix-2b}) reduces to
           \begin{eqnarray}\label{eq:3mix-2b-2}
                  {\rm i}\hbar\p{\eta}{t} &=& \dfrac{1}{\cos\vartheta\,2\,m_{\rm e}} \Biggl\{
                  {\cos^2}\vartheta\,\left[\vec{\Pi}^2 \,\openone - \left(1 -\tan\vartheta\,\dfrac{m_{\rm e}\,c}{p} \right)\dfrac{q\,\hbar}{c} (\vec{\sigma}\cdot\vec{B})\right]\nonumber\\
                  &&-\left[ {\sin^2}\vartheta m_{\rm e}^2 c^2 \frac{1}{p}- \sin\vartheta\,\cos\vartheta\, m_{\rm e}\,c\right]\dfrac{q\,\hbar}{c} \dfrac{1}{p^3}(\vec{p}\cdot\vec{B})(
                                    \vec{\sigma}\cdot\vec{p})
                   +{\sin^2}\vartheta m_{\rm e}^2 c^2 \openone \Biggr\}\eta +q\,\phi\,\eta,
            \end{eqnarray}
            which is Eq. (\ref{eq:EtaHeHb1v1}) for the $H_{\rm e}$-$H_{\rm b}^{\rm I}$ mixing.

            (iii) When $\vartheta\neq 0$, $\varphi=\pi/2$, Eq. (\ref{eq:3mix-2b}) reduces to
           \begin{eqnarray}\label{eq:3mix-2b-3}
           {\rm i}\hbar\p{\eta}{t} &=& \frac{1}{\cos\vartheta}\Bigg[\dfrac{1}{2\,m_{\rm e}} \vec{\Pi}^2
                  - \dfrac{q\,\hbar}{2\,m_{\rm e}\,c} (\vec{\sigma}\cdot\vec{B})
                        \Bigg]\eta
            +\frac{{\sin^2}\vartheta}{\cos\vartheta\,2\,m_{\rm e}}\left[-\dfrac{q\,\hbar}{c} \dfrac{m_{\rm e}^2 c^2}{p^4} (
                        \vec{p}\cdot\vec{B})(\vec{\sigma}\cdot\vec{p})
                  +\openone\, m_{\rm e}^2 c^2\right]\eta+q\,\phi\,\eta,
           \end{eqnarray}
            which is Eq. (\ref{eq:EtaHeHb2v1}) for the $H_{\rm e}$-$H_{\rm b}^{\rm II}$ mixing.
            $\blacksquare$
           \end{remark}

           Eq. (\ref{eq:3mix-2b}) can be recast to
           \begin{eqnarray}\label{eq:3mix-2c}
            {\rm i}\hbar\p{\eta}{t}
            &=& \dfrac{(\cos^2\vartheta+{\sin^2}\vartheta\,
                              {\sin^2}\varphi)}{\cos\vartheta} \left\{\dfrac{\vec{\Pi}^2}{2\,m_{\rm e}} -\left(1- \dfrac{m_{\rm e}c}{p}\dfrac{\cos\vartheta \sin\vartheta\,\cos\varphi}{\cos^2\vartheta+{\sin^2}\vartheta\,
                              {\sin^2}\varphi }\right)\dfrac{q\,\hbar}{2\,m_{\rm e}c} (\vec{\sigma}\cdot\vec{B}) \right\}\eta\nonumber\\
           &&+\dfrac{1}{\cos\vartheta} \left\{
           -\dfrac{m_{\rm e}^2c^2}{p^4} {\sin^2}\vartheta +\dfrac{m_{\rm e}c}{p^3}\cos\vartheta \sin\vartheta\,\cos\varphi \right\}\dfrac{q\,\hbar}{2\,m_{\rm e}c} (\vec{p}\cdot\vec{B})(
                                    \vec{\sigma}\cdot\vec{p})\eta+ \dfrac{1}{\cos\vartheta}
           {\sin^2}\vartheta \frac{m_{\rm e} c^2}{2} \openone\,\eta +q\,\phi\,\eta\nonumber\\
            &=& \dfrac{(\cos^2\vartheta+{\sin^2}\vartheta\,
                              {\sin^2}\varphi)}{\cos\vartheta} \left\{\dfrac{\vec{p}^2}{2\,m_{\rm e}} -\dfrac{q}{2\,m_{\rm e}c}\left[\vec{\ell}+2\left(1- \dfrac{\cos\vartheta \sin\vartheta\,\cos\varphi}{\cos^2\vartheta+{\sin^2}\vartheta\,
                              {\sin^2}\varphi }\dfrac{m_{\rm e}c}{p}\right)\vec{S}\right] \cdot\vec{B} \right\}\eta\nonumber\\
           &&+\dfrac{1}{\cos\vartheta} \left\{
           -\dfrac{m_{\rm e}^2c^2}{p^4} {\sin^2}\vartheta +\dfrac{m_{\rm e}c}{p^3}\cos\vartheta \sin\vartheta\,\cos\varphi \right\}\dfrac{q\,\hbar}{2\,m_{\rm e}c} (\vec{p}\cdot\vec{B})(
                                    \vec{\sigma}\cdot\vec{p})\eta+ \dfrac{1}{\cos\vartheta}
           {\sin^2}\vartheta \frac{m_{\rm e} c^2}{2} \openone\,\eta +q\,\phi\,\eta.
           \end{eqnarray}

            Based on (\ref{eq:3mix-2c}), we have the spin-Land{\' q} factor $g_{\rm s}$ (the ratio $g_{\rm spin}/g_{\rm orbit}$) as \begin{eqnarray}\label{eq:pauli-d2}
                 g_{\rm s}=\dfrac{g_{\rm spin}}{g_{\rm orbit}}=2\left(1-\dfrac{\cos\vartheta \sin\vartheta\,\cos\varphi}{\cos^2\vartheta+{\sin^2}\vartheta\,
                              {\sin^2}\varphi }\dfrac{m_{\rm e}c}{p}\right).
            \end{eqnarray}
            If we view $g_{\rm orbit}$ as 1, then the $g$ factor for the $H_{\rm e}$-${H}_{\rm b}^{\rm I}$-${H}_{\rm b}^{\rm II}$ mixing system is given by
            \begin{eqnarray}\label{eq:pauli-d3}
                 g=2\left(1- \dfrac{\cos\vartheta \sin\vartheta\,\cos\varphi}{\cos^2\vartheta+{\sin^2}\vartheta\,
                              {\sin^2}\varphi }\dfrac{m_{\rm e}c}{p}\right).
            \end{eqnarray}
             This means that the mixture of $H_{\rm e}$-${H}_{\rm b}^{\rm I}$-${H}_{\rm b}^{\rm II}$ can indeed alter the $g$ factor.

%
%
%
%
%
%
%
%
%
%
%
%
%
%